\documentclass[10pt]{behroozarxiv}

\usepackage{amsmath,amsfonts,bm, amssymb}
\usepackage{bbm}

\def\eqref#1{equation~\ref{#1}}

\def\1{\bm{1}}

\DeclareMathAlphabet{\mathsfit}{\encodingdefault}{\sfdefault}{m}{sl}
\SetMathAlphabet{\mathsfit}{bold}{\encodingdefault}{\sfdefault}{bx}{n}

\usepackage{xstring}
\usepackage{comment}
\usepackage{textcomp}
\usepackage{enumitem}

\usepackage{mathrsfs}
\usepackage{relsize}
\usepackage{stmaryrd}
\usepackage{bbm}
\usepackage{xfrac}

\begingroup\sffamily\endgroup
\begingroup\rmfamily\endgroup
\DeclareFontShape{T1}{lmss}{bx}{it}{<-> ec-lmssbo10}{}
\DeclareFontShape{T1}{lmss}{bx}{sc}{<-> ec-lmcsc10}{}
\DeclareFontShape{T1}{lmr}{m}{scit}{<-> ec-lmcsco10}{}

\usepackage{array}
\usepackage{tabularx}
\usepackage{longtable}
\usepackage{makecell}

\usepackage{algorithm}
\usepackage{algpseudocode}

\usepackage{pgffor}
\usepackage{wrapfig}
\usepackage{float} 
\usepackage{tikz}
\usetikzlibrary{calc}

\makeatletter
\newcommand{\AppendixOnlyTOC}{%
  \section*{Appendix Contents}%
  \begingroup
    \footnotesize
    \setlength{\parskip}{0pt}%
    \@starttoc{atoc}%
  \endgroup
}
\makeatother

\newcommand{\appsection}[1]{%
  \section{#1}%
  \addcontentsline{atoc}{section}{%
    \protect\numberline{\thesection}#1%
  }%
}

\newcommand{\appsubsection}[1]{%
  \subsection{#1}%
  \addcontentsline{atoc}{subsection}{%
    \protect\numberline{\thesubsection}#1%
  }%
}

\theoremstyle{plain}
\newtheorem{theorem}{Theorem}[section]
\newtheorem{lemma}[theorem]{Lemma}
\newtheorem{proposition}[theorem]{Proposition}
\newtheorem{corollary}[theorem]{Corollary}

\theoremstyle{definition}
\newtheorem{definition}[theorem]{Definition}

\theoremstyle{remark}
\newtheorem{remark}[theorem]{Remark}

\def\BibTeX{{\rm B\kern-.05em{\sc i\kern-.025em b}\kern-.08em
  T\kern-.1667em\lower.7ex\hbox{E}\kern-.125emX}}

\usepackage{bibunits}
\defaultbibliographystyle{tmlr}
\defaultbibliography{references}

\title{ScoreShield: Differentially Private Release of Similarity Scores}

\DeclareRobustCommand{\orcidiconbehrooz}{%
  \href{https://orcid.org/0000-0001-9568-4166}{%
    \raisebox{-0.2ex}{\includegraphics[height=1.6ex]{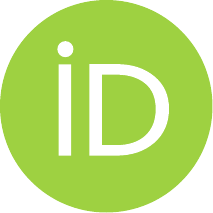}}%
  }%
}

\DeclareRobustCommand{\orcidiconparsa}{%
  \href{https://orcid.org/0000-0001-7927-268X}{%
    \raisebox{-0.2ex}{\includegraphics[height=1.6ex]{orcid.pdf}}%
  }%
}

\author[1]{Behrooz Razeghi~\orcidiconbehrooz}
\author[2]{Parsa Rahimi~\orcidiconparsa}
\affiliation[1]{School of Engineering and Applied Sciences, Harvard University}
\affiliation[2]{School of Engineering, École Polytechnique Fédérale de Lausanne (EPFL)}

\abstract{
A growing number of applications, such as biometrics and retrieval-augmented generation (RAG), rely on cosine similarity scores computed between vector embeddings of text, images, or audio. These systems return similarity scores through their APIs for ranking and verification. However, such releases can leak information about individual records and enable membership inference attacks. While differential privacy (DP) provides a principled metric for quantifying attack risks, na\"ive application of DP mechanisms---such as adding i.i.d. Gaussian noise to vector entries---leads to excessive distortion (i.e., low utility) at a given privacy constraint that scales poorly with the number of released scores. We propose \textsc{ScoreShield}, a perturb-then-project mechanism that adds Gaussian noise calibrated to global sensitivity of the chosen score release regime and then projects the result onto the feasibility set of valid cosine objects. \textsc{ScoreShield} satisfies $(\varepsilon,\delta)$--DP for releasing similarity score vectors and Gram matrices. We provide utility guarantees for the exact Frobenius metric projection used in the risk analysis, and prove convergence to feasibility for the practical averaged alternating-projection solver used for large-scale Gram releases. For full pairwise cosine Gram release under record-level replacement adjacency, the exact-projection bound improves the $n$-dependence of squared Frobenius risk from $\Theta(n^3)$ for the na\"ive Gaussian baseline to $\mathcal{O}(n^2)$ for fixed privacy parameters, with sharper local bounds at low-rank Grams. We evaluate the mechanism across RAG, face recognition, semantic retrieval, image similarity, and recommender-system tasks.
\vspace{-10pt}
}

\preprint{arXiv preprint, ScoreShield}
\date{\today}
\correspondence{\email{behroozrazeghi@seas.harvard.edu}}
\codeurl{https://github.com/BehroozRazeghi/scoreshield}

\begin{document}

\begin{bibunit}


\maketitle




\vspace{-5pt}

\section{Introduction}
\label{Sec:introduction}

\vspace{-5pt}

Similarity scores drive retrieval, verification and search \citep{reimers2019sentencebert, phan2022deepface, muennighoff2022mteb}. In RAG and enterprise semantic search, a service that maintains an indexed corpus returns, for each query, a ranked list of the top-$k$ matched records together with a per-record similarity (or relevance) score computed using a similarity metric \citep{lewis2020retrieval}. In biometric verification, standard matcher interfaces likewise return a scalar match score to the calling application. Recent work demonstrates black-box membership inference against RAG datastores, in which an adversary decides whether a target passage is contained in the retrieval database by issuing queries and analyzing the observable outputs of the system (retrieved context and/or generated responses) \citep{li2025budgetleak, li2025generating, naseh2025riddle, anderson2024my}. Moreover, similarity-score statistics can be leveraged to detect and localize membership-inference attempts, since they re-use the similarity scores already computed during top-$k$ selection \citep{choi2025safeguarding}. The biometrics literature has long noted that exposing match scores enables score-guided hill-climbing attacks \citep{maiorana2014hill, galbally2010vulnerability}.

\definecolor{dpgcolor}{HTML}{82B366}
\definecolor{nondpgcolor}{HTML}{B85450}
\begin{figure}
  \centering
  \includegraphics[width=0.5\linewidth]{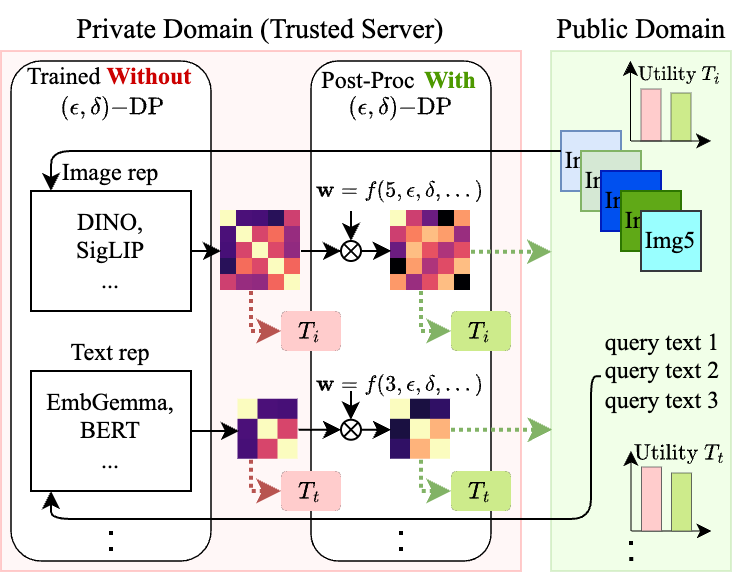}
    \caption{
      \textsc{ScoreShield} is a general, modality-agnostic post-processing DP wrapper for releasing similarity scores under central-model differential privacy.
      Given a downstream task $T$ that consumes similarity scores, the curator serves a non-private model and applies a calibrated Gaussian perturbation followed by projection onto the feasibility set, yielding a released vector/matrix that satisfies $(\varepsilon,\delta)$-DP with a quantified privacy--utility tradeoff.
      \textcolor{nondpgcolor}{Red dotted arrow} denotes the standard (non-private) release of similarity scores, while \textcolor{dpgcolor}{green dotted arrow} denotes releasing the same score objects through ScoreShield mechanism.
    }
  \label{fig:idea}
\end{figure}

A direct way to protect these score releases is to treat them as real-valued query outputs and apply a standard DP mechanism, for example by adding Gaussian noise calibrated to the global sensitivity of the released vector or matrix.
This baseline is model-agnostic and gives a formal $(\varepsilon,\delta)$-DP guarantee.
However, it treats cosine-score vectors and cosine Gram matrices as unconstrained Euclidean arrays.
For a fixed unit-norm query $\mathbf{q}$ and stored unit-norm embeddings $\mathbf{e}_1,\ldots,\mathbf{e}_n$, the released score vector $(\langle \mathbf{e}_1,\mathbf{q}\rangle,\ldots,\langle \mathbf{e}_n,\mathbf{q}\rangle)$ must lie in $[-1,1]^n$.
For full pairwise release, the score matrix $\mathbf{S}=\mathbf{E}\mathbf{E}^{\top}$ must be symmetric, positive semidefinite, and have unit diagonal.
Entrywise Gaussian perturbation does not preserve these constraints.
Thus, the na\"ive Gaussian mechanism is formally private but geometrically mismatched: it can add noise in infeasible directions and produce unnecessarily distorted inputs for ranking, thresholding, retrieval, clustering, verification, or recommendation.

\begin{figure*}[!t]
    \centering
    \includegraphics[width=0.95\linewidth]{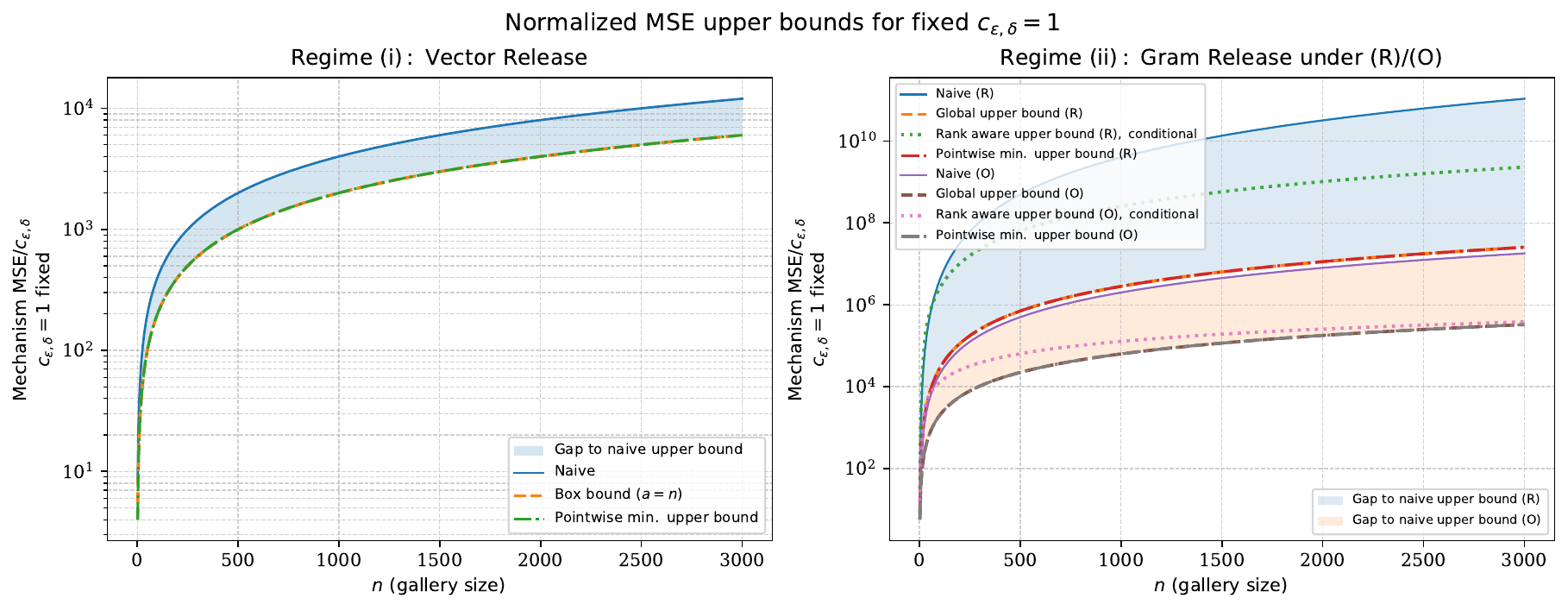}
    \vspace{-8pt}
    \caption{
    \textbf{Normalized MSE Scaling of Na\"ive Gaussian vs. ScoreShield Mechanisms.}
    \emph{Left (regime~(i), vector release):} releasing a query-to-collection cosine score vector.
    \emph{Right (regime~(ii), matrix release):} releasing the full pairwise cosine Gram matrix.
    }
    \vspace{-15pt}
    \label{fig:mse-scaling-main-body}
\end{figure*}

We study one-shot release of similarity scores under record-level $(\varepsilon,\delta)$-DP in the trusted-curator model \citep{dwork2014algorithmic}. The curator does not release embeddings; it releases either (i) a privatized score vector between one external query/probe embedding and all stored embeddings, used in verification, nearest-neighbor search, or re-ranking \citep{ijbc, whitelam2017iarpaijbb, johnson2019billion, zhong2017re, chum2007total, deng2019arcface, meng2021magface}, or (ii) a privatized pairwise cosine Gram matrix of the stored collection, used in non-interactive analytics such as de-duplication, clustering, or bias auditing \citep{ng2001spectral, shi2000normalized, mehrabi2021survey, wu2023face, slyman2024fairdedup, abbas2023semdedup}. \textsc{ScoreShield} applies the Gaussian mechanism to the chosen score statistic, with variance $\sigma^2_{\varepsilon,\delta}=c_{\varepsilon,\delta}\Delta^2$ calibrated to its global $\ell_2$ sensitivity, where $c_{\varepsilon,\delta}=2\log(2/\delta)/\varepsilon^2$, and then applies a privacy-preserving post-processing map that enforces the corresponding cosine feasibility constraints; see Figure~\ref{fig:idea}. Figure~\ref{fig:mse-scaling-main-body} summarizes the main risk behavior of \textsc{ScoreShield}. The stated Gram-risk bounds are for the exact Frobenius metric projection, which is DP-preserving post-processing and cannot increase squared distance to the non-private feasible Gram matrix.

\vspace{-5pt}

\paragraph{Main Contributions.}

\begin{enumerate}[leftmargin=1.38em,itemsep=3pt]
\item 
\textbf{Quantitative Improvement over Na\"ive Gaussian DP:}
For Gram-matrix release, the na\"ive Gaussian mechanism has expected squared Frobenius error $\Theta(n^{3}c_{\varepsilon,\delta})$ under record-level replacement adjacency and $\Theta(n^{2}\mathsf{\Delta}_{\mathsf{G}}^{2}c_{\varepsilon,\delta})$ under $\mathsf{\Delta}_{\mathsf{G}}$-Gram adjacency. For the exact Frobenius metric projection onto the cosine-Gram feasible set, the global Gaussian-complexity bound for \textsc{ScoreShield} gives $\mathcal{O}(n^{2}\sqrt{c_{\varepsilon,\delta}})$ and $\mathcal{O}(n^{3/2}\mathsf{\Delta}_{\mathsf{G}}\sqrt{c_{\varepsilon,\delta}})$, respectively. At rank-$r$ Gram matrices satisfying the local Gram-smoothness condition, the local tangent-cone bound gives $\mathcal{O}(n^{2}r c_{\varepsilon,\delta})$ and $\mathcal{O}(n r \mathsf{\Delta}_{\mathsf{G}}^{2}c_{\varepsilon,\delta})$, respectively. For vector releases, projection never increases squared-error risk and preserves the $\Theta(nc_{\varepsilon,\delta})$ worst-case scaling of na\"ive Gaussian DP, with possible constant-factor gains.\vspace{-3pt}
\item
\textbf{Utility Guarantees for Regime~(i):}
For query-to-collection vector releases calibrated to $(\varepsilon, \delta)$-DP: 
(i) We characterize the probability that perturbation noise flips the accept/reject decision for any interior threshold $\tau \in (-1,1)$.
(ii) We prove existence and uniqueness of a threshold re-calibration that restores any attainable operating point $(\mathsf{FPR}, \mathsf{TPR})$ achievable by thresholding the privatized scores, and show that in the small-noise regime $\sigma \rightarrow 0$ the required offset scales as $\Theta (\sigma^2)$ with a coefficient determined by the score density at $\tau$ (closed form for Gaussian impostors).
(iii) We establish uniform ROC/AUC stability: under $L$-Lipschitz score CDFs, both ROC coordinates and AUC deviate by at most $\mathcal{O}(\sigma)$. This $\mathcal{O}(\sigma)$ rate is first-order tight. The same $\mathcal{O}(\sigma)$ scaling holds for EER and partial-AUC. 
(iv) For small $\varepsilon$, we bound the sensitivity of the false match rate (FMR) function across neighboring galleries by $\mathcal{O} (\varepsilon + \delta)$ under one-shot $(\varepsilon, \delta)$-DP release.\vspace{-3pt}
\item
\textbf{A Fast, Scalable Projection Method with Guarantees:}
For regime~(ii), we use an averaged alternating-projection (AAP) algorithm for cosine Grams that alternates between the PSD cone $\mathcal{K}_n^+$ and the entrywise constraint set $\mathcal{C}_{\mathrm{unit}}^n$. We prove $R$-linear convergence to feasibility under bounded linear regularity; the AAP limit is not, in general, the exact Frobenius metric projection.\vspace{-2pt} 
\item 
\textbf{Evaluation:}
We evaluate regime~(i) for two applications: (a) differentially private retrieval-augmented generation (\textbf{DP-RAG}); (b)  differentially private face-recognition (\textbf{DP-FR}). For DP-RAG, we evaluate regime~(i) on the Google FRAMES benchmark for multi-hop RAG, employing multiple LLM generators and an LLM-based evaluator, and multiple text embedders for retrieval. For DP-FR, we evaluate regime~(i) on seven public FR benchmarks (\emph{e.g.,} LFW, IJB-C ) using three backbones across various privacy settings on an $(\varepsilon,\delta)$ grid and report several utility metrics. We evaluate regime~(ii) on \textbf{vision}/\textbf{NLP}/\textbf{recommender} tasks that consume only the private cosine Gram (CIFAR-10/100 and Oxford-IIIT Pets, STS-B, and MovieLens-100K).
\end{enumerate}

\vspace{-4pt}

\paragraph{Positioning.} 
To our knowledge, we provide the first systematic central-model $(\varepsilon, \delta)$-DP analysis of match-score releases in deep face recognition systems that connects threshold recalibration to decision-level guarantees and validates on standard public FR benchmarks.

\vspace{-4pt}

\paragraph{Notation.}
For $\mathbf{e} \! \in \! \mathbb{R}^{d}$, the Euclidean norm is denoted by ${\Vert \mathbf{e} \Vert}_2 \! = \! \big( \! \sum_{i=1}^d \! e_i^2 \big)^{\! 1/2}$, and the unit sphere in $\mathbb{R}^d$ is the set $\mathbb{S}^{d-1} \! \coloneqq \! \{\mathbf{e} \! \in \! \mathbb{R}^{d} \!\! : \! {\Vert \mathbf{e} \Vert}_2 \! = \! 1\}$. For a matrix $\mathbf{A} \! \in  \! \mathbb{R}^{n \times d}$, the Frobenius norm is ${\Vert \mathbf{A}\Vert}_{\mathrm{F}} \! = \! \sqrt{\mathrm{tr} (\mathbf{A}^{\!\top} \! \mathbf{A})}$.
The spectral norm of $\mathbf{A}$, denoted as ${\Vert \mathbf{A} \Vert}_2$, is the supremum $\sup_{\mathbf{e} \in \mathbb{S}^{d-1}} {\Vert \mathbf{A} \mathbf{e} \Vert}_2$ and equals its largest singular value. 
The nuclear norm ${\Vert \mathbf{A} \Vert}_\ast$ is the sum of the singular values of $\mathbf{A}$, i.e., $\sum_{k=1}^{\min{(n, d)}} \! \sigma_k (\mathbf{A})$. 
The maximum entry norm of $\mathbf{A}$ is defined as ${\Vert \mathbf{A} \Vert}_{\mathrm{max}} \!\! = \! \max_{1 \leq i \leq n, 1 \leq j \leq d} {\vert A_{ij} \vert}$. 
A symmetric matrix $\mathbf{A} \! \in \! \mathbb{R}^{n \times n}$ is PSD, denoted $\mathbf{A} \! \succeq 0$, if $\mathbf{v}^\top \! \mathbf{A} \mathbf{v} \! \geq 0, \, \forall \mathbf{v} \! \in \! \mathbb{R}^n$. 
For any integer $n \! \ge \! 1$, $\mathbf{I}_{n\times n}$ denotes the $n \! \times \! n$ identity matrix.
The standard deviation parameter in the Gaussian mechanism is denoted by $\sigma$. 
The Euclidean projection operator onto a closed convex set $\mathcal{C}$ is written as $\mathsf{proj}_{\mathcal{C}}(\cdot)$.

\vspace{-4pt}

\paragraph{Related Work.}

Releasing pairwise statistics with central DP has a long history \citep{blocki2012johnson, yang2017privacy}. Direct output perturbation adds noise calibrated to the global sensitivity of the released object; under record-level replacement, releasing a full $n\times n$ Gram modifies an entire row/column, so the expected squared Frobenius error can scale as $\Theta(n^3)$ (up to log factors in $\delta$) unless additional structure is used \citep{ji2024less}. These approaches do not guarantee that a released cosine Gram remains PSD with unit diagonal and bounded entries, nor do they provide guarantees for decision-level verification metrics.
Perturb-and-Project (PnP) \citep{cohen2024perturb} improves utility for cosine-similarity release by adding Gaussian noise and then projecting onto an admissible set of Gram matrices. Their analysis controls error via the Gaussian complexity of the feasibility set, yielding tighter Frobenius-risk bounds than na\"ive perturbation.
We differ in four respects: (i) we calibrate to record-level replacement for two disclosure regimes used in practice, and we also analyze the alternative $\mathsf{\Delta}_{\mathsf{G}}$-Gram (output-space) adjacency used in prior work; (ii) we enforce cosine-specific feasibility (PSD with unit diagonal and entrywise bounds) to guarantee validity of the released object; (iii) we provide decision-level utility guarantees (threshold recalibration, flip probability, ROC/AUC stability) that are not captured by norm-only bounds on $\ell_2$ or Frobenius error, and (iv) we prove convergence guarantees for a scalable alternating-projection algorithm.

A complementary line of work privatizes embeddings rather than scores. Johnson–Lindenstrauss (JL) transforms with calibrated Gaussian noise achieve vector-level DP while approximately preserving distances \cite{kenthapadi2012privacy}. Under suitable spectral assumptions, the Gaussian JL transform itself can satisfy $(\varepsilon,\delta)$-DP without an additional additive-noise step \cite{blocki2012johnson}. These methods protect the published vectors, but they do not by themselves bound leakage when the release object is an entire similarity vector or the full Gram matrix, which is our focus.

\textit{DP-FR:}
Regime~(i) matches the standard deep FR pipelines that operate on cosine similarity scores. Prior privacy-preserving FR research focuses on (i) local-DP or image/feature obfuscation \citep{chamikara2020privacy, ji2022privacy} and (ii) cryptographic inference (HE/MPC) \citep{bai2023cryptomask}. To our knowledge, prior work has not studied central-model $(\varepsilon,\delta)$-DP for the non-interactive publication of semantically valid FR similarity vectors together with decision-level verification guarantees. Perturb–and–project releases for cosine similarities do not provide verification decision-metric guarantees \citep{cohen2024perturb}. 
Regime~(i) therefore addresses a gap in the FR literature.

\textit{DP-RAG:}
Recent work studies differentially private RAG, where the goal is to protect an indexed corpus against leakage under (possibly adaptive) querying and generation. Existing approaches vary by what is privatized: some privatize retrieval outcomes (e.g., DP selection of IDs/ranks/scores) and treat subsequent processing as DP post-processing of the privatized retrieval output, while others spend privacy budget in the generation stage to make the final answer text DP, or privatize the corpus once via a DP synthetic proxy dataset \citep{wu2025private,koga2024privacy,grislain2025rag,mori2025differentially}. The multi-query regime is nontrivial because per-query privacy costs compose; recent mechanisms exploit relevance sparsity via screening and per-record privacy accounting to answer many queries under a fixed total budget \citep{wu2025beyond}. Our regime~(i) provides a one-shot central-DP primitive for releasing a bounded cosine score vector, and hence supports DP rankings/top-$k$/thresholding via post-processing, when the corpus is accessed only through these privatized scores. This is complementary to multi-query accountants and to end-to-end DP generation mechanisms.

For extended discussion, see Appendix~\ref{app:sec:extended-introduction} and Appendix~\ref{app:sec:extended-relatedwork}.

\vspace{-3pt}

\section{Preliminaries}
\label{Sec:preliminaries}

For completeness, Appendix~\ref{app:sec:extended-preliminaries} collects extended preliminaries and proofs. The proofs of all lemmas and corollaries in this section are deferred to the appendix.

\vspace{-3pt}

\paragraph{Setup.}
Let $\mathcal{D}= \{\mathbf{x}_1,\dots,\mathbf{x}_n\}$ be a dataset of $n$ records (e.g., images, text chunks, user profiles), and let $\phi:\mathcal{X}\to\mathbb{R}^d$ be a (fixed) feature encoder.
We assume $\mathbf{e}_i \coloneqq \phi(\mathbf{x}_i)$ and $\|\mathbf{e}_i\|_2 = 1, \forall i\in[n]$.
Stacking row vectors gives $\mathbf{E}=[\mathbf{e}_1^\top,\dots,\mathbf{e}_n^\top]^\top\in\mathbb{R}^{n\times d}$.
We study the differentially private release of cosine-similarity statistics derived from $\mathbf{E}$ under record-level replacement adjacency on $\mathcal{D}$.
Our two disclosure regimes are:
\textit{(i)} a query-to-collection score vector $\mathbf{s} \coloneq\mathbf{E}\mathbf{q}\in[-1,1]^n$ for a public query embedding $\mathbf{q}\in\mathbb{R}^d$ with $\|\mathbf{q}\|_2 = 1$; and
\textit{(ii)} the full pairwise cosine Gram matrix $\mathbf{S}\coloneqq \mathbf{E}\mathbf{E}^\top\in\mathbb{R}^{n\times n}$ with entries $S_{ij}=\langle \mathbf{e}_i,\mathbf{e}_j\rangle\in[-1,1]$.
Accordingly, the score-vector feasibility set is $\mathcal{C}_{\mathsf{query}} \coloneqq \{\mathbf{s}\in\mathbb{R}^{n}:\ |s_i|\le 1,\ \forall i\in[n]\} = [-1,1]^n$.
Since $\|\mathbf{e}_i\|_2= 1$, $\mathbf{S}$ is a correlation matrix: $\mathbf{S}\succeq 0$ and $\mathrm{diag}(\mathbf{S})=\mathbf{1}$; accordingly we define the feasible Gram set $\mathcal{C}_{\mathrm{coll}}\coloneqq\{\mathbf{S}\in\mathbb{R}^{n\times n}:\! \mathbf{S}\succeq0,\ \mathrm{diag}(\mathbf{S})=\mathbf{1},\ |S_{ij}|\le 1\ (i\neq j)\}$, and note $\mathrm{rank}(\mathbf{S})\le \min\{n,d\}$.
We instantiate this setup in face recognition, retrieval, and other similarity-based tasks.

\paragraph{Differential Privacy.}

\begin{definition}[Record-level Adjacency]
\label{def:recordlevel-adjacency-mainbody}
Datasets $\mathcal{D}, \mathcal{D}'$ (or equivalently, their embedding matrices $\mathbf{E}, \mathbf{E}'$) are \textit{adjacent}, denoted by $\mathcal{D} \sim \mathcal{D}'$, if they differ in at most one record (and thus in one embedding), i.e., $|\,\{i\in[n]: \mathbf{x}_{i}\neq \mathbf{x}_{i}'\}\,| = 1$.
\end{definition}

\begin{remark}
Under a fixed backbone model $\phi_{\boldsymbol{\theta}}$, adjacency implies that the embedding matrices $\mathbf{E}$ and $\mathbf{E}'$ differ in at most one row. 
Therefore, there exists at most one index $i$ such that $\mathbf{e}_j = \mathbf{e}_j' , \forall j \neq i$, and by our normalization assumption $\mathbf{e}_i \neq \mathbf{e}_i' \in \mathbb{R}^d$ satisfy $\| \mathbf{e}_i \|_2 = \| \mathbf{e}_i' \|_2 =1$.
\end{remark}

\begin{definition}[Output–space  (Gram-matrix) Adjacency]
\label{def:gram-adjacency-mainbody}
Two collections with embeddings $\mathbf{E},\mathbf{E}'$ are \textit{adjacent at radius} $\mathsf{\Delta}_{\mathsf{G}} > 0$ if
%
$\|\mathbf{E}\mathbf{E}^\top-\mathbf{E}'\mathbf{E}'^\top\|_{\mathrm{F}} \;\le\; \mathsf{\Delta}_{\mathsf{G}}$.
%
\end{definition}

\begin{definition}[$(\varepsilon,\delta)$--DP]
A (possibly randomized) mechanism $\mathcal{M}:\mathcal{E}\to\mathcal{O}$
satisfies $(\varepsilon,\delta)$--DP if for all measurable $\mathcal{T}\subseteq\mathcal{O}$
and all adjacent $\mathbf{E}\sim\mathbf{E}'$,
$\mathsf{Pr}\!\left[\mathcal{M}(\mathbf{E})\in\mathcal{T}\right] \;\le\;
e^{\varepsilon}\,\mathsf{Pr}\!\left[\mathcal{M}(\mathbf{E}')\in\mathcal{T}\right]+\delta$,
where $\varepsilon > 0$ and $\delta \in (0, 1)$ are privacy parameters.
\end{definition}

\begin{definition}[$\ell_2$-Sensitivity]
\label{def:sensitivity-mainbody}
For a function $f: \mathcal{E} \to \mathbb{R}^n$, the $\ell_2$-sensitivity is $\Delta_{f,2} = \sup_{\mathbf{E} \sim \mathbf{E}'} \| f(\mathbf{E}) - f(\mathbf{E}') \|_2$. 
We use the notation $\Delta$ for brevity. 
\end{definition}

\begin{lemma}[Gaussian Mechanism]
\label{lem:gaussian-mainbody}
Let $f:\mathcal{E} \to  \mathbb{R}^{n}$ have $\ell_2$–sensitivity $\Delta_{f, 2}= \sup_{\mathbf{E} \sim \mathbf{E}'} \|f(\mathbf{E}) - f(\mathbf{E}')\|_2$. Define $\mathcal{M} (\mathbf{E}) \! = \! f(\mathbf{E}) + \mathbf{w}$,
$\mathbf{w} \sim \mathcal{N} \bigl( \mathbf{0}, \sigma^2 \mathbf{I}_{n}\bigr)$,
$\sigma^2 \geq c_{\varepsilon,\delta}\,\Delta^{2}_{f,2}$ with $c_{\varepsilon,\delta} \coloneq \frac{2\log (2/\delta)}{\varepsilon^2}$.
Then $\mathcal{M}$ satisfies $(\varepsilon, \delta)$-DP.
\end{lemma}

\begin{corollary}[Gaussian Mechanism for Matrix-Valued Outputs]
\label{cor:gaussian-matrix-mainbody}
Let $f \! : \! \mathcal{E} \! \to \! \mathbb{R}^{n\times n}$ and define the Frobenius sensitivity
$\Delta_{f,\mathrm{F}} \! \coloneqq \! \sup_{\mathbf{E}\sim\mathbf{E}'} \|f(\mathbf{E}) \! - \! f(\mathbf{E}')\|_{\mathrm{F}}$.
Let $\mathbf{W}\in\mathbb{R}^{n\times n}$ have i.i.d.\ entries $W_{ij} \! \sim\! \mathcal{N}(0,\sigma^2)$, and define
$\mathcal{M}(\mathbf{E}) \coloneqq f(\mathbf{E})+\mathbf{W}$.
If $\sigma^2 \geq c_{\varepsilon,\delta}\,\Delta_{f,\mathrm{F}}^{2}$, with $c_{\varepsilon,\delta}=2\log(2/\delta)/\varepsilon^2$, then $\mathcal{M}$ is $(\varepsilon,\delta)$-DP.
\end{corollary}

\begin{lemma}[Post-processing]\label{lem:post-processing-mainbody}
Let $\mathcal{M}: \mathcal{E} \to \mathcal{O}$ be a mechanism that satisfies $(\varepsilon, \delta)$--differential privacy. For any measurable function $g: \mathcal{O} \to \mathcal{O}'$, the composed mechanism $g \circ \mathcal{M}: \mathcal{E} \to \mathcal{O}'$ also satisfies $(\varepsilon, \delta)$--differential privacy.
\end{lemma}

\vspace{-3pt}

\paragraph{Convex‑Geometry.}

\begin{definition}[Euclidean Projection Onto a Closed Convex Set]
\label{def:projection-mainbody}
Let $m\ge 1$ and let $\mathcal{C}\subset\mathbb{R}^{m}$ be non–empty, closed and convex. The Euclidean projection operator $\mathsf{proj}_{\mathcal{C}}:\mathbb{R}^{m}\to\mathcal{C}$ is defined for every $\mathbf{s} \in \mathbb{R}^{m}$ by $\mathsf{proj}_{\mathcal{C}}(\mathbf{s})  \coloneqq \operatorname*{arg\,min}_{\mathbf{y} \in \mathcal{C}} \| \mathbf{s} - \mathbf{y} \|_{2}$.
\end{definition}

\begin{definition}[Tangent Cone]
\label{def:tangent-cone-mainbody}
Let $\mathcal{C} \subset \mathbb{R}^{m}$ be  non‑empty, closed, and convex and fix a point $\mathbf{s} \in \mathcal{C}$. The tangent cone to $\mathcal{C}$ at $\mathbf{s}$ is
$\mathsf{T}_{\mathbf{s}} \left( \mathcal{C}  \right)  \coloneqq 
\mathsf{cl}\bigl\{ \lambda\,( \mathbf{y} - \mathbf{s}) \; : \; \lambda \ge 0,\; \mathbf{y} \in\mathcal{C} \bigr\}  \;\subset\; \mathbb{R}^{m}$,
where ``$\mathsf{cl}$'' denotes the Euclidean closure. Geometrically, $\mathsf{T}_{\mathbf{s}} \left( \mathcal{C}  \right) $ contains all velocity directions of feasible curves that start at $\mathbf{s}$ and remain inside $\mathcal{C}$.
\end{definition}

\begin{definition}[Gaussian Complexity of a Bounded Set]
\label{def:gaussian-complexity-mainbody}
Let $\mathcal{C}\subset\mathbb{R}^{m}$ be bounded. Its Gaussian complexity is
$\mathsf{GC}(\mathcal{C})  \coloneqq  \mathbb{E}_{\mathbf{w} \sim \mathcal{N}( \mathbf{0}, \mathbf{I}_m)} \left[ \, \sup_{\mathbf{s} \in \mathcal{C}} \, \langle \mathbf{w}, \mathbf{s} \rangle \, \right]$.
If $\mathcal{C}$ is unbounded we set $\mathsf{GC}(\mathcal{C})\;  \coloneqq  \; + \infty$ by convention.
\end{definition}

\begin{lemma}[Gaussian Complexity of $\mathcal{C}_{\mathrm{query}}$]
\label{lem:gc-box-mainbody}
Let $\mathcal{C}_{\mathrm{query}} \coloneqq \{\mathbf{s}\in\mathbb{R}^{n}:\ |s_i|\le 1,\ \forall i\in[n]\} \;=\; [-1,1]^n$ and $\mathbf{z} \sim \mathcal{N}(\mathbf{0},\mathbf{I}_n)$. Consider the Gaussian complexity definition \ref{def:gaussian-complexity}. Then $\mathsf{GC}( \mathcal{C}_{\mathrm{query}} ) = n\sqrt{\frac{2}{\pi}}= \Theta (n)$.
\end{lemma}

\begin{lemma}[Gaussian Complexity of $\mathcal{C}_{\mathrm{coll}}$]
\label{lem:gc-elliptope-mainbody}
Let $\mathcal{C}_{\mathrm{coll}}\coloneqq\{\mathbf{S}\in\mathbb{R}^{n\times n}:\! \mathbf{S}\succeq0,\ \mathrm{diag}(\mathbf{S})=\mathbf{1},\ |S_{ij}|\le 1\ (i\neq j)\}$. Then $\mathsf{GC}(\mathcal{C}_{\mathrm{coll}}) = \Theta(n^{3/2})$.
\end{lemma}

\begin{lemma}[Firm Non‑expansiveness of Euclidean Projections]
\label{lem:proj-firm-mainbody}
Let $\mathcal{C} \subset \mathbb{R}^{n}$ be non‑empty, closed and convex and define the Euclidean projector $\mathsf{proj}_{\mathcal{C}}(\mathbf{s}) \;=\; \operatorname*{arg\,min}_{\mathbf{s}'\in\mathcal{C}}\| \mathbf{s} - \mathbf{s}' \|_{2}$.    
Then for all $\mathbf{s},\mathbf{s}'\in \mathbb{R}^{n}$ we have $\|\mathsf{proj}_{\mathcal{C}}(\mathbf{s}) - \mathsf{proj}_{\mathcal{C}}(\mathbf{s}') \|_{2}^{2} \;\le\; \| \mathbf{s}- \mathbf{s}' \|_{2}^{2}$.
\end{lemma}

\vspace{-3pt}

\paragraph{Design Objective.}
Given private unit-norm embeddings, the task is to release cosine similarities without releasing the embeddings. The released object is either a vector of similarities between a public query and the database, or a matrix of pairwise similarities among database records. The mechanism must satisfy record-level differential privacy and output an element of the corresponding cosine-feasible set.
The objective is to minimize distortion relative to the non-private similarities.

\vspace{-3pt}

\paragraph{Threat Model.}
A curator holds the database (equivalently, the embedding matrix $\mathbf{E}$) and makes a single, non-interactive release of a privatized similarity object (either a score vector in regime~(i) or a Gram matrix in regime~(ii)).
The adversary is an external analyst who observes only the released output, knows the mechanism and all public parameters, including $(\varepsilon,\delta)$, and has unbounded computational power.
We allow arbitrary auxiliary information, including knowledge of all records except the one that may differ under the adjacency relation.



\section{ScoreShield Framework}
\label{sec:scoreshield}

\subsection{Query-to-Collection Similarity Score Vector Release}
\label{subsec:probe-vector}

\noindent
\textbf{Operational~Scenario.}
At run time, a client (or sensor) produces a fresh $\ell_{2}$-normalized probe embedding $\mathbf{q} \in  \mathbb{R}^{d}$ and submits it for matching. The server stores a fixed gallery of unit-norm enrollment embeddings $\mathbf{E}=[\mathbf{e}_{1}^{\top};\dots;\mathbf{e}_{n}^{\top}]\in \mathbb{R}^{n\times d}$ and computes the cosine-similarity vector $\mathbf{s} \;=\; \mathbf{E}\,\mathbf{q} = (\,\mathbf{e}_{1}^{\top}\mathbf{q},\dots, \mathbf{e}_{n}^{\top}\mathbf{q}) \in  \! [-1,1]^{n}$. This vector drives the downstream decisions (e.g., thresholding for $1{:}1$ verification or top-$k$ ranking for open-set search). Formally, the function to be privatized is $f_{\mathsf{query}} (\mathbf{E}, \mathbf{q}) \! = \! \mathbf{E} \, \mathbf{q} \! \in \! \mathbb{R}^n$. We omit the subscript in $f_{\mathsf{query}}$ when the context is clear.

\begin{wrapfigure}{r}{0.26\linewidth}
  \vspace{-\baselineskip}
  \centering
  \includegraphics[width=\linewidth]{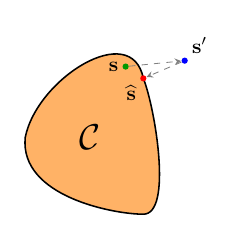}
   \caption{\textsc{ScoreShield} Visualization.}
  \label{fig:scoreshield-visualization}
  \vspace{-0.8\baselineskip}
\end{wrapfigure}

\noindent
\textbf{Sensitivity.}  
Fix any public $\mathbf{q}\in\mathbb{S}^{d-1}$. Denote $\boldsymbol{\delta} \! \coloneqq \! \mathbf{e}_{i}-\mathbf{e}'_{i}$ and note that $\|\boldsymbol\delta\|_{2}\le 2$ (e.g., $\mathbf{e}_i = \mathbf{q}$, $\mathbf{e}_i' = - \mathbf{q}$). Then under $\mathbf{E} \! \sim \! \mathbf{E}'$ differing in row $i$, $f(\mathbf{E},\mathbf{q}) \! - \! f(\mathbf{E}',\mathbf{q})
=\mathbf{E}\mathbf{q}- \mathbf{E}'\mathbf{q}
=\bigl(0,\dots,0, \boldsymbol\delta^{\top}\!\mathbf{q}, 0,\dots,0\bigr)$, hence, the inner product is bounded by Cauchy–Schwarz as follows $\|f(\mathbf{E},\mathbf{q})-f(\mathbf{E}',\mathbf{q})\|_{2}
=|\boldsymbol\delta^{\top}\mathbf{q}| \le \|\boldsymbol\delta\|_{2}\,\|\mathbf{q}\|_{2}
\le 2$. The global $\ell_{2}$-sensitivity is therefore the constant $\Delta_{f, 2} \eqqcolon \Delta_{\mathsf{query}} = 2$.

\noindent
\textbf{DP Mechanism.}
The DP query-to-collection similarity vector release algorithm is provided in Algorithm~\ref{alg:query-to-collection}. A conceptual illustration is shown in Figure~\ref{fig:scoreshield-visualization}.

\begin{theorem}[Privacy Guarantee of Query-to-Collection Similarity Score Vector Release]
\label{app:thm:probe-privacy_mainbody}
Let $\mathcal{M}_{\mathsf{query}}$ be the mechanism returned by Algorithm~\ref{alg:query-to-collection} with parameters $\varepsilon>0,\;\delta\in(0,1)$. Then for every pair of neighboring embedding matrices $\mathbf{E}\sim\mathbf{E}'$ and every measurable set $S\subseteq[-1,1]^n$, 
%
$\mathsf{Pr}\, \bigl[\mathcal{M}_{\mathsf{query}}(\mathbf{E}) \in S\bigr] \;\le\;e^{\varepsilon}\,
\mathsf{Pr}\, \bigl[\mathcal{M}_{\mathsf{query}}(\mathbf{E}') \in S\bigr]+\delta$ .
%
Hence $\mathcal{M}_{\mathsf{query}}$ is $(\varepsilon,\delta)$–DP.
\end{theorem}

\begin{proof}
See Appendix Theorem~\ref{app:thm:probe-privacy} in Appendix~\ref{app:sec:supplementary-regime1-theory}. 
\end{proof}

\begin{algorithm}[t]
\caption{DP Query-to-Collection Similarity Score Vector Release (regime~(i))}
\label{alg:query-to-collection}
\begin{algorithmic}[1]  
\State \strut \textbf{Input:} $\mathbf{E} \in \mathbb{R}^{n \times d}$, $\mathbf{q} \in \mathbb{R}^d$, $\varepsilon > 0$, $\delta \in (0, 1)$
\State \textbf{Output:} $\widehat{\mathbf{s}} \in \mathbb{R}^n$
\State Compute $\mathbf{s} = \mathbf{E} \, \mathbf{q}$
\State Set $\sigma_{\varepsilon, \delta} \;\gets\; \Delta_{\mathsf{query}} \sqrt{2\log(2/\delta)}/\varepsilon$
\State Sample noise $\mathbf{w} \sim \mathcal{N}\big( \mathbf{0}, \sigma_{\varepsilon, \delta}^2 \, \mathbf{I}_{n}\big)$
\State Compute $\mathbf{s}' = \mathbf{s} + \mathbf{w}$
\State Compute $\widehat{\mathbf{s}}~=~\mathsf{proj}_{\mathcal{C}} (\mathbf{s}')$,
\Statex
where~$(\mathsf{proj}_{\mathcal{C}}(\mathbf{s}^\prime))_{i} = \max\bigl(-1,\min(1, s_{i} ^\prime)\bigr)$ 
\State \textbf{Return} $\widehat{\mathbf{s}}$
\end{algorithmic}
\end{algorithm}

\paragraph{Risk Scaling: Na\"ive Gaussian vs. ScoreShield Mechanism}
\begin{lemma}[Global ScoreShield Bound via Gaussian Complexity]
\label{lem:global-gc-scoreshield-mainbody}
Let $m\ge 1$ and let $\mathcal{C}_{\mathsf{query}}\subset \mathbb{R}^{m}$ be nonempty, closed, convex, and bounded. Fix $\mathbf{s}\in\mathcal{C}_{\mathsf{query}}$ and let $\mathbf{w} \sim \mathcal{N} (\mathbf{0}, \sigma^{2}\mathbf{I}_{m})$. Consider ScoreShield projector $\widehat{\mathbf{s}} = \mathsf{proj}_{\mathcal{C}_{\mathsf{query}}}(\mathbf{s}+\mathbf{w})$. Then
\begin{equation}
\label{eq:global-gc-scoreshield-mainbody}
\mathbb{E}\,\|\widehat{\mathbf{s}}-\mathbf{s}\|_{2}^{2}
\;\le\; 4\, \sigma\,\mathsf{GC}(\mathcal{C}_{\mathsf{query}}),
\end{equation}
where the expectation is over $\mathbf{w}$ and for any bounded $\mathcal{A}\subset \mathbb{R}^{m}$, $\mathsf{GC}(\mathcal{A}) \; \coloneqq\; \mathbb{E}_{\mathbf{w} \sim \mathcal{N}(\mathbf{0},\mathbf{I}_{m})} \big[ \sup_{\mathbf{a}\in\mathcal{A}}\langle \mathbf{w},\mathbf{a} \rangle \big]$.
\end{lemma}

\begin{proof}
See Lemma~\ref{lem:global-gc-scoreshield} in Appendix~\ref{app:sec:naive-vs-scoreshield}. 
\end{proof}

\begin{lemma}[Local ScoreShield Bound via Tangent Cones]
\label{lem:local-tc-scoreshield-mainbody}
Let $m\ge 1$ and let $\mathcal{C}_{\mathsf{query}}\subset\mathbb{R}^{m}$ be nonempty, closed, and convex. Fix $\mathbf{s}\in\mathcal{C}_{\mathsf{query}}$ and let $\mathbf{w} \sim\mathcal{N}(\mathbf{0}, \sigma^{2}\mathbf{I}_{m})$. Let $\widehat{\mathbf{s}} = \mathsf{proj}_{\mathcal{C}_{\mathsf{query}}}(\mathbf{s}+ \mathbf{w})$. Then
\begin{equation}
\label{eq:local-tc-bound-mainbody}
\mathbb{E}\, \| \widehat{\mathbf{s}} - \mathbf{s} \|_{2}^{2}
\; \le \; \sigma^{2}\, \delta \! \left( \mathsf{T}_{\mathbf{s}}(\mathcal{C}_{\mathsf{query}}) \right),
\end{equation}
where $\mathsf{T}_{\mathbf{s}}(\mathcal{C}_{\mathsf{query}})$ is the tangent cone of $\mathcal{C}_{\mathsf{query}}$ at $\mathbf{s}$ and
\begin{equation}
\delta(\mathcal{K}) \coloneqq  \mathbb{E}\bigl[\|\mathsf{proj}_{\mathcal{K}}(\mathbf{z})\|_{2}^{2}\bigr],
\qquad \mathbf{z}\sim\mathcal{N}(\mathbf{0},\mathbf{I}_{m}),
\end{equation}
denotes the statistical dimension of a closed convex cone $\mathcal{K} \subset \mathbb{R}^{m}$.
\end{lemma}

\begin{proof}
See Lemma~\ref{lem:local-tc-scoreshield} in Appendix~\ref{app:sec:naive-vs-scoreshield}. 
\end{proof}

\noindent
\textbf{Na\"ive Gaussian Mechanism.}
Release $\mathbf{s}'= \mathbf{s}+\mathbf{w}$, $\mathbf{w}\sim\mathcal{N}( \mathbf{0},\sigma^2 \mathbf{I}_n)$ with $\sigma^2=c_{\varepsilon,\delta}\Delta_{\mathsf{query}}^2=4c_{\varepsilon,\delta}$. Then we have
\begin{equation}
\makebox[0pt][c]{\tcboxmath[
  enhanced,
  colback=gray!8,
  colframe=gray!40,
  boxrule=0.5pt,
  arc=2pt,
  boxsep=0pt,
  left=6pt,right=6pt,top=4pt,bottom=4pt,
  grow to left by=10mm,
  grow to right by=10mm
]{%
  \displaystyle
    \mathbb{E}\,\| \mathbf{s}' - \mathbf{s} \|_2^2 \, = \,  \mathbb{E}\|\mathbf{w}\|_2^2 \, =\, n\,\sigma^2 \; = \; 4\, c_{\varepsilon,\delta}\, n   
}}
\end{equation}

\noindent
\textbf{ScoreShield Mechanism.}
Let $\widehat{\mathbf{s}}= \mathsf{proj}_{\mathcal{C}_{\mathsf{query}}}(\mathbf{s} + \mathbf{w})$.
We have:

\noindent
\textit{Local (Pointwise) Risk Bound via the Tangent Cone.}
For the constraint set $\mathcal{C}_{\mathsf{query}} =[-1,1]^n$, the tangent cone at $\mathbf{s}$ is a product cone whose statistical dimension depends only on the active set. Let $a(\mathbf{s})$ denote the number of coordinates of $\mathbf{s}$ on the boundary $\{\pm1\}$. Then $\delta (\mathsf{T}_\mathbf{s} ( \mathcal{C}_{\mathsf{query}}) )= n-\tfrac12 a(\mathbf{s})$ (see Lemma~\ref{cor:polyhedral-cube-exact} for a proof), and the local conic denoising bound (Lemma~\ref{lem:local-tc-scoreshield-mainbody}) yields
\begin{equation}
\label{eq:ScoreShield-regime1-mainbody}
\makebox[0pt][c]{\tcboxmath[
  enhanced,
  colback=gray!8,
  colframe=gray!40,
  boxrule=0.5pt,
  arc=2pt,
  boxsep=0pt,
  left=6pt,right=6pt,top=4pt,bottom=4pt,
  grow to left by=10mm,
  grow to right by=10mm
]{%
  \displaystyle
    \mathbb{E}\|\widehat{\mathbf{s}}-\mathbf{s}\|_2^2
    \;\le\; \sigma^2\Bigl(n-\tfrac12 a(\mathbf{s})\Bigr)
    \;=\; 4\,c_{\varepsilon,\delta}\Bigl(n-\tfrac12 a(\mathbf{s})\Bigr). 
}}
\end{equation}
If no coordinate is active, $a(\mathbf{s})=0$ and $\delta(\mathsf{T}_{\mathbf{s}})=n$, i.e., the bound matches the na\"ive risk. If many coordinates are saturated (large $a(\mathbf{s})$), projection can reduce the bound by up to a factor $\frac{1}{2}$ on those coordinates.
Since $\Delta_{\mathsf{query}}$ is constant, the per-coordinate MSE is $\mathcal{O}(1)$ for both mechanisms.

\noindent
\textit{Global Risk Bound via the Gaussian Complexity.}
Let $\mathbf{z} \sim \mathcal{N} (\mathbf{0}, \mathbf{I}_n)$. For a universal constant $C > 0$, Lemma~\ref{lem:gc-box} and Lemma~\ref{lem:global-gc-scoreshield}) give the uniform estimate
\begin{equation}
\label{eq:global-gc-regime1-mainbody}
\makebox[0pt][c]{\tcboxmath[
  enhanced,
  colback=gray!8,
  colframe=gray!40,
  boxrule=0.5pt,
  arc=2pt,
  boxsep=0pt,
  left=6pt,right=6pt,top=4pt,bottom=4pt,
  grow to left by=10mm,
  grow to right by=10mm
]{%
  \displaystyle
    \mathbb{E}\|\widehat{\mathbf{s}}-\mathbf{s}\|_2^2
    \;\le\; C\,\sigma\,\mathsf{GC}(\mathcal{C}_{\mathsf{query}})
    \;=\; C\,\sigma\,n\sqrt{\frac{2}{\pi}}.
}}
\end{equation}
%

\noindent
\textbf{Adversary Gain.}
We analyze attacker reconstruction under three knowledge regimes:
(K0) no side information about $\mathbf{E}$;
(K1) full knowledge of $\mathbf{E}$ (so $\mathbf{s}=\mathbf{E}\mathbf{q}\in\mathrm{col}(\mathbf{E})$);
(K2) DP-admissible side information revealing $\mathbf{E}_{-i}$ (all rows except the single row $i$ that differs under adjacency), so $\mathbf{s}_{-i}\in\mathrm{col}(\mathbf{E}_{-i})$.
In (K1), under the no-clipping Gaussian model the optimal linear denoiser projects onto $\mathrm{col}(\mathbf{E})$, yielding average per-entry MSE $\sigma^2_{\varepsilon,\delta}r/n$ where $r= \mathrm{rank}(\mathbf{E})$.
In (K2), the attacker can denoise the unchanged coordinates via projection onto $\mathrm{col}(\mathbf{E}_{-i})$, but the changed coordinate remains at MSE $\sigma^2_{\varepsilon, \delta}$.

Extended discussion and additional results are provided in Appendix~\ref{ssec:attacker_regime1}.

\subsection{Full Pairwise Similarity Score Matrix Release}
\label{subsec:all-pairs}

\noindent
\textbf{Operational Scenario.}
Batch analytics workflows such as unsupervised clustering, multi-camera trajectory association, or demographic bias auditing require the complete pairwise similarity structure of the embeddings. To support these non-interactive tasks the curator makes a one-shot disclosure of the cosine Gram matrix $\mathbf{S} = \mathbf{E} \,\mathbf{E}^{\top} \;\in[-1,1]^{n\times n}$, where $S_{ij}= \mathbf{e}_{i}^{\top}\mathbf{e}_{j}$. The statistic to be privatized is therefore $f_{\mathsf{full}} (\mathbf{E}) =  \mathbf{E}\mathbf{E}^{\top} \in \mathbb{R}^{n \times n}$. We drop subscript for when the context is clear.

\noindent
\textbf{Sensitivity.}
Under output-space (Gram-matrix) adjacency, the global Frobenius sensitivity is $\Delta_{f,\mathrm{F}}=\Delta_{\mathsf G}$ by definition (for a fixed radius $\Delta_{\mathsf G}$); see Appendix~\ref{app:sec:output-space-adjacency}.%
\footnote{If $\Delta_{\mathsf G}$ is instead chosen to dominate record-level replacement, then necessarily $\Delta_{\mathsf G}\ge 2\sqrt{2(n-1)}$, hence $\Delta_{\mathsf G}=\Theta(\sqrt{n})$; see Appendix~\ref{app:sec:output-space-adjacency}.}
Under record-level adjacency, $\mathbf{E}$ and $\mathbf{E}'$ differ in one row $i$, so $\mathbf{E}'=\mathbf{E}+\boldsymbol{\Delta}$ where $\boldsymbol{\Delta}$ has a single nonzero row $\boldsymbol{\delta}^\top$ with $\|\boldsymbol{\delta}\|_2\le 2$.
Let $\mathbf{S}=\mathbf{E}\mathbf{E}^\top$ and $\mathbf{S}'=\mathbf{E}'\mathbf{E}'^\top$.
Then $\mathbf{S}'-\mathbf{S}$ is supported only on row/column $i$, and the diagonal satisfies $(\mathbf{S}'-\mathbf{S})_{ii}=0$ since $\|\mathbf{e}_i\|_2=\|\mathbf{e}'_i\|_2=1$.
Writing $v_j\coloneqq \boldsymbol{\delta}^\top \mathbf{e}_j$, we obtain
$\|\mathbf{S}'-\mathbf{S}\|_{\mathrm{F}}^{2} = 2\sum_{j\ne i} v_j^{2}
\le 2\sum_{j\ne i}\|\boldsymbol{\delta}\|_2^{2}\,\|\mathbf{e}_j\|_2^{2} = 2(n-1)\|\boldsymbol{\delta}\|_2^{2}
\le 8(n-1)$,
and therefore the global Frobenius sensitivity is
$\Delta_{f,F}  \eqqcolon \sup_{\mathbf{E}\sim\mathbf{E}'} \|\mathbf{S}' - \mathbf{S}\|_{\mathrm{F}} = 2\sqrt{2(n-1)}  \eqqcolon \Delta_{\mathrm{full}}$.

\noindent
\textbf{DP Mechanism.}
The DP full pairwise similarity score matrix release algorithm is provided in Algorithm~\ref{alg:all-pairs-regime2-mainbody}.

\begin{theorem}[Privacy Guarantee of Full Pairwise Similarity Score Matrix Release]
\label{app:thm:gram-dp-guarantee-mainbody}
Let $f_{\mathsf{full}}:\mathcal{E}\to \mathbb{R}^{n\times n}$ be
$f_{\mathsf{full}}(\mathbf{E})=\mathbf{S}=\mathbf{E}\mathbf{E}^\top$.
Fix an adjacency relation $\sim$ on $\mathcal{E}$ and assume that for some $\Delta>0$, $\|f_{\mathsf{full}}(\mathbf{E})-f_{\mathsf{full}}(\mathbf{E}')\|_{\mathrm{F}} \le \Delta,\forall \, \mathbf{E}\sim\mathbf{E}'$.
Let $\varepsilon>0$, $\delta\in(0,1)$, $c_{\varepsilon,\delta}\coloneqq 2\log(2/\delta)/\varepsilon^2$, and set $\sigma^2 \coloneqq c_{\varepsilon,\delta}\Delta^2$. Let $\mathbf{W}\in\mathbb{R}^{n\times n}$ have i.i.d. entries $W_{ij}\sim\mathcal{N}(0,\sigma^2)$. Let $\mathsf{proj}_{\mathcal{C}}$ be any (possibly randomized) measurable post-processing that depends on $\mathbf{E}$ only through its input argument. Define the release $\widehat{\mathbf{S}}= \mathcal{M}_{\mathsf{coll}} (\mathbf{E})  \;\coloneqq\; \mathsf{proj}_{\mathcal{C}}\!\left(f_{\mathsf{full}}(\mathbf{E})+\mathbf{W}\right)$. Then the mechanism $\mathbf{E}\mapsto \widehat{\mathbf{S}}$ is $(\varepsilon,\delta)$-DP with respect to $\sim$, i.e.,
\begin{equation}
\mathsf{Pr} \left[ \mathcal{M}_{\mathsf{coll}}(\mathbf{E})\in S \right]
\;\le\; e^{\varepsilon}\,\mathsf{Pr}\!\left[ \mathcal{M}_{\mathsf{coll}}(\mathbf{E}') \in S \right] + \delta,
\qquad \forall\,\mathbf{E}\sim \mathbf{E}'.
\end{equation}
\end{theorem}

\begin{proof}
See Theorem~\ref{app:thm:gram-dp-guarantee} in Appendix~\ref{app:sec:supplementary-regime2-theory}. 
\end{proof}

\begin{algorithm}[t]
\caption{DP All-Pairs Similarity Matrix Release (regime~(ii))}
\label{alg:all-pairs-regime2-mainbody}
\begin{algorithmic}[1]
\State \textbf{Input:} $\mathbf{E} \in \mathbb{R}^{n \times d}$, $\varepsilon > 0$, $\delta \in (0,1)$, sensitivity $\Delta > 0$
\State \textbf{Output:} $\widehat{\mathbf{S}} \in \mathcal{C}_{\mathsf{coll}} \subseteq \mathbb{R}^{n \times n}$
\State Construct: $\mathbf S\gets\mathbf E\mathbf E^\top$.
\State Gaussian calibration: set $\sigma^2_{\varepsilon, \delta} \;\gets\;  c_{\varepsilon, \delta} \Delta^2$
\State Sample noise $\mathbf{W} \sim \mathcal{N}\left( \mathbf{0}, \sigma^2 \mathbf{I}_{n \times n}\right)$
\State Perturb with noise: $\mathbf{S}' = \mathbf{S} + \mathbf{W} $
\State Project: $\widehat{\mathbf{S}}  =   \mathsf{proj}_{\mathcal{C}_{\mathsf{coll}}} (\mathbf{S}')  \!= \! \arg\min_{\mathbf{A} \in  \mathcal{C}_{\mathsf{coll}}} \! \|\mathbf{A} \! - \!  \mathbf{S}'\|_{\mathrm{F}}^2$
\State \textbf{Return} $\widehat{\mathbf{S}}$
\end{algorithmic}
\end{algorithm}

\paragraph{Fast Projection via Averaged Alternating Projections.}
Let $\mathbf{S}'= \mathbf{S} +\mathbf{W}\in \mathbb{R}^{n\times n}$ be a perturbed Gram matrix, where $\mathbf{S}= \mathbf{E}\mathbf{E}^{\top}$ is the clean cosine Gram matrix and ${\Vert \mathbf{e}_i\Vert}_2=1 , \forall i$. Our goal is to project $\mathbf{S}'$ onto the cosine-Gram feasibility set $\mathcal{C}_{\mathsf{coll}}  \coloneqq  \bigl\{ \mathbf{S}\in \mathbb{R}^{n\times n}: \, \mathbf{S}\succeq0,\; S_{ii}=1\;(1 \le  i \le  n),\; |S_{ij}|\le 1\;(i \ne  j) \bigr\}\subset \mathbb{R}^{n\times n}$. The exact projection onto the cosine-Gram feasible set $\mathcal{C}_{\mathsf{coll}}$ under the Frobenius norm is the metric projection onto the elliptope and in general requires solving an SDP.
Instead we compute an approximately feasible point by iterating a Krasnosel'ski\u{\i}-Mann averaged projector \citep{mann1953mean, ma1955two}, and hence replace the direct projection by alternating projections onto two closed convex sets, each admitting a closed-form projector. We decompose $\mathcal{C}_{\mathsf{coll}} = \mathcal{K}_{+}^{\,n} \cap \mathcal{C}_{\mathrm{unit}}^n$ where 
\begin{equation}
\mathcal{K}_{+}^{\,n}  = \{\mathbf{S} \in \mathbb{R}^{n \times n}  \mid \mathbf{S}\succeq0\}, 
\qquad
\mathcal{C}_{\mathrm{unit}}^n = \{ \mathbf{S} \in \mathbb{R}^{n \times n}  \mid \; S_{ii}=1, \, \vert S_{ij} \vert \le 1, (i\ne j)\}. 
\end{equation}
Starting from $\widehat{\mathbf{S}}_0\coloneqq \mathbf{S}'$, we iterate the equal-weights averaged map
\begin{equation}
\label{eq:aap-update-mainbody}
\widehat{\mathbf{S}}_{t+1}
\;=\; \frac{1}{2}\Bigl( \mathsf{proj}_{\mathcal{K}_+^{\,n}}\bigl(\mathsf{sym}(\widehat{\mathbf{S}}_{t})\bigr) \;+\; \mathsf{proj}_{\mathcal{C}_{\mathrm{unit}}^{\,n}} \bigl(\mathsf{sym}(\widehat{\mathbf{S}}_{t})\bigr),
\qquad \mathsf{sym}(\mathbf{Y})\coloneqq \tfrac12(\mathbf{Y}+\mathbf{Y}^\top).
\end{equation}
This feasibility map depends only on the privatized matrix $\mathbf{S}'$ (and any auxiliary randomness used internally), hence it is post-processing and does not affect the privacy guarantee. 
The fast projection method is provided in Algorithm~\ref{alg:fast_projection-mainbody}.
%

\begin{algorithm}[H]
\small
\caption{Fast Alternating Projection onto $\mathcal{C}_{\mathsf{coll}}$}
\label{alg:fast_projection-mainbody}
\begin{algorithmic}[1]
  \State \textbf{Input:} $\mathbf{E} \in \mathbb{R}^{n \times d}$, $\varepsilon > 0$, $\delta \in (0, 1)$, sensitivity $\Delta>0$, tolerance $\tau>0$ \textbf{or} maximum iterations $T\in\mathbb{N}$
  \State \textbf{Output:} $\widehat{\mathbf{S}} \in \mathcal{C}_{\mathsf{coll}} \subseteq \mathbb{R}^{n \times n}$ with $(\varepsilon, \delta)$--DP guarantee \, s.t. \,
      $\bigl\|\widehat{\mathbf{S}} - \mathsf{proj}_{\mathcal{C}_{\mathsf{coll}}}(\mathbf{S} + \mathbf{W})\bigr\|_{\mathrm{F}}\le\tau$
\vspace{10pt}
  \State \textbf{Construct:} $\mathbf{S}\gets\mathbf{E}\mathbf{E}^\top$.
  \State \textbf{DP noise:} set $\sigma^2 \;\gets\; c_{\varepsilon, \delta} \Delta^2$ 
  \State Sample noise $\mathbf{W} \sim \mathcal{N}\left( \mathbf{0}, \sigma^2 \, \mathbf{I}_{n \times n }\right)$
  \State $\mathbf{S}' \gets \mathbf{S} + \frac{1}{2} \left( \mathbf{W} + \mathbf{W}^\top \right)$
  \State Initialize $\widehat{\mathbf{S}}_{(0)} \gets \mathbf{S}'$

  \For{$t=0$ \textbf{to} $T-1$}

      \State \textsf{/* symmetrize current iterate */}
      \State $\mathbf{Y}_t = \frac{1}{2} \big(\widehat{\mathbf{S}}_t  + \widehat{\mathbf{S}}_t^\top \big)$
      
      \State \textsf{/* projection onto PSD Cone $\mathcal{K}_{+}^{\,n}$ */}
      \State Compute eigen-decomposition $\mathbf{Y}_t = \mathbf{U} \mathsf{diag}(\lambda_1, \dots, \lambda_n) \mathbf{U}^{\top}$
      \State $\lambda^+_k \gets \max\{0,\lambda_k\}$, $\forall k \in [n]$
      \State $\mathbf{P}_t \gets \mathsf{proj}_{\mathcal{K}_{+}^{\,n}}(\mathbf{Y}_t) = \mathbf{U} \mathsf{diag}(\lambda^+_1, \dots, \lambda_n^+)\mathbf{U}^\top$

      \If{Frobenius-ball constraint is enforced and $\|\mathbf{P}_t\|_{\mathrm{F}} > n$}
        \State $\mathbf{P}_t \gets (n/\|\mathbf{P}_t\|_{\mathrm{F}})\,\mathbf{P}_t$  \hfill\quad\% radial projection onto $\mathcal{K}_+^n\cap\mathcal{B}_{\mathrm{F}}^n$
      \EndIf
      
      \State \textsf{/* projection onto Unit-Hyper-Cube $\displaystyle\mathcal{C}_{\mathrm{unit}}^{\,n}$*/} 
      \State  $\mathbf{Q}_{t} \gets \mathsf{proj}_{\mathcal{C}_{\mathrm{unit}}^{\,n}}(\mathbf{Y}_t)$  \hfill \quad\% $(\mathbf{Q}_t)_{ii}=1$;\ $(\mathbf{Q}_t)_{ij}=\mathrm{clip}((\mathbf{Y}_t)_{ij},-1,1)$ for $i\neq j$
      
      \State \textsf{/* averaged update */}
      \State $\widehat{\mathbf{S}}_{t+1} \gets \tfrac12\bigl(\mathbf{P}_t+\mathbf{Q}_t\bigr)$

      \State \textsf{/* stopping tests */}
      \State $r_{\mathrm{chg}}\gets \|\widehat{\mathbf{S}}_{t+1}-\widehat{\mathbf{S}}_t\|_{\mathrm{F}}$
      \State $r_{\mathrm{psd}}\gets \|\mathbf{Y}_t-\mathbf{P}_t\|_{\mathrm{F}}$;\quad $r_{\mathrm{box}}\gets \|\mathbf{Y}_t-\mathbf{Q}_t\|_{\mathrm{F}}$
      \If{$r_{\mathrm{chg}}\le\tau$ \textbf{and} $\max\{r_{\mathrm{psd}},r_{\mathrm{box}}\}\le\tau$}
         \State \textbf{break}
      \EndIf
  \EndFor
  
  \State \textsf{/* PSD on return */}
  \State $\mathbf{S}_{\mathrm{avg}} \gets \widehat{\mathbf{S}}_{t+1}$
  \State $\widehat{\mathbf{S}}\gets \mathsf{proj}_{\mathcal{K}_+^n}(\tfrac12(\mathbf{S}_{\mathrm{avg}}+\mathbf{S}_{\mathrm{avg}}^\top))$
         
  \State \textbf{if} $\min_i \widehat{S}_{ii}\le 0$ \textbf{then}  
       $\widehat{\mathbf{S}} \gets \widehat{\mathbf{S}} + \mu \, \mathbf{I}$ \text{ with small } $\mu>0$ \text{ (e.g.,} $\mu=10^{-8}\|\widehat{\mathbf{S}}\|_{\mathrm{F}}/n)$
  \State $\mathbf{D} \gets \mathsf{diag}(\, \widehat{\mathbf{S}} \,)^{1/2}$, \quad  
  $\widehat{\mathbf{S}}\gets \mathbf{D}^{-1}\,\widehat{\mathbf{S}}\,\mathbf{D}^{-1}$   

  \State \textbf{Output:} $\widehat{\mathbf{S}}$ 
\end{algorithmic}
\end{algorithm}

\noindent
\textbf{Projection onto the PSD Cone.}
Given a symmetric iterate $\mathbf{Y} \!= \! \mathbf{Y}^{\top} \! = \! \tfrac{1}{2} \bigl(\widehat{\mathbf{S}}_{t}+\widehat{\mathbf{S}}_{t}^{\top}\bigr) \! \in \! \mathtt{S}^{n}$, $\widehat{\mathbf{S}}_{0} \! \coloneqq \! \mathbf{S} \! +\! \mathbf{W}$, we compute its spectral decomposition $\mathbf{Y}= \mathbf{U}\,\mathsf{diag}(\lambda_{1},\dots,\lambda_{n})\mathbf{U}^{ \top}$, costing $\mathcal{O}(n^{3})$. All subsequent steps are performed in the eigen-basis $\mathbf{U}$.

The Frobenius–orthogonal projector onto the PSD cone solves $\mathop{\min}_{\mathbf{S} \succeq 0} {\Vert \mathbf{S} - \mathbf{Y} \Vert}_{\mathrm{F}}^2$ and is obtained as follows. Define $\lambda_{k}^{+}\coloneqq \max\{0, \lambda_{k}\},\; k = 1,\dots, n$, and let $\boldsymbol\lambda^{+}\coloneqq (\lambda_{1}^{+},\dots,\lambda_{n}^{+})$. The orthogonal projector onto the PSD cone is
\begin{equation}
\label{eq:PSD-proj-mainbody}
\mathsf{proj}_{\mathcal{K}_{+}^{\,n}}(\mathbf{Y}) =
\mathbf{U}\, \mathsf{diag}(\lambda_{1}^{+},\dots,\lambda_{n}^{+})\, \mathbf{U}^{\top}.
\end{equation}
This map is a firmly non-expansive projector (see Lemma~\ref{lem:proj-firm}).

\noindent
\textbf{Projection Onto the Unit-Diagonal Box.}
The set $ \mathcal{C}_{\mathrm{unit}}^{\,n}  \coloneqq \bigl\{\mathbf{S}\in \mathbb{R}^{n\times n}:\; |S_{ij}|\le 1\;(i\neq j), \; S_{ii}=1 \; (1\leq i \le n ) \bigr\}$ is an axis-aligned hyper-box with fixed diagonal. Because the Frobenius norm decouples over coordinates, the orthogonal projector is entry-wise. For any $\mathbf{Y}\in \mathbb{R}^{n\times n}$ the projector onto $\mathcal{C}_{\mathrm{unit}}^{n}$ decouples entry-wise as 
\begin{eqnarray}
\bigl[\mathsf{proj}_{\mathcal{C}_{\mathrm{unit}}^{\,n}}(\mathbf{Y})\bigr]_{ij}\;=\;
\begin{cases}
1, & i=j,\\[6pt]
\mathsf{clip}(Y_{ij},-1,1), & i\neq j,
\end{cases}
\end{eqnarray}
where $\mathsf{clip}(y,-1,1) \coloneqq \max\{-1,\min\{1,y\}\}$.
Note that for each off-diagonal coordinate the convex problem $\mathop{\min}_{|z|\leq 1} {\left( z - Y_{ij} \right)}^2$ yields the clip operator. Moreover, the diagonal constraint is enforced exactly. This operation is firmly non-expansive (see Lemma~\ref{lem:proj-firm}), costs $\mathcal{O}(n^2)$ arithmetic operations and can be implemented in-place (with $\Theta(n^2)$ storage for the matrix itself).

\paragraph{Risk Scaling: Na\"ive Gaussian vs. ScoreShield Mechanism}
Consider the ScoreShield projector $\widehat{\mathbf{S}} = \mathsf{proj}_{\mathcal{C}_{\mathsf{coll}}}(\mathbf{S} + \mathbf{W})$, where $\mathsf{proj}_{\mathcal{C}_{\mathsf{coll}}}$ is the Frobenius (metric) projection onto $\mathcal{C}_{\mathsf{coll}}$. Then
\begin{equation}
\mathbb{E} \, \|\widehat{\mathbf{S}} - \mathbf{S} \|_{\mathrm{F}}^{2}
\; \le \; 4 \, \sigma \, \mathsf{GC}(\mathcal{C}_{\mathsf{coll}}),
\end{equation}
where $\mathsf{GC}(\mathcal{A}) = \mathbb{E}\big[\sup_{\mathbf{A}\in\mathcal{A}}\langle \mathbf{Z},\mathbf{A}\rangle_{\mathrm{F}}\big]$ (see Corollary~\ref{cor:global-gc-scoreshield-matrix} for more details).

Under record-level adjacency (\textsc{R}) the exact Frobenius sensitivity is $\Delta_{f,F} = 2\sqrt{2(n-1)}=\Theta(\sqrt n)$, hence $\sigma^2=\Theta(n)$.
Under output-space (Gram) adjacency (\textsc{O}) we have $\|\mathbf{S}-\mathbf{S}'\|_F \le \mathsf{\Delta}_{\mathsf{G}}$ with $\mathsf{\Delta}_{\mathsf{G}} = \Theta(1)$ independent of $n$, hence $\sigma^2 = \Theta(1)$.

\noindent
\textbf{Na\"ive Gaussian Mechanism.}
In our practical algorithm, we sample $\mathbf{W}$ with i.i.d. entries $W_{ij}\sim\mathcal{N}(0,\sigma^2)$ (Gaussian mechanism), and then apply the deterministic symmetrization post-processing $\mathbf{G}\coloneqq \tfrac12(\mathbf{W}+\mathbf{W}^\top)$. Since $\mathbf{S}$ is symmetric, the symmetrized release $\mathbf{S}' \coloneqq \tfrac12 \big( (\mathbf{S}+\mathbf{W})+ (\mathbf{S} +\mathbf{W})^\top \big) =\mathbf{S} + \mathbf{G}$ is a post-processing of $\mathbf{S}+\mathbf{W}$ and therefore preserves $(\varepsilon, \delta)$-DP\footnote{Equivalently, sampling $\mathbf{G}$ directly as a symmetric Gaussian with $\mathrm{Var}(G_{ii}) = \sigma^2$ and $\mathrm{Var}(G_{ij}) = \sigma^2/2, \forall i \neq j$, yields the same distribution as $\frac{1}{2} (\mathbf{W} + \mathbf{W}^\top)$.} (with $\sigma$ calibrated to the sensitivity of $\mathbf{S}$ under the chosen adjacency). We report utility for this symmetric pre-projection matrix $\mathbf{S}'$.
A direct variance calculation gives
\begin{equation}\label{eq:naive-frob}
\mathbb{E}\,\| \mathbf{S}' -\mathbf{S}\|_F^2
=\mathbb{E}\,\|\mathbf{G}\|_F^2
=\underbrace{n\sigma^2}_{\text{diagonal}}
+\underbrace{n(n-1) \frac{\sigma^2}{2}}_{\text{off-diagonal}}
= \frac{n^2+n}{2}\,\sigma^2 
= \Theta(n^2\sigma^2).
\end{equation}
Therefore, for the two adjacency definitions we have:
\begin{enumerate}[leftmargin=2em]
\item[(a)] Record-level adjacency (\textsc{R}): Using $\sigma^2=c_{\varepsilon,\delta}\Delta_{f,F}^2=\Theta \big(\frac{n\log(2/\delta)}{\varepsilon^2}\big)$,\vspace{-5pt}
\begin{equation}
\makebox[0pt][c]{\tcboxmath[
  enhanced,
  colback=gray!8,
  colframe=gray!40,
  boxrule=0.5pt,
  arc=2pt,
  boxsep=0pt,
  left=6pt,right=6pt,top=4pt,bottom=4pt,
  grow to left by=10mm,
  grow to right by=10mm
]{%
  \displaystyle
  \text{na\"ive + (\textsc{R})}:\qquad
  \mathbb{E}\,\| \mathbf{S}' - \mathbf{S} \|_{\mathrm{F}}^2
  = \Theta \Big( n^2 \, c_{\varepsilon,\delta}\, \Delta_{F,\mathrm{rec}}^2 \Big)
  = \Theta \Big( \frac{n^{3}\log(2/\delta)}{\varepsilon^2} \Big).
}}
\end{equation}
\item[(b)] Output-space adjacency (\textsc{O}): With $\sigma^2=c_{\varepsilon,\delta}\mathsf{\Delta}_{\mathsf{G}}^2$, 
\begin{equation}
\makebox[0pt][c]{\tcboxmath[
  enhanced,
  colback=gray!8,
  colframe=gray!40,
  boxrule=0.5pt,
  arc=2pt,
  boxsep=0pt,
  left=6pt,right=6pt,top=4pt,bottom=4pt,
  grow to left by=10mm,
  grow to right by=10mm
]{%
  \displaystyle
    \text{na\"ive + (\textsc{O})}:\qquad
    \mathbb{E} \,\| \mathbf{S}' - \mathbf{S} \|_{\mathrm{F}}^2
    = \Theta \Big( n^2\,c_{\varepsilon,\delta}\, \mathsf{\Delta}_{\mathsf{G}}^2 \Big)
    = \Theta \Big( \frac{n^{2} \mathsf{\Delta}_{\mathsf{G}}^2\log(2/\delta)}{\varepsilon^2} \Big).
}}
\end{equation}
\end{enumerate}

\noindent
\textit{Global Risk Bound via the Gaussian Complexity.}
Let $\widehat{\mathbf{S}}=\mathsf{proj}_{\mathcal{C}_{\mathsf{coll}}}(\mathbf{S}+\mathbf{G})$ for the regime (ii) Algorithm~\ref{alg:all-pairs-regime2}. Using Corollary~\ref{cor:global-gc-scoreshield-matrix} we have
\begin{equation}\label{eq:global-GC}
\mathbb{E}\|\widehat{\mathbf{S}}-\mathbf{S}\|_F^2
\;\le\; C\; \sigma \; \mathsf{GC}(\mathcal{C}_{\, \mathsf{coll}}\,)
\;\le\; \widetilde{C} \, \sigma \,  n^{3/2}\, ,
\end{equation}
where we used $\mathsf{GC}(\mathcal{C}_{\mathsf{coll}})=\Theta(n^{3/2})$.

\begin{enumerate}[leftmargin=2em]
\item[(a)] Record-level adjacency (\textsc{R}): With $\sigma = \Theta \big( \frac{ \sqrt{n\log(2/\delta)}}{\varepsilon}\big)$,\vspace{-5pt}
\begin{equation}
\makebox[0pt][c]{\tcboxmath[
  enhanced,
  colback=gray!8,
  colframe=gray!40,
  boxrule=0.5pt,
  arc=2pt,
  boxsep=0pt,
  left=6pt,right=6pt,top=4pt,bottom=4pt,
  grow to left by=10mm,
  grow to right by=10mm
]{%
  \displaystyle
    \text{ScoreShield + (\textsc{R})}:\quad
    \mathbb{E}\|\widehat{\mathbf{S}}-\mathbf{S}\|_F^2
    \le \widetilde{C}\,\frac{n^{2} \sqrt{\log(2/\delta)}}{\varepsilon}
    = \mathcal{O} \left( \frac{n^{2} \sqrt{\log(2/\delta)}}{\varepsilon} \right).
}}
\end{equation} 
\item[(b)] Output-space adjacency (\textsc{O}): With $\sigma = \frac{\sqrt{2\log(2/\delta)}}{\varepsilon} \mathsf{\Delta}_{\mathsf{G}}$,\vspace{-5pt}
\begin{eqnarray}
\makebox[0pt][c]{\tcboxmath[
  enhanced,
  colback=gray!8,
  colframe=gray!40,
  boxrule=0.5pt,
  arc=2pt,
  boxsep=0pt,
  left=6pt,right=6pt,top=4pt,bottom=4pt,
  grow to left by=10mm,
  grow to right by=10mm
]{%
  \displaystyle
    \text{ScoreShield + (\textsc{O})}: \quad
    \mathbb{E} \| \widehat{\mathbf{S}} - \mathbf{S} \|_F^2
    \le  \widetilde{C} \, \frac{n^{3/2} \mathsf{\Delta}_{\mathsf{G}} \sqrt{\log(2/\delta)}}{\varepsilon}
     =  \mathcal{O} \left( \frac{n^{3/2} \mathsf{\Delta}_{\mathsf{G}} \sqrt{\log(2/\delta)}}{\varepsilon} \right).
}}
\end{eqnarray}
\end{enumerate}

We refer the reader to Appendix~\ref{subsec:regime2-compare} for extended discussion.

\noindent
\textbf{Adversary Gain.}
We analyze attacker reconstruction under two knowledge regimes: (i) no side information and (ii) side information where the attacker knows the gallery embeddings $\{\mathbf{e}_j\}_{j\neq i}$ (one-row unknown).
Extended discussion and additional results are provided in Appendix~\ref{ssec:attacker_regime2}.

\vspace{-4pt}

\section{Experiments}
\label{sec:experiments}

\vspace{-4pt}

We evaluate \textsc{ScoreShield} in the two release regimes studied in the paper. In regime~(i), the mechanism releases a single privatized score vector. We instantiate this regime in face recognition and in single-query retrieval-augmented generation. In regime~(ii), the mechanism releases a single privatized cosine Gram matrix. We evaluate downstream tasks that consume only the released similarity matrix or deterministic transformations of it. Across both regimes, we compare the utility of the non-private release, the Gaussian mechanism applied directly to the score object, and the proposed perturb-then-project release. Full experimental details, extended grids, and ablations are deferred to App.~\ref{app:sec:supplementary-regime1-experiments-DP-RAG}, App.~\ref{app:sec:supplementary-regime1-experiments-DP-FR}, and App.~\ref{sec:supplementary-regime2-experiments}.

For regime~(i), all guarantees are for a single released score vector under central-model record-level replacement. The face-recognition study uses three representative LFW score sets for operating-point analysis and seven public FR benchmarks for aggregate evaluation. The DP-RAG study uses the \textsc{FRAMES} benchmark, EG300M retrieval embeddings, and Gemma-family generators. For regime~(ii), we evaluate CIFAR-10/100, Oxford-IIIT Pets, STS-B, and MovieLens-100K. In every matrix benchmark we compare the clean (non-private) Gram matrix $\mathbf{S}$, the symmetrized noisy release $\mathbf{S}'$, and the projected release $\widehat{\mathbf{S}}$; an SDP projector is included only where it is computationally tractable.

\vspace{-2pt}

\subsection{Regime (i): Similarity Score Vector Release}
\label{sec:exp-regime1}

\vspace{-2pt}

\begin{figure*}[t]
  \centering

  \begin{subfigure}[t]{0.32\textwidth}
    \includegraphics[width=\linewidth]
    {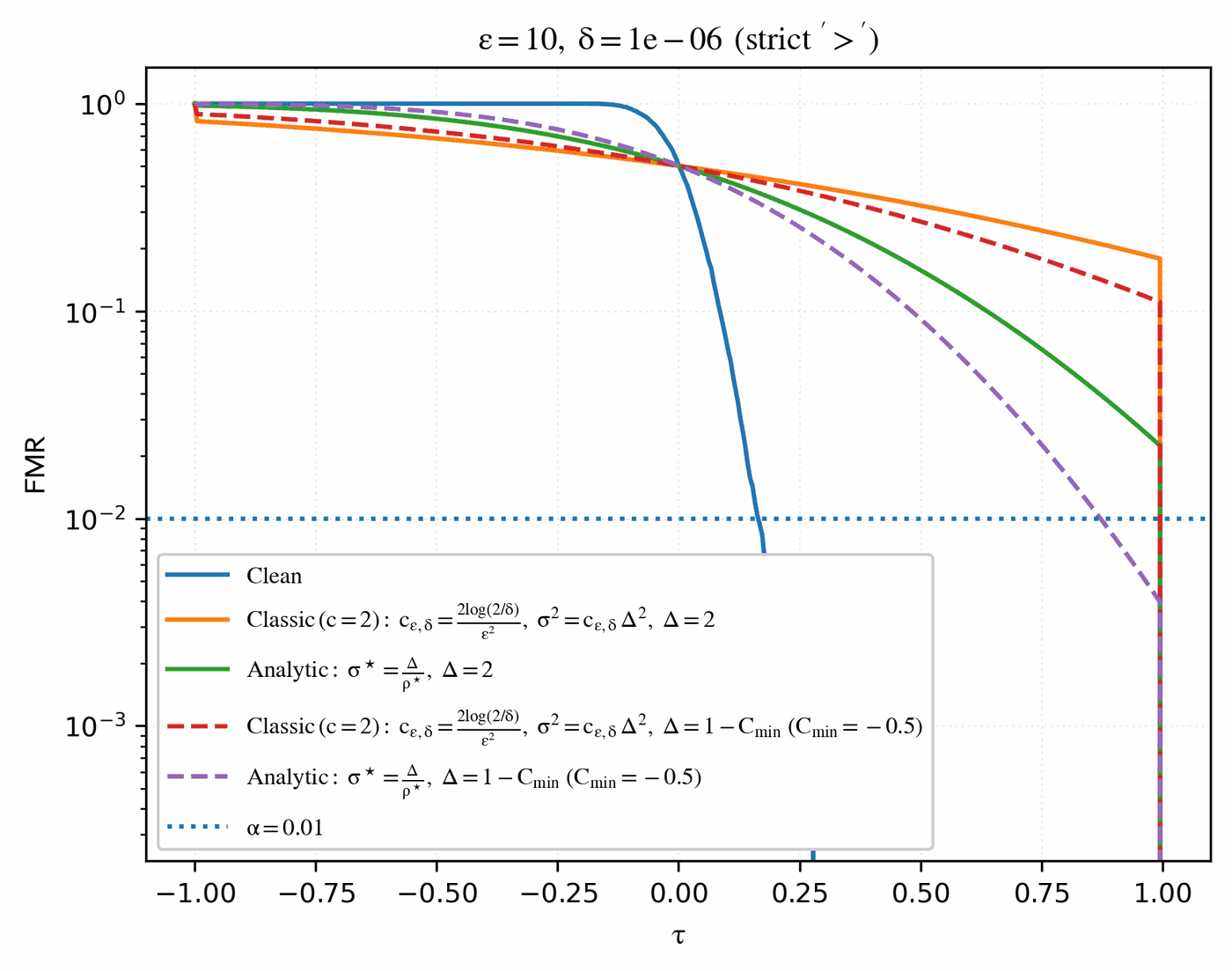}
    \caption{FMR curves at $\varepsilon=10$}
  \end{subfigure}
  \hfill
  \begin{subfigure}[t]{0.32\textwidth}
    \includegraphics[width=\linewidth]
    {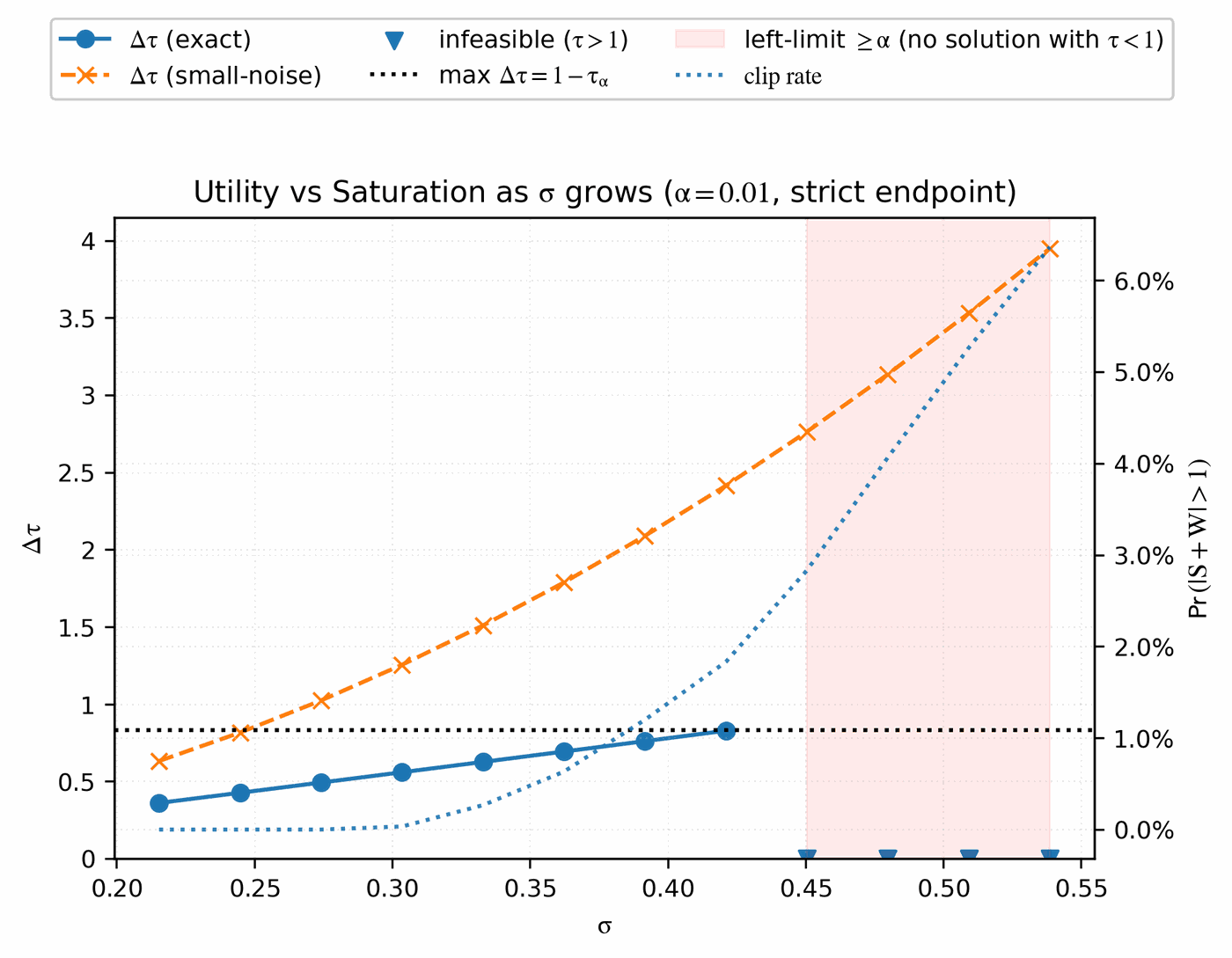}
    \caption{$\Delta\tau$ versus $\sigma$}
  \end{subfigure}
  \hfill
  \begin{subfigure}[t]{0.32\textwidth}
    \includegraphics[width=\linewidth]
    {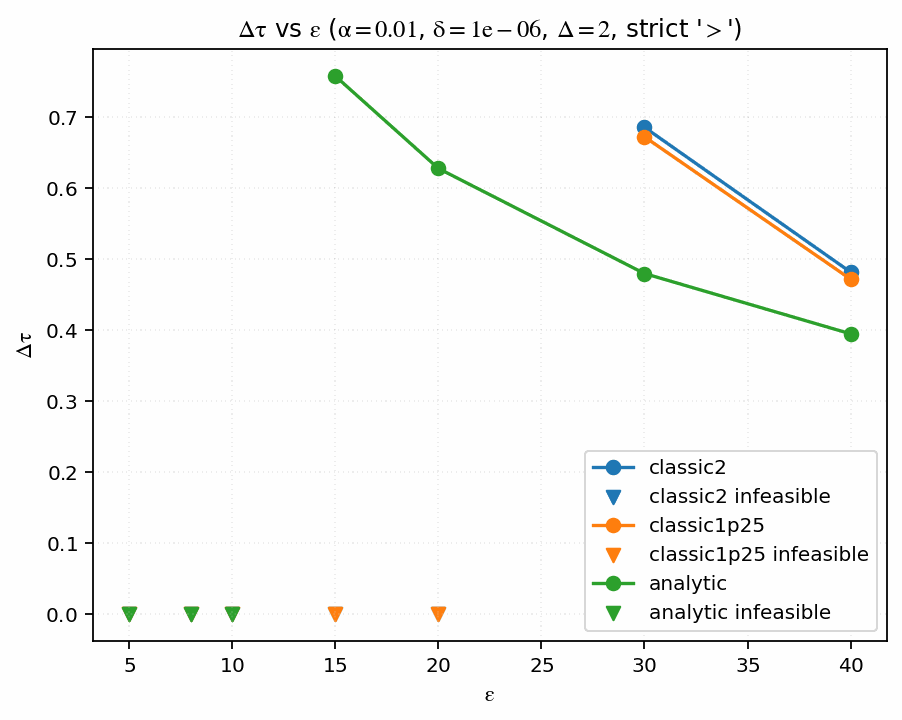}
    \caption{$\Delta\tau$ versus $\varepsilon$}
  \end{subfigure}
\caption{
\textbf{DP-FR operating-point behavior under score-vector release.}
All panels use LFW ArcFace-101/WebFace4M scores with target FMR $\alpha=10^{-2}$ under strict endpoint semantics.
(a) Post-privacy FMR curves at $\delta=10^{-6}$ and $\varepsilon=10$, comparing conservative and analytic Gaussian calibration under worst-case sensitivity $\Delta=2$, and the public-margin model $\Delta=1-c_{\min}=1.5$ with $c_{\min}=-0.5$.
(b) Threshold correction $\Delta\tau(\sigma)$ versus Gaussian noise scale; shading marks strict-endpoint infeasibility, where no threshold $\tau<1$ attains the target FMR.
(c) The corresponding $\Delta\tau$--$\varepsilon$ curve for analytic Gaussian calibration with $\Delta=2$ and $\delta=10^{-6}$
; filled markers denote feasible thresholds and inverted markers infeasible budgets.
}
  \label{fig:dpfr-main}
\end{figure*}

\paragraph{Regime~(i): DP-FR.}
For face recognition, average score distortion does not by itself determine deployment utility.  The relevant question is whether a target false-match operating point remains attainable after privatization.  We therefore evaluate utility at the decision level.  We use seven public benchmarks: LFW, CFP-FP, CALFW, CPLFW, AgeDB, IJB-B, and IJB-C.  For LFW, CFP-FP, CALFW, CPLFW, and AgeDB, we report verification accuracy.  For IJB-B and IJB-C, we report TAR at $\mathrm{FPR} \in\{10^{-6},10^{-5},10^{-4}\}$.  We evaluate WebFace4M-trained IR101 and ViT-Base backbones.  Full benchmark tables, calibration sweeps, synthetic score experiments, and operating-point analyses are reported in App.~\ref{app:sec:supplementary-regime1-experiments-DP-FR}.
Figure~\ref{fig:dpfr-main} illustrates how privacy noise affects the verification operating point.
At a fixed privacy budget, analytic Gaussian calibration injects less noise than the conservative sufficient calibration, and therefore gives a lower post-privacy FMR near the target threshold.
The public-margin curves show the possible gain under the condition $c_{\min}=-0.5$; the worst-case formal calibration remains $\Delta=2$.
Panels~(b)--(c) show the same effect through the required threshold correction: as the noise scale increases, the target FMR may become infeasible under strict endpoint semantics.

\begin{table*}[h]
\centering
\scriptsize
\setlength{\tabcolsep}{3.2pt}
\renewcommand{\arraystretch}{0.98}
\caption{
\textbf{Representative DP face-recognition results at \(\delta=10^{-4}\).}
For LFW we report verification accuracy. B-\(10^{-6}\), B-\(10^{-5}\), and
B-\(10^{-4}\) denote TAR on IJB-B at the corresponding FPR; C-\(10^{-6}\),
C-\(10^{-5}\), and C-\(10^{-4}\) are defined analogously for IJB-C.
``Avg.'' is the macro-average over IJB-B and IJB-C TAR at
\(\mathrm{FPR}\in\{10^{-6},10^{-5},10^{-4}\}\), and verification accuracy on
LFW, AgeDB, CFP-FP, CALFW, and CPLFW. The clean row is non-private. Entries
equal to \(0.00\) in the IJB columns indicate that the requested operating point
is not attained under the strict endpoint convention used in the appendix.
Full grids over \(\varepsilon\), \(\delta\), backbones, and benchmarks are
reported in App.~\ref{app:sec:supplementary-regime1-experiments-DP-FR}.
}
\label{tab:dpfr-main}
\resizebox{0.9\textwidth}{!}{
\begin{tabular}{llcccccccc}
\toprule
Backbone
& \((\varepsilon,\delta)\)
& LFW \(\uparrow\)
& B-\(10^{-6}\) \(\uparrow\)
& B-\(10^{-5}\) \(\uparrow\)
& B-\(10^{-4}\) \(\uparrow\)
& C-\(10^{-6}\) \(\uparrow\)
& C-\(10^{-5}\) \(\uparrow\)
& C-\(10^{-4}\) \(\uparrow\)
& Avg. \(\uparrow\) \\
\midrule
\multicolumn{10}{l}{\texttt{IR101/WebFace4M}}\\
IR101 & clean
& 99.70 & 89.46 & 93.07 & 95.52 & 43.49 & 89.07 & 93.72 & 89.74 \\
IR101 & \((100,10^{-4})\)
& 92.05 & 73.18 & 83.68 & 89.87 & 29.36 & 76.48 & 86.85 & 78.23 \\
IR101 & \((70,10^{-4})\)
& 86.72 & 53.82 & 66.66 & 78.60 & 20.35 & 60.17 & 75.10 & 67.81 \\
IR101 & \((35,10^{-4})\)
& 71.52 & 0.00 & 0.00 & 18.01 & 0.00 & 0.00 & 17.60 & 33.43 \\
IR101 & \((20,10^{-4})\)
& 61.67 & 0.00 & 0.00 & 0.00 & 0.00 & 0.00 & 0.00 & 26.69 \\
\midrule
\multicolumn{10}{l}{\texttt{ViT-Base/WebFace4M}}\\
ViT-Base & clean
& 99.80 & 87.12 & 94.54 & 96.89 & 38.62 & 90.51 & 95.39 & 90.01 \\
ViT-Base & \((100,10^{-4})\)
& 93.43 & 74.98 & 86.14 & 91.97 & 29.72 & 78.45 & 88.94 & 80.34 \\
ViT-Base & \((70,10^{-4})\)
& 87.07 & 52.67 & 67.86 & 80.93 & 35.75 & 60.49 & 77.05 & 70.72 \\
ViT-Base & \((35,10^{-4})\)
& 71.82 & 0.00 & 0.00 & 18.26 & 0.00 & 0.00 & 17.05 & 34.10 \\
ViT-Base & \((20,10^{-4})\)
& 62.30 & 0.00 & 0.00 & 0.00 & 0.00 & 0.00 & 0.00 & 27.05 \\
\bottomrule
\end{tabular}
}
\end{table*}

Table~\ref{tab:dpfr-main} reports a fixed-$\delta$ slice of the full DP-FR benchmark grid.  At $\delta=10^{-4}$, utility improves as $\varepsilon$ increases, but the recovery depends strongly on the operating point. For IR101, increasing $\varepsilon$ from $70$ to $100$ raises LFW accuracy from $86.72\%$ to $92.05\%$, while IJB-B TAR at FPR $10^{-6}$ increases from $53.82$ to $73.18$.  At the less stringent IJB-B FPR $10^{-4}$, the same change increases TAR from $78.60$ to $89.87$.  The ViT-Base rows show the same qualitative pattern. At $\varepsilon\in\{20,35\}$, the strict $10^{-6}$ and $10^{-5}$ IJB operating points are not attained in several cases under the strict endpoint convention, although LFW accuracy remains nonzero. The added Gaussian noise first affects
the most stringent low-FPR decisions, whereas verification accuracy on LFW, CFP-FP, CALFW, CPLFW, and AgeDB, and less stringent IJB thresholds, degrade more gradually.

\paragraph{Regime~(i): DP-RAG.}
We evaluate score-vector release as a retrieval primitive on \textsc{FRAMES}. For each query, the retriever computes cosine scores over a Wikipedia chunk index. Clean RAG applies thresholded top-$k$ retrieval to $\mathbf{s}(\mathbf{q})$, whereas DP-RAG applies the same rule to $\widehat{\mathbf{s}}(\mathbf{q})$. The generator, prompt format, decoding parameters, and LLM judge are fixed across all conditions.
The privacy guarantee applies only to the released score vector and its post-processing.
Here the source corpus is public Wikipedia text, so retrieved identifiers, thresholded retrieval sets, and answers generated from retrieved text are treated as post-processing of the privatized scores.
If the corpus text were private under the same adjacency relation, releasing chunks or generated answers would require an additional privacy mechanism.

\begin{table*}[h]
\centering
\scriptsize
\setlength{\tabcolsep}{4.0pt}
\renewcommand{\arraystretch}{0.98}
\caption{\textbf{Representative small $\delta$ DP-RAG results on \textsc{FRAMES}.}
``Base'' denotes generation without retrieved context. ``Full'' uses the linked Wikipedia pages and serves as an evidence upper bound. ``RAG'' uses thresholded top-$k$ retrieval. Clean rows retrieve from $\mathbf{s}(\mathbf{q})$, while DP rows retrieve from $\widehat{\mathbf{s}}(\mathbf{q})$.
``Gain'' is RAG accuracy minus the corresponding no-context baseline. 
The protocol is shown in App.~Fig.~\ref{fig:dprag-experimental-protocol}; full grids are reported in App.~\ref{app:sec:supplementary-regime1-experiments-DP-RAG}.
}
\label{tab:dprag-main}
\resizebox{0.9\textwidth}{!}{
\begin{tabular}{llccccccl}
\toprule
Generator & Embedding & Scores & \((\varepsilon,\delta)\) & \((\tau,k)\)
& Base Acc. & Full Acc. & RAG Acc. & Gain \\
\midrule
Gemma3-12B & EG300M & clean & -- & \((0.25,50)\)
& 45.87 & 72.09 & 55.46 & \(+9.59\) \\
Gemma3-12B & EG300M & DP & \((1,10^{-5})\) & \((0.25,50)\)
& 46.36 & 71.48 & 39.44 & \(-6.92\) \\
\midrule
Gemma3-12B & Q3VL-E2B & clean & -- & \((0.25,10)\)
& 46.36 & 71.48 & 56.92 & \(+10.56\) \\
Gemma3-12B & Q3VL-E2B & DP & \((1,10^{-6})\) & \((0.25,10)\)
& 46.36 & 71.48 & 42.23 & \(-4.13\) \\
Gemma3-12B & Q3VL-E2B & DP & \((10,10^{-6})\) & \((0.25,10)\)
& 46.36 & 71.48 & 41.75 & \(-4.61\) \\
\midrule
Gemma3-27B & EG300M & clean & -- & \((0.35,20)\)
& 40.17 & 72.82 & 60.92 & \(+20.75\) \\
Gemma3-27B & EG300M & DP & \((1,10^{-5})\) & \((0.35,20)\)
& 40.17 & 72.82 & 39.68 & \(-0.49\) \\
Gemma3-27B & EG300M & DP & \((1,10^{-6})\) & \((0.35,20)\)
& 40.17 & 72.82 & 37.86 & \(-2.31\) \\
\midrule
Gemma3-8B & Q3VL-E2B & clean & -- & \((0.25,10)\)
& 75.85 & 91.02 & 80.58 & \(+4.73\) \\
Gemma3-8B & Q3VL-E2B & DP & \((1,10^{-5})\) & \((0.25,10)\)
& 75.85 & 91.02 & 74.27 & \(-1.58\) \\
Gemma3-8B & Q3VL-E2B & DP & \((10,10^{-6})\) & \((0.25,10)\)
& 75.85 & 91.02 & 76.46 & \(+0.61\) \\
\bottomrule
\end{tabular}
}
\end{table*}

Table~\ref{tab:dprag-main} reports representative small-$\delta$ DP-RAG settings on \textsc{FRAMES}.
Clean retrieval improves accuracy in all displayed settings, with gains from $4.73$ to $20.75$ points over the no-context baseline.
Under DP score release, accuracy depends on whether the privatized scores preserve the thresholded top-$k$ retrieval set.
The $k=50$ G3-12B/EG300M row shows that larger retrieval depth alone does not make RAG robust to score perturbation.
The G3-27B/EG300M rows remain close to the no-context baseline at $\varepsilon=1$, while Q3-8B/Q3VL-E2B gives the strongest displayed small-$\delta$ result, slightly exceeding the baseline at $(\varepsilon,\delta)=(10,10^{-6})$.

\subsection{Regime (ii): Similarity Score Matrix Release}
\label{sec:exp-regime2}

Given normalized embeddings $\mathbf{E}\in\mathbb{R}^{n\times d}$, the non-private score object is the cosine Gram matrix $\mathbf{S}=\mathbf{E}\mathbf{E}^{\top}$. We compare the clean matrix $\mathbf{S}$, the symmetrized Gaussian perturbation $\mathbf{S}'$, and the projected release $\widehat{\mathbf{S}}$.  The projected matrix is obtained by projecting the perturbed matrix onto the cosine-Gram feasible set $\mathcal{C}_{\mathsf{coll}}$. Since this projection is post-processing of the noisy release, it preserves the same $(\varepsilon,\delta)$-DP guarantee.
For every benchmark, the downstream method receives only $\mathbf{M}\in\{\mathbf{S},\mathbf{S}',\widehat{\mathbf{S}}\}$, or a fixed deterministic transformation of $\mathbf{M}$. The encoder, sample, split, graph construction, prediction rule, and evaluation metric are fixed within each experiment. Thus, utility differences are attributable to the matrix supplied to the downstream method.

\vspace{-4pt}

\paragraph{Tasks.}
We evaluate \textsc{ScoreShield} on three classes of matrix-based downstream tasks. For image similarity, we use frozen DINOv2-B/14 embeddings and form cosine Gram matrices on CIFAR-10/100 and Oxford-IIIT Pets. Pairwise verification uses $M_{ij}$ as the score for the label $\mathbbm{1}\{y_i=y_j\}$ and reports ROC--AUC. Nearest-neighbor classification assigns each image the majority label among the $5$ largest off-diagonal entries in its row of $\mathbf{M}$ and reports top-1 accuracy. Instance retrieval ranks candidates $j\neq i$ by decreasing $M_{ij}$ and reports Recall@1. Spectral clustering forms the affinity $\mathbf{A}=(\mathbf{M}+\mathbf{1})/2$ with zero diagonal and reports NMI against the ground-truth labels.  For semantic textual similarity, we embed the unique STS-B sentences with SBERT; each labeled pair $(u,v)$ is scored by the released entry $M_{uv}$, and utility is Spearman correlation with human similarity scores.  For recommendation systems, we form a MovieLens-100K user--user cosine Gram matrix from mean-centered rating vectors and predict held-out ratings by a similarity-weighted neighborhood estimator; utility is RMSE.

\vspace{-4pt}

\begin{figure*}[t]
  \centering

  \begin{subfigure}[t]{0.32\textwidth}
    \includegraphics[
      trim={20pt 25pt 0pt 80pt},
      clip,
      width=\linewidth
    ]{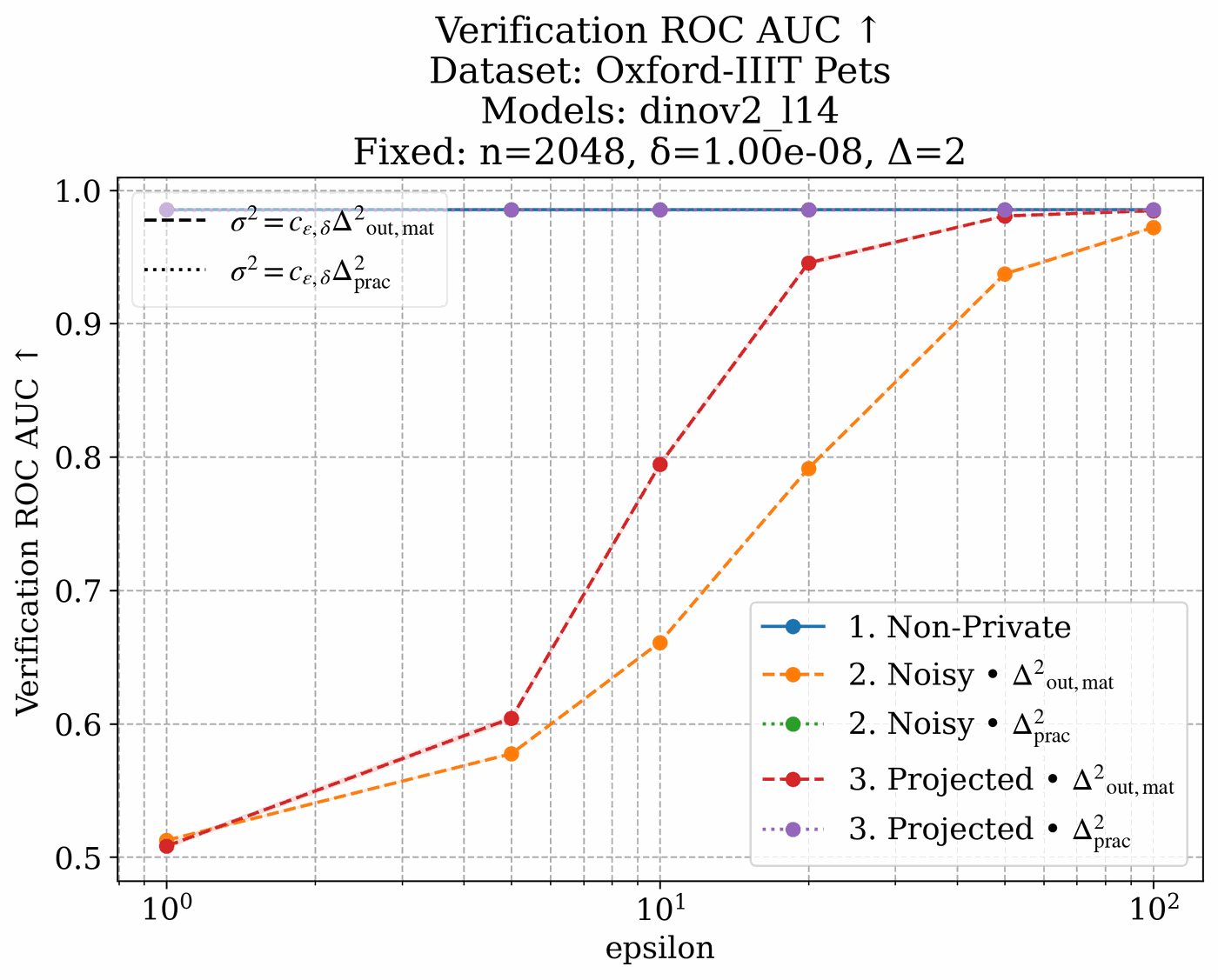}
    \caption{Pets verification AUC}
  \end{subfigure}
  \hfill
  \begin{subfigure}[t]{0.32\textwidth}
    \includegraphics[
      trim={20pt 25pt 0pt 80pt},
      clip,
      width=\linewidth
    ]{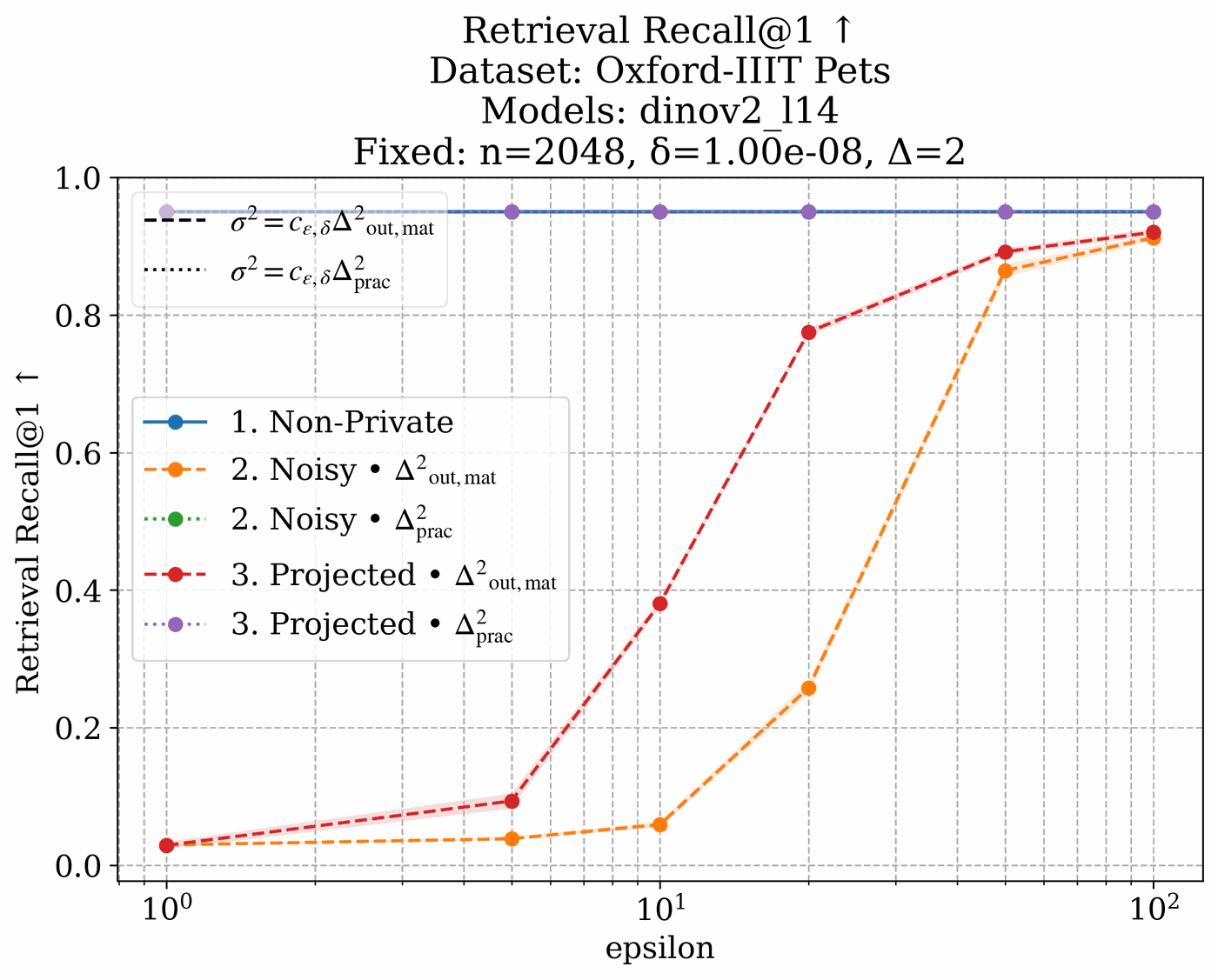}
    \caption{Pets Recall@1}
  \end{subfigure}
  \hfill
  \begin{subfigure}[t]{0.32\textwidth}
    \includegraphics[
      trim={20pt 25pt 0pt 80pt},
      clip,
      width=\linewidth
    ]{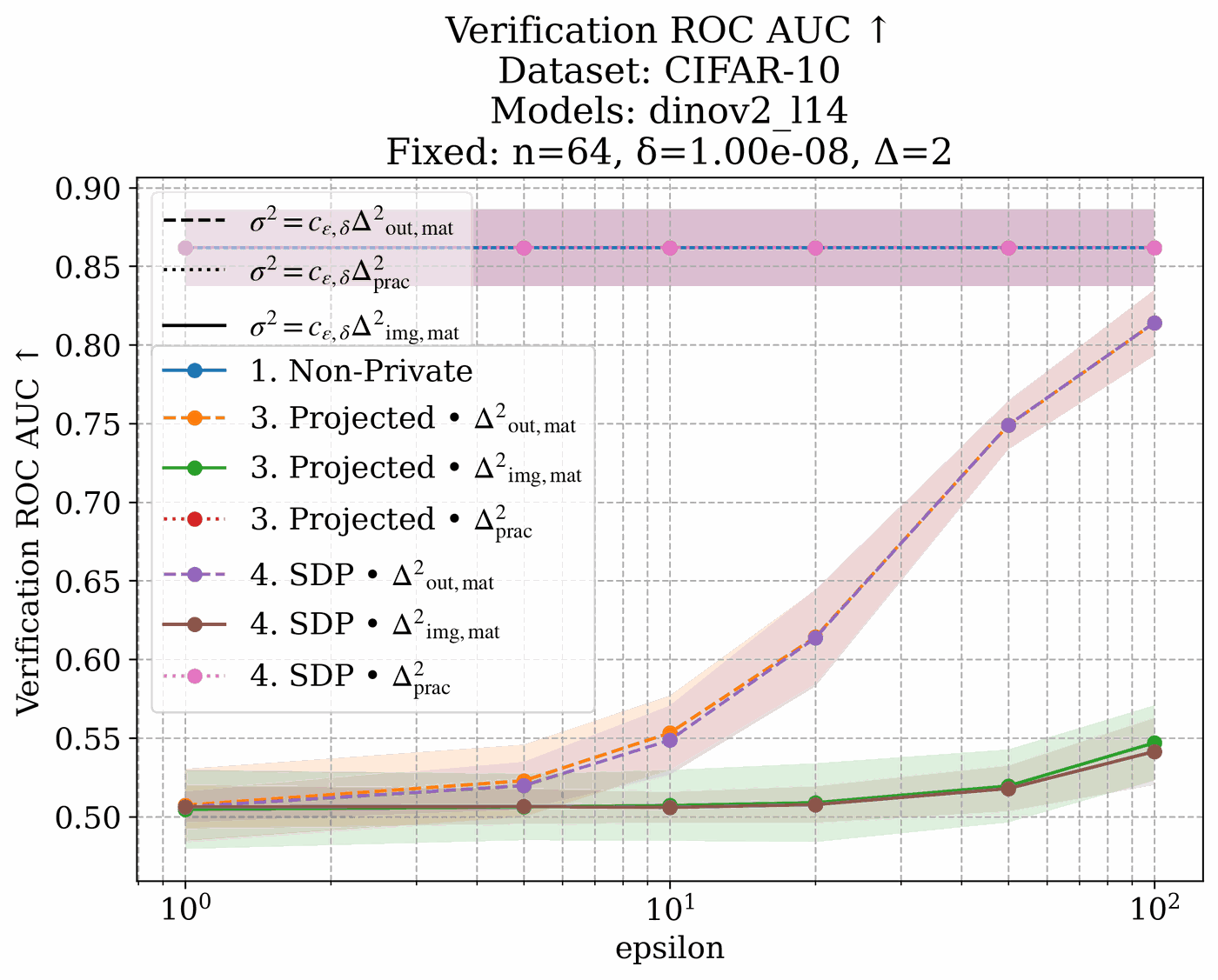}
    \caption{CIFAR-10 verification AUC}
  \end{subfigure}

  \vspace{0.5em}

  \begin{subfigure}[t]{0.32\textwidth}
    \includegraphics[
      trim={20pt 25pt 0pt 80pt},
      clip,
      width=\linewidth
    ]{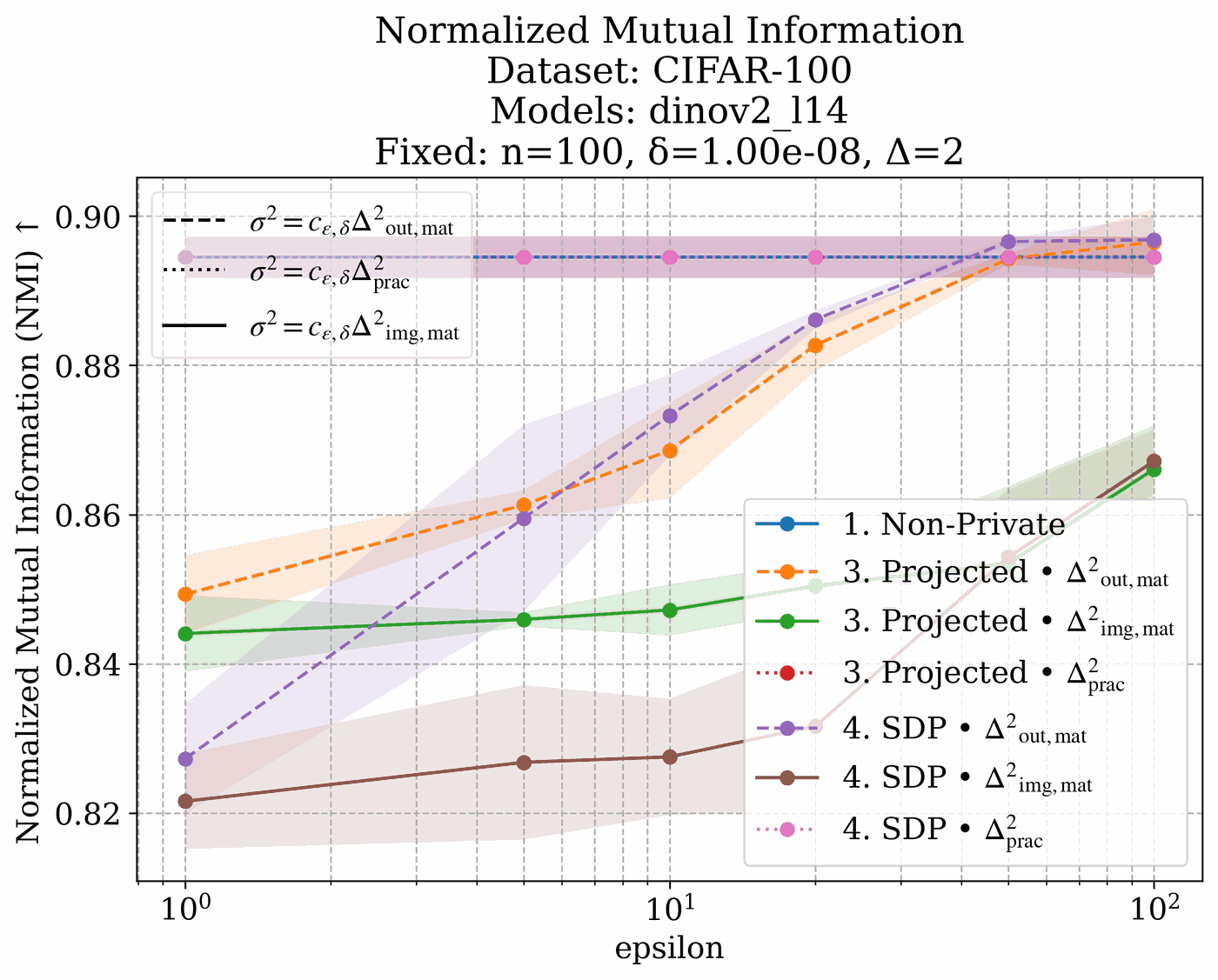}
    \caption{CIFAR-100 clustering NMI}
  \end{subfigure}
  \hfill
  \begin{subfigure}[t]{0.32\textwidth}
    \includegraphics[
      trim={20pt 25pt 0pt 80pt},
      clip,
      width=\linewidth
    ]{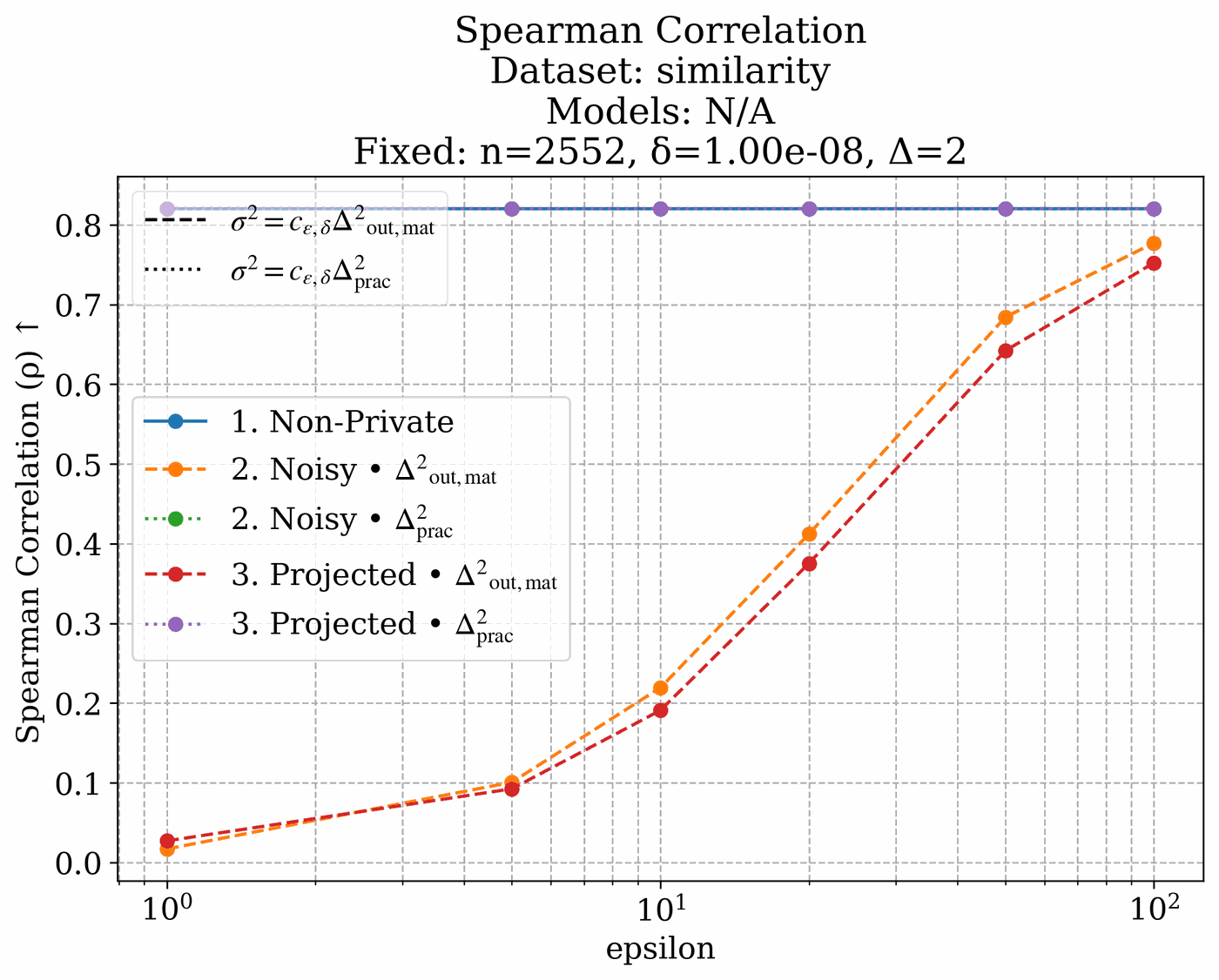}
    \caption{STS-B Spearman $\rho$}
  \end{subfigure}
  \hfill
  \begin{subfigure}[t]{0.32\textwidth}
    \includegraphics[
      trim={20pt 25pt 0pt 80pt},
      clip,
      width=\linewidth
    ]{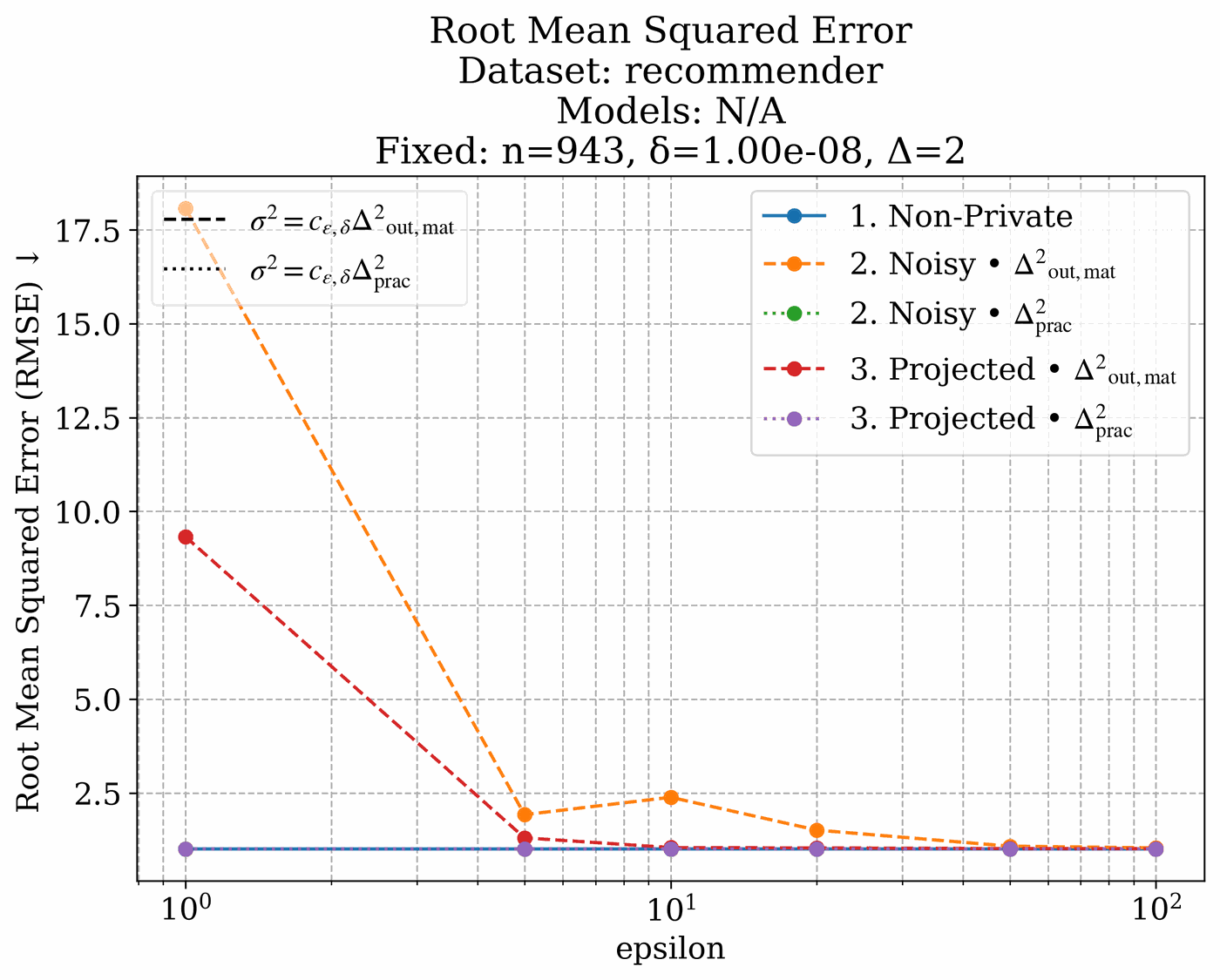}
    \caption{MovieLens RMSE}
  \end{subfigure}
\vspace{-2pt}
  \caption{
  \textbf{Regime~(ii): representative utilities for DP cosine-Gram release.}
  Each downstream task uses only the released matrix
  $\mathbf{M}\in\{\mathbf{S},\mathbf{S}',\widehat{\mathbf{S}}\}$ or a fixed
  deterministic transformation of it. The panels show representative results
  for image verification, image retrieval, spectral clustering, semantic
  textual similarity, and collaborative filtering. All displayed panels use
  $\delta=10^{-8}$ and $\Delta=2$, the main setting used in the paper;
  complete grids over $\Delta$, $\delta$, datasets, and metrics are reported in
  App.~\ref{sec:supplementary-regime2-experiments}. Higher is better for AUC,
  Recall@1, NMI, and Spearman correlation; lower is better for RMSE.
  Projection improves the raw Gaussian release by restoring the PSD,
  unit-diagonal, and entrywise cosine constraints required of a valid cosine
  Gram matrix.
  }
  \vspace{-6pt}
  \label{fig:regime2-main}
\end{figure*}

\paragraph{Representative results.}
Figure~\ref{fig:regime2-main} reports representative results at $\delta=10^{-8}$ and $\Delta=2$, the main calibration setting used for the displayed panels. The panels cover image verification on Oxford-IIIT Pets and CIFAR-10, image retrieval on Oxford-IIIT Pets, spectral clustering on CIFAR-100, semantic textual similarity on STS-B, and collaborative filtering on MovieLens-100K.  Full extended results over $\Delta$, $\delta$, datasets, and additional metrics are reported in App.~\ref{sec:supplementary-regime2-experiments}.
Gaussian perturbation alone can reduce utility because $\mathbf{S}'$ is not guaranteed to remain a valid cosine Gram matrix: it can be indefinite, its diagonal can differ from one, and its entries can lie outside $[-1,1]$. These violations affect the row-wise rankings, neighborhoods, and graph spectra used by the downstream methods. Projecting onto $\mathcal{C}_{\mathsf{coll}}$ enforces the PSD, unit-diagonal, and entrywise cosine constraints, and improves utility over $\mathbf{S}'$ at the same privacy level in the displayed settings. The improvement is most visible in tasks that depend on relative similarity structure, including retrieval, clustering, semantic similarity, and recommendation. For very small $\varepsilon$, the injected noise dominates and post-processing cannot in general recover the clean ordering.  As $\varepsilon$ increases, $\widehat{\mathbf{S}}$ moves toward the non-private baseline while preserving the same DP guarantee as the noisy release.

\paragraph{Projection scalability.}
We also compare AAP with an SDP projector for the nearest feasible cosine-Gram problem.  The SDP baseline is useful at small matrix sizes, but becomes time- or memory-limited for the larger matrices used in STS-B, MovieLens-100K, and Oxford-IIIT Pets. AAP remains tractable at these sizes because each iteration uses a closed-form unit-diagonal box projection and one PSD projection. In the small-$n$ cases where both projectors are run, AAP gives comparable downstream utility; at the larger scales used in the main experiments, AAP is the practical projection method.

\vspace{-4pt}

\section{Conclusion}

\vspace{-5pt}

We introduced \textsc{ScoreShield}, a perturb--then--project framework for the central-model $(\varepsilon,\delta)$-differentially private release of cosine similarity score vectors and cosine Gram matrices.
The mechanism adds Gaussian noise calibrated to the sensitivity of the chosen release regime and then projects the perturbed output onto the corresponding feasibility set of valid cosine objects. This projection is privacy-preserving by post-processing and enforces the structural constraints required by the released object. 
For vector release, \textsc{ScoreShield} preserves the privacy guarantee, does not increase squared error relative to the unprojected Gaussian release, and allows downstream thresholding, ranking, and top-$k$ selection rules applied to the privatized score vector to inherit the same differential-privacy guarantee by post-processing.
For full Gram release, \textsc{ScoreShield} exploits cosine-Gram geometry to improve Frobenius mean-squared error scaling over the unprojected Gaussian baseline, and we introduced a scalable averaged alternating-projection solver with convergence guarantees for feasibility projection.
Experiments on retrieval-augmented generation and face recognition evaluate the vector-release regime, while Gram-only evaluations on semantic textual similarity, image similarity and clustering, and recommender-system tasks evaluate the Gram-release regime; across these settings, the results support the predicted privacy--utility behavior across modalities and quantify the benefit of feasibility enforcement.
These results show that constraint-aware privatization of similarity scores is a useful primitive for similarity-based systems.  
%

\vspace{-6pt}

\section{Limitations and Future Works}
\vspace{-4pt}
Our results characterize a single release of cosine scores for fixed normalized embeddings under the stated central-DP adjacency relation. They do not address repeated or adaptive score queries, for which privacy losses compose and tight composition accounting is required. Nor do they characterize utility-optimal noise distributions for cosine-score release under joint $(\varepsilon,\delta)$-DP and cosine-feasibility constraints; the Gaussian mechanisms analyzed here satisfy the privacy guarantee but are not proved minimax- or instance-optimal.


\section*{Acknowledgments}
This work was supported by the Swiss National Science Foundation (SNSF) under Grant No.~222339. The authors thank Dr.~Flavio~P.~Calmon, Dr.~S\'{e}bastien Marcel, Dr.~Shahab Asoodeh, and Dr.~Hatef Otroshi Shahreza for insightful discussions and helpful suggestions that helped improve this work.


\putbib
\end{bibunit}

\appendix
\clearpage

\AppendixOnlyTOC
 

\begin{bibunit}
\counterwithin{table}{section}
\counterwithin{figure}{section}

\clearpage
\appsection{Extended Introduction: Differentially Private Deep Face Recognition}
\label{app:sec:extended-introduction}


Deep face-recognition (FR) systems are deployed in consumer devices, border-control gates, CCTV networks and large-scale photo grouping services. In these settings decision are made from cosine similarities: an image is mapped to an $\ell_{2}$-normalized embedding, and the dot product with reference embeddings is used for ranking and threshold-based match decisions. Although raw images often remain on device, similarity scores are often stored (e.g., for auditing and analytics) or shared across services. Releasing similarity scores without protection can enable membership inference about whether a specific enrolled record is in the gallery.

\vspace{-1pt}

\paragraph{Motivation.}
In regime (i), we study central-model $(\varepsilon,\delta)$-DP for releasing FR cosine-similarity scores. Due to three obstacles, this problem not directly treated in prior work:
\begin{enumerate}[label=(\roman*), font=\upshape, itemsep=1pt, leftmargin=1.58em]
\item 
\textit{What is released and how it scales.} 
The common release objects in FR are either \textit{embeddings} or their \textit{pairwise similarities}. Under standard record-level adjacency, the global sensitivity of the matrix releases grows with the gallery size, so na\"ively adding Gaussian noise degrades accuracy at strict operating points (very small FMR).
\item 
\textit{Accuracy in the low–false-match regime.} 
FR is tuned at thresholds that target very small impostor rates. Norm-based DP error bounds do not translate into decision-level guarantees (e.g., threshold shift, false-match/false-non-match changes, ROC/AUC), making it unclear how to translate Euclidean error bounds into guarantees on FMR/FNMR at a target threshold.
\item 
\textit{Geometry and computation.} Valid similarity vectors/matrices form a structured set (PSD, unit diagonal, bounded off-diagonals). Exploiting this geometry via projection can improve utility relative to unconstrained entrywise noise, but proving guarantees and obtaining scalable algorithms is nontrivial. Choose of adjacency (image-level, identity-level, or Gram-bounded) and privacy accounting under repeated probe releases also  require explicit composition analysis.
\end{enumerate}

\vspace{-2pt}

Consequently, prior work either applies local/instance-level perturbations to inputs/features or relies on HE/MPC without DP for released scores, leaving a gap: central-model $(\epsilon, \delta)$-DP guarantees for released similarity scores are rarely provided.

\vspace{-2pt}

\paragraph{Training-time DP (DP-SGD).}
A natural alternative is to train the encoder with training-time DP guarantee (e.g., DP-SGD). We view this as orthogonal to our goal: training-time DP protects membership with respect to the \textit{training corpus}, whereas our risk concerns the privacy of \textit{released similarity scores} over galleries and probes that may include individuals never seen during training. In practice, deploying DP-SGD within modern FR pipelines (large-batch schedules, margin-based losses) requires per-example clipping, careful noise calibration, and full retraining. Accordingly, we treat training-time DP as complementary: if similarity scores are released, release-time mechanisms remain necessary to certify central-model $(\varepsilon,\delta)$-DP for the scores themselves, regardless of how the encoder was trained.

\vspace{-2pt}

\paragraph{Our resolution.}
Two ingredients make this setting analyzable.
First, projection-based mechanisms for pairwise similarities \citep{cohen2024perturb} show that adding one Gaussian perturbation per release and projecting onto the feasible set can improve utility (in Frobenius/MSE scaling) relative to unconstrained noise by exploiting structure.
Second, modern FR backbones exhibit calibrated score distributions that allow threshold-level (decision-level) analysis at strict FMR targets. Building on these, we (a) formalize central, record-level adjacencies for three practical FR-release settings; (b) derive closed-form global sensitivities and the corresponding Gaussian standard deviations; and (c) couple projection with decision-level utility bounds at low FMR.
This connects generic central-DP releases of pairwise statistics to the needs of FR deployments.

\vspace{-2pt}

\paragraph{Adjacency for central model DP deep FR.}
Prior FR papers labeled ``DP'' typically act in the local model and perturb inputs or early features, without defining central-model neighbors for score releases. Prior work on DP release of cosine similarity matrices \citep{cohen2024perturb} defines a $\mathsf{\Delta}_{\mathsf{G}}$-bounded Gram adjacency that is task-agnostic. We define FR-specific central-model adjacencies at the image and identity levels and tie them to enrollment and verification procedures, enabling sensitivity-calibrated Gaussian noise and operating-point analysis for query (probe) vectors and Gram matrices of similarities.

\clearpage
\appsection{Extended Related Work}
\label{app:sec:extended-relatedwork}

\appsubsection{Differential Privacy for Vector, Covariance, and Gram Release Mechanisms}
\label{app:ssec:dp-vector-cov-gram-relatedwork}

\paragraph{Early vector-level random projections for privacy.}
A classical approach uses Johnson-Lindenstrauss (JL) embeddings to release \textit{vectors} with small distortion while approximately preserving pairwise distances. Blocki \textit{et al.} \citep{blocki2012johnson} showed that under specific spectral/conditioning assumptions (rank-1 neighbor change; all singular values above a threshold), a randomized JL transform \textit{itself} can satisfy differential privacy (``an old dog performs new tricks'') and applied this to edge-DP graph cut queries and directional variance/covariance release with distortion/estimation error bounds that does not scale with ambient dimension (with high probability). Building on this tool, JL with calibrated noise was then specialized to collaborative filtering: Yang \textit{et al.} \citep{yang2017privacy} proposed JLCF, proving $\varepsilon$--DP for a linear JL transform that approximately preserves user–user distances and, under margin conditions, the neighbors ($k$-NN structure), hence giving strong utility for $k$-NN recommenders. While effective for vector publication, these methods do not, by themselves, bound leakage when full similarity vectors or the entire Gram are disclosed.

\paragraph{Private ERM without heavy projections.}
A concurrent line in differentially private learning focuses on optimization rather than projecting outputs. Talwar~\textit{et~al.} \citep{talwar2015nearly} gave a nearly-optimal $(\varepsilon,\delta)$-DP LASSO via a private Frank-Wolfe method with the exponential mechanism, achieving excess risk $\tilde{\mathcal{O}}((n\varepsilon)^{-2/3})$ with only logarithmic dependence on dimension and is optimal up to logarithmic factors under standard assumptions. This direction is distinct from releasing pairwise similarities: it privatizes model training, whereas our setting privatizes a structured output (a similarity vector or matrix).

\paragraph{Covariance release revisited.}
Recent central-model work highlights how clipping and trace-dependent calibration affect utility for matrix-valued queries. Dong~\textit{et al.} propose \textit{AdaptiveCov} \citep{dong2022differentially}, which DP-selects clipping thresholds to balance bias and variance. On heavy-tailed data, AdaptiveCov achieves lower estimation error than both isotropic Gaussian and coordinatewise baselines, motivating structure-aware calibration that we later adapt to Gram release.

\paragraph{Structured output noise beyond full-rank Gaussians.}
Ji \textit{et al.} \citep{ji2024less} identify a ``curse of full-rank covariance'': for Gaussian mechanisms in their setting, the expected squared error depends on $\mathsf{tr}(\boldsymbol{\Sigma}_{\text{noise}})$, yielding matching lower bounds at fixed $(\varepsilon,\delta)$. They introduce a rank-$1$ singular multivariate Gaussian (R1SMG) mechanism whose covariance is random and rank-1, achieving $(\varepsilon,\delta)$-DP with a different risk scaling in the number of released coordinates.
This is complementary to projection-based post-processing: one can \textit{shape} the noise before projecting.

\paragraph{Perturb–and–Project (PnP) for structured releases.}
Cohen–Addad~\textit{et al.} \citep{cohen2024perturb} formalize a central-model perturb-and-project template for releasing structured objects: add Gaussian noise calibrated to the global $\ell_2$ sensitivity $\Delta$, then perform Euclidean projection onto a compact convex feasibility set. By post-processing, the mechanism is $(\varepsilon, \delta)$-DP. 
Their utility guarantees are governed by the Gaussian complexity of the feasibility set (rather than directly by the ambient dimension) and bound the expected squared error to the true projection of the input onto that set. 
For releasing a cosine-similarity matrix, they apply this template to the dataset’s Gram matrix and assume a Gram-space adjacency condition: neighboring datasets are those whose induced Gram matrices differ by at most a prescribed Frobenius-norm radius, i.e., the total squared change across all pairwise similarities is bounded.

\appsubsection{Differentially Private Retrieval-Augmented Generation}
\label{app:ssec:dp-rag-relatedwork}

\paragraph{Setup and Protection Goal.}
In retrieval-augmented generation (RAG), a server answers prompts by retrieving relevant corpus items and conditioning a generator on the retrieved context.
DP-RAG is commonly formalized as corpus-level $(\varepsilon,\delta)$-DP for the (possibly interactive) transcript produced under adaptively chosen prompts.
Across this literature, the privacy unit is often add/remove (document-deletion) at the document level (or a specified privacy unit), motivated by membership-inference and data extraction attacks in external knowledge bases \citep{wu2025private, mori2025differentially, koga2024privacy}.
\textsc{ScoreShield} regime~(i) instead adopts chunk-level replacement adjacency and treats prompts as public input in the curator model.

\paragraph{Privatized Objects.}
DP-RAG proposals differ primarily in (i) the adjacency relation and (ii) which intermediate or final objects are made private:
\begin{enumerate}[leftmargin=1.25em, itemsep=2pt,topsep=2pt]
\item \textit{Retrieval-stage privatization} (IDs/ranks/thresholds/scores):
the mechanism releases a DP version of the retrieval outcome (e.g., a DP top-$k$ set, a DP threshold, or a DP score signal), and subsequent steps are treated as post-processing \textit{of that released retrieval signal}.\footnote{In this category, post-processing applies only insofar as downstream computation depends on the corpus exclusively through the released DP retrieval signal (and public inputs/internal randomness), i.e., without additional access to sensitive record contents.}
\item \textit{Generation-stage privatization} (DP decoding / private prediction):
the system uses a DP mechanism during token generation (often via prompting an LLM on multiple documents and aggregating with a DP rule), so that the final text output satisfies DP w.r.t. the corpus.
\item \textit{One-time dataset privatization} (DP synthetic corpus):
the corpus is privatized once into a DP proxy dataset, enabling unlimited downstream (non-private) retrieval and generation by post-processing.
\end{enumerate}
These design points target different release goals and are therefore not generally interchangeable. That is privatizing a retrieval signal is often sufficient when the \textit{public release} is restricted to retrieval metadata (IDs/ranks/scores), whereas end-to-end DP for the \textit{generated answer text} generally
requires additional structure (generation-stage DP or dataset privatization) if the generator can access sensitive corpus text.

\paragraph{Single-query DP-RAG.}

A first line of work targets the single-query setting ($T=1$), where one prompt is answered under a fixed privacy budget. Representative mechanisms achieve end-to-end DP for the answer text by spending privacy budget in the generation stage (e.g., private-prediction / aggregated-generation style approaches such as \textsc{DPSparseVoteRAG}) \citep{koga2024privacy}. 
A complementary single-query strategy privatizes retrieval identifiers rather than releasing raw retrieved text.
%
%
For example, in \citep{grislain2025rag} the DP-RAG system designs a DP procedure to obtain a top-$k$ retrieval set while preserving the feasibility of a downstream DP generation step.
Concretely, it (i) DP-selects a similarity threshold $\theta$ (via an exponential-mechanism step),
(ii) forms a candidate set above $\theta$, and (iii) samples document indices under a DP distribution before invoking generation. On the generation side, the same work follows a DP aggregated generation / DP in-context learning template: it queries the LLM separately on each retrieved document and then aggregates token distributions with a DP mechanism to produce the final output. This class of methods targets end-to-end DP for the answer text by spending privacy budget during token selection.

\paragraph{Multi-Query DP-RAG is Nontrivial.}

In realistic deployments, an adversary can issue many (adaptive) prompts against the same corpus. If one applies a single-query DP-RAG mechanism independently to each query, under basic sequential composition, the privacy cost accumulates by sequential composition and can become very large after modest $T$ (e.g., $\varepsilon\approx 1000$ for $T=100$ queries at $\varepsilon_q\approx 10$) \citep{wu2025beyond}. This motivates mechanisms that exploit \textit{sparsity of relevance}, the empirical observation that each query typically touches only a small subset of the corpus, rather than paying as if all records participate in every query \citep{wu2025beyond}.

\paragraph{Multi-Query DP-RAG via Relevance Screening and Individual Privacy Filters (MURAG).}

The ``Beyond per-question privacy'' framework develops DP-RAG algorithms explicitly designed for the multi-query regime, introducing \textsc{MURAG} and \textsc{MURAG-ADA} \citep{wu2025beyond}.
Their core mechanism combines (i) \textit{relevance screening}, which restricts which records are eligible to be used for a given query, with (ii) \textit{individual privacy accounting} via R\'{e}nyi privacy filters, which track and halt a record's participation
once its ex-ante privacy budget is exhausted \citep{wu2025beyond}.
At a high level, \textsc{MURAG} maintains an active subset by screening records whose relevance score exceeds a fixed threshold $\tau$, answers the query using a single-query DP-RAG subroutine on the screened set, and decrements per-record budgets after each
use \citep{wu2025beyond}.
\textsc{MURAG-ADA} replaces a static $\tau$ by a query-adaptive private threshold calibrated to the (private) top-$K$ boundary: it discretizes similarity scores into bins and releases noisy prefix sums with Laplace noise until the cumulative count exceeds
$K$, spending a dedicated budget $\varepsilon_{\mathrm{thr}}$ for this threshold-release step \citep{wu2025beyond}.
%
They prove an overall corpus-level DP guarantee by ensuring each record participates only while its per-record privacy filter remains below a target budget \citep{wu2025beyond}, and empirically they report answering many queries (e.g., $T=100$) under total budgets such as $\varepsilon=10$ \citep{wu2025beyond}. This line of research addresses \textit{multi-query accounting} rather than proposing a new one-shot privatization of retrieval scores.

\paragraph{One-time dataset privatization for RAG via DP synthetic corpora (DP-SynRAG).}
A different strategy avoids additional per-query privacy expenditure by privatizing the corpus \textit{once} and then running standard (non-private) retrieval and generation over a DP proxy dataset. DP-SynRAG explicitly targets this regime, noting that enforcing DP directly on LLM outputs in RAG consumes privacy budget per query and can degrade rapidly as queries accumulate \citep{mori2025differentially}.
It adopts a private-prediction (subsample-and-aggregate) paradigm to generate synthetic text, emphasizing that RAG requires \textit{locality preservation} (query-relevant fine structure) rather than only global distributional similarity \citep{mori2025differentially}. Provided downstream retrieval/generation accesses only the DP proxy corpus, by producing a DP synthetic corpus once, subsequent retrieval and answering incur no additional privacy cost by post-processing, yielding a fixed total privacy budget that scales favorably with the number of queries \citep{mori2025differentially}, so the total privacy budget is fixed (independent of the number of queries $T$). Their experiments report lower attack success under repeated querying (under their threat model/metrics) than per-query DP baselines \citep{mori2025differentially}.

\paragraph{Relation to \textsc{ScoreShield}.}

\textsc{ScoreShield} regime~(i) can be viewed as a DP retrieval primitive that privatizes the score interface itself. Given a normalized prompt embedding $\mathbf{q}$ and chunk embeddings $\mathbf{x}_i$, it forms the cosine score vector $\mathbf{s}(\mathbf{q}) \coloneqq \big(\langle \mathbf{q},\mathbf{x}_1\rangle,\dots,\langle \mathbf{q},\mathbf{x}_n\rangle\big)\in[-1,1]^n$,
and releases
$\widehat{\mathbf{s}}(\mathbf{q}) \;=\; \mathsf{proj}_{[-1,1]^n} \big( \mathbf{s} (\mathbf{q}) + \mathbf{w}\big), \mathbf{w}\sim\mathcal{N}(\mathbf{0},\sigma^2 \mathbf{I}_n)$.
Under chunk-level replacement adjacency and clipping, at most one coordinate can change and its magnitude is bounded, yielding global $\ell_2$-sensitivity $\Delta_2=2$ and hence closed-form Gaussian calibration for $(\varepsilon,\delta)$-DP.
Importantly, the DP guarantee applies directly to the released retrieval signal $\widehat{\mathbf{s}}(\mathbf{q})$ and to any objects that are deterministic/randomized functions of $\widehat{\mathbf{s}}(\mathbf{q})$ and internal randomness only, such as rankings, top-$k$ sets, or thresholded retrieval \textit{identifiers} (by post-processing).
This places regime~(i) closest in spirit to retrieval-stage privatization methods (selection/IDs/thresholds), and it is complementary to multi-query layers (e.g., \textsc{MURAG}/\textsc{MURAG-ADA}) that manage privacy over $T \gg 1$ queries \citep{wu2025beyond}.
By contrast, providing end-to-end DP for the \textit{generated answer text} when the generator can access \textit{sensitive corpus content} typically requires additional mechanisms beyond score release (e.g., generation-stage DP aggregation or one-time corpus
privatization), and is the explicit target of private-prediction DP-RAG and DP-SynRAG-style approaches. 
In our experiments the retrieved corpus is public (Wikipedia), so returning retrieved text is not treated as a private release; the DP claim concerns the retrieval chunk IDs/ranks induced by it.
Empirically, \textsc{ScoreShield} evaluates regime~(i) as DP retrieval for multi-hop RAG on Google FRAMES using EmbeddingGemma for retrieval and Gemma~3--12B as generator/judge \citep{krishna2025fact, embeddinggemma2025, gemma2025gemma3}.

\appsubsection{Differentially Private Face Recognition}

\paragraph{Local/instance-level FR perturbations.}
In FR, several prior works adopt a local/instance adjacency and perturb inputs or early features \citep{chamikara2020privacy, ji2022privacy}. For example, \citep{ji2022privacy} maps images to frequency space (block-DCT), removes the DC (zero-frequency) component, and injects DP noise with learnable per-frequency budgets before feeding a standard FR backbone. This choice is motivated by the observation that visualization-critical and identification-critical information separate in frequency. The adjacency is also reframed in a learned representation (``secret'') space and target instance-level protection, rather than central-model DP guarantees for public score releases.

\clearpage
\appsection{Extended Preliminaries}
\label{app:sec:extended-preliminaries}

\vspace{-2pt}

\appsubsection{Notation}

\vspace{-2pt}

We use three dimension parameters:
(i) $n$ denotes the number of enrolled records,
(ii) $d$ denotes the embedding dimension,
(iii) $N \coloneqq \dim(\mathtt{S}^n)=\frac{n(n+1)}{2}$ denotes the ambient Euclidean dimension of symmetric $n\times n$ matrices $\mathtt{S}^n$ under Frobenius vectorization.
For any integer $m\ge 1$, the unit sphere in $\mathbb{R}^m$ is $\mathbb{S}^{m-1} \coloneqq \{\mathbf{u}\in\mathbb{R}^m:\|\mathbf{u}\|_2=1\}$.
For $\mathbf{e}\in \mathbb{R}^{d}$, the Euclidean norm is ${\Vert \mathbf{e} \Vert}_2 = \big( \sum_{i=1}^d e_i^2 \big)^{1/2}$.
For any integer $m \ge 1$, $\mathbf{I}_{m}$ denotes the $m\times m$ identity matrix.
For a matrix $\mathbf{A} \in \mathbb{R}^{n \times d}$, the Frobenius norm is ${\Vert \mathbf{A}\Vert}_{\mathrm{F}} = \big( \sum_{i=1}^n \sum_{j=1}^{d} A_{ij}^2\big)^{1/2} = \mathrm{tr} \big( \mathbf{A}^\top \mathbf{A} \big)^{1/2}$. The spectral norm of $\mathbf{A}$, denoted as ${\Vert \mathbf{A} \Vert}_2$, is the supremum $\sup_{\mathbf{e} \in \mathbb{S}^{d-1}} {\Vert \mathbf{A} \mathbf{e} \Vert}_2$ and equals its largest singular value. The nuclear norm ${\Vert \mathbf{A} \Vert}_\ast$ is the sum of the singular values of $\mathbf{A}$, i.e., $\sum_{k=1}^{\min{(n, d)}} \sigma_k (\mathbf{A})$. The maximum entry norm of $\mathbf{A}$ is defined as ${\Vert \mathbf{A} \Vert}_{\mathrm{max}} = \max_{1 \leq i \leq n, 1 \leq j \leq d} {\vert A_{ij} \vert}$. A symmetric matrix $\mathbf{A} \in \mathbb{R}^{n \times n}$ is positive semi-definite (PSD), denoted $\mathbf{A} \succeq 0$, if $\mathbf{e}^\top \mathbf{A} \mathbf{e} \geq 0, \, \forall \mathbf{e} \in \mathbb{R}^n$. The standard deviation parameter in the Gaussian mechanism is denoted by $\sigma$, and the Euclidean projection operator onto a closed convex set $\mathcal{C}$ is written as $\mathsf{proj}_{\, \mathcal{C}}(\cdot)$.
For sequences $a_n, b_n>0$, we write $a_n = \mathcal{O}(b_n)$ if there exist constants
$C<\infty$ and $n_0$ such that $a_n \le C\,b_n$ for all $n\ge n_0$.
We write $a_n = \Omega(b_n)$ if $b_n = \mathcal{O} (a_n)$, and $a_n = \Theta(b_n)$ if both $a_n=O(b_n)$ and $a_n= \Omega (b_n)$ hold.
Unless stated otherwise, asymptotics are with $n\to\infty$ and $(\varepsilon,\delta)$ treated as fixed.

\vspace{-2pt}

\appsubsection{Face Recognition}

\vspace{-2pt}

\paragraph{Setup.}
Let $\mathcal{D}=\{\mathbf{x}_{1},\dots, \mathbf{x}_{n}\}$ denote a dataset of face images. A pre-trained backbone model $\phi_{\boldsymbol{\theta}}$ maps each image $\mathbf{x}_i$ to an $\ell_{2}$-normalized embedding vector $\mathbf{e}_{i}=\phi_{\boldsymbol{\theta}}(\mathbf{x}_{i})\in\mathbb{R}^{d}$ with $\|\mathbf{e}_{i}\|_{2}=1$. The embeddings are aggregated into a dataset matrix $\mathbf{E} \in \mathbb{R}^{n \times d}$, where each row corresponds to the transpose of an embedding, i.e., $\mathbf{E}=[\mathbf{e}_{1}^{\top},\dots,\mathbf{e}_{n}^{\top}]^{\top}\in \mathbb{R}^{n\times d}$. For fixed $(n,d)$ define the admissible embedding space $\mathcal{E}\;\coloneqq\;\bigl\{\mathbf{E}\in\mathbb{R}^{n\times d}:\ \|\mathbf{e}_i\|_2=1, \forall i\in[n]\bigr\}$, where $\mathbf{e}_i^\top$ denotes row $i$ of $\mathbf{E}$.
The Gram matrix (cosine similarity matrix) is defined as $\mathbf{S}  \coloneqq  \mathbf{E} \mathbf{E}^{\top}\in\mathbb{R}^{n\times n}$, with entries $S_{ij}=\langle \mathbf{e}_{i}, \mathbf{e}_{j}\rangle \in [-1, 1]$. 
By construction (i) $\mathbf{S}$ is PSD ($\mathbf{S} \succeq 0$), as Gram matrices satisfy $\mathbf{v}^\top \mathbf{S} \mathbf{v} = {\Vert  \mathbf{E}^\top \mathbf{v} \Vert}_2^2 \geq 0$ for all $\mathbf{v} \in \mathbb{R}^n$, (ii) the diagonal entries satisfy $S_{ii} = 1, \forall i$ due to ${\Vert \mathbf{e}_i \Vert}_2 = 1$, (iii) the off-diagonal entries satisfy $\vert S_{ij} \vert \le1 \; , \forall i\neq j$. 
%
%
Hence, we define the set of valid cosine Gram matrices (correlation matrices) as
\begin{eqnarray}
\mathcal{C}_{\mathsf{coll}} \;  \coloneqq  \; \bigl\{ \mathbf{S}\in \mathbb{R}^{n\times n}: \mathbf{S}\succeq 0,\; S_{ii}=1 \; (1 \leq i \leq n) ,\; |S_{ij}| \le 1 \; (j\neq i) \bigr\}.
\end{eqnarray}
The rank of $\mathbf{S}$ is bounded by $\mathrm{rank} (\mathbf{S}) \leq \min (n, d)$, since the rank of a matrix product cannot exceed the rank of either factor.

\vspace{-2pt}

\paragraph{Similarity Scores in Face-Recognition Systems.}
FR systems perform tasks like verification, identification, and clustering by measuring how similar face embeddings are to one another.
For any pair $(i,j) \in \left[ n \right]^2$
\begin{equation}
S_{ij}\;  \coloneqq  \;\langle \mathbf{e}_i,\mathbf{e}_j \rangle \;=\;1-\tfrac12\|\mathbf{e}_i-\mathbf{e}_j\|_2^{2},
\end{equation}
so cosine similarity and squared Euclidean distance are affine transforms of each other, hence, they induce identical rankings of pairs. 
Decision thresholds translate via $\tau_{\mathrm{euc}} = 2 (1 - \tau_{\mathrm{cos}} )$ or $\tau_{\cos}=1-\tfrac12\tau_{\text{euc}}$, where $\tau_{\mathrm{cos}} \in [-1, 1]$, and $\tau_{\mathrm{euc}} \in [0, 4]$.

Verification, open-set identification, closed-set search, and even graph-based clustering all consume either (i)~cosine similarity scores $S_{ij} = \langle\mathbf{e}_i,\mathbf{e}_j\rangle$, or (ii)~their monotone surrogate $\|\mathbf{e}_i-\mathbf{e}_j\|_2^{2}$. 
For verification (1:1) a single score $S_{ij}$ is compared with a threshold~$\tau$ to decide a \textit{match} vs \textit{non-match}.
For closed-set identification (1:N) the probe is \textit{known} to correspond to one of the $n$ enrolled identities. Writing the noisy probe as $\mathbf{q}_i = \mathbf{e}_i + \mathbf{z}$ (for some unknown index~$i$), where $\mathbf{z}$ is nuisance noise, the backend must compare it with every gallery item, i.e., consume the entire $i$-th row $ \mathbf{s}_i=(S_{i1},\dots,S_{in})=\mathbf{E}\,\mathbf{e}_i$, in order to return $\arg\max_{j}S_{ij}$.
For open-set identification the probe may stem from an unseen identity, no row of~$\mathbf{S}$
is predetermined. The system instead forms the similarity vector $\mathbf{s}=\mathbf{E}\,\mathbf{q} \in \mathbb{R}^{n}$ and declares either the nearest neighbor or ``no match'' according to an open-set rule.
Clustering/multi-target tracking algorithms, such as spectral clustering, single-linkage, or Louvain, typically require the full pairwise affinity matrix $\mathbf{S}$ itself or a monotone transform such as $\exp(\alpha\mathbf{S})$, to define the graph's edge weights.

\appsubsection{Differential Privacy}

\begin{definition}[Record-level Adjacency]
\label{def:recordlevel-adjacency}
Datasets $\mathcal{D}, \mathcal{D}'$ (or equivalently, their embedding matrices $\mathbf{E}, \mathbf{E}'$) are \textit{adjacent}, denoted by $\mathcal{D} \sim \mathcal{D}'$, if they differ in at most one record (and thus in one embedding), i.e., $|\,\{i\in[n]: \mathbf{x}_{i}\neq \mathbf{x}_{i}'\}\,| = 1$.
\end{definition}

\begin{remark}
Under a fixed backbone model $\phi_{\boldsymbol{\theta}}$, adjacency implies that the embedding matrices $\mathbf{E}$ and $\mathbf{E}'$ differ in at most one row. 
Therefore, there exists at most one index $i$ such that $\mathbf{e}_j = \mathbf{e}_j' , \forall j \neq i$, and by our normalization assumption $\mathbf{e}_i \neq \mathbf{e}_i' \in \mathbb{R}^d$ satisfy $\| \mathbf{e}_i \|_2 = \| \mathbf{e}_i' \|_2 =1$.
\end{remark}

\begin{definition}[Output–space  (Gram-matrix) Adjacency]
\label{def:gram-adjacency}
Two collections with embeddings $\mathbf{E},\mathbf{E}'$ are \textit{adjacent at radius} $\mathsf{\Delta}_{\mathsf{G}} > 0$ if
%
$\|\mathbf{E}\mathbf{E}^\top-\mathbf{E}'\mathbf{E}'^\top\|_{\mathrm{F}} \;\le\; \mathsf{\Delta}_{\mathsf{G}}$.
%
\end{definition}

\begin{definition}[$(\varepsilon,\delta)$--DP]
A (possibly randomized) mechanism $\mathcal{M}:\mathcal{E}\to\mathcal{O}$
satisfies $(\varepsilon,\delta)$--DP if for all measurable $\mathcal{T}\subseteq\mathcal{O}$
and all adjacent $\mathbf{E}\sim\mathbf{E}'$,
$\mathsf{Pr}\!\left[\mathcal{M}(\mathbf{E})\in\mathcal{T}\right] \;\le\;
e^{\varepsilon}\,\mathsf{Pr}\!\left[\mathcal{M}(\mathbf{E}')\in\mathcal{T}\right]+\delta$,
where $\varepsilon > 0$ and $\delta \in (0, 1)$ are privacy parameters.
\end{definition}

\begin{definition}[$\ell_2$-Sensitivity]
\label{def:sensitivity}
For a function $f: \mathcal{E} \to \mathbb{R}^n$, the $\ell_2$-sensitivity is $\Delta_{f,2} = \sup_{\mathbf{E} \sim \mathbf{E}'} \| f(\mathbf{E}) - f(\mathbf{E}') \|_2$. 
We use the notation $\Delta$ for brevity. 
\end{definition}

\begin{lemma}[Gaussian Mechanism]
\label{lem:gaussian}
Let $f:\mathcal{E} \to  \mathbb{R}^{n}$ have $\ell_2$–sensitivity $\Delta_{f, 2}= \sup_{\mathbf{E} \sim \mathbf{E}'} \|f(\mathbf{E}) - f(\mathbf{E}')\|_2$. Define $\mathcal{M} (\mathbf{E}) \! = \! f(\mathbf{E}) + \mathbf{w}$,
$\mathbf{w} \sim \mathcal{N} \bigl( \mathbf{0}, \sigma^2 \mathbf{I}_{n}\bigr)$,
$\sigma^2 \geq c_{\varepsilon,\delta}\,\Delta^{2}_{f,2}$ with $c_{\varepsilon,\delta} \coloneq \frac{2\log (2/\delta)}{\varepsilon^2}$.
Then $\mathcal{M}$ satisfies $(\varepsilon, \delta)$-DP.
\end{lemma}


\begin{lemma}[Post-processing]\label{lem:post-processing}
Let $\mathcal{M}: \mathcal{E} \to \mathcal{O}$ be a mechanism that satisfies $(\varepsilon, \delta)$--differential privacy. For any measurable function $g: \mathcal{O} \to \mathcal{O}'$, the composed mechanism $g \circ \mathcal{M}: \mathcal{E} \to \mathcal{O}'$ also satisfies $(\varepsilon, \delta)$--differential privacy.
\end{lemma}
 

\begin{remark}
After adding Gaussian noise to the similarity scores to achieve $(\varepsilon, \delta)$-DP, we project the noisy output onto the closed convex set of valid similarity scores. Since projection is independent of the private dataset, the post-processing lemma ensures that this step does not compromise the privacy guarantee established by the perturbation.
\end{remark}


\begin{corollary}[Gaussian mechanism for matrix-valued outputs]
\label{cor:gaussian-matrix}
Let $f:\mathcal{E}\to \mathbb{R}^{n\times n}$ and define the Frobenius sensitivity
\begin{equation}
\Delta_{f,\mathrm{F}} \coloneqq \sup_{\mathbf{E}\sim\mathbf{E}'} \|f(\mathbf{E})-f(\mathbf{E}')\|_{\mathrm{F}}.
\end{equation}
Let $\mathbf{W}\in\mathbb{R}^{n\times n}$ have i.i.d.\ entries $W_{ij}\sim\mathcal{N}(0,\sigma^2)$, and define
\begin{equation}
\mathcal{M}(\mathbf{E}) \coloneqq f(\mathbf{E})+\mathbf{W}.
\end{equation}
If $\sigma^2 = c_{\varepsilon,\delta}\,\Delta_{f,\mathrm{F}}^{2}$ (with $c_{\varepsilon,\delta}=2\log(2/\delta)/\varepsilon^2$), then $\mathcal{M}$ is $(\varepsilon,\delta)$--differentially private.
\end{corollary}

\begin{proof}
Equip $\mathbb{R}^{n\times n}$ with the Frobenius norm $\|\cdot\|_{\mathrm{F}}$ and identify it with $\mathbb{R}^{n^2}$ via the vectorization map $\mathrm{vec}$, which is a linear isometry: $\|\mathbf{A}\|_{\mathrm{F}}=\|\mathrm{vec}(\mathbf{A})\|_2$. Therefore $\mathrm{vec}\circ f$ has $\ell_2$-sensitivity $\Delta_{f,\mathrm{F}}$.
Moreover, since $W_{ij}\stackrel{\mathrm{i.i.d.}}{\sim}\mathcal{N}(0,\sigma^2)$, we have $\mathrm{vec}(\mathbf{W})\sim \mathcal{N}(0,\sigma^2\mathbf{I}_{n^2})$. Applying Lemma~\ref{lem:gaussian} to the $\mathbb{R}^{n^2}$-valued mechanism $\mathrm{vec}(\mathcal{M}(\mathbf{E}))=\mathrm{vec}(f(\mathbf{E}))+\mathrm{vec}(\mathbf{W})$ yields $(\varepsilon,\delta)$--DP for $\mathcal{M}$.
\end{proof}

\begin{corollary}[Symmetric Gaussian noise via averaging]
\label{cor:gaussian-symmetric}
In the setting of Corollary~\ref{cor:gaussian-matrix}, assume $f(\mathbf{E})\in\mathtt{S}^n$ for all $\mathbf{E}$ and define
\begin{equation}
\mathcal{M}_{\mathrm{sym}}(\mathbf{E}) \coloneqq f(\mathbf{E}) + \tfrac12(\mathbf{W}+\mathbf{W}^\top).
\end{equation}
If $\sigma^2 = c_{\varepsilon,\delta}\,\Delta_{f,\mathrm{F}}^{2}$, then $\mathcal{M}_{\mathrm{sym}}$ is $(\varepsilon,\delta)$--DP.
Moreover, writing $\mathbf{G}\coloneqq\tfrac12(\mathbf{W}+\mathbf{W}^\top)$, the collection of upper-triangular entries $\{G_{ii}: i\in[n]\}\cup\{G_{ij}:1\le i<j\le n\}$ is mutually independent with $G_{ii}\sim\mathcal{N}(0,\sigma^2)$ and $G_{ij}\sim\mathcal{N}(0,\sigma^2/2)$ for $i<j$.
\end{corollary}

\begin{proof}
Define $g:\mathbb{R}^{n\times n}\to\mathtt{S}^n$ by $g(\mathbf{A})\coloneqq\tfrac12(\mathbf{A}+\mathbf{A}^\top)$.
Let $\mathcal{M}(\mathbf{E})\coloneqq f(\mathbf{E})+\mathbf{W}$ with i.i.d.\ $W_{ij}\sim\mathcal{N}(0,\sigma^2)$.
By Corollary~\ref{cor:gaussian-matrix}, if $\sigma^2=c_{\varepsilon,\delta}\Delta_{f,\mathrm{F}}^2$ then $\mathcal{M}$ is $(\varepsilon,\delta)$-DP.
Since $f(\mathbf{E})\in\mathtt{S}^n$, we have $g(f(\mathbf{E}))=f(\mathbf{E})$ and hence
\begin{equation}
(g\circ\mathcal{M})(\mathbf{E})
= g(f(\mathbf{E})+\mathbf{W})
= f(\mathbf{E})+g(\mathbf{W})
= f(\mathbf{E})+\tfrac12(\mathbf{W}+\mathbf{W}^\top)
= \mathcal{M}_{\mathrm{sym}}(\mathbf{E}).
\end{equation}
Thus $\mathcal{M}_{\mathrm{sym}}$ is a deterministic post-processing of $\mathcal{M}$, and is $(\varepsilon,\delta)$--DP by Lemma~\ref{lem:post-processing}.

For the distributional claim, note that $G_{ii}= W_{ii}\sim\mathcal{N}(0,\sigma^2)$. For $i<j$, $G_{ij}=\tfrac12(W_{ij}+W_{ji})$ is the average of two independent $\mathcal{N}(0,\sigma^2)$ variables, hence $G_{ij}\sim\mathcal{N}(0,\sigma^2/2)$. Independence across the upper-triangular collection follows because each $G_{ij}$ depends only on the disjoint set $\{W_{ij},W_{ji}\}$ (and each $G_{ii}$ only on $W_{ii}$), and the entries of $\mathbf{W}$ are mutually independent.
\end{proof}

\begin{remark}[Orthonormal-basis view]
Let $\{\mathbf{B}_k\}_{k=1}^N$ be any Frobenius-orthonormal basis of $\mathtt{S}^n$
(e.g., $\mathbf{e}_i\mathbf{e}_i^\top$ and $(\mathbf{e}_i\mathbf{e}_j^\top+\mathbf{e}_j\mathbf{e}_i^\top)/\sqrt{2}$ for $i<j$).
Then $\bigl(\langle \mathbf{G},\mathbf{B}_k\rangle_{\mathrm{F}}\bigr)_{k=1}^N$ are i.i.d.\ $\mathcal{N}(0,\sigma^2)$.
\end{remark}


\begin{theorem}[Analytic Gaussian Calibration \citep{balle2018improving}]
\label{thm:analytic-gaussian}
Let $\varepsilon>0$, $\delta\in(0,1)$, and write $\rho\coloneqq  \Delta/\sigma$. Define
\begin{equation}
\delta_{\mathrm{AG}}(\varepsilon,\rho)
\;=\; \Phi \Bigl(-\frac{\varepsilon}{\rho}+\frac{\rho}{2}\Bigr)
\;-\; e^{\varepsilon}\,\Phi \Bigl(-\frac{\varepsilon}{\rho}-\frac{\rho}{2}\Bigr), 
\end{equation}
where $\Phi$ is the standard normal cdf.
Then $\mathcal{M}_{\sigma}$ is $(\varepsilon,\delta)$--DP if and only if
\begin{equation}
\delta \;\;\ge\;\; \delta_{\mathrm{AG}}(\varepsilon,\Delta/\sigma).
\end{equation}
Equivalently, the minimal noise is obtained by the unique $\rho^\star>0$ solving
$\delta_{\mathrm{AG}}(\varepsilon,\rho^\star)=\delta$, with
\begin{equation}
\sigma^\star \;=\; \frac{\Delta}{\rho^\star}.
\end{equation}
Moreover, any measurable post–processing $g$ (e.g., projection onto $[-1,1]^n$) preserves $(\varepsilon,\delta)$ by DP post–processing.
\end{theorem}

\begin{proof}
A proof is beyond the scope of this paper; see \citep{balle2018improving} for a complete derivation and proof.
\end{proof}

\paragraph{Identity–Adjacency.}
\label{app:identity-adj}

We used the classical one–row/column notion $\mathbf{E}\;\sim_{\textsf{img}}\;\mathbf{E}' \quad\Longleftrightarrow\quad \exists\,i\in[n]:\; \mathbf{e}_i\neq\mathbf{e}_i',\; \mathbf{e}_j=\mathbf{e}_j'\;\forall j\neq i$. With unit–norm embeddings this gives $\Delta_{\textsf{img}}^{\mathsf{query}} = 2$ for the probe vector $f_{\mathsf{query}}(\mathbf{E}, \mathbf{q})=\mathbf{E} \mathbf{q}$. 
Now suppose each image (embedding) carries an identity label $y_i\in\{1,\dots,K\}$. Let $I_k=\{i:\,y_i=k\}$ for the index set of identity $k$ and cardinality $g_k \coloneqq  |I_k|$. We say $\mathbf{E} \sim_{\textsf{id}} \mathbf{E}'$ iff there exists exactly one identity $k^\star$ such that $\mathbf{e}_i\neq\mathbf{e}_i' \;\Longleftrightarrow\; i\in I_{k^\star}$. 
We say 
\begin{equation}
\mathbf{E}\; \sim_{\textsf{id}}\;\mathbf{E}'
\;\Longleftrightarrow\;
\exists\,k^\star:\; \mathbf{e}_i\neq\mathbf{e}_i'
\;\Longleftrightarrow\; i\in I_{k^\star}.
\end{equation}
Thus all embeddings belonging to at most one identity $k^\star$ may change (any or all of them can be added, removed, replaced); every other identity's rows stay fixed.

Let aggregate each identity to a unit–norm centroid
\begin{equation}
\mathbf{c}_k \; \coloneqq  \;\frac{1}{g_k}\sum_{i\in I_k}\mathbf{e}_i, \qquad \|\mathbf{c}_k\|_2 \leq 1,
\end{equation}
and publish the $K$ dimensional vector
\begin{equation}
\tilde f_{\mathrm{query}}(\mathbf{E},\mathbf{q}) \; = \; \bigl(\langle\mathbf{c}_1,\mathbf{q}\rangle,\dots, \langle\mathbf{c}_K,\mathbf{q}\rangle\bigr)^{\!\top} \in\mathbb{R}^{K}.
\end{equation}
In this case, only the centroid $k^\star$ may move 
\begin{equation}
\bigl\|\widetilde f_{\mathsf{query}}(\mathbf{E},\mathbf{q}) -\widetilde f_{\mathsf{query}}(\mathbf{E}',\mathbf{q})\bigr\|_2
= |\langle\mathbf{c}_{k^\star}-\mathbf{c}_{k^\star}',\mathbf{q}\rangle| \le 2 .
\end{equation}
Hence $\Delta_{\textsf{id}}^{\text{centroid}} = 2$. That is we have the same constant as image-adjacency definition.

Publishing the original $n$-vector $f_{\mathsf{query}}(\mathbf{E},\mathbf{q})=\mathbf{E}\mathbf{q}$ under identity-adjacency changes up to $g_{k^\star}$ coordinates:
\begin{equation}
\Delta_{\textsf{id}}^{\mathsf{query}} = 2\sqrt{g_{k^\star}}\; > 2 .
\end{equation}
Sensitivity grows only with $\sqrt{\vert I_{k^\star} \vert}$, but is larger than the centroid case.

Therefore, identity-level privacy is free ($\Delta=2$) \textit{iff} the mechanism aggregates all images of a person into one coordinate before noise is added. Otherwise the cost is the factor $\sqrt{g_{k^\star}}$.

\appsubsection{Convex Geometry}

\begin{definition}[Euclidean Projection Onto a Closed Convex Set]
\label{def:projection}
Let $m\ge 1$ and let $\mathcal{C}\subset\mathbb{R}^{m}$ be non–empty, closed and convex. The Euclidean projection operator $\mathsf{proj}_{\mathcal{C}}:\mathbb{R}^{m}\to\mathcal{C}$ is defined for every $\mathbf{s} \in \mathbb{R}^{m}$ by
\begin{equation}
\mathsf{proj}_{\mathcal{C}}(\mathbf{s})  \coloneqq 
\operatorname*{arg\,min}_{\mathbf{y} \in \mathcal{C}} \| \mathbf{s} - \mathbf{y} \|_{2}.
\end{equation}
\end{definition}

\begin{definition}[Tangent Cone]
\label{def:tangent-cone}
Let $\mathcal{C} \subset \mathbb{R}^{m}$ be  non‑empty, closed, and convex and fix a point $\mathbf{s} \in \mathcal{C}$. The tangent cone to $\mathcal{C}$ at $\mathbf{s}$ is
\begin{equation}
\mathsf{T}_{\mathbf{s}} \left( \mathcal{C}  \right)  \coloneqq 
\mathsf{cl}\Bigl\{ \lambda\,( \mathbf{y} - \mathbf{s}) \; : \; \lambda \ge 0,\; \mathbf{y} \in\mathcal{C} \Bigr\}  \;\subset\; \mathbb{R}^{m},
\end{equation}
where ``$\mathsf{cl}$'' denotes the Euclidean closure. Geometrically, $\mathsf{T}_{\mathbf{s}} \left( \mathcal{C}  \right) $ contains all velocity directions of feasible curves that start at $\mathbf{s}$ and remain inside $\mathcal{C}$.
\end{definition}

\begin{definition}[Gaussian Width of a Cone]
\label{def:gaussian-width}
Let $\mathcal{K}\subset\mathbb{R}^{m}$ be a non–empty, closed cone (i.e., $\lambda\, \mathcal{K}=\mathcal{K}$ for all $\lambda\ge 0$). Its Gaussian width is
\begin{equation}
\mathsf{GW}(\mathcal{K})
\coloneqq \mathbb{E}_{\mathbf{w} \sim \mathcal{N}(\mathbf{0}, \mathbf{I}_{m})}
\left[ \sup_{\mathbf{s}\in \mathcal{K}\cap \mathbb{S}^{m-1}} \langle \mathbf{w}, \mathbf{s}\rangle \right].
\end{equation}
\end{definition}

\begin{definition}[Gaussian Complexity of a Bounded Set]
\label{def:gaussian-complexity}
Let $\mathcal{C}\subset\mathbb{R}^{m}$ be bounded. Its Gaussian complexity is
\begin{equation}
\mathsf{GC}(\mathcal{C})  \coloneqq  \mathbb{E}_{\mathbf{w} \sim \mathcal{N}( \mathbf{0}, \mathbf{I}_m)} \left[ \, \sup_{\mathbf{s} \in \mathcal{C}} \, \langle \mathbf{w}, \mathbf{s} \rangle \, \right].
\end{equation}
If $\mathcal{C}$ is unbounded we set $\mathsf{GC}(\mathcal{C})\;  \coloneqq  \; + \infty$ by convention. When $\mathcal{C}$ is itself a cone intersected with the unit sphere, $\mathsf{GC}(\mathcal{C})$ coincides with the Gaussian width of that cone, i.e., $ \mathsf{GC}(\mathcal{C}) = \mathsf{GW} \bigl(\mathsf{cone}(\mathcal{C})\bigr)$.
\end{definition}

\begin{remark}[Statistical Dimension vs. Gaussian Width]
\label{rem:sd-vs-gw}
For a closed convex cone $\mathcal{K}\subset\mathbb{R}^{m}$, define its statistical dimension by
\begin{equation}
\delta(\mathcal{K}) \coloneqq \mathbb{E}\bigl[\|\mathsf{proj}_{\mathcal{K}}(\mathbf{z})\|_{2}^{2}\bigr],
\qquad \mathbf{z}\sim \mathcal{N}(\mathbf{0},\mathbf{I}_{m}).
\end{equation}
Then $\delta(\mathcal{K})$ is tightly comparable to the Gaussian width (see Lemma~\ref{lem:stability-proj} for a proof):
\begin{equation}
\label{eq:sd-gw-sandwich}
\mathsf{GW}(\mathcal{K})^{2}\ \le\ \delta(\mathcal{K})\ \le\ \mathsf{GW}(\mathcal{K})^{2}+1.
\end{equation}
Therefore any risk bound stated in terms of $\delta(\mathsf{T}_{\mathbf{s}}(\mathcal{C}))$ can equivalently
be expressed using $\mathsf{GW}(\mathsf{T}_{\mathbf{s}}(\mathcal{C}))$  (or $\mathsf{GC}(\mathsf{T}_{\mathbf{s}}(\mathcal{C})\cap \mathbb{S}^{m-1})$).
\end{remark}

\vspace{3pt}

\begin{remark}
In the convex-geometry definitions above, $m$ denotes the ambient Euclidean dimension.
In our applications:
(i) for score vectors $\mathbf{s}\in\mathbb{R}^{n}$ we have $m=n$ and the sphere is $\mathbb{S}^{n-1}$;
(ii) for Gram matrices $\mathbf{S}\in\mathtt{S}^n$ we identify $\mathtt{S}^n$ with $\mathbb{R}^{N}$ under the Frobenius inner product, where $N=\dim(\mathtt{S}^n)=n(n+1)/2$, and the sphere is $\mathbb{S}^{N-1}$.
\end{remark}

\vspace{3pt}

\begin{remark}[Matrix Gaussian Complexity under Frobenius Geometry]
\label{rem:matrix-gc-frobenius}
Definition~\ref{def:gaussian-complexity} is stated for subsets of $\mathbb{R}^m$. When $\mathcal{C}\subseteq \mathtt{S}^n$ is a set of symmetric matrices, we view $\mathtt{S}^n$ as a Euclidean space with Frobenius inner product $\langle \mathbf{A},\mathbf{B} \rangle \coloneqq \mathrm{Tr}(\mathbf{A}^\top \mathbf{B})$ and dimension $N=\dim(\mathtt{S}^n)=n(n+1)/2$. Equivalently, one may identify $\mathtt{S}^n$ with $\mathbb{R}^{N}$ via any fixed linear isometry (e.g., vectorization of the upper triangle with the appropriate $\sqrt{2}$ scaling on off-diagonal entries). Accordingly, for $\mathcal{C}\subseteq \mathtt{S}^n$ we write
\begin{equation}
\mathsf{GC}(\mathcal{C})
\;=\; \mathop{\mathbb{E}}_{\mathbf{W} \sim \mathcal{N}(\mathbf{0}, \mathbf{I}_{n \times n})} \Big[ \, \sup_{\mathbf{S}\in\mathcal{C}} \, \langle \mathbf{W},\mathbf{S} \rangle \, \Big],   
\end{equation}
where $\mathbf{W}$ is a standard Gaussian in this ambient Euclidean space.
\end{remark}

\begin{lemma}[Symmetrization Invariance of Matrix Gaussian Complexity]
\label{lem:gc-symmetrization}
Let $\mathbf{W}\in\mathbb{R}^{n\times n}$ have i.i.d. $\mathcal{N}(0,1)$ entries. For any $\mathcal{C}\subseteq \mathtt{S}^n$,
\begin{equation}
\mathsf{GC}(\mathcal{C})
= \mathop{\mathbb{E}}_{\mathbf{W} \sim \mathcal{N}(\mathbf{0}, \mathbf{I}_{n \times n})} \, \Big[ \, \sup_{\mathbf{S}\in\mathcal{C}} \, \langle \mathbf{W},\mathbf{S}\rangle\, \Big]
= \mathop{\mathbb{E}}_{\mathbf{W} \sim \mathcal{N}(\mathbf{0}, \mathbf{I}_{n \times n})} \Big[ \, \sup_{\mathbf{S}\in\mathcal{C}}\, \Big\langle \frac{\mathbf{W}+\mathbf{W}^\top}{2},\,\mathbf{S}\Big\rangle \, \Big].   
\end{equation}
Moreover, $\mathbf{G} \coloneqq (\mathbf{W}+\mathbf{W}^\top)/2$ satisfies
$G_{ii} \sim \mathcal{N}(0,1)$ and $G_{ij} \sim \mathcal{N}(0,\tfrac{1}{2})$ for $i<j$.
\end{lemma}

\begin{proof}
For any symmetric $\mathbf{S}$, $\langle \mathbf{W},\mathbf{S}\rangle = \mathrm{Tr} \left( \mathbf{W}^\top\mathbf{S} \right) =\mathrm{Tr} \left( \frac{\mathbf{W} + \mathbf{W}^\top}{2}\,\mathbf{S} \right)$. The distributional claims follow by direct computation.
\end{proof}

\begin{lemma}[Gaussian Complexity of $\mathcal{C}_{\mathrm{query}}$]
\label{lem:gc-box}
Let $\mathcal{C}_{\mathrm{query}} \coloneqq \{\mathbf{s}\in\mathbb{R}^{n}:\ |s_i|\le 1,\ \forall i\in[n]\} \;=\; [-1,1]^n$ and $\mathbf{z} \sim \mathcal{N}(\mathbf{0},\mathbf{I}_n)$. Consider the Gaussian complexity definition \ref{def:gaussian-complexity}. Then
\begin{equation}
\label{eq:gc-box-lemma}
\mathsf{GC}( \mathcal{C}_{\mathrm{query}} )
\;=\; \mathbb{E} \Big[ \sup_{\mathbf{x} \in [{-}1,1]^n}\langle \mathbf{z},\mathbf{x} \rangle \Big] \;=\; \mathbb{E} \Big[ \sum_{i=1}^n |z_i| \Big] \;=\; n\sqrt{\frac{2}{\pi}} = \Theta (n).
\end{equation}
\end{lemma}

\begin{proof}
Fix $\mathbf{z} \in \mathbb{R}^n$. For any $\mathbf{x} \in [{-}1,1]^n$, 
\begin{equation}
\langle \mathbf{z},\mathbf{x}\rangle
=\sum_{i=1}^n z_i x_i
\le \sum_{i=1}^n |z_i|\,|x_i|
\le \sum_{i=1}^n |z_i|.
\end{equation}
Equality is achieved by choosing $x_i = \mathrm{sign}(z_i)$ (with any value in $[-1, 1]$ when $z_i = 0$). Hence 
\begin{equation}
\sup_{\mathbf{x} \in [{-}1,1]^n} \langle \mathbf{z}, \mathbf{x} \rangle
=\sum_{i=1}^n |z_i|.
\end{equation}
Taking expectations gives the first two equalities in Eq.~\ref{eq:gc-box-lemma}. Since the coordinates of $\mathbf{z}$ are i.i.d. $\mathcal{N} (0,1)$ and $\mathbb{E}|Z| = \sqrt{2/\pi}$ for $Z \sim \mathcal{N}(0,1)$,
\begin{equation}
\mathbb{E}\Big[\sum_{i=1}^n |z_i|\Big]
= \sum_{i=1}^n \mathbb{E}|z_i|
= n\sqrt{\frac{2}{\pi}}.
\end{equation}
\end{proof}

\begin{lemma}[Elliptope Equivalence]
\label{lem:elliptope-equivalence}
Let $\mathcal{E}_n \coloneqq \{\mathbf{S}\in\mathtt{S}^n: \mathbf{S} \succeq \mathbf{0}, \mathrm{diag}(\mathbf{S})=\mathbf{1}\}$.
Then every $\mathbf{S}\in\mathcal{E}_n$ satisfies $|S_{ij}|\le 1$ for all $i\neq j$. Consequently, $\mathcal{C}_{\mathsf{coll}} = \bigl\{ \mathbf{S}\in\mathbb{R}^{n\times n}:  \mathbf{S}\succeq \mathbf{0},  \mathrm{diag}(\mathbf{S})=\mathbf{1}, |S_{ij}| \le 1\ (i\neq j) \bigr\} = \mathcal{E}_n$.
\end{lemma}

\begin{proof}
Fix $\mathbf{S}\in\mathcal{E}_n$. Since $\mathbf{S}\succeq\mathbf{0}$, there exist vectors $\{\mathbf{e}_i\}_{i=1}^n$ such that $S_{ij}= \langle \mathbf{e}_i,\mathbf{e}_j\rangle$ and $S_{ii}=\|\mathbf{e}_i\|_2^2$. Because $\mathrm{diag}(\mathbf{S})= \mathbf{1}$, we have $\|\mathbf{e}_i\|_2=1$ for all $i$. Thus for $i\neq j$,
$|S_{ij}|=|\langle \mathbf{e}_i,\mathbf{e}_j\rangle|\le \|\mathbf{e}_i\|_2\|\mathbf{e}_j\|_2=1$.
\end{proof}

\begin{lemma}[Gaussian Complexity of $\mathcal{C}_{\mathsf{coll}}$]
\label{lem:gc-elliptope}
Let $\mathcal{E}_n \;\coloneqq\; \{\mathbf{S}\in\mathtt{S}^n:\ \mathbf{S}\succeq \mathbf{0},\ \mathrm{diag}(\mathbf{S})=\mathbf{1}\}$. There exist universal constants $0<c\le C<\infty$ such that
\begin{equation}
c\,n^{3/2} \; \le \; \mathsf{GC}(\mathcal{E}_n) \; \le \; C\,n^{3/2}.    
\end{equation}
Equivalently, $\mathsf{GC}(\mathcal{C}_{\mathsf{coll}})= \Theta(n^{3/2})$.
\end{lemma}

\begin{proof}
By Lemma~\ref{lem:gc-symmetrization}, with $\mathbf{G}=(\mathbf{W}+ \mathbf{W}^\top)/2$, $\mathsf{GC}(\mathcal{E}_n) = \mathbb{E} \Big[ \sup_{\mathbf{S}\in\mathcal{E}_n} \langle \mathbf{G},\mathbf{S} \rangle \Big]$, where the expectation is over the randomness of $\mathbf{W}$ (equivalently, $\mathbf{G}$).

\noindent
\textit{Upper bound:}
For $\mathbf{S}\in \mathcal{E}_n$, $\mathrm{Tr}(\mathbf{S})=n$, hence
\begin{equation}
\sup_{\mathbf{S}\in\mathcal{E}_n}\langle \mathbf{G},\mathbf{S}\rangle
\le \sup_{\substack{\mathbf{S}\succeq\mathbf{0}\\ \mathrm{Tr}(\mathbf{S})=n}} \langle \mathbf{G},\mathbf{S}\rangle
= n \, \lambda_{\max}(\mathbf{G}),    
\end{equation}
where the equality holds because the right-hand side is achieved by $\mathbf{S}=n\,\mathbf{v} \mathbf{v}^\top$ for any unit top-eigenvector $\mathbf{v}$ of $\mathbf{G}$.
Therefore, taking expectations gives 
\begin{equation}\label{eq:gc-elliptope-upper}
\mathsf{GC}(\mathcal{E}_n)\le n\, \mathop{\mathbb{E}}[ \lambda_{\max} (\mathbf{G}) ].
\end{equation}
Next, note that $\mathbf{G}$ has independent Gaussian entries (up to symmetry) with $G_{ii}\sim \mathcal{N}(0,1)$ and $G_{ij}\sim \mathcal{N}(0,\tfrac12)$ for $i<j$. 
Standard spectral-norm bounds for such Wigner matrices imply $\mathbb{E} \big[ \lambda_{\max}(\mathbf{G}) \big] \le \mathbb{E} \big[\|\mathbf{G}\|_{\mathrm{op}} \big] \le C_0 \sqrt{n}$ for a universal constant $C_0<\infty$, and therefore Eq.~\ref{eq:gc-elliptope-upper} yields $\mathsf{GC} (\mathcal{E}_n) \le C \, n^{3/2}$.

\noindent
\textit{Lower bound:}
For any $\mathbf{s}\in\{\pm 1\}^n$, the rank-one matrix $\mathbf{S}=\mathbf{s}\mathbf{s}^\top$ satisfies $\mathbf{S}\succeq\mathbf{0}$ and $\mathrm{diag}(\mathbf{S})=\mathbf{1}$, hence $\mathbf{s}\mathbf{s}^\top\in\mathcal{E}_n$. Thus 
\begin{equation}
\sup_{\mathbf{S}\in\mathcal{E}_n}\langle \mathbf{G},\mathbf{S}\rangle
\ge \max_{\mathbf{s}\in\{\pm 1\}^n}\langle \mathbf{G},\mathbf{s}\mathbf{s}^\top\rangle
= \max_{\mathbf{s}\in\{\pm 1\}^n}\mathbf{s}^\top \mathbf{G}\mathbf{s}.
\end{equation}
Define the centered Gaussian process $X_{\mathbf{s}} \coloneqq \mathbf{s}^\top \mathbf{G} \mathbf{s}$ indexed by $\mathbf{s} \in \{\pm1\}^n$.

We first choose a subset $T\subseteq\{\pm1\}^n$ whose pairwise Hamming distances are bounded both from below and from above. Specifically, there exists a universal constant $c_0>0$ and a set $T\subseteq\{\pm1\}^n$ such that
\begin{equation}
\label{eq:balanced-code}
|T|\ge \exp(c_0 n),
\qquad
\frac{n}{4}\le d_{\mathrm H}(\mathbf{s},\mathbf{t})\le \frac{3n}{4},
\quad \forall \mathbf{s}\neq \mathbf{t}\in T .
\end{equation}
To see this, draw $M$ independent codewords
$\mathbf{s}^{(1)},\ldots,\mathbf{s}^{(M)}$ uniformly from $\{\pm1\}^n$.
For any fixed pair $a\neq b$, the Hamming distance
$d_{\mathrm H}(\mathbf{s}^{(a)},\mathbf{s}^{(b)})$ has distribution
$\mathrm{Binomial}(n,1/2)$. By Chernoff's bound, for a universal constant
$c_1>0$,
\begin{equation}
\Pr\!\left[
d_{\mathrm H}(\mathbf{s}^{(a)},\mathbf{s}^{(b)})<\frac n4
\ \text{or}\
d_{\mathrm H}(\mathbf{s}^{(a)},\mathbf{s}^{(b)})>\frac{3n}{4}
\right]
\le 2e^{-c_1 n}.
\end{equation}
Taking $M=\lfloor e^{c_0 n}\rfloor$ with $2c_0<c_1$, the union bound gives
\begin{equation}
\Pr\!\left[
\exists\,a<b:\,
d_{\mathrm H}(\mathbf{s}^{(a)},\mathbf{s}^{(b)})\notin[n/4,3n/4]
\right]
\le
M^2\,2e^{-c_1 n}
<1
\end{equation}
for all sufficiently large $n$. Hence a deterministic set $T$ satisfying
Eq.~\ref{eq:balanced-code} exists. In particular,
$\log |T|=\Omega(n)$; finite values of $n$ can be absorbed into the universal constants.

Recall $\mathbf{G}$ is symmetric with independent entries $\{G_{ij}: i\le j\}$, where $\mathrm{Var}(G_{ii}) = 1$ and $\mathrm{Var}(G_{ij}) = 1/2$ for $i< j$. For any $\mathbf{s} \in \{ \pm1 \}^n$, we have
\begin{equation}
X_{\mathbf{s}} = \sum_{i=1}^n G_{ii} s_i^2 + 2\sum_{1 \le i<j \le n} G_{ij} s_i s_j
= \sum_{i=1}^n G_{ii} + 2\sum_{i<j} G_{ij} s_i s_j.
\end{equation}
Hence the diagonal term cancels in differences, and for $\mathbf{s}, \mathbf{t} \in \{\pm1\}^n$, we have $X_{\mathbf{s}} - X_{\mathbf{t}}  = 2\sum_{i<j} G_{ij} \bigl(s_i s_j - t_i t_j \bigr)$. Using independence and $\mathrm{Var}(G_{ij})= 1/2$ for $i<j$, 
\begin{subequations}
\label{eq:xsxt-distance}    
\begin{align}
\mathbb{E} \bigl[ (X_{\mathbf{s}} - X_{\mathbf{t}})^2 \bigr]
& = 4\sum_{i<j} \mathrm{Var}(G_{ij})\, (s_i s_j - t_i t_j)^2\\
&= 4\cdot\frac12\sum_{i<j}(s_i s_j - t_i t_j)^2 = 2\sum_{i<j}( s_i s_j - t_i t_j)^2.
\end{align}
\end{subequations}
Now $(s_i s_j - t_i t_j) \in \{0,\pm2\}$, so $(s_i s_j - t_i t_j)^2 = 4 \cdot \mathbf{1}\{s_i s_j \neq t_i t_j\}$. Let $D = \{i:\ s_i \neq t_i\}$ with $d = |D| = d_\mathrm{H}(\mathbf{s},\mathbf{t})$. Then $s_i s_j \neq t_i t_j$ iff exactly one of $\{i, j\}$ lies in $D$, so the number of such pairs is $| \{(i,j): i<j,\ s_is_j\neq t_it_j\} | =d (n -d)$. Therefore $\mathbb{E} \bigl[ (X_{\mathbf{s}}-X_{\mathbf{t}})^2 \bigr] = 8 \, d (n-d)$. For distinct $\mathbf{s},\mathbf{t}\in T$, Eq.~\ref{eq:balanced-code} gives
$d\in[n/4,3n/4]$. Consequently, $d(n-d) \ge \frac n4\cdot\frac{3n}{4}$, and Eq.~\ref{eq:xsxt-distance} yields
\begin{equation}
\mathbb{E} \bigl[ (X_{\mathbf{s}} - X_{\mathbf{t}})^2 \bigr]
\; \ge \; 8\cdot \frac{n}{4} \cdot \frac{3n}{4} = \frac{3}{2}\, n^2.
\end{equation}
Thus the canonical metric \citep{adler2007random} $d_X (\mathbf{s}, \mathbf{t}) \coloneqq \sqrt{ \mathbb{E}[(X_{\mathbf{s}} - X_{\mathbf{t}})^2] }$ satisfies
\begin{equation}
\inf_{\mathbf{s} \neq \mathbf{t} \in T} d_X(\mathbf{s},\mathbf{t}) \; \ge\; \sqrt{\frac{3}{2}} \, n .
\end{equation}

Sudakov's minoration inequality \citep{talagrand1992sudakov, chu2025talagrand} for Gaussian processes yields
\begin{equation}
\mathbb{E} \Big[ \max_{\mathbf{s} \in T} X_{\mathbf{s}} \Big] \; \ge \;
c\, \Big( \inf_{\mathbf{s} \neq \mathbf{t}\in T} d_X(\mathbf{s},\mathbf{t}) \Big) \, \sqrt{\log|T|}
\; \ge \; c' \, n \sqrt{n} = c' \, n^{3/2},
\end{equation}
for universal constants $c,c'>0$ (using $\log|T|=\Omega(n)$). Since $\max_{\mathbf{s} \in \{\pm1\}^n} X_{\mathbf{s}} \ge \max_{\mathbf{s}\in T} X_{\mathbf{s}}$, 
\begin{equation}
\mathbb{E} \Big[ \max_{\mathbf{s} \in \{\pm1\}^n} \mathbf{s}^\top \mathbf{G} \mathbf{s} \Big] \; \ge \; c'\, n^{3/2},
\end{equation}
and therefore $\mathsf{GC}( \mathcal{E}_n ) \ge c'\, n^{3/2}$.
\end{proof}

\begin{lemma}[Rank-$r$ Gram-manifold Tangent Space and its relation to the Elliptope Tangent Cone]
\label{lem:elliptope_tangent_manifold_vs_cone}
Let $\mathbf{E}= [\mathbf{e}_1^\top; \dots; \mathbf{e}_n^\top] \in \mathbb{R}^{n\times d}$ have unit-norm rows $\|\mathbf{e}_i\|_2=1$, and let $\mathbf{S} = \mathbf{E}\mathbf{E}^\top\in\mathtt{S}^n$ be the associated (cosine) Gram matrix with rank $r\coloneqq \mathrm{rank}(\mathbf{S}) \le \min \{n,d\}$. Consider the elliptope (correlation-matrix set) $\mathcal{E}_n = \bigl\{\mathbf{Y}\in \mathtt{S}^n: \ \mathbf{Y} \succeq \mathbf{0},\ \mathrm{diag}(\mathbf{Y}) = \mathbf{1} \bigr\}$.
Equivalently, $\mathcal{C}_{\mathsf{coll}} = \mathcal{E}_n$ since $|Y_{ij}| \le 1$ for $i\neq j$ is implied by $\mathbf{Y} \succeq \mathbf{0}$ and $\mathrm{diag} (\mathbf{Y})= \mathbf{1}$.
Let $\mathbf{F} \in \mathbb{R}^{n\times r}$ be any rank factor such that $\mathbf{S}= \mathbf{F}\mathbf{F}^\top$, and write $\mathbf{f}_i^\top$ for the $i$-th row of $\mathbf{F}$ (so $\|\mathbf{f}_i\|_2^2=S_{ii}=1$). Define the row-orthogonality constraint set
\begin{equation}
\mathcal{M} \; \coloneqq \; \Bigl\{ \Delta \in \mathbb{R}^{n\times r}:\ \langle \mathbf{f}_i,\Delta_i \rangle =0 , \ \forall i\in[n] \Bigr\},   
\end{equation}
and the associated linear image (the rank-$r$ Gram-manifold tangent space at $\mathbf{S}$)
\begin{equation}
\mathcal{T}_{\mathrm{man}} (\mathbf{S})
\; \coloneqq \; \Bigl\{\mathbf{F} \Delta^\top + \Delta\mathbf{F}^\top :\ \Delta \in \mathcal{M} \Bigr\}
\, \subseteq\ \mathtt{S}^n.
\end{equation}
Then:
\begin{enumerate}[label=(\roman*),leftmargin=2.2em]
\item 
%
%
$\mathcal{T}_{\mathrm{man}}(\mathbf{S})$ is a linear subspace and admits the representation
\begin{equation}
\label{eq:Tman_rep}
\mathcal{T}_{\mathrm{man}}(\mathbf{S}) = \Bigl\{\mathbf{F}\Delta^\top+\Delta\mathbf{F}^\top:\ \Delta\in\mathbb{R}^{n\times r},\ \langle \mathbf{f}_i,\Delta_i\rangle=0, \forall i \in [n] \Bigr\},
\end{equation}
with
\begin{equation}
\label{eq:Tman_dim}
\dim \bigl(\mathcal{T}_{\mathrm{man}}(\mathbf{S})\bigr) \;=\; n(r-1)\;-\;\frac{r(r-1)}{2}.
\end{equation}
\item
%
%
Let $\mathsf{T}_{\mathbf{S}}(\mathcal{E}_n)$ denote the contingent tangent cone of $\mathcal{E}_n$ at $\mathbf{S}$. Then
\begin{equation}
\label{eq:Tman_subset_Tcone}
\mathcal{T}_{\mathrm{man}}(\mathbf{S}) \ \subseteq\
\mathsf{T}_{\mathbf{S}}(\mathcal{E}_n) \ \subseteq\
\bigl\{ \mathbf{H}\in\mathtt{S}^n:\ \mathrm{diag}(\mathbf{H})=\mathbf{0} \bigr\}.
\end{equation}
Consequently, for the statistical dimension $\delta (\cdot)$,
\begin{equation}
\label{eq:statdim_lower}
\delta \bigl( \mathsf{T}_{\mathbf{S}}(\mathcal{E}_n) \bigr) \;\ge\;
\delta \bigl( \mathcal{T}_{\mathrm{man}}(\mathbf{S}) \bigr) \;=\;
\dim \bigl( \mathcal{T}_{\mathrm{man}}(\mathbf{S}) \bigr)
\; = \; n(r-1) \; - \;\frac{r(r-1)}{2}.
\end{equation}
\item 
%
%
If $r= n$ (equivalently, $\mathbf{S}\succ \mathbf{0}$), then $\mathbf{S}$ is an interior point of the PSD constraint and
\begin{equation}
\label{eq:full_rank_cone}
\mathsf{T}_{\mathbf{S}}(\mathcal{E}_n) \; = \; \bigl\{ \mathbf{H}\in\mathtt{S}^n:\ \mathrm{diag}(\mathbf{H})=\mathbf{0} \bigr\},
\qquad \delta \bigl( \mathsf{T}_{\mathbf{S}} (\mathcal{E}_n) \bigr)
\;=\; \frac{n(n-1)}{2} = \Theta (n^2).
\end{equation}
In particular, when $r = n$ one has $\mathcal{T}_{\mathrm{man}}(\mathbf{S})=\mathsf{T}_{\mathbf{S}}(\mathcal{E}_n)$.
\end{enumerate}
\end{lemma}

\begin{proof}
%
Note that if $\mathbf{Y}\succeq \mathbf{0}$ and $\mathrm{diag}(\mathbf{Y})=\mathbf{1}$, then for all $i\neq j$, $|Y_{ij}|\le \sqrt{Y_{ii}Y_{jj}}=1$ by Cauchy--Schwarz for PSD matrices. Hence $\mathcal{C}_{\mathsf{coll}}=\mathcal{E}_n$.

\noindent
\textbf{(i)}
%
%
Consider a smooth perturbation of the factor $\mathbf{F}(t) \coloneqq \mathbf{F}+ t \Delta$
with $\Delta\in\mathbb{R}^{n\times r}$. Then
\begin{equation}
\mathbf{F}(t) \mathbf{F}(t)^\top
= \mathbf{S} + t \,(\mathbf{F}\Delta^\top + \Delta \mathbf{F}^\top) + t^2 \, \Delta \Delta^\top,    
\end{equation}
so the first-order variation of $\mathbf{S}$ induced by $\Delta$ is $\mathbf{H}= \mathbf{F} \Delta^\top +\Delta \mathbf{F}^\top$. Moreover, for each $i \in [n]$,
\begin{equation}
\frac{\mathrm{d}}{\mathrm{d}t} \Big|_{t=0}\ \| \mathbf{f}_i(t) \|_2^2 = \frac{\mathrm{d}}{\mathrm{d}t} \Big|_{t=0} \ \|\mathbf{f}_i + t \Delta_i\|_2^2 = 2 \, \langle \mathbf{f}_i, \Delta_i \rangle.
\end{equation}
Thus the unit-diagonal constraint $\mathrm{diag}( \mathbf{F}(t) \mathbf{F}(t)^\top )= \mathbf{1}$ holds to first order if and only if $\langle \mathbf{f}_i,\Delta_i\rangle = 0$ for all $i$, yielding Eq.~\ref{eq:Tman_rep} and showing that $\mathcal{T}_{\mathrm{man}}(\mathbf{S})$ is a linear subspace.
To compute its dimension, note that the constraints $\langle \mathbf{f}_i,\Delta_i\rangle=0$ are $n$ independent row-wise linear constraints, each reducing the $r$ degrees of freedom in $\Delta_i$ by one because $\|\mathbf{f}_i\|_2=1$. Hence
\begin{equation}
\dim(\mathcal{M}) = n(r-1).
\end{equation}
Define the linear map $\mathcal{L}: \mathcal{M} \to \mathtt{S}^n$ by $\mathcal{L}(\Delta)\coloneqq \mathbf{F} \Delta^\top + \Delta \mathbf{F}^\top$, so $\mathrm{Im}(\mathcal{L}) = \mathcal{T}_{\mathrm{man}}(\mathbf{S})$. Its kernel is
\begin{equation}
\ker(\mathcal{L}) = \bigl\{ \mathbf{F} \boldsymbol{\Omega}:\ \boldsymbol{\Omega}^\top= - \boldsymbol{\Omega} \bigr\}.
\end{equation}
Indeed, if $\Delta= \mathbf{F}\boldsymbol{\Omega}$ with $\boldsymbol{\Omega}^\top = - \boldsymbol{\Omega}$, then $\mathcal{L}(\Delta)= \mathbf{F}(\boldsymbol{\Omega} + \boldsymbol{\Omega}^\top) \mathbf{F}^\top = \mathbf{0}$. Conversely, if $\mathcal{L}(\Delta) = \mathbf{0}$, decompose $\Delta = \mathbf{F} \boldsymbol{\Omega} + \Delta_\perp$ where the columns of $\Delta_\perp$ lie in $\mathrm{range}(\mathbf{F})^\perp$. Then $\mathbf{F}\Delta_\perp^\top + \Delta_\perp\mathbf{F}^\top = \mathbf{0}$ forces $\Delta_\perp=\mathbf{0}$, and $\mathbf{F}(\boldsymbol{\Omega} + \boldsymbol{\Omega}^\top)\mathbf{F}^\top = \mathbf{0}$ implies $\boldsymbol{\Omega}^\top = - \boldsymbol{\Omega}$ since $\mathbf{F}$ has full column rank. Therefore
\begin{equation}
\dim\ker(\mathcal{L}) = \dim\{\boldsymbol{\Omega} \in \mathbb{R}^{r\times r}: \boldsymbol{\Omega}^\top = -\boldsymbol{\Omega}\} = r(r-1)/2,
\end{equation}
and rank--nullity gives Eq.~\ref{eq:Tman_dim}.

\noindent
\textbf{(ii)}
%
%
Fix $\Delta\in\mathcal{M}$ and let $\mathbf{H} \coloneqq \mathbf{F}\Delta^\top + \Delta\mathbf{F}^\top \in \mathcal{T}_{\mathrm{man}}(\mathbf{S})$.
Define the PSD curve
\begin{equation}
\mathbf{X}(t) \; \coloneqq\; (\mathbf{F}+ t\, \Delta)(\mathbf{F}+ t \,\Delta)^\top \; \succeq\; \mathbf{0}.
\end{equation}
As above, $\mathbf{X}(t)=\mathbf{S} + t \, \mathbf{H} + \mathcal{O}(t^2)$ in Frobenius norm. Its diagonal satisfies, for each $i$,
\begin{equation}
\mathrm{diag}(\mathbf{X}(t))_i = \|\mathbf{f}_i + t\Delta_i\|_2^2 = 1+ t^2 \|\Delta_i\|_2^2,
\end{equation}
using $\langle \mathbf{f}_i, \Delta_i \rangle = 0$. Thus $\mathrm{diag}(\mathbf{X}(t))$ is entrywise positive for all $t$, and we may define the diagonal scaling
\begin{equation}
\mathbf{D}(t) \;\coloneqq\; \mathrm{Diag} \big( \mathrm{diag}(\mathbf{X}(t)) \big)^{-1/2}.
\end{equation}
Set the diagonally normalized curve
\begin{equation}
\widetilde{\mathbf{X}}(t) \;\coloneqq\; \mathbf{D}(t)\, \mathbf{X}(t)\, \mathbf{D}(t).
\end{equation}
Then $\widetilde{\mathbf{X}}(t) \succeq \mathbf{0}$ and $\mathrm{diag}(\widetilde{\mathbf{X}}(t)) = \mathbf{1}$ for all $t$, hence $\widetilde{\mathbf{X}}(t) \in \mathcal{E}_n$ for all $t$. Moreover, since $\mathbf{D}(t)= \mathbf{I}+ \mathcal{O} (t^2)$ entrywise and $\mathbf{X}(t) = \mathbf{S}+t\mathbf{H} + \mathcal{O} (t^2)$, we have $\widetilde{\mathbf{X}}(t) = \mathbf{S} + t \, \mathbf{H} + \mathcal{O}(t^2)$.
By the definition of the contingent tangent cone, this implies $\mathbf{H} \in \mathsf{T}_{\mathbf{S}}(\mathcal{E}_n)$, proving the left inclusion in Eq.~\ref{eq:Tman_subset_Tcone}.
For the right inclusion in Eq.~\ref{eq:Tman_subset_Tcone}, note that $\mathrm{diag}(\mathbf{Y}) = \mathbf{1}$ is an affine constraint on $\mathbf{Y}$; hence any feasible first-order velocity at $\mathbf{S}$ must satisfy $\mathrm{diag}(\mathbf{H}) = \mathbf{0}$.
Finally, Eq.~\ref{eq:statdim_lower} follows from (a) monotonicity of statistical dimension under set inclusion and (b) $\delta(\mathcal{U}) = \dim(\mathcal{U})$ for any linear subspace $\mathcal{U}$.

\noindent
\textbf{(iii)}
%
%
If $r = n$, then $\mathbf{S} \succ \mathbf{0}$ is an interior point of the PSD cone, so $\mathsf{T}_{\mathbf{S}}(\mathtt{S}^n_+) = \mathtt{S}^n$. Intersecting with the affine constraint $\mathrm{diag}(\mathbf{Y})=\mathbf{1}$ yields $\mathsf{T}_{\mathbf{S}}(\mathcal{E}_n) = \{\mathbf{H}\in \mathtt{S}^n:\mathrm{diag}(\mathbf{H})= \mathbf{0} \}$.
This is a linear subspace of dimension $n(n-1)/2$, hence its statistical dimension equals $n(n-1)/2$, establishing Eq.~\ref{eq:full_rank_cone}.
\end{proof}

\begin{lemma}[Rank-aware Upper Bound on $\delta \big(\mathsf{T}_{\mathbf{S}}(\mathcal{E}_n)\big)$ Under Gram-Smoothness]
\label{lem:rank_aware_statdim_upper}
Let $\mathbf{S} \in \mathcal{E}_n$ be a correlation matrix with rank $r\le n$, where $\mathcal{E}_n \;\coloneqq\; \{\mathbf{X}\in\mathtt{S}^n:\ \mathbf{X} \succeq \mathbf{0},\ \mathrm{diag}(\mathbf{X}) =\mathbf{1}\}$. Let $\mathbf{S}= \mathbf{F}\mathbf{F}^\top$ for some $\mathbf{F}\in \mathbb{R}^{n\times r}$ with rows $\mathbf{f}_i^\top$ satisfying $\|\mathbf{f}_i\|_2^2 = S_{ii}=1$. Consider the rank-$r$ Gram-manifold tangent space in Eq.~\ref{eq:Tman_rep}. Assume that the Bouligand tangent cone of $\mathcal{E}_n$ at $\mathbf{S}$ coincides with rank-$r$ Gram manifold tangent space (i.e., assume local Gram-smoothness at $\mathbf{S}$):
\begin{equation}
\label{eq:gram_smoothness}
\mathsf{T}_{\mathbf{S}}(\mathcal{E}_n)\; = \; \mathcal{T}_{\mathrm{man}}(\mathbf{S}).
\end{equation}
Then
\begin{equation}
\label{eq:rank_aware_statdim_upper}
\delta \bigl(\mathsf{T}_{\mathbf{S}}(\mathcal{E}_n)\bigr) \; = \; n(r-1)\; - \; \frac{r(r-1)}{2}
\; \le \; nr.
\end{equation}
\end{lemma}

\begin{proof}
Under Eq.~\ref{eq:gram_smoothness}, we have $\delta(\mathsf{T}_{\mathbf{S}}(\mathcal{E}_n)) = \delta(\mathcal{T}_{\mathrm{man}}(\mathbf{S}))$. By Lemma~\ref{lem:elliptope_tangent_manifold_vs_cone}(i), $\mathcal{T}_{\mathrm{man}}(\mathbf{S})$ is a linear subspace with
\begin{equation}
\dim\bigl(\mathcal{T}_{\mathrm{man}}(\mathbf{S})\bigr) = n(r-1)\;-\;\frac{r(r-1)}{2}.
\end{equation}
For any linear subspace $\mathcal{U}$, $\delta(\mathcal{U})= \dim(\mathcal{U})$, hence the first equality in Eq.~\ref{eq:rank_aware_statdim_upper} follows. For the upper bound, note that
\begin{equation}
n(r-1)-\frac{r(r-1)}{2} = (r- 1 ) (n - \frac{r}{2}) \le rn,
\end{equation}
which completes the proof.
\end{proof}

\begin{remark}[Scope of the rank-aware bound]
\label{rem:rank-aware-scope}
Lemma~\ref{lem:rank_aware_statdim_upper} is conditional on the local Gram-smoothness assumption in Eq.~\ref{eq:gram_smoothness}. It is not a general upper bound on $\delta(\mathsf T_{\mathbf S}(\mathcal E_n))$ for every rank-$r$ point of the elliptope. At singular boundary points, the elliptope tangent cone can be strictly larger than the rank-$r$ Gram-manifold tangent space. For example, at a rank-one extreme point $\mathbf S=\mathbf v\mathbf v^\top$ with $\mathbf v\in\{\pm1\}^n$, $\mathcal T_{\mathrm{man}}(\mathbf S)=\{\mathbf 0\}$, while $\mathsf T_{\mathbf S}(\mathcal E_n)$ contains nontrivial zero-diagonal feasible directions. Thus the rank-aware risk bound used in our analysis should be interpreted only on the local smooth stratum where Eq.~\ref{eq:gram_smoothness} holds.
\end{remark}

\appsubsection{Non‑Expansive Operators: Clipping and Euclidean Projection}

\begin{lemma}[Firm Non‑expansiveness of Euclidean Projections]
\label{lem:proj-firm}
Let $\mathcal{C} \subset \mathbb{R}^{n}$ be non‑empty, closed and convex and define the Euclidean projector
\begin{equation}
\mathsf{proj}_{\mathcal{C}}(\mathbf{s}) \;=\; \operatorname*{arg\,min}_{\mathbf{s}'\in\mathcal{C}}\| \mathbf{s} - \mathbf{s}' \|_{2}.    
\end{equation}
Then for all $\mathbf{s},\mathbf{s}'\in \mathbb{R}^{n}$ we have
\begin{equation}\label{eq:firm}
\|\mathsf{proj}_{\mathcal{C}}(\mathbf{s}) - \mathsf{proj}_{\mathcal{C}}(\mathbf{s}') \|_{2}^{2}
\;\le\; \bigl \langle \mathsf{proj}_{\mathcal{C}}(\mathbf{s}) - \mathsf{proj}_{\mathcal{C}}(\mathbf{s}'), \, \mathbf{s}- \mathbf{s}' \bigr \rangle \;\le\; \| \mathbf{s}- \mathbf{s}' \|_{2}^{2}.
\end{equation}
Consequences:

(a) 1‑Lipschitzness.  
Taking square‑roots in Eq.~\ref{eq:firm} gives $\|\mathsf{proj}_{\mathcal{C}}(\mathbf{s}) - \mathsf{proj}_{\mathcal{C}}(\mathbf{s}')\|_{2}\le\|\mathbf{s} - \mathbf{s}' \|_{2}$.

(b) Difference bound. 
For every $\mathbf{w} \in \mathbb{R}^{n}$, $\|\mathsf{proj}_{\mathcal{C}}(\mathbf{s} + \mathbf{w})- \mathsf{proj}_{\mathcal{C}}(\mathbf{s}) \|_{2}\le\|\mathbf{w} \|_{2}$.

(c) Second moment bound.
If $\mathbf{w} \sim \mathcal{N}( \mathbf{0},\sigma^{2} \mathbf{I}_{n})$ then $\mathbb{E}\|\mathsf{proj}_{\mathcal{C}}( \mathbf{s}+ \mathbf{w})- \mathsf{proj}_{\mathcal{C}}(\mathbf{s}) \|_{2}^{2}\le n \,\sigma^{2}$.
\end{lemma}

\begin{proof}
%
Optimality of the projection implies
\begin{equation}
  \langle \mathbf{s} - \mathsf{proj}_{\mathcal{C}}(\mathbf{s}),\mathbf{v} - \mathsf{proj}_{\mathcal{C}}(\mathbf{s}) \rangle \le 0,
  \qquad 
  \langle \mathbf{s}'- \mathsf{proj}_{\mathcal{C}}(\mathbf{s}'), \mathbf{v} - \mathsf{proj}_{\mathcal{C}}(\mathbf{s}') \rangle \le 0,
  \quad \forall \, \mathbf{v}\in \mathcal{C}.
\end{equation}
Choose $\mathbf{v} = \mathsf{proj}_{\mathcal{C}}(\mathbf{s}')$ in the first and $\mathbf{v} = \mathsf{proj}_{\mathcal{C}}(\mathbf{s})$ in the second. 
Adding the two gives
\begin{eqnarray}
      \langle \mathbf{s} - \mathsf{proj}_{\mathcal{C}}(\mathbf{s}), \mathsf{proj}_{\mathcal{C}}(\mathbf{s}') - \mathsf{proj}_{\mathcal{C}}(\mathbf{s}) \rangle
  +\langle \mathbf{s}'- \mathsf{proj}_{\mathcal{C}}(\mathbf{s}') , \mathsf{proj}_{\mathcal{C}}(\mathbf{s}) - \mathsf{proj}_{\mathcal{C}}(\mathbf{s}') \rangle
  \;\le\; 0 .
\end{eqnarray}
Hence
\begin{eqnarray}  
  \langle \mathsf{proj}_{\mathcal{C}}(\mathbf{s}) - \mathsf{proj}_{\mathcal{C}}(\mathbf{s}') ,\mathbf{s} - \mathbf{s}' \rangle
  \;\ge\;
  \| \mathsf{proj}_{\mathcal{C}}(\mathbf{s}) - \mathsf{proj}_{\mathcal{C}}(\mathbf{s}') \|_{2}^{2},
\end{eqnarray}
which is the first inequality in Eq.~\ref{eq:firm}. The second follows by Cauchy–Schwarz.  

Part (a) is immediate; (b) is (a) with $\mathbf{s}' = \mathbf{s}+ \mathbf{w}$; (c) squares (b) and uses $\mathbb{E} \| \mathbf{w} \|_{2}^{2}= n \sigma^{2}$.
\end{proof}

\begin{lemma}[Euclidean Projection Onto the Hypercube]
\label{lem:box-equals-clip}
Let $\mathcal{C}=[-1,1]^n$. The Euclidean projector onto $\mathcal{C}$ satisfies
\begin{equation}
\bigl(\mathsf{proj}_{\mathcal{C}}(\mathbf{s})\bigr)_i = \max\{-1,\min\{1,s_i\}\},\qquad i=1,\dots,n.
\end{equation}
\end{lemma}

\begin{proof}
Since $\mathcal{C}$ is a Cartesian product of intervals, the minimization $\min_{\mathbf{x} \in \mathcal{C}}\|\mathbf{s} - \mathbf{x} \|_2^2 =\min_{\mathbf{x} \in \mathcal{C}} \sum_{i=1}^n (s_i-x_i)^2$ separates across coordinates. Thus each $x_i$ is the minimizer of $\min_{x \in [-1,1]}(s_i-x)^2$, which is the scalar clip of $s_i$ onto $[-1,1]$.
\end{proof}

\begin{corollary}[Pythagorean Inequality for Euclidean Projections]%
\label{cor:pythagoras}
Fix $\mathbf{s} \in \mathbb{R}^{n}$. For every $\mathbf{y} \in \mathcal{C}$,
\begin{equation}
\label{eq:pythagoras-id}
\| \mathbf{s} - \mathbf{y} \|_{2}^{2}
=
\|\mathbf{s} - \mathsf{proj}_{\mathcal{C}}(\mathbf{s})\|_{2}^{2}
+
\| \mathsf{proj}_{\mathcal{C}}(\mathbf{s}) - \mathbf{y} \|_{2}^{2}
+
2\,\bigl\langle
\mathbf{s}- \mathsf{proj}_{\mathcal{C}}(\mathbf{s}),
\mathsf{proj}_{\mathcal{C}}(\mathbf{s})-\mathbf{y}
\bigr\rangle .
\end{equation}
Moreover,
\begin{equation}
\label{eq:cross-ge0}
\bigl\langle
\mathbf{s} - \mathsf{proj}_{\mathcal{C}}(\mathbf{s}),
\mathsf{proj}_{\mathcal{C}}(\mathbf{s}) - \mathbf{y}
\bigr\rangle
\ge 0 .
\end{equation}
Consequently,
\begin{equation}
\label{eq:pythagoras-ineq}
\| \mathsf{proj}_{\mathcal{C}}(\mathbf{s}) -\mathbf{y}\|_{2}^{2}
\le
\|\mathbf{s} - \mathbf{y}\|_{2}^{2}
-
\|\mathbf{s}- \mathsf{proj}_{\mathcal{C}}(\mathbf{s}) \|_{2}^{2},
\qquad \forall \, \mathbf{y} \in\mathcal{C}.
\end{equation}
\end{corollary}

\begin{proof}
The identity Eq.~\ref{eq:pythagoras-id} is the elementary expansion
$\|\mathbf{a}+\mathbf{b}\|_2^2=\|\mathbf{a}\|_2^2+\|\mathbf{b}\|_2^2+2\langle\mathbf{a},\mathbf{b}\rangle$
with
\begin{equation}
\mathbf{a}\coloneqq \mathbf{s}-\mathsf{proj}_{\mathcal{C}}(\mathbf{s}),
\qquad
\mathbf{b}\coloneqq \mathsf{proj}_{\mathcal{C}}(\mathbf{s})-\mathbf{y}.
\end{equation}
Also, $\mathbf{a}+\mathbf{b}=\mathbf{s}-\mathbf{y}$.
The variational inequality for Euclidean projections onto a closed convex set gives
\begin{equation}
\label{eq:proj-vi-pythagoras}
\bigl\langle
\mathbf{s}-\mathsf{proj}_{\mathcal{C}}(\mathbf{s}),
\mathbf{v}-\mathsf{proj}_{\mathcal{C}}(\mathbf{s})
\bigr\rangle
\le 0,
\qquad \forall\,\mathbf{v}\in\mathcal{C}.
\end{equation}
Choosing $\mathbf{v}=\mathbf{y}$ in Eq.~\ref{eq:proj-vi-pythagoras} gives
\begin{equation}
\bigl\langle
\mathbf{s}-\mathsf{proj}_{\mathcal{C}}(\mathbf{s}),
\mathbf{y}-\mathsf{proj}_{\mathcal{C}}(\mathbf{s})
\bigr\rangle
\le 0,
\end{equation}
or equivalently Eq.~\ref{eq:cross-ge0}. Substituting Eq.~\ref{eq:cross-ge0} into
Eq.~\ref{eq:pythagoras-id} gives
\begin{equation}
\|\mathbf{s}-\mathbf y\|_2^2
\ge \|\mathbf{s}-\mathsf{proj}_{\mathcal{C}}(\mathbf{s})\|_2^2
+ \|\mathsf{proj}_{\mathcal{C}}(\mathbf{s})-\mathbf y\|_2^2,
\end{equation}
which is equivalent to Eq.~\ref{eq:pythagoras-ineq}.
\end{proof}

\clearpage
\appsection{Regime~(iii): Per-record Similarity Score Vector Release}
\label{sec:reference-row-regime3}

\paragraph{Operational Scenario.}
In closed-set $1{:}N$ identification at e-gates or corporate turnstiles, duplicate-enrollment audits in civil-ID databases, or when a watch-list entry must be shipped to an air-gapped or mobile device, the back-end needs the \textit{entire similarity profile} of one enrolled identity.
Fix a public index $i \in [n]$ and define
\begin{equation}
f_{\mathsf{ref},i}(\mathbf E)
\coloneqq
\mathbf E\mathbf e_i
=
(\mathbf e_1^\top\mathbf e_i,\dots,\mathbf e_n^\top\mathbf e_i)
\in[-1,1]^n,
\end{equation}
where all embeddings satisfy $\|\mathbf e_j\|_2=1$.
Publishing $f_{\mathsf{ref},i}(\mathbf E)$ once lets downstream systems run top-$k$ searches, threshold checks, or quality audits without further access to the raw embeddings.
We drop the subscript $i$ when the reference index is clear.

\paragraph{Threat Model.}
We consider one-shot, non-interactive release of the $i$-th similarity row
$\mathbf r=f_{\mathsf{ref},i}(\mathbf E)=\mathbf E\mathbf e_i\in[-1,1]^n$ with $r_i=1$.
We consider central $(\varepsilon,\delta)$-DP under record-level adjacency where identity $i$ may change, so the $n-1$ off-diagonal similarities involving identity $i$ may change while the diagonal entry remains equal to one.
The adversary sees only the noisy vector $\widehat{\mathbf r}$, knows the mechanism and all hyperparameters $(\varepsilon,\delta,\sigma,n,i)$, and is computationally unbounded.
We consider: (i) \textit{no side information} about $\mathbf E$; and (ii) \textit{full gallery known}, enabling linear subspace denoising; the diagonal constraint is public ($r_i=1$).
See App.~\ref{subsec:regime3-reference} and App.~\ref{ssec:attacker_regime1}.

\paragraph{Sensitivity.}
Let $\mathbf E'$ differ from $\mathbf E$ only in row $i$, replacing $\mathbf e_i$ by $\mathbf e_i'$. Write $\boldsymbol\delta \coloneqq \mathbf e_i-\mathbf e_i'$, $\|\boldsymbol\delta\|_2\le 2$.
The neighboring outputs are $f_{\mathsf{ref},i}(\mathbf E)=\mathbf E\mathbf e_i$, $f_{\mathsf{ref},i}(\mathbf E')=\mathbf E'\mathbf e_i'$.
For $j\neq i$, the $j$-th component of their difference is $\mathbf e_j^\top\mathbf e_i-\mathbf e_j^\top\mathbf e_i' = \mathbf e_j^\top\boldsymbol\delta$. For the diagonal component,
\begin{equation}
\bigl(f_{\mathsf{ref},i}(\mathbf E)\bigr)_i
- \bigl(f_{\mathsf{ref},i}(\mathbf E')\bigr)_i
= \mathbf e_i^\top\mathbf e_i - {\mathbf e_i'}^\top\mathbf e_i'
= 1-1
= 0.    
\end{equation}
Therefore,
\begin{equation}
f_{\mathsf{ref},i}(\mathbf E)-f_{\mathsf{ref},i}(\mathbf E')
= (\mathbf e_1^\top\boldsymbol\delta,\dots,\mathbf e_{i-1}^\top\boldsymbol\delta,0,
\mathbf e_{i+1}^\top\boldsymbol\delta,\dots,\mathbf e_n^\top\boldsymbol\delta).
\end{equation}
Hence
\begin{equation}
\|f_{\mathsf{ref},i}(\mathbf E)-f_{\mathsf{ref},i}(\mathbf E')\|_2^2
= \sum_{j\neq i}(\mathbf e_j^\top\boldsymbol\delta)^2 
\le \sum_{j\neq i}\|\mathbf e_j\|_2^2\|\boldsymbol\delta\|_2^2 
\le 4(n-1).
\end{equation}
Thus the global $\ell_2$-sensitivity satisfies
\begin{equation}
\Delta_{f,2}\coloneqq \Delta_{\mathsf{ref}}\le 2\sqrt{n-1}.
\end{equation}
This bound is tight in the worst case, so $\Delta_{\mathsf{ref}}=2\sqrt{n-1}$.

\paragraph{Adversary Gain \& Risk Scaling.}
With $\Delta_{\mathrm{ref}}=2\sqrt{n-1}$, the Gaussian mechanism uses per–coordinate noise
$\sigma \ge \Delta_{\mathrm{ref}} \sqrt{c_{\varepsilon,\delta}})$, with $c_{\varepsilon,\delta}= 2\log (2/\delta)/\varepsilon^2$, hence $\sigma=\Theta(\sqrt n)$. For any off–diagonal entry $j\neq i$,
\begin{equation}
\mathrm{SNR}_j \;\coloneqq \; \frac{|\mathbf{e}_i^\top \mathbf{e}_j|}{\sigma}
\;\le\; \frac{1}{\sigma}
\;\le\; \frac{1}{2\sqrt{(n-1) \, c_{\varepsilon,\delta}}}
= \mathcal{O} \Bigl(\frac{\varepsilon}{\sqrt{n \log(1/\delta)}}\Bigr).
\end{equation}
That is each off-diagonal coordinate is perturbed by $\mathcal{N}(0,\sigma^2)$ with per--coordinate SNR at most $\mathcal{O}(1/\sqrt n)$ (since $|\mathbf{e}_i^\top \mathbf{e}_j|\le 1$). If $\mathbf{E}$ is unknown to the attacker, a na\"ive entrywise estimator would then incur per–entry MSE $\sigma^2=\Theta(n)$.

\noindent
\textit{Jointly optimal denoising under auxiliary knowledge.}
Given $\mathbf E$, the clean row lies in the rank-$r$ subspace $\mathsf{col}(\mathbf E)\subseteq\mathbb R^n$. Let $\mathbf r_i'=\mathbf r_i+\mathbf w_i$ denote the noisy row. The affine-unbiased oracle subspace estimator is
$\widehat{\mathbf r}_i^\star=\mathbf P\mathbf r_i'$ with
$\mathbf P\coloneqq \mathbf E(\mathbf E^\top\mathbf E)^\dagger\mathbf E^\top\in\mathbb R^{n\times n}$. Its error law is $\widehat{\mathbf{r}}_i^\star - \mathbf{r}_i \sim \mathcal{N}(0,\sigma^2 P )$, so the $j$-th entry has $\mathsf{MSE}_j = \sigma^2 P_{jj} = \sigma^2 \ell_j$, where $\ell_j\in[0,1]$ are statistical leverage scores of column $j$, with $\sum_{j=1}^n \ell_j = \mathsf{tr} (\mathbf{P} ) = r$. Hence, noting that $\sigma^2 = \Theta (n)$, the average per–entry MSE equals
\begin{equation}
\frac{1}{n}\sum_{j=1}^n \mathsf{MSE}_j
= \frac{\sigma^2}{n}\mathsf{tr}(\mathbf{P})
= \sigma^2 \frac{r}{n}
= \Theta(r), 
\end{equation}
which is constant in $n$ for fixed embedding dimension $d$ (full rank $r=d$), while at most $\mathcal{O}(r)$ high–leverage indices can reach the na\"ive $\Theta(n)$ level. 
Note that each $\mathsf{MSE}_j = \sigma^2 \ell_j$ can be as large as $\sigma^2 = \Theta (n)$ if $\ell_j \approx 1$ (a high-leverage index), but only $\mathcal{O}(r)$ entries can be large because $\sum_j \ell_j = r$. In particular, the median per-entry MSE is $\mathcal{O} (\sigma^2 r / n) = \Theta (r)$.
If face embeddings of $\mathbf{E}$ are approximately isotropic, which is typical in practice for large $n$ with unit-norm embeddings, then $\ell_j \approx r/n, \forall j$ (concentration of leverage scores). In that case, $\mathsf{MSE}_j \approx \sigma^2 \frac{r}{n} = \Theta(r)$, uniformly over $j$, hence, again independent of $n$ for fixed $r$. Therefore, given knowledge of embeddings dataset $\mathbf{E}$, the attacker can denoise jointly and recover the whole row to constant per-entry accuracy (scaling with 
$r$, not  $n$). The na\"ive $\Theta(n)$ per-entry MSE only applies if the attacker ignores structure.
The indistinguishability guarantee for membership inference remains governed by $(\varepsilon,\delta)$
independently of $n$.

\begin{lemma}[Orthogonal subspace denoising]
\label{lem:projection-subspace}
Let $\mathbf r'=\mathbf r+\mathbf w$, $\mathbf w\sim \mathcal N(\mathbf 0,\sigma^2\mathbf I_m)$, where the unknown deterministic vector satisfies $\mathbf r\in\mathcal S$ for a known linear subspace $\mathcal S\subseteq\mathbb R^m$. Let $\mathbf P$ denote the orthogonal projector onto $\mathcal S$, and define $\widehat{\mathbf r}^{\star} \coloneqq \mathbf P\mathbf r' $. Then, for every realization $\mathbf r'$, $\widehat{\mathbf r}^{\star}$ is the unique solution of $\min_{\mathbf z\in\mathcal S} \|\mathbf z-\mathbf r'\|_2^2$ .
Moreover, $\widehat{\mathbf r}^{\star}-\mathbf r = \mathbf P\mathbf w$, and therefore $\mathbb E\|\widehat{\mathbf r}^{\star}-\mathbf r\|_2^2  = \sigma^2\operatorname{tr}(\mathbf P) = \sigma^2\dim(\mathcal S)$. For each coordinate $j$, $\mathbb E\bigl[(\widehat r^{\star}_j-r_j)^2\bigr] =\sigma^2 P_{jj}$. Among all affine unbiased estimators $\widehat{\mathbf r}=\mathbf A\mathbf r'+\mathbf b$ that take values in $\mathcal S$ and satisfy $\mathbb E_{\mathbf r}[\widehat{\mathbf r}]=\mathbf r$ for every $\mathbf r\in\mathcal S$, $\widehat{\mathbf r}^{\star}=\mathbf P\mathbf r'$ is the unique minimizer of the mean squared error.
\end{lemma}

\begin{proof}
Since $\mathcal S$ is a closed linear subspace of $\mathbb R^m$, every $\mathbf r'\in\mathbb R^m$ admits the orthogonal decomposition $\mathbf r' = \mathbf P\mathbf r' + (\mathbf I-\mathbf P)\mathbf r'$, where $\mathbf P\mathbf r'\in\mathcal S$ and $(\mathbf I-\mathbf P)\mathbf r'\in\mathcal S^\perp$.
Thus, for any $\mathbf z\in\mathcal S$, $\|\mathbf z-\mathbf r'\|_2^2
=\|\mathbf z-\mathbf P\mathbf r'\|_2^2 + \|(\mathbf I-\mathbf P)\mathbf r'\|_2^2$.
The second term is independent of $\mathbf z$, and the first term is uniquely minimized at $\mathbf z=\mathbf P\mathbf r'$. Hence
$\widehat{\mathbf r}^{\star}=\mathbf P\mathbf r'$ is the unique Euclidean projection of $\mathbf r'$ onto $\mathcal S$.

Since $\mathbf r\in\mathcal S$, we have $\mathbf P\mathbf r=\mathbf r$. Therefore, $\widehat{\mathbf r}^{\star}-\mathbf r =
\mathbf P(\mathbf r+\mathbf w)-\mathbf r = \mathbf P\mathbf w$.
Consequently,
\begin{equation}
\mathbb E\|\widehat{\mathbf r}^{\star}-\mathbf r\|_2^2
= \mathbb E\|\mathbf P\mathbf w\|_2^2
= \sigma^2\operatorname{tr}(\mathbf P^2)
= \sigma^2\operatorname{tr}(\mathbf P)
= \sigma^2\dim(\mathcal S),    
\end{equation}
where we used $\mathbf P^2=\mathbf P$ and $\operatorname{tr}(\mathbf P)=\dim(\mathcal S)$.
Also, $\operatorname{Cov}(\mathbf P\mathbf w) = \sigma^2\mathbf P$,
which gives $\mathbb E\bigl[(\widehat r^{\star}_j-r_j)^2\bigr] = \sigma^2 P_{jj}$.
It remains to prove the affine-unbiased optimality statement.
Let $\widehat{\mathbf r}=\mathbf A\mathbf r'+\mathbf b$ be affine, take values in $\mathcal S$, and be unbiased for every $\mathbf r\in\mathcal S$.
Unbiasedness at $\mathbf r=\mathbf 0$ gives $\mathbf b=\mathbf 0$.
Unbiasedness for all $\mathbf r\in\mathcal S$ gives $\mathbf A\mathbf r=\mathbf r, \forall \mathbf r\in\mathcal S$, equivalently $\mathbf A\mathbf P=\mathbf P$.
Since the estimator takes values in $\mathcal S$, we also have $\mathbf P\mathbf A=\mathbf A$.
Hence $\mathbf A = \mathbf A\mathbf P+\mathbf A(\mathbf I-\mathbf P)
= \mathbf P+\mathbf A(\mathbf I-\mathbf P)$.
The two terms are orthogonal in Frobenius inner product, so $\|\mathbf A\|_F^2 = \|\mathbf P\|_F^2 + \|\mathbf A(\mathbf I-\mathbf P)\|_F^2 \ge \|\mathbf P\|_F^2$, with equality if and only if $\mathbf A(\mathbf I-\mathbf P)=\mathbf 0$, i.e., $\mathbf A=\mathbf P$. For any such affine unbiased estimator, $\mathbb E\|\widehat{\mathbf r}-\mathbf r\|_2^2 = \sigma^2\|\mathbf A\|_F^2$.
Therefore the unique minimizer is $\mathbf A=\mathbf P$, which gives
$\widehat{\mathbf r}^{\star}=\mathbf P\mathbf r'$.
\end{proof}

\vspace{20pt}

\begin{algorithm}
\caption{DP Reference-Identity Similarity Row Release (Regime~(iii))}
\begin{algorithmic}
\State \strut \textbf{Input:} $\mathbf{E} \in \mathbb{R}^{n \times d}$, $\mathbf{e} \in \mathbb{R}^d$, $\varepsilon > 0$, $\delta \in (0, 1)$
\State \textbf{Output:} $\widehat{\mathbf{s}} \in \mathbb{R}^n$
\State Compute $\mathbf{s} = \mathbf{E} \, \mathbf{e}$
\State \textbf{DP noise:} set $\sigma_{\varepsilon, \delta}^2 \;\gets\;  c_{\varepsilon, \delta} \Delta^2$ 
\State Sample $\mathbf{w} \sim \mathcal{N}\left( \mathbf{0}, \, \sigma_{\varepsilon, \delta}^2 \, \mathbf{I}_{n}\right)$
\State Compute $\mathbf{s}' = \mathbf{s} + \mathbf{w}$
\State Compute $\widehat{\mathbf{s}} = \mathsf{proj}_{\mathcal{C}_\mathsf{ref}} (\mathbf{s}')$, \quad
 where ${\big( \mathsf{proj}_{{[-1, 1]}^n} (\mathbf{s}') \big)}_i = \max(-1, \min(1, s'_i))$
\State \textbf{Return} $\widehat{\mathbf{s}}$
\end{algorithmic}
\label{alg:reference-row}
\end{algorithm}

\vspace{20pt}

We do not analyze regime~(iii) further and focus on the more general regime~(ii) (Sec.~\ref{subsec:all-pairs}).

\clearpage
\appsection{Output–Space Adjacency for DP Face-Recognition}
\label{app:sec:output-space-adjacency}

This appendix formalizes an \textit{output–space} (Gram-level) adjacency as an alternative to the \textit{image-adjacency} used in our framework\footnote{Note that  one record corresponds to one image (equivalently, one embedding), so record-level replacement adjacency coincides with image-level adjacency.}. We show what privacy it delivers, how to calibrate the adjacency radius $\Delta$ to operational threat models, and how it compares, case by case, to our record–level adjacency guarantees and to the na\"ive Gaussian mechanism.

\paragraph{Setting and Notation.}
Let $\mathcal{D}=\{x_1,\dots,x_n\}$ be the enrollment (collection) set, $f_{\boldsymbol\theta}$ a fixed backbone, and
$\mathbf{E}=[\mathbf{e}_1^\top,\ldots,\mathbf{e}_n^\top]^\top \in \mathbb{R}^{n\times d}$ the unit-normalized embeddings ($\|\mathbf{e}_i\|_2=1$). Let $\mathbf{S}=\mathbf{E}\mathbf{E}^\top\in\mathbb{R}^{n\times n}$ be the cosine Gram. Define the feasible set $\mathcal{C}_{\mathsf{coll}}
\;\coloneqq \; \bigl\{ \mathbf{X}\in\mathbb{R}^{n\times n}:\; \mathbf{X}\succeq 0,\;\; X_{ii}=1,\;\; |X_{ij}|\le 1\ (i\neq j) \bigr\}$. We write $c_{\varepsilon,\delta}\coloneqq  2\log(2/\delta)/\varepsilon^2$. In our paper adjacency was \textit{record-adjacency}: $\mathcal{D}\sim\mathcal{D}'$ iff the embeddings $\mathbf{E},\mathbf{E}'$ differ in at most one row.

\begin{definition}[Output–Space (Gram-Matrix) Adjacency]
Two collections with embeddings $\mathbf{E},\mathbf{E}'$ are \textit{adjacent at radius} $\mathsf{\Delta}_{\mathsf{G}} > 0$ if
\begin{equation}
\label{eq:gram-adj}
\|\mathbf{E}\mathbf{E}^\top-\mathbf{E}'\mathbf{E}'^\top\|_{\mathrm{F}} \;\le\; \mathsf{\Delta}_{\mathsf{G}}.
\end{equation}
\end{definition}

A randomized mechanism $\mathcal{M}$ mapping $\mathbf{E}$ to an output in $\mathbb{R}^{n \times n}$ is $(\varepsilon,\delta)$-DP at Gram radius $\mathsf{\Delta}_{\mathsf{G}}$ if for all $\mathbf{E},\mathbf{E}'$ satisfying Eq.~\ref{eq:gram-adj} and all measurable sets $\mathcal{A}$, $\mathsf{Pr} [ \mathcal{M}(\mathbf{E}) \in \mathcal{A}] \le e^{\varepsilon} \mathsf{Pr}[ \mathcal{M}(\mathbf{E}') \in \mathcal{A}] + \delta$.

\begin{remark}
By the Gaussian mechanism on the Euclidean space $\left( \mathtt{S}^n, {\Vert \cdot \Vert}_F \right)$, adding i.i.d. Gaussian noise calibrated to Frobenius sensitivity $\mathsf{\Delta}_{\mathsf{G}}$ yields $(\varepsilon, \delta)$-DP. We therefore set $\sigma = \mathsf{\Delta}_{\mathsf{G}} \sqrt{2 \log (2/\delta)} / \varepsilon$ (equivalently, $\sigma^2 = c_{\varepsilon, \delta} \mathsf{\Delta}_{\mathsf{G}}^2$). Post-processing (projection) preserves DP. 
\end{remark}

\paragraph{ScoreShield under $\mathsf{\Delta}_{\mathsf{G}}$-Gram Adjacency.}
ScoreShield performs \textit{perturb–then–project}:
\begin{equation}
\label{eq:score-shield-gram}
\widehat{\mathbf{S}} \;=\; \mathsf{proj}_{\mathcal{C}_{\mathsf{coll}}}
\! \left(\mathbf{S}+\tfrac12(\mathbf{W}+\mathbf{W}^\top)\right),
\qquad W_{ij}\stackrel{\mathrm{i.i.d.}}{\sim}\mathcal{N}(0,\sigma^2),\;
\sigma^2= c_{\varepsilon,\delta}\Delta^2.
\end{equation}

\paragraph{Semantic Appropriateness of \(\mathsf{\Delta}_{\mathsf{G}}\)-Gram DP.}
\label{app:semantics}
$\mathsf{\Delta}_{\mathsf{G}}$-Gram DP is natural for non-interactive analytics whose object is the \textit{pairwise similarity structure} (e.g., clustering, link-mining, bias auditing). 
This provides $(\varepsilon, \delta)$-DP for any two collections whose Gram matrices are within Frobenius distance $\mathsf{\Delta}_{\mathsf{G}}$, i.e., $\| \mathbf{S} - \mathbf{S}' \|_F \leq \mathsf{\Delta}_{\mathsf{G}}$. 
It is not a person-/image-level guarantee unless $\mathsf{\Delta}_{\mathsf{G}}$ is chosen to upper bound the Gram change induced by any single-image replacement (see Sec.~\ref{app:rel-to-record}).

\appsubsection{%
  Calibrating
  \texorpdfstring{\(\mathsf{\Delta}_{\mathsf{G}}\)}{Delta G}
  to Operational Scenarios%
}
\label{app:calibration}

We provide simple sufficient calibrations with explicit $n, k, m$ dependence, linking $\mathsf{\Delta}_{\mathsf{G}}$ to interpretable threat parameters.

\begin{lemma}[Sparse pairwise perturbations]
\label{lem:pairwise}
Suppose at most $m$ \textit{undirected} pairs $(i,j)$, $i < j$ change and for each such pair
$| \mathit{\Delta} S_{ij}|=|S'_{ij}-S_{ij}|\le \tau$.
Then $\|\mathbf{S}'-\mathbf{S}\|_{\mathrm{F}}^2 \le 2m\,\tau^2$ and hence a sufficient choice is
\begin{equation}
\label{eq:cal-A}
\mathsf{\Delta}_{\mathsf{G}} \;\ge\; \tau\,\sqrt{2m}\, .
\end{equation}
\end{lemma}
\begin{proof}
Each undirected pair contributes two symmetric entries, so $\|\mathbf{S}' - \mathbf{S}\|_{\mathrm{F}}^2 = \sum_{i\ne j}(\mathit{\Delta} S_{ij})^2 \le 2m\tau^2$.
\end{proof}

\begin{lemma}[At most $k$ images drift by at most $\eta$ in $\ell_2$]
\label{lem:k-drift}
Let $\mathit{\Delta} \mathbf{E} = \mathbf{E}'-\mathbf{E}$ have at most $k$ nonzero rows and $\| \mathit{\Delta} \mathbf{e}_i \|_2 \le \eta$ for each changed row. Then
\begin{equation}
\label{eq:cal-B}
\|\mathbf{S}'-\mathbf{S}\|_{\mathrm{F}} \;\le\; 2\eta\sqrt{nk}\;+\;\eta^2 k ,
\end{equation}
hence a sufficient choice is
\begin{equation}
\mathsf{\Delta}_{\mathsf{G}} \;\ge\; 2 \eta \sqrt{nk} \;+\; \eta^2 k .
\end{equation}
\end{lemma}
\begin{proof}
$\mathbf{S}'-\mathbf{S}= \mathbf{E} \mathit{\Delta} \mathbf{E}^\top + \mathit{\Delta} \mathbf{E}\,\mathbf{E}^\top + \Delta\mathbf{E}\, \mathit{\Delta} \mathbf{E}^\top$. Using $\|\mathbf{A} \mathbf{B}^\top\|_{\mathrm{F}}\le \| \mathbf{A} \|_2\|\mathbf{B} \|_{\mathrm{F}}$, $\|\mathbf{E}\|_2\le \|\mathbf{E}\|_{\mathrm{F}}= \sqrt{n}$, and
$\| \mathit{\Delta} \mathbf{E}\|_{\mathrm{F}}\le \eta \,\sqrt{k}$ we have
\begin{equation}
\|\mathbf{S}' - \mathbf{S}\|_{\mathrm{F}}
\le 2\, \| \mathbf{E}\|_2\, \| \mathit{\Delta} \mathbf{E} \|_{\mathrm{F}} + \| \mathit{\Delta} \mathbf{E} \|_{\mathrm{F}}^2 \le 2\, \eta\sqrt{nk} + \eta^2 k,   
\end{equation}
which yields Eq.~\ref{eq:cal-B}. Note that for small $\eta$, the linear term dominates.
\end{proof}

\begin{lemma}[Row-sparse single-image effect]
\label{lem:row-sparse}
If a single image changes but only $m$ of its similarities move by at most $\tau$, then $\|\mathbf{S}' - \mathbf{S}\|_{\mathrm{F}}\le \tau\sqrt{2m}$ and a sufficient choice is $\mathsf{\Delta}_{\mathsf{G}} \ge \tau \sqrt{2m}$.
\end{lemma}

\begin{proof}
This is immediate from Lemma~\ref{lem:pairwise} by taking the set of changed undirected pairs.
\end{proof}

\noindent 
These bounds express the Gram-adjacency radius $\mathsf{\Delta}_{\mathsf{G}}$ as an explicit function of interpretable threat parameters (either $(m, \tau)$ for sparse score changes or $(k, \eta)$ for $k$-record embedding drift).

\appsubsection{Bridging to Image Adjacency}
\label{app:rel-to-record}

Our main setup adopts \text{image-adjacency}: $\mathbf{E},\mathbf{E}'$ differ in a single row. In that case the Gram sensitivity is exact:
\begin{equation}
\label{eq:Delta-img}
\Delta_{F,\mathrm{img}}
\;\coloneqq \; \sup_{\substack{\mathbf{E} \sim \mathbf{E}'}}
\|\mathbf{E}\mathbf{E}^\top-\mathbf{E}'\mathbf{E}'^\top\|_{\mathrm{F}} \;=\; 2\sqrt{2(n-1)} \;=\; \Theta(\sqrt{n}).
\end{equation}

\begin{proposition}[Dominating image-adjacency with Gram-adjacency]
\label{prop:dominate}
If one wishes $\mathsf{\Delta}_{\mathsf{G}}$-Gram DP to imply image-level DP, it is necessary and sufficient to take
\begin{equation}
\label{eq:Delta-necessary}
\mathsf{\Delta}_{\mathsf{G}} \;\ge\; \Delta_{F, \mathrm{img}} \;=\; 2 \,\sqrt{2\,(n-1)}.
\end{equation}
Consequently, the minimum Gaussian variance required to dominate image-adjacency is $\sigma^2_{\min}=c_{\varepsilon,\delta}\Delta_{F,\mathrm{img}}^2 = 8\,c_{\varepsilon,\delta}(n-1)=\Theta(n)$, and any larger $\mathsf{\Delta}_{\mathsf{G}}$ yields proportionally larger variance.
\end{proposition}

\noindent 
Conversely, fixing $\mathsf{\Delta}_{\mathsf{G}} = \mathcal{O}(1)$ yields a distinct guarantee that does not cover arbitrary single-image changes. It protects only changes that only pairs $(\mathbf{E}, \mathbf{E}')$ satisfying $\Vert \mathbf{S} - \mathbf{S}' \Vert_\mathrm{F} \leq \mathsf{\Delta}_{\mathsf{G}}$ (see Lemmas \ref{lem:pairwise}–\ref{lem:row-sparse}).

\clearpage
\appsection{Na\"ive Gaussian vs. ScoreShield Mechanism Under Different Release Regimes}
\label{app:sec:naive-vs-scoreshield}

We compare mechanisms for privatizing similarity statistics derived from unit-normalized face embeddings $\mathbf{E}= [\mathbf{e}_1^\top,\dots,\mathbf{e}_n^\top]^\top\in\mathbb{R}^{n\times d}$ with $\|\mathbf{e}_i\|_2=1$. We use the central model $(\varepsilon,\delta)$-DP and calibrate the Gaussian mechanism to the global $\ell_2$-sensitivity of the (vectorized) released statistic. We denote $c_{\varepsilon,\delta}\coloneqq 2\log(2/\delta)/\varepsilon^2$, and choose the noise variance as $\sigma^2 = c_{\varepsilon,\delta}\,\Delta^2$ (any $\sigma^2\ge c_{\varepsilon,\delta}\,\Delta^2$ is valid).
Unless stated otherwise, all asymptotics are as $n\to\infty$ with $(\varepsilon,\delta)$ fixed.
In particular, $c_{\varepsilon,\delta}=2\log(2/\delta)/\varepsilon^2$ is treated as $\Theta(1)$ in $n$.

\appsubsection{ScoreShield Projection Risk Bounds}
\label{app:ssec:ScoreShield-projection-risk-bounds}

\paragraph{Global Projection Risk via Gaussian Complexity.}
Consider ScoreShield with isotropic Gaussian noise in the ambient Euclidean space. A uniform (set-level) bound controls the squared projection error by the Gaussian complexity of the feasible set as
\begin{equation}
\label{eq:global-risk-gc-intro}
\mathbb{E}\,\| \widehat{\mathbf{S}} - \mathbf{S} \|_{F}^{2} \; \le\; C\,\sigma \,\mathsf{GC}(\mathcal{C}),
\end{equation}
for a universal constant $C>0$; see Lemma~\ref{lem:global-gc-scoreshield}.\footnote{%
Using $\sigma^2= c_{\varepsilon,\delta} \Delta^2$ gives $\sigma= \sqrt{c_{\varepsilon,\delta}}\Delta$, hence
$\mathbb{E} \|\widehat{\mathbf{S}}-\mathbf{S}\|_F^2 \le C\sqrt{c_{\varepsilon,\delta}}\,\Delta\, \mathsf{GC}(\mathcal C) = C\frac{\sqrt{2\log(2/\delta)}}{\varepsilon}\,\Delta\,\mathsf{GC}(\mathcal C)$.}
This guarantee is global (uniform over all $\mathbf{S} \in \mathcal{C}$), does not rely on a small-noise regime, and depends only on the set geometry through $\mathsf{GC}(\mathcal{C})$ (and boundedness of $\mathcal{C}$).

\begin{lemma}[Global ScoreShield Bound via Gaussian Complexity]
\label{lem:global-gc-scoreshield}
Let $m\ge 1$ and let $\mathcal{C}\subset \mathbb{R}^{m}$ be nonempty, closed, convex, and bounded. Fix $\mathbf{s}\in\mathcal{C}$ and let $\mathbf{w} \sim \mathcal{N} (\mathbf{0}, \sigma^{2}\mathbf{I}_{m})$. Consider ScoreShield projector $\widehat{\mathbf{s}} = \mathsf{proj}_{\mathcal{C}}(\mathbf{s}+\mathbf{w})$. Then
\begin{equation}
\label{eq:global-gc-scoreshield}
\mathbb{E}\,\|\widehat{\mathbf{s}}-\mathbf{s}\|_{2}^{2}
\;\le\; 2\, \sigma\,\mathsf{GC}(\mathcal{C}-\mathcal{C})
\;\le\; 4\, \sigma\,\mathsf{GC}(\mathcal{C}),
\end{equation}
where the expectation is over $\mathbf{w}$, $\mathcal{C} -\mathcal{C} \coloneqq \{\mathbf{u}-\mathbf{v}:\mathbf{u},\mathbf{v}\in\mathcal{C}\}$, and for any bounded $\mathcal{A}\subset \mathbb{R}^{m}$, $\mathsf{GC}(\mathcal{A}) \; \coloneqq\; \mathbb{E}_{\mathbf{z} \sim \mathcal{N}(\mathbf{0},\mathbf{I}_{m})} \big[ \sup_{\mathbf{a}\in\mathcal{A}}\langle \mathbf{z},\mathbf{a} \rangle \big]$.
\end{lemma}

\begin{proof}
Let $\widehat{\mathbf{s}} = \mathsf{proj}_{\mathcal{C}}(\mathbf{s} + \mathbf{w})$ and set $\boldsymbol{\Delta}\coloneqq \widehat{\mathbf{s}}-\mathbf{s}$. By optimality of Euclidean projection, for every $\mathbf{y}\in\mathcal{C}$,
\begin{equation}
\label{eq:proj-opt}
\|\mathbf{s}+ \mathbf{w} - \widehat{\mathbf{s}}\|_2^{2}
\; \le  \; \|\mathbf{s} + \mathbf{w}- \mathbf{y}\|_2^{2}.
\end{equation}
Choosing $\mathbf{y} =\mathbf{s}\in\mathcal{C}$ yields
\begin{equation}
\|\mathbf{s}+\mathbf{w} - \widehat{\mathbf{s}}\|_2^{2}
\; \le \; \|\mathbf{w}\|_2^{2}.    
\end{equation}
Expanding the left-hand side,
\begin{equation}
\|\mathbf{s}+\mathbf{w}-\widehat{\mathbf{s}}\|_2^{2}
= \|( \mathbf{s} - \widehat{\mathbf{s}} ) + \mathbf{w}\|_2^{2}
= \|\boldsymbol{\Delta}\|_2^{2}+ \|\mathbf{w}\|_2^{2}
-2\langle \mathbf{w},\boldsymbol{\Delta} \rangle.
\end{equation}
Cancel $\|\mathbf{w}\|_2^{2}$ from both sides to obtain the deterministic inequality
\begin{equation}
\label{eq:key-det}
\|\boldsymbol{\Delta}\|_2^{2}
\;\le\; 2\langle \mathbf{w},\boldsymbol{\Delta}\rangle.
\end{equation}
Since $\widehat{\mathbf{s}},\mathbf{s}\in\mathcal{C}$, we have $\boldsymbol{\Delta} \in \mathcal{C}- \mathcal{C}$, hence
\begin{equation}
\langle \mathbf{w},\boldsymbol{\Delta}\rangle
\;\le\; \sup_{\mathbf{u}\in \mathcal{C}- \mathcal{C}}\langle \mathbf{w},\mathbf{u} \rangle.
\end{equation}
Combining with Eq.~\ref{eq:key-det} and taking expectations gives
\begin{equation}
\label{eq:exp-step}
\mathbb{E} \, \|\widehat{\mathbf{s}}- \mathbf{s}\|_2^{2}
\;\le\; 2\,\mathbb{E}\Big[\sup_{\mathbf{u}\in\mathcal{C} - \mathcal{C}}\langle \mathbf{w}, \mathbf{u}\rangle \Big].
\end{equation}
Write $\mathbf{w} = \sigma \mathbf{z}$ with $\mathbf{z} \sim \mathcal{N}(\mathbf{0},\mathbf{I}_{m})$. Then
\begin{equation}
\mathbb{E}\Big[\sup_{\mathbf{u} \in\mathcal{C}- \mathcal{C}}\langle \mathbf{w},\mathbf{u}\rangle\Big]
= \sigma\, \mathbb{E}\Big[\sup_{\mathbf{u} \in \mathcal{C}- \mathcal{C}} \langle \mathbf{z},\mathbf{u} \rangle\Big]
= \sigma\, \mathsf{GC}(\mathcal{C}-\mathcal{C}),
\end{equation}
which together with Eq.~\ref{eq:exp-step} proves the first inequality in Eq.~\ref{eq:global-gc-scoreshield}.

For the second inequality, for any fixed $\mathbf{z}$,
\begin{equation}
\sup_{\mathbf{u}\in\mathcal{C}-\mathcal{C}}\langle \mathbf{z}, \mathbf{u}\rangle
= \sup_{\mathbf{a},\mathbf{b}\in \mathcal{C}} \langle \mathbf{z},\mathbf{a}- \mathbf{b} \rangle \le \sup_{\mathbf{a}\in \mathcal{C}}\langle \mathbf{z},\mathbf{a}\rangle + \sup_{\mathbf{b}\in \mathcal{C}}\langle -\mathbf{z},\mathbf{b} \rangle.
\end{equation}
Taking expectations and using $-\mathbf{z} \stackrel{d}{=} \mathbf{z}$ yields $\mathsf{GC}(\mathcal{C} -\mathcal{C}) \le 2 \, \mathsf{GC}(\mathcal{C})$.
\end{proof}

\begin{corollary}[Global ScoreShield Bound for Gram Matrices]
\label{cor:global-gc-scoreshield-matrix}
Let $\mathcal{H}\coloneqq(\mathbb{R}^{n\times n},\langle\cdot,\cdot\rangle_F)$ with Frobenius norm $\|\cdot\|_F$, and let $\mathcal{C}\subset \mathbb{R}^{n\times n}$ be nonempty, closed, convex, and bounded.
Fix $\mathbf{S}\in\mathcal{C}$ and let $\mathbf{W}\in\mathbb{R}^{n\times n}$ satisfy $\mathrm{vec}(\mathbf{W}) \sim \mathcal{N}(\mathbf{0},\sigma^{2}\mathbf{I}_{n^{2}})$. Consider the ScoreShield projector $\widehat{\mathbf{S}} = \mathsf{proj}_{\mathcal{C}}(\mathbf{S}+\mathbf{W})$, where $\mathsf{proj}_{\mathcal{C}}$ is the Frobenius (metric) projection onto $\mathcal{C}$.
Then
\begin{equation}
\mathbb{E} \, \|\widehat{\mathbf{S}} - \mathbf{S} \|_{F}^{2}
\; \le \; 2 \sigma\,\mathsf{GC}(\mathcal{C}-\mathcal{C})
\; \le \; 4 \sigma \,\mathsf{GC}(\mathcal{C}),
\end{equation}
where $\mathcal{C}-\mathcal{C}\coloneqq\{\mathbf{U}-\mathbf{V}:\mathbf{U},\mathbf{V}\in\mathcal{C}\}$ and, for any bounded $\mathcal{A}\subset\mathbb{R}^{n\times n}$, we have 
\begin{equation}
\mathsf{GC}(\mathcal{A})
\;\coloneqq\;
\mathbb{E}\Big[\sup_{\mathbf{A}\in\mathcal{A}}\langle \mathbf{Z},\mathbf{A}\rangle_F\Big],
\qquad \mathrm{vec}(\mathbf{Z})\sim\mathcal{N}(\mathbf{0},\mathbf{I}_{n^{2}}).
\end{equation}
\end{corollary}

\begin{proof}
Define the linear isometry $\mathrm{vec}:\left(\mathbb{R}^{n\times n},\|\cdot\|_F\right)\to\left(\mathbb{R}^{n^{2}},\|\cdot\|_2\right)$.
Let $\widetilde{\mathcal{C}}\coloneqq \mathrm{vec}(\mathcal{C})\subset\mathbb{R}^{n^{2}}$ and $\widetilde{\mathbf{s}}\coloneqq\mathrm{vec}(\mathbf{S})$.
Since $\mathrm{vec}$ is an isometry and $\mathcal{C}$ is closed convex, we have
\begin{equation}
\mathrm{vec}\!\left(\mathsf{proj}_{\mathcal{C}}(\mathbf{Y})\right)
=\mathsf{proj}_{\widetilde{\mathcal{C}}}\!\left(\mathrm{vec}(\mathbf{Y})\right),
\qquad \forall\,\mathbf{Y}\in\mathbb{R}^{n\times n}.
\end{equation}
Moreover, by assumption $\widetilde{\mathbf{w}}\coloneqq \mathrm{vec}(\mathbf{W})\sim \mathcal{N}(\mathbf{0},\sigma^{2}\mathbf{I}_{n^{2}})$.
Therefore $\mathrm{vec}(\widehat{\mathbf{S}}) =\mathsf{proj}_{\widetilde{\mathcal{C}}}(\widetilde{\mathbf{s}}+\widetilde{\mathbf{w}})$.

Apply Lemma~\ref{lem:global-gc-scoreshield} with $m=n^{2}$ to the convex set $\widetilde{\mathcal{C}}\subset\mathbb{R}^{n^{2}}$. Finally, use $\|\widehat{\mathbf{S}}-\mathbf{S}\|_{F}=\|\mathrm{vec}(\widehat{\mathbf{S}})-\mathrm{vec}(\mathbf{S})\|_{2}$ and note that gaussian complexities are equal by the identity $\langle \mathrm{vec}(\mathbf{Z}),\mathrm{vec}(\mathbf{A})\rangle = \langle \mathbf{Z},\mathbf{A}\rangle_F$. This yields the stated bound.
\end{proof}

\paragraph{Projection Risk via Tangent Cones.}
We next state a point-dependent (instance-dependent) bound in terms of the tangent cone geometry at the true point. Let $m\ge 1$ and let $\mathcal{C} \subset \mathbb{R}^{m}$ be nonempty, closed, and convex. Fix $\mathbf{s} \in \mathcal{C}$ and let $\mathbf{w} \sim \mathcal{N}(\mathbf{0}, \sigma^{2} \mathbf{I}_{m})$. Let $\widehat{\mathbf{s}}=\mathsf{proj}_{\mathcal{C}}(\mathbf{s}+ \mathbf{w})$. Then the squared error is controlled by the statistical dimension of the tangent cone $\mathsf{T}_{\mathbf{s}}(\mathcal{C})$:
\begin{equation}
\label{eq:local-risk-tcone-intro}
\mathbb{E} \,\| \widehat{\mathbf{s}} - \mathbf{s} \|_{2}^{2}
\; \le \; \sigma^{2}\,\delta \! \left(\mathsf{T}_{\mathbf{s}}(\mathcal{C}) \right),
\end{equation}
as formalized in Lemma~\ref{lem:local-tc-scoreshield}.
Unlike the global Gaussian-complexity bound, Eq.~\ref{eq:local-risk-tcone-intro} is local and can be substantially smaller when $\mathsf{T}_{\mathbf{s}}(\mathcal{C})$ is low-dimensional (e.g., due to rank structure).

\begin{lemma}[Local ScoreShield Bound via Tangent Cones]
\label{lem:local-tc-scoreshield}
Let $m\ge 1$ and let $\mathcal{C}\subset\mathbb{R}^{m}$ be nonempty, closed, and convex. Fix $\mathbf{s}\in\mathcal{C}$ and let $\mathbf{w} \sim\mathcal{N}(\mathbf{0}, \sigma^{2}\mathbf{I}_{m})$. Let $\widehat{\mathbf{s}} = \mathsf{proj}_{\mathcal{C}}(\mathbf{s}+ \mathbf{w})$. Then
\begin{equation}
\label{eq:local-tc-bound}
\mathbb{E}\, \| \widehat{\mathbf{s}} - \mathbf{s} \|_{2}^{2}
\; \le \; \sigma^{2}\, \delta \! \left( \mathsf{T}_{\mathbf{s}}(\mathcal{C}) \right),
\end{equation}
where $\mathsf{T}_{\mathbf{s}}(\mathcal{C})$ is the tangent cone of $\mathcal{C}$ at $\mathbf{s}$ (see Definition~\ref{def:tangent-cone}) and
\begin{equation}
\delta(\mathcal{K}) \coloneqq  \mathbb{E}\bigl[\|\mathsf{proj}_{\mathcal{K}}(\mathbf{z})\|_{2}^{2}\bigr],
\qquad \mathbf{z}\sim\mathcal{N}(\mathbf{0},\mathbf{I}_{m}),
\end{equation}
denotes the statistical dimension of a closed convex cone $\mathcal{K} \subset \mathbb{R}^{m}$.
\end{lemma}

\begin{proof}
Let $\boldsymbol{\Delta} \coloneqq \widehat{\mathbf s}-\mathbf s$ and define the translated set
$\mathcal D \coloneqq \mathcal C-\mathbf s = \{\mathbf y-\mathbf s:\mathbf y\in\mathcal C\}$.
Then $\mathcal D$ is closed and convex, $\mathbf 0\in\mathcal D$, and by translation invariance of Euclidean projection,
$\boldsymbol{\Delta} = \mathsf{proj}_{\mathcal D}(\mathbf w)$.
Let $\mathcal K \coloneqq \mathsf T_{\mathbf 0}(\mathcal D)$. Since $\mathcal D$ is convex and contains $\mathbf 0$, its tangent cone at the origin is
\begin{equation}
\mathcal K = \mathsf T_{\mathbf 0}(\mathcal D) = \mathsf{cl}\bigl(\mathsf{cone}(\mathcal D)\bigr),    
\end{equation}
and hence $\mathcal D\subseteq\mathcal K$. Moreover, $\mathcal K=\mathsf T_{\mathbf s}(\mathcal C)$ by translation invariance of tangent cones.

We next prove the deterministic inequality
\begin{equation}
\label{eq:projD-projK-deterministic}
\|\mathsf{proj}_{\mathcal D}(\mathbf w)\|_2
\le
\|\mathsf{proj}_{\mathcal K}(\mathbf w)\|_2 .
\end{equation}
Set $\boldsymbol{\Delta}=\mathsf{proj}_{\mathcal D}(\mathbf w)$. Since $\mathbf 0\in\mathcal D$, the variational inequality for Euclidean projection gives
\begin{equation}
\langle \mathbf w-\boldsymbol{\Delta},\mathbf 0-\boldsymbol{\Delta}\rangle \le 0.
\end{equation}
Equivalently,
\begin{equation}
\label{eq:delta-vi-norm}
\|\boldsymbol{\Delta}\|_2^2 \le \langle \mathbf w,\boldsymbol{\Delta}\rangle .
\end{equation}
Because $\boldsymbol{\Delta}\in\mathcal D\subseteq\mathcal K$, if $\boldsymbol{\Delta}\neq\mathbf 0$, then
$\boldsymbol{\Delta}/\|\boldsymbol{\Delta}\|_2\in\mathcal K\cap\mathbb S^{m-1}$. Therefore
\begin{equation}
\label{eq:wdelta-support}
\langle \mathbf w,\boldsymbol{\Delta}\rangle
\le
\|\boldsymbol{\Delta}\|_2
\sup_{\mathbf v\in\mathcal K,\ \|\mathbf v\|_2\le1}
\langle \mathbf w,\mathbf v\rangle .
\end{equation}
We now identify the support function of $\mathcal K\cap\mathbb B_2$. Let
\begin{equation}
\mathcal K^\circ
\coloneqq
\{\mathbf q\in\mathbb R^m:\langle \mathbf q,\mathbf v\rangle\le0,\ \forall \mathbf v\in\mathcal K\}
\end{equation}
denote the polar cone. By Moreau's decomposition for closed convex cones \citep{moreau1962decomposition, bauschke2020correction},
\begin{equation}
\label{eq:moreau-decomposition}
\mathbf w = \mathsf{proj}_{\mathcal K}(\mathbf w)
+ \mathsf{proj}_{\mathcal K^\circ}(\mathbf w),
\qquad
\left\langle \mathsf{proj}_{\mathcal K}(\mathbf w), \mathsf{proj}_{\mathcal K^\circ}(\mathbf w) \right\rangle = 0 .
\end{equation}
Since $\mathsf{proj}_{\mathcal K^\circ}(\mathbf w)\in\mathcal K^\circ$, for every $\mathbf v\in\mathcal K$ with $\|\mathbf v\|_2\le1$ we have
\begin{equation}
\langle \mathbf w,\mathbf v\rangle
= \langle \mathsf{proj}_{\mathcal K}(\mathbf w),\mathbf v\rangle
+ \langle \mathsf{proj}_{\mathcal K^\circ}(\mathbf w),\mathbf v\rangle
\le \langle \mathsf{proj}_{\mathcal K}(\mathbf w),\mathbf v\rangle
\le \|\mathsf{proj}_{\mathcal K}(\mathbf w)\|_2 .
\end{equation}
If $\mathsf{proj}_{\mathcal K}(\mathbf w)\neq\mathbf 0$, equality is attained by
$\mathbf v=\mathsf{proj}_{\mathcal K}(\mathbf w)/\|\mathsf{proj}_{\mathcal K}(\mathbf w)\|_2$. If $\mathsf{proj}_{\mathcal K}(\mathbf w)=\mathbf 0$, the supremum is zero by the preceding inequality. Hence
\begin{equation}
\label{eq:support-cone-ball}
\sup_{\mathbf v\in\mathcal K,\ \|\mathbf v\|_2\le1}
\langle \mathbf w,\mathbf v\rangle
= \|\mathsf{proj}_{\mathcal K}(\mathbf w)\|_2 .
\end{equation}
Combining Eqs.~\eqref{eq:delta-vi-norm}, \eqref{eq:wdelta-support}, and \eqref{eq:support-cone-ball} gives
\begin{equation}
\|\boldsymbol{\Delta}\|_2^2
\le \|\boldsymbol{\Delta}\|_2
\|\mathsf{proj}_{\mathcal K}(\mathbf w)\|_2 .
\end{equation}
Thus
\begin{equation}
\|\boldsymbol{\Delta}\|_2
\le \|\mathsf{proj}_{\mathcal K}(\mathbf w)\|_2 ,
\end{equation}
with the conclusion trivial when $\boldsymbol{\Delta}=\mathbf 0$. Hence Eq.~\eqref{eq:projD-projK-deterministic} holds, and
\begin{equation}
\|\widehat{\mathbf s}-\mathbf s\|_2^2
= \|\boldsymbol{\Delta}\|_2^2
\le \|\mathsf{proj}_{\mathcal K}(\mathbf w)\|_2^2 .
\end{equation}

Taking expectations over $\mathbf w\sim\mathcal N(\mathbf 0,\sigma^2\mathbf I_m)$ gives
\begin{equation}
\mathbb E\|\widehat{\mathbf s}-\mathbf s\|_2^2
\le \mathbb E\|\mathsf{proj}_{\mathcal K}(\mathbf w)\|_2^2 .
\end{equation}
Writing $\mathbf w=\sigma\mathbf z$ with
$\mathbf z\sim\mathcal N(\mathbf 0,\mathbf I_m)$ and using positive homogeneity of projection onto a cone,
$\mathsf{proj}_{\mathcal K}(\sigma\mathbf z)=\sigma\,\mathsf{proj}_{\mathcal K}(\mathbf z)$ for $\sigma\ge0$, we obtain
\begin{equation}
\mathbb E\|\mathsf{proj}_{\mathcal K}(\mathbf w)\|_2^2
= \sigma^2
\mathbb E\|\mathsf{proj}_{\mathcal K}(\mathbf z)\|_2^2
= \sigma^2\delta(\mathcal K).
\end{equation}
Finally, since $\mathcal K=\mathsf T_{\mathbf s}(\mathcal C)$, we get
\begin{equation}
\mathbb E\|\widehat{\mathbf s}-\mathbf s\|_2^2
\le \sigma^2\delta\bigl(\mathsf T_{\mathbf s}(\mathcal C)\bigr).
\end{equation}
\end{proof}

\begin{corollary}[Local ScoreShield bound for matrix inputs under i.i.d. Gaussian noise]
\label{cor:local-tc-scoreshield-matrix}
Let $\mathcal{H}\coloneqq(\mathbb{R}^{n\times n},\langle\cdot,\cdot\rangle_F)$ with Frobenius norm $\|\cdot\|_F$, and let $\mathcal{C}\subset\mathbb{R}^{n\times n}$ be nonempty, closed, and convex.
Fix $\mathbf{S}\in\mathcal{C}$ and let $\mathbf{W}\in\mathbb{R}^{n\times n}$ satisfy $\mathrm{vec}(\mathbf{W}) \sim \mathcal{N}(\mathbf{0},\sigma^{2}\mathbf{I}_{n^{2}})$. Consider the ScoreShield projector $\widehat{\mathbf{S}}\;\coloneqq\;\mathsf{proj}_{\mathcal{C}}(\mathbf{S}+\mathbf{W})$, where $\mathsf{proj}_{\mathcal{C}}$ denotes the metric projection in Frobenius norm. Then
\begin{equation}
\mathbb{E}\,\|\widehat{\mathbf{S}}-\mathbf{S}\|_{\mathrm{F}}^{2}
\;\le\; \sigma^{2}\,\delta \left(\mathsf{T}_{\mathbf{S}}(\mathcal{C})\right),
\end{equation}
where $\mathsf{T}_{\mathbf{S}}(\mathcal{C})$ is the tangent cone of $\mathcal{C}$ at $\mathbf{S}$ in $\mathcal{H}$, and for any closed convex cone $\mathcal{K}\subset\mathbb{R}^{n\times n}$,
\begin{equation}
\delta(\mathcal{K}) = \mathbb{E}\bigl[\|\mathsf{proj}_{\mathcal{K}}(\mathbf{Z})\|_{\mathrm{F}}^{2}\bigr],
\qquad \mathrm{vec}(\mathbf{Z})\sim\mathcal{N}(\mathbf{0},\mathbf{I}_{n^{2}}).
\end{equation}
\end{corollary}

\begin{proof}
Let $L\coloneqq \mathrm{vec}:(\mathbb{R}^{n\times n},\|\cdot\|_F)\to(\mathbb{R}^{n^{2}},\|\cdot\|_2)$ be the canonical linear isometry. Set $\widetilde{\mathcal{C}}\coloneqq L(\mathcal{C})\subset\mathbb{R}^{n^{2}}$,
$\widetilde{\mathbf{s}}\coloneqq L(\mathbf{S})$, and $\widetilde{\mathbf{w}}\coloneqq L(\mathbf{W})$.
By assumption, $\widetilde{\mathbf{w}}\sim\mathcal{N}(\mathbf{0},\sigma^{2}\mathbf{I}_{n^{2}})$.
Since $L$ is an isometry and $\mathcal{C}$ is closed convex, metric projections commute with $L$:
\begin{equation}
L\!\left(\mathsf{proj}_{\mathcal{C}}(\mathbf{Y})\right)
= \mathsf{proj}_{\widetilde{\mathcal{C}}}\!\left(L(\mathbf{Y})\right),
\qquad \forall\,\mathbf{Y}\in\mathbb{R}^{n\times n}.
\end{equation}
Hence
\begin{equation}
L(\widehat{\mathbf{S}}) = \mathsf{proj}_{\widetilde{\mathcal{C}}}(\widetilde{\mathbf{s}}+\widetilde{\mathbf{w}}).
\end{equation}
Apply Lemma~\ref{lem:local-tc-scoreshield} with $m=n^{2}$ to the convex set $\widetilde{\mathcal{C}}\subset\mathbb{R}^{n^{2}}$:
\begin{equation}
\mathbb{E}\,\|L(\widehat{\mathbf{S}})-L(\mathbf{S})\|_{2}^{2}
\;\le\; \sigma^{2}\,\delta\!\left(\mathsf{T}_{\widetilde{\mathbf{s}}}(\widetilde{\mathcal{C}})\right).
\end{equation}
Using $\|L(\widehat{\mathbf{S}})-L(\mathbf{S})\|_2=\|\widehat{\mathbf{S}}-\mathbf{S}\|_F$, it remains to relate tangent cones and statistical dimensions under $L$.
By the definition $\mathsf{T}_{\mathbf{S}}(\mathcal{C})=\mathrm{cl}\,\mathrm{cone}(\mathcal{C}-\mathbf{S})$ and linearity of $L$,
\begin{equation}
L\!\left(\mathsf{T}_{\mathbf{S}}(\mathcal{C})\right)
= \mathsf{T}_{L(\mathbf{S})}(L(\mathcal{C}))
= \mathsf{T}_{\widetilde{\mathbf{s}}}(\widetilde{\mathcal{C}}).
\end{equation}
Moreover, for $\mathbf{Z}$ with $\mathrm{vec}(\mathbf{Z})\sim\mathcal{N}(\mathbf{0},\mathbf{I}_{n^{2}})$, we have
\begin{equation}
\delta\!\left(\mathsf{T}_{\widetilde{\mathbf{s}}}(\widetilde{\mathcal{C}})\right)
= \mathbb{E}\Big[\big\|\mathsf{proj}_{L(\mathsf{T}_{\mathbf{S}}(\mathcal{C}))}(L(\mathbf{Z}))\big\|_2^2\Big]
= \mathbb{E}\Big[\big\|\mathsf{proj}_{\mathsf{T}_{\mathbf{S}}(\mathcal{C})}(\mathbf{Z})\big\|_F^2\Big]
= \delta \left(\mathsf{T}_{\mathbf{S}}(\mathcal{C})\right),
\end{equation}
where the middle equality uses that $L$ is an isometry and projections commute with $L$ on closed convex cones. Substituting completes the proof.
\end{proof}

\vspace{3pt}
We now use these tools in our three regimes.

\appsubsection{Regime (i): Query-to-Collection Similarity Score Vector}
\label{subsec:regime1-compare}

%
Given a fixed query $\mathbf{q}\in\mathbb{R}^d$ with $\|\mathbf{q}\|_2=1$, the released vector is $\mathbf{s}=f_{\mathsf{query}}(\mathbf{E},\mathbf{q})=\mathbf{E}\mathbf{q}\in[-1,1]^n$. Under record–level replacement adjacency on $\mathbf{E}$), only the corresponding coordinate may change, and the global $\ell_2$-sensitivity is constant $\Delta_{\mathsf{query}}=2$.
We therefore calibrate the Gaussian mechanism with variance $\sigma^2 = c_{\varepsilon,\delta}\,\Delta_{\mathsf{query}}^2 = 4\,c_{\varepsilon,\delta}$.

\vspace{2pt}

\paragraph{Na\"ive Gaussian Mechanism.}
Release $\mathbf{s}'= \mathbf{s}+\mathbf{w}$, $\mathbf{w}\sim\mathcal{N}( \mathbf{0},\sigma^2 \mathbf{I}_n)$ with $\sigma^2=c_{\varepsilon,\delta}\Delta_{\mathsf{query}}^2=4c_{\varepsilon,\delta}$. Then we have
\begin{equation}\label{eq:naive-regime1}
\mathbb{E}\,\| \mathbf{s}' - \mathbf{s} \|_2^2 \, = \,  \mathbb{E}\|\mathbf{w}\|_2^2 \, =\, n\,\sigma^2 \; = \; 4\, c_{\varepsilon,\delta}\,n.
\end{equation}

\paragraph{ScoreShield Mechanism.}

Let $\widehat{\mathbf{s}}= \mathsf{proj}_{\mathcal{C}_{\mathsf{query}}}(\mathbf{s} + \mathbf{w})$. We have:

\noindent
\textbf{Local (Pointwise) Risk Bound via the Tangent Cone.}
For the constraint set $\mathcal{C}_{\mathsf{query}} =[-1,1]^n$, the tangent cone at $\mathbf{s}$ is a product cone whose statistical dimension depends only on the active set. Let $a(\mathbf{s})$ denote the number of coordinates of $\mathbf{s}$ on the boundary $\{\pm1\}$. Then $\delta (\mathsf{T}_\mathbf{s} ( (\mathcal{C})) )=n-\tfrac12 a(\mathbf{s})$ (see Lemma~\ref{cor:polyhedral-cube-exact} for a proof), and the local conic denoising bound (Lemma~\ref{lem:local-tc-scoreshield}) yields
\begin{equation}
\label{eq:ScoreShield-regime1}
\mathbb{E}\|\widehat{\mathbf{s}}-\mathbf{s}\|_2^2
\;\le\; \sigma^2\Bigl(n-\tfrac12 a(\mathbf{s})\Bigr)
\;=\; 4\,c_{\varepsilon,\delta}\Bigl(n-\tfrac12 a(\mathbf{s})\Bigr).    
\end{equation}
If no coordinate is active, $a(\mathbf{s})=0$ and $\delta(\mathsf{T}_{\mathbf{s}})=n$, i.e., the bound matches the na\"ive risk in Eq.~\ref{eq:naive-regime1}. If many coordinates are saturated (large $a(\mathbf{s})$), projection can reduce the bound by up to a factor $\frac{1}{2}$ on those coordinates.
Since $\Delta_{\mathsf{query}}$ is constant, the per-coordinate MSE is $\mathcal{O}(1)$ for both
mechanisms.

\noindent
\textbf{Global Risk Bound via the Gaussian Complexity.}
Let $\mathbf{z} \sim \mathcal{N} (\mathbf{0}, \mathbf{I}_n)$, then using Lemma~\ref{lem:gc-box}, we have
\begin{equation}
\label{eq:gc-box}
\mathsf{GC}(\mathcal{C}_{\mathsf{query}})
=\mathbb{E} \Big[\sup_{\mathbf{x} \in[-1,1]^n} \langle \mathbf{z}, \mathbf{x} \rangle\Big] 
= n\sqrt{\frac{2}{\pi}} .
\end{equation}
Applying the global ScoreShield bound (Lemma~\ref{lem:global-gc-scoreshield}) gives the uniform estimate
\begin{equation}
\label{eq:global-gc-regime1}
\mathbb{E}\|\widehat{\mathbf{s}}-\mathbf{s}\|_2^2
\;\le\; C\,\sigma\,\mathsf{GC}(\mathcal{C}_{\mathsf{query}})
\;=\; C\,\sigma\,n\sqrt{\frac{2}{\pi}},
\end{equation}
for a universal constant $C > 0$. 
This bound is uniform in $\mathbf{s} \in [-1,1]^n$ but does not exploit the local active-set geometry captured by the tangent-cone bound in Eq.~\ref{eq:ScoreShield-regime1}. 
Because $\mathsf{GC}(\mathcal{C}_{\mathsf{query}}) = \Theta(n)$, the global bound scales as $\Theta(\sigma \,n)$, whereas the local tangent-cone bound scales as $\Theta( \sigma^2 n)$.
Since $\sigma$ is fixed by privacy calibration, either inequality may be numerically smaller depending on the regime.
In particular, for the common high-privacy regime where $\sigma$ is not small, the two bounds are of comparable order up to constants, but only the tangent-cone bound captures the reduction by $a(\mathbf{s})$.

\appsubsection{Regime (iii): Per-Record Similarity Score Vector}
\label{subsec:regime3-reference}

%
Fix $i\in[n]$ and released the similarity vector $\mathbf{r} = f_{\text{ref}} (\mathbf{E},\mathbf{e}_i)=\mathbf{E}\mathbf{e}_i\in[-1,1]^n$, which satisfies $r_i = 1$. Under record-level replacement adjacency, only the off-diagonal coordinates can change, and $\Delta_{\text{ref}} = 2\sqrt{n-1} = \Theta(\sqrt{n})$. See the exact derivation in Sec.~\ref{sec:reference-row-regime3}. Therefore $\sigma^2= c_{\varepsilon,\delta}\Delta_{\text{ref}}^2 = 4c_{\varepsilon,\delta}(n-1)$.

\paragraph{Na\"ive Gaussian Mechanism.}
Release $\mathbf{r}' = \mathbf{r} +\mathbf{w}$, $\mathbf{w} \sim\mathcal{N} ( \mathbf{0},\sigma^2 \mathbf{I}_n)$, $\sigma^2 = c_{\varepsilon,\delta}\, \Delta_{\text{ref}}^2 = 4c_{\varepsilon,\delta}(n-1)$. Then
\begin{equation}
\label{eq:naive-regime3}
\mathbb{E}\,\| \mathbf{r}' - \mathbf{r} \|_2^2 \;=\; n\,\sigma^2 \;=\; 4\,c_{\varepsilon,\delta}\,n\,(n-1)
\;=\; \Theta \big(c_{\varepsilon,\delta}\,n^2\big).
\end{equation}

\paragraph{ScoreShield Mechanism.}
Let $\widehat{\mathbf{r}}= \mathsf{proj}_{\mathcal{C}_{\mathsf{ref}}}(\mathbf{r} + \mathbf{w})$. We have:

\noindent
\textbf{Local (Pointwise) Risk Bound via the Tangent Cone.}
Project onto $\mathcal{C}_{\text{ref}} \coloneqq \{\mathbf{x}\in[-1,1]^n:\ x_i=1\}$. The equality constraint fixes the $i$th (self) coordinate and removes one free direction. If $a_{\neg i}(\mathbf{r})$ of the remaining coordinates are on the boundary $\{\pm 1\}$, then $\delta(\mathsf{T}_{\mathbf{r}} (\mathcal{C}_{\text{ref}}) )= (n-1) - \frac{1}{2} a_{\neg i}(\mathbf{r})$, and therefore Lemma~\ref{lem:local-tc-scoreshield} gives
\begin{equation}
\label{eq:ScoreShield-regime3}
\mathbb{E}\,\left[ \|\widehat{\mathbf{r}} - \mathbf{r} \|_2^2 \right] \;\le\; \sigma^2 \Bigl( (n-1) - \tfrac{1}{2} a_{\neg i}(\mathbf{r})\Bigr)
\;=\; 4\,c_{\varepsilon,\delta}\,(n-1)\,\Bigl((n-1) - \tfrac{1}{2}a_{\neg i}(\mathbf{r})\Bigr),
\end{equation}
where $a_{\neg i}(\mathbf{r})$ counts boundary activations among the coordinates $j \neq i$.
Thus ScoreShield yields data-dependent constant-factor improvement (up to roughly a factor $\frac{1}{2}$ on activated coordinates) relative to the fully interior case, while the overall scaling remains $\mathcal{O}(n^2)$ because $\sigma^2 = \Theta(n)$.
Equivalently, over the $(n-1)$ free coordinates, the average per-coordinate MSE scales as $\Theta(n)$ (reduced by up to a factor $2$ on boundary-active coordinates)\footnote{%
Compared to the fully interior case $a_{\neg i}(\mathbf{r}) = 0$, boundary-active coordinates reduce the tangent-cone statistical dimension by $\frac{1}{2}$ each, so the bound drops from $\sigma^2 (n-1)$ to $\sigma^2 ((n-1) - a_{\neg i}(\mathbf{r})/2)$. In the extreme case $a_{\neg i}(\mathbf{r}) = n -1$, this is a $2\times$ reduction, while the overall order remains $\Theta (n^2)$ since $\sigma^2 = \Theta(n)$.}, and the fixed coordinate incurs zero error after projection due to hard constraint $r_i = 1$.

\noindent
\textbf{Global Risk Bound via the Gaussian Complexity.}
The feasible set is the affine slice of the box $\mathcal{C}_{\mathsf{ref}} \coloneqq \bigl\{ \mathbf{x} \in [-1,1]^n:\ x_i=1\bigr\}$. Since $\mathcal{C}_{\mathsf{ref}}$ differs from $\mathcal{C}_{\mathsf{query}}$ only by fixing one coordinate, it follows immediately that $\mathsf{GC}(\mathcal{C}_{\mathsf{ref}})=\Theta(n)$ as in Eq.~\ref{eq:gc-box}. Below we compute the exact constant. 

Applying Lemma~\ref{lem:global-gc-scoreshield} with $\mathcal{C}=\mathcal{C}_{\mathsf{ref}}$ yields the uniform bound
\begin{equation}
\label{eq:global-gc-regime3}
\mathbb{E} \, \| \widehat{\mathbf{r}}-\mathbf{r} \|_2^2
\; \le \; 2 \sigma \, \mathsf{GC} \bigl(\mathcal{C}_{\mathsf{ref}} - \mathcal{C}_{\mathsf{ref}}\bigr)
\; \le \; 4\sigma \, \mathsf{GC}(\mathcal{C}_{\mathsf{ref}}) ,
\qquad \widehat{\mathbf{r}}=\mathsf{proj}_{\mathcal{C}_{\mathsf{ref}}}(\mathbf{r} + \mathbf{w}),
\;\;  \mathbf{w}\sim\mathcal{N}(\mathbf{0}, \sigma^2 \mathbf{I}_n).
\end{equation}
We can compute $\mathsf{GC}(\mathcal{C}_{\mathsf{ref}})$ in closed form. Let $\mathbf{z}\sim\mathcal{N}(\mathbf{0},\mathbf{I}_n)$. Then 
\begin{equation}
\sup_{\mathbf{x} \in \mathcal{C}_{\mathsf{ref}}} \langle \mathbf{z}, \mathbf{x} \rangle
= z_i \cdot 1 \;+\; \sum_{j\neq i} \sup_{x_j\in[-1,1]} z_j x_j
= z_i \;+\; \sum_{j\neq i} |z_j|.
\end{equation}
Taking expectation and using $\mathbb{E}[z_i]=0$ and $\mathbb{E}|z_j|=\sqrt{2/\pi}$ gives
\begin{equation}
\label{eq:gc-cref}
\mathsf{GC}(\mathcal{C}_{\mathsf{ref}}) = \mathbb{E}\Big[ \sup_{\mathbf{x} \in \mathcal{C}_{\mathsf{ref}}} \langle \mathbf{z},\mathbf{x} \rangle \Big] = (n-1)\sqrt{\frac{2}{\pi}} = \Theta(n).
\end{equation}
Moreover,
\begin{equation}
\mathcal{C}_{\mathsf{ref}}-\mathcal{C}_{\mathsf{ref}}
= \bigl\{\mathbf{u}\in\mathbb{R}^n:\ u_i=0,\ |u_j|\le 2\ \forall j\neq i\bigr\}
= 2\Bigl\{\mathbf{u}\in\mathbb{R}^n:\ u_i=0,\ |u_j|\le 1\ \forall j\neq i\Bigr\},
\end{equation}
hence by the scaling law $\mathsf{GC}(a \mathcal{A}) = a\, \mathsf{GC}(\mathcal{A})$,
\begin{equation}
\label{eq:gc-diff-cref}
\mathsf{GC}\!\bigl(\mathcal{C}_{\mathsf{ref}}-\mathcal{C}_{\mathsf{ref}}\bigr)
= 2(n-1)\sqrt{\frac{2}{\pi}} = \Theta(n).
\end{equation}
Substituting Eq.~\ref{eq:gc-diff-cref} into Eq.~\ref{eq:global-gc-regime3} yields the explicit global bound
\begin{equation}
\label{eq:global-gc-regime3-explicit}
\mathbb{E}\,\| \widehat{\mathbf{r}} - \mathbf{r} \|_2^2 \;\le\; 2 \sigma \cdot 2(n-1) \sqrt{\frac{2}{\pi}}
= 4 \sigma (n-1) \sqrt{\frac{2}{\pi}}
= \mathcal{O}(\sigma n).
\end{equation}
Using $\sigma^2 = 4c_{\varepsilon,\delta}(n-1)$, we have $\sigma = 2\sqrt{c_{\varepsilon,\delta}}\sqrt{n-1}$, so 
\begin{equation}
\label{eq:global-gc-regime3-dp}
\mathbb{E}\,\| \widehat{\mathbf{r}} - \mathbf{r} \|_2^2 \; \le\; 
8 \sqrt{\frac{2}{\pi}}\; \sqrt{c_{\varepsilon,\delta}} \; (n-1)^{3/2}
= \mathcal{O} \bigl(\sqrt{c_{\varepsilon, \delta}}\, n^{3/2}\bigr).
\end{equation}
This global bound is uniform over $\mathbf{r} \in \mathcal{C}_{\mathsf{ref}}$ but does not exploit the local active-set geometry captured by the tangent-cone bound in Eq.~\ref{eq:ScoreShield-regime3}. Since the local bound scales as $\Theta(\sigma^2 n) = \Theta(c_{\varepsilon, \delta}n^2)$ while the global bound scales as $\Theta(\sigma n) = \Theta(\sqrt{c_{\varepsilon,\delta}}n^{3/2})$, either inequality can be numerically smaller depending on the privacy noise level (via $\sigma$) and the active set size $a_{\neg i}(\mathbf{r})$.

\paragraph{Summary for Regimes (i) \& (iii).}
%
Under the same record-level adjacency model and Gaussian calibration, ScoreShield projection is non-expansive and therefore cannot increase the squared error relative to the unprojected Gaussian release.
Moreover, it can yield data-dependent constant-factor improvements governed by the number of active (saturated) coordinates. 
In particular, the local tangent-cone bounds give:
\begin{center}
\begin{tabular}{lcc}
\toprule
\textbf{Regime} & \textbf{Na\"ive Mechanism MSE} & \textbf{ScoreShield Mechanism MSE} \\
\midrule
Regime~(i) ($\Delta_{\mathsf{query}}=2$)
& $\,4\,c_{\varepsilon,\delta}\,n$
& $\le 4\,c_{\varepsilon,\delta}\Bigl(n-\tfrac{1}{2}a(\mathbf{s})\Bigr)$ \\
Regime~(iii) ($\Delta_{\mathsf{ref}}=2\sqrt{n-1}$)
& $\,4\,c_{\varepsilon,\delta}\,n(n-1)$
& $\le 4\,c_{\varepsilon,\delta}\,(n-1)\Bigl((n-1)-\tfrac{1}{2}a_{\neg i}(\mathbf{r})\Bigr)$ \\
\bottomrule
\end{tabular}
\end{center}

In regimes~(i) and (iii), the feasible sets are (products of) intervals (with an additional equality constraint in regime~(iii)), so projection yields constant-factor improvements and does not change the exponent in the leading $n$-scaling under the local tangent-cone bound. 
In the following, we address the attacker's reconstruction error.

\vspace{10pt}

\appsubsection{Attacker Reconstruction Error for Regime (i) \& (iii)}
\label{ssec:attacker_regime1}

\paragraph{ScoreShield release model.}
For any released score vector we use $\mathbf{s}' \;=\; \mathbf{s}+\mathbf{w},\qquad \mathbf{w}\sim\mathcal{N}(\mathbf{0},\sigma^2 \mathbf{I}_n)$, $\widehat{\mathbf{s}} \;=\; \mathsf{proj}_{\mathcal{C}}(\mathbf{s}')$. 
For regime~(i), $\mathcal{C} = \mathcal{C}_{\mathsf{query}} = \{ \mathbf{s} \in [-1,1]^n \}$ and the projection is coordinate-wise clipping
\[
\bigl(\mathsf{proj}_{[-1,1]^n}(\mathbf{s}')\bigr)_j =
\max(-1,\min(1,s'_j)).
\]
For regime~(iii), $\mathcal{C}= \mathcal{C}_{\mathrm{ref}} \coloneqq \{\mathbf{s} \in [-1,1]^n:\ s_i=1 \}$ and the projection onto $\mathcal{C}_{\mathrm{ref}}$ sets the reference coordinate to one and clips the remaining coordinates:
\[
\bigl(\mathsf{proj}_{\mathcal{C}_{\mathrm{ref}}}(\mathbf{s}')\bigr)_i=1,
\qquad
\bigl(\mathsf{proj}_{\mathcal{C}_{\mathrm{ref}}}(\mathbf{s}')\bigr)_j
= \max(-1,\min(1,s'_j)), 
\quad j\neq i.
\]
For the box-clipping coordinates, the scalar clipping map is $1$-Lipschitz, so
$\mathbb{E}[(\widehat s_j-s_j)^2]\le\sigma^2$.
For the fixed reference coordinate in regime~(iii), $\widehat s_i=s_i=1$, so the MSE is zero.

\paragraph{Adjacency and altered index set.}
Under record-level replacement, $\mathbf{E}$ and $\mathbf{E}'$ differ in exactly one row $i$. Let $\mathcal{I}$ denote the set of altered indices under adjacency. For regime~(i) (query-to-collection), only coordinate $i$ of $\mathbf{s}=\mathbf{E}\mathbf{q}$ may change, hence $\mathcal{I}_1 \coloneqq \{i\}$.
For regime~(iii) (reference-to-gallery), the altered set is the off-diagonals $\mathcal{I}_2(i) \coloneqq \{ j\neq i \}$ of size $n{-}1$.

\paragraph{Risk Functionals.}
For any index $j$, define the coordinate MSE
\begin{equation}
\mathsf{MSE}_j \; \coloneqq  \; \mathbb{E} \big[(\widehat{s}_j - s_j)^2\big],    
\end{equation}
and for an altered index set $\mathcal{I}$ define the restricted risk
\begin{equation}
\mathcal{R} (\mathcal{I}) \; \coloneqq  \; \mathbb{E} \big[\| (\widehat{\mathbf{s}} - \mathbf{s})_{\mathcal I} \|_2^2\big]
\; = \; \sum_{j\in\mathcal{I}} \mathsf{MSE}_j.
\end{equation}

\paragraph{Auxiliary-information regimes.}
The reconstruction calculations below concern estimation error from the noisy released score vector; they are separate from the differential-privacy guarantee. We distinguish knowledge of the score-generating values from knowledge of the subspace in which the clean score vector lies. If an attacker knows all quantities that determine the clean score vector, for example both $\mathbf E$ and a public query $\mathbf q$ in regime~(i), then $\mathbf s=\mathbf E\mathbf q$ is computable exactly and the reconstruction MSE is zero. The linear denoising bounds below are therefore stated for a subspace-knowledge model, in which the attacker knows the relevant column space but not the latent coefficient vector generating the released scores.

We consider three regimes:
\begin{enumerate}[label=(K\arabic*)]
\item \textbf{No side information:} the attacker knows neither $\mathbf E$ nor the relevant column space $\mathrm{col}(\mathbf E)$.

\item \textbf{Knows $\mathrm{col}(\mathbf E)$, but not the latent score generator:}
the attacker knows the column space $\mathrm{col}(\mathbf E)$, equivalently the orthogonal projector
\begin{equation}
\mathbf P\coloneqq \mathbf E(\mathbf E^\top\mathbf E)^\dagger\mathbf E^\top\in\mathbb R^{n\times n},
\end{equation}
but does not know the vector that generates the clean score vector within this subspace. Let
$r\coloneqq\mathrm{rank}(\mathbf E)$.

\item \textbf{Knows $\mathbf{E}_{-i}$:} the attacker knows all rows except the single differing row $i$, i.e.,
$\mathbf E_{-i}\in\mathbb R^{(n-1)\times d}$.
Let $r_{-i}\coloneqq\mathrm{rank}(\mathbf E_{-i})$ and let
\begin{equation}
\mathbf P_{-i}\coloneqq
\mathbf E_{-i}(\mathbf E_{-i}^\top\mathbf E_{-i})^\dagger\mathbf E_{-i}^\top
\in\mathbb R^{(n-1)\times(n-1)}
\end{equation}
be the orthogonal projector onto $\mathrm{col}(\mathbf E_{-i})$.
\end{enumerate}

\paragraph{Baseline (coordinatewise upper bounds).}
Without using any structure beyond the released values, a natural estimator for $s_j$ is $\widehat{s}_j$. By non-expansiveness,
\begin{equation}
\mathsf{MSE}_j\le \sigma^2, \quad\forall j\in[n], 
\end{equation}
with equality on coordinates that do not clip (in particular, exactly under the no-clipping Gaussian model $\widehat{\mathbf{s}}= \mathbf{s}'=\mathbf{s}+\mathbf{w}$).

\paragraph{Linear denoising under subspace knowledge.}
Assume the no-clipping Gaussian model $\widehat{\mathbf s}=\mathbf s'=\mathbf s+\mathbf w$. The following formulas apply when the clean score vector is an unknown deterministic vector in the known subspace. They do not apply to the case where the attacker knows all score-generating inputs; in particular, in regime~(i), if both $\mathbf E$ and the public query $\mathbf q$ are known to the attacker, then $\mathbf s=\mathbf E\mathbf q$ is known exactly and the reconstruction MSE is zero.

\textit{(K2) Knows $\mathrm{col}(\mathbf E)$, but not the latent generator.}
In the subspace-knowledge model, the clean vector is an unknown element of $\mathrm{col}(\mathbf E)$; for example, in regime~(i), $\mathbf s=\mathbf E\mathbf q\in\mathrm{col}(\mathbf E)$ when $\mathbf q$ is not disclosed to the attacker. Under additive Gaussian noise, the affine-unbiased least-squares estimator is the orthogonal projection
\begin{equation}
\widetilde{\mathbf{s}}_{\mathrm{LS}}=\mathbf P\widehat{\mathbf{s}}.
\end{equation}
Its error covariance is
\begin{equation}
\mathrm{Cov}(\widetilde{\mathbf{s}}_{\mathrm{LS}}-\mathbf{s})=\sigma^2\mathbf P,
\end{equation}
and hence
\begin{equation}
\mathsf{MSE}_j=\sigma^2P_{jj},
\qquad 
\frac{1}{n}\sum_{j=1}^n\mathsf{MSE}_j
= \frac{\sigma^2}{n}\mathrm{tr}(\mathbf P)
= \frac{\sigma^2 r}{n}.
\end{equation}
The coordinatewise risk is governed by the leverage score $P_{jj}\in[0,1]$.

\textit{(K3) Knows $\mathbf E_{-i}$.}
In the subspace-knowledge model, the attacker can denoise the subvector indexed by $j\neq i$ using
$\mathrm{col}(\mathbf E_{-i})$, but cannot use $\mathbf E_{-i}$ to denoise the altered coordinate $i$.
A natural affine-unbiased estimator is
\begin{equation}
\widetilde{s}_i=\widehat{s}_i,
\qquad
\widetilde{\mathbf{s}}_{-i}=\mathbf P_{-i}\widehat{\mathbf{s}}_{-i}.
\end{equation}
Under no clipping, for $j\neq i$,
\begin{equation}
\mathsf{MSE}_j=\sigma^2(\mathbf P_{-i})_{jj},
\end{equation}
and
\begin{equation}
\sum_{j\neq i}\mathsf{MSE}_j
= \sigma^2\mathrm{tr}(\mathbf P_{-i})
= \sigma^2 r_{-i},
\qquad
\frac{1}{n-1}\sum_{j\neq i}\mathsf{MSE}_j
= \frac{\sigma^2 r_{-i}}{n-1}.
\end{equation}
The altered coordinate remains at the no-denoising level, $\mathsf{MSE}_i=\sigma^2$. If, in regime~(i), the query $\mathbf q$ is public and $\mathbf E_{-i}$ is known, then the unchanged coordinates
$\mathbf s_{-i}=\mathbf E_{-i}\mathbf q$ are known exactly; in that case the above subspace-denoising bound is not the relevant reconstruction model for those coordinates.

\paragraph{Attacker's reconstruction error for regime~(i) (query vector).}
In regime~(i), $\sigma^2=4c_{\varepsilon,\delta}$ and $\mathcal{I}_1=\{i\}$.
Therefore, the restricted risk equals the altered-coordinate risk,
$\mathcal{R}(\mathcal{I}_1)=\mathsf{MSE}_i$. Under the no-clipping Gaussian model, in the subspace-knowledge setting described above,
\begin{align*}
\text{(K1)~~~No side information:}\quad &
\mathsf{MSE}_i \le \sigma^2,\qquad \mathcal{R}(\mathcal{I}_1)\le\sigma^2.\\[2pt]
\text{(K2)~~~Knows $\mathrm{col}(\mathbf E)$:}\quad &
\mathsf{MSE}_i=\sigma^2 P_{ii}\in[0,\sigma^2],\qquad
\mathcal{R}(\mathcal{I}_1)=\sigma^2 P_{ii}.\\[2pt]
\text{(K3)~~~Knows $\mathbf{E}_{-i}$:}\quad &
\mathsf{MSE}_i=\sigma^2,\qquad \mathcal{R}(\mathcal{I}_1)=\sigma^2.
\end{align*}
In (K2), the average per-entry MSE over all coordinates equals $\sigma^2 r/n$ under no clipping. This is an average over $j\in[n]$ and is not the risk on the single altered coordinate unless the altered index is randomized. If the query $\mathbf q$ is public and the attacker knows $\mathbf E$, then $\mathbf s=\mathbf E\mathbf q$ is known exactly and the reconstruction MSE is zero; the K2 formulas above apply only to the subspace-knowledge model.

\paragraph{Attacker's reconstruction error for regime~(iii) (reference vector).}
In regime~(iii), $\sigma^2=4c_{\varepsilon,\delta}(n-1)$ and
$\mathcal{I}_2(i)=\{j\neq i\}$.
Under the no-clipping Gaussian model, in the subspace-knowledge setting described above,
\begin{align*}
\text{(K1)~~No side information:}\;\; &
\mathsf{MSE}_j\le\sigma^2\ (j\neq i),\quad
\mathcal{R}(\mathcal{I}_2(i))\le (n-1)\sigma^2.\\[2pt]
\text{(K2)~~Knows $\mathrm{col}(\mathbf E)$:}\;\; &
\mathsf{MSE}_j=\sigma^2P_{jj}\ (j\neq i),\quad
\mathcal{R}(\mathcal{I}_2(i))=\sigma^2 \sum_{j\neq i}P_{jj}
=\sigma^2(r-P_{ii})\le \sigma^2 r.\\[2pt]
\text{(K3)~~Knows $\mathbf{E}_{-i}$:}\;\; &
\mathsf{MSE}_j=\sigma^2(\mathbf{P}_{-i})_{jj}\ (j\neq i),\quad
\mathcal{R}(\mathcal{I}_2(i))=\sigma^2 r_{-i}.
\end{align*}
If the attacker knows the full gallery $\mathbf E$, then the reference row
$\mathbf r_i=\mathbf E\mathbf e_i$ is computable exactly and the reconstruction MSE is zero. Therefore the K2 formulas above should be read as subspace-denoising benchmarks, not as full-gallery-knowledge risks.

The equalities involving $\sigma^2P_{jj}$ and $\sigma^2(\mathbf P_{-i})_{jj}$ are no-clipping Gaussian benchmarks. With clipping, the universal coordinate-wise bound $\mathsf{MSE}_j\le\sigma^2$ still holds for box-clipped coordinates, and the total nonexpansive bound
\[
\mathbb E\|\widehat{\mathbf s}-\mathbf s\|_2^2\le n\sigma^2
\]
continues to hold. Clipping may reduce the error on saturated coordinates, but the exact leverage-score covariance identities need not hold after clipping.
%
Moreover, note that these bounds are attacker estimation errors given the one-shot release $\widehat{\mathbf{s}}$ and the specified auxiliary-information regime. They do not alter the $(\varepsilon,\delta)$ indistinguishability guarantee, which is defined at the dataset level under the adopted adjacency.
We summarize the attacker's reconstruction error bounds in the following table.
\begin{center}
\begin{tabular}{@{}llccc@{}}
\toprule
\textbf{Regime} & \textbf{Quantity} & \textbf{(K1) none} & \textbf{(K2) knows $\mathrm{col}(\mathbf E)$} & \textbf{(K3) knows $\mathbf E_{-i}$} \\
\midrule
\multirow{2}{*}{(i) $\ \mathcal{I}_1=\{i\}$}
& $\mathsf{MSE}_i$ & $\le\sigma^2$ & $\sigma^2 P_{ii}$ & $\sigma^2$ \\
& $\mathcal{R}(\mathcal{I}_1)$ & $\le\sigma^2$ & $\sigma^2 P_{ii}$ & $\sigma^2$ \\
\midrule
\multirow{2}{*}{(iii) $\ \mathcal{I}_2(i)=\{j\neq i\}$}
& $\frac{1}{n-1}\sum_{j\neq i}\mathsf{MSE}_j$ & $\le\sigma^2$ & $\frac{\sigma^2(r-P_{ii})}{n-1}$ & $\frac{\sigma^2 r_{-i}}{n-1}$ \\
& $\mathcal{R}(\mathcal{I}_2(i))$ & $\le (n-1)\sigma^2$ & $\sigma^2(r-P_{ii})$ & $\sigma^2 r_{-i}$ \\
\bottomrule
\end{tabular}
\end{center}

\emph{Note.} K2 is a subspace-knowledge model. If all score-generating values are known, such as both $\mathbf E$ and a public $\mathbf q$ in regime~(i), or the full gallery $\mathbf E$ in regime~(iii), then the corresponding clean score vector is exactly computable and the reconstruction MSE is zero.

\vspace{30pt}
 
\appsubsection{Regime (ii): Full Pairwise Similarity Score Matrix}
\label{subsec:regime2-compare}

%
We release $\mathbf{S}= \mathbf{E}\mathbf{E}^\top\in[-1,1]^{n\times n}$ with $\mathbf{S} \in \mathcal{C}_{\mathsf{coll}}  \coloneqq  \bigl\{ \mathbf{S}\in \mathbb{R}^{n\times n}: \, \mathbf{S}\succeq0,\; S_{ii}=1\;(1 \le  i \le  n),\; |S_{ij}|\le 1\;(i \ne  j) \bigr\}\subset \mathbb{R}^{n\times n}$. Our ScoreShield mechanism under \textit{record-level adjacency} on $\mathbf{E}$ has the exact Frobenius sensitivity $\Delta_{f, F} \eqqcolon \Delta_{F,\mathrm{rec}}$
\begin{equation}
\Delta_{F,\mathrm{rec}} = 2 \sqrt{2(n-1)}
= \Theta(\sqrt{n}),
\end{equation}
since changing record $i$ changes only the $2(n-1)$ off-diagonal entries in row/column $i$.

We analyze both adjacency notions. In each regime we set $\sigma^2 \;=\; c_{\varepsilon, \delta}\, \Delta^2$,
$c_{\varepsilon, \delta}\coloneqq \frac{2\log(2/\delta)}{\varepsilon^2}$.
\begin{itemize}[leftmargin=*]
\item 
\textbf{(\textsc{R})} Record-level adjacency on $\mathbf{E}$: exact Frobenius sensitivity $\Delta_{f,F} = 2\sqrt{2(n-1)}=\Theta(\sqrt n)$, hence $\sigma^2=\Theta(n)$.
\item 
\textbf{(\textsc{O})} Output-space adjacency on $\mathbf{S}$ \citep{cohen2024perturb}:
$\|\mathbf{S}-\mathbf{S}'\|_F \le \mathsf{\Delta}_{\mathsf{G}}$ with $ \mathsf{\Delta}_{\mathsf{G}} = \Theta(1)$ independent of $n$, hence $\sigma^2 = \Theta(1)$.
\end{itemize}

\paragraph{Na\"ive Gaussian Mechanism.}
In our practical algorithm, we sample $\mathbf{W}$ with i.i.d. entries $W_{ij}\sim\mathcal{N}(0,\sigma^2)$ (Gaussian mechanism), and then apply the deterministic symmetrization post-processing $\mathbf{G}\coloneqq \tfrac12(\mathbf{W}+\mathbf{W}^\top)$.
Since $\mathbf{S}$ is symmetric, the symmetrized release $\mathbf{S}' \coloneqq \tfrac12\big((\mathbf{S}+\mathbf{W})+(\mathbf{S}+\mathbf{W})^\top\big) =\mathbf{S}+\mathbf{G}$ is a post-processing of $\mathbf{S}+\mathbf{W}$ and therefore preserves $(\varepsilon,\delta)$-DP\footnote{Equivalently, sampling $\mathbf{G}$ directly as a symmetric Gaussian with $\mathrm{Var}(G_{ii}) = \sigma^2$ and $\mathrm{Var}(G_{ij}) = \sigma^2/2, \forall i \neq j$, yields the same distribution as $\frac{1}{2} (\mathbf{W} + \mathbf{W}^\top)$.} (with $\sigma$ calibrated to the sensitivity of $\mathbf{S}$ under the chosen adjacency). We report utility for this symmetric pre-projection matrix $\mathbf{S}'$.
A direct variance calculation gives
\begin{equation}\label{eq:naive-frob-appx}
\mathbb{E}\,\| \mathbf{S}' -\mathbf{S}\|_F^2
=\mathbb{E}\,\|\mathbf{G}\|_F^2
=\underbrace{n\sigma^2}_{\text{diagonal}}
+\underbrace{n(n-1) \frac{\sigma^2}{2}}_{\text{off-diagonal}}
=\frac{n^2+n}{2}\,\sigma^2 
=\Theta(n^2\sigma^2).
\end{equation}
Therefore, for the two adjacency definitions we have:
\begin{enumerate}[leftmargin=2em]
\item[(a)] Record-level adjacency \textbf{(\textsc{R})}: Using $\sigma^2=c_{\varepsilon,\delta}\Delta_{f,F}^2=\Theta \big(\frac{n\log(2/\delta)}{\varepsilon^2}\big)$,
\begin{equation}
\text{na\"ive + (\textsc{R})}:\qquad
\mathbb{E}\,\| \mathbf{S}' - \mathbf{S} \|_{\mathrm{F}}^2
= \Theta \Big( n^2 \, c_{\varepsilon,\delta}\, \Delta_{F,\mathrm{rec}}^2 \Big)
= \Theta \Big( \frac{n^{3}\log(2/\delta)}{\varepsilon^2} \Big).
\end{equation}
\item[(b)] Output-space adjacency \textbf{(\textsc{O})}: With $\sigma^2=c_{\varepsilon,\delta}\mathsf{\Delta}_{\mathsf{G}}^2$,
\begin{equation}
\text{na\"ive + (\textsc{O})}:\qquad
\mathbb{E} \,\| \mathbf{S}' - \mathbf{S} \|_{\mathrm{F}}^2
= \Theta \Big( n^2\,c_{\varepsilon,\delta}\, \mathsf{\Delta}_{\mathsf{G}}^2 \Big)
= \Theta \Big( \frac{n^{2} \mathsf{\Delta}_{\mathsf{G}}^2\log(2/\delta)}{\varepsilon^2} \Big).
\end{equation}
\end{enumerate}

\begin{remark}
If instead we perturb only the upper triangle and reflect, Eq.~\ref{eq:naive-frob-appx} still yields $\Theta(n^2\sigma^2)$, while constants differ but the exponent is unchanged.    
\end{remark}

\paragraph{ScoreShield Mechanism.}
Let $\widehat{\mathbf{S}}=\mathsf{proj}_{\mathcal{C}_{\mathsf{coll}}}(\mathbf{S}+\mathbf{G})$ denote the exact Frobenius metric-projection release. The risk bounds in this paragraph apply to $\widehat{\mathbf{S}}$. The AAP feasibility solver used in large-scale experiments is a different post-processing map and is not the object of these exact-projection risk bounds.

\noindent
\textbf{Global Risk Bound via the Gaussian Complexity (uniform in $\mathbf{S}$).}
Using Corollary~\ref{cor:global-gc-scoreshield-matrix} we have
\begin{equation}
\label{eq:global-GC-appx}
\mathbb{E}\|\widehat{\mathbf{S}}-\mathbf{S}\|_F^2
\;\le\; C\; \sigma \; \mathsf{GC}(\mathcal{C}_{\, \mathsf{coll}}\,)
\;\le\; \widetilde{C} \, \sigma \,  n^{3/2}\, ,
\end{equation}
where we used $\mathsf{GC}(\mathcal{C}_{\mathsf{coll}}) = \Theta(n^{3/2})$.

\begin{enumerate}[leftmargin=2em]
\item[(a)] Record-level adjacency \textbf{(\textsc{R})}: With $\sigma = \Theta \left( \frac{ \sqrt{n\log(2/\delta)}}{\varepsilon}\right)$,
\begin{equation}
\text{ScoreShield + (\textsc{R})}:\qquad
\mathbb{E}\|\widehat{\mathbf{S}}-\mathbf{S}\|_F^2
\;\le\; \widetilde{C}\,\frac{n^{2} \sqrt{\log(2/\delta)}}{\varepsilon}
\;=\; \mathcal{O} \left( \frac{n^{2} \sqrt{\log(2/\delta)}}{\varepsilon} \right).
\end{equation}
\item[(b)] Output-space adjacency \textbf{(\textsc{O})}: With $\sigma = \frac{\sqrt{2\log(2/\delta)}}{\varepsilon} \mathsf{\Delta}_{\mathsf{G}}$,
\begin{equation}
\text{ScoreShield + (\textsc{O})}: \qquad
\mathbb{E} \| \widehat{\mathbf{S}} - \mathbf{S} \|_F^2
\; \le \; \widetilde{C} \, \frac{n^{3/2} \mathsf{\Delta}_{\mathsf{G}} \sqrt{\log(2/\delta)}}{\varepsilon}
\; = \; \mathcal{O} \left( \frac{n^{3/2} \mathsf{\Delta}_{\mathsf{G}} \sqrt{\log(2/\delta)}}{\varepsilon} \right).
\end{equation}
\end{enumerate}

\noindent
\textbf{Local Risk Bound via Rank-Aware Tangent-Cone (under local Gram-smoothness).}
Let $r \coloneqq  \mathrm{rank}(\mathbf{S})\le d$ and $\mathsf{T}_{\mathbf{S}} \coloneqq \mathsf{T}_{\mathbf{S}} (\mathcal{C}_{\mathsf{coll}})$ denote the contingent tangent cone of the elliptope at $\mathbf{S}$. For the projected Gaussian estimator $\widehat{\mathbf{S}} = \mathsf{proj}_{\mathcal{C}_{\mathsf{coll}}} \left( \mathbf{S} + \mathbf{G} \right)$, the conic denoising bound gives
\begin{equation}
\mathbb{E} \| \widehat{\mathbf{S}} - \mathbf{S} \|_F^2 \leq \sigma^2 \,  \delta (\mathsf{T}_{\mathbf{S}}).
\end{equation}
By the rank-aware upper bound (Lemma~\ref{lem:rank_aware_statdim_upper}), under local Gram-smoothness $\mathsf T_{\mathbf S}(\mathcal E_n)=\mathcal T_{\mathrm{man}}(\mathbf S)$, we have
%
$\delta(\mathsf{T}_{\mathbf{S}}) \le nr$.
%
The conic bound gives
\begin{equation}\label{eq:local-cone-risk}
\mathbb{E}\|\widehat{\mathbf{S}}-\mathbf{S}\|_F^2 \;\le\; \sigma^2\,\delta(\mathsf{T}_{\mathbf{S}})
\;\le\; \widetilde{C}\,\sigma^2\,n r = \mathcal{O} (\sigma^2 n r).
\end{equation}
\begin{enumerate}[leftmargin=2em]
\item[(a)] Record-level adjacency \textbf{(\textsc{R})}: With $\sigma^2 = \Theta \big(\frac{n\log(2/\delta)}{\varepsilon^2}\big)$,
\begin{equation}
\text{ScoreShield + (\textsc{R})}:\qquad
\mathbb{E} \| \widehat{\mathbf{S}} - \mathbf{S} \|_F^2 \; \leq  \;
\mathcal{O} \Big( \frac{n^2 \, r \,\log(2/\delta)}{\varepsilon^2} \Big).
\end{equation}
\item[(b)] Output-space adjacency \textbf{(\textsc{O})}: With $\sigma^2 = \frac{2\log(2/\delta)}{\varepsilon^2} \mathsf{\Delta}_{\mathsf{G}}^2$,
\begin{equation}
\text{ScoreShield + (\textsc{O})}:\qquad
\mathbb{E} \| \widehat{\mathbf{S}} - \mathbf{S} \|_F^2 \leq \;
\mathcal{O} \Big( \frac{n\, r \, \mathsf{\Delta}_{\mathsf{G}}^2\log(2/\delta)}{\varepsilon^2} \Big) .
\end{equation}
\end{enumerate}

\paragraph{Summary.}

We summarize the mechanism reconstruction error bounds in the following table.

\begin{center}
\renewcommand{\arraystretch}{2}
\begin{tabular}{lcc}
\toprule
\textbf{Mechanism} & \textbf{Adjacency} & 
\textbf{Mechanism MSE}\; $\mathbb{E}\|\,\widehat{\mathbf{S}} - \mathbf{S} \,\|_{\mathrm{F}}^2$ \\
\midrule
Na\"ive Gaussian & (\textsc{R}): replace one row 
& $\Theta \big( n^{3} c_{\varepsilon,\delta} \big)$ \\

\textbf{ScoreShield} (global) & (\textsc{R}): replace one row 
& $\mathcal{O} \big( n^{2}\sqrt{c_{\varepsilon,\delta}} \big)$ \\

\textbf{ScoreShield} (rank-aware) & (\textsc{R}): replace one row 
& $\mathcal{O} \big( n^{2} r \, c_{\varepsilon,\delta} \big)$ \\
\hline

Na\"ive Gaussian & (\textsc{O}): 
$\| \mathbf{S}-\mathbf{S}' \|_{\mathrm{F}} \le \mathsf{\Delta}_{\mathsf{G}}$ 
& $\Theta \big( n^{2} \mathsf{\Delta}_{\mathsf{G}}^2 c_{\varepsilon,\delta} \big)$ \\ 

\textbf{ScoreShield} (global) & (\textsc{O}): 
$\| \mathbf{S}-\mathbf{S}' \|_{\mathrm{F}} \le \mathsf{\Delta}_{\mathsf{G}}$ 
& $\mathcal{O} \big( n^{3/2} \mathsf{\Delta}_{\mathsf{G}} \,\sqrt{c_{\varepsilon,\delta}} \big)$ \\

\textbf{ScoreShield} (rank-aware) & (\textsc{O}): 
$\| \mathbf{S}-\mathbf{S}' \|_{\mathrm{F}} \le \mathsf{\Delta}_{\mathsf{G}}$ 
& $\mathcal{O} \big( n r \,\mathsf{\Delta}_{\mathsf{G}}^2 \, c_{\varepsilon,\delta} \big)$ \\
\bottomrule
\end{tabular}
\end{center}

\vspace{10pt}

\appsubsection{Attacker Reconstruction Error for Regime (ii)}
\label{ssec:attacker_regime2}

\paragraph{ScoreShield release model.}
Let $\mathbf{W}\in\mathbb{R}^{n\times n}$ have i.i.d. entries $W_{ij}\sim\mathcal{N}(0,\sigma^2)$ and apply the deterministic symmetrization post-processing $\mathbf{G}\coloneqq \tfrac12(\mathbf{W}+\mathbf{W}^\top)$. Then $G_{ii}\sim\mathcal{N}(0,\sigma^2)$ and for $i<j$, $G_{ij}=G_{ji}\sim\mathcal{N}(0,\sigma^2/2)$. We release $\widehat{\mathbf{S}}=\mathsf{proj}_{\mathcal{C}_{\mathsf{coll}}}(\mathbf{S}+\mathbf{G})$ where
$\mathbf{S}=\mathbf{E}\mathbf{E}^\top\in\mathcal{C}_{\mathsf{coll}}\coloneqq \{\mathbf X\in\mathtt S^n: \mathbf X\succeq0,\ \mathrm{diag}(\mathbf X)=\mathbf 1,\ |X_{ij}|\le 1\ (i\neq j)\}$. Note that our practical AAP feasibility solver used in large-scale experiments is a different post-processing map and is not analyzed by the coordinatewise formulas below. 
Euclidean projection in Frobenius norm is firmly non-expansive (and therefore $1$-Lipschitz). Hence
\begin{equation}
\|\widehat{\mathbf{S}}-\mathbf{S}\|_{\mathrm F}\le \|\mathbf{G}\|_{\mathrm F},\qquad
\mathbb{E} \left[ \|\widehat{\mathbf{S}}-\mathbf{S}\|_{\mathrm F}^2\ \right] \le\ \mathbb{E} \left[ \|\mathbf{G}\|_{\mathrm F}^2\ \right] = \frac{n^2+n}{2}\,\sigma^2 \leq  n^2\sigma^2.   
\end{equation}
Per–coordinate inequalities like $\mathbb{E}[(\widehat S_{ij}-S_{ij})^2]\le \mathrm{Var}(G_{ij})$ need not hold for the elliptope due to PSD coupling (unlike coordinatewise clipping). They do hold for coordinate–separable clipping.

\paragraph{Risk functionals.}
Under record–level replacement of row $i$ in $\mathbf{E}$, the affected Gram entries are the off–diagonal row/column $\mathcal{J}_i^{\mathrm{row}}\coloneqq\{(i,j):j\neq i\}$, and, if counted, the symmetric strip $\mathcal{J}_i^{\mathrm{strip}}\coloneqq\{(i,j):j\neq i\}\cup\{(j,i):j\neq i\}$.
Because $S_{ij}=S_{ji}$ and $G_{ij}\equiv G_{ji}$\footnote{Since $\mathbf{S}=\mathbf{E}\mathbf{E}^\top$ is symmetric and we use $\mathbf{S}'=\mathbf{S}+\tfrac12(\mathbf{W}+\mathbf{W}^\top)$, both the pre- and post-projection matrices are symmetric. The projector also preserves symmetry.}, the two strips are redundant. Therefore, $\mathcal{R}_{\mathrm{strip}}(i)=2\,\mathcal{R}_{\mathrm{row}}(i)$ holds exactly. Define the coordinate risk $\mathsf{MSE}_{ij}\coloneqq\mathbb{E}[(\widehat S_{ij}-S_{ij})^2]$ and the restricted risks
\begin{equation}
\mathcal{R}_{\mathrm{row}}(i)\coloneqq \sum_{j\neq i}\mathsf{MSE}_{ij},\qquad
\mathcal{R}_{\mathrm{strip}}(i)\coloneqq 2\,\mathcal{R}_{\mathrm{row}}(i).   
\end{equation}

\paragraph{Knowledge regimes.}
We consider two auxiliary-information regimes: (i) no side information about the embeddings; and (ii) one-row-unknown side information, where the attacker knows the gallery embeddings
$\{\mathbf e_j\}_{j\neq i}$ but not the changed row $\mathbf e_i$.

\begin{enumerate}[label=(\roman*),leftmargin=1.4em]
\item \textbf{No side information on embeddings.}
For the symmetric additive pre-projection model
$\mathbf S'=\mathbf S+\mathbf G$, we have, for every $j\neq i$,
\begin{equation}
\mathsf{MSE}_{ij}
= \mathbb E[(S'_{ij}-S_{ij})^2]
= \mathrm{Var}(G_{ij})
= \frac{\sigma^2}{2}.
\end{equation}
Hence
\begin{equation}
\mathcal R_{\mathrm{row}}(i)
= \frac{(n-1)\sigma^2}{2},
\qquad
\mathcal R_{\mathrm{strip}}(i)
= (n-1)\sigma^2
\end{equation}
for the symmetric additive pre-projection model.

For the exact metric-projection release
$\widehat{\mathbf S}
= \mathsf{proj}_{\mathcal C_{\mathsf{coll}}}(\mathbf S+\mathbf G)$,
PSD coupling can redistribute error across entries. Therefore, coordinatewise inequalities such as $\mathbb E[(\widehat S_{ij}-S_{ij})^2]\le \mathrm{Var}(G_{ij})$ are not guaranteed. However, the row-restricted risk is bounded by the total Frobenius risk:
\begin{equation}
\mathcal R_{\mathrm{row}}(i)
= \mathbb E\sum_{j\neq i}(\widehat S_{ij}-S_{ij})^2
\le \mathbb E\|\widehat{\mathbf S}-\mathbf S\|_F^2 .
\end{equation}
Since $\mathrm{diag}(\widehat{\mathbf S})=\mathrm{diag}(\mathbf S)=\mathbf 1$, the diagonal error is zero. Moreover, the diagonal part of $\mathbf G$ contributes only an additive constant to the Frobenius projection objective over
$\mathcal C_{\mathsf{coll}}$, because every feasible matrix has unit diagonal. Thus the exact projection depends only on the off-diagonal perturbation for the purpose of the minimizer, and non-expansiveness gives
\begin{equation}
\mathbb E\|\widehat{\mathbf S}-\mathbf S\|_F^2
\le \mathbb E\|\mathbf G_{\mathrm{off}}\|_F^2
= \frac{n(n-1)}{2}\sigma^2 ,
\end{equation}
where $\mathbf G_{\mathrm{off}}$ denotes the off-diagonal part of $\mathbf G$.
Consequently,
\begin{equation}
\mathcal R_{\mathrm{row}}(i)
\le \frac{n(n-1)}{2}\sigma^2,
\qquad
\mathcal R_{\mathrm{strip}}(i) \le n(n-1)\sigma^2 .
\end{equation}
\item \textbf{Knows embeddings $\{\mathbf{e}_j\}_{j\neq i}$ (one-row unknown).}
Let $\mathbf E_{-i}\in\mathbb R^{(n-1)\times d}$ stack the rows $\{\mathbf e_j^\top\}_{j\neq i}$, let
$r_{-i}\coloneqq\mathrm{rank}(\mathbf E_{-i})$, and let
\begin{equation}
\mathbf P_{-i} \coloneqq
\mathbf E_{-i}(\mathbf E_{-i}^\top\mathbf E_{-i})^\dagger\mathbf E_{-i}^\top
\in\mathbb R^{(n-1)\times(n-1)}
\end{equation}
be the orthogonal projector onto $\mathrm{col}(\mathbf E_{-i})$.
Under the symmetric additive pre-projection row model,
\begin{equation}
\mathbf y = \mathbf E_{-i}\mathbf e_i+\boldsymbol\eta,
\qquad
\boldsymbol\eta\sim\mathcal N\!\left(\mathbf 0,\frac{\sigma^2}{2}\mathbf I_{n-1}\right),
\end{equation}
the clean off-diagonal score vector $\mathbf E_{-i}\mathbf e_i$ lies in $\mathrm{col}(\mathbf E_{-i})$. The affine-unbiased least-squares estimator of this score vector is
\begin{equation}
\widetilde{\mathbf y}_{\mathrm{LS}}
= \mathbf P_{-i}\mathbf y.
\end{equation}
Its error covariance is
\begin{equation}
\mathrm{Cov}\!\left(\widetilde{\mathbf y}_{\mathrm{LS}}-\mathbf E_{-i}\mathbf e_i\right)
= \frac{\sigma^2}{2}\mathbf P_{-i}.
\end{equation}
Therefore, for $j\neq i$,
\begin{equation}
\mathsf{MSE}_{ij}^{\mathrm{LS}} =
\frac{\sigma^2}{2}(\mathbf P_{-i})_{jj},
\end{equation}
and
\begin{equation}
\sum_{j\neq i}\mathsf{MSE}_{ij}^{\mathrm{LS}}
= \frac{\sigma^2}{2}\operatorname{tr}(\mathbf P_{-i})
= \frac{\sigma^2 r_{-i}}{2},
\qquad
\frac{1}{n-1}\sum_{j\neq i}\mathsf{MSE}_{ij}^{\mathrm{LS}}
= \frac{\sigma^2 r_{-i}}{2(n-1)}.
\end{equation}
Moreover, since $\mathbf P_{-i}$ is an orthogonal projector,
\begin{equation}
\frac{\sigma^2 r_{-i}}{2(n-1)}
\le \max_{j\neq i}\mathsf{MSE}_{ij}^{\mathrm{LS}}
\le \frac{\sigma^2}{2}.
\end{equation}
For the released projected matrix $\widehat{\mathbf S}$, these LS formulas are pre-projection additive benchmarks. They are not coordinatewise guarantees after projection, because projection onto $\mathcal C_{\mathsf{coll}}$ can redistribute error across entries.
\end{enumerate}

\paragraph{Calibration under (\textsc{R}) and (\textsc{O}): benchmarks vs. guarantees.}
Let $\sigma^2=c_{\varepsilon,\delta}\Delta^2$ with $c_{\varepsilon,\delta}=2\log(2/\delta)/\varepsilon^2$.
\begin{itemize}[leftmargin=1.2em]
\item \textbf{(\textsc{R}) Record–level adjacency:} $\Delta=\Delta_{F,\mathrm{rec}}=2\sqrt{2(n-1)}=\Theta(\sqrt n)$ so $\sigma^2=\Theta \big(\frac{n\log(2/\delta)}{\varepsilon^2}\big)$.
\[
\begin{aligned}
\text{No side info:}\quad &
\text{avg.\ per off–diagonal }=\Theta \Big(\frac{n\log(2/\delta)}{\varepsilon^2}\Big),\quad 
\mathcal{R}_{\mathrm{row}}(i)=\Theta \Big(\frac{n^2\log(2/\delta)}{\varepsilon^2}\Big).\\
\text{Knows }\mathbf E_{-i}:\quad &
\text{avg. per off--diagonal }=
\Theta\left(\frac{r_{-i}\log(2/\delta)}{\varepsilon^2}\right),
\quad
\mathcal R_{\mathrm{row}}(i)=
\Theta\left(\frac{n r_{-i}\log(2/\delta)}{\varepsilon^2}\right).
\end{aligned}
\]
With fixed $r$, the adversary's \textit{average} per–entry error is $\mathcal{O}(1)$ in $n$, while a few high-leverage coordinates may be larger.
\item \textbf{(\textsc{O}) Output–space adjacency:} $\mathsf{\Delta}_{\mathsf{G}} = \Theta(1)$ and $\sigma^2 =  \frac{2\log(2/\delta)}{\varepsilon^2} \mathsf{\Delta}_{\mathsf{G}}^2 =  \Theta(1)$.
\[
\begin{aligned}
\text{No side info:}\quad &
\text{avg.\ per off–diagonal }=\Theta\Big(\frac{ \mathsf{\Delta}_{\mathsf{G}}^2 \log(2/\delta)}{\varepsilon^2}\Big),\quad 
\mathcal{R}_{\mathrm{row}}(i)= \Theta \Big(\frac{n \, \mathsf{\Delta}_{\mathsf{G}}^2 \, \log(2/\delta)}{\varepsilon^2}\Big).\\
\text{Knows } \mathbf{E}_{-i}:\quad &
\text{avg.\ per off–diagonal }=\Theta \Big(\frac{r_{-i}\, \mathsf{\Delta}_{\mathsf{G}}^2 \, \log(2/\delta)}{n\,\varepsilon^2}\Big),\quad 
\mathcal{R}_{\mathrm{row}}(i)=\Theta \Big(\frac{r_{-i}\, \mathsf{\Delta}_{\mathsf{G}}^2 \, \log(2/\delta)}{\varepsilon^2}\Big).
\end{aligned}
\]
\end{itemize}

\paragraph{Summary of pre-projection row-risk benchmarks.}
Let $r_{-i}=\mathrm{rank}(\mathbf{E}_{-i})$. The following rates are for the symmetric additive pre-projection model $\mathbf S'=\mathbf S+\mathbf G$. They are not asserted as coordinatewise or row-restricted guarantees for the projected release $\widehat{\mathbf S}$, because projection onto $\mathcal C_{\mathsf{coll}}$ can redistribute error across entries.
\[
\begin{array}{@{}lcc@{}}
\toprule
\textbf{Knowledge regime} & \textbf{(\textsc{R}): } \mathcal{R}_{\mathrm{row}}(i) & \textbf{(\textsc{O}): } \mathcal{R}_{\mathrm{row}}(i) \\
\midrule
\text{No side info} & \Theta\Big( \dfrac{n^2\log(2/\delta)}{\varepsilon^2} \Big) & \Theta\Big( \dfrac{n\,\mathsf{\Delta}_{\mathsf{G}}^2\log(2/\delta)}{\varepsilon^2} \Big) \\
\text{Knows }E_{-i} & \Theta\Big( \dfrac{n\,r_{-i}\,\log(2/\delta)}{\varepsilon^2} \Big) & \Theta\Big( \dfrac{r_{-i}\,\mathsf{\Delta}_{\mathsf{G}}^2\log(2/\delta)}{\varepsilon^2} \Big) \\
\bottomrule
\end{array}
\]
For the symmetric two-strip count in the same pre-projection model, the row risks are multiplied by $2$.

\noindent\textit{Remarks.}
(i) If one perturbs only the upper triangle and mirrors, constants change but the rates above are unaffected.  
(ii) If the attacker knows all of $\mathbf{E}$, then $\mathbf{S}=\mathbf{E}\mathbf{E}^\top$ is already known and reconstruction is trivial. The one-row-unknown model matches record-level adjacency (\textsc{R}).  
(iii) Exact metric projection onto $\mathcal{C}_{\mathsf{coll}}$ cannot increase the overall Frobenius error relative to the symmetric additive input $\mathbf S+\mathbf G$. However, it can redistribute error across entries. Therefore, the pre-projection strip formulas are exact for the purely additive model, while post-projection row or strip risks should be interpreted through the global Frobenius bound unless additional structure is imposed.

\appsubsection{Visual Comparison of MSE Scaling Bounds}
\label{app:sec:mse-scaling-visual}

This subsection reports the analytical MSE bounds used to compare the na\"ive Gaussian mechanism with \textsc{ScoreShield}. We use two scalings of the same bounds. The first set of figures reports unnormalized MSE bounds at fixed privacy parameters $(\varepsilon,\delta)$. The second set reports the same bounds divided by $c_{\varepsilon,\delta} \coloneqq \frac{2\log(2/\delta)}{\varepsilon^2}$, $\sigma^2=c_{\varepsilon,\delta}\Delta^2$. Dividing by $c_{\varepsilon,\delta}$ removes the privacy-calibration factor only for bounds that are linear in $\sigma^2$. This applies to the na\"ive Gaussian risk and to the local rank-aware tangent-cone bound. It does not apply to the global Gaussian-complexity bound, which is linear in $\sigma$; after division by $c_{\varepsilon,\delta}$, that bound retains the factor $c_{\varepsilon,\delta}^{-1/2}$.

\paragraph{Regime~(i): vector release.}
For query-to-collection vector release, the global $\ell_2$-sensitivity is
$\Delta_{\mathsf{query}}=2$. The na\"ive Gaussian release satisfies $B_{\mathsf{naive}}^{(i)}(n) \coloneqq \mathbb E\|\mathbf s'-\mathbf s\|_2^2 =n\sigma^2 = n\,c_{\varepsilon,\delta}\Delta_{\mathsf{query}}^2$.
For projection onto $[-1,1]^n$, if $a(\mathbf s)$ coordinates are boundary-active, the local tangent-cone bound gives $B_{\mathsf{box}}^{(i)}(n,a) \coloneqq c_{\varepsilon,\delta}\Delta_{\mathsf{query}}^2 \left(n-\frac{a(\mathbf s)}{2}\right)$. Thus the projection changes the leading constant through boundary activity, but it does not change the $\Theta(n)$ dependence on $n$. In the vector-release panels, the curve labeled by $a=n$ corresponds to the maximal boundary-active reduction; the case $a=0$ coincides with the na\"ive Gaussian risk.

\paragraph{Regime~(ii): Gram release.}
For full pairwise Gram release, the symmetrized Gaussian perturbation is $\mathbf G=\frac12(\mathbf W+\mathbf W^\top)$, $W_{ij}\stackrel{\mathrm{i.i.d.}}{\sim}\mathcal N(0,\sigma^2)$.
Then $G_{ii}\sim\mathcal N(0,\sigma^2)$ and $G_{ij}=G_{ji}\sim\mathcal N(0,\sigma^2/2)$ for $i<j$. Hence the na\"ive symmetrized Gaussian risk is $B_{\mathsf{naive}}^{(ii)}(n,\Delta) \coloneqq \mathbb E\|\mathbf S'-\mathbf S\|_F^2 = \mathbb E\|\mathbf G\|_F^2 = \frac{n(n+1)}{2}\sigma^2 = \frac{n(n+1)}{2}c_{\varepsilon,\delta}\Delta^2$.

Under record-level adjacency \textbf{(\textsc{R})}, the exact Frobenius sensitivity satisfies $\Delta^2 = \Delta_{F,\mathrm{rec}}^2 = 8(n-1)$, and therefore $B_{\mathsf{naive}}^{(ii,\textsc{R})}(n) = \Theta(c_{\varepsilon,\delta}n^3)$.
Under output-space adjacency \textbf{(\textsc{O})}, $\Delta=\Delta_{\mathsf G}=\Theta(1)$, so $B_{\mathsf{naive}}^{(ii,\textsc{O})}(n) = \Theta(c_{\varepsilon,\delta}\Delta_{\mathsf G}^2 n^2)$.

For the exact Frobenius metric-projection release $\widehat{\mathbf S} = \mathsf{proj}_{\mathcal C_{\mathsf{coll}}}(\mathbf S+\mathbf G)$, the global Gaussian-complexity bound gives $B_{\mathsf{glob}}^{(ii)}(n,\Delta) \coloneqq \widetilde C_{\mathsf{glob}}\,
\sigma n^{3/2} = \widetilde C_{\mathsf{glob}}\, \sqrt{c_{\varepsilon,\delta}}\, \Delta n^{3/2}$.
Consequently,
\[
B_{\mathsf{glob}}^{(ii,\textsc{R})}(n)
= \mathcal O(\sqrt{c_{\varepsilon,\delta}}\,n^2),
\qquad
B_{\mathsf{glob}}^{(ii,\textsc{O})}(n)
= \mathcal O(\sqrt{c_{\varepsilon,\delta}}\,\Delta_{\mathsf G}n^{3/2}).
\]
After division by $c_{\varepsilon,\delta}$, these bounds become
\[
\frac{B_{\mathsf{glob}}^{(ii,\textsc{R})}(n)}{c_{\varepsilon,\delta}}
= \mathcal O\!\left(\frac{n^2}{\sqrt{c_{\varepsilon,\delta}}}\right),
\qquad
\frac{B_{\mathsf{glob}}^{(ii,\textsc{O})}(n)}{c_{\varepsilon,\delta}}
= \mathcal O\!\left(\frac{\Delta_{\mathsf G}n^{3/2}}{\sqrt{c_{\varepsilon,\delta}}}\right).
\]

The conditional rank-aware tangent-cone bound is
\[
B_{\mathsf{rank}}^{(ii)}(n,r,\Delta) \coloneqq \widetilde C_{\mathsf{rank}}\, \sigma^2 n r = \widetilde C_{\mathsf{rank}}\,
c_{\varepsilon,\delta}\Delta^2 n r .
\]
This bound assumes the local Gram-smoothness condition $\mathsf T_{\mathbf S}(\mathcal E_n) = \mathcal T_{\mathrm{man}}(\mathbf S)$. Under \textbf{(\textsc{R})},
\[
B_{\mathsf{rank}}^{(ii,\textsc{R})}(n,r) = \mathcal O(c_{\varepsilon,\delta}n^2r),
\]
whereas under \textbf{(\textsc{O})},
\[
B_{\mathsf{rank}}^{(ii,\textsc{O})}(n,r) = \mathcal O(c_{\varepsilon,\delta}\Delta_{\mathsf G}^2nr).
\]

\paragraph{Pointwise minimum of valid upper bounds.}
Individual analytical upper bounds can be looser than the na\"ive Gaussian risk for some values of $n$. For example, the rank-aware upper bound may lie above the na\"ive curve when the rank-aware estimate is not active. This should not be interpreted as an increase in the exact Frobenius risk after projection. Since $\mathbf S\in\mathcal C_{\mathsf{coll}}$ and Euclidean projection is non-expansive, $\bigl\| \mathsf{proj}_{\mathcal C_{\mathsf{coll}}}(\mathbf S+\mathbf G)-\mathbf S \bigr\|_F \le \|\mathbf G\|_F$. Therefore the na\"ive Gaussian risk is also a valid upper bound on the exact metric-projection risk. Hence the pointwise minimum of the displayed valid upper bounds is itself a valid upper bound:
\[
B_{\mathsf{min}}(n)
\coloneqq
\min\{
B_{\mathsf{naive}}(n),
B_{\mathsf{glob}}(n),
B_{\mathsf{rank}}(n)
\}.
\]
The curve $B_{\mathsf{min}}$ is not a lower bound and is not an empirical risk estimate. It is the smallest among the displayed valid analytical upper bounds at each $n$. If the local Gram-smoothness assumption required for $B_{\mathsf{rank}}$ is not invoked, then the corresponding pointwise minimum is $\min\{B_{\mathsf{naive}}(n), B_{\mathsf{glob}}(n) \}$.
The shaded regions in the figures indicate the difference between the na\"ive Gaussian upper bound and the pointwise minimum of the displayed valid upper bounds.

\paragraph{Global--rank crossover.}
The crossover between the global and rank-aware bounds is obtained from
\[
\widetilde C_{\mathsf{rank}}\,c_{\varepsilon,\delta}\Delta^2nr
\le \widetilde C_{\mathsf{glob}}\sqrt{c_{\varepsilon,\delta}}\Delta n^{3/2}.
\]
Equivalently,
\[
r \le
\frac{\widetilde C_{\mathsf{glob}}}{\widetilde C_{\mathsf{rank}}}
\frac{\sqrt n}{\Delta\sqrt{c_{\varepsilon,\delta}}}.
\]
Under record-level adjacency \textbf{(\textsc{R})}, $\Delta=\Delta_{F,\mathrm{rec}}=\sqrt{8(n-1)}$, so the condition becomes
\[
r \le \frac{\widetilde C_{\mathsf{glob}}}{\widetilde C_{\mathsf{rank}}}
\frac{\sqrt n}{\sqrt{8(n-1)c_{\varepsilon,\delta}}}.
\]
For large $n$, this is a constant-rank condition:
\[
r = \mathcal O\!\left(c_{\varepsilon,\delta}^{-1/2}\right).
\]
Under output-space adjacency \textbf{(\textsc{O})}, $\Delta=\Delta_{\mathsf G}$ is independent of $n$, and the condition becomes
\[
r = \mathcal O\!\left( \frac{\sqrt n}{\Delta_{\mathsf G}\sqrt{c_{\varepsilon,\delta}}} \right).
\]
Thus the crossover $r=\mathcal O(\sqrt n)$ applies only under output-space adjacency with fixed $\Delta_{\mathsf G}$ and fixed privacy parameters.

\begin{figure}[t]
  \centering
  \includegraphics[width=\linewidth]{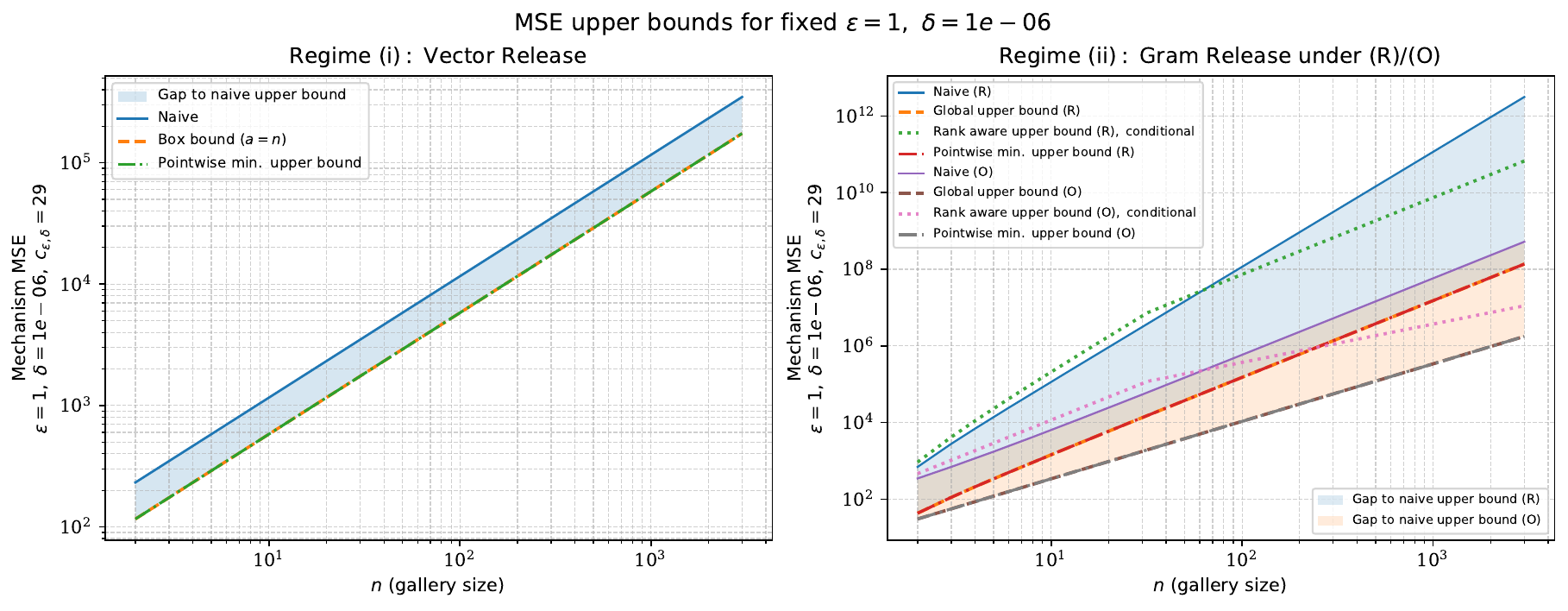}
  \caption{
  \textbf{MSE upper bounds at fixed $(\varepsilon,\delta)$ on a log--log scale.}
  The left panel reports regime~(i), where the na\"ive vector-release risk is $n\,c_{\varepsilon,\delta}\Delta_{\mathsf{query}}^2$ and the box-projection bound with $a=n$ boundary-active coordinates changes only the leading constant.
  The right panel reports regime~(ii) under record-level adjacency \textbf{(\textsc{R})} and output-space adjacency \textbf{(\textsc{O})}.
  The global and conditional rank-aware bounds are shown separately.
  The dash-dotted curve reports the pointwise minimum of the displayed valid upper bounds,
  $\min\{B_{\mathsf{naive}},B_{\mathsf{glob}},B_{\mathsf{rank}}\}$.
  The shaded region indicates the difference between $B_{\mathsf{naive}}$ and this pointwise minimum.
  }
  \label{fig:app:mse-fixed-loglog}
\end{figure}

\begin{figure}[t]
  \centering
  \includegraphics[width=\linewidth]{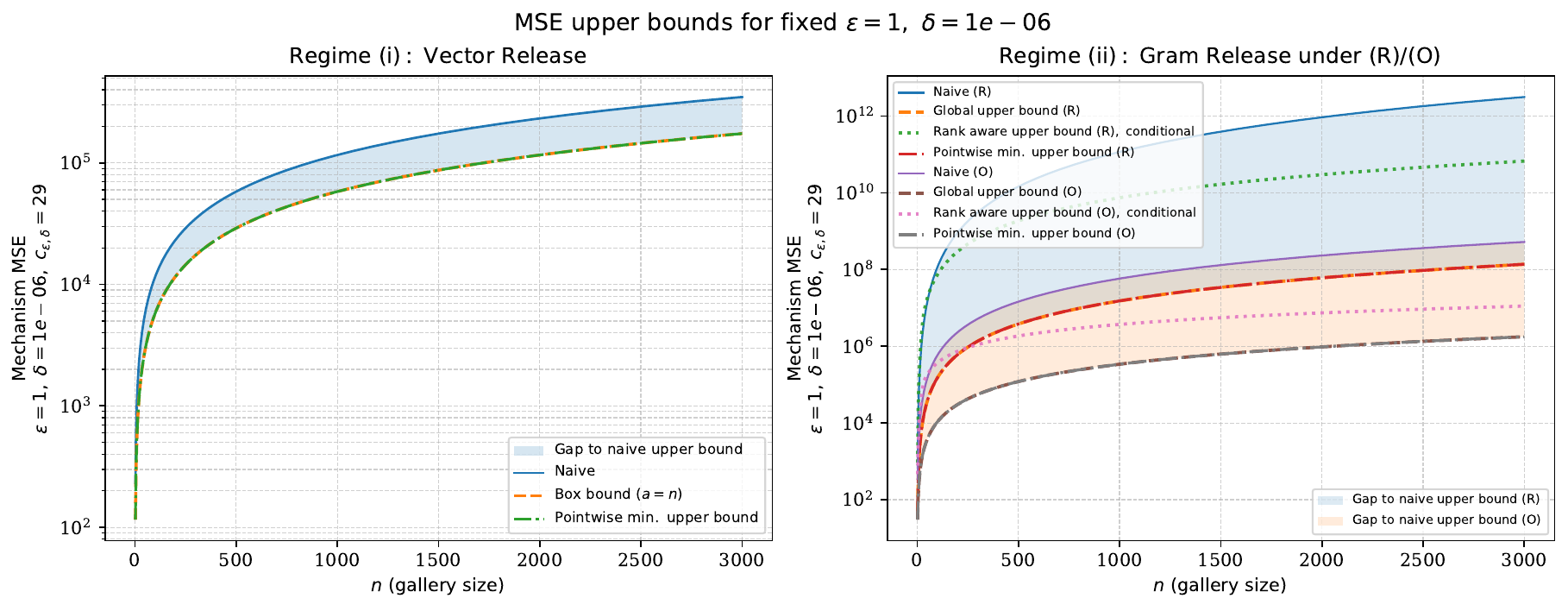}
  \caption{
  \textbf{MSE upper bounds at fixed $(\varepsilon,\delta)$ on a semi-log scale.}
  The quantities are the same as in Fig.~\ref{fig:app:mse-fixed-loglog}.
  The semi-log scale separates the finite-$n$ values of the displayed upper bounds.
  The rank-aware curve can lie above the na\"ive curve when the rank-aware upper bound is loose.
  The dash-dotted curve reports the pointwise minimum of the displayed valid upper bounds.
  }
  \label{fig:app:mse-fixed-semilogy}
\end{figure}

\begin{figure}[t]
  \centering
  \includegraphics[width=\linewidth]{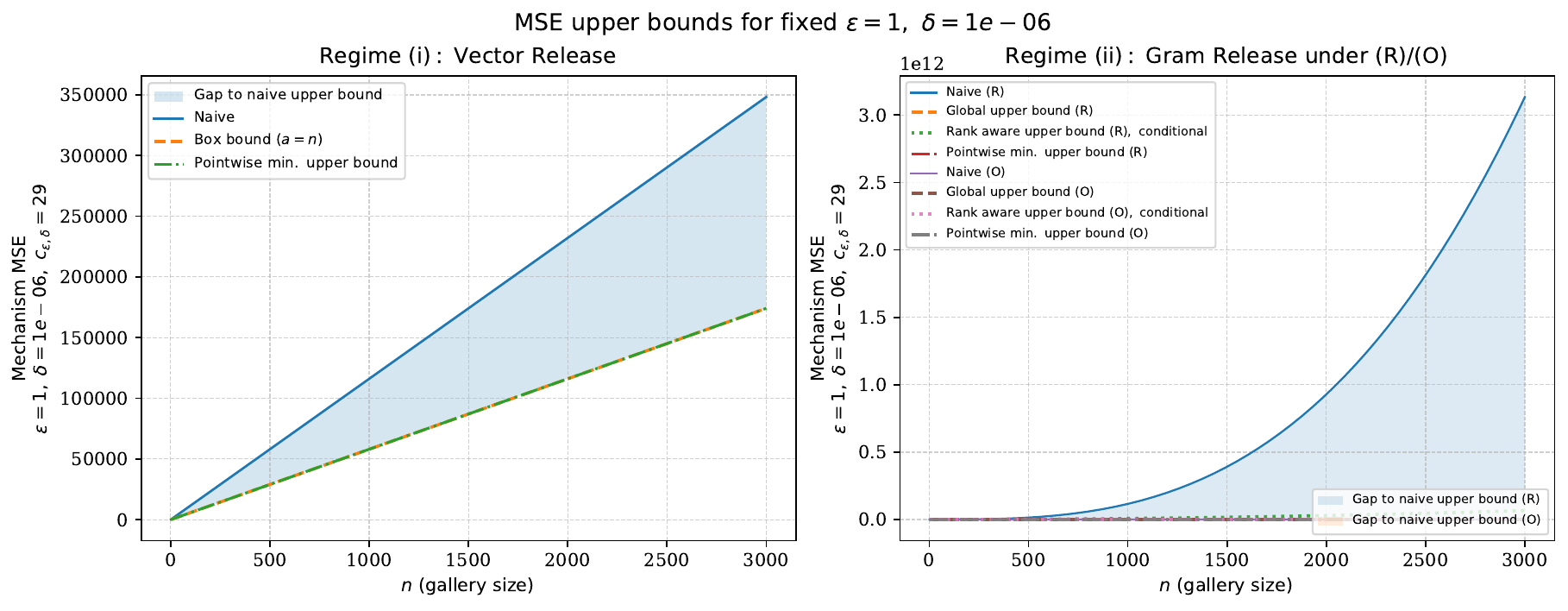}
  \caption{
  \textbf{MSE upper bounds at fixed $(\varepsilon,\delta)$ on a linear scale.}
  This panel reports the same bounds as Figs.~\ref{fig:app:mse-fixed-loglog}--\ref{fig:app:mse-fixed-semilogy} using a linear vertical scale.
  }
  \label{fig:app:mse-fixed-linear}
\end{figure}

\begin{figure}[t]
  \centering
  \includegraphics[width=0.6\linewidth]{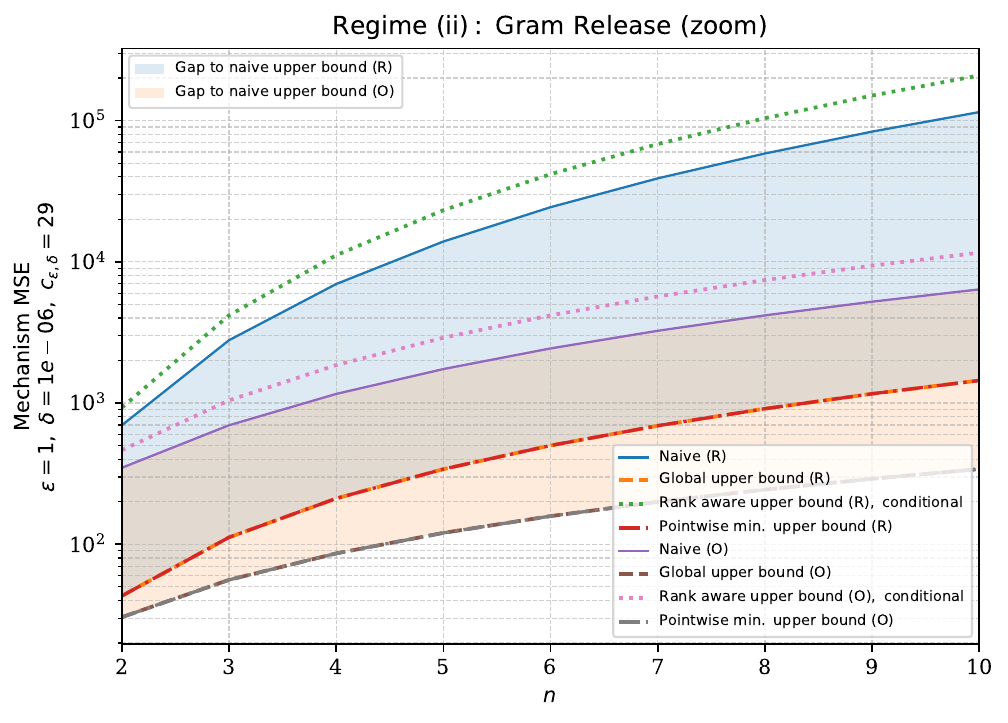}
  \caption{
  \textbf{Finite-$n$ view of unnormalized regime~(ii) MSE upper bounds at fixed $(\varepsilon,\delta)$.}
  This panel restricts the horizontal axis to the finite-$n$ range shown in the figure.
  The rank-aware curves are conditional local bounds and can exceed the na\"ive curve when they are not active.
  In such ranges, the pointwise minimum of the displayed valid upper bounds coincides with either the na\"ive bound or the global bound.
  }
  \label{fig:app:mse-fixed-zoom}
\end{figure}

\begin{figure}[t]
  \centering
  \includegraphics[width=\linewidth]{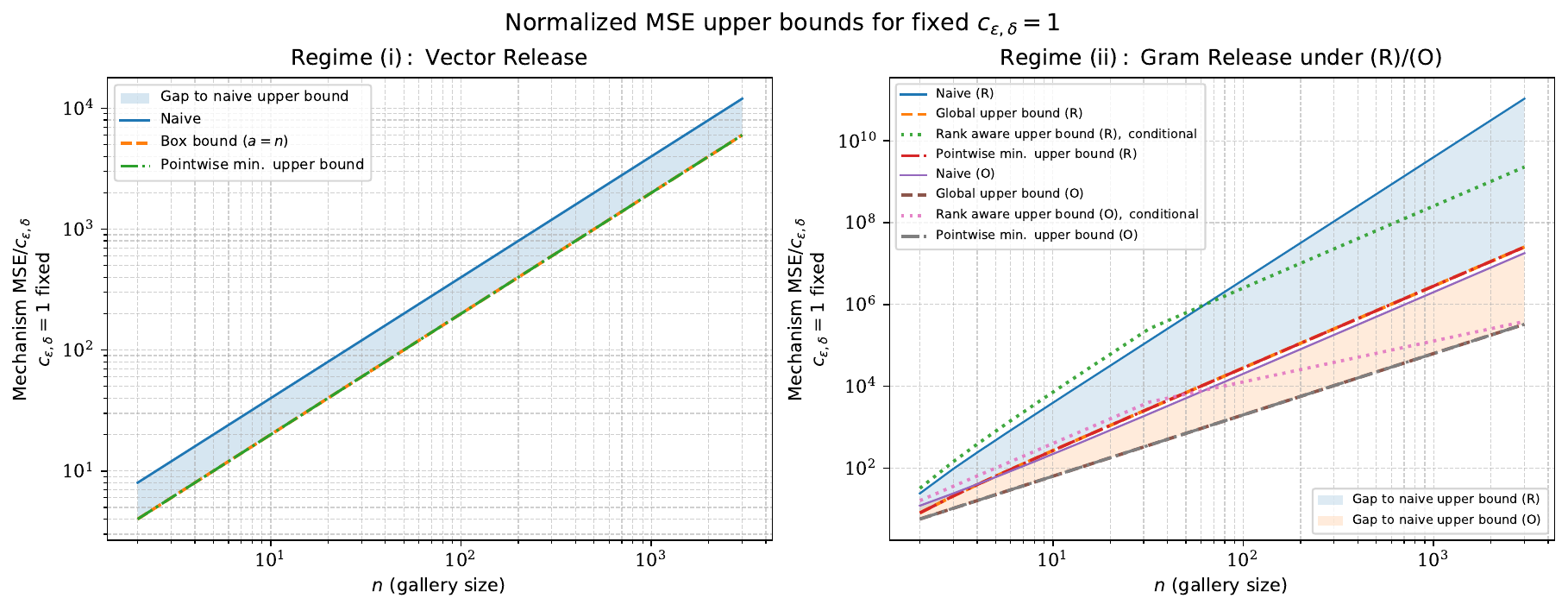}
  \caption{
  \textbf{MSE upper bounds divided by $c_{\varepsilon,\delta}$ on a log--log scale.}
  The displayed quantities are the bounds in Fig.~\ref{fig:app:mse-fixed-loglog} divided by a fixed value of $c_{\varepsilon,\delta}$.
  This division removes the privacy-calibration factor from the na\"ive and rank-aware bounds, which are linear in $\sigma^2$.
  It leaves a factor $c_{\varepsilon,\delta}^{-1/2}$ in the global bound, which is linear in $\sigma$.
  The dash-dotted curve reports the pointwise minimum of the displayed valid upper bounds after the same division by $c_{\varepsilon,\delta}$.
  }
  \label{fig:app:mse-normalized-loglog}
\end{figure}

\begin{figure}[t]
  \centering
  \includegraphics[width=\linewidth]{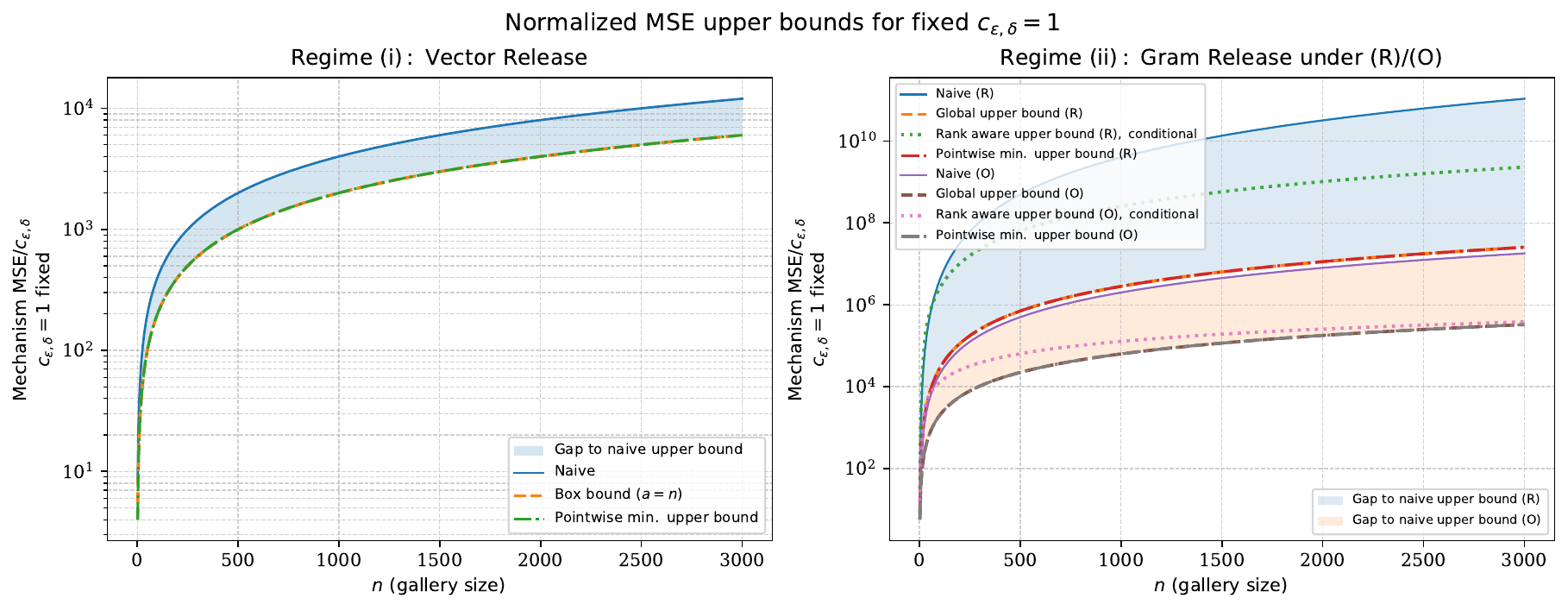}
  \caption{
  \textbf{MSE upper bounds divided by $c_{\varepsilon,\delta}$ on a semi-log scale.}
  The quantities are the same as in Fig.~\ref{fig:app:mse-normalized-loglog}.
  The rank-aware curves are shown even when they are not the smallest valid upper bound.
  }
  \label{fig:app:mse-normalized-semilogy}
\end{figure}

\begin{figure}[t]
  \centering
  \includegraphics[width=\linewidth]{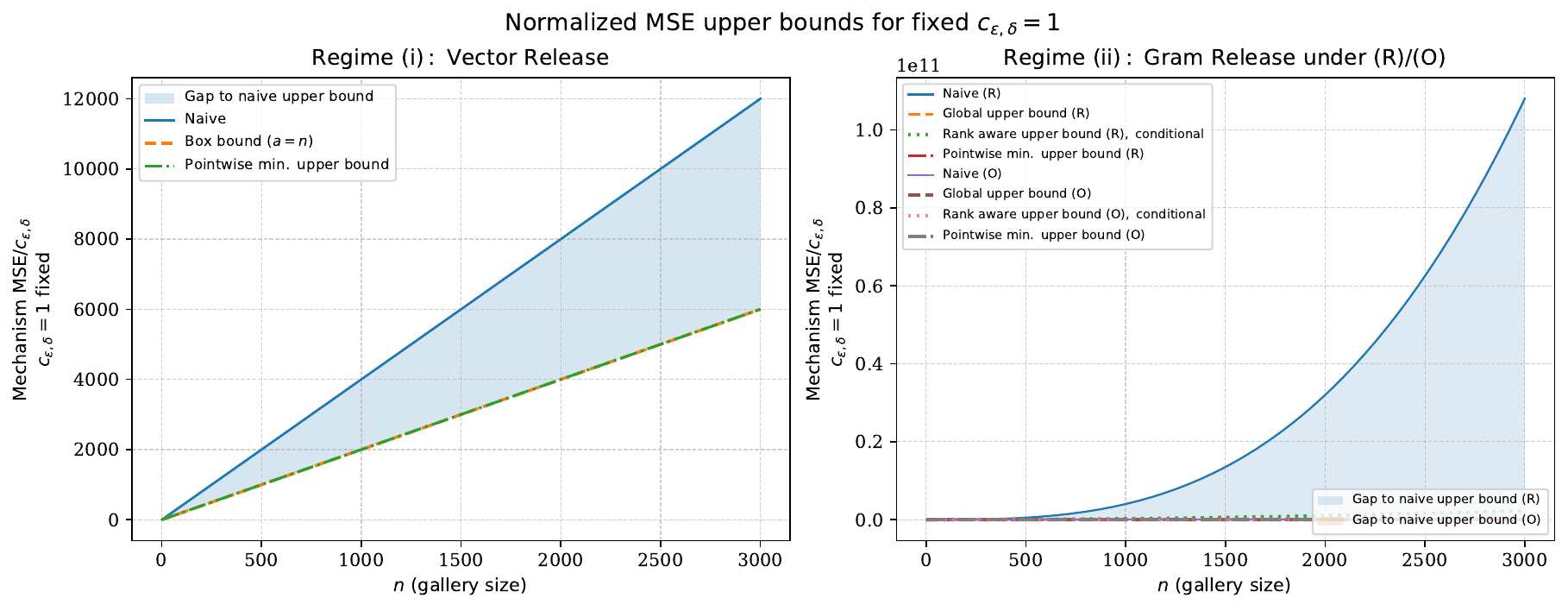}
  \caption{
  \textbf{MSE upper bounds divided by $c_{\varepsilon,\delta}$ on a linear scale.}
  This panel reports the same divided bounds as Figs.~\ref{fig:app:mse-normalized-loglog}--\ref{fig:app:mse-normalized-semilogy} using a linear vertical scale.
  The pointwise minimum curve is an analytical upper bound and is not an empirical MSE curve.
  }
  \label{fig:app:mse-normalized-linear}
\end{figure}

\begin{figure}[t]
  \centering
  \includegraphics[width=0.65\linewidth]{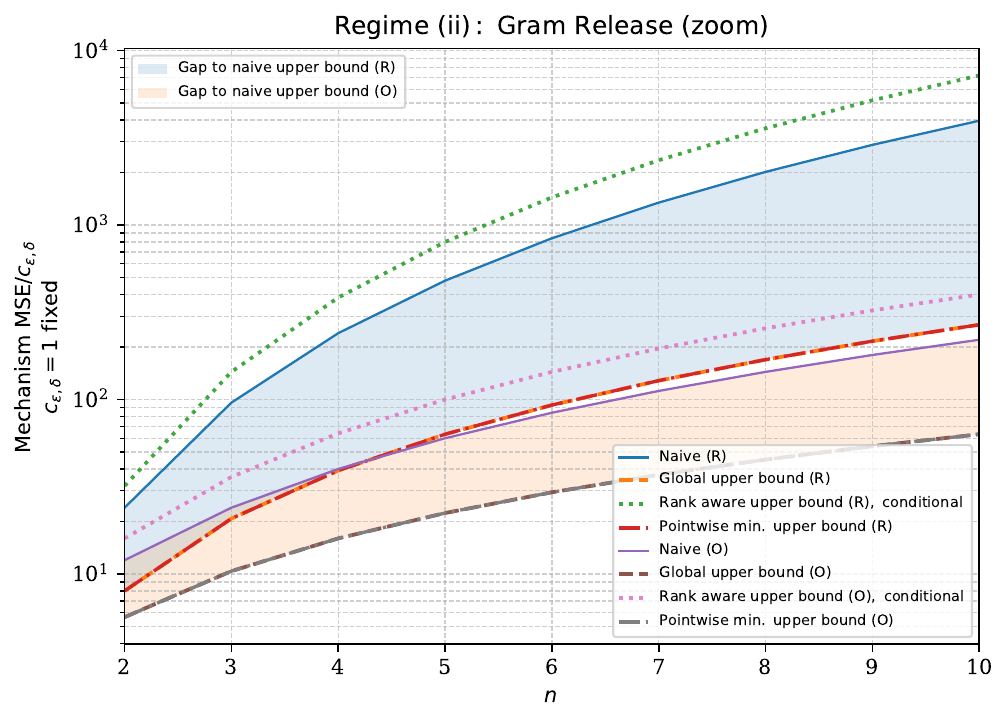}
  \caption{
  \textbf{Finite-$n$ view of regime~(ii) MSE upper bounds divided by $c_{\varepsilon,\delta}$.}
  This panel restricts the horizontal axis to the finite-$n$ range shown in the figure.
  Under record-level adjacency \textbf{(\textsc{R})}, the global--rank crossover is a constant-rank condition up to the constants in the two bounds and the factor $c_{\varepsilon,\delta}^{-1/2}$.
  Under output-space adjacency \textbf{(\textsc{O})}, the crossover satisfies
  $r=\mathcal O(\sqrt n/(\Delta_{\mathsf G}\sqrt{c_{\varepsilon,\delta}}))$.
  }
  \label{fig:app:mse-normalized-zoom}
\end{figure}

\vfill
        
\clearpage
\appsection{Supplementary Details for Regime~(i): Omitted Theorems, Propositions, Proofs and Lemmas}
\label{app:sec:supplementary-regime1-theory}

\appsubsection{Privacy Guarantee and Stability of Projections}

\begin{theorem}[Privacy Guarantee of Query-to-Collection Similarity Score Vector Release]
\label{app:thm:probe-privacy}
Let $\mathcal{M}_{\mathsf{query}}$ be the mechanism returned by Algorithm~\ref{alg:query-to-collection} with parameters $\varepsilon>0,\;\delta\in(0,1)$. Then for every pair of neighboring embedding matrices $\mathbf{E}\sim\mathbf{E}'$ (Def.~\ref{def:sensitivity}) and every measurable set $S\subseteq[-1,1]^n$, 
\begin{equation}
\mathsf{Pr}\, \bigl[\mathcal{M}_{\mathsf{query}}(\mathbf{E}) \in S\bigr] \;\le\;e^{\varepsilon}\,
\mathsf{Pr}\, \bigl[\mathcal{M}_{\mathsf{query}}(\mathbf{E}') \in S\bigr]+\delta .
\end{equation}
Hence $\mathcal{M}_{\mathsf{query}}$ is $(\varepsilon,\delta)$–DP.
\end{theorem}

\begin{proof}
Section~\ref{subsec:probe-vector} established the global $\ell_{2}$-sensitivity $\Delta_{\mathrm{query}}= 2$. The Gaussian mechanism (Lemma~\ref{lem:gaussian}) with scale  
$\sigma=\Delta_{\mathrm{query}}\sqrt{2\log(2/\delta)}/\varepsilon =2\sqrt{2\log(2/\delta)}/\varepsilon$ is therefore $(\varepsilon,\delta)$–DP. Projection onto the fixed convex set $\mathcal{C}=[-1,1]^n$ depends only on the noisy output, so by the post-processing lemma (Lemma~\ref{lem:post-processing}) the composite mechanism remains $(\varepsilon,\delta)$–DP.
\end{proof}

\begin{definition}[Normal cone]
\label{def:normal-cone}
Let $\mathcal{C}\subset \mathbb{R}^n$ be non-empty, closed, and convex, and let $\mathbf{x} \in \mathcal{C}$.
The normal cone to $\mathcal{C}$ at $\mathbf{x}$ is
\begin{equation}
\mathsf{N}_{\mathbf{x}}(\mathcal{C}) \;\coloneqq\; \bigl\{\mathbf{v}\in\mathbb{R}^n:\ \langle \mathbf{v},\mathbf{y}-\mathbf{x}\rangle\le 0,\ \forall\,\mathbf{y}\in\mathcal{C}\bigr\}.    
\end{equation}
\end{definition}

\begin{definition}[Critical cone]
\label{def:critical-cone}
Let $\mathcal{C}\subset\mathbb{R}^n$ be non-empty, closed, and convex. Fix $\mathbf{s}\in\mathbb{R}^n$ and define $\mathbf{x}\coloneqq\mathsf{proj}_{\mathcal{C}}(\mathbf{s})$ and $\mathbf{u}\coloneqq \mathbf{s}-\mathbf{x}\in\mathsf{N}_{\mathbf{x}}(\mathcal{C})$ (see Lemma~\ref{lem:proj-vi}). The critical cone at $(\mathbf{x},\mathbf{u})$ is
\begin{equation}
\mathcal{K}(\mathbf{x},\mathbf{u})
\;\coloneqq\; \mathsf{T}_{\mathbf{x}}(\mathcal{C})\cap \mathbf{u}^{\perp},
\qquad \mathbf{u}^{\perp}\coloneqq\{\mathbf{v}\in\mathbb{R}^n:\langle \mathbf{v},\mathbf{u}\rangle=0\}.
\end{equation}
\end{definition}

\begin{lemma}[Variational Inequality for Euclidean Projection]
\label{lem:proj-vi}
Let $\mathcal{C}\subset\mathbb{R}^n$ be non-empty, closed, and convex. For any $\mathbf{z}\in\mathbb{R}^n$ and $\mathbf{p}\coloneqq \mathsf{proj}_{\mathcal{C}}(\mathbf{z})$, one has
\begin{equation}
\label{eq:vi}
\langle \mathbf{z}-\mathbf{p},\,\mathbf{y}-\mathbf{p}\rangle \le 0, \qquad \forall\,\mathbf{y}\in\mathcal{C}.
\end{equation}
Equivalently, $\mathbf{z}-\mathbf{p}\in \mathsf{N}_{\mathbf{p}}(\mathcal{C})$.
\end{lemma}

\begin{proof}
The point $\mathbf{p}$ minimizes the convex function $\phi(\mathbf{y})\coloneqq \tfrac12\|\mathbf{z}-\mathbf{y}\|_2^2$ over the closed convex set $\mathcal{C}$. For any $\mathbf{y}\in\mathcal{C}$ and $\tau\in(0,1)$, the point $\mathbf{p}_{\tau}\coloneqq (1-\tau)\mathbf{p}+\tau\mathbf{y}\in\mathcal{C}$, hence
$\phi(\mathbf{p})\le \phi(\mathbf{p}_{\tau})$. Expanding $\phi(\mathbf{p}_{\tau})$ and dividing by $\tau$ then letting $\tau \downarrow 0$ yields Eq.~\ref{eq:vi}. The normal-cone equivalence is exactly Definition~\ref{def:normal-cone}.
\end{proof}

\begin{lemma}[Monotonicity of the Normal Cone Mapping]
\label{lem:normal-monotone}
Let $\mathcal{C}\subset\mathbb{R}^n$ be non-empty, closed, and convex. If $\mathbf{v}\in\mathsf{N}_{\mathbf{x}}(\mathcal{C})$ and $\mathbf{v}'\in\mathsf{N}_{\mathbf{x}'}(\mathcal{C})$, then
\begin{equation}
\label{eq:normal-monotone}
\langle \mathbf{v}-\mathbf{v}',\,\mathbf{x}-\mathbf{x}'\rangle \ge 0.
\end{equation}
\end{lemma}

\begin{proof}
Since $\mathbf{v} \in \mathsf{N}_{\mathbf{x}}(\mathcal{C})$ and $\mathbf{x}'\in \mathcal{C}$, we have
$\langle \mathbf{v}, \mathbf{x}'- \mathbf{x}\rangle\le 0$. Similarly, $\mathbf{v}'\in \mathsf{N}_{\mathbf{x}'}(\mathcal{C})$ and $\mathbf{x} \in \mathcal{C}$ imply $\langle \mathbf{v}', \mathbf{x} - \mathbf{x}' \rangle \le 0$. Adding the two inequalities gives Eq.~\ref{eq:normal-monotone}.
\end{proof}

\begin{lemma}[Directional derivative of the projector onto a polyhedral convex set]
\label{lem:proj-directional-derivative-polyhedral}
Let $\mathcal C\subset\mathbb R^m$ be a nonempty closed convex polyhedron.
Fix $\mathbf s\in\mathbb R^m$ and define $\mathbf x\coloneqq \mathsf{proj}_{\mathcal C}(\mathbf s)$, $\mathbf u\coloneqq \mathbf s-\mathbf x\in \mathsf N_{\mathbf x}(\mathcal C)$. Let $\mathcal K \coloneqq \mathsf T_{\mathbf x}(\mathcal C)\cap \mathbf u^\perp$ be the critical cone. Then, for every direction $\mathbf h\in\mathbb R^m$, the one-sided directional derivative
\begin{equation}
\mathrm D\mathsf{proj}_{\mathcal C}(\mathbf s)(\mathbf h)
\coloneqq \lim_{t\downarrow0}
\frac{\mathsf{proj}_{\mathcal C}(\mathbf s+t\mathbf h)-\mathsf{proj}_{\mathcal C}(\mathbf s)}{t}
\end{equation}
exists and satisfies
\begin{equation}
\mathrm D\mathsf{proj}_{\mathcal C}(\mathbf s)(\mathbf h)
= \mathsf{proj}_{\mathcal K}(\mathbf h).
\end{equation}
Equivalently,
\begin{equation}
\mathsf{proj}_{\mathcal C}(\mathbf s+t\mathbf h) = \mathbf x
+ t\,\mathsf{proj}_{\mathcal K}(\mathbf h)
+ o(t),
\qquad t\downarrow0.
\end{equation}
\end{lemma}

\begin{proof}
Since $\mathcal C$ is polyhedral, write $\mathcal C = \{\mathbf z\in\mathbb R^m:\mathbf A\mathbf z\le \mathbf b,\ \mathbf B\mathbf z=\mathbf c\}$.
Let $I(\mathbf x)\coloneqq\{i:\mathbf a_i^\top\mathbf x=b_i\}$ be the active inequality set at $\mathbf x$. Then the tangent cone is
\begin{equation}
\mathsf T_{\mathbf x}(\mathcal C)
= \{\mathbf d:\mathbf a_i^\top\mathbf d\le0\ \forall i\in I(\mathbf x),\ \mathbf B\mathbf d=0\}.
\end{equation}
For $t>0$, set $\mathbf x_t\coloneqq \mathsf{proj}_{\mathcal C}(\mathbf s+t\mathbf h)$, $\mathbf d_t\coloneqq \frac{\mathbf x_t-\mathbf x}{t}$.

By nonexpansiveness of the projector,
\begin{equation}
\|\mathbf d_t\|_2 = \frac{\|\mathsf{proj}_{\mathcal C}(\mathbf s+t\mathbf h)-\mathsf{proj}_{\mathcal C}(\mathbf s)\|_2}{t}
\le \|\mathbf h\|_2,
\end{equation}
so $(\mathbf d_t)$ is bounded.

Because inactive inequalities at $\mathbf x$ have a positive slack, for every bounded set of directions and all sufficiently small $t$, the condition
$\mathbf x+t\mathbf d\in\mathcal C$ is equivalent to
$\mathbf d\in\mathsf T_{\mathbf x}(\mathcal C)$.
Hence, for all sufficiently small $t$, $\mathbf d_t$ is the unique minimizer over $\mathsf T_{\mathbf x}(\mathcal C)$ of $\frac12 \|\mathbf x+t\mathbf d-(\mathbf s+t\mathbf h)\|_2^2$.

Since $\mathbf s=\mathbf x+\mathbf u$, this is equivalent, after removing constants and dividing by $t>0$, to
\begin{equation}
\mathbf d_t = \arg\min_{\mathbf d\in\mathsf T_{\mathbf x}(\mathcal C)}
\left\{ -\langle \mathbf u,\mathbf d\rangle
+ \frac t2\|\mathbf d-\mathbf h\|_2^2 \right\}.
\end{equation}
Because $\mathbf u\in\mathsf N_{\mathbf x}(\mathcal C)$, we have  $\langle \mathbf u,\mathbf d\rangle\le0$, $\forall \mathbf d\in\mathsf T_{\mathbf x}(\mathcal C)$.

Thus the first term is nonnegative and vanishes exactly on $\mathcal K = \mathsf T_{\mathbf x}(\mathcal C)\cap\mathbf u^\perp$.
Let $\mathbf v\in\mathcal K$. By optimality of $\mathbf d_t$,
\begin{equation}
\label{eq:dt-optimality}
-\langle \mathbf u,\mathbf d_t\rangle +
\frac t2\|\mathbf d_t-\mathbf h\|_2^2
\le \frac t2\|\mathbf v-\mathbf h\|_2^2 .
\end{equation}
Since $-\langle \mathbf u,\mathbf d_t\rangle\ge0$, Eq.~\ref{eq:dt-optimality} implies
\begin{equation}
\label{eq:dt-distance-ineq}
\|\mathbf d_t-\mathbf h\|_2^2
\le
\|\mathbf v-\mathbf h\|_2^2,
\qquad \forall \mathbf v\in\mathcal K .
\end{equation}
The same inequality also implies
\begin{equation}
\label{eq:normal-decay}
0
\le
-\langle \mathbf u,\mathbf d_t\rangle
\le
\frac t2\|\mathbf v-\mathbf h\|_2^2,
\qquad \forall \mathbf v\in\mathcal K .
\end{equation}
Let $t_k\downarrow0$ be any sequence such that $\mathbf d_{t_k}\to\mathbf d$.
Since $\mathsf T_{\mathbf x}(\mathcal C)$ is closed, $\mathbf d\in\mathsf T_{\mathbf x}(\mathcal C)$.
Taking the limit in Eq.~\ref{eq:normal-decay} gives
$\langle \mathbf u,\mathbf d\rangle=0$; hence $\mathbf d\in\mathcal K$.
Taking the limit in Eq.~\ref{eq:dt-distance-ineq} gives
\begin{equation}
\|\mathbf d-\mathbf h\|_2^2
\le \|\mathbf v-\mathbf h\|_2^2,
\qquad \forall \mathbf v\in\mathcal K.
\end{equation}
Therefore $\mathbf d=\mathsf{proj}_{\mathcal K}(\mathbf h)$.
Every cluster point of $(\mathbf d_t)$ is the same vector, so $\mathbf d_t\to \mathsf{proj}_{\mathcal K}(\mathbf h)$ as $t\downarrow0$.
This proves the claimed directional derivative.
\end{proof}

\begin{lemma}[Directional derivative of the PSD-cone projector]
\label{lem:psd-projector-directional-derivative}
Let $\mathbf A\in\mathbb S^n$ and let
\begin{equation}
\mathbf A =
\mathbf U
\begin{bmatrix}
    \boldsymbol\Lambda_{+} & 0 & 0\\
    0 & 0 & 0\\
    0 & 0 & \boldsymbol\Lambda_{-}
\end{bmatrix}
\mathbf U^\top
\end{equation}
be an eigendecomposition, where $\boldsymbol\Lambda_{+}\succ0$ contains the positive eigenvalues,
$\boldsymbol\Lambda_{-}\prec0$ contains the negative eigenvalues, and the middle block corresponds to
the zero eigenspace. For $\mathbf H\in\mathbb S^n$, write
\begin{equation}
\widetilde{\mathbf H}
= \mathbf U^\top \mathbf H\mathbf U
=
\begin{bmatrix}
    \widetilde{\mathbf H}_{++} & \widetilde{\mathbf H}_{+0} & \widetilde{\mathbf H}_{+-}\\
    \widetilde{\mathbf H}_{0+} & \widetilde{\mathbf H}_{00} & \widetilde{\mathbf H}_{0-}\\
    \widetilde{\mathbf H}_{-+} & \widetilde{\mathbf H}_{-0} & \widetilde{\mathbf H}_{--}
\end{bmatrix}.
\end{equation}
Define the matrix $\boldsymbol\Gamma\in\mathbb R^{|\!+\!|\times|\!-\!|}$ by
\begin{equation}
\Gamma_{ij} =
\frac{\lambda_i}{\lambda_i-\lambda_j},
\qquad
\lambda_i>0,\ \lambda_j<0.
\end{equation}
Then the projector $\mathsf{proj}_{\mathbb S^n_+}$ is directionally differentiable at $\mathbf A$, and
\begin{equation}
\label{eq:psd-proj-derivative}
\mathrm D\mathsf{proj}_{\mathbb S^n_+}(\mathbf A)(\mathbf H)
= \mathbf U
\begin{bmatrix}
    \widetilde{\mathbf H}_{++}
    &
    \widetilde{\mathbf H}_{+0}
    &
    \boldsymbol\Gamma\circ \widetilde{\mathbf H}_{+-}
    \\
    \widetilde{\mathbf H}_{0+}
    &
    \mathsf{proj}_{\mathbb S^{|0|}_+}(\widetilde{\mathbf H}_{00})
    &
    0
    \\
    \boldsymbol\Gamma^\top\circ \widetilde{\mathbf H}_{-+}
    &
    0
    &
    0
\end{bmatrix}
\mathbf U^\top ,
\end{equation}
where $\circ$ denotes the Hadamard product. Empty blocks are omitted.
\end{lemma}

\begin{proof}
The PSD projection is the spectral operator associated with the scalar function $f(\lambda)=\max\{\lambda,0\}$:
\begin{equation}
\mathsf{proj}_{\mathbb S^n_+}(\mathbf A)
=\mathbf U\,\mathrm{diag}(f(\lambda_1),\ldots,f(\lambda_n))\,\mathbf U^\top.
\end{equation}
For eigenvalue pairs away from zero, the directional derivative of a spectral operator is governed by the first divided differences of $f$:
\begin{equation}
f^{[1]}(\lambda_i,\lambda_j) =
\begin{cases}
\dfrac{f(\lambda_i)-f(\lambda_j)}{\lambda_i-\lambda_j},
& \lambda_i\neq\lambda_j,\\[1.2ex]
f'(\lambda_i),
& \lambda_i=\lambda_j,\ \lambda_i\neq0.
\end{cases}
\end{equation}
Thus the positive-positive block has coefficient $1$, the negative-negative block has coefficient $0$,
the positive-zero block has coefficient $1$, the zero-negative block has coefficient $0$, and the
positive-negative block has coefficient
\begin{equation}
    \frac{\lambda_i}{\lambda_i-\lambda_j},
    \qquad \lambda_i>0,\ \lambda_j<0.
\end{equation}
On the zero eigenspace, $f$ is not differentiable as a scalar function. The directional derivative of
the spectral operator restricted to this block is therefore the spectral operator generated by the
one-sided directional derivative of $f$ at zero, namely $f'_+(0;\mu)=\max\{\mu,0\}$.

Applied to the zero-eigenspace compression $\widetilde{\mathbf H}_{00}$, this gives $\mathsf{proj}_{\mathbb S^{|0|}_+}(\widetilde{\mathbf H}_{00})$.
Combining these block contributions gives Eq.~\ref{eq:psd-proj-derivative}, which is the standard directional-derivative formula for the spectral projection onto the PSD cone.
\end{proof}

\begin{lemma}[Stability of ScoreShield Projections]
\label{lem:stability-proj}
Let $\mathcal{C}\subset\mathbb{R}^n$ be non-empty, closed, and convex, and let $\mathsf{proj}_{\mathcal{C}}$ denote the Euclidean projector as defined in Definition~\ref{def:projection}. Fix $\sigma>0$ and let $\mathbf{w}\sim\mathcal{N}(\mathbf{0},\sigma^2\mathbf{I}_n)$. For any $\mathbf{s}\in\mathbb{R}^n$, define $\mathbf{x}\coloneqq \mathsf{proj}_{\mathcal{C}}(\mathbf{s})$, $\boldsymbol{\Delta}\coloneqq \mathsf{proj}_{\mathcal{C}}(\mathbf{s}+\mathbf{w})-\mathsf{proj}_{\mathcal{C}}(\mathbf{s})$. Then:

\begin{enumerate}[label=\textup{\textbf{(\alph*)}}, itemsep=5pt, wide=0pt]
\item 
\textit{Coarse Global Bound.}
\begin{equation}
\mathbb{E}\|\boldsymbol{\Delta}\|_2^2 \;\le\; \mathbb{E}\|\mathbf{w}\|_2^2 \;=\; n\sigma^2. 
\end{equation}    
\item 
\textit{Global Gaussian‑Complexity Bound (non-asymptotic).}  
If $\mathcal{C}$ is bounded, then
\begin{equation}
\mathbb{E}\|\boldsymbol{\Delta}\|_2^2 \;\le\;  4 \, \sigma\,\mathsf{GC}(\mathcal{C}),
\end{equation}
%
%
\item 
\textit{Small-noise Local Limit (geometry-aware).}
Let $\mathbf{z} \sim \mathcal{N}(\mathbf{0}, \mathbf{I}_n)$ and define $\mathbf{u} \coloneqq \mathbf{s} - \mathbf{x} \in \mathsf{N}_{\mathbf{x}}(\mathcal{C})$. Let $\mathcal{K} \coloneqq \mathcal{K}(\mathbf{x}, \mathbf{u})$ be the critical cone (Definition~\ref{def:critical-cone}). Then
\begin{equation}
\lim_{t\downarrow 0}\frac{1}{t^2} \, \mathbb{E}\big\|\mathsf{proj}_{\mathcal{C}}(\mathbf{s} + t \, \sigma \, \mathbf{z}) - \mathbf{x}\big\|_2^2 \;=\; \sigma^2\,\delta(\mathcal{K}), \qquad \delta(\mathcal{K})\coloneqq\mathbb{E}\|\mathsf{proj}_{\mathcal{K}}(\mathbf{z})\|_2^2 .
\end{equation}
Moreover, $\delta(\mathcal{K}) \le n$ and
\begin{equation}
\mathsf{GW}(\mathcal{K})^2 \;\le\; \delta(\mathcal{K}) \; \le \; \mathsf{GW} (\mathcal{K})^2+1.
\end{equation}
\end{enumerate}
\end{lemma}

\begin{proof}
\leavevmode

\noindent
\textbf{(a)}  
By firm non-expansiveness of $\mathsf{proj}_{\mathcal{C}}$ (Lemma~\ref{lem:proj-firm}), one has $\|\boldsymbol{\Delta}\|_2 \le \|\mathbf{w}\|_2$. Squaring and taking expectations gives $\mathbb{E}\|\boldsymbol{\Delta}\|_2^2 \le \mathbb{E}\|\mathbf{w}\|_2^2 = n\sigma^2$.

\textbf{(b)}  
See Lemma~\ref{lem:global-gc-scoreshield} for a complete proof.

\textbf{(c)}  
Define $\mathbf{x}= \mathsf{proj}_{\mathcal{C}}(\mathbf{s})$ and $\mathbf{u} = \mathbf{s}- \mathbf{x}$, and let $\mathcal{K} = \mathcal{K}(\mathbf{x}, \mathbf{u})$. For each fixed realization of $\mathbf{z}$, Lemma~\ref{lem:proj-directional-derivative-polyhedral} yields
\begin{equation}
\label{eq:ptwise}
\frac{\mathsf{proj}_{\mathcal{C}}(\mathbf{s}+ t \, \sigma \,\mathbf{z})-\mathbf{x}}{t}
\; \xrightarrow[t\downarrow 0]{}\;
\sigma\,\mathsf{proj}_{\mathcal{K}}(\mathbf{z}).
\end{equation}
Moreover, by $1$-Lipschitzness of $\mathsf{proj}_{\mathcal{C}}$,
\begin{equation}
\Big\|\frac{\mathsf{proj}_{\mathcal{C}}(\mathbf{s}+ t \, \sigma \,\mathbf{z})-\mathbf{x}}{t}\Big\|_2
\;\le\; \sigma \, \|\mathbf{z}\|_2, \qquad \forall\,t>0.
\end{equation}
Since $\mathbb{E}\|\mathbf{z}\|_2^2<\infty$, the family $\big\|\big(\mathsf{proj}_{\mathcal{C}}(\mathbf{s}+t\sigma\mathbf{z}) - \mathbf{x}\big)/t\big\|_2^2$ is dominated by $\sigma^2 \|\mathbf{z}\|_2^2$. Applying dominated convergence to Eq.~\ref{eq:ptwise} gives
\begin{equation}
\lim_{t\downarrow 0}\frac{1}{t^2}\, \mathbb{E}\big\|\mathsf{proj}_{\mathcal{C}}(\mathbf{s}+t\sigma\mathbf{z})-\mathbf{x}\big\|_2^2 = \sigma^2\,\mathbb{E}\|\mathsf{proj}_{\mathcal{K}}(\mathbf{z})\|_2^2 = \sigma^2\,\delta(\mathcal{K}).
\end{equation}
The bound $\delta(\mathcal{K})\le n$ follows from $\| \mathsf{proj}_{\mathcal{K}}(\mathbf{z}) \|_2\le \|\mathbf{z}\|_2$ and $\mathbb{E}\| \mathbf{z} \|_2^2 = n$. The inequalities $\mathsf{GW}(\mathcal{K})^2 \le \delta(\mathcal{K}) \le \mathsf{GW}(\mathcal{K})^2 + 1$ are the standard relationship between statistical dimension and squared Gaussian width for closed convex cones.
\end{proof}

\begin{corollary}[Piecewise-affine structure of polyhedral projections in \textsc{ScoreShield}]
\label{cor:polyhedral-cube-exact}
Let $\mathcal C\subset\mathbb R^n$ be a nonempty closed convex polyhedron.
Fix $\mathbf s\in\mathbb R^n$ and define $\mathbf x\coloneqq \mathsf{proj}_{\mathcal C}(\mathbf s)$, $\mathbf u\coloneqq \mathbf s-\mathbf x\in \mathsf N_{\mathbf x}(\mathcal C)$. Let $\mathcal K \coloneqq \mathsf T_{\mathbf x}(\mathcal C)\cap \mathbf u^\perp$ denote the critical cone at $(\mathbf x,\mathbf u)$.

\begin{enumerate}[label=\textup{\textbf{(\alph*)}}, itemsep=6pt, wide=0pt]

\item
\textit{Locally exact first-order expansion.}
There exist finitely many polyhedral cones $\{\mathcal Q_j\}_{j=1}^J$ whose union is $\mathbb R^n$ and whose relative interiors are disjoint such that, for every $j$ and every $\mathbf h\in\mathrm{ri}(\mathcal Q_j)$, there exists $t_0(\mathbf h)>0$ for which
\begin{equation}
\mathsf{proj}_{\mathcal C}(\mathbf s+t\mathbf h)
= \mathbf x + t\,\mathrm D\mathsf{proj}_{\mathcal C}(\mathbf s)(\mathbf h),
\qquad
\forall t\in(0,t_0(\mathbf h)).
\end{equation}
Moreover, by Lemma~\ref{lem:proj-directional-derivative-polyhedral}, $\mathrm D\mathsf{proj}_{\mathcal C}(\mathbf s)(\mathbf h) = \mathsf{proj}_{\mathcal K}(\mathbf h)$.

Thus the first-order expansion has zero remainder along all directions in the relative interiors of the cones of this partition. In particular, if $\mathbf h$ has an absolutely continuous distribution, for example $\mathbf h\sim\mathcal N(\mathbf 0,\mathbf I_n)$, then this locally exact regime holds with probability one.

\item
\textit{Explicit critical cone and statistical dimension for the box.}
Let $\mathcal C=[-1,1]^n$ and let $\mathbf x=\mathsf{proj}_{\mathcal C}(\mathbf s)$ be the componentwise clipping of $\mathbf s$. Define
\begin{equation}
I_0 \coloneqq \{i:\ |x_i|<1\},\qquad
I_+ \coloneqq \{i:\ x_i=1\},\qquad
I_- \coloneqq \{i:\ x_i=-1\}.
\end{equation}
Refine the active sets according to complementarity:
\begin{equation}
I_+^{>0}\coloneqq \{i\in I_+:\ u_i>0\},\qquad
I_+^{0}\coloneqq \{i\in I_+:\ u_i=0\},
\end{equation}
and
\begin{equation}
I_-^{<0}\coloneqq \{i\in I_-:\ u_i<0\},\qquad
I_-^{0}\coloneqq \{i\in I_-:\ u_i=0\}.
\end{equation}
Then
\begin{equation}
\mathcal K =
\Bigl\{
\mathbf v\in\mathbb R^n:
v_i\in\mathbb R\ \forall i\in I_0,\ 
v_i\le 0\ \forall i\in I_+^{0},\
v_i\ge 0\ \forall i\in I_-^{0},\
v_i=0\ \forall i\in I_+^{>0}\cup I_-^{<0}
\Bigr\}.
\end{equation}
Consequently, for $\mathbf z\sim\mathcal N(\mathbf 0,\mathbf I_n)$,
\begin{equation}
\delta(\mathcal K) =
\mathbb E\|\mathsf{proj}_{\mathcal K}(\mathbf z)\|_2^2
= |I_0| + \frac12\bigl(|I_+^{0}|+|I_-^{0}|\bigr).
\end{equation}
In particular, if $I_+^{0}=I_-^{0}=\emptyset$, $\delta(\mathcal K)=|I_0|$ and, if $\mathbf u=\mathbf 0$, equivalently $\mathbf s\in[-1,1]^n$ and $\mathbf x=\mathbf s$, then
\begin{equation}
\delta(\mathcal K)
= |I_0| + \frac12(|I_+|+|I_-|)
= n-\frac12(|I_+|+|I_-|).
\end{equation}

\end{enumerate}
\end{corollary}

\begin{proof}
\leavevmode

\noindent
\textbf{(a)}
Since $\mathcal C$ is polyhedral, the Euclidean projector
$\mathsf{proj}_{\mathcal C}$ is piecewise affine. Hence there exist finitely many polyhedra $\{\mathcal R_j\}_{j=1}^J$ covering $\mathbb R^n$ and affine maps $\mathbf z\mapsto \mathbf A_j\mathbf z+\mathbf b_j$ such that $\mathsf{proj}_{\mathcal C}(\mathbf z) = \mathbf A_j\mathbf z+\mathbf b_j$, $ \forall \mathbf z\in\mathcal R_j$. Fix $\mathbf s$ and define the corresponding cones of directions
\begin{equation}
\mathcal Q_j \coloneqq
\left\{ \,
\mathbf h\in\mathbb R^n: \exists\,t_0>0  \text{ such that }
\mathbf s+t\mathbf h\in\mathcal R_j,\;\forall t\in(0,t_0)
\, \right\}.
\end{equation}
After discarding empty cones and refining overlaps if necessary, these sets form a finite polyhedral conic partition of $\mathbb R^n$.

If $\mathbf h\in\mathrm{ri}(\mathcal Q_j)$, then there exists $t_0(\mathbf h)>0$ such that
$\mathbf s+t\mathbf h\in\mathcal R_j$ for all $t\in(0,t_0(\mathbf h))$. Therefore
\begin{subequations}
\begin{eqnarray}
\mathsf{proj}_{\mathcal C}(\mathbf s+t\mathbf h)
&=& \mathbf A_j(\mathbf s+t\mathbf h)+\mathbf b_j  \\
&=& \mathsf{proj}_{\mathcal C}(\mathbf s)+t\mathbf A_j\mathbf h,
\qquad \forall t\in(0,t_0(\mathbf h)).    
\end{eqnarray}
\end{subequations}
By Lemma~\ref{lem:proj-directional-derivative-polyhedral},
\begin{equation}
\mathbf A_j\mathbf h = \mathrm D\mathsf{proj}_{\mathcal C}(\mathbf s)(\mathbf h) = \mathsf{proj}_{\mathcal K}(\mathbf h).
\end{equation}
This proves the locally exact expansion. The boundary of a finite polyhedral conic partition is a finite union of lower-dimensional polyhedral cones and therefore has Lebesgue measure zero. Hence any absolutely continuous direction belongs to the union of the relative interiors with probability one.

\noindent
\textbf{(b)}
For $\mathcal C=[-1,1]^n$, the tangent cone at $\mathbf x$ is coordinatewise:
\begin{equation}
v_i\in\mathbb R\quad (i\in I_0),\qquad
v_i\le0\quad (i\in I_+),\qquad
v_i\ge0\quad (i\in I_-).
\end{equation}
The normal vector $\mathbf u=\mathbf s-\mathbf x$ has the corresponding signs
\begin{equation}
u_i=0\quad (i\in I_0),\qquad
u_i\ge0\quad (i\in I_+),\qquad
u_i\le0\quad (i\in I_-).
\end{equation}
Therefore, for every $\mathbf v\in\mathsf T_{\mathbf x}(\mathcal C)$, $u_i v_i\le 0$, $i=1,\ldots,n$.
The condition $\mathbf v\in\mathbf u^\perp$ is $\sum_{i=1}^n u_i v_i=0$.

Since every summand is nonpositive, the sum can equal zero only if $u_i v_i=0$, $i=1,\ldots,n$.
Thus, on an active coordinate with strict complementarity, namely
$u_i>0$ on $I_+$ or $u_i<0$ on $I_-$, we must have $v_i=0$. On active coordinates with $u_i=0$, the one-sided tangent restriction remains. This gives the stated product-form description of $\mathcal K$.

Since $\mathcal K$ is a Cartesian product of coordinate cones, its Euclidean projection acts coordinatewise. A free coordinate contributes $\mathbb E[z_i^2]=1$. A half-line coordinate contributes $\mathbb E[(z_i)_+^2] = \mathbb E[(z_i)_-^2] = \frac12$, by symmetry of the standard normal distribution. A fixed-zero coordinate contributes zero. Summing the coordinate contributions yields
\begin{equation}
\label{eq:statdim-box-critical-cone}
\delta(\mathcal K) = |I_0| + \frac12\bigl(|I_+^{0}|+|I_-^{0}|\bigr).
\end{equation}
If strict complementarity holds on all active coordinates, then
$I_+^{0}=I_-^{0}=\emptyset$, and Eq.~\ref{eq:statdim-box-critical-cone} gives
$\delta(\mathcal K)=|I_0|$.
If $\mathbf u=\mathbf 0$, equivalently $\mathbf s\in[-1,1]^n$ and $\mathbf x=\mathbf s$, then
$I_+^{>0}=I_-^{<0}=\emptyset$, and Eq.~\ref{eq:statdim-box-critical-cone} gives
\[
\delta(\mathcal K)
=
|I_0|
+
\frac12(|I_+|+|I_-|)
=
n-\frac12(|I_+|+|I_-|).
\]
\end{proof}

\appsubsection{Impact on Verification Thresholds}
\label{ssec:threshold-impact}

Fix an operating verification threshold $\tau \in [-1,1]$. 
The \textit{clean} verification decision for a pair $(i,j)$ is $\textsf{accept} \;\Longleftrightarrow \; S_{ij}\;>\;\tau$, where $S_{ij}=\langle\mathbf{e}_i,\mathbf{e}_j\rangle$ is the cosine-similarity score. 
Let $\mathbf{s}\in[-1,1]^n$ be the vector of query-to-collection similarities for a given probe, one entry per gallery identity. \textsc{ScoreShield} perturbs and then projects scores coordinatewise:
\[
\mathbf{s}' \;=\; \mathbf{s}+\mathbf{w},\qquad \mathbf{w}\sim\mathcal{N}(\mathbf{0},\sigma^2 \mathbf{I}_n),
\]
\[
\widehat{\mathbf{s}} \;=\; \mathsf{proj}_{\mathcal{C}_{\mathsf{query}}}(\mathbf{s}'),\qquad
\mathcal{C}_{\mathsf{query}} =[-1,1]^n,\qquad
\bigl(\mathsf{proj}_{\mathcal{C}_{\mathsf{query}}}(\mathbf{s}')\bigr)_i=\max \bigl(-1,\min(1,s'_i)\bigr).
\]

\paragraph{Projection monotonicity at interior thresholds.}
The scalar projection $\mathsf{proj}_{[-1,1]}$ is nondecreasing and fixes every point in $(-1,1)$. Hence, for any scalar score $S$ and any $\tau \in [-1, 1)$,
\begin{equation}
\label{eq:proj-monotone-interior}
\mathbf{1}  \left\{\,\mathsf{proj}_{[-1,1]}(S+W)>\tau\,\right\}
 \;=\; \mathbf{1} \left\{\,S+W>\tau\,\right\}.  
\end{equation}
Indeed, if $S{+}W\le\tau$ then $\mathsf{proj}(S{+}W)\le\tau$, if $\tau<S{+}W \le 1$ then $\mathsf{proj}(S{+}W)= S{+}W > \tau$, and if $S{+}W>1$ then $\mathsf{proj}(S{+}W)=1>\tau$ since $\tau<1$. The only exceptional case is $\tau=1$, where $\mathbf{1}\{\mathsf{proj}(S{+}W)>1\}\equiv 0$ but $\mathbf{1}\{S{+}W>1\}$ may be $1$.

Therefore, for any interior threshold $\tau\in(-1,1)$, clipping does not affect the decision rule:
$\mathbf{1}\{\widehat{S}>\tau\}=\mathbf{1}\{S+W>\tau\}$.
Hence post-privacy FMR depend only on the additive Gaussian mechanism noise $W$ and hold for an arbitrary distribution of the clean score $S$.

\newcommand{\Q}[1]{\mathsf{Q} \left(#1\right)}   

\begin{definition}[Complementary Gaussian Tail]
\label{def:gaussian-tail}
For a standard normal variable $Z \sim \mathcal{N}(0,1)$ we denote
\begin{equation}
\Q{x} \;=\; \mathsf{Pr}\,[Z \ge x] \;=\; \frac{1}{\sqrt{2\pi}}\int_{x}^{\infty} e^{-u^{2}/2}\, \mathrm{d}u. 
\end{equation}
The lower-tail CDF is $\Phi(x)=  1- \mathsf{Q}(x)$.
\end{definition}

\begin{definition}[Decision Flip]
\label{def:decision-flip}
Let $S_{ij}\in[-1,1]$ be the clean cosine-similarity score for a pair $(i,j)$, let $\widehat{S}_{ij}$ be its  perturbed-and-projected version, and let $\tau\in[-1,1]$ be a fixed verification threshold. A \textit{decision flip} occurs when the private verdict disagrees with the clean verdict, i.e.,
\begin{equation}
(S_{ij}-\tau) \, (\widehat{S}_{ij}-\tau) < 0 .
\end{equation}
%
Ties are not counted as a flip.
\end{definition}

\begin{proposition}[Exact flip probability at interior thresholds]
\label{prop:flip-exact-interior}
Let $S\in[-1,1]$, $\tau\in(-1,1)$ and $W\sim\mathcal{N}(0,\sigma^{2})$. Then
\begin{equation}
\mathsf{Pr}[ \, \textnormal{flip}\mid S=s \, ] \;=\; \Q{\frac{|s-\tau|}{\sigma}}.    
\end{equation}
\end{proposition}

\begin{proof}
By Eq.~\ref{eq:proj-monotone-interior}, for every $\tau<1$ we have
$\mathbf{1} \{\widehat{S} > \tau \} = \mathbf{1} \{ S+W > \tau \}$, so a flip occurs iff $S$ and $S{+}W$ lie on different sides of $\tau$, i.e., iff $W< -d$ when $d \coloneqq  s-\tau > 0$, or $W > -d$ when $d<0$. Each case has probability $\Q{|d|/\sigma}$ by symmetry.
\end{proof}

\begin{remark}[Endpoint behavior at $\tau=1$]
\label{rem:endpoint-flips}
With strict rule ``$\, > \,$'', $\mathbf{1} \{\widehat{S} > 1\} \equiv 0$ for all $S \le 1$, so flips cannot occur at $\tau=1$. The flip probability equals $0$ even though $\mathsf{Pr}[S{+}W>1]$ may be positive. That is
\[
\mathsf{Pr}[\textnormal{flip}\mid S=s]=0
\qquad\text{and}\qquad
\mathsf{Pr}[\widehat{S}>1]=0.
\]
This is the only threshold at which clipping alters the decision rule. For every interior threshold $\tau\in(-1,1)$, clipping does not affect decisions and the flip law is given by Proposition~\ref{prop:flip-exact-interior}.
This endpoint effect does not preclude threshold re-calibration. Under $W\sim\mathcal{N}(0,\sigma^2)$,
the random variable $S+W$ admits a continuous density. Hence, for any fixed (possibly random) $S\in[-1,1]$, the map $\tau\mapsto \mathsf{Pr}[\widehat{S}>\tau]=\mathsf{Pr}[S+W>\tau]$ is continuous and strictly decreasing on $(-1,1)$, and with the strict rule ``$>$'' we have $\mathsf{Pr}[\widehat{S}>1]=0$.
One can potentially raise the threshold from any $\tau_\alpha \in(-1,1)$ to a unique $\tau_\alpha + \Delta\tau\le1$ that restores any target $\alpha \in(0,1)$. 
We formalize this in the following. 
\end{remark}

\begin{remark}[Exponential decay and Mills' bound]
\label{rem:mills}
For $k > 0$,
\begin{equation}
\mathsf{Pr}[\textnormal{flip}\mid S=s] \;=\; \Q{k}\;\le\;\frac{e^{-k^2/2}}{k\sqrt{2\pi}} ,    
\end{equation}
so the flip probability in Proposition~\ref{prop:flip-exact-interior} decays like $\exp(-k^2/2)$ in the normalized verification margin $k = |s-\tau|/ \sigma$. The bound is conservative but tightens as $k$ grows.
\end{remark}

\paragraph{Acceptance probabilities.}
For any $\tau<1$, Eq.~\ref{eq:proj-monotone-interior} yields 
\begin{equation}
\label{eq:acc-prob-conv}
\mathsf{Pr}[\widehat{S} > \tau]
\;=\; \mathbb{E}\bigl[\mathbf{1}\{S+W>\tau\}\bigr]
\;=\; \mathbb{E}\Bigl[\Q{\frac{\tau-S}{\sigma}}\Bigr],
\end{equation}
where the expectation is taken over the (arbitrary) distribution of the clean score $S$ and $W\sim\mathcal{N}(0,\sigma^2)$ is independent. At the endpoint, $\mathsf{Pr}[\widehat{S}>1]=0$ under the strict rule ``$>$'' (Remark~\ref{rem:endpoint-flips}).
Because $W$ has a continuous density, $\mathsf{Pr}[S+W=\tau]=0$ for every $\tau <1$, so the strict ``$>$'' and non-strict ``$\ge$'' rules coincide on $(-1,1)$.

\paragraph{Endpoint behavior at $\tau=1$.}
With the strict rule ``$\widehat{S} > \tau$'', projection saturates at $1$ but $1\not>1$, hence
\[
\mathsf{Pr}[\widehat{S}>1]=0, \qquad \forall \, S \le 1,
\]
which is the only threshold where clipping alters the decision rule. For every $\tau<1$ the projection is monotone and fixes $(-1,1)$, hence ``$>$'' and ``$\ge$'' coincide and $\mathbf{1}\{\widehat{S}>\tau\}=\mathbf{1}\{S+W>\tau\}$. If the upper endpoint rule is instead non-strict (i.e., ``$\widehat{S} \ge \tau$''), then
\begin{equation}
\mathsf{Pr}[ \widehat{S} \ge 1 ] = \mathsf{Pr}[ S+W \ge 1 ] = \mathbb{E} \Bigl[ \Q{\frac{1-S}{\sigma}} \Bigr] > 0,
\end{equation}
so the minimum achievable acceptance over $\tau \le 1$ is strictly positive in that semantics. (For every $\tau<1$, ``$>$'' and ``$\ge$'' remain identical.)

\paragraph{Recalibration guarantee and feasibility conditions.}
The map $\tau\mapsto \mathsf{Pr}[\widehat{S}>\tau]$ is continuous and strictly decreasing on $(-1,1)$. 
From Eq.~\ref{eq:acc-prob-conv} we have
\begin{equation}
\frac{\mathrm{d}}{\mathrm{d}\tau}\mathsf{Pr}[\widehat{S}>\tau]= -\frac{1}{\sigma}\,\mathbb{E}\!\left[\varphi\!\left(\frac{\tau-S}{\sigma}\right)\right]
\;<\;0,
\end{equation}
Thus $\frac{\mathrm{d}}{\mathrm{d}\tau}\mathsf{Pr}[\widehat{S}>\tau]<0$ for $\tau\in(-1,1)$. At $\tau=1$ the strict rule induces a jump discontinuity, so this derivative statement applies only to the interior.
Consequently, starting from any interior operating point $\tau_0\in(-1,1)$ there exists a  \textit{unique} offset $\Delta\tau\in[0,1-\tau_0]$ such that
\[
\mathsf{Pr}[\widehat{S} > \tau_0 + \Delta \tau]= \alpha,
\]
where $\alpha$ is a desired target utility level. 
In the following sections we will instantiate this with impostor scores (yielding a recalibration that restores a desired operating rate).
By contrast, under the non-strict endpoint rule one encounters a genuine \textit{feasibility frontier}: targets below $\mathbb{E}[\Q{(1-S)/\sigma}]$ are unattainable unless $\sigma$ is reduced (equivalently, $\varepsilon$ increased).  
We defer the precise statement and computation to the next section.


\begin{definition}[Clean False–Match Rate ($\mathsf{FMR}$)]
\label{def:clean-fmr}
Let $S_{ij}$ denote an \textit{impostor} cosine–similarity score ($i\neq j$). For any threshold $\tau \in [-1,1]$ define the \textit{clean} false–match rate\footnote{For impostor comparisons (negative class), the false match rate used in biometrics coincides with the false positive rate used in ROC analysis:
\[
\mathsf{FMR} (\tau) = \mathsf{Pr} \bigl[ \, S_{ij} > \tau \, \mid \, i \neq j \, \bigr] = 1 - F_i (\tau) = \mathsf{FPR} (\tau).
\]
We use FMR when discussing verification operating points and FPR when parameterizing ROC curves.
}
\begin{equation}
\mathsf{FMR}_{\mathrm{clean}}(\tau) \coloneqq 
\mathsf{Pr} \bigl[ \, S_{ij} > \tau \, \mid \, i \neq j \, \bigr].   
\end{equation}
An operational FR system fixes a target $\mathsf{FMR}$ level $\alpha \in (0,1)$ (e.g., $\alpha=10^{-2}$ or $10^{-3}$) and chooses
\begin{equation}
\tau_{\alpha} \coloneqq 
\inf \bigl\{ \, \tau: \mathsf{FMR}_{\mathrm{clean}}(\tau) \le \alpha \, \bigr\}.
\end{equation}
When the impostor CDF is continuous at $\tau_\alpha$ one has $\mathsf{FMR}_{\mathrm{clean}}(\tau_\alpha)=\alpha$. Otherwise $\mathsf{FMR}_{\mathrm{clean}}(\tau_\alpha)\le\alpha$ by construction.
\end{definition}

\begin{remark}[Endpoint behavior of $\mathsf{FMR}_{\mathrm{clean}}$]
\label{rem:endpoint-clean}
For $S\in[-1,1]$, $\mathsf{FMR}_{\mathrm{clean}}(\tau)=\mathsf{Pr}[S>\tau]$ is right–continuous and non–increasing on $[-1,1]$, with
\[
\mathsf{FMR}_{\mathrm{clean}}(1)=0,\qquad
\lim_{\tau\uparrow 1}\mathsf{FMR}_{\mathrm{clean}}(\tau)=\mathsf{Pr}[S=1].
\]
Thus there is no jump at $\tau=1$ iff $\mathsf{Pr}[S=1]=0$ (e.g., when the impostor score law is continuous at $1$).
\end{remark}

\paragraph{Exact post-privacy FMR at interior thresholds.}
For $\tau\in(-1,1)$, projection monotonicity (Eq.~\ref{eq:proj-monotone-interior}) implies
\begin{equation}
\label{eq:fmr-convolution}
\mathsf{FMR}_{\mathrm{priv}}(\tau)
\;\coloneqq \; \mathsf{Pr} \left[\widehat{S}_{ij}>\tau \,|\,i\neq j \right]
\;=\; \mathbb{E}\!\left[\Phi\!\left(\frac{S_{ij}-\tau}{\sigma}\right) \,|\,i\neq j \right]
\;=\; \mathbb{E}\!\left[\Q{\frac{\tau-S_{ij}}{\sigma}} \,|\,i\neq j\right],
\end{equation}
i.e., the clean survivor function smoothed by a Gaussian kernel. Evaluating at $\tau= \tau_\alpha \in (-1,1)$ gives the exact post-privacy FMR at the clean operating point.

\begin{remark}[Endpoint behavior of $\mathsf{FMR}_{\mathrm{priv}}$ at $\tau=1$]
\label{rem:endpoint-fmr}
With the strict rule ``$>$'', $\mathsf{FMR}_{\mathrm{priv}}(\tau)=\mathbb{E}[\mathbf{1}\{S+W>\tau\}]$ is continuous and strictly decreasing on $(-1,1)$, and
\begin{equation}
\lim_{\tau\uparrow 1}\mathsf{FMR}_{\mathrm{priv}}(\tau)=\mathbb{E}\!\left[\mathsf{Q}\!\left(\frac{1-S}{\sigma}\right)\right] \eqqcolon L(\sigma) \;>\;0.
\end{equation}
At the endpoint the strict rule forces $\mathsf{FMR}_{\mathrm{priv}}(1)=0$ because $1\not>1$, so there is a jump discontinuity of size $L(\sigma)$ at $\tau=1$. Consequently, a target level $\alpha\in(0,1)$ can be attained by thresholding below $1$ iff $\alpha\ge L(\sigma)$. The targets $\alpha\in(0,L(\sigma))$ are unattainable in strict semantics (the endpoint itself attains only $\alpha=0$).

In contrast, with the non–strict rule ``$\ge$'' we have $\mathsf{FMR}_{\mathrm{priv}}(1)=L(\sigma)$ and the map is continuous at $\tau=1$. The minimum achievable acceptance over $\tau\le 1$ is exactly $L(\sigma)>0$, which induces a genuine feasibility frontier.
\end{remark}

\begin{corollary}[Feasible threshold shifts and uniqueness]
\label{cor:feasible-shift}
Fix $\sigma>0$ and an interior operating point $\tau_0\in(-1,1)$ with target level $\alpha\in(0,1)$.

\textbf{Strict endpoint (``$>$'').} The map $\tau\mapsto \mathsf{FMR}_{\mathrm{priv}}(\tau)$ is continuous and strictly decreasing on $(-1,1)$ with
\[
\lim_{\tau\uparrow 1}\mathsf{FMR}_{\mathrm{priv}}(\tau)=L(\sigma)>0
\quad\text{and}\quad
\mathsf{FMR}_{\mathrm{priv}}(1)=0.
\]
Hence the equation $\mathsf{FMR}_{\mathrm{priv}}(\tau_0+\Delta\tau)=\alpha$ has a \textit{unique} interior solution $\Delta\tau\in[0,1-\tau_0)$ iff
\[
\alpha\in\bigl[L(\sigma),\,\mathsf{FMR}_{\mathrm{priv}}(\tau_0)\bigr].
\]
The boundary value $\tau_0+\Delta\tau=1$ attains $\alpha=0$. Targets $\alpha\in(0,L(\sigma))$ are unattainable by any $\tau\le 1$.

\textbf{Non–strict endpoint (``$\ge$'').} The map is continuous on $[-1,1]$ with minimum $\mathsf{FMR}_{\mathrm{priv}}(1)=L(\sigma)>0$, so a (unique) solution exists iff $\alpha\in\bigl[L(\sigma),\,\mathsf{FMR}_{\mathrm{priv}}(\tau_0)\bigr]$.
\end{corollary}

\vspace{7pt}

\begin{proposition}[Threshold-crossing identity and small–noise expansion at $\tau_\alpha$]
\label{prop:fmr-decomp}
Assume the additive mechanism noise is $W=\sigma Z$ with $Z\sim\mathcal{N}(0,1)$ independent of the clean impostor score $S_{ij}$.
Let $\tau_\alpha\in(-1,1)$ satisfy $\mathsf{FMR}_{\mathrm{clean}}(\tau_\alpha)=\Pr[S_{ij}>\tau_\alpha\mid i\neq j]=\alpha$.
For $\tau\in(-1,1)$ define
\begin{equation}
\mathsf{FMR}_{\mathrm{priv}}(\tau)\coloneqq \Pr[S_{ij}+W>\tau \mid i\neq j]
=\mathbb{E}\!\left[\mathsf{Q}\!\left(\frac{\tau-S_{ij}}{\sigma}\right)\middle|\, i\neq j\right].
\end{equation}
Then, at $\tau=\tau_\alpha$,
\begin{align}
\mathsf{FMR}_{\mathrm{priv}}(\tau_\alpha)
&= \alpha
- \mathbb{E}\!\left[\mathbf{1}\{S_{ij}>\tau_\alpha\}\; \mathsf{Q}\!\left(\frac{S_{ij}-\tau_\alpha}{\sigma}\right)\middle|\, i\neq j\right]
+ \mathbb{E}\!\left[\mathbf{1}\{S_{ij}\le\tau_\alpha\}\; \mathsf{Q}\!\left(\frac{\tau_\alpha-S_{ij}}{\sigma}\right)\middle|\, i\neq j\right].
\label{eq:flip-plus-minus-correct}
\end{align}
Moreover, suppose the conditional (impostor) law of $S_{ij}\mid(i\neq j)$ admits a density $f_{\mathrm{i}}$ in a neighborhood of $\tau_\alpha$,
and $f_{\mathrm{i}}$ is continuously differentiable at $\tau_\alpha$.
Then, as $\sigma\to 0$,
\begin{equation}
\label{eq:fmr-small-sigma-correct}
\mathsf{FMR}_{\mathrm{priv}}(\tau_\alpha)
=\alpha-\frac{\sigma^2}{2}\,f_{\mathrm{i}}'(\tau_\alpha)+o(\sigma^2).
\end{equation}
Equivalently, since $F_{\mathrm{i}}'=f_{\mathrm{i}}$, if $F_{\mathrm{i}}''(\tau_\alpha)$ exists then
\begin{equation}
\mathsf{FMR}_{\mathrm{priv}}(\tau_\alpha)
=\alpha-\frac{\sigma^2}{2}\,F_{\mathrm{i}}''(\tau_\alpha)+o(\sigma^2).
\end{equation}
\end{proposition}

\begin{proof}
\textit{Flip decomposition.}
Condition on $S_{ij}=s$. For $\tau\in(-1,1)$,
\begin{equation}
\Pr[S_{ij}+W>\tau\mid S_{ij}=s]=\Pr[W>\tau-s]=\mathsf{Q}\!\left(\frac{\tau-s}{\sigma}\right).
\end{equation}
If $s>\tau$, then $\mathsf{Q}((\tau-s)/\sigma)=1-\mathsf{Q}((s-\tau)/\sigma)$, so
\begin{equation}
\mathbf{1}\{s>\tau\}\,\mathsf{Q}\!\left(\frac{\tau-s}{\sigma}\right)
=\mathbf{1}\{s>\tau\}-\mathbf{1}\{s>\tau\}\,\mathsf{Q}\!\left(\frac{s-\tau}{\sigma}\right).
\end{equation}
If $s\le \tau$, then $\mathsf{Q}((\tau-s)/\sigma)=\mathsf{Q}((\tau-s)/\sigma)$ as written.
Taking expectations (conditional on $i\neq j$) and using $\Pr[S_{ij}>\tau_\alpha\mid i\neq j]=\alpha$
gives Eq.~\ref{eq:flip-plus-minus-correct}.

\textit{Rewrite the smoothing bias.}
Fix $\tau\in(-1,1)$ and write (still under $i\neq j$)
\begin{equation}
\mathsf{FMR}_{\mathrm{priv}}(\tau)
=\int \mathsf{Q}\!\left(\frac{\tau-s}{\sigma}\right)\,f_{\mathrm{i}}(s)\, \mathrm{d}s.
\end{equation}
Introduce the odd correction kernel
\begin{equation}
q(u)\coloneqq \mathsf{Q}(u)-\mathbf{1}\{u<0\}.
\end{equation}
Then $\mathsf{Q}(u)=\mathbf{1}\{u<0\}+q(u)$, hence
\begin{equation}
\mathsf{FMR}_{\mathrm{priv}}(\tau)
=\underbrace{\int \mathbf{1}\{s>\tau\}f_{\mathrm{i}}(s)\,\mathrm{d}s}_{=\ \mathsf{FMR}_{\mathrm{clean}}(\tau)}
+\int q\!\left(\frac{\tau-s}{\sigma}\right)f_{\mathrm{i}}(s)\, \mathrm{d}s.
\end{equation}
At $\tau=\tau_\alpha$ the first term equals $\alpha$. It remains to expand the second term.

\textit{Change variables and use a first-order Taylor remainder.}
Let $u=(\tau-s)/\sigma$ so $s=\tau-\sigma u$ and $ds=-\sigma du$. Then
\begin{equation}
\int q\!\left(\frac{\tau-s}{\sigma}\right)f_{\mathrm{i}}(s)\,\mathrm{d}s
=\sigma\int_{\mathbb{R}} q(u)\,f_{\mathrm{i}}(\tau-\sigma u)\,\mathrm{d}u.
\end{equation}
Because $f_{\mathrm{i}}$ is differentiable at $\tau$, write
\begin{equation}
f_{\mathrm{i}}(\tau-\sigma u)=f_{\mathrm{i}}(\tau)-\sigma u f_{\mathrm{i}}'(\tau)+\sigma u\,\varepsilon_\sigma(u),
\end{equation}
where $\varepsilon_\sigma(u)\to 0$ pointwise as $\sigma\to 0$ (by the definition of derivative).
Plugging in:
\begin{equation}
\sigma\int q(u)f_{\mathrm{i}}(\tau-\sigma u)\,\mathrm{d}u
=\sigma f_{\mathrm{i}}(\tau)\int q(u)\,\mathrm{d}u
-\sigma^2 f_{\mathrm{i}}'(\tau)\int u q(u)\,\mathrm{d}u
+\sigma^2\int u q(u)\,\varepsilon_\sigma(u)\,\mathrm{d}u.
\end{equation}

\textit{Evaluate the kernel moments and bound the remainder.}
First, $q$ is odd because for $u>0$, $q(u)=\mathsf{Q}(u)$ while for $u<0$,
$q(u)=\mathsf{Q}(u)-1=-(1-\mathsf{Q}(u))=-\mathsf{Q}(-u)$, hence $q(-u)=-q(u)$.
Therefore $\int q(u)\,du=0$.

Next, $u q(u)$ is even and integrable, and
\begin{equation}
\int_{\mathbb{R}} u q(u)\,\mathrm{d}u
=2\int_0^\infty u\,\mathsf{Q}(u)\,\mathrm{d}u.
\end{equation}
Using the identity for $Z\sim\mathcal{N}(0,1)$,
\begin{equation}
\mathbb{E}[(Z_+)^2]=\int_0^\infty \Pr(Z_+^2>t)\, \mathrm{d}t
=\int_0^\infty \Pr(Z>\sqrt t)\, \mathrm{d}t
=\int_0^\infty 2u\,\mathsf{Q}(u)\, \mathrm{d}u,
\end{equation}
and $\mathbb{E}[(Z_+)^2]=\tfrac12\mathbb{E}[Z^2]=\tfrac12$, we obtain
$\int_0^\infty u\,\mathsf{Q}(u)\, \mathrm{d}u=\tfrac14$, hence
\[
\int_{\mathbb{R}} u q(u)\, \mathrm{d}u= \frac12.
\]

Finally, since $\int |u q(u)|\, \mathrm{d}u<\infty$ and $\varepsilon_\sigma(u)\to 0$ pointwise with $|\varepsilon_\sigma(u)|$ bounded for small $\sigma$ (from local boundedness of $f_{\mathrm{i}}'$ near $\tau$),
dominated convergence gives
\begin{equation}
\int u q(u)\,\varepsilon_\sigma(u)\, \mathrm{d}u= o(1)\qquad(\sigma\to 0).
\end{equation}
Combining the above,
\begin{equation}
\int q\!\left(\frac{\tau-s}{\sigma}\right)f_{\mathrm{i}}(s)\, \mathrm{d}s
= -\sigma^2 f_{\mathrm{i}}'(\tau)\cdot \frac12 + o(\sigma^2).
\end{equation}
Setting $\tau=\tau_\alpha$ and adding the clean term $\alpha$ yields Eq.~\ref{eq:fmr-small-sigma-correct}.
The equivalence with $F_{\mathrm{i}}''(\tau_\alpha)$ follows from $F_{\mathrm{i}}'=f_{\mathrm{i}}$.
\end{proof}

\vspace{7pt}

\begin{corollary}[One-sided flip bounds (interior thresholds)]
\label{cor:fmr-bound}
For $\tau_\alpha\in(-1,1)$,
\begin{equation}
\alpha\;-\;\mathbb{E}\!\left[\mathbf{1}\{S_{ij}>\tau_\alpha\}\; \Q{\frac{S_{ij}-\tau_\alpha}{\sigma}}\right]
\;\le\; \mathsf{FMR}_{\mathrm{priv}}(\tau_\alpha)
\;\le\; \alpha\;+\;\mathbb{E}\!\left[\mathbf{1}\{S_{ij}\le\tau_\alpha\}\; \Q{\frac{\tau_\alpha-S_{ij}}{\sigma}}\right].
\end{equation}
Applying Mills' inequality $\Q{z}\le e^{-z^{2}/2}/(z\sqrt{2\pi})$ yields closed-form, data-dependent upper/lower bounds.
\end{corollary}

\begin{definition}[Feasibility frontier under non-strict endpoint]
\label{def:feasibility}
For fixed $(\alpha,\delta)$ let $\sigma_{\varepsilon,\delta}$ be the Gaussian DP scale. Define
\begin{equation}
\varepsilon_{\min}(\alpha,\delta)
\;=\;\inf\left\{\varepsilon>0:\;
\mathbb{E}\!\left[\Q{\frac{1-S_{ij}}{\sigma_{\varepsilon,\delta}}}\right]\le\alpha\right\}.
\end{equation}
Because the integrand is strictly decreasing in $\varepsilon$, $\varepsilon_{\min}$ is well-defined and can be found by 1-D root finding.
\end{definition}

\paragraph{Threshold re-calibration.}
Let $F_{\mathrm{i}}$ be the CDF of the \textit{clean} impostor scores and $f_{\mathrm{i}}= F'_{\mathrm{i}}$ its density. Let $\tau_{0}$ be the operating threshold that achieves the desired pre-privacy target FMR $1-F_{\mathrm{i}}(\tau_{0}) = \mathsf{FMR}_{0}$. By adding isotropic Gaussian noise with standard deviation $\sigma$, the \textit{ScoreShield} mechanism smooths the (impostor) score distribution by convolution, $\widetilde F_{\mathrm{i}} \;=\; F_{\mathrm{i}} * g_{\sigma}$, where $g_{\sigma}(x)= \sigma^{-1}\, \varphi \bigl(x/\sigma\bigr)$.
To preserve the target FMR after smoothing, raise the threshold by $\Delta\tau$ so that
\begin{equation}\label{eq:FMR-equality}
   \widetilde F_{\mathrm{i}}(\tau_{0}+\Delta\tau)
   \;=\;
   F_{\mathrm{i}}(\tau_{0}).
\end{equation}

\vspace{7pt}

\textit{Small-noise regime ($\sigma \ll 1$).}  
Assume $F_{\mathrm{i}}\in C^{2}$ in a neighborhood of $\tau_{0}$ (i.e., $F_{\mathrm{i}}$ is twice continuously differentiable at $\tau_{0}$) with density $f_{\mathrm{i}}(\tau_{0}) > 0$. 
Gaussian smoothing admits the local expansion
\begin{equation}
( F_{\mathrm{i}}*g_{\sigma} )(t) \;=\; F_{\mathrm{i}}(t)
\;+\;\frac{\sigma^{2}}{2}\,F_{\mathrm{i}}''(t) \;+\;r_{\sigma}(t), \qquad
r_{\sigma}(t)=o(\sigma^{2}) \;\; \text{as }\sigma \to 0,    
\end{equation}
uniformly for $t$ near $\tau_{0}$. Evaluate at $t = \tau_0 + \Delta \tau$ and Taylor–expand in $\Delta \tau$:
\begin{subequations}
\begin{eqnarray}
F_{\mathrm{i}} (\tau_0 + \Delta \tau) &=& F_{\mathrm{i}} (\tau_0) + f_{\mathrm{i}} (\tau_0) \Delta \tau + \frac{1}{2} F_{\mathrm{i}}^{''} (\tau_0) \Delta \tau^2 + o (\Delta \tau^2),\\
F_{\mathrm{i}}^{''} (\tau_0 + \Delta \tau) &=& F_{\mathrm{i}}^{''} (\tau_0) + o (1).
\end{eqnarray}
\end{subequations}
Plug in to Eq.~\ref{eq:FMR-equality} gives
\begin{eqnarray}
F_{\mathrm{i}}(\tau_{0}) + f_{\mathrm{i}} (\tau_{0})\, \Delta \tau + \frac{1}{2} F_{\mathrm{i}}''(\tau_{0})\,\Delta \tau^{2} + \frac{\sigma^{2}}{2} F_{\mathrm{i}}''(\tau_{0}) + o(\sigma^2) + o (\Delta \tau^2) \;=\; F_{\mathrm{i}}( \tau_{0} ).
\end{eqnarray}
Solving for $\Delta \tau$ yields
\begin{equation}
\Delta\tau\;=\;-\frac{\sigma^{2}}{2}\,\frac{F_{\mathrm{i}}''(\tau_{0})}{f_{\mathrm{i}}(\tau_{0})}
\; \;+\; o(\sigma^{2})=\;-\frac{\sigma^{2}}{2}\,\frac{f'_{\mathrm{i}}(\tau_{0})}{f_{\mathrm{i}}(\tau_{0})}
\;+\; o(\sigma^{2}),
\label{eq:delta-tau}
\end{equation}
and hence $\Delta\tau=\Theta(\sigma^{2})$.
If, additionally, $F_{\mathrm{i}}\in C^{4}$ near $\tau_{0}$, the remainder sharpens to $\mathcal{O}(\sigma^{4})$.

\vspace{7pt}

\begin{remark}
Define the log-pdf slope $\ell(t)=\frac{\mathrm{d}}{\mathrm{d}t}\log f_{\mathrm{i}}(t)=f_{\mathrm{i}}'(t)/f_{\mathrm{i}}(t)$. For right-tail operating thresholds of a unimodal density we have $\ell(\tau_\alpha) < 0$, so $\Delta\tau > 0$, matching the intuition that the operating point moves to the right.\footnote{The threshold must move \textit{to the right}, i.e, become stricter, as noise spreads the scores.}
\end{remark}

\vspace{7pt}

\begin{remark}[Gaussian impostor example (exact calibration)]
Assume the clean impostor score is standard normal, $S_{ij} \sim \mathcal{N}(0,1)$, and the added noise is $W \sim \mathcal{N} (0,\sigma^{2})$. For any interior threshold $\tau<1$ under the strict rule ``$>$'', clipping does not affect the decision. The post-privacy impostor tail at threshold $\tau$ is $\mathsf{Pr}[ S+ W > \tau]= \mathsf{Q} \big(\tau/\sqrt{1+\sigma^{2}}\big)$. If the pre-privacy operating point satisfies $\alpha= \mathsf{Q}(\tau_{0})$, preserving this tail after noise requires
\begin{equation}
\mathsf{Q} \Big(\frac{\tau_{0}+\Delta\tau}{\sqrt{1+\sigma^{2}}}\Big)=\mathsf{Q}(\tau_{0}).
\end{equation}
Since $\mathsf{Q}$ is strictly decreasing, it follows that
\begin{eqnarray}
\Delta\tau=\tau_{0}\big(\sqrt{1+\sigma^{2}}-1\big)\;.
\end{eqnarray}
Expanding $\sqrt{1+\sigma^{2}}=1+\tfrac{\sigma^{2}}{2}-\tfrac{\sigma^{4}}{8} + \mathcal{O} (\sigma^{6})$ yields
\begin{equation}
\Delta\tau=\tfrac{\sigma^{2}}{2}\,\tau_{0} - \tfrac{\sigma^{4}}{8}\,\tau_{0}+ \mathcal{O} (\sigma^{6})
= \tfrac{\sigma^{2}}{2}\,\tau_{0} + \mathcal{O}(\sigma^{4}).
\end{equation}
This matches the small-noise regime $\Delta\tau=-(\sigma^{2}/2)\,f_{\mathrm{i}}'(\tau_{0})/f_{\mathrm{i}}(\tau_{0}) + o(\sigma^{2})$ from Eq.~\ref{eq:delta-tau}, because for the standard normal $f_{\mathrm{i}}' / f_{\mathrm{i}} = - \tau_{0}$.
\end{remark}

\vspace{7pt}

\begin{remark}[Extreme–FMR scaling and tail shape]
\label{rem:extreme-fmr}
Recall the small-noise calibration formula
$\Delta\tau= -\frac{\sigma^{2}}{2}\,\frac{f_{\mathrm{i}}'(\tau_{0})}{f_{\mathrm{i}}(\tau_{0})} + o(\sigma^{2})$ from Eq.~\ref{eq:delta-tau}. For a \textit{fixed} target $\alpha$ (hence fixed $\tau_{0}\in(-1,1)$), the coefficient $|f'_{\mathrm{i}}(\tau_{0})/f_{\mathrm{i}}(\tau_{0})|$ is a finite constant and $\Delta\tau=\Theta(\sigma^{2})$. When $\alpha\downarrow 0$ (extreme right tail), the coefficient's growth is determined entirely by the impostor tail:
\begin{itemize}[leftmargin=*, itemsep=2pt]
\item 
\textit{Quadratic log–tail (Gaussian/sub-Gaussian) regime.} 
If $\log f_{\mathrm{i}}(t)=-\psi(t)+ \mathcal{O}(1)$ with $\psi'(t)=\Theta(t)$ as $t \to +\infty$
(e.g., the standard normal, where $\psi(t)=t^{2}/2$ and $f'_{\mathrm{i}}/f_{\mathrm{i}}= -t$),
then $\bigl|f'_{\mathrm{i}}(\tau_{0})/f_{\mathrm{i}}(\tau_{0})\bigr|= \Theta(|\tau_{0}|)$ and
\begin{equation}
\Delta\tau=\Theta \bigl(\sigma^{2}\,|\tau_{0}|\bigr).
\end{equation}
For the standard normal, $\tau_{0}=\Phi^{-1}(1-\alpha)\sim \sqrt{2\log(1/\alpha)}$, hence
\begin{equation}
\Delta\tau=\Theta \bigl(\sigma^{2}\sqrt{\log(1/\alpha)}\bigr).
\end{equation}
\item 
\textit{Bounded support with algebraic vanishing near the endpoint.}
If scores live in $[-1,1]$ and $f_{\mathrm{i}}(t)\asymp C(1-t)^{\beta}$ as $t\uparrow 1$ for some $\beta > 0$, then $f'_{\mathrm{i}}(t)/f_{\mathrm{i}}(t)\sim -\beta/(1-t)$ and $1-F_{\mathrm{i}}(\tau_{0}) = \alpha \asymp C'(1-\tau_{0})^{\beta+1}$. Consequently,
\begin{equation}
\Delta\tau\sim \frac{\sigma^{2}}{2}\,\frac{\beta}{1-\tau_{0}}
=\Theta \bigl(\sigma^{2}\,\alpha^{-1/(\beta+1)}\bigr).
\end{equation}
\end{itemize}
In all cases the dependence on the privacy noise scale remains quadratic in $\sigma$.  Only the multiplicative constant changes with how extreme the operating point is, via the local tail shape encoded by $f'_{\mathrm{i}}(\tau_{0})/f_{\mathrm{i}}(\tau_{0})$.
\end{remark}

\begin{proposition}[Exact re-calibration: existence and uniqueness]
\label{prop:delta-tau-equation}
Let $\tau_{\alpha}\in(-1,1)$ satisfy $\mathsf{FMR}_{\mathrm{clean}}(\tau_{\alpha})=\alpha$ as in Definition~\ref{def:clean-fmr}. For $\eta\in[0,\,1-\tau_{\alpha}]$ define
\begin{equation}
G(\eta) \;\coloneqq \; \mathsf{FMR}_{\mathrm{priv}}(\tau_{\alpha} + \eta),
\end{equation}
where, for $\tau\in(-1,1)$,
\begin{equation}
\label{eq:fmr-convolution-recall}
\mathsf{FMR}_{\mathrm{priv}}(\tau)
=\mathbb{E}\!\left[\mathsf{Q}\!\left(\frac{\tau-S_{ij}}{\sigma}\right)\middle|\,i\neq j\right].
\end{equation}
Then for any $\sigma>0$:
\begin{enumerate}[leftmargin=2.1em,label=(\alph*)]
\item \textit{Strict endpoint (``$>$'').} $G$ is continuous and strictly decreasing on $[0,1-\tau_{\alpha})$ with $\lim_{\eta\uparrow 1-\tau_{\alpha}}G(\eta)=L(\sigma)>0$ and $G(1-\tau_{\alpha})=0$ (at the endpoint). There exists a unique interior solution $\eta^\star\in[0,1-\tau_{\alpha})$ to $G(\eta)=\alpha$ iff $\alpha\in[L(\sigma),G(0)]$. 
In other words,
\begin{equation} 
\exists \;\Delta\tau\in[0,\,1-\tau_{\alpha}] \;\text{~with~}\; \mathsf{FMR}_{\mathrm{priv}}(\tau_{\alpha}+\Delta\tau)= \alpha 
\quad\Longleftrightarrow\quad G(0)\;\ge\;\alpha.
\end{equation}
In particular, if $G(0)>\alpha$ then $\Delta\tau\in(0,\,1-\tau_{\alpha})$. 
If $\alpha=0$, the unique solution is $\eta^\star=1-\tau_{\alpha}$ (i.e., $\tau=1$). If $\alpha<L(\sigma)$ there is no solution with $\tau\le 1$ in strict semantics.
\item \textit{Non–strict endpoint (``$\ge$'').} $G$ is continuous and strictly decreasing on $[0,1-\tau_{\alpha}]$ with $G(1-\tau_{\alpha})=L(\sigma)>0$. There exists a unique solution $\eta^\star\in[0,1-\tau_{\alpha}]$ iff $\alpha\in[L(\sigma),G(0)]$, where $\alpha = L(\sigma)$ is attained at $\tau=1$.
\end{enumerate}
\end{proposition}

\begin{proof}
For $\tau\in(-1,1)$, projection monotonicity and independence of $W\sim\mathcal{N}(0,\sigma^2)$ implies $\mathsf{Pr}[\widehat{S}_{ij}>\tau\,|\,i\neq j]= \mathsf{Pr}[S_{ij} + W>\tau\,|\,i\neq j] = \mathbb{E}[\,\mathsf{Q}((\tau-S_{ij})/\sigma)\mid i\neq j\,]$. Fix $\sigma>0$. Since $\frac{\partial}{\partial\tau}\mathsf{Q}((\tau-s)/\sigma) = -\sigma^{-1}\varphi((\tau-s)/\sigma)$ and $S_{ij}\in[-1,1]$ a.s., dominated convergence yields
\begin{equation}
\frac{\mathrm{d}}{\mathrm{d}\tau}\,\mathsf{FMR}_{\mathrm{priv}}(\tau)
= -\frac{1}{\sigma}\,\mathbb{E}\!\left[\varphi\!\left(\frac{\tau-S_{ij}}{\sigma}\right)\middle|\,i\neq j\right] < 0, \qquad \tau\in(-1,1),
\end{equation}
so $\mathsf{FMR}_{\mathrm{priv}}$ is continuous and strictly decreasing on $(-1,1)$ and continuous up to $\tau=1$. With the strict ``$>$'' rule and $S_{ij} \leq 1$, $\mathsf{FMR}_{\mathrm{priv}}(1)=0$, hence $G$ is continuous up to $\tau = 1$ and strictly decreasing on $[0,1-\tau_{\alpha})$ with $G(1-\tau_{\alpha})=0$. By the intermediate value theorem, the equation $G(\eta)=\alpha$ has a unique solution in $[0,1-\tau_{\alpha}]$ iff $G(0)\ge\alpha$. Uniqueness follows from strict decrease.
\end{proof}

\begin{definition}[Recalibration offset]
The recalibration offset $\Delta\tau$ is the unique $\eta\in[0,1-\tau_\alpha]$ satisfying
\begin{equation}
\label{eq:delta-tau-root}
\mathsf{FMR}_{\mathrm{priv}}(\tau_{\alpha}+\eta)=\alpha .
\end{equation}
By Proposition~\ref{prop:delta-tau-equation}, this is well-defined whenever $G(0)\ge\alpha$.
\end{definition}

\vspace{5pt}

\begin{remark}[Small-noise Law]
\label{rem:delta-tau-small-noise}
Under the regularity assumptions stated in the re-calibration section (twice differentiable impostor CDF with $f_{\mathrm{i}}(\tau_\alpha)>0$), the exact shift obeys the previously derived expansion
\begin{equation}
\Delta\tau \;=\; -\frac{\sigma^{2}}{2}\,\frac{f'_{\mathrm{i}}(\tau_{\alpha})}{f_{\mathrm{i}}(\tau_{\alpha})}
\;+\;o(\sigma^{2}) ,
\end{equation}
see Eq.~\ref{eq:delta-tau}. Thus exact re-calibration cancels the $\Theta(\sigma^2)$ inflation at $\tau_\alpha$ with a shift only $\Theta(\sigma^{2})$.
\end{remark}

\begin{remark}[Small-noise Regime (When is $G(0)>\alpha$?)]
\label{rem:G0gtalpha}
If $f_{\mathrm{i}}(\tau_\alpha)>0$ and $F_{\mathrm{i}}$ is $C^2$ near the clean operating point $\tau_{\alpha}$, the small-noise expansion
\begin{equation}
\mathsf{FMR}_{\mathrm{priv}}(\tau_{\alpha})
=\alpha+\frac{\sigma\,f_{\mathrm{i}}(\tau_{\alpha})}{\sqrt{2\pi}}+ o(\sigma),
\qquad(\sigma\to 0)   ,
\end{equation}
implies $G(0)>\alpha$ for all sufficiently small $\sigma>0$. In that case the unique solution satisfies $\Delta\tau\in(0,\,1-\tau_{\alpha})$.
\end{remark}

\begin{remark}[Conservative sufficient offset]
\label{rem:sufficient-upper}
Define, for $\eta\ge0$,
\begin{equation}
\mathcal{I}_{\mathrm{up}}(\eta,\sigma) \;\coloneqq \;
\mathbb{E}\!\left[\mathbf{1}\{S_{ij}\le\tau_{\alpha}+\eta\}\;
\mathsf{Q}\!\left(\frac{\tau_{\alpha}+\eta-S_{ij}}{\sigma}\right)\middle|\,i\neq j\right].
\end{equation}
Applying the upper bound from Corollary~\ref{cor:fmr-bound} at $\tau=\tau_\alpha+\eta$ gives
\begin{equation}
\mathsf{FMR}_{\mathrm{priv}}(\tau_\alpha+\eta) \;\le\;
\mathsf{FMR}_{\mathrm{clean}}(\tau_\alpha+\eta)\;+\;\mathcal{I}_{\mathrm{up}}(\eta,\sigma).
\end{equation}
Hence any $\eta\ge0$ satisfying $\mathsf{FMR}_{\mathrm{clean}}(\tau_\alpha+\eta)+\mathcal{I}_{\mathrm{up}}(\eta,\sigma)\le\alpha$ is a \textit{sufficient} (conservative) recalibration offset. 
%
\end{remark}

\vspace{5pt}

\begin{remark}[Operational Interpretation]
\label{rem:operational}
Keeping $\tau=\tau_{\alpha}$ induces additive inflation of order $\Theta(\sigma^2)$ (Proposition~\ref{prop:fmr-decomp}). Exact re-calibration finds the unique $\Delta\tau$ that restores the target and, by Remark~\ref{rem:delta-tau-small-noise}, requires a shift only $\Theta(\sigma^{2})$. When $\sigma\,f_{\mathrm{i}}(\tau_\alpha)/\sqrt{2\pi}$ is already below tolerance, re-calibration can be skipped, otherwise compute $\Delta\tau$ via the root condition Eq.~\ref{eq:delta-tau-root} (or use the small-noise approximation Eq.~\ref{eq:delta-tau} when appropriate. See Remark~\ref{rem:delta-tau-small-noise}).
\end{remark}

\appsubsection{Effect on the ROC Curve and AUC}
\label{ssec:auc-perturbation}

Let $F_{\mathrm{g}}$ and $F_{\mathrm{i}}$ denote the CDFs of the genuine and impostor scores prior to perturbation. We view them as CDFs on $\mathbb{R}$ by extending constantly outside $[-1,1]$ (i.e., $F(t)=0$ for $t<-1$ and $F(t)=1$ for $t>1$). Adding zero-mean Gaussian noise with variance $\sigma^{2}$ smooths both distributions by convolution\footnote{Because $\mathbf{1}\{\mathsf{proj}(S{+}W)>\tau\}=\mathbf{1} \{S{+}W > \tau\}$ for $\tau\in(-1,1)$, both genuine and impostor CDFs are perturbed by Gaussian convolution.}:
\[
\widetilde F_{\mathrm{g}}=F_{\mathrm{g}} *g_{\sigma},
\qquad
\widetilde F_{\mathrm{i}}=F_{\mathrm{i}} *g_{\sigma},
\qquad
g_{\sigma}(x)=\tfrac{1}{\sigma}\,\varphi(x/\sigma),
\quad \varphi(x)=\tfrac{e^{-x^{2}/2}}{\sqrt{2\pi}}.
\]

For a decision threshold $\tau \in[-1,1]$ we write
\begin{equation}
\label{eq:TPR_FPR_def}
\mathsf{TPR}_{F_{\mathrm{g}}}( \tau) = 1-F_{\mathrm{g}}( \tau),\qquad
\mathsf{FPR}_{F_{\mathrm{i}}}( \tau)=1-F_{\mathrm{i}}( \tau),
\end{equation}
and analogously $\mathsf{TPR}_{\widetilde{F}_{\mathrm{g}}}$, $\mathsf{FPR}_{\widetilde{F}_{\mathrm{i}}}$.

\begin{theorem}[Uniform perturbation bound on the ROC]
\label{thm:roc-bound}
Assume that for $\ell\in\{\mathrm{g},\mathrm{i}\}$ the CDF $F_{\ell}$ is $L$-Lipschitz on $\mathbb{R}$ (equivalently, its density $f_\ell$ exists a.e. and satisfies $0 \le f_\ell\le L $), and let $\widetilde{F}_{\ell} = F_{\ell} * g_{\sigma}$ be its Gaussian smoothing. Then for every threshold $\tau \in [-1,1]$,
\begin{subequations}
\begin{align}
\bigl|\mathsf{TPR}_{\widetilde F_{\mathrm{g}}}(\tau) -\mathsf{TPR}_{F_{\mathrm{g}}}(\tau)\bigr|
&\;\le\;L\sigma\sqrt{\tfrac{2}{\pi}},\\
\bigl|\mathsf{FPR}_{\widetilde F_{\mathrm{i}}}(\tau) -\mathsf{FPR}_{F_{\mathrm{i}}}(\tau)\bigr|
&\;\le\;L\sigma\sqrt{\tfrac{2}{\pi}}.
\end{align}  
\end{subequations}
Consequently, for the Hausdorff distance induced by the $\ell_\infty$ metric on $[0,1]^2$,
\begin{equation}
d_{\mathsf{H}} \bigl( \, \mathsf{ROC}( \, \widetilde F_{\mathrm{g}},\widetilde F_{\mathrm{i}}), \,
\mathsf{ROC}(F_{\mathrm{g}},F_{\mathrm{i}} \,) \, \bigr) \;\le\;L\sigma\sqrt{\tfrac{2}{\pi}}.
\end{equation}
\end{theorem}

\begin{proof}
For any $L$-Lipschitz CDF $F$ on $\mathbb{R}$,
\begin{equation}
\sup_{\tau\in\mathbb{R}}\bigl| F(\tau) - (F * g_{\sigma})(\tau) \bigr|
=\sup_{\tau} \bigl| \mathbb{E}\!\left[F(\tau-\sigma Z)-F(\tau)\right] \bigr|
\le L \,\mathbb{E} \left[ \, |\sigma Z| \,\right] = L \, \sigma\sqrt{\tfrac{2}{\pi}}, 
\end{equation}
$Z\sim \mathcal{N}(0,1)$. Since $\mathsf{TPR}= 1-F_{\mathrm{g}}$ and $\mathsf{FPR}= 1-F_{\mathrm{i}}$, the same bound holds for both coordinates. For the ROC Hausdorff bound, at each $\tau$ the two ROC points differ by at most this amount in each coordinate, so the $\ell_\infty$ distance between the two curves is at most the same bound.
\end{proof}

\vspace{5pt}

\begin{remark}[Lipschitz on \text{[-1, 1]} via constant extension]
If $F$ is $L$-Lipschitz on $[-1, 1]$ (equivalently, its density exists a.e. on $[-1, 1]$ with $0 \leq f \leq 1$) and we extend $F$ constantly outside $[-1, 1]$ by setting $F(t) = 0$ for $t < -1$ and $F(t) = 1$ for $t>1$, then the extension is globally $L$-Lipschitz on $\mathbb{R}$. Hence the proof of Theorem \ref{thm:roc-bound} (which applies convolution on $\mathbb{R}$) goes through unchanged.
\end{remark}

\begin{remark}[Hausdorff metric choice]
The theorem states the bound using the Hausdorff distance induced by the $\ell_{\infty}$ metric on ${[0,1]}^2$. If we instead measure Hausdorff distance with the Euclidean metric, we need to multiply the constant by $\sqrt{2}$. 
\end{remark}

\paragraph{AUC Functional.}
For CDFs $F_{\mathrm{g}},F_{\mathrm{i}}$ supported on $[-1,1]$,
\begin{equation}
\mathsf{AUC}(F_{\mathrm{g}},F_{\mathrm{i}}) \;=\;
\int_{-1}^{1}\bigl(1-F_{\mathrm{g}}(\tau)\bigr)\,\mathrm{d} F_{\mathrm{i}}(\tau)
\;=\;\int_{-1}^{1}\bigl(1-F_{\mathrm{g}}(\tau)\bigr)\,f_{\mathrm{i}}(\tau)\,\mathrm{d}\tau   ,
\end{equation}
whenever $F_{\mathrm{i}}$ is absolutely continuous with density $f_{\mathrm{i}}$. Equivalently, if $(S_{\mathrm{g}},S_{\mathrm{i}})$ has a continuous joint law (no ties), $\mathsf{AUC}= \mathsf{Pr}[S_{\mathrm{g}}>S_{\mathrm{i}}]$. In general $\mathsf{AUC}= \mathsf{Pr}[S_{\mathrm{g}}>S_{\mathrm{i}}] + \tfrac12\mathsf{Pr}[S_{\mathrm{g}}=S_{\mathrm{i}}]$ \citep{bamber1975area, muschelli2020roc}.

\vspace{5pt}


\begin{theorem}[AUC Perturbation Bound]
\label{thm:auc-perturb}
Let $F_{\mathrm{g}}, F_{\mathrm{i}}$ be CDFs supported on $[-1,1]$. Extend $F_{\mathrm{g}}$ to $\mathbb{R}$ by setting $F_{\mathrm{g}}(t)=0$ for $t<-1$ and $F_{\mathrm{g}}(t)=1$ for $t>1$, and assume that this extension is $L$-Lipschitz. Let $\widetilde{F}_{\mathrm{g}}= F_{\mathrm{g}} * g_{\sigma}$ and $\widetilde{F}_{\mathrm{i}} = F_{\mathrm{i}} * g_{\sigma}$ with $g_{\sigma}(x)= \sigma^{-1} \varphi(x/ \sigma)$, $\varphi(x) = e^{-x^{2}/2} / \sqrt{2\pi}$. Define $\eta \coloneqq  L\,\sigma\sqrt{2/\pi}$. Then
\begin{equation}
\bigl| \mathsf{AUC}(\widetilde F_{\mathrm{g}},\widetilde F_{\mathrm{i}})
-\mathsf{AUC}(F_{\mathrm{g}},F_{\mathrm{i}}) \bigr| \;\le\; 2\,\eta. 
\end{equation}
In particular,
\begin{equation}
\bigl| \mathsf{AUC}(\widetilde{F}_{\mathrm{g}}, \widetilde{F}_{\mathrm{i}})
- \mathsf{AUC}(F_{\mathrm{g}}, F_{\mathrm{i}}) \bigr|
= \mathcal{O}(\sigma),
\end{equation}
with constant $2L\sqrt{2/\pi}$.
\end{theorem}

\begin{proof}
Let $Y\sim F_{\mathrm{i}}$ and couple $\widetilde{Y}\sim \widetilde{F}_{\mathrm{i}}$ as
$\widetilde{Y}=Y+\sigma Z$, where $Z\sim\mathcal{N}(0,1)$ is independent of $Y$.
Using the Stieltjes form,
\begin{equation}
\mathsf{AUC}(F_{\mathrm{g}}, F_{\mathrm{i}})
= \mathbb{E}\bigl[1-F_{\mathrm{g}}(Y)\bigr],
\qquad \mathsf{AUC}(\widetilde{F}_{\mathrm{g}}, \widetilde{F}_{\mathrm{i}}) = \mathbb{E}\bigl[1-\widetilde{F}_{\mathrm{g}}(\widetilde{Y})\bigr].
\end{equation}
Set
\begin{equation}
\Delta
\coloneqq
\mathsf{AUC}(F_{\mathrm{g}},F_{\mathrm{i}})
- \mathsf{AUC}(\widetilde F_{\mathrm{g}},\widetilde F_{\mathrm{i}}).
\end{equation}
Adding and subtracting $\mathbb{E}[1-F_{\mathrm{g}}(\widetilde Y)]$ gives
\begin{equation}
\Delta_{\mathrm{AUC}} =
\underbrace{\mathbb{E}\!\left[\widetilde F_{\mathrm{g}}(\widetilde Y)-F_{\mathrm{g}}(\widetilde Y)\right]}_{\mathrm{(A)}}
+ \underbrace{\Bigl(\mathbb{E}[1-F_{\mathrm{g}}(Y)]-\mathbb{E}[1-F_{\mathrm{g}}(\widetilde Y)]\Bigr)}_{\mathrm{(B)}}.
\end{equation}
For term $\mathrm{(A)}$, the global $L$-Lipschitz property of $F_{\mathrm{g}}$ gives
\begin{equation}
\|F_{\mathrm{g}}-\widetilde F_{\mathrm{g}}\|_{\infty}
= \sup_t \left|F_{\mathrm{g}}(t)-\mathbb{E}F_{\mathrm{g}}(t-\sigma Z)\right|
\le L\sigma\mathbb{E}|Z|
= \eta,
\end{equation}
and hence $|\mathrm{(A)}|\le \eta$.
For term $\mathrm{(B)}$, the function $h(t)=1-F_{\mathrm{g}}(t)$ is also $L$-Lipschitz, so
\begin{equation}
|\mathrm{(B)}|
= \left|\mathbb{E}h(Y)-\mathbb{E}h(\widetilde Y)\right|
\le L\,\mathbb{E}|Y-\widetilde Y|
= L\sigma\mathbb{E}|Z|
= \eta.
\end{equation}
Therefore $|\Delta_{\mathrm{AUC}}| \le |\mathrm{(A)}|+|\mathrm{(B)}| \le 2\eta$, which proves the claim.
\end{proof}

\begin{corollary}[Privacy–Utility Rate]
\label{cor:auc-scaling}
Let $\mathsf{AUC}_{0} \coloneqq  \mathsf{AUC}(F_{\mathrm{g}}, F_{\mathrm{i}})$ and $\mathsf{AUC}_{\sigma} \coloneqq  \mathsf{AUC}(\widetilde{F}_{\mathrm{g}}, \widetilde{F}_{\mathrm{i}})$, and set $\eta = L\,\sigma \sqrt{2/\pi}$. Then
\begin{equation}
\bigl| \mathsf{AUC}_{\sigma} - \mathsf{AUC}_{0} \bigr|
\;\le\; 2\,\eta
\;=\; 2\,L\,\sigma \sqrt{\tfrac{2}{\pi}}
\;=\; \Theta(\sigma).  
\end{equation}
If the $(\varepsilon,\delta)$--DP Gaussian mechanism uses $\sigma = c(\delta)/\varepsilon$
(e.g., $c(\delta)=2\sqrt{2\log(2/\delta)}$ in our setting), then $\bigl| \mathsf{AUC}_{\sigma}-\mathsf{AUC}_{0} \bigr| = \mathcal{O} (1/\varepsilon)$. Moreover, Theorem~\ref{thm:roc-bound} implies the same $\Theta(\sigma)$ rate holds \textit{pointwise} along the ROC: for every $\tau \in [-1,1]$,
\begin{equation}
\bigl|\mathsf{TPR}_{\widetilde F_{\mathrm{g}}}(\tau)-\mathsf{TPR}_{F_{\mathrm{g}}}(\tau)\bigr|
\;\le\;\eta,\qquad
\bigl|\mathsf{FPR}_{\widetilde F_{\mathrm{i}}}(\tau)-\mathsf{FPR}_{F_{\mathrm{i}}}(\tau)\bigr|
\;\le\;\eta.
\end{equation}
\end{corollary}

\begin{corollary}[EER Perturbation]
\label{cor:eer-bound}
Let
\begin{eqnarray}
\mathrm{EER}_0  & \coloneqq   &  \inf_{\tau\in[-1,1]}
\max \bigl\{\mathsf{FPR}_{F_{\mathrm{i}}}(\tau),\,1-\mathsf{TPR}_{F_{\mathrm{g}}}(\tau)\bigr\},
\\
\mathrm{EER}_\sigma & \coloneqq  &  \inf_{\tau\in[-1,1]}
\max \bigl\{\mathsf{FPR}_{\widetilde{F}_{\mathrm{i}}}(\tau),\, 1-\mathsf{TPR}_{\widetilde{F}_{\mathrm{g}}}(\tau)\bigr\},
\end{eqnarray}
be the equal–error rates before and after smoothing. Under the hypotheses of Theorem~\ref{thm:roc-bound}, with $\eta = L\,\sigma\sqrt{2/\pi}$,
\begin{equation}
\bigl|\mathrm{EER}_\sigma-\mathrm{EER}_0\bigr| \;\le\; \eta.
\end{equation}
\end{corollary}

\begin{proof}
By Theorem~\ref{thm:roc-bound}, for every $\tau \in [-1,1]$, $| \mathsf{FPR}_{\widetilde{F}_{\mathrm{i}}}(\tau)- \mathsf{FPR}_{F_{\mathrm{i}}}(\tau) | \le \eta$ and $| \mathsf{TPR}_{\widetilde{F}_{\mathrm{g}}} (\tau) - \mathsf{TPR}_{F_{\mathrm{g}}} (\tau) |\le \eta$. Hence, setting $e_0(\tau) = \max\{\mathsf{FPR}_{F_{\mathrm{i}}}(\tau), \, 1-\mathsf{TPR}_{F_{\mathrm{g}}}(\tau)\}$ and $e_\sigma(\tau)= \max\{\mathsf{FPR}_{\widetilde{F}_{\mathrm{i}}}(\tau), \, 1-\mathsf{TPR}_{\widetilde{F}_{\mathrm{g}}}(\tau)\}$, we have for each $\tau$:
\begin{equation}
\vert e_\sigma(\tau) - e_0(\tau) \vert \;\le\; \eta,
\end{equation}
because the max is $1$-Lipschitz in the $\ell_\infty$ norm. Taking infima over $\tau$ on both sides gives $\mathrm{EER}_\sigma = \inf_\tau e_\sigma(\tau) \le \inf_\tau(e_0(\tau)+ \eta) = \mathrm{EER}_0 + \eta$, and symmetrically $\mathrm{EER}_0 \le \mathrm{EER}_\sigma + \eta$. Combining the two inequalities complete the proof.
\end{proof}

In face recognition, utility is often evaluated at extremely low false-positive rates (FPR), where the global AUC can mask performance in the left tail of the ROC. We therefore study the ROC over and FPR-restricted range $[0, \alpha]$ for $\alpha \in (0, 1)$. For a threshold $\tau \in [-1, 1]$, define $\mathsf{TPR}_{F_\mathrm{g}} (\tau)$ and $\mathsf{FPR}_{F_\mathrm{i}} (\tau)$ as in Eq.~\ref{eq:TPR_FPR_def}. Since the set-valued ROC may contain vertical segments, we work with the standard ROC upper envelope
\begin{equation}
\label{eq:roc-envelope}
R (u;F_{\mathrm{g}},F_{\mathrm{i}}) \; \coloneqq \; \sup \Bigl\{ \mathsf{TPR}_{F_{\mathrm{g}}}(\tau): \mathsf{FPR}_{F_{\mathrm{i}}}(\tau) \le u\Bigr\}, \qquad u\in [0,1].
\end{equation}
The (unnormalized) partial area under the ROC curve \citep{dodd2003partial, yang2019two, yang2021all} up to $\alpha$ is
\begin{equation}
\label{eq:pauc-def}
\mathrm{pAUC}(\alpha;F_{\mathrm{g}},F_{\mathrm{i}})
\;\coloneqq\; \mathrm{AUC}_{[0,\alpha]}(F_{\mathrm{g}},F_{\mathrm{i}}) = 
\int_{0}^{\alpha} R(u;F_{\mathrm{g}},F_{\mathrm{i}})\,\mathrm{d}u.
\end{equation}

\begin{corollary}[Partial-AUC perturbation]
\label{cor:pauc}
Assume the hypotheses of Theorem~\ref{thm:roc-bound} and set $\eta \coloneqq L\,\sigma\sqrt{2/\pi}$.
Then, for every $\alpha\in(0,1)$,
\begin{equation}
\label{eq:pauc-bound}
\Bigl| \mathrm{pAUC}( \alpha; \widetilde{F}_{\mathrm{g}}, \widetilde{F}_{\mathrm{i}})
-\mathrm{pAUC}(\alpha;F_{\mathrm{g}},F_{\mathrm{i}}) \Bigr|
\;\le\; \alpha\,\eta \;+\; \min\{\alpha,\eta\} \;\le\; (\alpha+1)\,\eta.
\end{equation}
In particular,
\begin{equation}
\Bigl| \mathrm{pAUC}(\alpha;\widetilde F_{\mathrm{g}}, \widetilde F_{\mathrm{i}})
- \mathrm{pAUC}(\alpha;F_{\mathrm{g}}, F_{\mathrm{i}}) \Bigr| \;= \;\mathcal{O}(\sigma),
\end{equation}
with constant $(\alpha + 1) L \sqrt{2/\pi}$.
\end{corollary}

\begin{proof}
Write $R_0 (u) \coloneqq R(u;F_{\mathrm{g}},F_{\mathrm{i}})$ and $R_\sigma(u) \coloneqq R(u; \widetilde F_{\mathrm{g}}, \widetilde{F}_{\mathrm{i}})$. By Theorem~\ref{thm:roc-bound}, for every $\tau \in [-1,1]$,
\begin{equation}
\label{eq:tauwise-bounds}
\bigl|\mathsf{TPR}_{\widetilde F_{\mathrm{g}}}(\tau)-\mathsf{TPR}_{F_{\mathrm{g}}}(\tau)\bigr|\le \eta,
\qquad \bigl|\mathsf{FPR}_{\widetilde F_{\mathrm{i}}}(\tau)-\mathsf{FPR}_{F_{\mathrm{i}}}(\tau)\bigr|\le \eta.
\end{equation}
We claim that for all $u \in [0,1]$,
\begin{equation}
\label{eq:envelope-sandwich}
R_\sigma(u) \; \le \; R_0(\min\{1 ,u +\eta\}) + \eta,
\qquad R_0(u)\;\le\; R_\sigma(\min\{1 ,u + \eta\}) + \eta.
\end{equation}
To prove the first inequality, fix $u$ and any $\tau$ such that $\mathsf{FPR}_{\widetilde{F}_{\mathrm{i}}}(\tau) \le u$. Then by Eq.~\ref{eq:tauwise-bounds} we have $\mathsf{FPR}_{F_{\mathrm{i}}}(\tau) \le u + \eta$, and also $\mathsf{TPR}_{\widetilde{F}_{\mathrm{g}}}(\tau) \le \mathsf{TPR}_{F_{\mathrm{g}}}(\tau) + \eta$. Taking the supremum over such $\tau$ yields $R_\sigma(u)\le R_0(\min\{1, u +\eta\})+ \eta$.
The second inequality follows symmetrically by swapping $(F_{\mathrm{g}}, F_{\mathrm{i}})$ with
$(\widetilde{F}_{\mathrm{g}}, \widetilde{F}_{\mathrm{i}})$.

Integrating Eq.~\ref{eq:envelope-sandwich} over $u\in[0,\alpha]$ gives
\begin{equation}
\int_0^\alpha R_\sigma(u)\,du
\le \int_0^\alpha R_0(\min\{1,u+\eta\})\,du
+ \alpha\eta .
\end{equation}
Define the clipped extension $\bar R_0(v)\coloneqq R_0(\min\{1,v\})$ for $v\ge 0$.
Then
\begin{equation}
\int_0^\alpha R_0(\min\{1,u+\eta\})\,du
= \int_\eta^{\alpha+\eta}\bar R_0(v)\,dv .
\end{equation}
Since $0\le \bar R_0\le 1$, we have
\begin{equation}
\int_\eta^{\alpha+\eta}\bar R_0(v)\,dv
\le \int_0^\alpha R_0(v)\,dv+\min\{\alpha,\eta\}.
\end{equation}
Indeed, if $\eta\le \alpha$, then
\begin{align}
\int_\eta^{\alpha+\eta}\bar R_0(v)\,dv
- \int_0^\alpha R_0(v)\,dv
&= \int_\alpha^{\alpha+\eta}\bar R_0(v)\,dv
- \int_0^\eta R_0(v)\,dv  \\
&\le \eta .
\end{align}
If $\eta>\alpha$, then the left integral is at most $\alpha$, and hence
\begin{equation}
\int_\eta^{\alpha+\eta}\bar R_0(v)\,dv
\le
\int_0^\alpha R_0(v)\,dv+\alpha .
\end{equation}
Combining the two cases gives
\begin{equation}
\int_0^\alpha R_0(\min\{1,u+\eta\})\,du
\le \mathrm{pAUC}(\alpha;F_{\mathrm g},F_{\mathrm i})+\min\{\alpha,\eta\}.
\end{equation}
Therefore,
\begin{equation}
\mathrm{pAUC}(\alpha;\widetilde F_{\mathrm g},\widetilde F_{\mathrm i})
\le \mathrm{pAUC}(\alpha;F_{\mathrm g},F_{\mathrm i})
+ \alpha\eta+\min\{\alpha,\eta\}.
\end{equation}
The reverse inequality follows by swapping $(F_{\mathrm g},F_{\mathrm i})$ and
$(\widetilde F_{\mathrm g},\widetilde F_{\mathrm i})$.
\end{proof}

\appsubsection{Rate-Optimal Upper and Matching Bound}

Consider the clean impostor false–match rate $\mathsf{FMR}_{\mathrm{clean}}(\tau) \coloneqq  \mathsf{Pr} \bigl[ \, S_{ij} > \tau \, \mid \, i \neq j \, \bigr]$. For a one-shot release $\widehat{\mathbf{s}}= \mathcal{M} (\mathbf{E}, \mathbf{q})\in[-1,1]^n$ on a fixed probe $\mathbf{q}$ and gallery $\mathbf{E}$ of size $n$ (with the probe not enrolled, hence all $n$ comparisons are impostors), the corresponding gallery–conditional FMR is
\begin{equation}
\label{eq:gallery-conditional-FMR}
\mathsf{FMR}^{\mathcal{M}}_{\mathbf{E}}(\tau) \;\coloneqq \; 
\mathsf{Pr} \left[ \widehat{S}_{IJ} > \tau  \mid I \neq J , \mathbf{E} \right] =
\frac{1}{n}\sum_{j=1}^n \mathsf{Pr} \big[ \widehat{s}_j > \tau \,\big| \, \mathbf{E} \big],
\end{equation}
where $\tau \in [-1, 1]$ is a verification threshold. We keep the strict rule ``$\, > \,$'' as we discussed before. At $\tau = 1$ the strict rule gives $\mathsf{Pr} \left[ \widehat{S} > 1\right] = 0$.
When $\mathbf{E} \sim \mathbf{E}'$ differ in exactly one enrollment (neighbors), consider $\mathsf{FMR}^{\mathcal{M}}_{\mathbf{E}'}(t)$ for the corresponding FMR under $\mathbf{E}'$.
Also note that according to the regime~(i) setup in Sec.~\ref{subsec:probe-vector}, we assume that the probe (query) is not enrolled (so all $n$ coordinates are impostor comparisons).

\begin{theorem}[Universal FMR perturbation bounds under DP]
\label{thm:fmr-upper}
Let $\mathcal{M}$ be \textit{any} one-shot $(\varepsilon, \delta)$--DP mechanism that releases, for a fixed probe $\mathbf{q}$, the full vector of clipped similarity scores $\widehat{\mathbf{s}} = \mathcal{M} \left( \mathbf{E}, \mathbf{q} \right) \in {[-1, 1]}^n$ for a gallery $\mathbf{E}$ of size $n$. For neighboring galleries $\mathbf{E} \sim \mathbf{E}'$, define the gallery-conditional FMR of $\mathcal{M}$ by Eq.~\ref{eq:gallery-conditional-FMR}. For neighboring galleries $\mathbf{E}\sim\mathbf{E}'$ the following hold.
\begin{enumerate}[label=(\alph*), leftmargin=2.1em]
\item 
\textit{Universal bound (no structural condition).}
\begin{equation}
\label{eq:universal-FMR-TV}
\sup_{\tau\in[-1,1]}
\bigl|\mathsf{FMR}^{\mathcal{M}}_{\mathbf{E}}(\tau)-\mathsf{FMR}^{\mathcal{M}}_{\mathbf{E}'}(\tau)\bigr|
\;\le\;
\mathsf{TV}\!\left(P_{\mathrm{vec}},Q_{\mathrm{vec}}\right)
\;\le\;\tanh \bigl(\varepsilon/2\bigr)+\delta
\;\le\;\frac{\varepsilon}{2}+\delta
\quad(\varepsilon\le 1),
\end{equation}
where $P_{\mathrm{vec}} \coloneqq  \mathcal{L}(\widehat{\mathbf{s}}\mid \mathbf{E})$ and 
$Q_{\mathrm{vec}} \coloneqq  \mathcal{L} (\widehat{\mathbf{s}}\mid \mathbf{E}')$, and $\mathsf{TV}(P,Q)= \sup_{A} |P(A) - Q(A)| $ is total variation distance.
\item 
\textit{Sharpened bound under marginal stability.}
Assume the following marginal–stability condition:
If $\mathbf{E}\sim\mathbf{E}'$ differ only at index $i$, then for every $j\neq i$,
$\mathcal{L}(\widehat{s}_j\mid\mathbf{E})=\mathcal{L}(\widehat{s}_j\mid\mathbf{E}')$.
Let $P_i \coloneqq  \mathcal{L}(\widehat{s}_i\mid\mathbf{E})$ and $Q_i\coloneqq  \mathcal{L}(\widehat{s}_i\mid\mathbf{E}')$ are the one-dimensional marginals of the changed coordinate $\widehat{\mathbf{s}}_i$ under $\mathbf{E}$ and $\mathbf{E}'$, respectively. Then
\begin{equation}
\label{eq:sharpened-1overN}
\sup_{\tau\in[-1,1]}
\bigl|\mathsf{FMR}^{\mathcal{M}}_{\mathbf{E}}(\tau)-\mathsf{FMR}^{\mathcal{M}}_{\mathbf{E}'}(\tau)\bigr|
\;\le\;\frac{\mathsf{TV}(P_i,Q_i)}{n}
\;\le\;\frac{\tanh(\varepsilon/2)+\delta}{n}
\;\le\;\frac{\varepsilon/2+\delta}{n}
\quad(\varepsilon\le 1).
\end{equation}
\end{enumerate}
\end{theorem}

\begin{proof}
For $\tau\in[-1,1]$ define the coordinate exceedance sets $A_j^\tau\coloneqq \{x\in[-1,1]^n:\; x_j>\tau\}$, $j=1,\dots,n$. By definition in Eq.~\ref{eq:gallery-conditional-FMR},
\begin{equation}
\mathsf{FMR}^{\mathcal{M}}_{\mathbf{E}}(\tau) =\frac1n\sum_{j=1}^n P_{\mathrm{vec}}(A_j^\tau),
\qquad
\mathsf{FMR}^{\mathcal{M}}_{\mathbf{E}'}(\tau) =\frac1n\sum_{j=1}^n Q_{\mathrm{vec}}(A_j^\tau).
\end{equation}
Hence
\begin{subequations}
\begin{align}
\bigl|\mathsf{FMR}^{\mathcal{M}}_{\mathbf{E}}(\tau)-\mathsf{FMR}^{\mathcal{M}}_{\mathbf{E}'}(\tau)\bigr|
&= \Bigl|\frac1n\sum_{j=1}^n \bigl(P_{\mathrm{vec}}(A_j^\tau)-Q_{\mathrm{vec}}(A_j^\tau)\bigr)\Bigr|\\
&\le \frac1n\sum_{j=1}^n \bigl|P_{\mathrm{vec}}(A_j^\tau)-Q_{\mathrm{vec}}(A_j^\tau)\bigr|
\;\le\; \frac1n\sum_{j=1}^n \mathsf{TV}(P_{\mathrm{vec}},Q_{\mathrm{vec}})\\
&=\mathsf{TV}(P_{\mathrm{vec}},Q_{\mathrm{vec}}).
\end{align}
\end{subequations}
Taking $\sup_{\tau\in[-1,1]}$ preserves the bound, proving the first inequality in Eq.~\ref{eq:universal-FMR-TV}. Because $\mathcal{M}$ is $(\varepsilon, \delta)$–DP on neighboring inputs, the standard DP$\Rightarrow$TV conversion gives $\mathsf{TV}(P_{\mathrm{vec}},Q_{\mathrm{vec}})\le \tanh(\varepsilon/2)+\delta$\footnote{For $\delta=0$ (pure DP) the inequality $\mathsf{TV}\le\tanh(\varepsilon/2)$ is tight. For $(\varepsilon,\delta)$–DP a standard extension gives $\mathsf{TV}\le\tanh(\varepsilon/2)+\delta$.}. Finally, $\tanh(x)\le x$ for $x\ge 0$ yields the small–$\varepsilon$ simplification.

For part (B), let $i$ be the unique index where $\mathbf{E}$ and $\mathbf{E}'$ differ.
By marginal stability, $P_{\mathrm{vec}}(A_j^\tau)=Q_{\mathrm{vec}}(A_j^\tau)$ for all $j\neq i$, hence
\begin{equation}
\mathsf{FMR}^{\mathcal{M}}_{\mathbf{E}}(\tau)-\mathsf{FMR}^{\mathcal{M}}_{\mathbf{E}'}(\tau)
=\frac1n\Big(P_{\mathrm{vec}}(A_i^\tau)-Q_{\mathrm{vec}}(A_i^\tau)\Big)
=\frac1n\Big(P_i\big((\tau,1]\big)-Q_i\big((\tau,1]\big)\Big),
\end{equation}
where we used that $\widehat{s}_i\in[-1,1]$ implies $\{\widehat{s}_i>\tau\}=\{\widehat{s}_i\in(\tau,1]\}=\{\widehat{s}_i\in(\tau,\infty)\}$. Also note total variation is the supremum over all Borel sets, so taking the supremum over this particular subfamily of half-lines can only underestimate $\mathsf{TV}$. Therefore
\begin{equation}
\sup_{\tau\in[-1,1]} \bigl|\mathsf{FMR}^{\mathcal{M}}_{\mathbf{E}}(\tau)-\mathsf{FMR}^{\mathcal{M}}_{\mathbf{E}'}(\tau)\bigr|
\;\le\; \frac{1}{n}\,\sup_{\tau\in\mathbb{R}}\bigl|P_i\big((\tau,\infty)\big)-Q_i\big((\tau,\infty)\big)\bigr| \;\le\; \frac{\mathsf{TV}(P_i,Q_i)}{n}.
\end{equation}
Since $\widehat{s}_i$ is a measurable post–processing of the DP output $\widehat{\mathbf{s}}$, the pair $(P_i,Q_i)$ also satisfies $(\varepsilon,\delta)$–DP, hence $\mathsf{TV}(P_i,Q_i)\le \tanh(\varepsilon/2)+\delta$ and Eq.~\ref{eq:sharpened-1overN} follows.
\end{proof}

\begin{remark}[Marginal Stability]
\label{rem:marginal-stability}
The condition does \textit{not} require independence across coordinates and allows shared internal randomness. A sufficient formulation is: there exist measurable maps $K_j$ and a data–independent random seed $U$ such that  $\widehat{s}_j = K_j(s_j,U)$ for each $j$. Then changing one enrollment (hence only one clean $s_i$ for a non–enrolled probe) leaves the marginals of $\widehat{s}_j$ unchanged for all $j\neq i$. Mechanisms such as per–coordinate additive noise followed by clipping (e.g., Gaussian noise) satisfy this property.
\end{remark}

\begin{remark}[Half-lines]
\label{rem:halflines}
Since outputs are clipped to $[-1,1]$, the exceedance event $\{\widehat{s}_j>\tau\}$ equals $\{\widehat{s}_j\in(\tau,1]\}$, which is the same as $\{\widehat{s}_j\in(\tau,\infty)\}$ because there is no mass above $1$. Thus using $(\tau,\infty)$ is a notational convenience for survivor sets and does not affect probabilities or suprema.
\end{remark}


\begin{remark}[ROC-level Stability]
The bounds hold \textit{uniformly} over $\tau\in[-1,1]$, so the entire impostor ROC (equivalently, the false–match rate curve across thresholds) changes by at most $\mathcal{O}( \varepsilon+  \delta )$ between neighboring galleries \textit{without} structure, and by at most $\mathcal{O}( (\varepsilon+\delta)/n )$ under marginal stability.
\end{remark}

\vspace{5pt}

\begin{proposition}[Achievability by the Gaussian ScoreShield mechanism]
\label{prop:fmr-gauss-proj}
Consider regime~(i) with a fixed (non-enrolled) probe $\mathbf{q} \in \mathbb{R}^{d}$ and gallery $\mathbf{E} \in\mathbb{R}^{n\times d}$. Let $\mathbf{s} \;=\; \mathbf{E} \mathbf{q} \in[-1,1]^n$ be the clean probe-to-enrollment similarity vector, and let the DP mechanism be
$\mathbf{s}' \;=\; \mathbf{s} + \mathbf{w}$, $\mathbf{w} \sim \mathcal{N} (\mathbf{0}, \sigma^2 \mathbf{I}_n)$, $\widehat{\mathbf{s}}\;=\;\mathsf{proj}_{[-1,1]^n}(\mathbf{s}')$. Assume the Gaussian calibration for $(\varepsilon,\delta)$–DP with regime~(i) sensitivity $\Delta_{f,2}=2$, namely $\sigma \;=\; \frac{2\sqrt{2\log(2/\delta)}}{\varepsilon}$. Fix neighboring galleries $\mathbf{E} \sim \mathbf{E}'$ that differ only at index $i=1$ and satisfy
\begin{equation}
s^{(\mathbf{E})}_1= +1, \qquad s^{(\mathbf{E}')}_1= -1, \qquad
s^{(\mathbf{E})}_j= s^{(\mathbf{E}')}_j = a_j \in[-1,1] \quad (j= 2, \dots, n), 
\end{equation}
with arbitrary constants $a_j$. Denote the gallery–conditional impostor FMR of this mechanism by $\mathsf{FMR}^{\mathcal{M}}_{\mathbf{E}}(\tau)$. Then the FMR gap admits the \textit{exact identity}
\begin{equation}
\label{eq:gauss-proj-exact}
\sup_{\tau\in[-1,1]}\,
\bigl|\mathsf{FMR}^{\mathcal{M}}_{\mathbf{E}}(\tau)-\mathsf{FMR}^{\mathcal{M}}_{\mathbf{E}'}(\tau)\bigr|
\;=\;\frac{1}{n}\Bigl(2\,\Phi(1/\sigma)-1\Bigr)
\;=\;\frac{1}{n}\,\mathrm{erf} \Bigl(\frac{1}{\sigma\sqrt{2}}\Bigr).
\end{equation}
Moreover, for $\varepsilon\le1$ and $\delta\in(0,0.1]$ there exist absolute constants
$0 < c < C < \infty$ such that
\begin{equation}
\label{eq:gauss-proj-theta}
\frac{c}{n} \cdot \frac{\varepsilon}{\sqrt{\log(2/\delta)}}
\;\le\;
\sup_{\tau\in[-1,1]}\,
\bigl| \mathsf{FMR}^{\mathcal{M}}_{\mathbf{E}}(\tau) - \mathsf{FMR}^{\mathcal{M}}_{\mathbf{E}'}(\tau) \bigr|
\;\le\;
\frac{C}{n}\cdot\frac{\varepsilon}{\sqrt{\log(2/\delta)}},
\end{equation}
and hence, for any fixed $\delta\in(0,0.1]$ the gap is $\Theta( \varepsilon/n )$, matching the DP upper bound in Theorem~\ref{thm:fmr-upper} up to constants.
\end{proposition}

\vspace{5pt}

\begin{proof}
%
Because $\mathbf{E}$ and $\mathbf{E}'$ differ only at index $1$ and the mechanism is \textit{coordinatewise} perturbation plus \textit{coordinatewise} projection, the coordinates $j \ge 2$ have identical distributions under $\mathbf{E}$ and $\mathbf{E}'$ and thus cancel in the FMR difference, irrespective of the values $a_j$. Therefore, for every threshold $\tau\in[-1,1]$,
\begin{subequations}
\begin{align}
\mathsf{FMR}^{\mathcal{M}}_{\mathbf{E}}(\tau)-\mathsf{FMR}^{\mathcal{M}}_{\mathbf{E}'}(\tau)
& = \frac{1}{n} \sum_{j=1}^n \Bigl( \mathsf{Pr}[\widehat{s}^{(\mathbf{E})}_j > \tau] - \mathsf{Pr}[ \widehat{s}^{(\mathbf{E}')}_j > \tau] \Bigr)\\
& = \frac{1}{n} \Bigl( \mathsf{Pr}[\widehat{s}^{\, (\mathbf{E})}_1 > \tau] - \mathsf{Pr}[\widehat{s}^{\,(\mathbf{E}')}_1 > \tau] \Bigr).
\end{align}
\end{subequations}
%
By projection monotonicity (see Eq.~\ref{eq:proj-monotone-interior}), for any \textit{interior} threshold $\tau < 1$ and any scalar $S \in [-1,1]$, $\mathbf{1}\{\mathsf{proj}_{[-1,1]}(S+W)>\tau\}=\mathbf{1}\{S+W>\tau\}$. The optimizer we will obtain satisfies $\tau^\star = 0 < 1$, hence clipping does not alter the exceedance events at the optimizer and we may analyze the \textit{unclipped} Gaussian
convolutions.
Under $\mathbf{E}$ and $\mathbf{E}'$, the first coordinate laws (before projection) are $\mathcal{N}(+1, \sigma^2)$ and $\mathcal{N} (-1,\sigma^2)$, respectively. Thus, for any $\tau$,
\begin{equation}
\Delta_1 (\tau) \;\coloneqq \; \mathsf{Pr}[\widehat{s}^{\,(\mathbf{E})}_1 > \tau]-\mathsf{Pr}[ \widehat{s}^{\, (\mathbf{E}')}_1 > \tau]
=\Phi \Bigl( \frac{1-\tau}{\sigma}\Bigr) - \Phi \Bigl(\frac{-1-\tau}{\sigma} \Bigr). 
\end{equation}
By differentiation we have:
\begin{equation}
\Delta_1'(\tau) =\frac{1}{\sigma}\Bigl(\phi \bigl(\tfrac{-1-\tau}{\sigma}\bigr)-\phi \bigl(\tfrac{1-\tau}{\sigma}\bigr)\Bigr),
\end{equation}
with $\phi$ the standard normal pdf (even and strictly decreasing on $[0,\infty)$). Hence $\Delta_1'(0) = 0$ and $\Delta_1'(\tau) > 0$ for $\tau < 0$ while $\Delta_1'(\tau) < 0$ for $\tau > 0$, so $\Delta_1 (\tau)$ is maximized at $\tau^\star= 0$. Therefore
\begin{equation}
\sup_{\tau\in[-1,1]}\Delta_1(\tau) = \Delta_1(0) = \Phi(1/\sigma) - \Phi(-1/\sigma)
= 2\,\Phi(1/\sigma)-1 = \mathrm{erf} \Bigl(\frac{1}{\sigma\sqrt{2}}\Bigr).
\end{equation}
Dividing by $n$ gives the exact identity Eq.~\ref{eq:gauss-proj-exact}.
%
For $x\ge0$,
\begin{equation}
\frac{2}{\sqrt{\pi}}\, x\,e^{-x^{2}} \;\le\; \mathrm{erf}(x) \;\le\; \frac{2}{\sqrt{\pi}}\, x. 
\end{equation}
With $x= 1/(\sigma\sqrt{2})$ and $\varepsilon \le 1$, $\delta \le 0.1$ imply $\sigma \ge 1$, yielding
\begin{equation}
\frac{\sqrt{2}}{\sqrt{\pi}}\cdot\frac{e^{-1/(2\sigma^{2})}}{\sigma}
\;\le\; \mathrm{erf} \Bigl(\frac{1}{\sigma\sqrt{2}}\Bigr)
\;\le\; \frac{1}{\sqrt{2\pi}}\cdot\frac{2}{\sigma} \, .   
\end{equation}
Using $e^{-1/(2\sigma^2)}\ge e^{-1/2}$ for $\sigma \ge 1$ and substituting $\sigma = 2\sqrt{2\log(2/\delta)}/\varepsilon$ gives
\begin{equation}
\frac{e^{-1/2}}{2\sqrt{\pi}}\cdot\frac{1}{n}\cdot
\frac{\varepsilon}{\sqrt{\log(2/\delta)}}
\;\le\;
\sup_{\tau\in[-1,1]}
\bigl|\mathsf{FMR}^{\mathcal{M}}_{\mathbf{E}}(\tau)-\mathsf{FMR}^{\mathcal{M}}_{\mathbf{E}'}(\tau)\bigr|
\;\le\;
\frac{1}{2\sqrt{\pi}}\cdot\frac{1}{n}\cdot
\frac{\varepsilon}{\sqrt{\log(2/\delta)}}.
\end{equation}
For any fixed $\delta\in(0,0.1]$, the factor $1/\sqrt{\log(2/\delta)}$ is a constant, hence the rate is $\Theta( \varepsilon/n )$.
\end{proof}

\vspace{5pt}

\begin{remark}[Consistency with the DP upper bound and marginal stability]
\label{rem:gauss-proj-consistency}
The perturb–then–project release in Proposition~\ref{prop:fmr-gauss-proj} is \textit{coordinatewise}
(add noise and project per coordinate), hence it satisfies the marginal–stability assumption of
Theorem~\ref{thm:fmr-upper}~(b). Combining Eq.~\ref{eq:sharpened-1overN} with the exact identity
Eq.~\ref{eq:gauss-proj-exact} yields the sandwich
\begin{equation}
\underbrace{\frac{1}{n}\,\mathrm{erf}  \Bigl(\frac{1}{\sigma\sqrt{2}}\Bigr)}_{\text{Prop.~\ref{prop:fmr-gauss-proj}}}
\; \leq \;
\sup_{\tau\in[-1,1]}
\bigl|\mathsf{FMR}^{\mathcal{M}}_{\mathbf{E}}(\tau)-\mathsf{FMR}^{\mathcal{M}}_{\mathbf{E}'}(\tau)\bigr|
\;\le\;
\underbrace{\frac{\tanh(\varepsilon/2)+\delta}{n}}_{\text{Thm.~\ref{thm:fmr-upper}~(b)}}.
\end{equation}
With $\sigma=\tfrac{2\sqrt{2\log(2/\delta)}}{\varepsilon}$ and $\varepsilon\le1$, $\delta\in(0,0.1]$,
the left-hand side equals $\Theta  \bigl(\varepsilon/(n\sqrt{\log(2/\delta)})\bigr)$ while the
right-hand side is $\mathcal{O}(\varepsilon/n)$. Thus, for fixed $\delta$, the marginal–stability upper bound
in Theorem~\ref{thm:fmr-upper}~(b) is \textit{rate-tight} up to constants. By contrast, without marginal stability, Theorem~\ref{thm:fmr-upper}~(a) gives the universal mechanism–independent bound
$\mathcal{O}( \varepsilon+ \delta)$ (no $1/n$ factor), which the Gaussian mechanism trivially satisfies but does
not saturate.
\end{remark}

\vspace{5pt}

\newcommand{\Gap}{{\mathrm{Gap}}}
\newcommand{\FMRgap}[3]{%
\sup_{\tau\in[-1,1]}\bigl|\mathsf{FMR}^{#1}_{#2}(\tau)-\mathsf{FMR}^{#1}_{#3}(\tau)\bigr|
}

\begin{remark}[Rate tightness and the minimax viewpoint]
\label{rem:rate-tight-minimax}
For a one–shot mechanism $\mathcal{M}$, define the worst–case FMR disturbance
\begin{equation}
\Gap(\mathcal{M})\;\coloneqq \;\sup_{\mathbf{E}\sim\mathbf{E}'}\ \sup_{\tau\in[-1,1]}\,
\bigl|\mathsf{FMR}^{\mathcal{M}}_{\mathbf{E}}(\tau)-\mathsf{FMR}^{\mathcal{M}}_{\mathbf{E}'}(\tau)\bigr|.    
\end{equation}
Theorem~\ref{thm:fmr-upper}~(a) gives the universal mechanism–independent bound $\Gap(\mathcal{M})\le\tanh(\varepsilon/2)+\delta=\mathcal{O}(\varepsilon+\delta)$ for $\varepsilon\le1$, and under marginal stability Theorem~\ref{thm:fmr-upper}~(b) sharpens this to $\Gap(\mathcal{M})\le(\tanh(\varepsilon/2)+\delta)/n=\mathcal{O}((\varepsilon+\delta)/n)$. Proposition~\ref{prop:fmr-gauss-proj} shows \textit{existential} rate tightness:
\begin{equation}
\Gap(\mathcal{M})=\frac{1}{n}\,\mathrm{erf} \Bigl(\frac{1}{\sigma\sqrt{2}}\Bigr)
=\Theta \Bigl(\frac{1}{n}\cdot\frac{\varepsilon}{\sqrt{\log(2/\delta)}}\Bigr),
\qquad \sigma=\frac{2\sqrt{2\log(2/\delta)}}{\varepsilon},    
\end{equation}
hence $\Gap(\mathcal{M})=\Theta(\varepsilon/n)$ for fixed $\delta$.
This shows that the marginal–stability upper bound in Theorem~\ref{thm:fmr-upper}~(b) is \textit{rate-tight} up to constants. Moreover, a per–coordinate randomized–response release (one bit per coordinate with $(\varepsilon,0)$–DP) also satisfies marginal stability and achieves $\Gap(\mathcal{M})= \tfrac{1}{n}\tanh(\varepsilon/2)=\Theta(\varepsilon/n)$ on a suitable neighboring pair, i.e., without the $\sqrt{\log}$ factor. In contrast, the universal part (a) without marginal stability provides only the mechanism–independent bound $\mathcal{O}(\varepsilon+\delta)$ (no $1/n$ factor), which the above mechanisms trivially satisfy but do not saturate.
\end{remark}

\vspace{5pt}

\begin{remark}[Absence of a mechanism–uniform lower bound]
\label{rem:no-universal-lb}
There is no positive lower bound that holds uniformly over all $(\varepsilon,\delta)$–DP one–shot
mechanisms. In particular,
\begin{equation}
\inf_{\mathcal{M} \text{$(\varepsilon, \delta)$--DP}} \Gap( \mathcal{M}) \;=\; 0.    
\end{equation}
Consider a data–independent mechanism that outputs a random vector with a fixed law (e.g., i.i.d. $\mathcal{N}(0,1)$), regardless of the input. Such a mechanism is $(0,0)$--DP and therefore $(\varepsilon, \delta)$--DP for all parameters. Its gallery–conditional FMRs coincide under any neighboring galleries, whence $\Gap( \mathcal{M})= 0$. Consequently, statements asserting $\Gap(\mathcal{M})\ge c\,\varepsilon/n$ for \textit{every} DP mechanism do not hold. By contrast,
Proposition~\ref{prop:fmr-gauss-proj} and the randomized–response example show that there \textit{exist} natural mechanisms and neighboring pairs achieving $\Gap (\mathcal{M})= \Theta(\varepsilon/n)$ (for fixed $\delta$), while Theorem~\ref{thm:fmr-upper}~(b) provides the matching $\mathcal{O}(\varepsilon/n)$ upper bound under marginal stability. These results characterize the achievable rate up to constant factors, but not via a mechanism–uniform lower bound.
\end{remark}

\clearpage
\appsection{Supplementary Details for Regime~(i): DP-FR}
\label{app:sec:supplementary-regime1-experiments-DP-FR}

\appsubsection{Experimental Setup}

\paragraph{Domain–restricted sensitivity.}
\label{app:domain–restricted-sensitivity}

Assume that all admissible record--query pairs satisfy the public margin constraint $\langle \mathbf{e}, \mathbf{q} \rangle \in [c_{\min},1]$, $c_{\min}\in(-1,1)$. Consider the score-vector release $f_{\mathsf{query}}(\mathbf{E},\mathbf{q})=\mathbf{E}\mathbf{q}$, under \emph{row-replacement adjacency}, where one row of $\mathbf{E}$ may change and $\|\mathbf{e}\|_2=\|\mathbf{q}\|_2=1$. Then neighboring databases differ in exactly one score coordinate, and the corresponding effective $\ell_2$ sensitivity satisfies
\begin{equation}
\Delta_{\mathrm{query}}^{\mathrm{eff}}\;\le\;
\sup_{a,b\in[c_{\min},1]}|a-b|
=\;1-c_{\min}. 
\end{equation}

In our audits across several face-recognition backbones and datasets, off-diagonal cosine similarities seldom took highly negative values: the minimum observed negative similarity was typically in the range $[-0.45,-0.40]$, and the mean over negative entries was approximately $-0.06$, with about one quarter of off-diagonal pairs being negative.\footnote{For a representative $100\mathrm{k}\times100\mathrm{k}$ similarity matrix, we observed approximately $25\%$ negative off-diagonal entries, mean over negatives approximately $-0.066$, and minimum approximately $-0.424$. Exact values vary across backbones and datasets. We therefore treat these numbers as descriptive rather than universal.}
Motivated by this empirical structure, we also study the reduced sensitivity
\begin{equation}
\Delta_{\mathrm{query}}^{\mathrm{eff}}=1-c_{\min} 
\end{equation}
as a utility model under the stated public margin restriction. 
For example, the conservative choice $c_{\min}=-0.5$ gives $\Delta_{\mathrm{query}}^{\mathrm{eff}}=1.5$, which is often more consistent with the observed privacy--utility degradation curves than the worst-case value $\Delta=2$. 
Unless the margin restriction is enforced as a public precondition of the mechanism, however, all formal DP guarantees in this paper remain calibrated with the worst-case global sensitivity $\Delta=2$.

\paragraph{Configurations compared (calibration $\times$ sensitivity).}
To separate the effect of the calibration rule from the effect of the sensitivity choice, we evaluate the following four configurations:
\begin{itemize}[leftmargin=*]
  \item 
  \textit{Conservative sufficient calibration with worst-case sensitivity}:
    the standard sufficient bound $\sigma=\Delta\sqrt{2\ln(2/\delta)}/\varepsilon$ with $\Delta=2$ (worst-case for our FR mechanism). 
  \item 
  \textit{Conservative sufficient calibration with domain-restricted sensitivity}:
    the same sufficient bound, but with $\Delta=1-c_{\min}$ under the public margin constraint $\langle \mathbf{e},\mathbf{q}\rangle \ge c_{\min}$. 
  \item 
  \textit{Analytic calibration~\citep{balle2018improving} with worst-case sensitivity}:
   $\sigma$ is the smallest solution of $\delta_{\mathrm{AG}}(\varepsilon,\Delta/\sigma)=\delta$ with $\Delta=2$. 
  \item 
  \textit{Analytic calibration~\citep{balle2018improving} with domain-restricted sensitivity}:
    the same analytic calibration, but with $\Delta=1-c_{\min}$ under the margin restriction. 
\end{itemize}
In the two domain-restricted configurations, both methods use the same reduced sensitivity $\Delta=1-c_{\min}$, so any performance difference isolates the effect of the calibration rule itself. Note that all formal privacy claims are calibrated with the worst-case value $\Delta=2$; the reduced-sensitivity results are reported to interpret utility under the stated margin model.
For some figures, we additionally plot a classical reference calibration with constant $c= 1.25$, alongside the conservative sufficient calibration and the analytic calibration.

\paragraph{Impostor--genuine score sets.}
We evaluate two types of impostor--genuine (imp--gen) score sets.

\begin{figure}[!t]
  \centering
  \begin{subfigure}[t]{0.32\linewidth}
    \includegraphics[width=\linewidth]{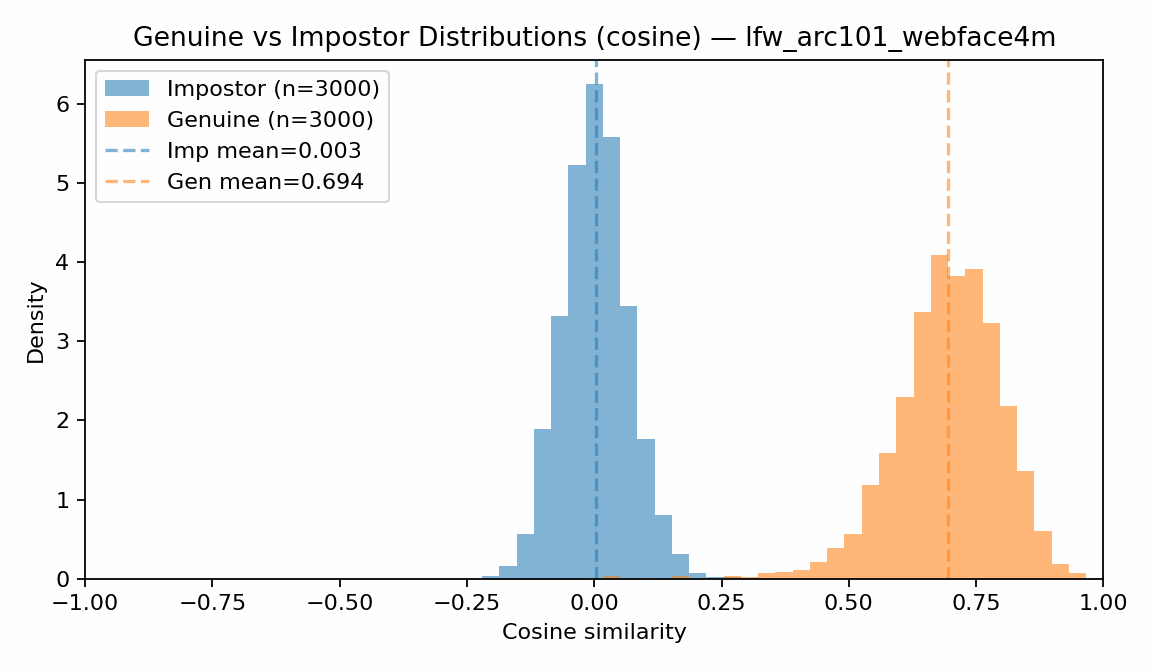}
    \caption{}\label{fig:impgen_lfw_arc101_webface4m}
  \end{subfigure}
   \begin{subfigure}[t]{0.32\linewidth}
    \includegraphics[width=\linewidth]{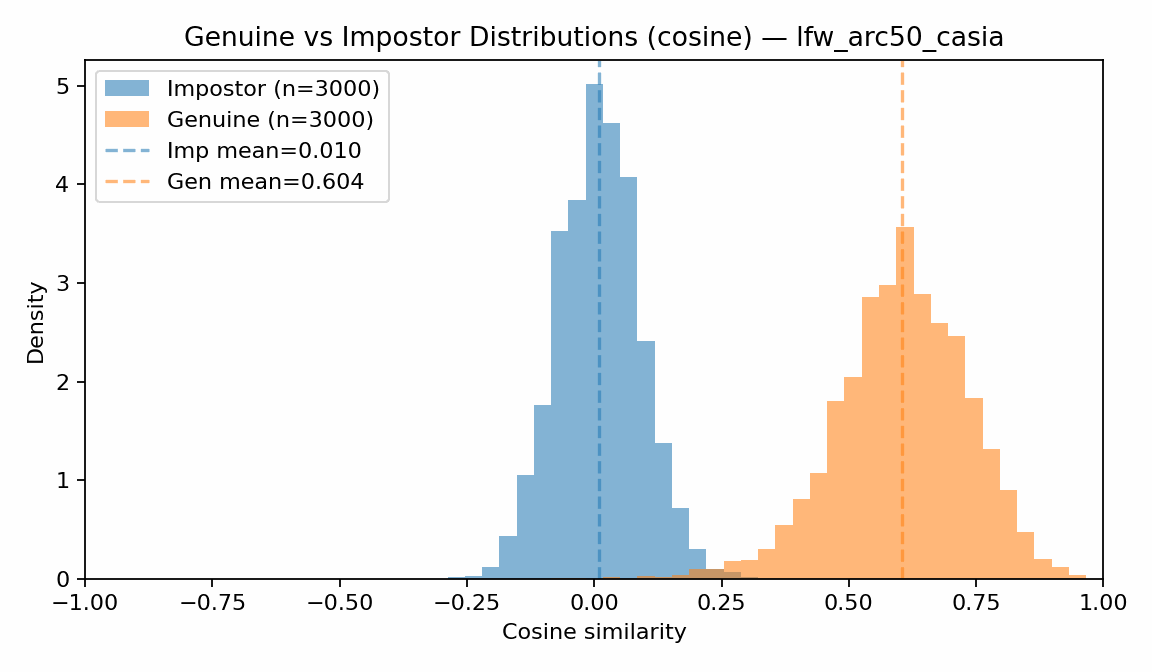}
    \caption{}\label{fig:impgen_lfw_arc50_casia}
  \end{subfigure}
  \begin{subfigure}[t]{0.32\linewidth}
    \includegraphics[width=\linewidth]{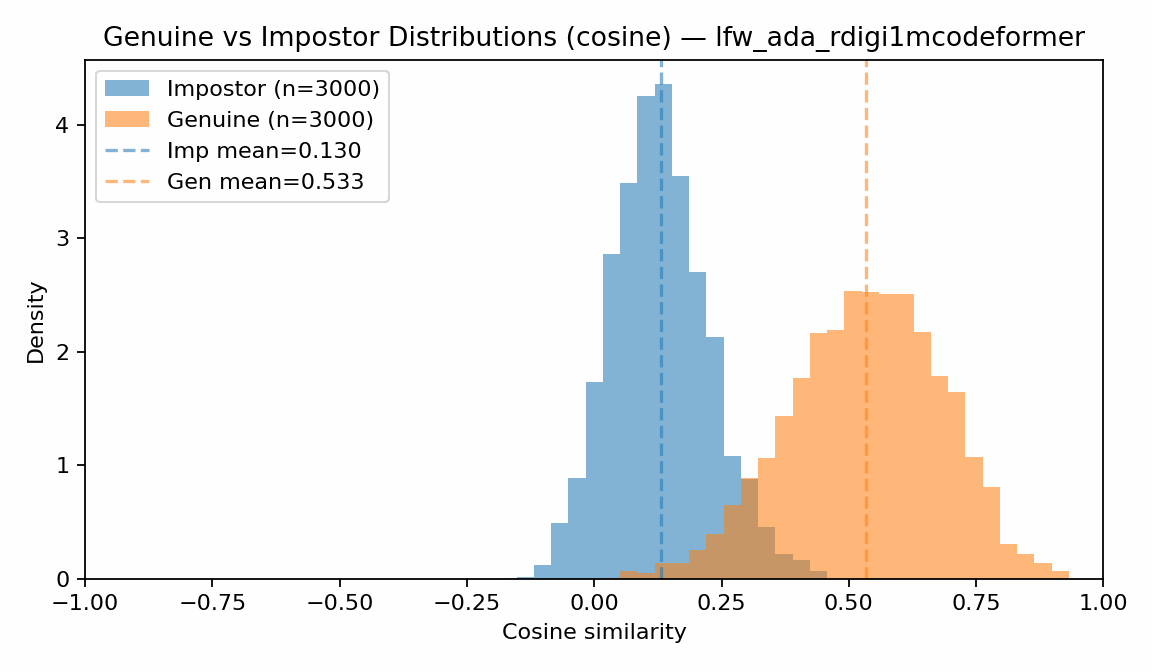}
    \caption{}\label{fig:impgen_lfw_ada_rdigi1mcodeformer}
  \end{subfigure}
  \vspace{-4pt}
  \caption{Real impostor–-genuine score distributions on LFW.
  Each panel overlays the impostor (blue) and genuine (orange) cosine-similarity histograms for one FR pipeline.
  (a) ArcFace with an IR-101 backbone pretrained on \textit{WebFace4M} (\texttt{lfw\_arc101\_webface4m}).
  (b) ArcFace with an IR-50 backbone pretrained on \textit{Casia} (\texttt{lfw\_arc50\_casia}).
  (c) AdaFace with an IR-101 backbone pretrained on \textit{DigiFace-1M}, with inputs restored by \textit{CodeFormer} before embedding (\texttt{lfw\_ada\_rdigi1mcodeformer}).
  Each run yields roughly $3{,}000$ impostor and $3{,}000$ genuine pairs.}
  \label{fig:real-impgen}
\end{figure}

\textit{(i) Real FR scores.}
We compute cosine-similarity scores on LFW using three off-the-shelf FR pipelines and use all available genuine and impostor comparisons from each run (about $3{,}000$ pairs per class in our setup):
\begin{enumerate}[label=(\alph*), leftmargin=2.2em]
  \item 
  \texttt{lfw\_arc101\_webface4m}: ArcFace with an IR-101 backbone pretrained on \textit{WebFace4M}.
  \item 
  \texttt{lfw\_arc50\_casia}: ArcFace with an IR-50 backbone pretrained on \textit{Casia}.
  \item 
  \texttt{lfw\_ada\_rdigi1mcodeformer}: AdaFace with an IR-101 backbone pretrained on \textit{DigiFace-1M}, with \textit{CodeFormer} restoration before embedding.
\end{enumerate}
Histograms of these real imp--gen score distributions are shown in Fig.~\ref{fig:real-impgen}.

\textit{(ii) Synthetic FR-like scores.}
To probe behavior under controlled distributional shapes, we generate matched imp--gen score sets with $N=200{,}000$ i.i.d. samples per side, using independent random seeds. To avoid boundary atoms, all scores are constrained to the open interval $(-1,1)$. Hence if $X$ is drawn from a base law $F$, we retain only draws in $(-1,1)$, i.e., $X_{\mathrm{tr}} \sim F(\,\cdot \mid -1<X<1\,)$,
implemented by accept--reject sampling. We report the rejection rate
\[
r_{\mathrm{rej}}
\coloneqq
1-\frac{n}{N_{\mathrm{prop}}},
\]
that is, the fraction of proposals discarded by truncation. This quantifies the mass that the untruncated law assigns outside $(-1,1)$ and ensures that the synthetic histograms have no spikes at $\pm1$.

\textit{Impostor families.} 
We use four synthetic impostor families, each truncated to $(-1,1)$:
\begin{enumerate}[label=(\alph*), leftmargin=1.6em]
  \item 
  Gaussian: $X\sim\mathcal{N}(0,\sigma)$.
  \item 
  Mixture: $X\sim (1-p)\,\mathcal{N}(0,\sigma_1)+p\,\mathcal{N}(\mu_2,\sigma_2)$.
  \item 
  Student-$t$: $X=s\,T_\nu$,  $s=\sigma\sqrt{(\nu-2)/\nu}$
    where $T_\nu$ is standard $t_\nu$, so that $\operatorname{sd}(X)=\sigma$ before truncation (for $\nu>2$).
  \item 
  Symmetric Beta: draw $Z\sim\mathrm{Beta}(a,a)$ on $[0,1]$ and map $X=2Z-1$. This law already lies in $[-1,1]$; we numerically nudge the samples into $(-1,1)$ to avoid exact endpoints.
\end{enumerate}

\textit{Matching genuine families.} 
For each impostor file, we generate a matched genuine file (same $N$) with mass shifted toward higher similarity while preserving the basic family shape: 
\begin{itemize}[leftmargin=1.6em, itemsep=1pt]
  \item 
  \textit{Right-shifted Gaussian:} 
  $G\sim\mathcal{N}(\mu,\sigma)$ with $\mu\in[0.6,0.7]$.
  \item 
  \textit{Right-shifted mixture:} 
    a dominant high-mean component plus a small low-mean tail to mimic difficult genuine pairs.
  \item 
  \textit{Shifted Student-$t$:} 
  $G=\mu + s\,T_\nu$ with the mass centered in the genuine region.
  \item 
  \textit{Skewed Beta:} 
  draw $Z\sim\mathrm{Beta}(a,b)$ with $a\ge b$, then map $G=2Z-1$, concentrating mass near $+1$.
\end{itemize}

\begin{figure}[!t]
  \centering
  \begin{subfigure}[t]{0.35\linewidth}
    \includegraphics[width=\linewidth]{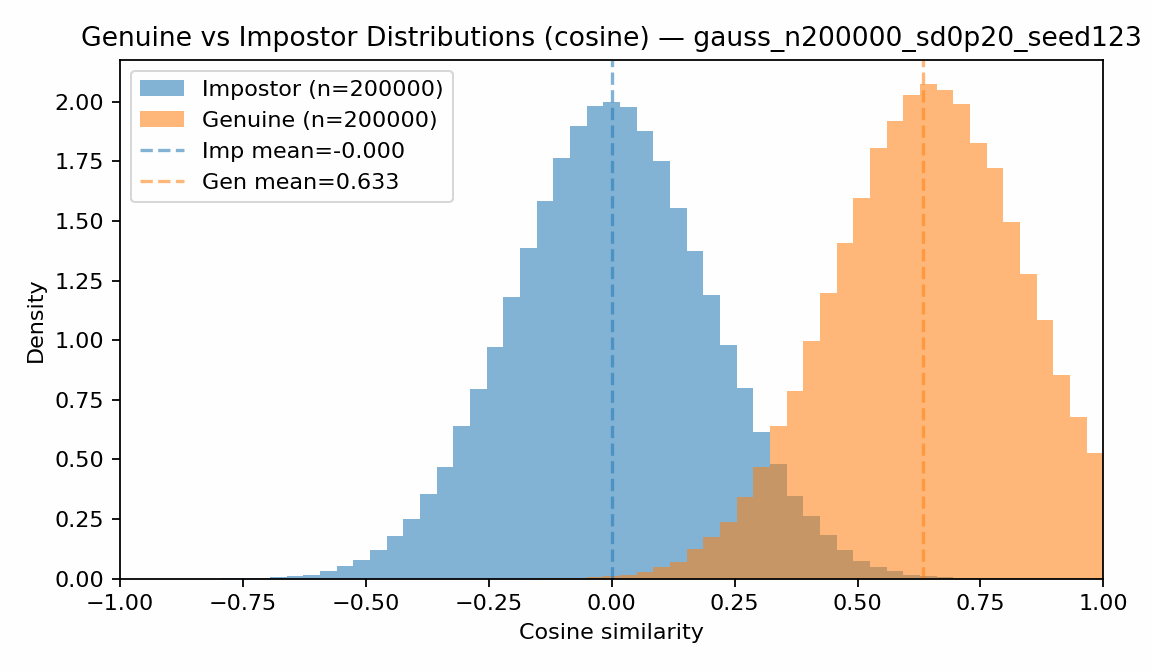}
    \caption{}\label{fig:impgen-synth-gauss}
  \end{subfigure}~~
  \begin{subfigure}[t]{0.35\linewidth}
    \includegraphics[width=\linewidth]{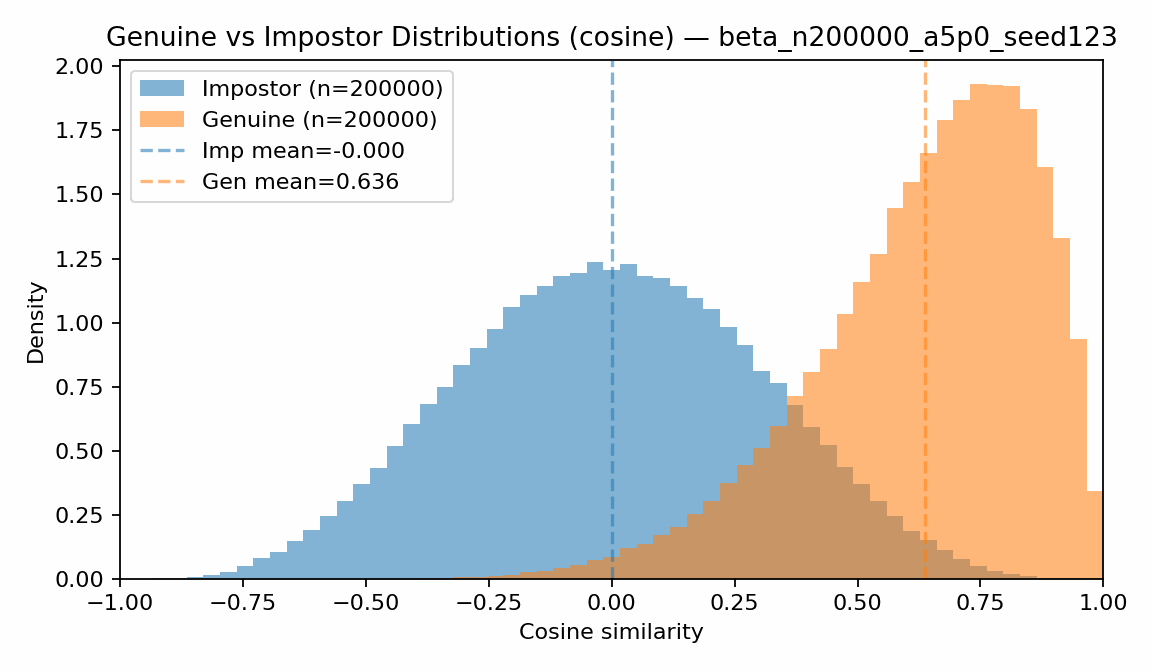}
    \caption{}\label{fig:impgen-synth-beta}
  \end{subfigure}

  \begin{subfigure}[t]{0.35\linewidth}
    \includegraphics[width=\linewidth]{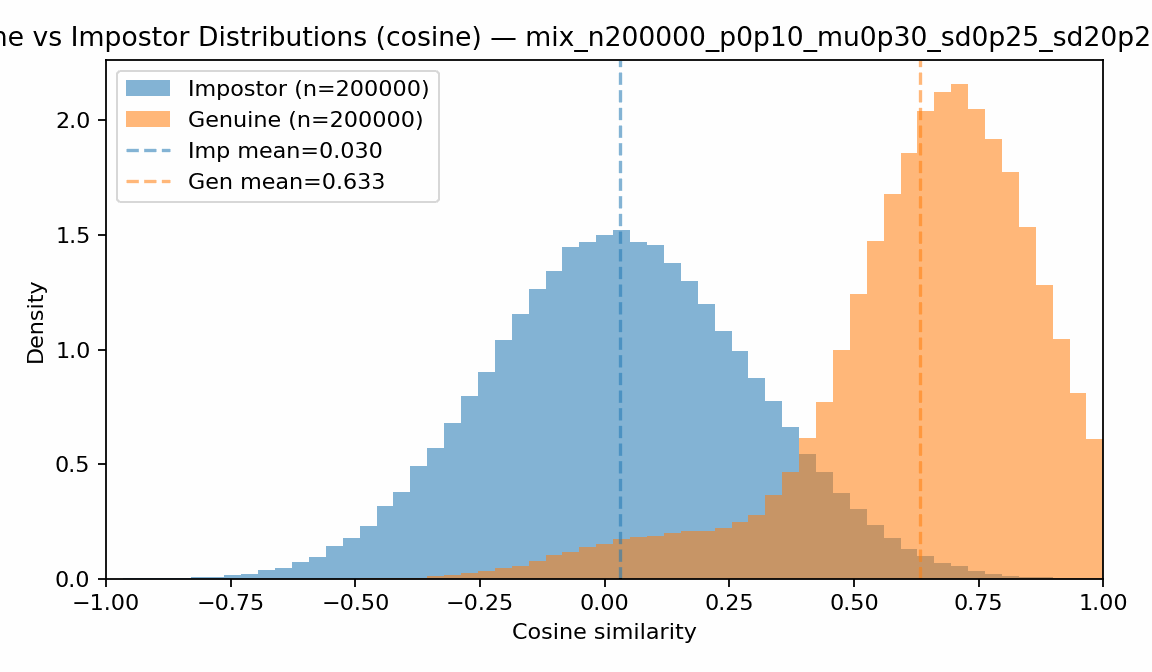}
    \caption{}\label{fig:impgen-synth-mix}
  \end{subfigure}~~
  \begin{subfigure}[t]{0.35\linewidth}
    \includegraphics[width=\linewidth]{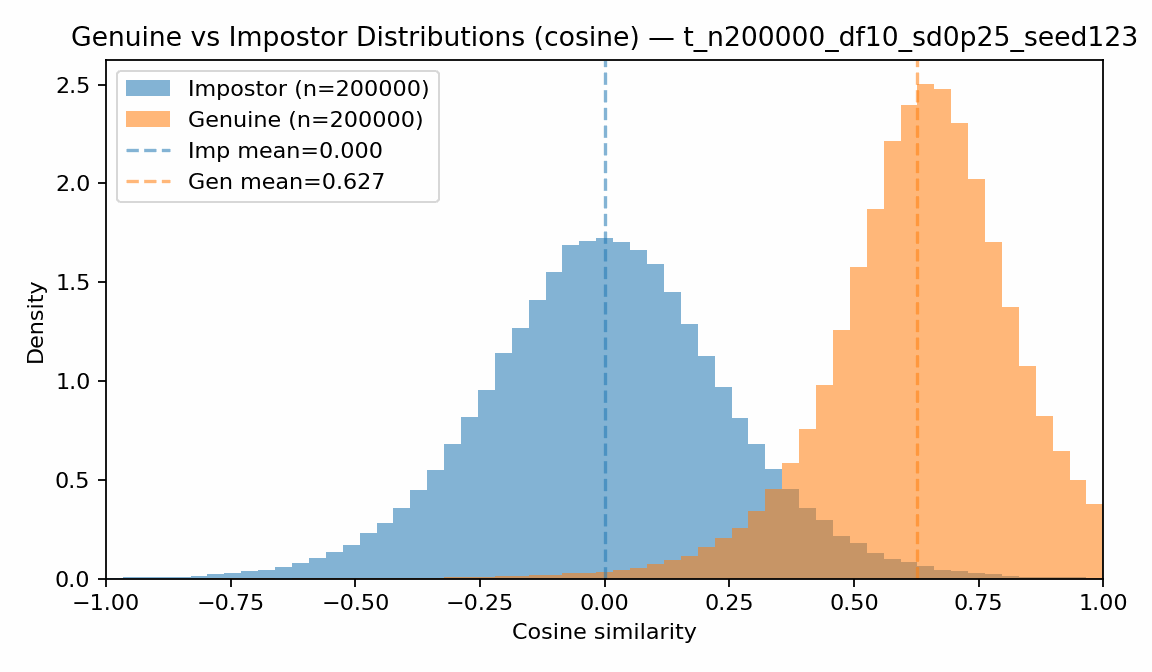}
    \caption{}\label{fig:impgen-synth-t}
  \end{subfigure}
  \vspace{-4pt}
  \caption{Synthetic impostor--genuine score distributions.
  Each panel overlays impostor (blue) and genuine (orange) histograms for $N=200{,}000$ samples per side, using independent random seeds.
  (a) Impostor: Gaussian with $\sigma=0.20$; Genuine: right-shifted Gaussian (default $\mu\approx0.65$, $\sigma\approx0.20$).
  (b) Impostor: symmetric Beta with $a=5$ mapped via $x=2z-1$; Genuine: skewed Beta (default $a=9$, $b=2$) mapped to concentrate mass near $+1$.
  (c) Impostor: mixture $(1-p)\,\mathcal{N}(0,\sigma_{1}) + p\,\mathcal{N}(\mu_{2},\sigma_{2})$ with $p=0.10$, $\mu_{2}=0.30$, $\sigma_{1}=0.25$, $\sigma_{2}=0.20$; Genuine: right-shifted mixture with a dominant high-mean component and a small low-mean tail. 
  (d) Impostor: Student-$t$ with $\nu=10$ and pre-truncation standard deviation $0.25$; Genuine: shifted Student-$t$. 
  The generator reports the rejection rate
  $r_{\mathrm{rej}} \coloneqq 1 - \frac{n}{N_{\mathrm{prop}}}$ for each file. In all displayed settings $r_{\mathrm{rej}}\ll 1$, indicating negligible boundary mass at $\pm 1$.}
  \label{fig:impgen-synth}
\end{figure}

Representative synthetic imp--gen histograms are shown in Figure~\ref{fig:impgen-synth}.

\paragraph{Operating points and endpoint semantics.}
A pre-privacy target $\mathsf{FMR}=\alpha\in\{10^{-2},10^{-3}\}$ 
determines the clean operating threshold $\tau_\alpha$ (Definition~\ref{def:clean-fmr}). We then apply \textsc{ScoreShield} with Gaussian scale $\sigma$, calibrated from $(\varepsilon,\delta)$ and the chosen sensitivity $\Delta$, and evaluate either strict endpoint semantics (`$>$') or non-strict endpoint semantics (`$\ge$') at $\tau=1$.

\paragraph{Privacy grid and reproducibility.}
We use $\delta\in\{10^{-8},10^{-6},10^{-5}\}$, $\varepsilon\in\{0.5,1,2,3, 5,8,10,15,20,30,40\}$.
All experiments use fixed random seeds, single-threaded numerical routines, and non-interactive rendering.

\appsubsection{Performance Analysis}

\begin{figure}[!t]
  \centering
  \begin{subfigure}[t]{0.32\linewidth}
    \includegraphics[width=\linewidth]{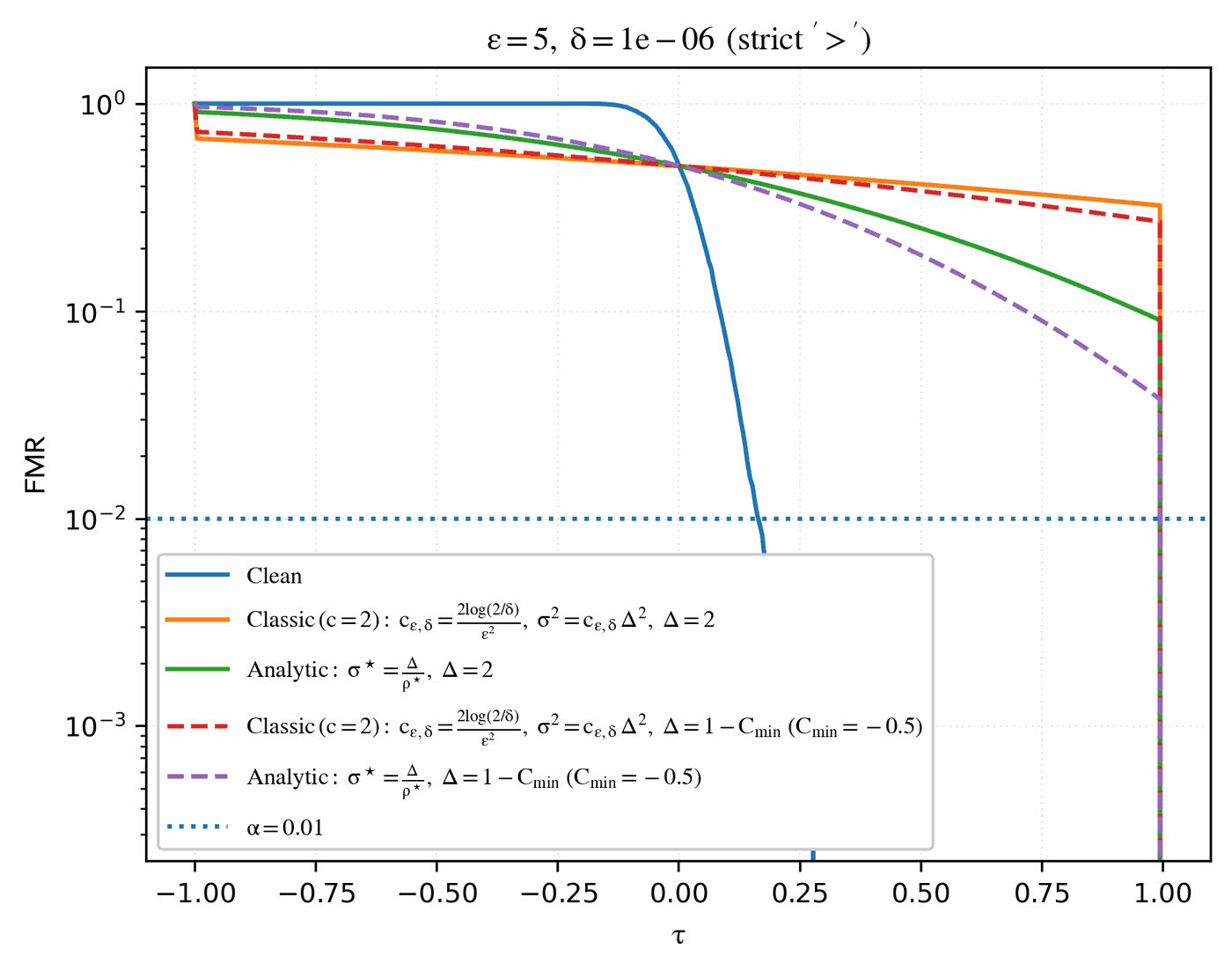}%
    \caption{\textbf{Real} (LFW, ArcFace-101, WebFace4M), $\varepsilon=5$}
    \label{fig:overlay_real_e5}
    \end{subfigure}
  \hfill
  \begin{subfigure}[t]{0.32\linewidth}
    \includegraphics[width=\linewidth]{appx_results/case1/fmr/fmr_overlay_alpha0.01_e10.0_d1e-06_strict_Delta2.0_1.5_real_lfw_arc101_webface4m.png}%
    \caption{\textbf{Real}, $\varepsilon=10$}
    \label{fig:overlay_real_e10}
    \end{subfigure}
  \hfill
  \begin{subfigure}[t]{0.32\linewidth}
    \includegraphics[width=\linewidth]{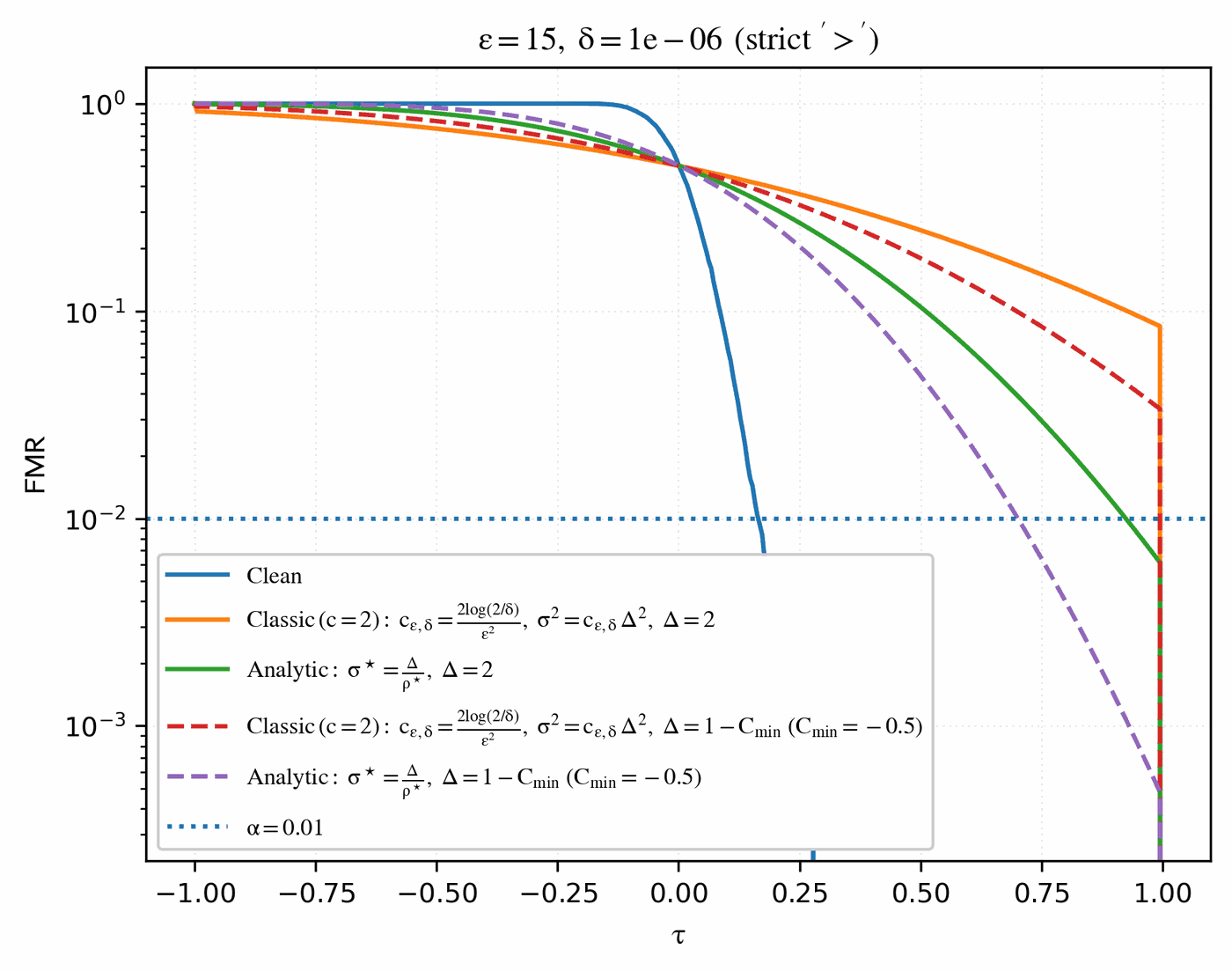}%
    \caption{\textbf{Real}, $\varepsilon=15$}
    \label{fig:overlay_real_e15}
    \end{subfigure}

  \begin{subfigure}[t]{0.32\linewidth}
    \includegraphics[width=\linewidth]{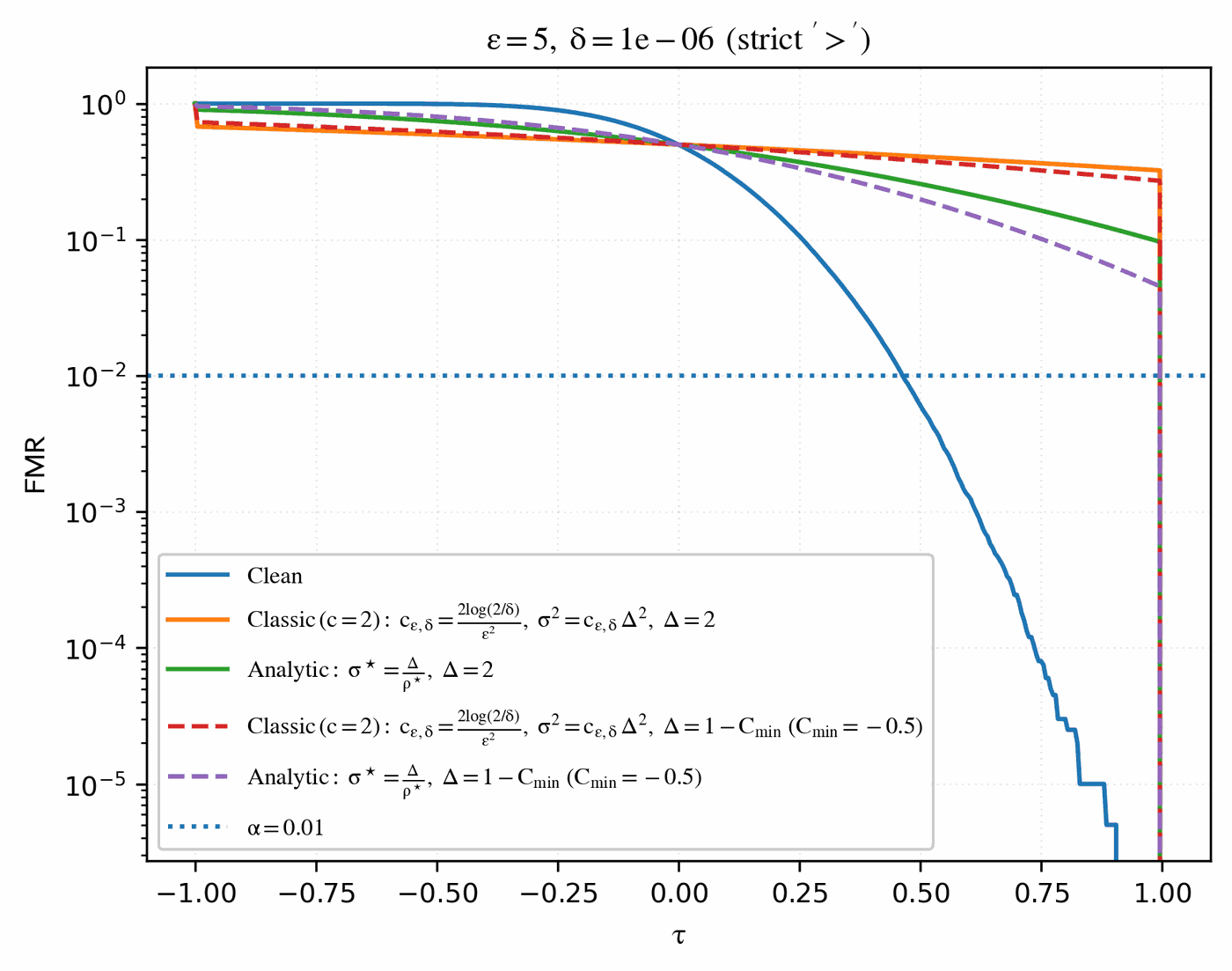}%
    \caption{\textbf{Synthetic} Gaussian, $\varepsilon=5$}
    \label{fig:overlay_gauss_e5}
    \end{subfigure}
  \hfill
  \begin{subfigure}[t]{0.32\linewidth}
    \includegraphics[width=\linewidth]{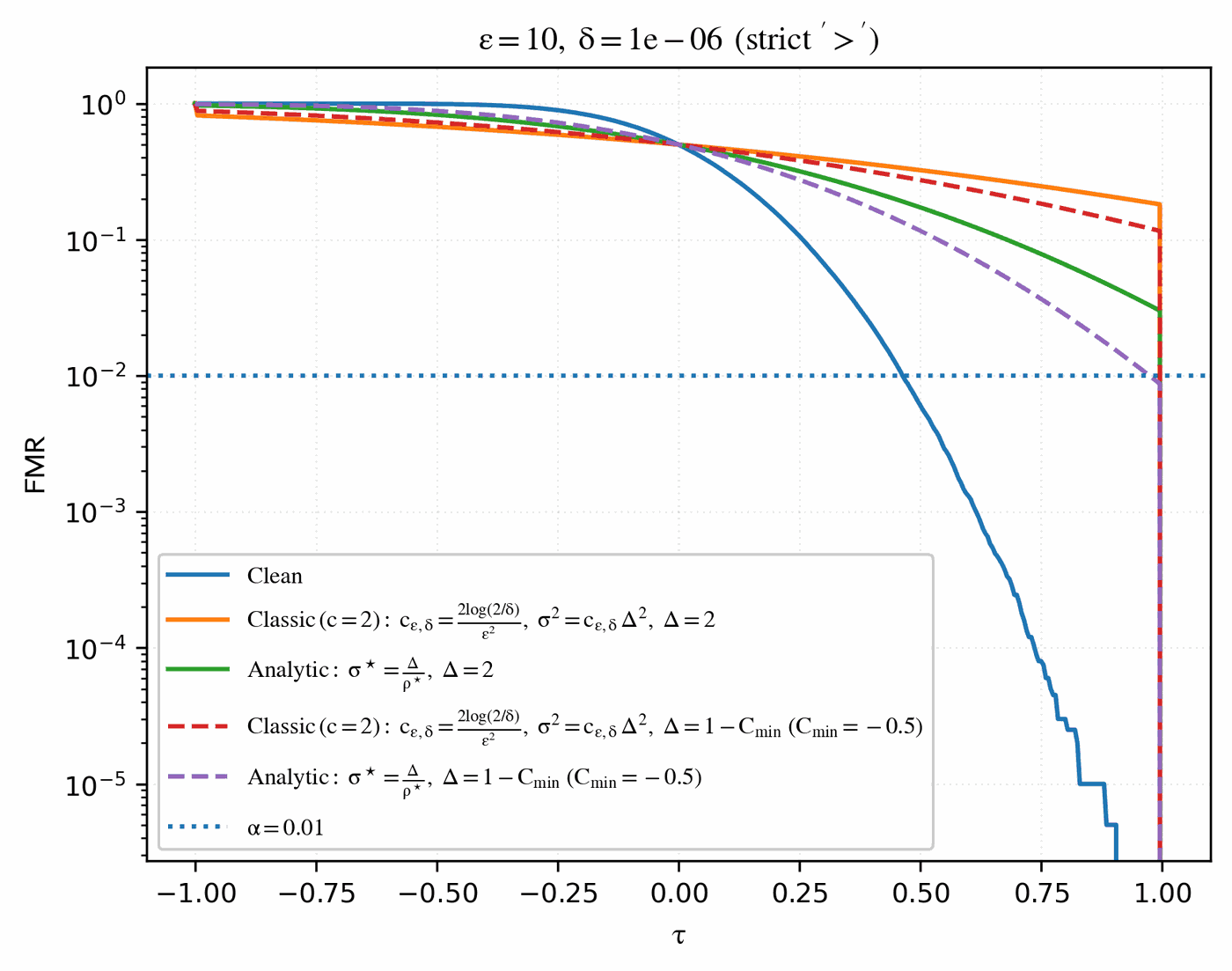}%
    \caption{\textbf{Synthetic} Gaussian, $\varepsilon=10$}
    \label{fig:overlay_gauss_e10}
  \end{subfigure}
  \hfill
  \begin{subfigure}[t]{0.32\linewidth}
    \includegraphics[width=\linewidth]{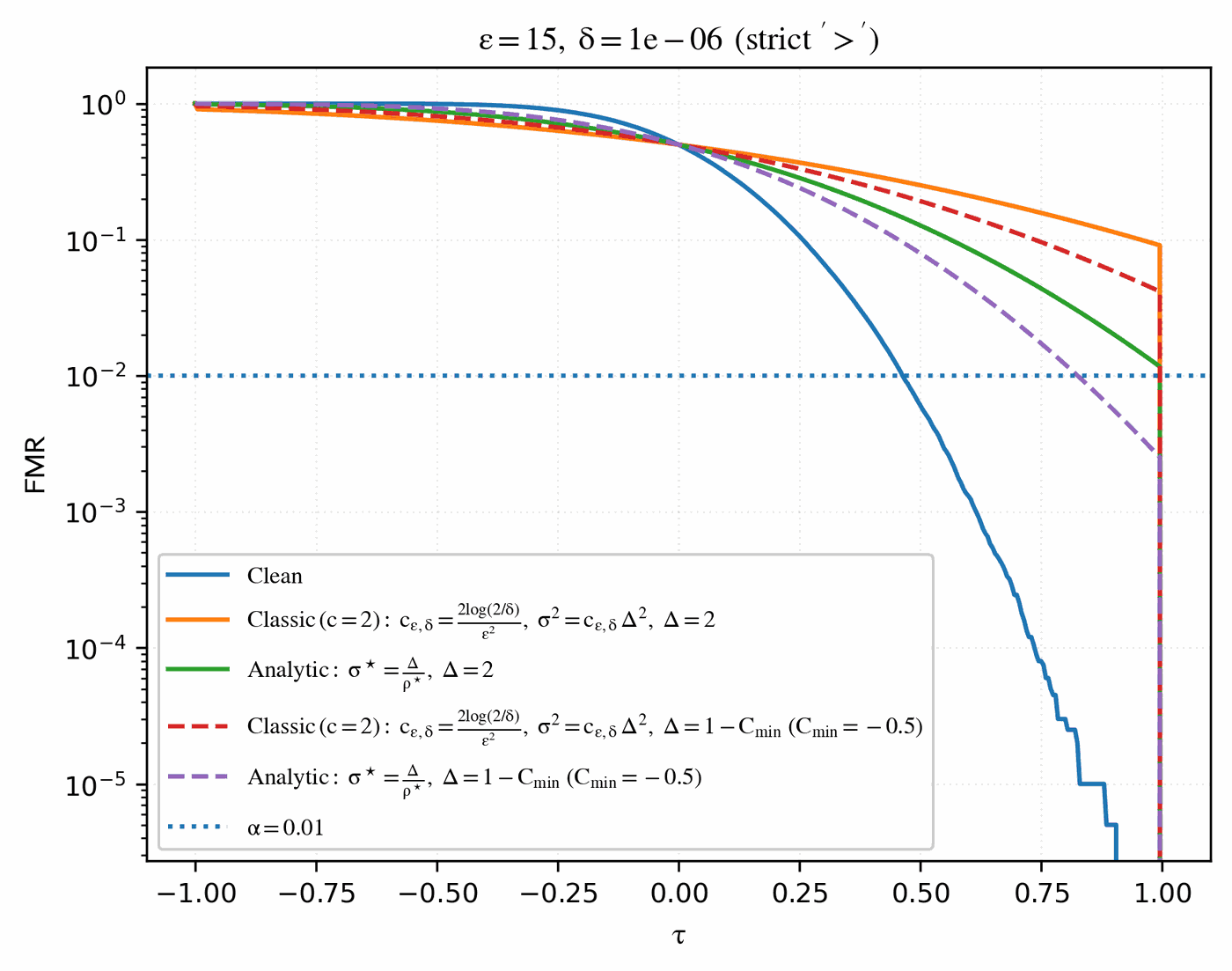}%
    \caption{\textbf{Synthetic} Gaussian, $\varepsilon=15$}
    \label{fig:overlay_gauss_e15}
  \end{subfigure}
  \begin{subfigure}[t]{0.32\linewidth}
    \includegraphics[width=\linewidth]{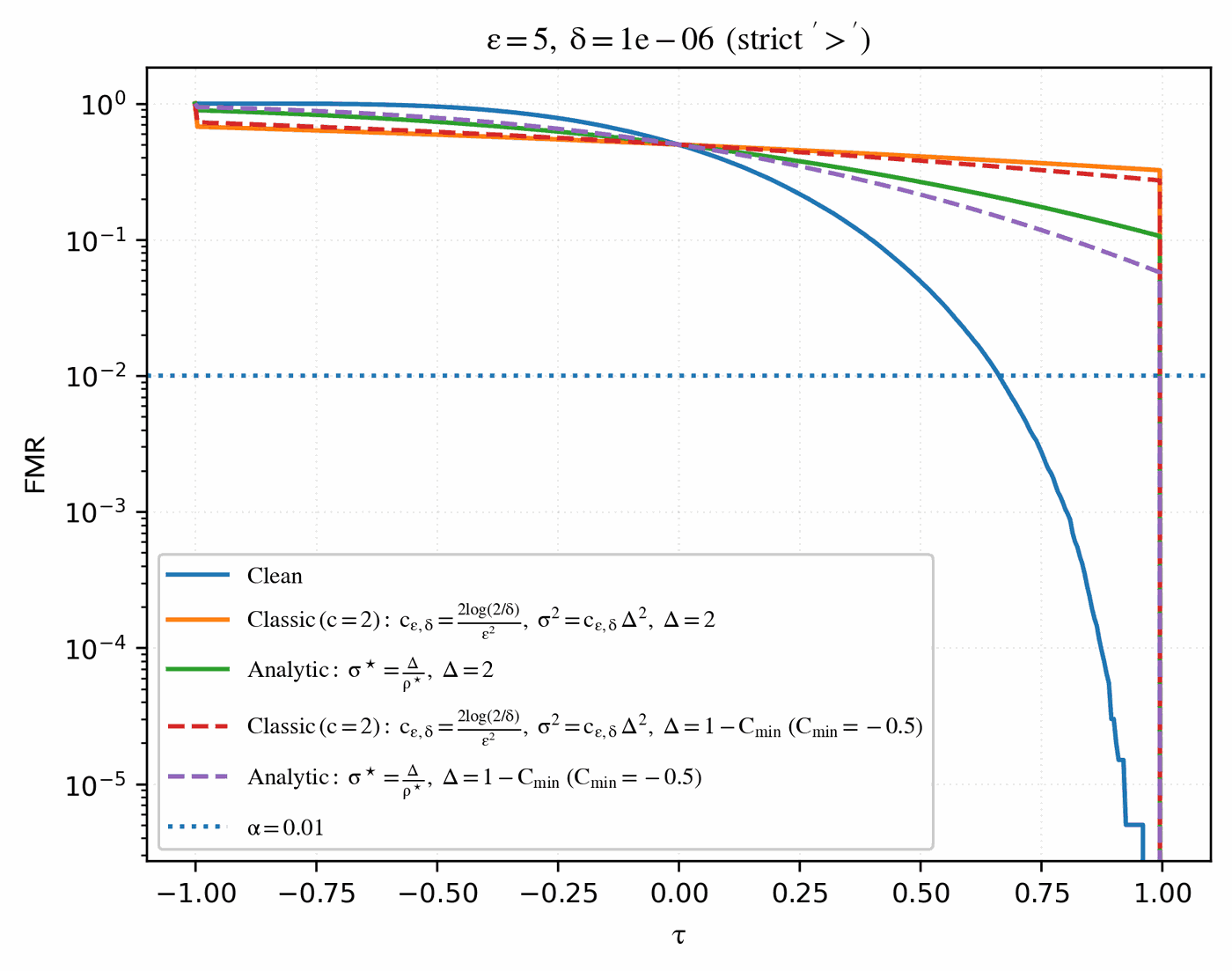}%
    \caption{\textbf{Synthetic} Beta, $\varepsilon=5$}
    \label{fig:overlay_beta_e5}
    \end{subfigure}
  \hfill
  \begin{subfigure}[t]{0.32\linewidth}
    \includegraphics[width=\linewidth]{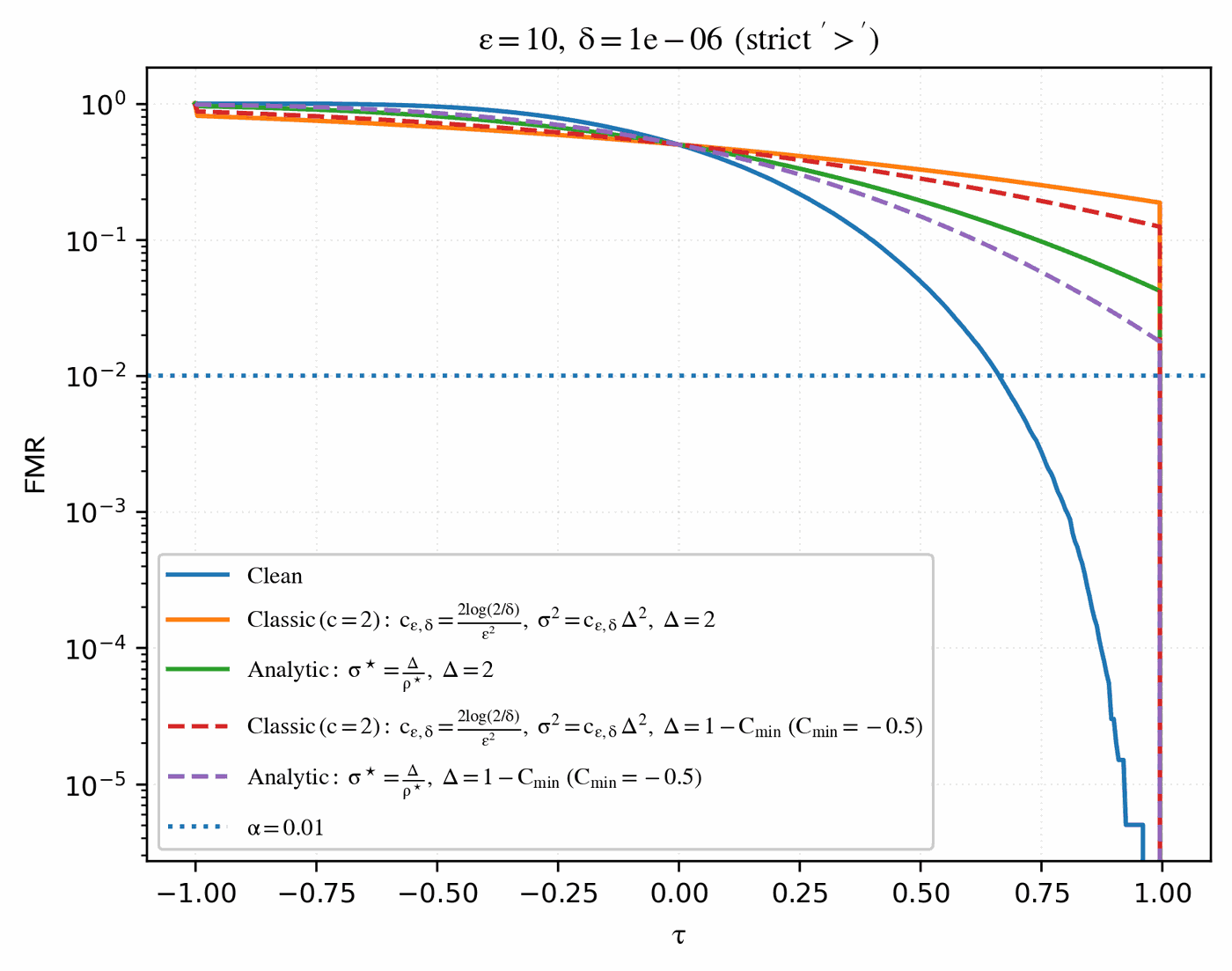}%
    \caption{\textbf{Synthetic} Beta, $\varepsilon=10$}
    \label{fig:overlay_beta_e10}
    \end{subfigure}
  \hfill
  \begin{subfigure}[t]{0.32\linewidth}
    \includegraphics[width=\linewidth]{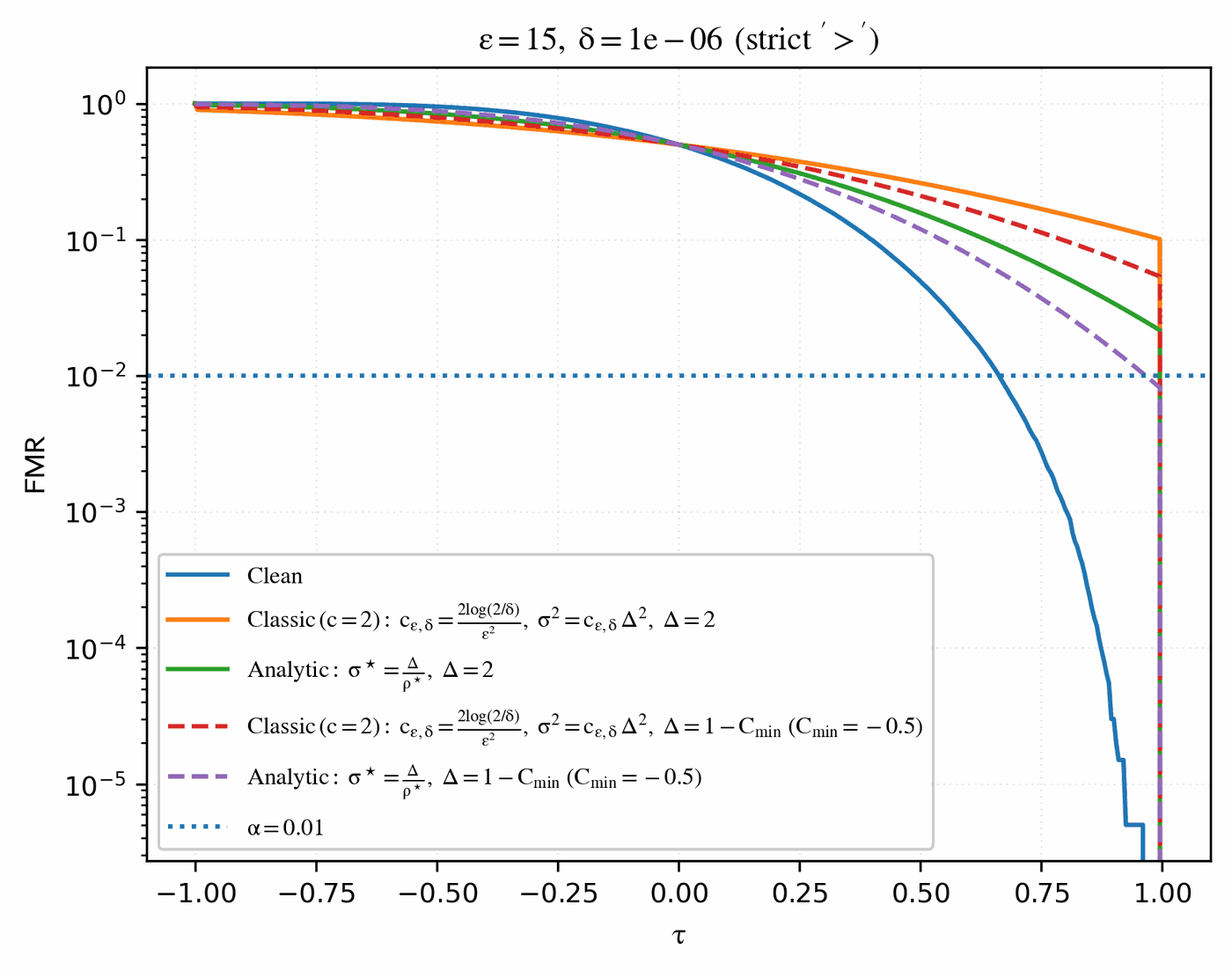}%
    \caption{\textbf{Synthetic} Beta, $\varepsilon=15$}
    \label{fig:overlay_beta_e15}
  \end{subfigure}

  \begin{subfigure}[t]{0.32\linewidth}
    \includegraphics[width=\linewidth]{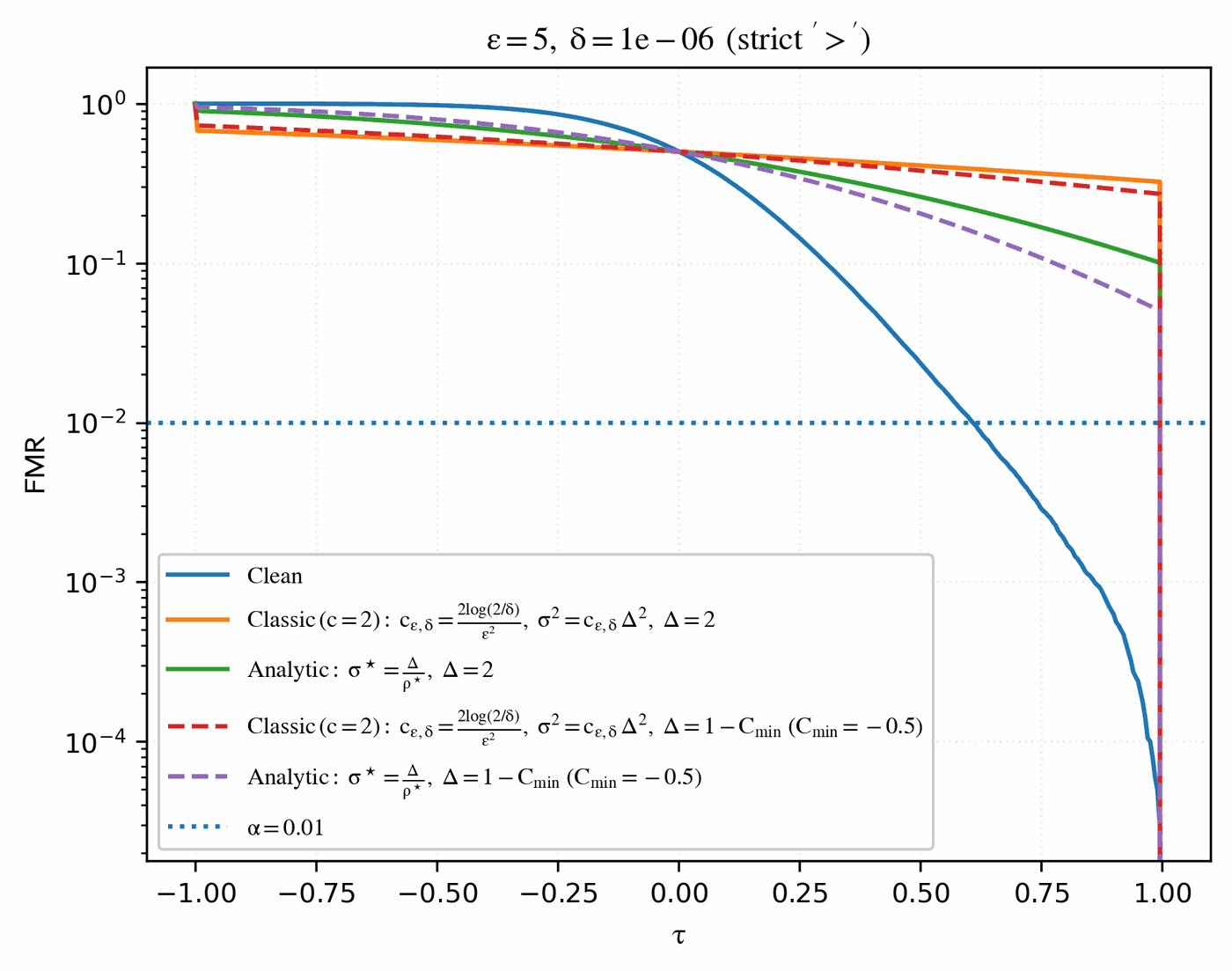}%
    \caption{\textbf{Synthetic} Student-$t$, $\varepsilon=5$}
    \label{fig:overlay_t_e5}
    \end{subfigure}
  \hfill
  \begin{subfigure}[t]{0.32\linewidth}
    \includegraphics[width=\linewidth]{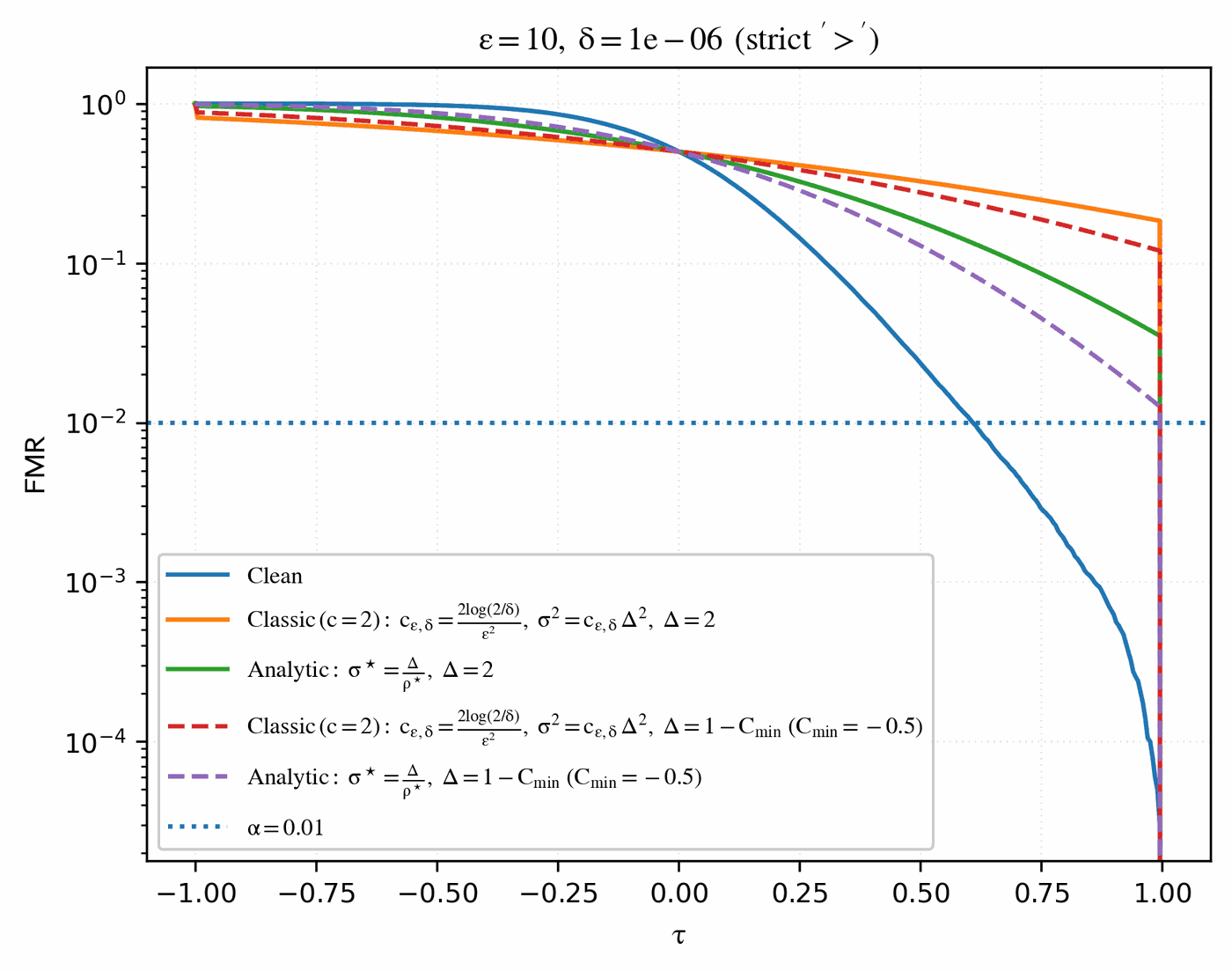}%
    \caption{\textbf{Synthetic} Student-$t$, $\varepsilon=10$}
    \label{fig:overlay_t_e10}
    \end{subfigure}
  \hfill
  \begin{subfigure}[t]{0.32\linewidth}
    \includegraphics[width=\linewidth]{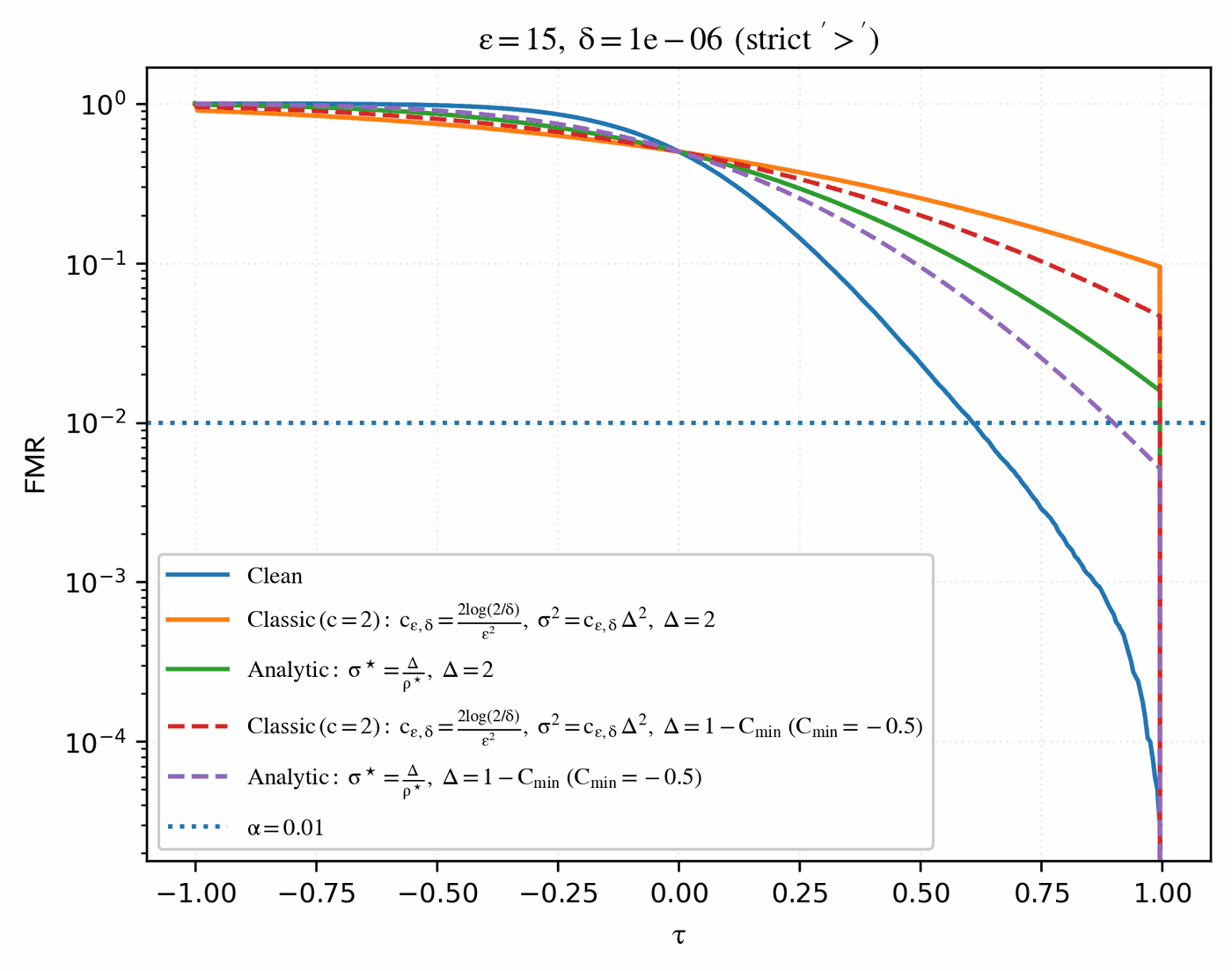}%
    \caption{\textbf{Synthetic} Student-$t$, $\varepsilon=15$}
    \label{fig:overlay_t_e15}
 \end{subfigure}
  \vspace{-4pt}
  \caption{\textbf{FMR vs threshold under different calibrations.}
  Each panel plots $\mathsf{FMR}(\tau)$ for fixed $\delta=10^{-6}$ and target level $\alpha=10^{-2}$ under the strict endpoint rule.
  Solid curves use the worst-case sensitivity $\Delta=2$; dashed curves use the domain-restricted sensitivity $\Delta=1-C_{\min}$ with $C_{\min}=-0.5$, hence $\Delta=1.5$;
  \textit{Classic} (orange/red) uses the sufficient Gaussian calibration $\sigma^2=c_{\varepsilon,\delta}\Delta^2$ with $c_{\varepsilon,\delta}=2\log(2/\delta)/\varepsilon^2$; 
  \textit{Analytic} (green/purple) uses the exact analytic Gaussian calibration ($\sigma^\star=\Delta/\rho^\star$).
  Columns sweep $\varepsilon\in\{5,10,15\}$; rows vary the underlying score distribution (one real dataset and three synthetic families).
  In the threshold range relevant to $\alpha=10^{-2}$, analytic calibration yields lower post-privacy FMR than the conservative sufficient bound, and the reduced-sensitivity model further improves the right-tail behavior.
  As $\varepsilon$ increases, all private curves move closer to the FMR curve.}
  \label{fig:fmr_overlays_eps_sweep_strict}
\end{figure}

\paragraph{Discussion of Figure~\ref{fig:fmr_overlays_eps_sweep_strict}.}
Figure~\ref{fig:fmr_overlays_eps_sweep_strict} compares $\mathsf{FMR}(\tau)$ under several privacy calibrations at fixed $\delta=10^{-6}$ and target level $\alpha=10^{-2}$, using strict endpoint semantics. For fixed $(\varepsilon,\delta,\Delta)$, the analytic calibration chooses the smallest Gaussian noise scale $\sigma$ that satisfies differential privacy, whereas the conservative sufficient bound uses a larger $\sigma$. Consequently, in the threshold range relevant to small-$\alpha$ operation, the analytic curves lie below the corresponding conservative curves. Replacing the worst-case sensitivity $\Delta=2$ by the domain-restricted value $\Delta=1-c_{\min}=1.5$ with $c_{\min}=-0.5$ reduces the required noise for both calibration rules and further improves the right-tail behavior.

As $\varepsilon$ increases from $5$ to $10$ to $15$, the noise scale decreases and the private curves approach the clean curve. The qualitative ordering remains the same across all displayed score families. Under strict endpoint semantics, feasibility is governed by the left limit $L(\sigma)\coloneqq \lim_{\tau\uparrow 1}\mathsf{FMR}_{\mathrm{priv}}(\tau)$. If $L(\sigma)\ge \alpha$, then no threshold $\tau<1$ attains the target level $\alpha$. Thus, for fixed $\delta$ and $\alpha$, feasibility is lost first under conservative calibration with worst-case sensitivity and retained longest under analytic calibration with reduced sensitivity. This effect is especially visible for the Student-$t$ impostor family, whose right tail is harder to suppress after Gaussian perturbation.

For a fixed privacy budget $(\varepsilon,\delta)$, analytic calibration and, when justified, domain-restricted sensitivity both improve the attainable operating region relative to the conservative worst-case baseline.

\begin{figure}[!t]
  \centering
  \begin{subfigure}[t]{0.48\linewidth}
    \includegraphics[width=\linewidth]{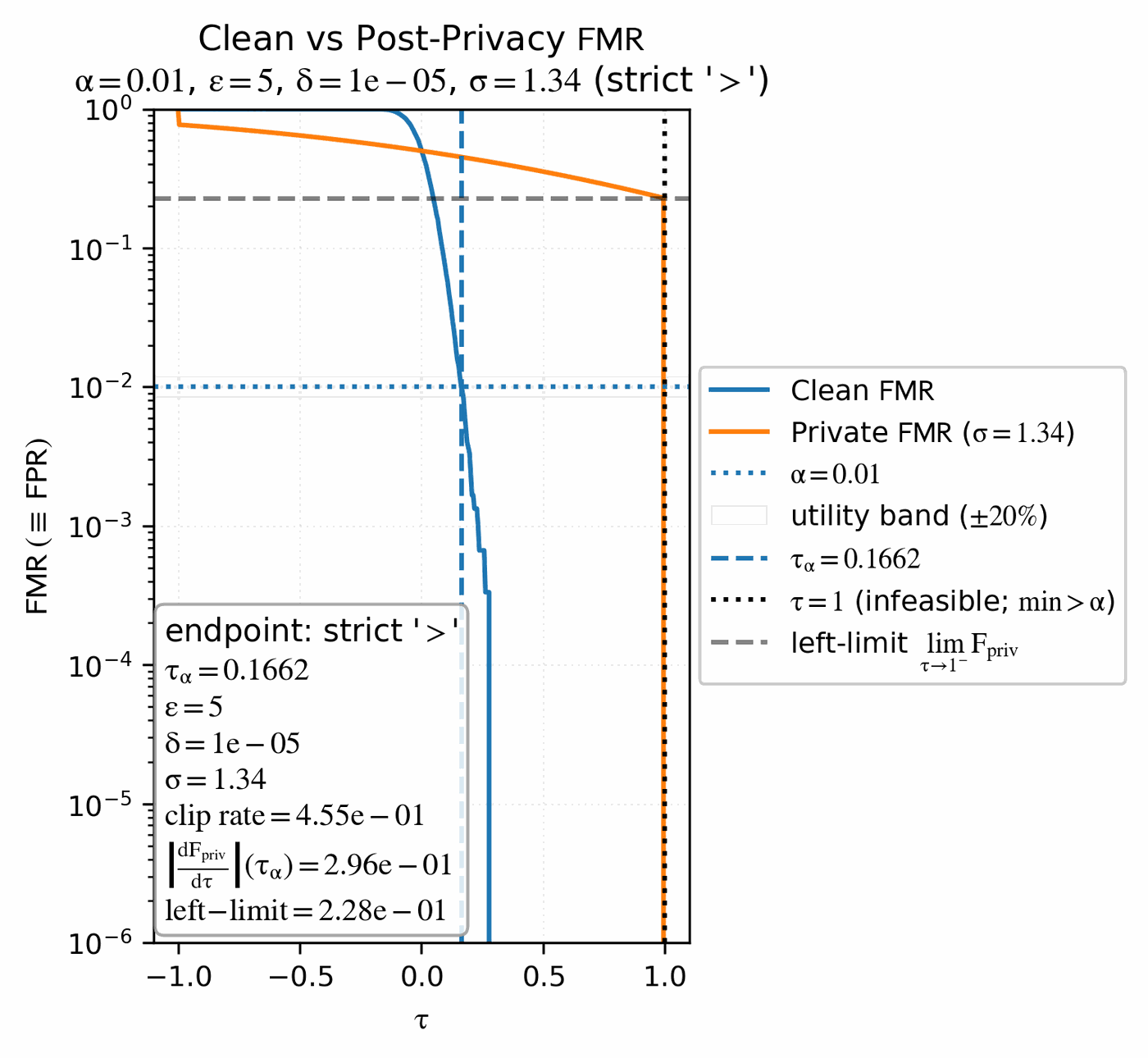}%
    \caption{$\varepsilon\!=\!5$}
    \label{fig:fmr-lfw-arc101-webface4m-e5}
    \end{subfigure}
  \hfill
  \begin{subfigure}[t]{0.48\linewidth}
    \includegraphics[width=\linewidth]{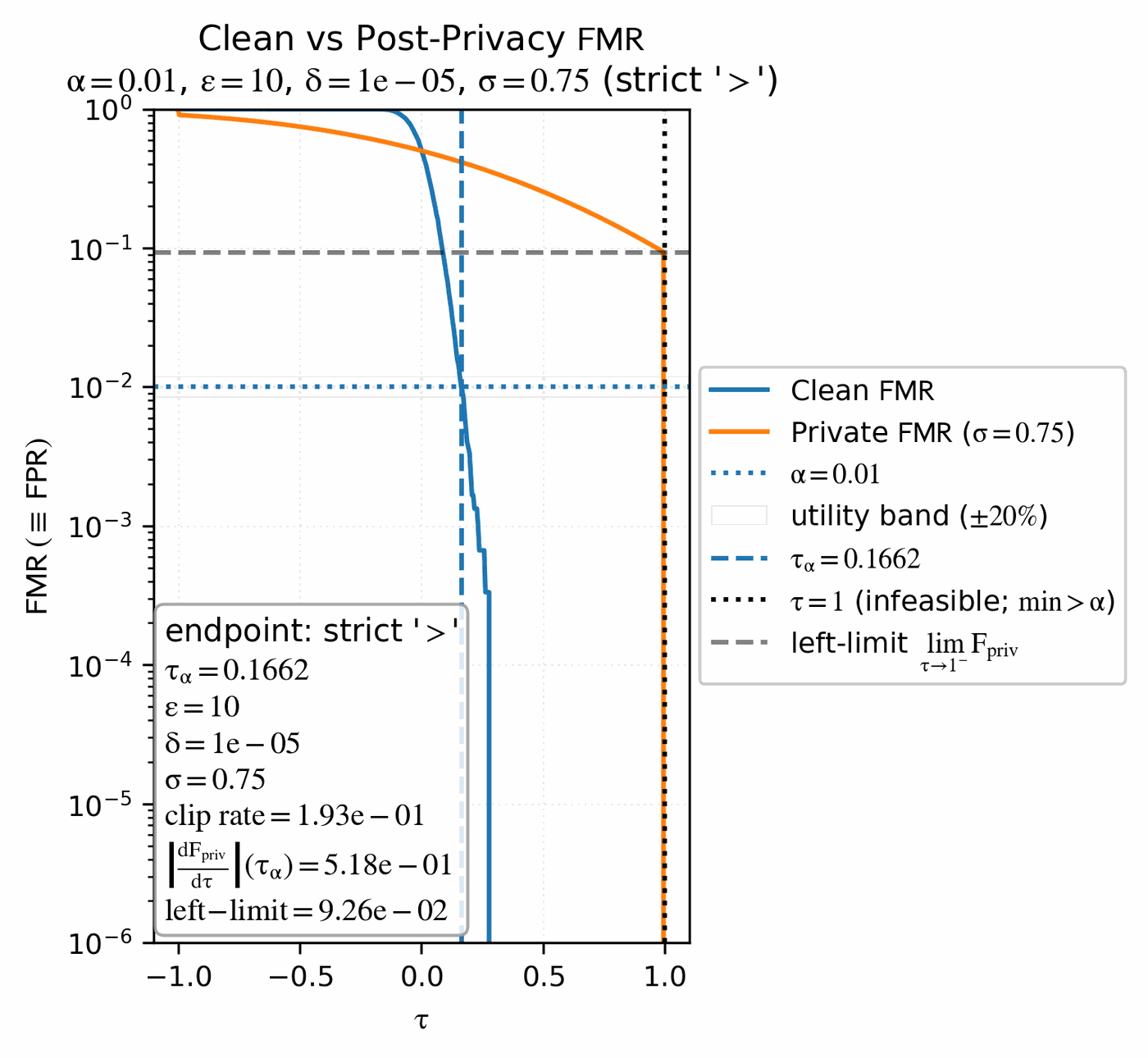}%
    \caption{$\varepsilon\!=\!10$}
    \label{fig:fmr-lfw-arc101-webface4m-e10}
    \end{subfigure}
 
  \begin{subfigure}[t]{0.48\linewidth}
    \includegraphics[width=\linewidth]{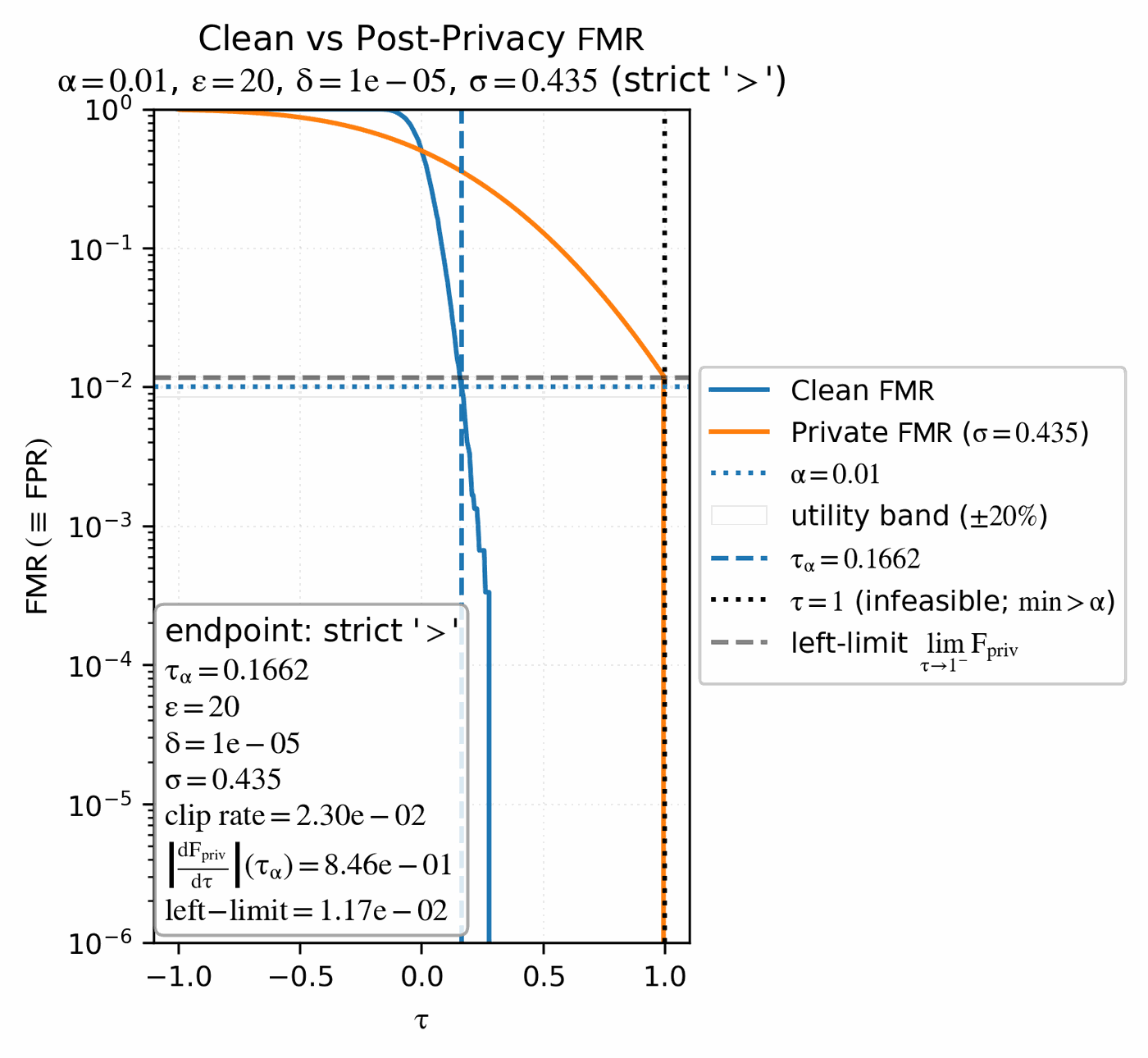}%
    \caption{$\varepsilon\!=\!20$}
    \label{fig:fmr-lfw-arc101-webface4m-e20}
    \end{subfigure}
  \hfill
  \begin{subfigure}[t]{0.48\linewidth}
    \includegraphics[width=\linewidth]{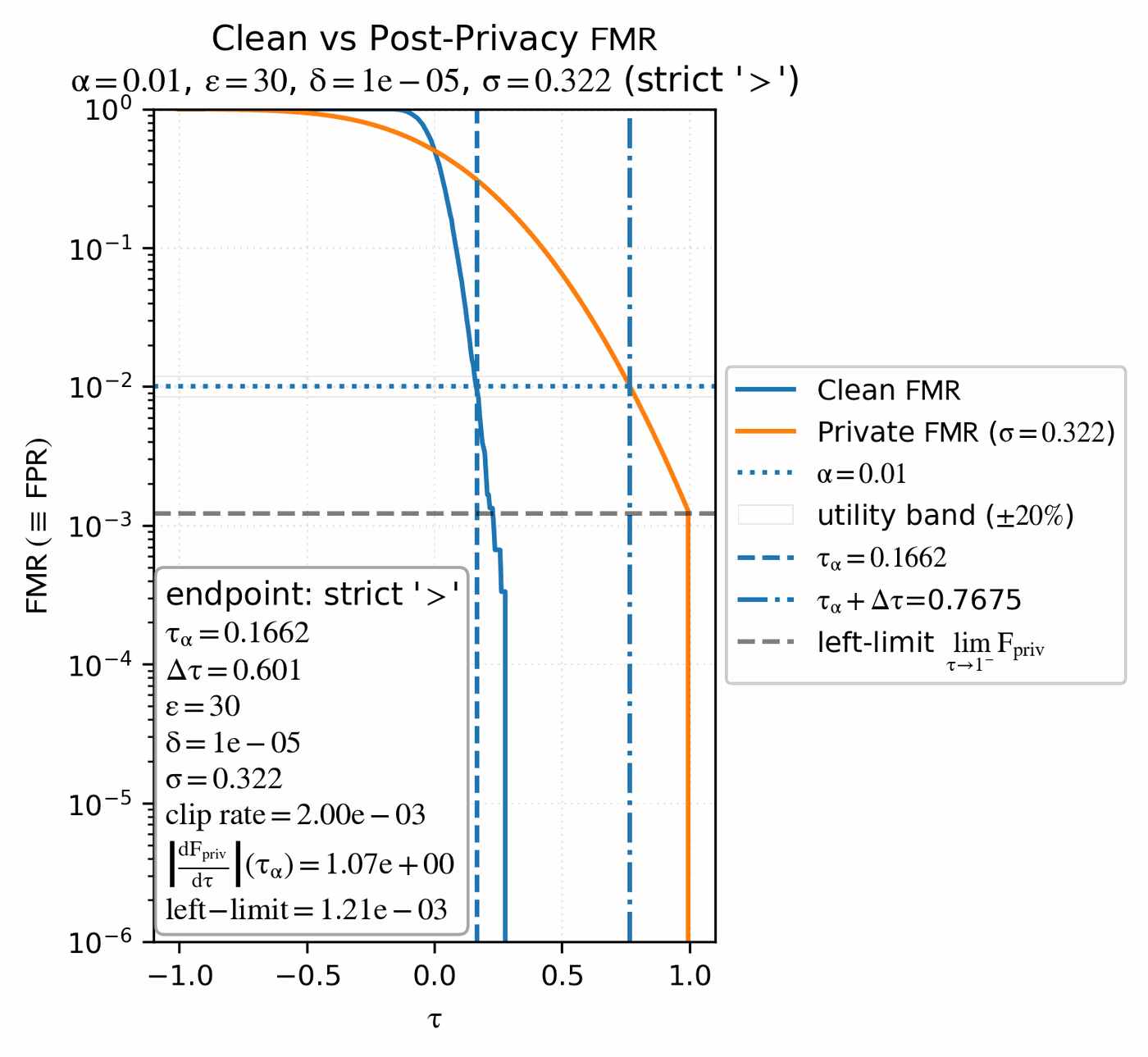}%
    \caption{$\varepsilon\!=\!30$}
    \label{fig:fmr-lfw-arc101-webface4m-e30}
    \end{subfigure}
  \vspace{-4pt}
  \caption{\textbf{LFW (ArcFace-101 trained on WebFace4M): clean vs. post-privacy FMR under analytic calibration with domain-restricted sensitivity.}
  Each panel shows the clean and private $\mathsf{FMR}$ curves (log-scale) for a fixed $\delta=10^{-5}$ and varying $\varepsilon\in\{5,10,20,30\}$, using the analytic Gaussian calibration with $\Delta = 1 - C_{\min}$, $C_{\min}=-0.5$, hence $\Delta=1.5$. The dotted horizontal line marks the target level $\alpha$. The dashed vertical line marks the clean operating threshold $\tau_\alpha$. The dash-dotted line marks the re-calibrated threshold $\tau_\alpha+\Delta\tau$ an interior solution exists. The shaded band indicates a $\pm 20\%$ tolerance region around $\alpha$. Endpoint semantics are strict (`$>$') at $\tau=1$.}
  \label{fig:fmr-lfw-arc101-webface4m-analytic-pr}
\end{figure}

\begin{figure}[!t]
  \centering
  \begin{subfigure}[t]{0.48\linewidth}
    \includegraphics[width=\linewidth]{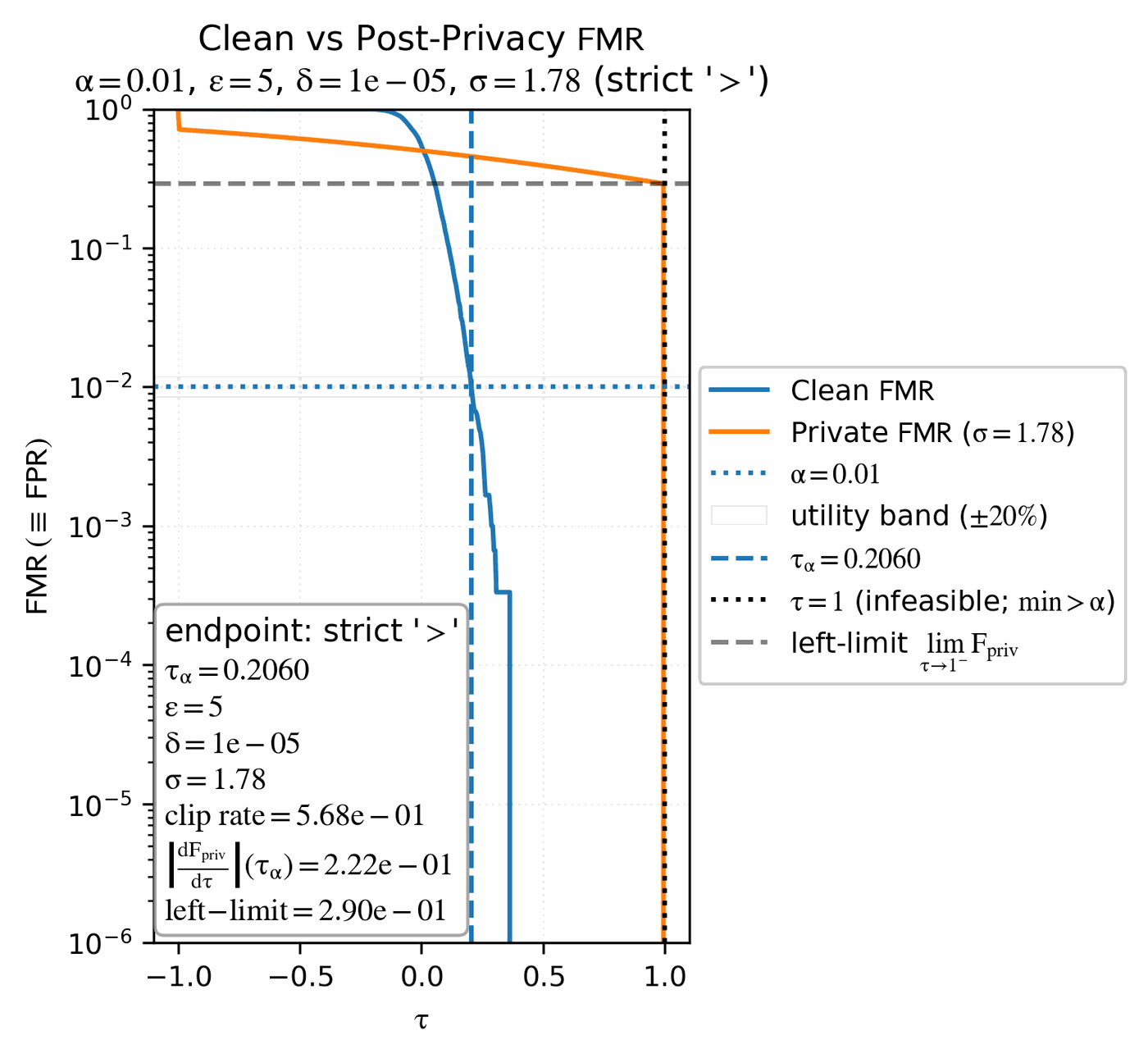}%
    \caption{$\varepsilon\!=\!5$}
    \label{fig:fmr-lfw-arc101-casia-e5}
    \end{subfigure}
  \hfill
  \begin{subfigure}[t]{0.48\linewidth}
    \includegraphics[width=\linewidth]{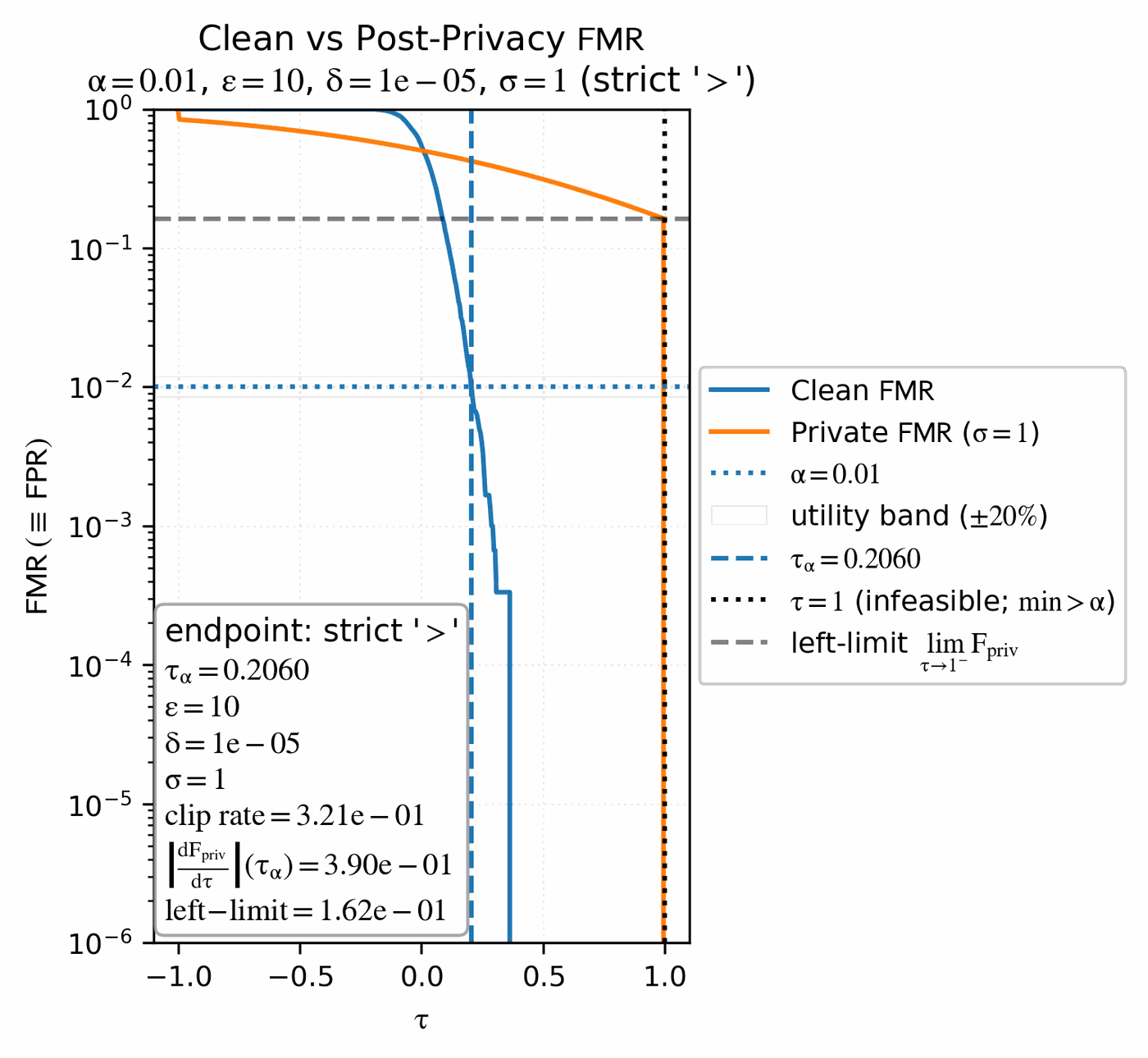}%
    \caption{$\varepsilon\!=\!10$}
    \label{fig:fmr-lfw-arc101-casia-e10}
    \end{subfigure}

  \begin{subfigure}[t]{0.48\linewidth}
    \includegraphics[width=\linewidth]{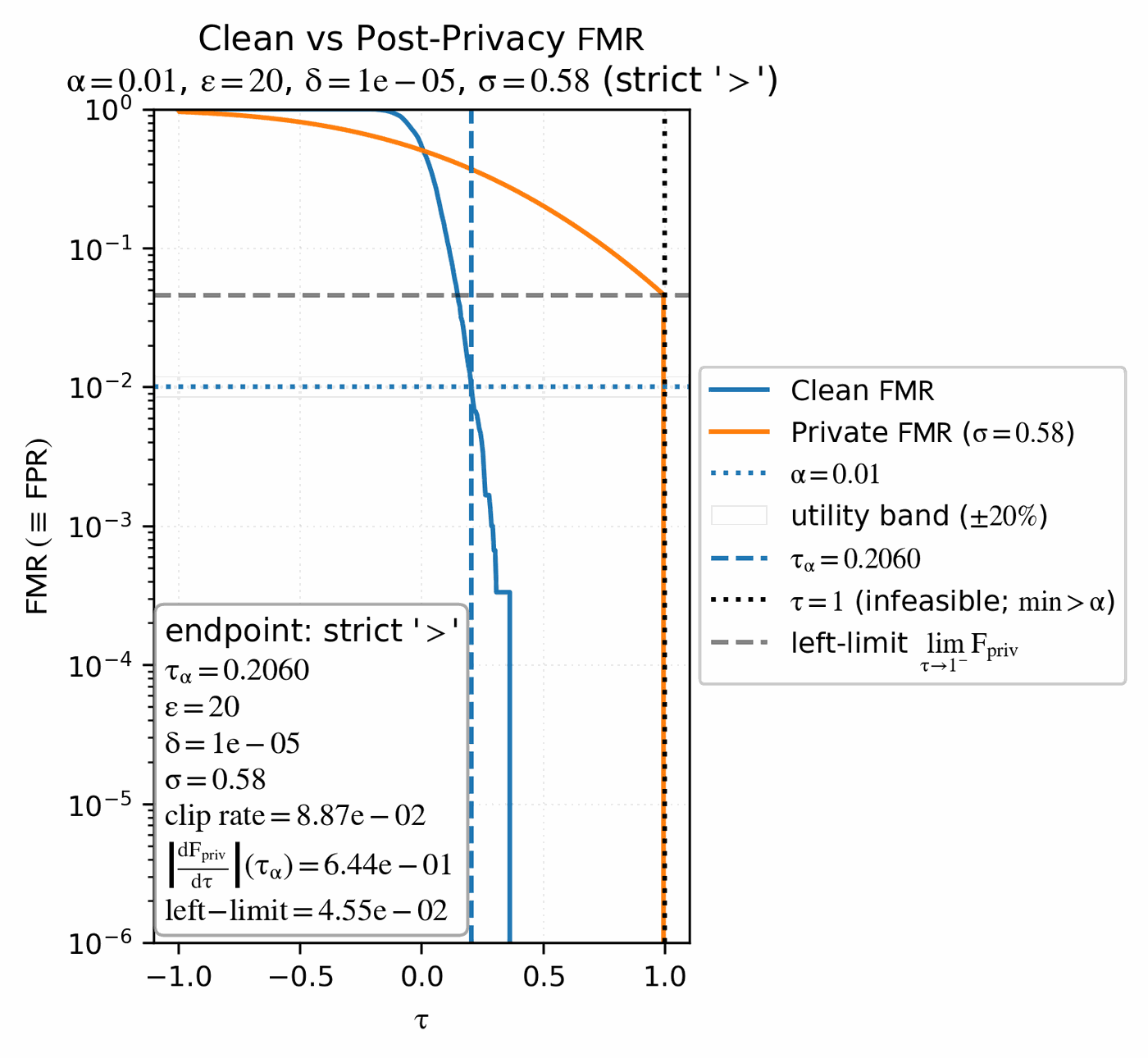}%
    \caption{$\varepsilon\!=\!20$}
    \label{fig:fmr-lfw-arc101-casia-e20}
  \end{subfigure}
  \hfill
  \begin{subfigure}[t]{0.48\linewidth}
    \includegraphics[width=\linewidth]{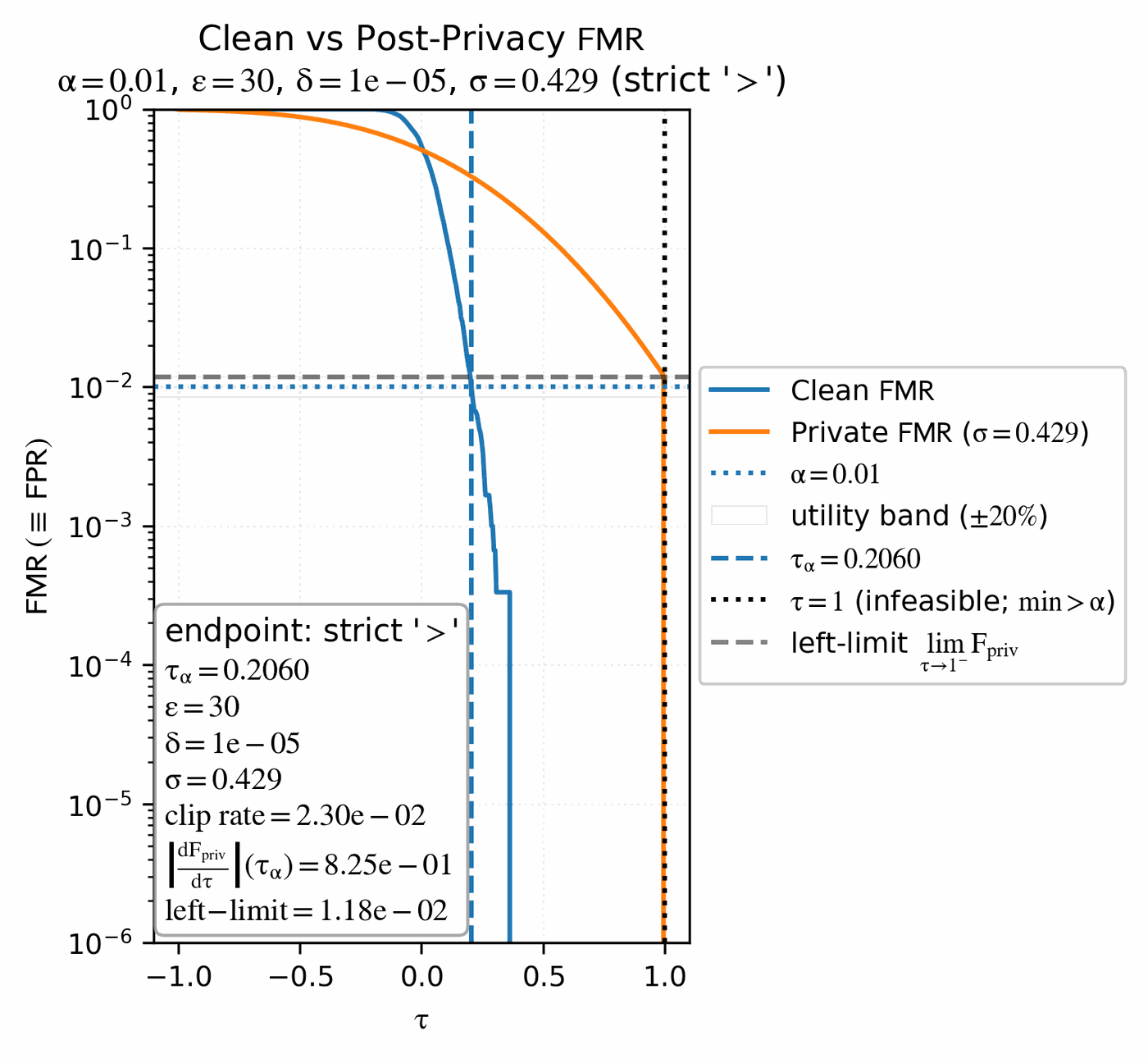}%
    \caption{$\varepsilon\!=\!30$}
    \label{fig:fmr-lfw-arc101-casia-e30}
  \end{subfigure}
  \vspace{-4pt}
  \caption{\textbf{LFW (ArcFace-50 trained on Casia): clean and post-privacy FMR under analytic calibration with worst-case sensitivity.}
  Same setup as Figure~\ref{fig:fmr-lfw-arc101-webface4m-analytic-pr}, except that the analytic calibration uses the worst-case sensitivity $\Delta=2$. We fix $\delta=10^{-5}$, vary $\varepsilon\in\{5,10,20,30\}$, and annotate the target $\alpha$ (dotted), the clean threshold $\tau_\alpha$ (dashed), and the re-calibrated threshold $\tau_\alpha+\Delta\tau$ (dash-dotted, when feasible). The shaded band indicates a $\pm20\%$ tolerance region around $\alpha$.
  Endpoint semantics are strict (`$>$') at $\tau=1$.}
  \label{fig:fmr-lfw-arc50-casia-analytic-wc}
\end{figure}

\begin{figure}[!t]
  \centering
  \begin{subfigure}[t]{0.48\linewidth}
    \includegraphics[width=\linewidth]{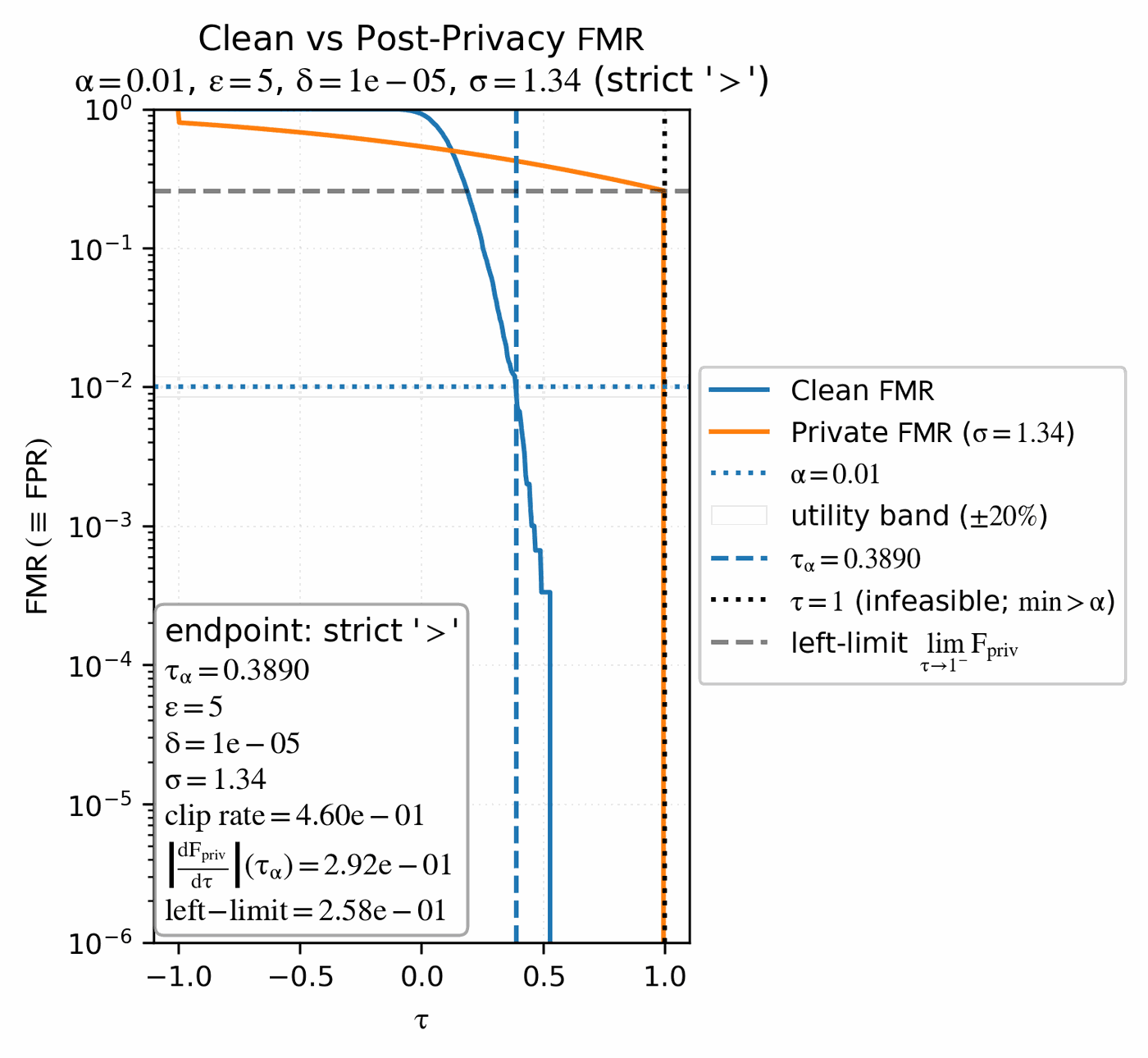}%
    \caption{$\varepsilon\!=\!5$}
    \label{fig:fmr-lfw-ada-e5}
  \end{subfigure}
  \hfill
  \begin{subfigure}[t]{0.48\linewidth}
    \includegraphics[width=\linewidth]{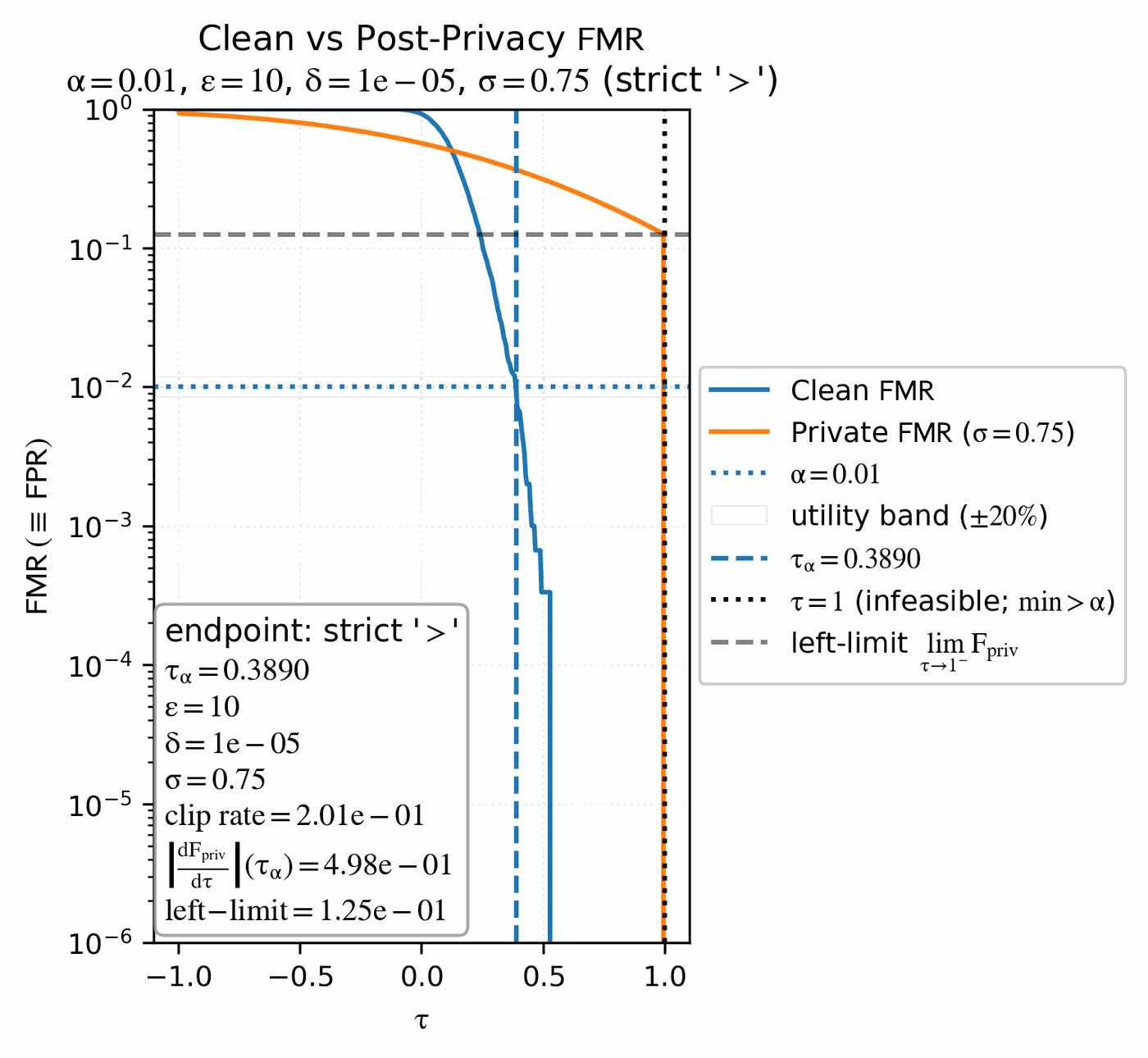}%
    \caption{$\varepsilon\!=\!10$}
    \label{fig:fmr-lfw-ada-e10}
  \end{subfigure}

  \begin{subfigure}[t]{0.48\linewidth}
    \includegraphics[width=\linewidth]{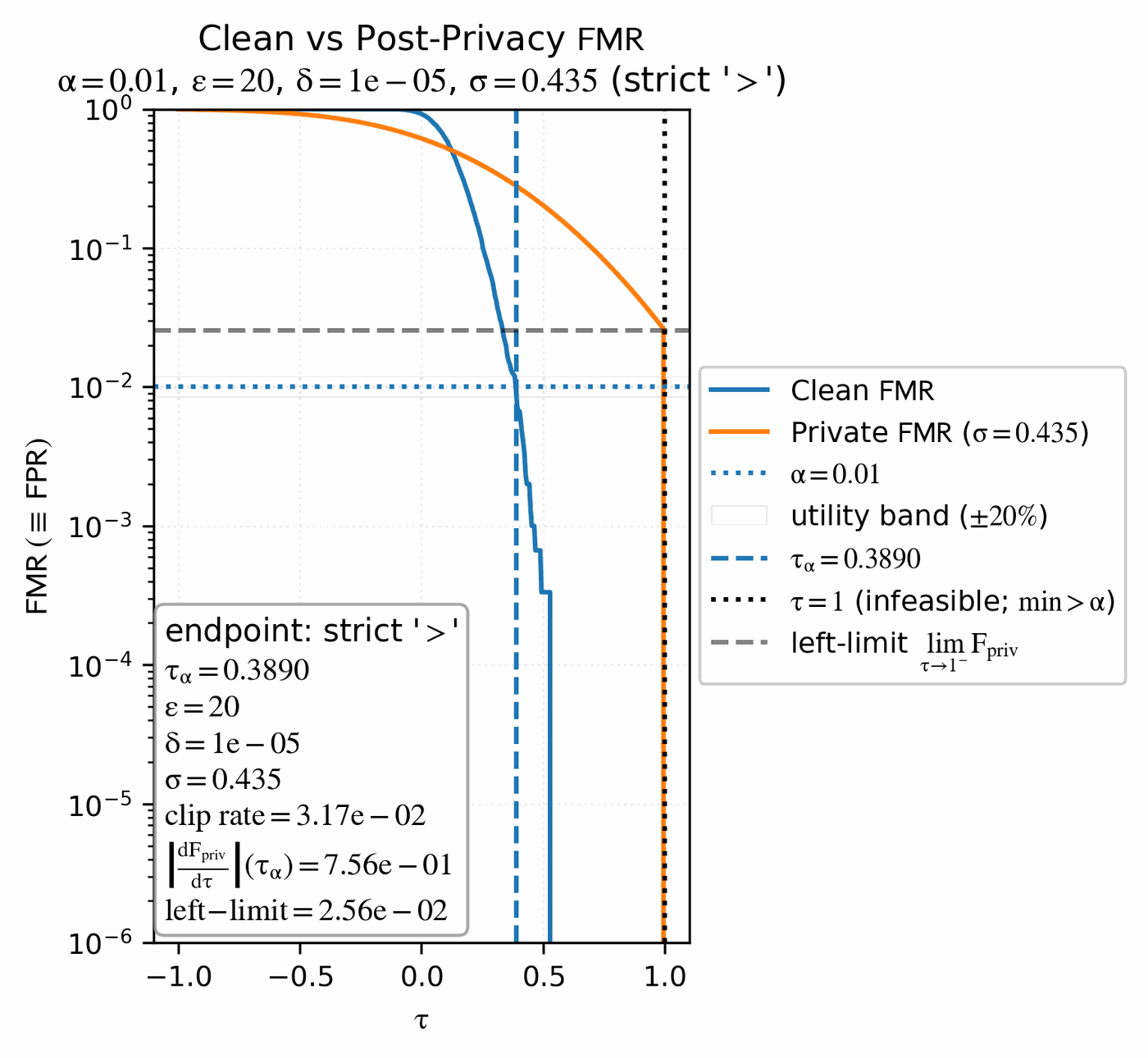}%
    \caption{$\varepsilon\!=\!20$}
    \label{fig:fmr-lfw-ada-e20}
    \end{subfigure}
  \hfill
  \begin{subfigure}[t]{0.48\linewidth}
    \includegraphics[width=\linewidth]{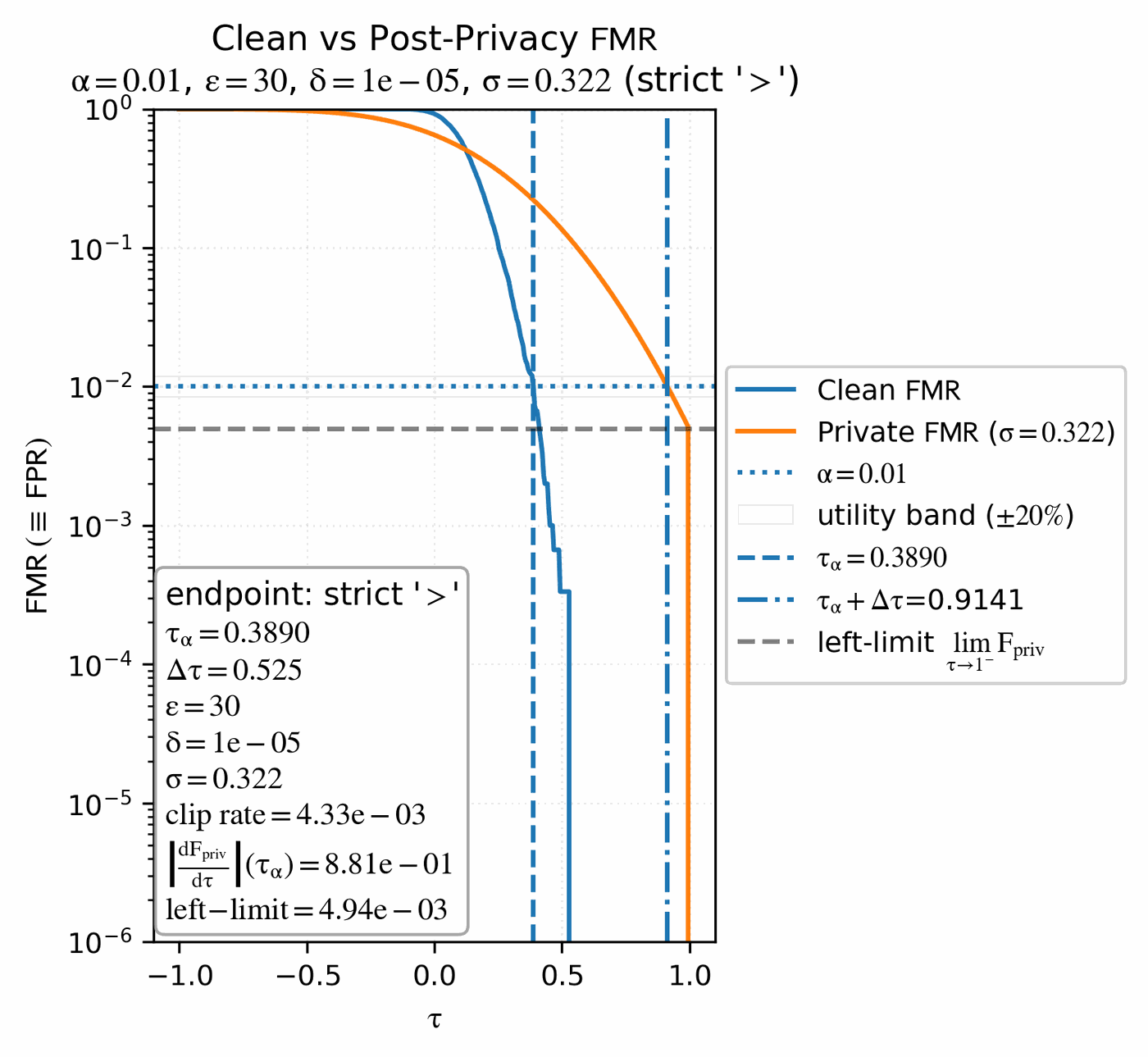}%
    \caption{$\varepsilon\!=\!30$}
    \label{fig:fmr-lfw-ada-e30}
  \end{subfigure}
  \vspace{-4pt}
  \caption{\textbf{LFW (AdaFace + RDigi1M/CodeFormer): clean and post-privacy FMR under analytic calibration with domain-restricted sensitivity.}
  Same setup as Figure~\ref{fig:fmr-lfw-arc101-webface4m-analytic-pr}, but using the AdaFace-101 pipeline with RDigi1M pretraining and CodeFormer restoration.
  The analytic calibration uses $\Delta=1-c_{\min}=1.5$, $\qquad c_{\min}=-0.5$.}
  \label{fig:fmr-lfw-ada-analytic-pr}
\end{figure}

\paragraph{Discussion of Figs.~\ref{fig:fmr-lfw-arc101-webface4m-analytic-pr}--\ref{fig:fmr-lfw-ada-analytic-pr}.}

Across the three figure sets we fix the target impostor rate at $\alpha=10^{-2}$ with $\delta=10^{-5}$ and sweep $\varepsilon\in\{5,10,20,30\}$.
The blue curves show the clean $\mathsf{FMR}(\tau)$, and the orange curves show the post-privacy $\mathsf{FMR}_{\mathrm{priv}}(\tau)$. The dashed vertical line marks the clean operating threshold $\tau_\alpha$, while the dash-dotted line marks the re-calibrated threshold $\tau_\alpha+\Delta\tau$ when an interior solution exists. All panels use strict endpoint semantics at $\tau=1$.

Increasing $\varepsilon$ reduces the Gaussian noise scale $\sigma$, so the private curve contracts monotonically toward the clean curve. Feasibility is controlled by $L(\sigma)\coloneqq \lim_{\tau\uparrow 1}\mathsf{FMR}_{\mathrm{priv}}(\tau)$. Under strict endpoint semantics, an interior threshold achieving $\alpha$ exists only when $L(\sigma)<\alpha$.
If $L(\sigma)\ge \alpha$, then no threshold $\tau<1$ attains the target.

Sensitivity modeling shifts this feasibility boundary. In Figs.~\ref{fig:fmr-lfw-arc101-webface4m-analytic-pr} and~\ref{fig:fmr-lfw-ada-analytic-pr}, the analytic calibration uses the domain-restricted sensitivity $\Delta=1-c_{\min}=1.5$, $c_{\min}=-0.5$, whereas Figure~\ref{fig:fmr-lfw-arc50-casia-analytic-wc} uses the worst-case sensitivity $\Delta=2$. The smaller $\Delta$ produces a smaller $\sigma$ at the same $(\varepsilon,\delta)$, and therefore reaches feasibility at a lower privacy cost. The backbone and training set mainly affect the clean operating threshold and the local shape of the impostor tail near $\tau_\alpha$; the primary determinant of feasibility remains the noise scale $\sigma(\varepsilon,\delta,\Delta)$.

For ArcFace-101/WebFace4M (Figure~\ref{fig:fmr-lfw-arc101-webface4m-analytic-pr}), the target is infeasible for $\varepsilon\in \{5,10,20\}$ and becomes feasible at $\varepsilon=30$. AdaFace/RDigi1M/CodeFormer (Figure~\ref{fig:fmr-lfw-ada-analytic-pr}) shows the same feasibility pattern. For ArcFace-50/Casia under worst-case sensitivity (Figure~\ref{fig:fmr-lfw-arc50-casia-analytic-wc}), the noise inflation is larger, and even at $\varepsilon=30$ the left limit remains slightly above $\alpha$, so the target is still infeasible. In that case, feasibility would require a larger $\varepsilon$, a larger $\delta$, or a less stringent target level $\alpha$.

When feasibility holds, the required correction $\tau_\alpha\mapsto \tau_\alpha+\Delta\tau$ remains away from the saturation boundary $\tau=1$, which helps preserve genuine-match utility.

\begin{figure}[!t]
  \centering
  \begin{subfigure}[t]{0.47\linewidth}
    \includegraphics[width=\linewidth]{appx_results/case1/sigma_sweep/sigma_sweep_a0p01_strict_real_lfw_arc101_webface4m.png}%
    \caption{ArcFace-R101 (WebFace4M)}
    \label{fig:sigma-sweep-lfw_arc101_webface4m}
    \end{subfigure}
  \hfill
  \begin{subfigure}[t]{0.47\linewidth}
    \includegraphics[width=\linewidth]{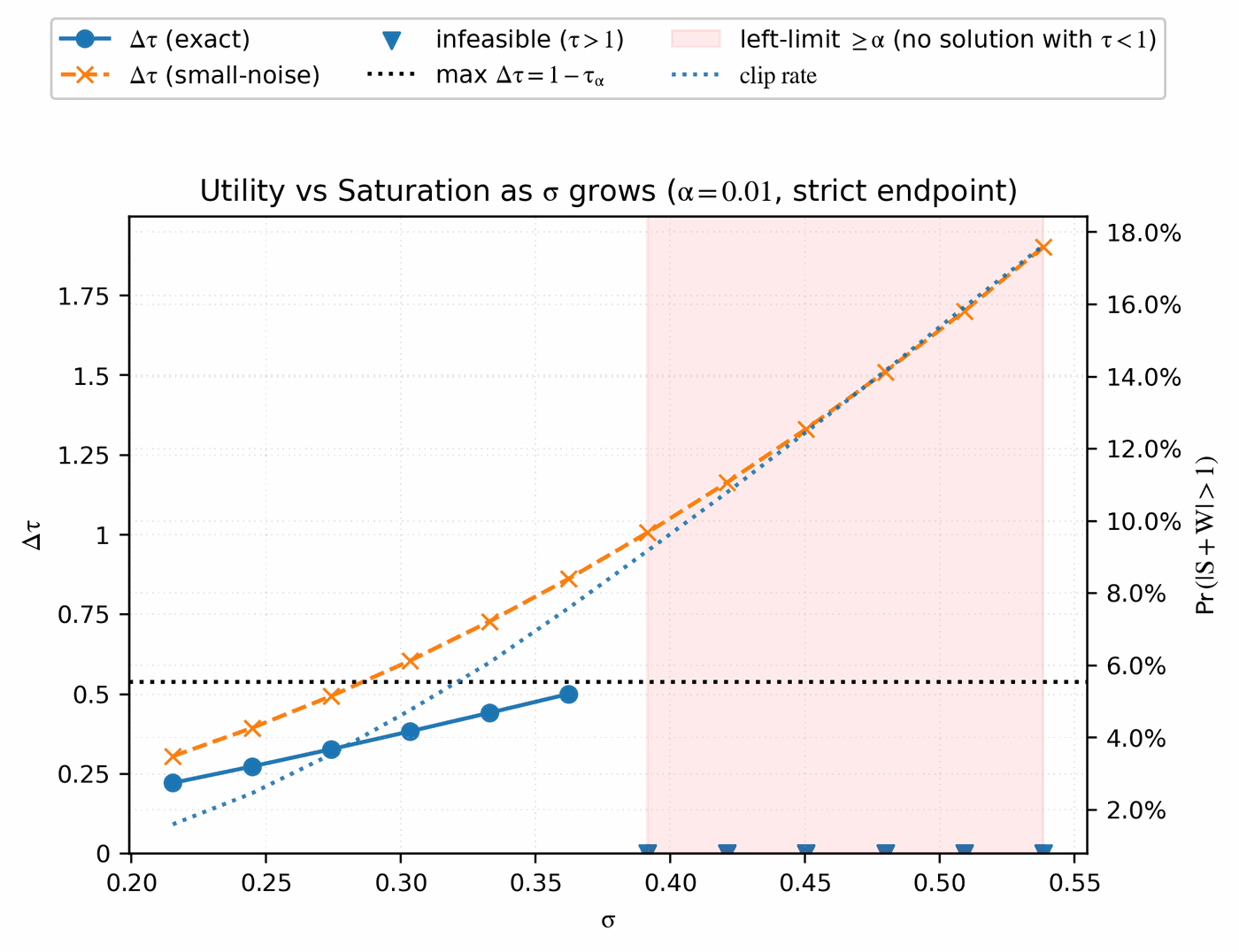}%
    \caption{Synthetic Gaussian}
    \label{fig:sweep-arc101-analytic-strict}
    \end{subfigure}

    \begin{subfigure}[t]{0.47\linewidth}
    \includegraphics[width=\linewidth]{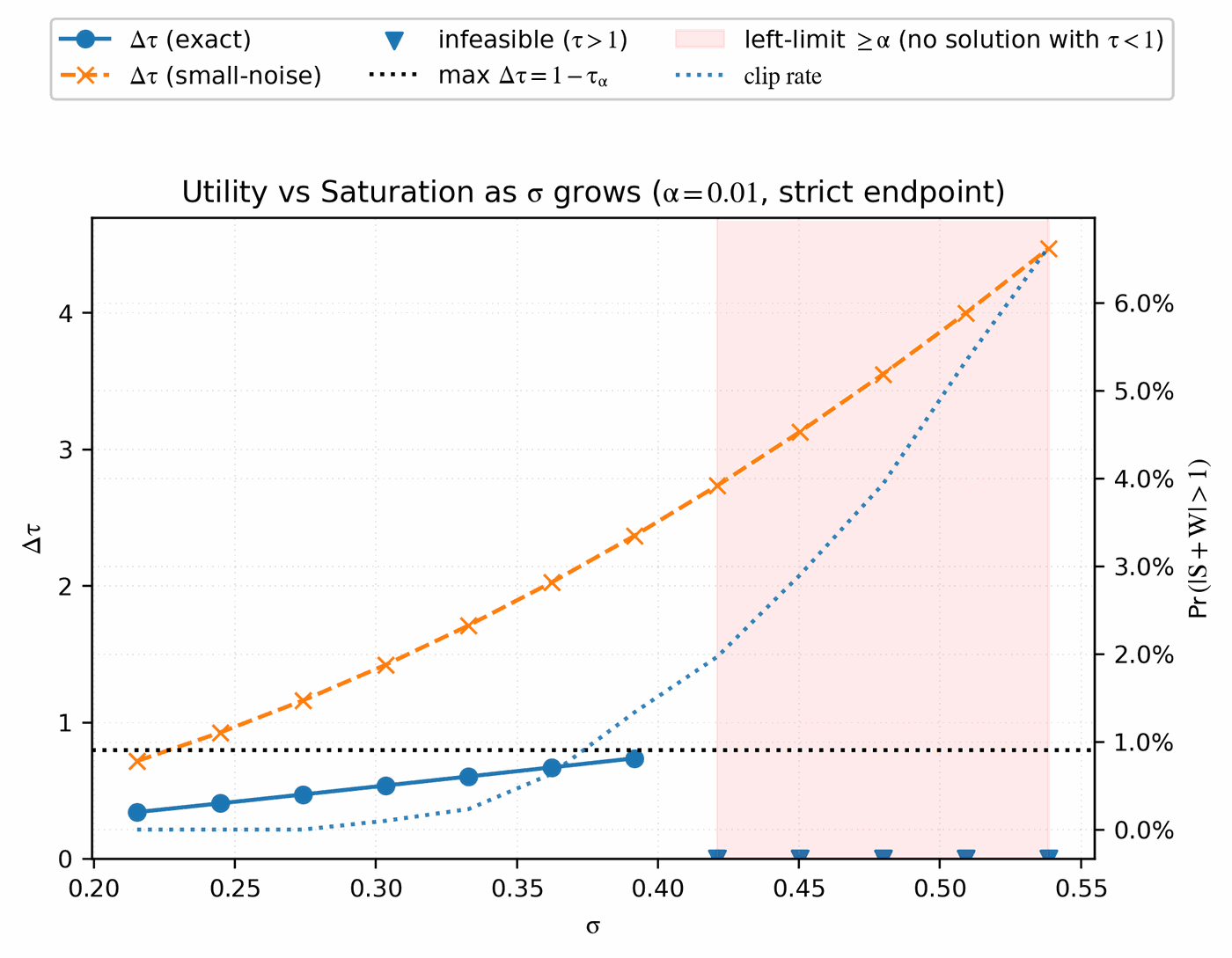}%
    \caption{ArcFace-R50 (Casia)}
    \label{fig:sigma-sweep-lfw_arc50_casia}
    \end{subfigure}
  \hfill
  \begin{subfigure}[t]{0.47\linewidth}
    \includegraphics[width=\linewidth]{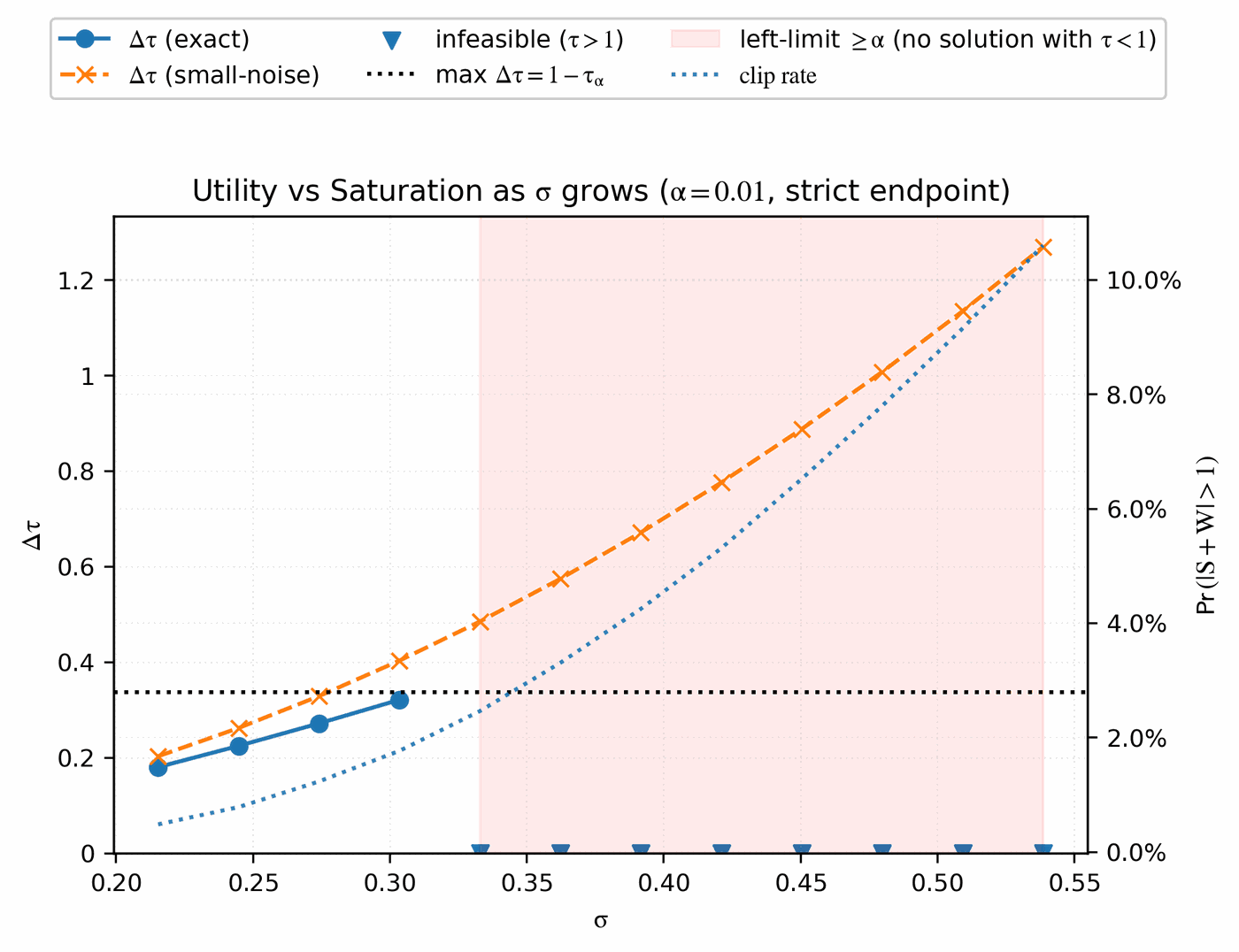}%
    \caption{Synthetic Symmetric Beta}
    \label{fig:sigma-sweep-synth_beta}
    \end{subfigure}

  \begin{subfigure}[t]{0.47\linewidth}      
    \includegraphics[width=\linewidth]{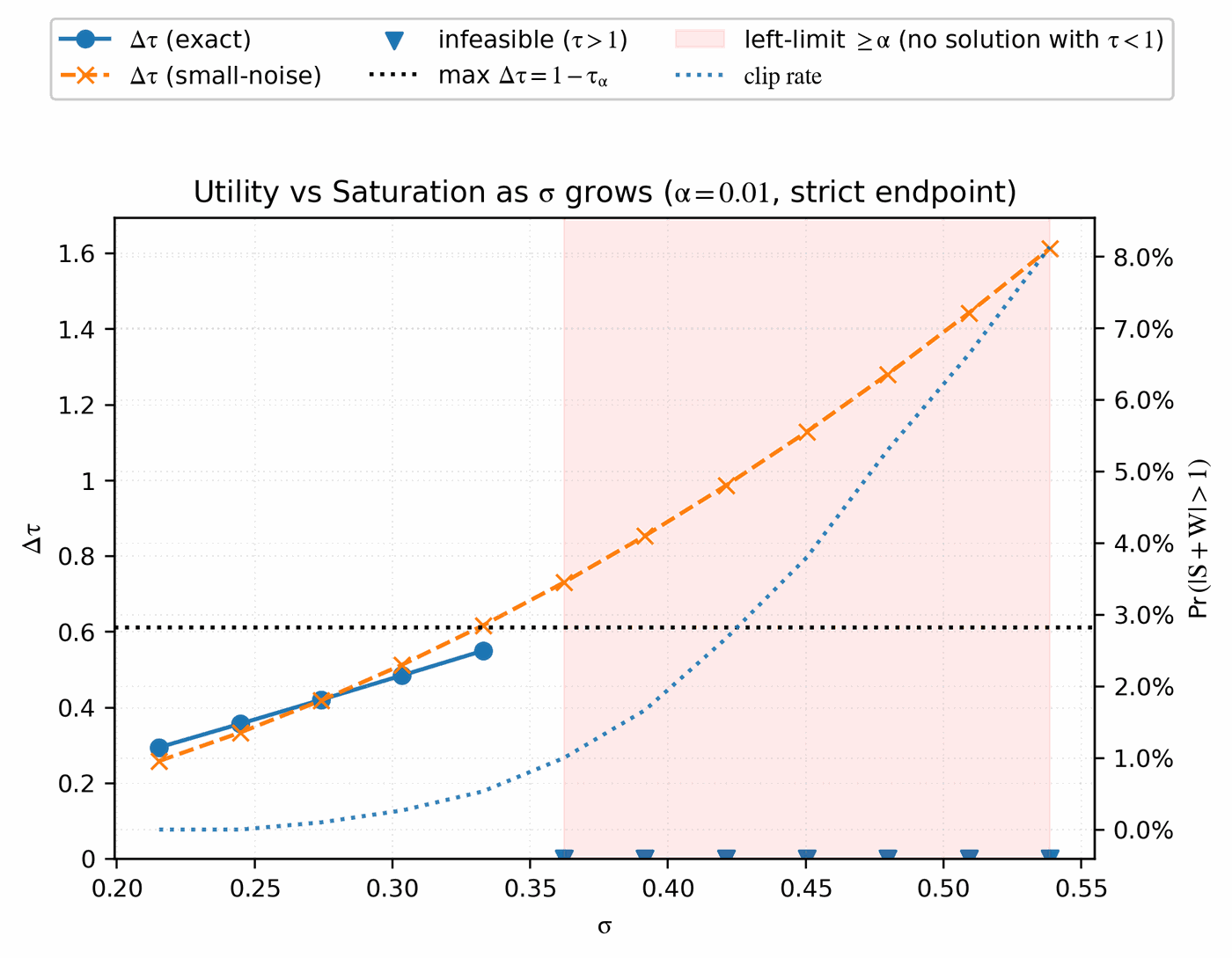}%
    \caption{AdaFace + RDigi1M/CodeFormer}
    \label{fig:sigma-sweep-lfw_ada}
    \end{subfigure}
  \hfill
  \begin{subfigure}[t]{0.47\linewidth}
    \includegraphics[width=\linewidth]{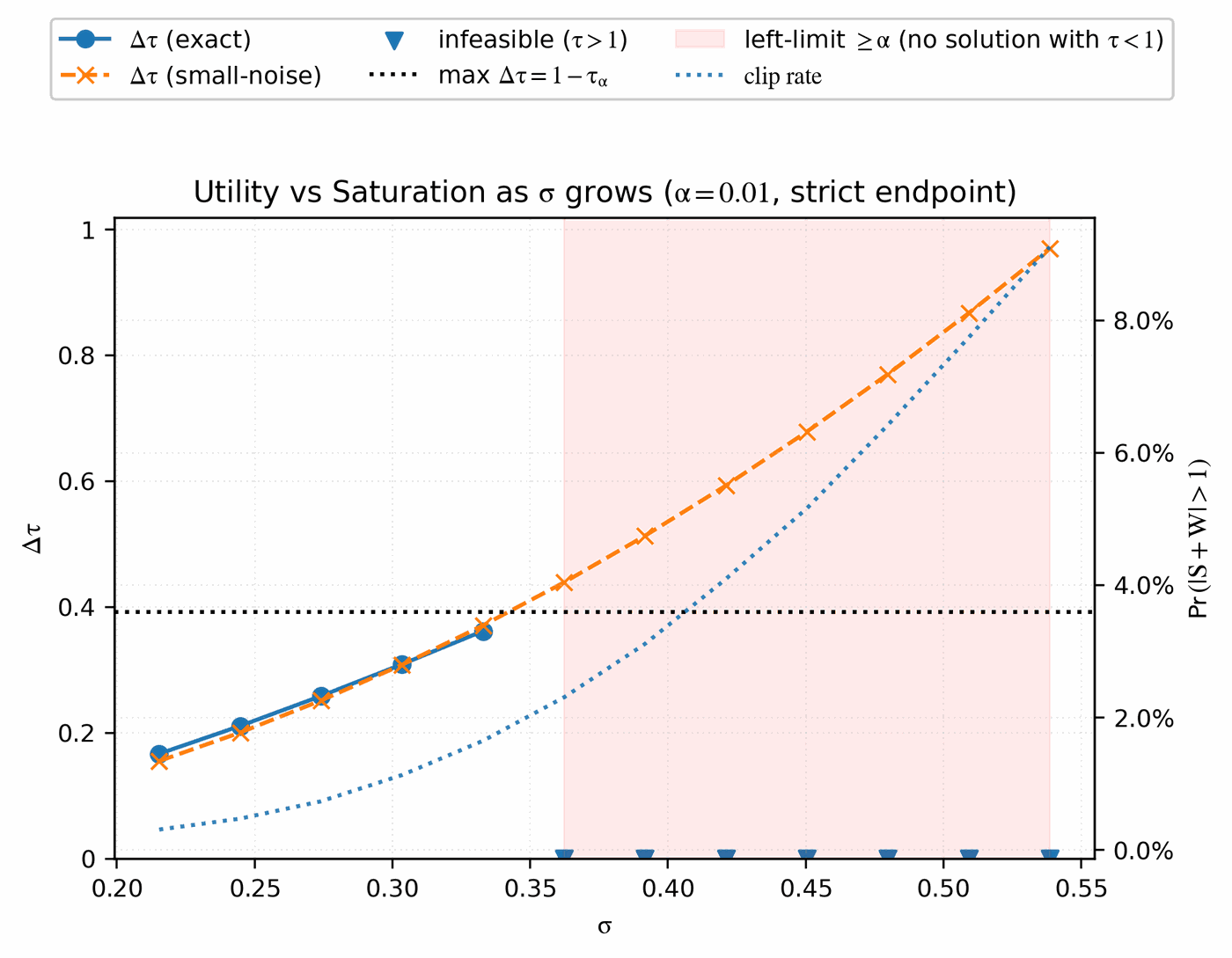}%
    \caption{Synthetic Student-t}
    \label{fig:sigma-sweep-synth_t}
   \end{subfigure}
  \vspace{-4pt}
  \caption{\textbf{Sigma sweep: threshold shift $\Delta\tau$ versus noise scale $\sigma$.}
  Each panel fixes the target impostor rate $\alpha=0.01$ and plots the calibration-agnostic threshold correction $\Delta\tau(\sigma)$ for the indicated score model.
  Panels (a,c,e) use real face-recognition score sets; panels (b,d,f) use synthetic score families.
  Shaded regions indicate strict-endpoint infeasibility: if $L(\sigma)=\lim_{\tau\uparrow 1}\mathsf{FMR}_{\mathrm{priv}}(\tau)\ge \alpha$, then no threshold $\tau<1$ attains the target.
  For small $\sigma$, $\Delta\tau(\sigma) = -\frac{\sigma^2}{2}\, \frac{f'_{\mathrm{i}}(\tau_\alpha)}{f_{\mathrm{i}}(\tau_\alpha)} +o(\sigma^2)$, so the leading curvature is determined by the impostor density near the clean operating point $\tau_\alpha$.
  A secondary top axis, when shown, maps $\sigma$ to $\varepsilon$ under a selected Gaussian calibration; it is included only for interpretation.}
\label{fig:sigma-sweep-real-and-synthetic}
\end{figure}

\paragraph{Discussion of Figure~\ref{fig:sigma-sweep-real-and-synthetic}.}

%
For fixed target level $\alpha$ and fixed endpoint semantics, the threshold correction $\Delta\tau(\sigma) = \tau_{\mathrm{priv}}(\alpha;\sigma)-\tau_{\mathrm{clean}}(\alpha)$ depends only on the clean score distribution and the Gaussian noise scale $\sigma$. Hence different privacy calibrations agree whenever they induce the same $\sigma$; they differ only through the map
$(\varepsilon,\delta,\Delta)\mapsto \sigma$.

Under strict endpoint semantics, $\mathsf{FMR}_{\mathrm{priv}}(\tau)$ is continuous and strictly decreasing on $(-1,1)$, with $L(\sigma)\coloneqq \lim_{\tau\uparrow 1}\mathsf{FMR}_{\mathrm{priv}}(\tau)>0$. An interior threshold achieving $\alpha$ exists if and only if $\alpha>L(\sigma)$. 
This yields the feasibility boundary $\sigma_{\mathrm{crit}}(\alpha) \coloneqq \inf\{\sigma: L(\sigma)\ge \alpha\}$. For $\sigma\ge \sigma_{\mathrm{crit}}(\alpha)$, the target $\alpha$ is infeasible under strict endpoint semantics.

In the small-noise regime, Proposition~\ref{prop:fmr-decomp} gives the local expansion $\Delta\tau(\sigma) = -\frac{\sigma^{2}}{2}\, \frac{f'_{\mathrm{i}}(\tau_\alpha)}{f_{\mathrm{i}}(\tau_\alpha)} + o(\sigma^{2})$, so the leading correction is quadratic in $\sigma$. Its magnitude is governed by the local log-derivative $-\frac{f'_{\mathrm{i}}(\tau_\alpha)}{f_{\mathrm{i}}(\tau_\alpha)}$ of the impostor density at the clean operating point, rather than by tail class alone. Across datasets, the qualitative shape is similar, but the vertical offset and the feasibility boundary are distribution-dependent.

Practically, we therefore report $\Delta\tau(\sigma)$ using (i) a shared $\sigma$-grid across panels, (ii) the clean threshold $\tau_\alpha$ as a reference, (iii) a shaded infeasible region where $\alpha\le L(\sigma)$, and, when useful, (iv) a secondary top axis mapping $\sigma$ to $\varepsilon$ under a selected calibration. That secondary axis is only interpretive; it does not alter the $\Delta\tau(\sigma)$ curve itself.

\begin{figure}[!t]
  \centering

  \begin{subfigure}[t]{0.32\linewidth}
    \includegraphics[width=\linewidth]{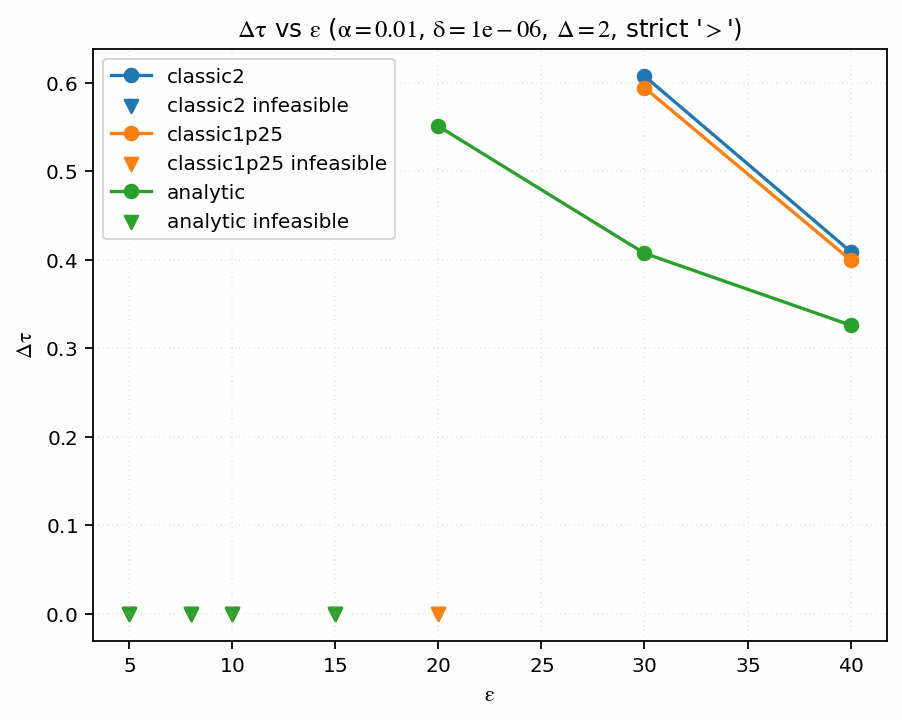}
    \caption{\textbf{LFW, AdaFace RDIGI1M+CodeFormer}, $\Delta=2$}
  \end{subfigure}
  ~
  \begin{subfigure}[t]{0.32\linewidth}
    \includegraphics[width=\linewidth]{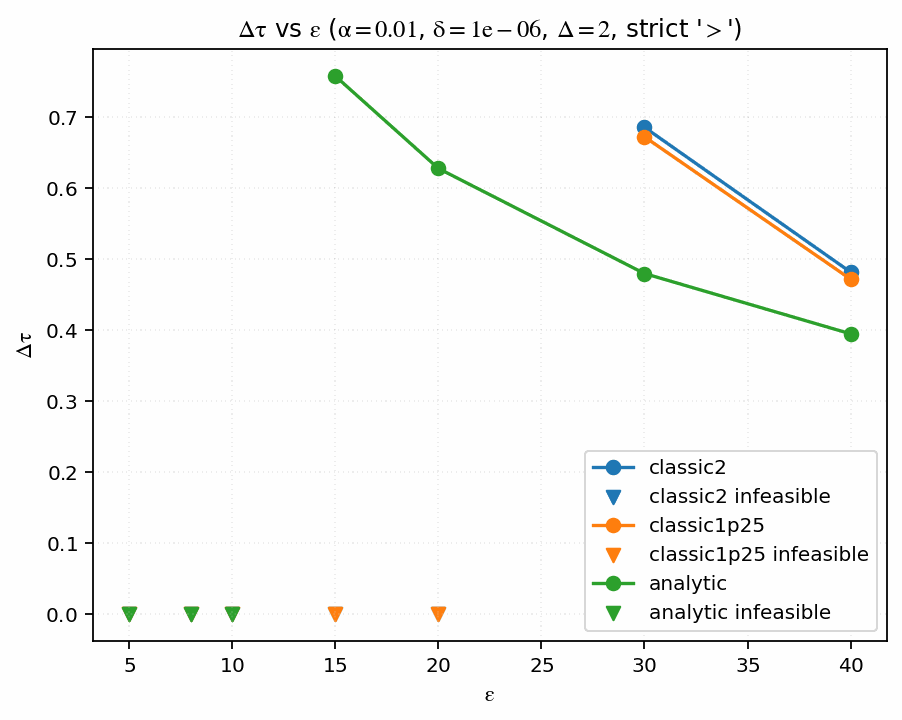}
    \caption{\textbf{LFW, ArcFace-101 WebFace4M}, $\Delta=2$}
  \end{subfigure}
  ~
  \begin{subfigure}[t]{0.32\linewidth}
    \includegraphics[width=\linewidth]{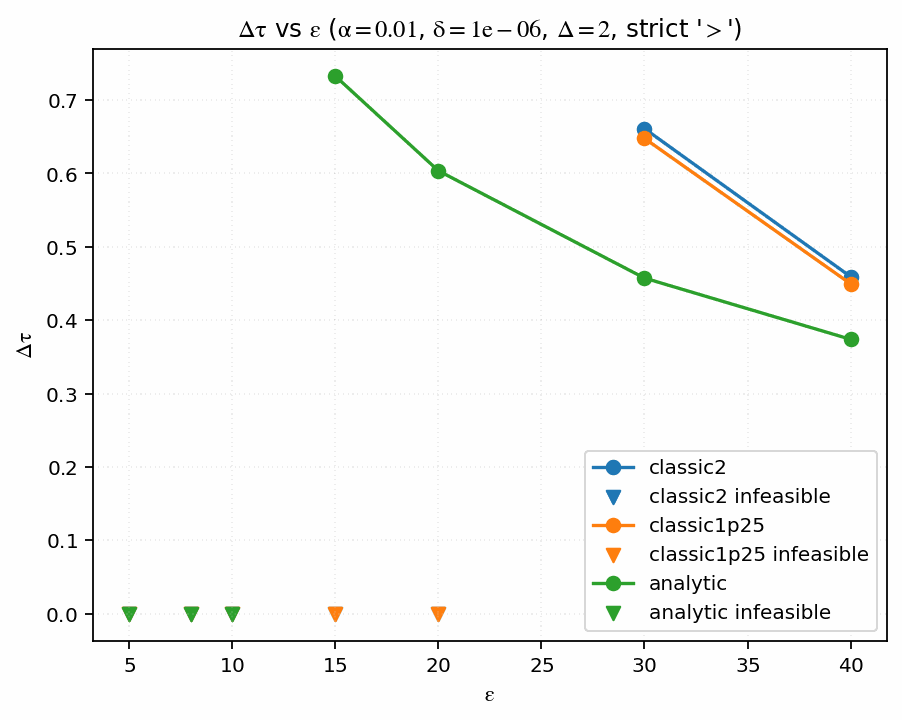}
    \caption{\textbf{LFW, ArcFace-50 CASIA}, $\Delta=2$}
  \end{subfigure}

  \vspace{4pt}

  \begin{subfigure}[t]{0.32\linewidth}
    \includegraphics[width=\linewidth]{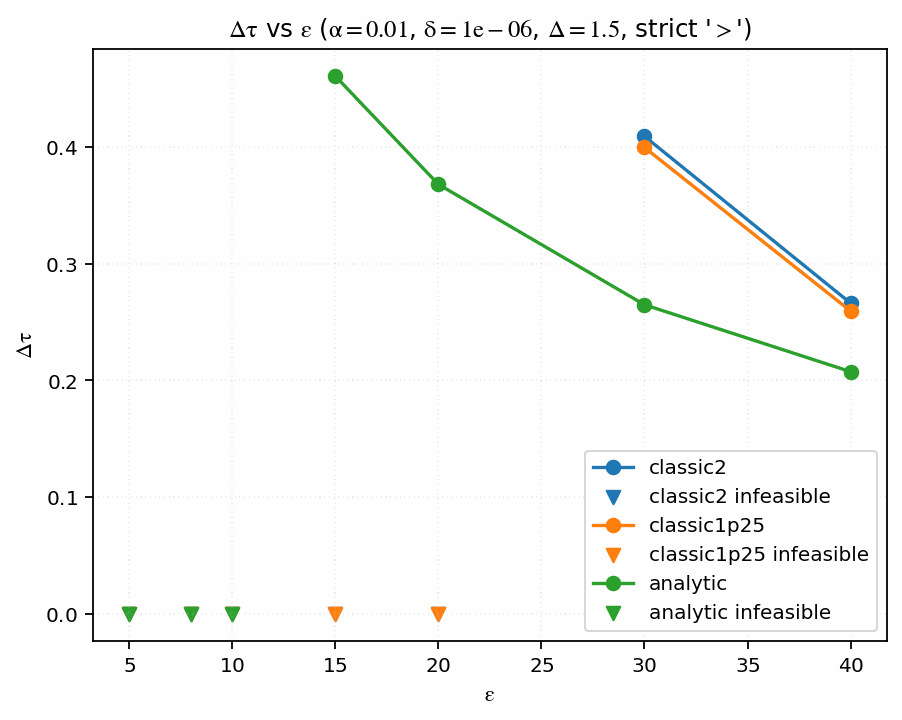}
    \caption{\textbf{LFW, AdaFace RDIGI1M+CodeFormer}, $\Delta=1.5$}
  \end{subfigure}
  ~
  \begin{subfigure}[t]{0.32\linewidth}
    \includegraphics[width=\linewidth]{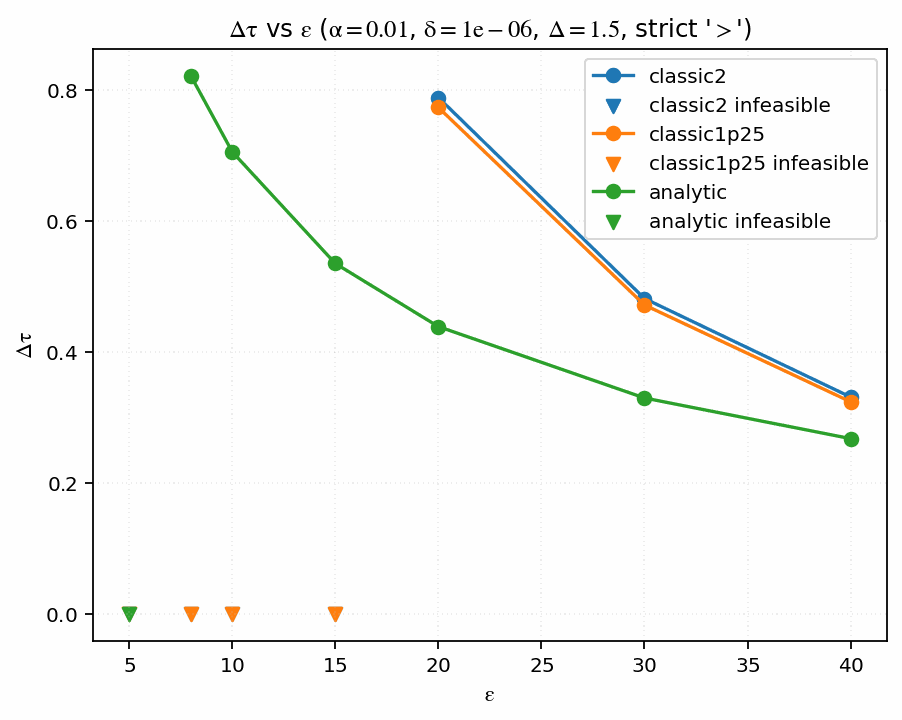}
    \caption{\textbf{LFW, ArcFace-101 WebFace4M}, $\Delta=1.5$}
  \end{subfigure}
  ~
  \begin{subfigure}[t]{0.32\linewidth}
    \includegraphics[width=\linewidth]{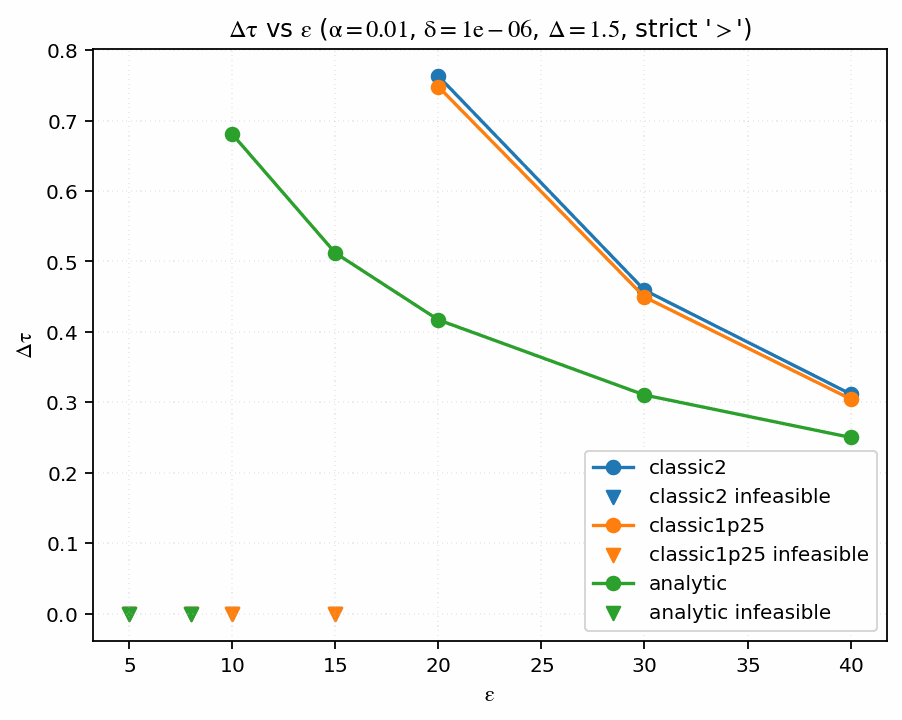}
    \caption{\textbf{LFW, ArcFace-50 CASIA}, $\Delta=1.5$}
  \end{subfigure}

  \vspace{-3pt}

  \caption{\textbf{Threshold correction $\Delta\tau$ versus privacy budget $\varepsilon$ on LFW-based score sets at fixed $\delta=10^{-6}$ under strict endpoint semantics.}
  Each panel reports the threshold correction $\Delta\tau$ required to attain $\alpha=10^{-2}$.
  The curves inside each panel correspond to the Gaussian calibration rules shown in the panel legend, using the same sensitivity value for that panel.
  The top row uses the worst-case score-vector sensitivity $\Delta=2$.
  The bottom row uses the domain-restricted sensitivity with $c_{\min}=-0.5$, hence $\Delta=1-c_{\min}=1.5$.
  Filled circles mark feasible operating points; inverted triangles mark infeasible budgets for which
  $\lim_{\tau\uparrow 1}\mathsf{FMR}_{\mathrm{priv}}(\tau)\ge \alpha$.
  }
  \label{fig:epssweep_real}
\end{figure}

\paragraph{Discussion of Figure~\ref{fig:epssweep_real} ($\varepsilon$-sweep of $\Delta\tau$).}

Figure~\ref{fig:epssweep_real} reports the threshold correction $\Delta\tau$ as a function of the privacy budget $\varepsilon$ for two sensitivity models. For each backbone, the top row uses the worst-case score-vector sensitivity $\Delta=2$, whereas the bottom row uses the domain-restricted sensitivity induced by $c_{\min}=-0.5$, namely $\Delta=1-c_{\min}=1.5$.

Across the three LFW-based score sets, $\Delta\tau$ decreases as $\varepsilon$ increases. This follows from the monotonic decrease of the Gaussian noise scale $\sigma(\varepsilon,\delta,\Delta)$ with $\varepsilon$ at fixed $(\delta,\Delta)$. Among feasible operating points at the same $(\varepsilon,\delta,\Delta)$, the analytic Gaussian calibration gives the smallest noise scale among the displayed calibrations and therefore the smallest required threshold correction. The conservative sufficient calibration gives the largest displayed correction, while the intermediate classical reference curve lies between these two calibrations.

Under strict endpoint semantics, feasibility at a target false-match rate $\alpha$ is determined by the limiting privatized false-match rate at the upper endpoint. If $\lim_{\tau\uparrow 1}\mathsf{FMR}_{\mathrm{priv}}(\tau)\ge \alpha$, then no admissible threshold $\tau<1$ attains the target. In Fig.~\ref{fig:epssweep_real}, such cases are marked by inverted triangles. The critical noise level for feasibility depends on the clean impostor-score distribution and on $\alpha$; the calibration rule determines which noise level corresponds to a given $(\varepsilon,\delta,\Delta)$.

Changing the sensitivity from $\Delta=2$ to $\Delta=1.5$ reduces the Gaussian noise scale $\sigma(\varepsilon,\delta,\Delta)$ by the factor $1.5/2=0.75$ for each displayed calibration rule, at fixed $(\varepsilon,\delta)$. Therefore, for the same backbone, calibration rule, and feasible value of $\varepsilon$, the domain-restricted model requires no larger threshold correction than the worst-case model. The bottom-row curves can nevertheless span a wider range of $\varepsilon$ values, because the smaller sensitivity makes some lower privacy budgets feasible that are infeasible when $\Delta=2$. These newly feasible points occur at smaller $\varepsilon$ and can still require relatively large threshold corrections. Differences across backbones are reflected in the level and curvature of the curves, which depend on the local impostor-score distribution near the clean operating threshold $\tau_\alpha$.


\clearpage

\appsubsection{Benchmarks}
\label{sec:complete-benchmark-results-regime1}

\paragraph{Dataset availability.}
IJB-B and IJB-C provide large-scale, template-based face-verification
protocols, including evaluation at stringent false-positive rates
\citep{ijbc}. NIST discontinued official distribution of the IJB
challenges on March 14, 2023.
We retain IJB-B/C because the publicly obtainable identity-style
benchmarks used in our evaluation---LFW, AgeDB, CFP-FP, CALFW, and
CPLFW---primarily use pairwise verification protocols and do not
provide scientifically equivalent large-scale, template-based
evaluation at false-positive rates of $10^{-6}$, $10^{-5}$, and
$10^{-4}$. Reproducing the IJB-B/C results therefore requires
previously authorized access to these datasets.

Before discussing the aggregate benchmark tables, we note that very large values of $\delta$ (e.g., $\delta\ge 10^{-1}$) are included only as \textit{utility ablations} to trace the dependence of performance on the Gaussian noise scale. They should not be interpreted as standard operational DP regimes.

\begin{figure*}[!h]
  \centering
  \begin{subfigure}[t]{0.48\linewidth}
    \includegraphics[width=\linewidth]{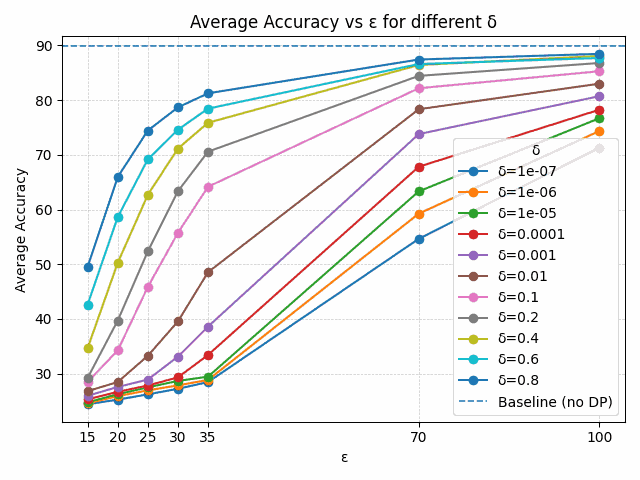}
    \caption{}\label{fig:avg_ir101_7bench}
  \end{subfigure}\hfill
  \begin{subfigure}[t]{0.48\linewidth}
    \includegraphics[width=\linewidth]{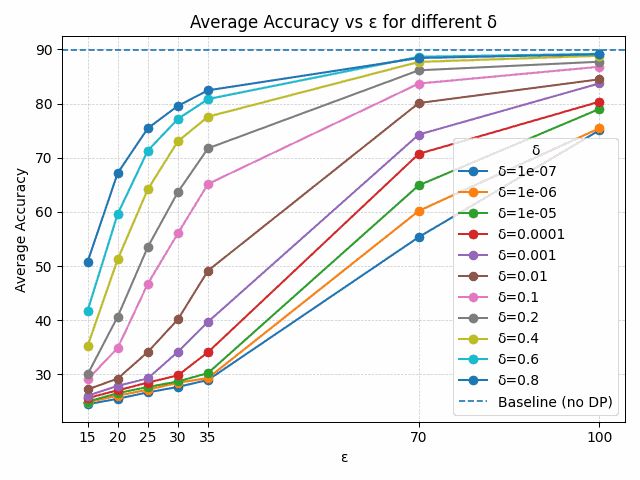}
    \caption{}\label{fig:avg_vit_7bench}
  \end{subfigure}
    \caption{Average performance across seven benchmarks when applying Algorithm~\ref{alg:query-to-collection} at different $(\delta,\varepsilon)$ values to (a) an IR101 backbone and (b) a ViT-Base backbone, both trained on WebFace4M.}
  \label{fig:avg_7bench_subfig}
\end{figure*}

\paragraph{Privacy-utility tradeoff.}
Figure~\ref{fig:avg_7bench_subfig} and Tables~\ref{tab:ir101}--\ref{tab:vit} show monotonic recovery of utility as the privacy budget $\varepsilon$ increases for fixed $\delta$. The curves are steepest for $\varepsilon$ in the range roughly $[15,35]$ and begin to saturate around $\varepsilon\approx 70$--$100$. For fixed $\varepsilon$, increasing $\delta$ improves utility because the Gaussian calibration $\sigma(\varepsilon,\delta)=\frac{\Delta\sqrt{2\log(2/\delta)}}{\varepsilon}$ decreases as $\delta$ increases.
For example, at $\varepsilon=70$, the IR101 macro average increases from $54.65\%$ at $\delta=10^{-7}$ to $86.40\%$ at $\delta=0.4$, while the ViT-Base macro average increases from $55.35\%$ to $87.69\%$. Beyond $\varepsilon\approx 70$, the gains become modest, indicating diminishing returns.

%
Most of the degradation is concentrated at the strict IJB-B/C operating points, especially at $\mathrm{FPR}=10^{-6}$.
The identity-style sets (LFW, AgeDB, CFP-FP, CALFW, CPLFW) remain much closer to their clean baselines, with LFW the most robust.
For IR101 at $(\varepsilon,\delta)=(70,0.4)$, LFW drops by $1.88$ percentage points (from $99.70\%$ to $97.82\%$), whereas IJB-B at $10^{-6}$ drops by $3.49$ percentage points (from $89.46\%$ to $85.97\%$).
ViT-Base shows the same qualitative pattern: LFW drops by $1.35$ percentage points (from $99.80\%$ to $98.45\%$), while IJB-B at $10^{-6}$ drops by $2.55$ percentage points (from $87.12\%$ to $84.57\%$).
This matches the decision-flip profile under Gaussian perturbation: the flip probability is smallest far from the decision threshold and largest in the extreme right tail.

\paragraph{Infeasible endpoints.}
Rows with zeros in the IJB-B/C@$10^{-6}$ columns correspond to infeasible operating points under the calibrated Gaussian noise scale, i.e., $\lim_{\tau\uparrow 1}\mathrm{FMR}_{\mathrm{priv}}(\tau) > 10^{-6}$. Feasibility can be restored by increasing $\varepsilon$, increasing $\delta$, or moving to a less stringent operating point such as $10^{-5}$ or $10^{-4}$.
For IR101, the $10^{-6}$ columns become nonzero around $(\varepsilon,\delta)\approx(35,10^{-2})$ or $(20,0.4)$, while the $10^{-5}$ and $10^{-4}$ targets are attainable at stricter privacy budgets.

\paragraph{Backbone comparison.}
Across most of the grid, ViT-Base is more robust than IR101 at matched $(\varepsilon,\delta)$, typically by about $0.8$--$2.5$ percentage points in macro average.
For example, at $(35,0.6)$, ViT-Base attains $80.83\%$ versus $78.43\%$ for IR101, and at $(100,0.6)$ it attains $89.18\%$ versus $87.68\%$.
A plausible explanation is that the ViT-Base score distribution places less impostor mass near the operational thresholds, which lowers the decision-flip probability under the same Gaussian noise scale.

\paragraph{Practical regimes.}
The benchmark sweeps suggest three coarse regimes:
\begin{enumerate}[leftmargin=*,itemsep=1pt,topsep=2pt]
\item \emph{Small-budget regime} ($\varepsilon\le 30$): utility is strongly degraded, and the IJB@$10^{-6}$ operating points are often infeasible.
\item \emph{Transition regime} ($35\le \varepsilon<70$): utility improves rapidly with either increasing $\varepsilon$ or relaxing $\delta$.
\item \emph{Near-saturation regime} ($\varepsilon\ge 70$): performance is close to the clean baseline, with diminishing gains from further increases in $\varepsilon$.
\end{enumerate}
Thus, for moderate $\delta$, the practical knee of the privacy--utility trade-off occurs around $\varepsilon\approx 70$ in these experiments. If the operating point can be relaxed from $10^{-6}$ to $10^{-5}$ or $10^{-4}$, useful accuracy extends to smaller $\varepsilon$.

\paragraph{Effect of $\delta$.}
At fixed $\varepsilon$, increasing $\delta$ improves utility because the Gaussian noise scale shrinks with $\delta$. For IR101 at $\varepsilon=100$, the macro average rises from $71.29\%$ at $\delta=10^{-7}$ to $80.70\%$ at $\delta=10^{-3}$ and to $88.44\%$ at $\delta=0.8$.
ViT-Base shows the same pattern, increasing from $75.07\%$ to $83.72\%$ and then to $89.10\%$.
When analytic calibration or domain-restricted sensitivity reduces the effective $\sigma$, the same $(\varepsilon,\delta)$ pair moves closer to the low-noise regime, shifting both the feasibility frontier and the practical knee toward stricter privacy.

\begin{table*}[!htb]
\tiny
\centering
\caption{
Results for the IR101 backbone trained on WebFace4M.
For LFW, CFP-FP, CPLFW, AgeDB, and CALFW, we report average verification accuracy. Columns labeled B-1e-6, B-1e-5, and B-1e-4 report TAR on IJB-B at the corresponding FPR; columns C-1e-6, C-1e-5, and C-1e-4 are defined analogously for IJB-C. For each $(\varepsilon,\delta)$ pair, the table reports performance after adding Gaussian noise to the released score vector in Algorithm~\ref{alg:query-to-collection}. The row with $\varepsilon=\mathrm{N/A}$ and $\delta=\mathrm{N/A}$ is the clean, non-private baseline.}
\resizebox{0.80\textwidth}{!}{ 
\begin{tabular}{c|c||c|c|c|c|c|c|c|c|c|c|c||c}
\toprule
$\varepsilon$ & $\delta$ & B-1e-6 & B-1e-5 & B-1e-4 & C-1e-6 & C-1e-5 & C-1e-4 & AgeDB & CALFW & CFPFP & CPLFW & LFW & Avg \\
\midrule
N/A & N/A & 89.46 & 93.07 & 95.52 & 43.49 & 89.07 & 93.72 & 96.35 & 95.57 & 98.37 & 92.82 & 99.70 & 89.74 \\
\midrule
100.00 & 1e-7 & 58.54 & 72.81 & 83.41 & 25.60 & 65.13 & 79.26 & 75.72 & 80.02 & 79.34 & 76.77 & 87.57 & 71.29 \\
100.00 & 1e-6 & 68.08 & 77.37 & 85.49 & 28.39 & 70.83 & 82.23 & 77.15 & 81.85 & 81.24 & 76.53 & 88.55 & 74.34 \\
100.00 & 1e-5 & 70.31 & 80.71 & 87.98 & 32.37 & 73.99 & 84.81 & 78.47 & 82.77 & 82.41 & 79.25 & 90.37 & 76.68 \\
100.00 & 1e-4 & 73.18 & 83.68 & 89.87 & 29.36 & 76.48 & 86.85 & 79.78 & 84.47 & 84.73 & 80.10 & 92.05 & 78.23 \\
100.00 & 1e-3 & 79.99 & 87.05 & 91.82 & 26.29 & 81.09 & 89.16 & 82.65 & 86.13 & 87.01 & 82.63 & 93.87 & 80.70 \\
100.00 & 1e-2 & 80.47 & 88.73 & 93.08 & 31.80 & 83.32 & 90.57 & 85.72 & 89.20 & 89.56 & 84.35 & 96.03 & 82.98 \\
100.00 & 0.10 & 86.35 & 90.88 & 94.32 & 29.66 & 85.65 & 92.20 & 88.70 & 91.77 & 92.71 & 87.48 & 98.12 & 85.26 \\
100.00 & 0.20 & 86.26 & 91.42 & 94.71 & 39.36 & 86.00 & 92.57 & 90.43 & 92.63 & 93.91 & 88.40 & 98.70 & 86.76 \\
100.00 & 0.40 & 88.17 & 92.07 & 94.86 & 43.25 & 87.72 & 92.92 & 91.83 & 93.33 & 95.27 & 89.92 & 99.17 & 88.05 \\
100.00 & 0.60 & 87.92 & 92.33 & 95.15 & 35.69 & 87.37 & 93.13 & 93.40 & 93.65 & 96.23 & 90.32 & 99.33 & 87.68 \\
100.00 & 0.80 & 88.81 & 92.25 & 95.26 & 39.64 & 87.97 & 93.35 & 93.88 & 94.38 & 96.59 & 91.12 & 99.53 & 88.44 \\
\midrule
70.00 & 1e-7 & 24.47 & 40.48 & 58.17 & 16.29 & 37.13 & 55.07 & 70.10 & 73.08 & 73.59 & 71.38 & 81.37 & 54.65 \\
70.00 & 1e-6 & 33.72 & 49.64 & 65.33 & 23.10 & 42.62 & 61.51 & 70.98 & 75.93 & 74.09 & 72.12 & 82.67 & 59.25 \\
70.00 & 1e-5 & 39.52 & 57.38 & 72.30 & 25.80 & 50.95 & 68.16 & 72.95 & 76.43 & 75.71 & 73.35 & 83.90 & 63.31 \\
70.00 & 1e-4 & 53.82 & 66.66 & 78.60 & 20.35 & 60.17 & 75.10 & 73.92 & 77.53 & 77.93 & 75.17 & 86.72 & 67.81 \\
70.00 & 1e-3 & 61.19 & 74.96 & 84.41 & 36.37 & 69.71 & 81.37 & 76.78 & 80.58 & 80.97 & 76.47 & 88.75 & 73.78 \\
70.00 & 1e-2 & 73.44 & 82.93 & 89.36 & 34.15 & 77.06 & 86.20 & 80.32 & 84.25 & 83.01 & 79.53 & 91.27 & 78.32 \\
70.00 & 0.10 & 81.22 & 88.55 & 92.73 & 27.91 & 82.78 & 90.05 & 84.72 & 87.87 & 88.57 & 83.72 & 95.55 & 82.15 \\
70.00 & 0.20 & 84.54 & 89.52 & 93.44 & 38.26 & 84.02 & 91.09 & 86.85 & 88.87 & 90.23 & 84.93 & 96.78 & 84.41 \\
70.00 & 0.40 & 85.97 & 90.91 & 94.30 & 44.22 & 85.58 & 91.85 & 89.35 & 90.98 & 92.73 & 86.72 & 97.82 & 86.40 \\
70.00 & 0.60 & 86.96 & 91.32 & 94.69 & 36.18 & 86.85 & 92.32 & 91.05 & 92.00 & 93.60 & 88.48 & 98.83 & 86.57 \\
70.00 & 0.80 & 87.40 & 91.92 & 94.87 & 38.26 & 86.93 & 92.78 & 91.95 & 93.53 & 95.29 & 89.25 & 99.20 & 87.40 \\
\midrule
35.00 & 1e-7 & 0.00 & 0.00 & 0.00 & 0.00 & 0.00 & 0.00 & 60.63 & 62.92 & 61.44 & 60.47 & 67.75 & 28.47 \\
35.00 & 1e-6 & 0.00 & 0.00 & 0.00 & 0.00 & 0.00 & 0.00 & 61.07 & 63.68 & 62.96 & 61.85 & 67.95 & 28.86 \\
35.00 & 1e-5 & 0.00 & 0.00 & 0.00 & 0.00 & 0.00 & 0.00 & 61.77 & 65.50 & 64.74 & 63.07 & 69.00 & 29.46 \\
35.00 & 1e-4 & 0.00 & 0.00 & 18.01 & 0.00 & 0.00 & 17.60 & 63.75 & 66.28 & 65.81 & 64.75 & 71.52 & 33.43 \\
35.00 & 1e-3 & 0.00 & 14.64 & 28.11 & 0.00 & 13.23 & 26.80 & 65.13 & 68.45 & 67.64 & 65.70 & 75.08 & 38.62 \\
35.00 & 1e-2 & 17.57 & 28.35 & 46.40 & 14.28 & 25.59 & 43.35 & 68.45 & 72.18 & 70.77 & 69.07 & 78.00 & 48.55 \\
35.00 & 0.10 & 42.13 & 58.61 & 73.00 & 26.07 & 52.52 & 69.58 & 72.73 & 77.13 & 76.34 & 73.43 & 84.23 & 64.16 \\
35.00 & 0.20 & 55.13 & 69.57 & 80.58 & 37.11 & 62.14 & 76.80 & 75.17 & 79.05 & 78.46 & 75.77 & 86.53 & 70.57 \\
35.00 & 0.40 & 69.21 & 79.14 & 87.12 & 30.30 & 73.36 & 84.23 & 78.65 & 82.52 & 81.96 & 78.17 & 89.68 & 75.85 \\
35.00 & 0.60 & 76.71 & 83.74 & 89.88 & 24.63 & 77.26 & 86.93 & 80.87 & 84.92 & 84.51 & 80.67 & 92.62 & 78.43 \\
35.00 & 0.80 & 77.07 & 86.53 & 91.85 & 35.02 & 81.17 & 88.79 & 83.07 & 86.52 & 86.64 & 82.63 & 94.22 & 81.23 \\
\midrule
30.00 & 1e-7 & 0.00 & 0.00 & 0.00 & 0.00 & 0.00 & 0.00 & 57.70 & 59.53 & 60.13 & 58.52 & 63.83 & 27.25 \\
30.00 & 1e-6 & 0.00 & 0.00 & 0.00 & 0.00 & 0.00 & 0.00 & 59.40 & 61.07 & 60.91 & 59.02 & 66.20 & 27.87 \\
30.00 & 1e-5 & 0.00 & 0.00 & 0.00 & 0.00 & 0.00 & 0.00 & 61.13 & 63.50 & 62.81 & 60.82 & 67.27 & 28.68 \\
30.00 & 1e-4 & 0.00 & 0.00 & 0.00 & 0.00 & 0.00 & 0.00 & 62.55 & 64.95 & 63.41 & 62.62 & 69.08 & 29.33 \\
30.00 & 1e-3 & 0.00 & 0.00 & 17.05 & 0.00 & 0.00 & 16.14 & 63.73 & 66.18 & 65.56 & 64.17 & 71.50 & 33.12 \\
30.00 & 1e-2 & 0.00 & 16.10 & 30.90 & 0.00 & 14.29 & 28.84 & 66.05 & 68.55 & 67.50 & 67.22 & 75.65 & 39.55 \\
30.00 & 0.10 & 27.65 & 41.93 & 60.11 & 19.44 & 38.86 & 56.23 & 70.57 & 73.98 & 71.77 & 70.80 & 81.65 & 55.73 \\
30.00 & 0.20 & 39.44 & 55.93 & 71.33 & 28.02 & 51.97 & 67.58 & 72.63 & 77.35 & 76.27 & 73.13 & 83.07 & 63.34 \\
30.00 & 0.40 & 54.00 & 70.91 & 81.70 & 36.80 & 65.96 & 78.37 & 74.38 & 79.47 & 78.23 & 75.15 & 87.50 & 71.13 \\
30.00 & 0.60 & 67.55 & 78.67 & 87.02 & 20.44 & 72.94 & 83.25 & 78.00 & 82.17 & 81.81 & 78.48 & 90.08 & 74.58 \\
30.00 & 0.80 & 74.49 & 83.36 & 90.06 & 31.31 & 77.54 & 86.76 & 79.72 & 84.47 & 84.71 & 80.62 & 92.18 & 78.66 \\
\midrule
25.00 & 1e-7 & 0.00 & 0.00 & 0.00 & 0.00 & 0.00 & 0.00 & 55.95 & 57.77 & 57.86 & 55.93 & 61.02 & 26.23 \\
25.00 & 1e-6 & 0.00 & 0.00 & 0.00 & 0.00 & 0.00 & 0.00 & 56.75 & 59.52 & 58.89 & 58.60 & 62.83 & 26.96 \\
25.00 & 1e-5 & 0.00 & 0.00 & 0.00 & 0.00 & 0.00 & 0.00 & 59.10 & 59.42 & 60.66 & 58.65 & 64.77 & 27.51 \\
25.00 & 1e-4 & 0.00 & 0.00 & 0.00 & 0.00 & 0.00 & 0.00 & 58.72 & 61.97 & 60.99 & 59.03 & 65.63 & 27.85 \\
25.00 & 1e-3 & 0.00 & 0.00 & 0.00 & 0.00 & 0.00 & 0.00 & 60.23 & 63.73 & 63.74 & 61.83 & 68.70 & 28.93 \\
25.00 & 1e-2 & 0.00 & 0.00 & 17.00 & 0.00 & 0.00 & 16.34 & 63.92 & 66.30 & 65.41 & 64.47 & 72.63 & 33.28 \\
25.00 & 0.10 & 12.80 & 24.78 & 41.07 & 11.58 & 21.43 & 38.88 & 67.53 & 70.53 & 70.04 & 67.85 & 78.02 & 45.86 \\
25.00 & 0.20 & 19.23 & 35.74 & 54.72 & 18.66 & 33.56 & 51.78 & 68.37 & 71.75 & 72.41 & 70.17 & 79.48 & 52.35 \\
25.00 & 0.40 & 40.68 & 55.13 & 71.30 & 25.35 & 50.67 & 66.89 & 72.15 & 75.65 & 75.61 & 72.50 & 83.57 & 62.68 \\
25.00 & 0.60 & 53.25 & 67.51 & 80.25 & 26.58 & 63.02 & 76.10 & 75.37 & 78.88 & 78.04 & 74.73 & 87.02 & 69.16 \\
25.00 & 0.80 & 64.33 & 76.95 & 85.64 & 34.54 & 69.76 & 82.26 & 77.58 & 81.00 & 80.40 & 77.17 & 89.08 & 74.43 \\
\midrule
20.00 & 1e-7 & 0.00 & 0.00 & 0.00 & 0.00 & 0.00 & 0.00 & 54.30 & 54.50 & 55.39 & 56.37 & 57.33 & 25.26 \\
20.00 & 1e-6 & 0.00 & 0.00 & 0.00 & 0.00 & 0.00 & 0.00 & 56.28 & 57.10 & 56.27 & 56.30 & 59.07 & 25.91 \\
20.00 & 1e-5 & 0.00 & 0.00 & 0.00 & 0.00 & 0.00 & 0.00 & 57.72 & 58.18 & 56.81 & 55.93 & 60.25 & 26.26 \\
20.00 & 1e-4 & 0.00 & 0.00 & 0.00 & 0.00 & 0.00 & 0.00 & 56.63 & 59.50 & 58.83 & 56.95 & 61.67 & 26.69 \\
20.00 & 1e-3 & 0.00 & 0.00 & 0.00 & 0.00 & 0.00 & 0.00 & 58.90 & 60.92 & 59.79 & 59.30 & 64.55 & 27.59 \\
20.00 & 1e-2 & 0.00 & 0.00 & 0.00 & 0.00 & 0.00 & 0.00 & 60.78 & 62.62 & 61.53 & 60.35 & 67.95 & 28.48 \\
20.00 & 0.10 & 0.00 & 0.00 & 20.15 & 0.00 & 0.00 & 19.95 & 63.87 & 67.82 & 67.16 & 64.87 & 72.97 & 34.25 \\
20.00 & 0.20 & 0.00 & 16.98 & 32.03 & 0.00 & 14.99 & 29.16 & 65.07 & 68.87 & 68.40 & 66.38 & 74.93 & 39.71 \\
20.00 & 0.40 & 18.99 & 31.37 & 50.28 & 13.84 & 29.79 & 46.65 & 68.88 & 71.35 & 71.83 & 70.13 & 79.78 & 50.26 \\
20.00 & 0.60 & 32.45 & 48.07 & 64.77 & 22.68 & 42.97 & 60.03 & 71.65 & 75.40 & 73.56 & 71.53 & 81.62 & 58.61 \\
20.00 & 0.80 & 45.84 & 61.56 & 75.12 & 29.11 & 56.03 & 72.10 & 73.05 & 77.35 & 75.96 & 73.90 & 85.17 & 65.93 \\
\midrule
15.00 & 1e-7 & 0.00 & 0.00 & 0.00 & 0.00 & 0.00 & 0.00 & 53.05 & 54.38 & 53.40 & 52.85 & 55.07 & 24.43 \\
15.00 & 1e-6 & 0.00 & 0.00 & 0.00 & 0.00 & 0.00 & 0.00 & 52.82 & 55.17 & 53.06 & 52.95 & 56.47 & 24.59 \\
15.00 & 1e-5 & 0.00 & 0.00 & 0.00 & 0.00 & 0.00 & 0.00 & 53.15 & 53.98 & 53.67 & 54.38 & 57.30 & 24.77 \\
15.00 & 1e-4 & 0.00 & 0.00 & 0.00 & 0.00 & 0.00 & 0.00 & 54.72 & 55.13 & 54.90 & 56.37 & 57.43 & 25.32 \\
15.00 & 1e-3 & 0.00 & 0.00 & 0.00 & 0.00 & 0.00 & 0.00 & 55.27 & 56.90 & 57.37 & 56.53 & 59.53 & 25.96 \\
15.00 & 1e-2 & 0.00 & 0.00 & 0.00 & 0.00 & 0.00 & 0.00 & 57.40 & 58.90 & 58.87 & 57.72 & 62.20 & 26.83 \\
15.00 & 0.10 & 0.00 & 0.00 & 0.00 & 0.00 & 0.00 & 0.00 & 61.37 & 62.82 & 61.96 & 60.37 & 67.02 & 28.50 \\
15.00 & 0.20 & 0.00 & 0.00 & 0.00 & 0.00 & 0.00 & 0.00 & 62.47 & 64.13 & 64.03 & 62.17 & 68.73 & 29.23 \\
15.00 & 0.40 & 0.00 & 0.00 & 22.47 & 0.00 & 0.00 & 21.60 & 64.25 & 67.48 & 67.36 & 64.82 & 73.43 & 34.67 \\
15.00 & 0.60 & 0.00 & 19.90 & 35.46 & 11.92 & 18.34 & 32.94 & 67.43 & 69.07 & 68.90 & 67.97 & 77.03 & 42.63 \\
15.00 & 0.80 & 16.10 & 31.52 & 49.03 & 11.43 & 28.21 & 46.55 & 68.78 & 71.92 & 71.81 & 70.37 & 79.12 & 49.53 \\
\bottomrule
\end{tabular}
}
\label{tab:ir101}
\end{table*}

\begin{table*}[!htb]
\centering
\tiny
\caption{
Results for the ViT-Base backbone trained on WebFace4M.
For LFW, CFP-FP, CPLFW, AgeDB, and CALFW, we report verification accuracy.
Columns labeled B-1e-6, B-1e-5, and B-1e-4 report TAR on IJB-B at the corresponding FPR; columns C-1e-6, C-1e-5, and C-1e-4 are defined analogously for IJB-C.
For each $(\varepsilon,\delta)$ pair, the table reports performance after adding Gaussian noise to the released score vector in Algorithm~\ref{alg:query-to-collection}.
The row with $\varepsilon=\mathrm{N/A}$ and $\delta=\mathrm{N/A}$ is the clean, non-private baseline.}
\resizebox{0.80\textwidth}{!}{ 
\begin{tabular}{c|c||c|c|c|c|c|c|c|c|c|c|c||c}
\toprule
$\varepsilon$ & $\delta$ & B-1e-6 & B-1e-5 & B-1e-4 & C-1e-6 & C-1e-5 & C-1e-4 & AgeDB & CALFW & CFPFP & CPLFW & LFW & Avg  \\
\midrule
N/A & N/A & 87.12 & 94.54 & 96.89 & 38.62 & 90.51 & 95.39 & 97.28 & 96.07 & 98.99 & 94.88 & 99.80 & 90.01 \\
\midrule
100.00 & 1e-7 & 60.84 & 74.62 & 84.93 & 43.50 & 68.40 & 82.13 & 78.92 & 81.62 & 81.90 & 79.40 & 89.50 & 75.07 \\
100.00 & 1e-6 & 67.04 & 79.05 & 87.57 & 21.67 & 72.85 & 84.29 & 79.32 & 83.58 & 83.33 & 81.08 & 90.08 & 75.44 \\
100.00 & 1e-5 & 71.69 & 82.05 & 89.73 & 34.94 & 77.93 & 86.71 & 81.13 & 84.92 & 85.61 & 82.23 & 92.10 & 79.00 \\
100.00 & 1e-4 & 74.98 & 86.14 & 91.97 & 29.72 & 78.45 & 88.94 & 83.82 & 85.22 & 87.04 & 84.07 & 93.43 & 80.34 \\
100.00 & 1e-3 & 79.64 & 88.92 & 93.72 & 40.61 & 83.51 & 91.08 & 84.98 & 87.40 & 89.74 & 86.20 & 95.15 & 83.72 \\
100.00 & 1e-2 & 81.42 & 91.16 & 94.83 & 28.18 & 85.53 & 92.46 & 88.17 & 89.93 & 92.29 & 88.27 & 97.02 & 84.48 \\
100.00 & 0.10 & 82.59 & 92.89 & 95.78 & 32.69 & 87.83 & 94.01 & 91.30 & 92.58 & 95.46 & 90.82 & 98.78 & 86.79 \\
100.00 & 0.20 & 86.81 & 93.16 & 95.91 & 33.00 & 88.86 & 94.35 & 92.60 & 93.28 & 96.10 & 91.83 & 99.08 & 87.73 \\
100.00 & 0.40 & 86.18 & 93.60 & 96.33 & 38.83 & 89.40 & 94.72 & 93.83 & 94.52 & 97.31 & 92.95 & 99.42 & 88.83 \\
100.00 & 0.60 & 85.99 & 93.78 & 96.49 & 39.81 & 89.60 & 94.89 & 94.75 & 94.88 & 97.61 & 93.60 & 99.55 & 89.18 \\
100.00 & 0.80 & 84.34 & 94.06 & 96.56 & 36.83 & 90.22 & 95.15 & 95.68 & 95.33 & 98.10 & 94.10 & 99.70 & 89.10 \\
\midrule
70.00 & 1e-7 & 25.13 & 39.46 & 59.06 & 16.29 & 35.87 & 55.32 & 72.38 & 74.48 & 75.47 & 73.92 & 81.42 & 55.35 \\
70.00 & 1e-6 & 29.96 & 47.21 & 66.67 & 26.94 & 44.26 & 61.70 & 73.67 & 76.38 & 77.04 & 74.52 & 83.63 & 60.18 \\
70.00 & 1e-5 & 37.29 & 56.93 & 73.36 & 30.94 & 52.83 & 70.12 & 74.73 & 76.95 & 78.31 & 76.37 & 86.17 & 64.91 \\
70.00 & 1e-4 & 52.67 & 67.86 & 80.93 & 35.75 & 60.49 & 77.05 & 76.82 & 79.87 & 80.27 & 79.12 & 87.07 & 70.72 \\
70.00 & 1e-3 & 58.04 & 76.45 & 86.15 & 28.89 & 69.46 & 83.19 & 79.35 & 82.68 & 83.14 & 79.62 & 90.03 & 74.27 \\
70.00 & 1e-2 & 71.37 & 84.52 & 91.01 & 37.24 & 78.48 & 88.21 & 82.23 & 84.83 & 86.67 & 83.58 & 92.90 & 80.10 \\
70.00 & 0.10 & 81.20 & 90.44 & 94.31 & 27.32 & 85.10 & 92.04 & 86.47 & 89.50 & 90.69 & 87.67 & 95.85 & 83.69 \\
70.00 & 0.20 & 81.98 & 91.31 & 95.22 & 39.81 & 87.16 & 92.89 & 89.13 & 90.80 & 93.13 & 88.87 & 97.25 & 86.14 \\
70.00 & 0.40 & 84.57 & 92.43 & 95.76 & 43.55 & 87.83 & 93.57 & 90.72 & 92.38 & 94.51 & 90.80 & 98.45 & 87.69 \\
70.00 & 0.60 & 85.98 & 93.09 & 96.05 & 44.46 & 88.53 & 94.15 & 92.82 & 93.15 & 95.96 & 91.82 & 99.02 & 88.64 \\
70.00 & 0.80 & 85.95 & 93.50 & 96.30 & 36.43 & 89.23 & 94.59 & 93.55 & 94.20 & 96.97 & 92.65 & 99.18 & 88.41 \\
\midrule
35.00 & 1e-7 & 0.00 & 0.00 & 0.00 & 0.00 & 0.00 & 0.00 & 61.77 & 62.90 & 63.66 & 62.80 & 67.53 & 28.97 \\
35.00 & 1e-6 & 0.00 & 0.00 & 0.00 & 0.00 & 0.00 & 0.00 & 61.88 & 65.47 & 62.86 & 63.23 & 69.05 & 29.32 \\
35.00 & 1e-5 & 0.00 & 0.00 & 0.00 & 0.00 & 0.00 & 0.00 & 62.98 & 65.78 & 65.73 & 65.77 & 72.38 & 30.24 \\
35.00 & 1e-4 & 0.00 & 0.00 & 18.26 & 0.00 & 0.00 & 17.05 & 65.17 & 68.28 & 67.66 & 66.87 & 71.82 & 34.10 \\
35.00 & 1e-3 & 0.00 & 15.92 & 28.03 & 0.00 & 15.03 & 25.80 & 67.77 & 69.98 & 70.50 & 68.40 & 75.33 & 39.71 \\
35.00 & 1e-2 & 17.05 & 28.95 & 46.23 & 11.94 & 24.58 & 43.20 & 70.03 & 73.35 & 73.51 & 71.85 & 79.95 & 49.15 \\
35.00 & 0.10 & 43.42 & 57.97 & 74.08 & 21.27 & 53.43 & 71.02 & 75.63 & 77.98 & 79.03 & 76.55 & 86.35 & 65.16 \\
35.00 & 0.20 & 53.35 & 69.76 & 82.13 & 35.91 & 63.36 & 78.75 & 76.92 & 80.73 & 81.24 & 78.98 & 88.00 & 71.74 \\
35.00 & 0.40 & 66.49 & 81.48 & 88.96 & 33.61 & 74.53 & 85.95 & 80.80 & 83.77 & 84.81 & 81.88 & 91.02 & 77.57 \\
35.00 & 0.60 & 74.62 & 86.18 & 91.84 & 35.38 & 79.31 & 89.05 & 82.83 & 85.53 & 87.81 & 83.53 & 93.00 & 80.83 \\
35.00 & 0.80 & 76.83 & 89.27 & 93.57 & 28.70 & 83.15 & 91.11 & 86.10 & 88.02 & 89.14 & 86.02 & 94.98 & 82.44 \\
\midrule
30.00 & 1e-7 & 0.00 & 0.00 & 0.00 & 0.00 & 0.00 & 0.00 & 58.53 & 60.90 & 60.01 & 59.88 & 65.15 & 27.68 \\
30.00 & 1e-6 & 0.00 & 0.00 & 0.00 & 0.00 & 0.00 & 0.00 & 59.98 & 62.13 & 62.17 & 62.30 & 66.40 & 28.45 \\
30.00 & 1e-5 & 0.00 & 0.00 & 0.00 & 0.00 & 0.00 & 0.00 & 61.05 & 62.30 & 63.80 & 62.22 & 66.22 & 28.69 \\
30.00 & 1e-4 & 0.00 & 0.00 & 0.00 & 0.00 & 0.00 & 0.00 & 62.53 & 66.40 & 65.77 & 64.20 & 69.13 & 29.82 \\
30.00 & 1e-3 & 0.00 & 0.00 & 19.02 & 0.00 & 0.00 & 18.45 & 65.42 & 67.35 & 66.91 & 66.03 & 72.70 & 34.17 \\
30.00 & 1e-2 & 0.00 & 15.56 & 30.67 & 0.00 & 14.62 & 28.80 & 66.80 & 69.33 & 70.10 & 69.48 & 76.43 & 40.16 \\
30.00 & 0.10 & 25.75 & 41.70 & 60.18 & 16.35 & 36.54 & 56.38 & 71.95 & 75.13 & 75.69 & 73.88 & 82.95 & 56.05 \\
30.00 & 0.20 & 38.42 & 55.71 & 72.13 & 21.90 & 51.15 & 68.57 & 75.65 & 77.73 & 78.61 & 75.55 & 84.77 & 63.65 \\
30.00 & 0.40 & 59.83 & 72.27 & 83.74 & 30.84 & 66.40 & 80.69 & 78.22 & 81.27 & 81.97 & 79.60 & 88.68 & 73.05 \\
30.00 & 0.60 & 66.94 & 80.96 & 88.84 & 28.65 & 73.95 & 85.82 & 80.67 & 84.07 & 85.20 & 82.95 & 90.68 & 77.16 \\
30.00 & 0.80 & 73.60 & 85.86 & 91.46 & 22.94 & 79.55 & 88.96 & 82.27 & 86.55 & 87.44 & 83.52 & 93.28 & 79.58 \\
\midrule
25.00 & 1e-7 & 0.00 & 0.00 & 0.00 & 0.00 & 0.00 & 0.00 & 57.07 & 58.17 & 58.10 & 58.27 & 61.85 & 26.68 \\
25.00 & 1e-6 & 0.00 & 0.00 & 0.00 & 0.00 & 0.00 & 0.00 & 58.38 & 59.48 & 59.61 & 58.90 & 62.43 & 27.16 \\
25.00 & 1e-5 & 0.00 & 0.00 & 0.00 & 0.00 & 0.00 & 0.00 & 59.57 & 60.43 & 60.44 & 59.35 & 64.57 & 27.67 \\
25.00 & 1e-4 & 0.00 & 0.00 & 0.00 & 0.00 & 0.00 & 0.00 & 60.83 & 62.28 & 61.84 & 61.65 & 66.67 & 28.48 \\
25.00 & 1e-3 & 0.00 & 0.00 & 0.00 & 0.00 & 0.00 & 0.00 & 63.37 & 64.32 & 63.20 & 62.57 & 68.33 & 29.25 \\
25.00 & 1e-2 & 0.00 & 0.00 & 18.45 & 0.00 & 0.00 & 17.88 & 65.35 & 67.45 & 68.09 & 65.48 & 72.75 & 34.13 \\
25.00 & 0.10 & 14.24 & 23.08 & 40.62 & 12.96 & 20.93 & 38.62 & 69.53 & 72.35 & 72.54 & 70.85 & 78.50 & 46.75 \\
25.00 & 0.20 & 22.15 & 36.31 & 54.91 & 16.75 & 32.40 & 50.93 & 71.02 & 74.47 & 75.67 & 72.68 & 81.53 & 53.53 \\
25.00 & 0.40 & 41.10 & 56.44 & 72.11 & 27.59 & 50.54 & 68.34 & 74.12 & 77.57 & 76.59 & 76.27 & 85.32 & 64.18 \\
25.00 & 0.60 & 50.97 & 69.69 & 81.44 & 33.58 & 63.65 & 77.88 & 77.55 & 81.08 & 81.23 & 79.17 & 88.30 & 71.32 \\
25.00 & 0.80 & 65.19 & 78.83 & 87.55 & 25.03 & 72.02 & 84.03 & 79.85 & 82.85 & 83.23 & 80.92 & 90.25 & 75.43 \\
\midrule
20.00 & 1e-7 & 0.00 & 0.00 & 0.00 & 0.00 & 0.00 & 0.00 & 55.65 & 56.50 & 55.63 & 54.63 & 57.98 & 25.49 \\
20.00 & 1e-6 & 0.00 & 0.00 & 0.00 & 0.00 & 0.00 & 0.00 & 56.48 & 56.65 & 56.80 & 55.88 & 60.85 & 26.06 \\
20.00 & 1e-5 & 0.00 & 0.00 & 0.00 & 0.00 & 0.00 & 0.00 & 58.45 & 58.95 & 56.91 & 57.40 & 60.05 & 26.52 \\
20.00 & 1e-4 & 0.00 & 0.00 & 0.00 & 0.00 & 0.00 & 0.00 & 58.22 & 59.73 & 58.71 & 58.60 & 62.30 & 27.05 \\
20.00 & 1e-3 & 0.00 & 0.00 & 0.00 & 0.00 & 0.00 & 0.00 & 59.73 & 60.23 & 61.29 & 60.93 & 64.37 & 27.87 \\
20.00 & 1e-2 & 0.00 & 0.00 & 0.00 & 0.00 & 0.00 & 0.00 & 62.47 & 64.55 & 64.11 & 63.50 & 66.87 & 29.23 \\
20.00 & 0.10 & 0.00 & 0.00 & 20.40 & 0.00 & 0.00 & 19.77 & 65.18 & 69.18 & 68.61 & 67.00 & 73.47 & 34.87 \\
20.00 & 0.20 & 0.00 & 16.54 & 31.76 & 0.00 & 14.90 & 29.18 & 67.50 & 70.62 & 70.89 & 69.93 & 76.05 & 40.67 \\
20.00 & 0.40 & 18.79 & 32.46 & 50.77 & 15.71 & 29.11 & 47.32 & 70.87 & 72.52 & 73.14 & 72.72 & 80.30 & 51.25 \\
20.00 & 0.60 & 29.03 & 47.20 & 65.23 & 22.12 & 44.73 & 62.32 & 72.43 & 77.15 & 76.93 & 75.07 & 83.72 & 59.63 \\
20.00 & 0.80 & 44.87 & 62.17 & 76.83 & 28.01 & 56.69 & 72.87 & 76.42 & 78.83 & 79.03 & 76.63 & 86.67 & 67.18 \\
\midrule
15.00 & 1e-7 & 0.00 & 0.00 & 0.00 & 0.00 & 0.00 & 0.00 & 52.90 & 53.72 & 54.21 & 53.53 & 55.03 & 24.49 \\
15.00 & 1e-6 & 0.00 & 0.00 & 0.00 & 0.00 & 0.00 & 0.00 & 54.80 & 54.78 & 54.76 & 52.73 & 56.12 & 24.84 \\
15.00 & 1e-5 & 0.00 & 0.00 & 0.00 & 0.00 & 0.00 & 0.00 & 53.28 & 54.90 & 54.26 & 56.48 & 55.60 & 24.96 \\
15.00 & 1e-4 & 0.00 & 0.00 & 0.00 & 0.00 & 0.00 & 0.00 & 55.27 & 57.28 & 55.64 & 55.75 & 57.97 & 25.63 \\
15.00 & 1e-3 & 0.00 & 0.00 & 0.00 & 0.00 & 0.00 & 0.00 & 55.78 & 57.53 & 57.21 & 56.88 & 59.33 & 26.07 \\
15.00 & 1e-2 & 0.00 & 0.00 & 0.00 & 0.00 & 0.00 & 0.00 & 58.13 & 60.35 & 59.59 & 59.15 & 62.07 & 27.21 \\
15.00 & 0.10 & 0.00 & 0.00 & 0.00 & 0.00 & 0.00 & 0.00 & 61.95 & 62.48 & 63.50 & 63.95 & 67.78 & 29.06 \\
15.00 & 0.20 & 0.00 & 0.00 & 0.00 & 0.00 & 0.00 & 0.00 & 64.67 & 65.00 & 65.56 & 64.52 & 70.27 & 30.00 \\
15.00 & 0.40 & 0.00 & 0.00 & 21.86 & 0.00 & 0.00 & 20.35 & 66.25 & 69.23 & 69.03 & 66.90 & 74.27 & 35.26 \\
15.00 & 0.60 & 0.00 & 18.03 & 33.69 & 0.00 & 16.59 & 32.91 & 67.80 & 71.97 & 70.96 & 69.78 & 77.80 & 41.78 \\
15.00 & 0.80 & 18.00 & 31.12 & 49.75 & 14.04 & 28.11 & 46.88 & 70.48 & 74.10 & 74.24 & 72.33 & 79.65 & 50.79 \\
\midrule
\bottomrule
\end{tabular}
}
\label{tab:vit}
\end{table*}

\clearpage

\vfill

\clearpage

%
\appsection{Supplementary Details for Regime~(i): DP-RAG}
\label{app:sec:supplementary-regime1-experiments-DP-RAG}

We evaluate a single-query ($T=1$) DP-RAG retrieval primitive. For each query, the mechanism releases a differentially private query-to-collection score vector and then applies standard thresholded top-$k$ retrieval to the privatized scores.

\appsubsection{Privacy Object, Adjacency, and Retrieval Scores}
\label{app:ssec:regime1_privacy_object}

In our experiments, the source corpus is public Wikipedia text. The DP guarantee therefore concerns the released similarity-score vector and index-level functions of that vector, such as rankings, top-$k$ indices, and thresholded retrieval decisions. 
It does not claim secrecy of the Wikipedia text itself.

\paragraph{Setting.}
We work in the central model, where a trusted server holds an indexed retrieval corpus and releases only the \textit{retrieval output} (a set of top-$k$ chunks) and the \textit{generated answer}. For each evaluation instance, we construct an instance-specific knowledge base $\mathsf{KB}=\{\mathtt{d}_1,\dots,\mathtt{d}_n\}$, where each record $\mathtt{d}_i\in\mathcal{D}$ is one indexed document chunk (raw text, i.e., a finite token/string sequence) and $n$ is the number of chunks produced by the scrape--chunk--index pipeline described in Sec.~\ref{app:ssec:regime1_corpus_construction}. The protected object is the indexed retrieval representation used to compute query--corpus similarity scores.

\paragraph{Adjacency.}
We use record-level replacement adjacency on chunks. Two knowledge bases $\mathsf{KB} \sim\mathsf{KB}'$ are neighboring if they differ in exactly one chunk, i.e., $\mathtt{d}_j \neq \mathtt{d}'_j$ for one $j\in[n]$, $\mathtt{d}_i=\mathtt{d}'_i$, for all $i \neq j$. 
The user query is treated as public input. Protecting the query itself would require a different threat model, such as local or distributed privacy, and is outside the scope of this experiment.

\paragraph{Embedding map and score vector.}
Let $\phi:\text{(text)}\to\mathbb{R}^d$ denote the document embedding map.
Each chunk $\mathtt{d}_i$ is embedded as $\mathbf{x}_i \coloneq \phi(\mathtt{d}_i) \in\mathbb{R}^d$ and then clipped to the unit ball as $\mathbf{x}_i \leftarrow \mathbf{x}_i / \max\{1,\| \mathbf{x}_i \|_2\}$, so $\| \mathbf{x}_i\|_2 \le 1, \forall i$.
Given a public prompt, we form a retrieval query embedding $\mathbf{q} \coloneq \phi_{\mathrm{qry}}(\text{prompt})$ and apply the same clipping rule as $\mathbf{q} \leftarrow \mathbf{q} / \max\{1,\| \mathbf{q} \|_2\}$, so that $\|\mathbf{q}\|_2\le 1$.
We define the retrieval score for chunk $i\in[n]$ as $s_i(\mathbf{q})\; \coloneqq\; \langle \mathbf{q} , \mathbf{x}_i\rangle \in [-1,1]$, and write $\mathbf{s} (\mathbf{q}) = (s_1(\mathbf{q}), \dots, s_n(\mathbf{q}))\in[-1,1]^n$.

\appsubsection{DP-RAG Mechanism and Privacy Calibration}
\label{app:ssec:regime1_mechanism_calibration}

\paragraph{\textsc{ScoreShield} Mechanism.}
We privatize the complete query-to-collection score vector using the Gaussian mechanism followed by entrywise clipping:
\begin{equation}
\widehat{\mathbf{s}}( \mathbf{q} )
\;\coloneqq\;
\mathsf{clip}_{[-1,1]^n} \bigl( \mathbf{s}(\mathbf{q})+ \mathbf{w}\bigr),
\qquad
\mathbf{w} \sim \mathcal{N}( \mathbf{0}, \sigma^2_{\varepsilon,\delta} \;\mathbf{I}_n),
\label{eq:dp_rag_scores_rewrite}
\end{equation}
where $\mathsf{clip}_{[-1,1]^n}$ denotes entrywise clamping to the interval $[-1,1]$.
Retrieval is then performed on $\widehat{\mathbf{s}}( \mathbf{q} )$ using the same thresholded top-$k$ rule as in the non-private pipeline. 
Finally, the generator receives the retrieved text and produces an answer.
Any index-level output computed from $\widehat{\mathbf{s}}(\mathbf{q})$, including rankings, top-$k$ index lists, and thresholded index sets, is an $(\varepsilon,\delta)$-DP post-processing of the privatized score vector.

If the retrieved text is not itself part of the private dataset, then the retrieved indices/text and the generated answer are measurable functions of $\widehat{\mathbf{s}}(\mathbf{q})$, the public prompt, and the mechanism's internal randomness, they are post-processings of $\widehat{\mathbf{s}}(\mathbf{q})$ and inherit the same $(\varepsilon,\delta)$-DP guarantee\footnote{DP post-processing allows arbitrary functions of the DP output that do not additionally access the private dataset. If the corpus text is private and differs across neighboring KBs, then $\mathtt{index} \mapsto \mathtt{chunk~text}$ is an additional dataset access, so \textsc{ScoreShield} lone does not imply end-to-end DP for revealed chunk text or for the final answer.}.

\begin{remark}[ScoreShield Scope of the DP guarantee]
\textsc{ScoreShield} guarantees DP for the released score vector and for index-level functions of that vector. In our DP-RAG experiments, this is sufficient because the underlying corpus text is public.
If the corpus text itself is private under the adjacency relation, then releasing raw retrieved chunks (or an answer generated from those chunks) generally requires additional DP mechanisms beyond score release (e.g., DP generation/aggregation or one-time corpus privatization). ScoreShield alone guarantees privacy for the released score object and index-level functions of it, but does not provide end-to-end DP for revealed chunk contents or for generated answers conditioned on those contents.
\end{remark}

\paragraph{Global $\ell_2$-sensitivity under record replacement.}
Fix a public query $\mathbf{q}$ with $\| \mathbf{q} \|_2\le 1$. For neighboring KBs differing only in record $j$, the score vector differs in exactly one coordinate $s_i( \mathbf{q} )= s_i'( \mathbf{q} )\;\; (i \neq j)$, $s_j( \mathbf{q} )-s_j'( \mathbf{q} )=\langle \mathbf{q} , \mathbf{x}_j - \mathbf{x}_j'\rangle$.
Hence
\begin{equation}
\| \mathbf{s}(\mathbf{q}) - \mathbf{s}'(\mathbf{q})\|_2
= |\langle \mathbf{q}, \mathbf{x}_j - \mathbf{x}_j' \rangle|
\le \|\mathbf{q}\|_2\,\|\mathbf{x}_j - \mathbf{x}_j'\|_2
\le 2,
\label{eq:dp_rag_sensitivity_rewrite}
\end{equation}
Thus the global $\ell_2$-sensitivity is at most $\Delta_{\mathsf{query}}=2$.

\paragraph{Gaussian calibration.}
Using Lemma~\ref{lem:gaussian}, we set $\sigma^2_{\varepsilon,\delta} \;=\; c_{\varepsilon,\delta}\,\Delta_{\mathsf{query}}^2$, $c_{\varepsilon,\delta}\coloneqq \frac{2\log(2/\delta)}{\varepsilon^2}$, with $\Delta_{\mathsf{query}} =2$. This gives $\sigma^2 = 4 c_{\varepsilon,\delta}$ and $\sigma_{\varepsilon,\delta} = 2\sqrt{2\log(2/\delta)}/\varepsilon$.

\appsubsection{Experimental Protocol}
\label{app:ssec:regime1_corpus_and_baselines}

We evaluate a single-query DP-RAG retrieval primitive on the FRAMES benchmark~\citep{krishna2025fact}. For each query, the retrieval module computes query--corpus similarity scores, optionally privatizes these scores, and then applies the same thresholded top-$k$
retrieval rule to either the clean or privatized score vector.

\subsubsection{Run-Level Corpus Construction}
\label{app:ssec:regime1_corpus_construction}

Each FRAMES example contains a set of Wikipedia pages in its \texttt{wiki\_links} field. For a fixed evaluation run, we collect the union of all linked Wikipedia pages appearing in the selected examples and construct a single run-level retrieval corpus. Thus, RAG retrieves from one
global chunk index for the run, rather than from a question-specific corpus.

For each page, we scrape the article text using the \texttt{wikipedia} API and convert the content to plain text using \texttt{BeautifulSoup}. The text is split into overlapping token windows of approximately 2000 tokens with 200-token overlap, where token counts are computed using the embedding tokenizer. Each chunk is embedded together with its page title using the
retrieval encoder and the document instruction \texttt{"Retrieval-document"}. We denote the resulting chunk embeddings by $\mathbf{x}_1,\dots,\mathbf{x}_n \in \mathbb{R}^d$, where $n$ is the number of chunks in the run-level corpus.

\subsubsection{%
  Retrieval Scores and Thresholded Top-\texorpdfstring{$k$}{k} Rule%
}
\label{app:ssec:regime1_retrieval_baselines}

Given a query $\mathbf{q}$, we embed it using the retrieval encoder and the query instruction \texttt{"Retrieval-query"}. The retrieval score for chunk $i$ is the cosine similarity $s_i(\mathbf{q}) = \langle \mathbf{q}, \mathbf{x}_i\rangle$ , $i\in[n]$, where the query and chunk embeddings are normalized before computing the inner product. We write $\mathbf{s}(\mathbf{q}) = (s_1(\mathbf{q}),\dots,s_n(\mathbf{q})) \in [-1,1]^n$ .

For a generic score vector $\mathbf{z}\in[-1,1]^n$, define the eligible
index set
\begin{equation}
\mathcal{I}_\tau(\mathbf{z})
\coloneqq
\{i\in[n]: z_i\ge \tau\}.
\end{equation}
Let $\prec_{\mathbf{z}}$ denote the strict total order on $\mathcal{I}_\tau(\mathbf{z})$ that sorts indices by decreasing score and breaks ties by smaller index. That is, for $i\neq j$,
\begin{equation}
i \prec_{\mathbf{z}} j
\quad\Longleftrightarrow\quad
\bigl(z_i > z_j\bigr)
\ \text{or}\
\bigl(z_i=z_j \ \text{and}\ i<j\bigr).
\end{equation}
Let $i_{(1)}(\mathbf{z}),\dots, i_{(|\mathcal{I}_\tau(\mathbf{z})|)}(\mathbf{z})$ be the elements of $\mathcal{I}_\tau(\mathbf{z})$ listed according to $\prec_{\mathbf{z}}$. The thresholded top-$k$ retrieval operator returns
\begin{equation}
\mathsf{TopK}_{k,\tau}(\mathbf{z})
\coloneqq
\bigl(i_{(1)}(\mathbf{z}),\dots,i_{(m)}(\mathbf{z})\bigr),
\qquad
m=\min\{k,|\mathcal{I}_\tau(\mathbf{z})|\}.
\end{equation}
We use $k= 20$ and $\tau= 0.35$.

\subsubsection{Evaluation Conditions}
\label{app:ssec:regime1_eval_conditions}

We compare the following conditions.

\begin{itemize}[leftmargin=*]
\item
\textbf{No-context baseline.}
The generator receives only the user question. No retrieved document text is included in the prompt.
\item
\textbf{Oracle linked-page context.}
The generator receives text from the Wikipedia pages listed in the current example's \texttt{wiki\_links} field. The concatenated article text is truncated to 10,000 generator tokens before being inserted into the prompt. This condition measures performance when the model is given the benchmark linked evidence, subject to the same context-length constraint.
\item
\textbf{RAG with clean retrieval.}
The retrieval module computes the clean score vector $\mathbf{s}(\mathbf{q})$ over the run-level corpus and returns $\mathsf{TopK}_{k,\tau}(\mathbf{s}(\mathbf{q}))$. The corresponding chunks are inserted into the generator prompt in retrieved order. If no chunk satisfies the threshold, the prompt states that no document was retrieved.
\item
\textbf{DP-RAG with noisy retrieval.}
The retrieval module first privatizes the score vector using
\begin{equation}
\widehat{\mathbf{s}}(\mathbf{q})
= \mathsf{clip}_{[-1,1]^n}
\bigl(\mathbf{s}(\mathbf{q})+\mathbf{w}\bigr),
\qquad
\mathbf{w}\sim
\mathcal{N}\!\left(\mathbf{0},
\sigma_{\varepsilon,\delta}^2 \mathbf{I}_n\right),
\end{equation}
where $\sigma_{\varepsilon,\delta} = \frac{2\sqrt{2\log(2/\delta)}}{\varepsilon}$. Retrieval then returns $\mathsf{TopK}_{k,\tau}(\widehat{\mathbf{s}}(\mathbf{q}))$. Thus, clean RAG and DP-RAG differ only in whether the retrieval rule is applied to $\mathbf{s}(\mathbf{q})$ or to $\widehat{\mathbf{s}}(\mathbf{q})$.
\end{itemize}

\subsubsection{Models and Evaluation}
\label{app:ssec:regime1_models_eval}

We use \texttt{vLLM} for generation and \texttt{sentence-transformers} utilities for embedding and dense retrieval. 
The retrieval backbone encoder is \texttt{Embedding Gemma}~\citep{embeddinggemma2025} (\emph{i.e.,} EG300M) or \textsc{Qwen3-VL-Embedding}~\citep{qwen3vlembedding} (\emph{i.e.,} Q3VL-E2B). 
The evaluated generators are listed in Table~\ref{tab:dprag-accuracy-summary}. 
Prompts are rendered using the chat template associated with each generator. Decoding hyperparameters, including temperature, top-$p$, and maximum generation length, are fixed across conditions for a given generator.

Because FRAMES contains open-ended answers, exact string matching is not reliable. We therefore use an LLM judge. For each example, the judge receives the question, the reference answer, and the generated answer, and returns a binary match decision. We report answer accuracy as the percentage of examples marked correct. Latency is measured only for answer generation;
it excludes corpus construction, embedding, retrieval-index construction, and judging.

\begin{figure}[!t]
    \centering
    \includegraphics[width=0.99\linewidth]{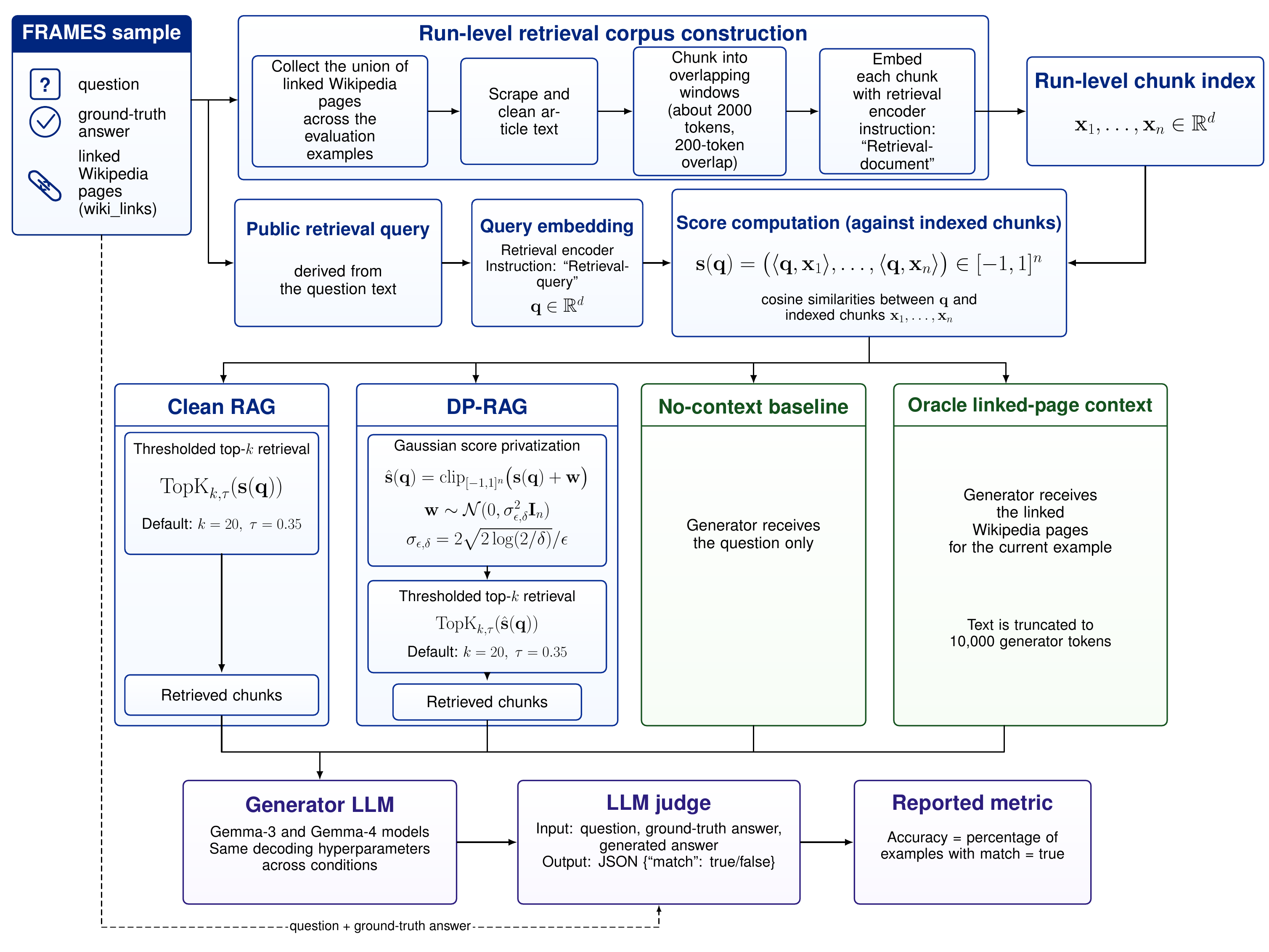}
    \caption{Experimental protocol for the FRAMES DP-RAG evaluation. A run-level retrieval corpus is constructed from the union of linked Wikipedia pages across the evaluation examples and embedded into a global chunk index $\mathbf{x}_1,\ldots,\mathbf{x}_n.$ For each public retrieval query $\mathbf{q}$, clean RAG applies thresholded top-\(k\) retrieval to the clean score vector \(\mathbf{s}(\mathbf{q})\), whereas DP-RAG first applies Gaussian score privatization and retrieves from $\widehat{\mathbf{s}}(\mathbf{q})$. The no-context baseline, clean RAG, DP-RAG, and oracle linked-page condition are evaluated with the same generator and LLM-judge protocol.}
\label{fig:dprag-experimental-protocol}
\end{figure}

Figure~\ref{fig:dprag-experimental-protocol} summarizes the experimental protocol used for the FRAMES evaluation.

\appsubsection{Empirical Results}

\paragraph{Evaluation protocol and retrieval database size.}
For each question in FRAMES, we extract the associated Wiki links and chunk the corresponding documents with a overlap based on the maximum context length supported by the embedding model. The resulting chunks define the retrieval database queried at inference time. Therefore, the number
of database entries depends on the embedding extractor: EG300M produces 22,880 entries, while Q3VL-E2B produces 6,892 entries (with a fixed number number of token overlaps), due to its larger supported input length. The privacy parameters $(\epsilon,\delta)$ only affect the noisy retrieval mechanism. They do not change the underlying language model, the embedding extractor, the question set, or the oracle/no-context baselines. Consequently, oracle and baseline accuracies remain fixed for a given model--embedding
configuration, up to small variation due to repeated runs or generation/evaluation seeds.

\paragraph{Retrieval improves utility in the non-private setting.}
Across all configurations where non-noisy RAG is evaluated, adding retrieved context improves accuracy over the no-context baseline. For example, Gemma3-27B improves from $40.17$ to $60.92$, Gemma4-26B-A4B improves from $26.09$ to $62.26$ at threshold $0.25$ and top-$k=50$, and Q3-8B improves from $75.85$ to $80.58$. In all cases, non-noisy RAG remains below the oracle setting, which provides only the relevant evidence. This gap indicates that retrieval is useful, but also that the retrieved context still contains irrelevant or incomplete information compared with the oracle evidence.

\paragraph{Effect of generation and evaluation seed.}
Small differences in oracle and baseline accuracy across otherwise similar settings are caused by
run-level variation rather than by the privacy parameters. For instance, the first two Gemma3-12B
EG300M rows show very similar oracle and baseline accuracies, $72.09/45.87$ versus
$71.48/46.36$. This suggests that the effect of the generation/evaluation seed is minor compared
with the effect of adding retrieval or injecting noise into the retrieval scores.

\paragraph{Noisy retrieval exhibits a sharp privacy--utility tradeoff.}
Injecting noise into the retrieval scores can substantially degrade utility, especially at stronger
privacy settings. For Gemma4-26B-A4B with EG300M, noisy RAG at $\epsilon\in\{1,10\}$ achieves only
$4.61$--$6.07$ accuracy, far below the $26.09$ baseline. Similarly, Gemma4-E4B remains below $2$
accuracy for several noisy settings at $\epsilon\in\{1,10\}$. These results show that, when the
noise is too large, the retrieved context can become misleading enough that RAG performs worse
than using no retrieved context at all.

\paragraph{Privacy--utility operating points.}
As $\epsilon$ and $\delta$ increase, the amount of injected noise decreases. The noisy retrieval
mechanism therefore approaches the behavior of standard non-private RAG, but with weaker privacy
guarantees. \colorbox{green!12}{We mark in green} the noisy retrieval settings whose RAG accuracy
exceeds the corresponding no-context baseline. For example, Gemma3-12B with EG300M reaches $54.98$
accuracy at $(\epsilon,\delta)=(100,0.01)$, compared with a $46.36$ baseline and $55.46$
non-noisy RAG accuracy. Gemma3-4B improves from a $49.51$ baseline to $52.43$ under $(100,0.01)$.
Gemma4-26B-A4B improves from $26.09$ to $50.24$ at $(100,0.01)$, and Gemma4-E4B improves from $10.19$
to $39.68$ under the same setting. These operating points demonstrate that noisy retrieval can
preserve part of the utility gain from RAG, but only under relatively weak privacy parameters in
the present runs.

\paragraph{Main empirical takeaway.}
The results support three observations. First, retrieval is consistently beneficial when no noise is
added. Second, strong privacy settings can destroy retrieval quality and may reduce accuracy below
the no-context baseline. Third, weaker privacy settings can recover a substantial fraction of the
non-private RAG gain, indicating a clear privacy--utility tradeoff. We therefore do not interpret
the highlighted rows as evidence that private RAG always improves performance, but rather as
empirical operating points where the proposed noisy retrieval mechanism remains useful.

\begin{table}[ht]
\centering
\caption{Accuracy summary by model and noise setting. Highlighted rows have noise enabled and RAG accuracy above baseline accuracy.}
\label{tab:dprag-accuracy-summary}
\resizebox{\textwidth}{!}{%
\begin{tabular}{ll|rrrrr|rrr}
\toprule
Model & Embedding & $\delta$ & $\epsilon$ & Top-$k$ & Threshold & DB Entries & Oracle Acc. & Baseline Acc. & RAG Acc. \\
\midrule
Gemma3-12B & EG300M & N/A & N/A & 50 & 0.25 & 22880 & 72.09 & 45.87 & 55.46 \\
Gemma3-12B & EG300M & 1e-05 & 1.0 & 50 & 0.25 & 22880 & 71.48 & 46.36 & 39.44 \\
Gemma3-12B & EG300M & 0.001 & 10.0 & 50 & 0.25 & 22880 & 71.48 & 46.36 & 40.41 \\
Gemma3-12B & EG300M & 0.01 & 10.0 & 50 & 0.25 & 22880 & 71.48 & 46.36 & 39.68 \\
\rowcolor{green!12}
Gemma3-12B & EG300M & 0.001 & 100.0 & 50 & 0.25 & 22880 & 71.48 & 46.36 & 52.31 \\
\rowcolor{green!12}
Gemma3-12B & EG300M & 0.01 & 100.0 & 50 & 0.25 & 22880 & 71.48 & 46.36 & 54.98 \\
\midrule
Gemma3-12B & Q3VL-E2B & N/A & N/A & 10 & 0.25 & 6892 & 71.48 & 46.36 & 56.92 \\
Gemma3-12B & Q3VL-E2B & 1e-05 & 1.0 & 10 & 0.25 & 6892 & 71.48 & 46.36 & 41.63 \\
Gemma3-12B & Q3VL-E2B & 1e-06 & 1.0 & 10 & 0.25 & 6892 & 71.48 & 46.36 & 42.23 \\
Gemma3-12B & Q3VL-E2B & 1e-05 & 10.0 & 10 & 0.25 & 6892 & 71.48 & 46.36 & 41.63 \\
Gemma3-12B & Q3VL-E2B & 1e-06 & 10.0 & 10 & 0.25 & 6892 & 71.48 & 46.36 & 41.75 \\
\midrule
Gemma3-27B & EG300M & N/A & N/A & 20 & 0.35 & 22880 & 72.82 & 40.17 & 60.92 \\
Gemma3-27B & EG300M & 1e-05 & 1.0 & 20 & 0.35 & 22880 & 72.82 & 40.17 & 39.68 \\
Gemma3-27B & EG300M & 1e-06 & 1.0 & 20 & 0.35 & 22880 & 72.82 & 40.17 & 37.86 \\
\midrule
Gemma3-4B & EG300M & 0.001 & 10.0 & 50 & 0.25 & 22880 & 65.66 & 49.51 & 46.60 \\
Gemma3-4B & EG300M & 0.01 & 10.0 & 50 & 0.25 & 22880 & 65.66 & 49.51 & 47.94 \\
\rowcolor{green!12}
Gemma3-4B & EG300M & 0.001 & 100.0 & 50 & 0.25 & 22880 & 65.66 & 49.51 & 52.18 \\
\rowcolor{green!12}
Gemma3-4B & EG300M & 0.01 & 100.0 & 50 & 0.25 & 22880 & 65.66 & 49.51 & 52.43 \\
\midrule
Gemma4-26B-A4B & EG300M & N/A & N/A & 20 & 0.35 & 22880 & 73.30 & 26.09 & 52.55 \\
Gemma4-26B-A4B & EG300M & 1e-05 & 1.0 & 20 & 0.35 & 22880 & 73.30 & 26.09 & 4.61 \\
Gemma4-26B-A4B & EG300M & 1e-06 & 1.0 & 20 & 0.35 & 22880 & 73.30 & 26.09 & 4.85 \\
Gemma4-26B-A4B & EG300M & 1e-05 & 10.0 & 20 & 0.35 & 22880 & 73.30 & 26.09 & 4.61 \\
Gemma4-26B-A4B & EG300M & 1e-06 & 10.0 & 20 & 0.35 & 22880 & 73.30 & 26.09 & 4.61 \\
Gemma4-26B-A4B & EG300M & N/A & N/A & 50 & 0.25 & 22880 & 73.30 & 26.09 & 62.26 \\
Gemma4-26B-A4B & EG300M & 1e-05 & 1.0 & 50 & 0.25 & 22880 & 73.30 & 26.09 & 5.34 \\
Gemma4-26B-A4B & EG300M & 1e-06 & 1.0 & 50 & 0.25 & 22880 & 73.30 & 26.09 & 4.73 \\
Gemma4-26B-A4B & EG300M & 0.01 & 10.0 & 50 & 0.25 & 22880 & 73.30 & 26.09 & 4.85 \\
Gemma4-26B-A4B & EG300M & 1e-06 & 10.0 & 50 & 0.25 & 22880 & 73.30 & 26.09 & 6.07 \\
\rowcolor{green!12}
Gemma4-26B-A4B & EG300M & 0.01 & 100.0 & 50 & 0.25 & 22880 & 73.30 & 26.09 & 50.24 \\
\rowcolor{green!12}
Gemma4-26B-A4B & EG300M & 1e-06 & 100.0 & 50 & 0.25 & 22880 & 73.30 & 26.09 & 36.04 \\
\midrule
Gemma4-31B & EG300M & 1e-05 & 1.0 & 20 & 0.35 & 22880 & 80.58 & 29.98 & 0.85 \\
Gemma4-31B & EG300M & N/A & N/A & 50 & 0.25 & 22880 & 80.58 & 29.98 & 72.57 \\
\midrule
Gemma4-E4B & EG300M & N/A & N/A & 20 & 0.35 & 22880 & 61.04 & 9.83 & 41.38 \\
Gemma4-E4B & EG300M & 1e-05 & 1.0 & 20 & 0.35 & 22880 & 61.04 & 9.83 & 0.97 \\
Gemma4-E4B & EG300M & 1e-06 & 1.0 & 20 & 0.35 & 22880 & 61.04 & 9.83 & 0.61 \\
Gemma4-E4B & EG300M & 1e-05 & 10.0 & 20 & 0.35 & 22880 & 61.04 & 9.83 & 1.46 \\
Gemma4-E4B & EG300M & 1e-06 & 10.0 & 20 & 0.35 & 22880 & 61.04 & 9.83 & 1.21 \\
Gemma4-E4B & EG300M & N/A & N/A & 50 & 0.25 & 22880 & 62.01 & 10.19 & 47.82 \\
Gemma4-E4B & EG300M & 0.01 & 10.0 & 50 & 0.25 & 22880 & 62.01 & 10.19 & 1.82 \\
Gemma4-E4B & EG300M & 1e-06 & 10.0 & 50 & 0.25 & 22880 & 62.01 & 10.19 & 1.70 \\
\rowcolor{green!12}
Gemma4-E4B & EG300M & 0.01 & 100.0 & 50 & 0.25 & 22880 & 62.01 & 10.19 & 39.68 \\
\rowcolor{green!12}
Gemma4-E4B & EG300M & 1e-06 & 100.0 & 50 & 0.25 & 22880 & 62.01 & 10.19 & 28.03 \\
\midrule
Q3-8B & Q3VL-E2B & N/A & N/A & 10 & 0.25 & 6892 & 91.02 & 75.85 & 80.58 \\
Q3-8B & Q3VL-E2B & 1e-05 & 1.0 & 10 & 0.25 & 6892 & 91.02 & 75.85 & 74.27 \\
Q3-8B & Q3VL-E2B & 1e-06 & 1.0 & 10 & 0.25 & 6892 & 91.02 & 75.85 & 73.54 \\
Q3-8B & Q3VL-E2B & 1e-05 & 10.0 & 10 & 0.25 & 6892 & 91.02 & 75.85 & 74.76 \\
\rowcolor{green!12}
Q3-8B & Q3VL-E2B & 1e-06 & 10.0 & 10 & 0.25 & 6892 & 91.02 & 75.85 & 76.46 \\
\bottomrule
\end{tabular}
}%
\end{table}

\clearpage
\appsection{Supplementary Details for Regime~(ii): Omitted Theorems, Propositions, Proofs and Lemmas}
\label{app:sec:supplementary-regime2-theory}

This appendix complements Sec.~\ref{subsec:all-pairs}.
It first derives the sensitivity for regime~(ii) and provides an extended presentation of the fast projection via alternating steps  algorithm.
Next, we provide the full statement and proof of the averaged–alternating-projection theorem used in Sec.~\ref{subsec:all-pairs}. We then contrast our projection scheme with the perturb-and-project method of Cohen-Addad \textit{et al.}~\citep{cohen2024perturb}, detailing  
(i) the respective feasible sets and adjacency models,  
(ii) the decomposition choices that make the projection steps cheap,  
(iii) the role of the unit-diagonal constraint, and  
(iv) complexity and convergence rates.

\appsubsection{Global Frobenius Sensitivity Under Record-Level Adjacency}
\label{app:ssec:global_Frobenius_sensitivity}

Under record-level adjacency model, $\mathbf{E}$ and $\mathbf{E}'$ differ in at most a single row $i$. Let $\boldsymbol\delta=\mathbf{e}'_{i}-\mathbf{e}_{i}$ denote the row–difference, with  $\|\boldsymbol{\delta} \|_{2}\le 2$, and embed it into an $n\times d$ matrix $\boldsymbol{\Delta} \; \coloneqq \; \bigl[\mathbf{0};\dots;\boldsymbol{\delta}^{ \top};\dots;\mathbf{0}\bigr]$.
Hence, we have $\mathbf{E}' \;=\; \mathbf{E} + \boldsymbol{\Delta}$.
Then 
\begin{align}
f_{\mathrm{full}}(\mathbf{E})-f_{\mathrm{full}}(\mathbf{E}')
=\mathbf{E}\mathbf{E}^{\top}
-(\mathbf{E}+\boldsymbol\Delta)(\mathbf{E}+\boldsymbol\Delta)^{\top}
= - \mathbf{E}\boldsymbol\Delta^{\top}
- \boldsymbol\Delta \mathbf{E}^{\top}
- \boldsymbol\Delta\boldsymbol\Delta^{\top}.
\label{eq:full-decomp}
\end{align}
Equivalently, $\mathbf{S}'-\mathbf{S} = \mathbf{E}'\mathbf{E}'^{\top}-\mathbf{E}\mathbf{E}^{\top}
= \mathbf{E}\boldsymbol\Delta^{\top}+\boldsymbol\Delta \mathbf{E}^{\top}+\boldsymbol\Delta\boldsymbol\Delta^{\top}$.
Because $\boldsymbol{\Delta}$ has exactly one non-zero row, 
above decomposition populates only row~$i$ and column~$i$ of the $n\times n$ difference matrix. Writing $v_{j}  \coloneqq  \boldsymbol{\delta}^{\top} \mathbf{e}_{j}$ and $\mathbf{v} \coloneqq (v_{1}, \dots, v_{n})^{\top}$, the diagonal cancels via $ (\mathbf{S}'-\mathbf{S})_{ii} =  2\,\mathbf{e}_{i}^{\top}\boldsymbol{\delta}+\|\boldsymbol{\delta}\|_{2}^{2} = 0$ (since $\|\mathbf{e}_i'\|_2^2=\|\mathbf{e}_i\|_2^2=1$).
Hence
$\mathbf{S}'-\mathbf{S} = \mathbf{v} \,\mathbf{e}_{(i)}^{ \top}
\;+\; \mathbf{e}_{(i)}\,\mathbf{v}^{ \top}$,
where $\mathbf{e}_{(i)}\in\mathbb{R}^{n}$ denotes the $i$-th canonical basis vector.
%
%
Only the off-diagonal pairs $(i,j)$ and $(j,i)$, $j\neq i$, survive. Hence
\begin{subequations}
\begin{align}
    \| \mathbf{S}'-\mathbf{S} \|_{F}^{2}
    &= 2\sum_{j\neq i} v_j^{2} = 2 \sum_{j \neq i} \left( \boldsymbol{\delta}^\top \mathbf{e}_j \right)^2\\
    &= 2  \boldsymbol{\delta}^\top \! \Big( \sum_{j \neq i} \mathbf{e}_j \mathbf{e}_j^\top \Big) \boldsymbol{\delta}
     \leq  2 (n-1) {\Vert \boldsymbol{\delta} \Vert}_2^2
     \le  8(n-1). 
\end{align}    
\end{subequations}
Therefore the global Frobenius sensitivity is $\Delta_{f,F}  \eqqcolon  \Delta_{\mathrm{full}} = 2\sqrt{2(n-1)}$.

\appsubsection{DP Release Mechanism for the Full Pairwise Similarity Score Matrix}

\begin{algorithm}[h]
\caption{DP All-Pairs Similarity Matrix Release (regime~(ii))}
\label{alg:all-pairs-regime2}
\begin{algorithmic}[1]
\State \textbf{Input:} $\mathbf{E} \in \mathbb{R}^{n \times d}$, $\varepsilon > 0$, $\delta \in (0,1)$, sensitivity $\Delta > 0$
\State \textbf{Output:} $\widehat{\mathbf{S}} \in \mathcal{C}_{\mathsf{coll}} \subseteq \mathbb{R}^{n \times n}$
\State Construct: $\mathbf{S}\gets\mathbf{E}\mathbf{E}^\top$.
\State Gaussian calibration: set $\sigma^2_{\varepsilon, \delta} \;\gets\;  c_{\varepsilon, \delta} \Delta^2$
\State Sample noise $\mathbf{W} \sim \mathcal{N}\left( \mathbf{0}, \sigma^2 \mathbf{I}_{n \times n}\right)$
\State Perturb with noise: $\mathbf{S}' = \mathbf{S} + \mathbf{W} $
\State Project: $\widehat{\mathbf{S}}  =   \mathsf{proj}_{\mathcal{C}_{\mathsf{coll}}} (\mathbf{S}')  \!= \! \arg\min_{\mathbf{A} \in  \mathcal{C}_{\mathsf{coll}}} \! \|\mathbf{A} \! - \!  \mathbf{S}'\|_{\mathrm{F}}^2$
\State \textbf{Return} $\widehat{\mathbf{S}}$
\end{algorithmic}
\end{algorithm}

\appsubsection{Privacy Guarantee of ScoreShield for Regime~(ii)}

\begin{theorem}[Privacy Guarantee of Full Pairwise Similarity Score Matrix Release]
\label{app:thm:gram-dp-guarantee}
Let $f_{\mathsf{full}}:\mathcal{E}\to \mathbb{R}^{n\times n}$ be
$f_{\mathsf{full}}(\mathbf{E})=\mathbf{S}=\mathbf{E}\mathbf{E}^\top$.
Fix an adjacency relation $\sim$ on $\mathcal{E}$ and assume that for some $\Delta>0$,
\begin{equation}
\|f_{\mathsf{full}}(\mathbf{E})-f_{\mathsf{full}}(\mathbf{E}')\|_{\mathrm{F}}
\le \Delta,
\qquad \forall \, \mathbf{E}\sim\mathbf{E}'.
\end{equation}
Let $\varepsilon > 0$, $\delta \in (0,1)$, $c_{\varepsilon, \delta} \coloneqq 2 \log(2/\delta)/ \varepsilon^2$, and set $\sigma^2 \coloneqq c_{\varepsilon,\delta}\Delta^2$. Let $\mathbf{W}\in\mathbb{R}^{n\times n}$ have i.i.d. entries $W_{ij}\sim\mathcal{N}(0,\sigma^2)$. Let $\mathsf{proj}_{\mathcal{C}}$ be any (possibly randomized) measurable post-processing that depends on $\mathbf{E}$ only through its input argument. Define the release $\widehat{\mathbf{S}}= \mathcal{M}_{\mathsf{coll}} (\mathbf{E})  \;\coloneqq\; \mathsf{proj}_{\mathcal{C}}\!\left(f_{\mathsf{full}}(\mathbf{E}) + \mathbf{W}\right)$. Then the mechanism $\mathbf{E}\mapsto \widehat{\mathbf{S}}$ is $(\varepsilon,\delta)$-DP with respect to $\sim$.
\end{theorem}

\begin{proof}
By assumption, the statistic $f_{\mathsf{full}}:\mathcal{E}\to\mathbb{R}^{n\times n}$ has global Frobenius sensitivity $\Delta_{f,\mathrm{F}} \;\coloneqq\; \sup_{\mathbf{E} \sim \mathbf{E}'}\, \| f_{\mathsf{full}}(\mathbf{E}) - f_{\mathsf{full}}(\mathbf{E}') \|_{\mathrm{F}} \;\le\; \Delta$.
Consider the additive Gaussian mechanism $\mathcal{M}_{0}(\mathbf{E}) \; \coloneqq \; f_{\mathsf{full}}(\mathbf{E})+\mathbf{W}$, $ W_{ij}\stackrel{\mathrm{i.i.d.}}{\sim}\mathcal{N}(0,\sigma^{2})$, $\sigma^{2}=c_{\varepsilon,\delta}\Delta^{2}$.
Viewing $\mathbb{R}^{n\times n}$ as a Euclidean space equipped with the Frobenius norm, $\mathcal{M}_0$
is exactly the Gaussian mechanism for matrix-valued outputs (Corollary~\ref{cor:gaussian-matrix}).
Therefore, for every pair $\mathbf{E}\sim \mathbf{E}'$ and every measurable set $T\subseteq\mathbb{R}^{n\times n}$,
\begin{equation}
\mathsf{Pr}\!\left[\mathcal{M}_0(\mathbf{E})\in T\right]
\;\le\; e^{\varepsilon}\,\mathsf{Pr}\!\left[\mathcal{M}_0(\mathbf{E}')\in T\right]+\delta .
\end{equation}
Now define the released mechanism $\widehat{\mathbf{S}}= \mathcal{M}_{\mathsf{coll}} (\mathbf{E})  \;\coloneqq\; \mathsf{proj}\!\left(\mathcal{M}_0(\mathbf{E})\right) \;=\; \mathsf{proj}\!\left(f_{\mathsf{full}}(\mathbf{E})+\mathbf{W}\right)$, where $\mathsf{proj}$ is any measurable map that depends on $\mathbf{E}$ only through its input argument (possibly using additional randomness independent of $\mathbf{E}$). By the post-processing property (Lemma~\ref{lem:post-processing}), for every measurable set
$S\subseteq\mathcal{Y}$,
\begin{equation}
\mathsf{Pr} \left[ \mathcal{M}_{\mathsf{coll}}(\mathbf{E})\in S \right]
\;\le\;
e^{\varepsilon}\,\mathsf{Pr}\!\left[ \mathcal{M}_{\mathsf{coll}}(\mathbf{E}') \in S \right] + \delta,
\qquad \forall\,\mathbf{E}\sim \mathbf{E}'.
\end{equation}
Hence the mechanism $\mathbf{E}\mapsto \widehat{\mathbf{S}}$ is $(\varepsilon,\delta)$-DP with respect to $\sim$.
\end{proof}

In particular, under output-space (Gram) adjacency (Def.~\ref{def:gram-adjacency-mainbody}), the global sensitivity bound is $\Delta = \Delta_{\mathsf{G}}$ by definition of the adjacency radius. Under record-level (single-record replacement) adjacency, the global Frobenius sensitivity is
$\Delta = \Delta_{\mathrm{full}}= 2 \sqrt{2(n-1)}$.

\appsubsection{Fast Projection via Averaged Alternating Projections}

Let $\mathbf{S}'= \mathbf{S} +\mathbf{W}\in \mathbb{R}^{n\times n}$ be a perturbed Gram matrix, where $\mathbf{S}= \mathbf{E}\mathbf{E}^{\top}$ is the clean cosine Gram matrix and ${\Vert \mathbf{e}_i\Vert}_2=1 , \forall i$. Our goal is to project $\mathbf{S}'$ onto the cosine-Gram feasibility set $\mathcal{C}_{\mathsf{coll}}  \coloneqq  \bigl\{ \mathbf{S}\in \mathbb{R}^{n\times n}: \, \mathbf{S}\succeq0,\; S_{ii}=1\;(1 \le  i \le  n),\; |S_{ij}|\le 1\;(i \ne  j) \bigr\}\subset \mathbb{R}^{n\times n}$.
The exact projection onto the cosine-Gram feasible set $\mathcal{C}_{\mathsf{coll}}$ under the Frobenius norm is the metric projection onto the elliptope and in general requires solving an SDP.
Instead we compute an approximately feasible point by iterating a Krasnosel'ski\u{\i}-Mann averaged projector \citep{mann1953mean, ma1955two},
%
and hence replace the direct projection by alternating projections onto two closed convex sets, each admitting a closed-form projector\footnote{%
In our mechanism definition and in the geometric analysis, $\mathsf{proj}_{\mathcal{C}_{\mathsf{coll}}}$ denotes the Euclidean projector (metric projection) onto $\mathcal{C}_{\mathsf{coll}}$. In implementation, however, we enforce feasibility using a Krasnosel'ski\u{\i}–Mann \citep{mann1953mean, ma1955two} averaged-projection iteration that returns an \emph{approximately feasible} point in $\mathcal{C}_{\mathsf{coll}}$ but is not, in general, the metric projection of the noisy input onto $\mathcal{C}_{\mathsf{coll}}$. Since this feasibility map depends only on the perturbed matrix (and any internal randomness is independent of the data), it is post-processing and therefore does not affect the privacy guarantee.
}. We decompose $\mathcal{C}_{\mathsf{coll}} = \mathcal{K}_{+}^{\,n} \cap \mathcal{C}_{\mathrm{unit}}^n$ where 
\begin{equation}
\mathcal{K}_{+}^{\,n}  = \{\mathbf{S} \in \mathbb{R}^{n \times n}  \mid \mathbf{S}\succeq0\}, 
\qquad \mathcal{C}_{\mathrm{unit}}^n = \{ \mathbf{S} \in \mathbb{R}^{n \times n}  \mid \; S_{ii}=1, \, \vert S_{ij} \vert \le 1, (i\ne j)\}. 
\end{equation}
Starting from $\widehat{\mathbf{S}}_0\coloneqq \mathbf{S}'$, we iterate the equal-weights averaged map
\begin{equation}
\label{eq:aap-update}
\widehat{\mathbf{S}}_{t+1}
\;=\; \frac{1}{2}\Bigl(
\mathsf{proj}_{\mathcal{K}_+^{\,n}} \bigl(\mathsf{sym}(\widehat{\mathbf{S}}_{t})\bigr)
\;+\; \mathsf{proj}_{\mathcal{C}_{\mathrm{unit}}^{\,n}}\bigl(\mathsf{sym}(\widehat{\mathbf{S}}_{t})\Bigr),
\qquad \mathsf{sym}(\mathbf{Y})\coloneqq \tfrac12(\mathbf{Y}+\mathbf{Y}^\top).
\end{equation}
%

Under bounded linear regularity of $(\mathcal{K}_+^{\,n},\mathcal{C}_{\mathrm{unit}}^{\,n})$ on a ball containing the iterates, Theorem~\ref{thm:aap} yields a geometric contraction of the feasibility gap: $\mathsf{dist}(\widehat{\mathbf{S}}_{t},\mathcal{C}_{\mathsf{coll}})\le \rho^{t}\mathsf{dist}(\widehat{\mathbf{S}}_{0},\mathcal{C}_{\mathsf{coll}})$ for some $\rho\in(0,1)$. Consequently, to reach $\mathsf{dist}(\widehat{\mathbf{S}}_{t},\mathcal{C}_{\mathsf{coll}})\le \tau$ it suffices that $t \ge \log(\mathsf{dist}(\widehat{\mathbf{S}}_{0},\mathcal{C}_{\mathsf{coll}})/\tau)/\log(1/\rho)
= \mathcal{O}(\log(1/\tau))$ (see Corollary~\ref{cor:iter}). Each iteration is dominated by an $n\times n$ eigendecomposition for the PSD projection ($\mathcal{O}(n^3)$), so the total cost is $\mathcal{O}(n^3\log(1/\tau))$ operations to tolerance.

\noindent
\textit{Symmetrization.}
Before projecting onto $\mathcal{K}_+^{\,n}$, we need to replace $\widehat{\mathbf{S}}_t$ by $\mathbf{Y}=\mathsf{sym}(\widehat{\mathbf{S}}_t)$. By Lemma~\ref{lem:symmetrization}, the PSD projection depends only on the symmetric part under $\|\cdot\|_{\mathrm{F}}$, so $\mathsf{proj}_{\mathcal{K}_+^{\,n}}(\widehat{\mathbf{S}}_t)=\mathsf{proj}_{\mathcal{K}_+^{\,n}}(\mathbf{Y})$. This removes antisymmetric numerical artifacts that would otherwise inflate $\|\widehat{\mathbf{S}}_t\|_{\mathrm{F}}$.

\subsubsection{Projection onto the PSD Cone}

\textit{Step (i): Spectral Decomposition.}
Given a symmetric iterate $\mathbf{Y} = \mathbf{Y}^{\top} = \tfrac{1}{2} \bigl(\widehat{\mathbf{S}}_{t}+\widehat{\mathbf{S}}_{t}^{\top}\bigr) \in \mathtt{S}^{n}$, $\widehat{\mathbf{S}}_{0}  \coloneqq  \mathbf{S} + \mathbf{W}$, we compute its spectral decomposition $\mathbf{Y}= \mathbf{U}\,\mathsf{diag}(\lambda_{1},\dots,\lambda_{n})\mathbf{U}^{ \top}$, costing $\mathcal{O}(n^{3})$. All subsequent steps are performed in the eigenbasis $\mathbf{U}$.

\noindent
\textit{Step (ii): PSD Truncation.}  
The Frobenius–orthogonal projector onto the PSD cone solves $\mathop{\min}_{\mathbf{S} \succeq 0} {\Vert \mathbf{S} - \mathbf{Y} \Vert}_{\mathrm{F}}^2$ and is obtained as follows. Define $\lambda_{k}^{+} \coloneqq \max\{0, \lambda_{k}\},\; k = 1,\dots, n$, and let $\boldsymbol\lambda^{+}\coloneqq (\lambda_{1}^{+},\dots,\lambda_{n}^{+})$. The orthogonal projector onto the PSD cone is
\begin{equation}\label{eq:PSD-proj}
\mathsf{proj}_{\mathcal{K}_{+}^{\,n}}(\mathbf{Y}) =
\mathbf{U}\, \mathsf{diag}(\lambda_{1}^{+},\dots,\lambda_{n}^{+})\, \mathbf{U}^{\top}.
\end{equation}
This map is a firmly non-expansive projector (see Lemma~\ref{lem:proj-firm}).

\subsubsection{Projection Onto the Unit-Diagonal Box} 
\label{ssec:projection-unit-hyper-cube}

The set $ \mathcal{C}_{\mathrm{unit}}^{\,n}  \coloneqq \bigl\{\mathbf{S}\in \mathbb{R}^{n\times n}:\; |S_{ij}|\le 1\;(i\neq j), \; S_{ii}=1 \; (1\leq i \le n ) \bigr\}$ is an axis-aligned hyper-box with fixed diagonal.
Because the Frobenius norm decouples over coordinates, the orthogonal projector is entry-wise. For any $\mathbf{Y}\in \mathbb{R}^{n\times n}$ the projector onto $\mathcal{C}_{\mathrm{unit}}^{n}$ decouples entry-wise as:
\begin{eqnarray}
\bigl[\mathsf{proj}_{\mathcal{C}_{\mathrm{unit}}^{\,n}}(\mathbf{Y})\bigr]_{ij}\;=\;
\begin{cases}
    1, & i=j,\\[6pt]
    \mathsf{clip}(Y_{ij},-1,1), & i\neq j,
\end{cases}
\end{eqnarray}
where $\mathsf{clip}(y,-1,1) \coloneqq \max\{-1,\min\{1,y\}\}$.
Note that for each off-diagonal coordinate the convex problem $\mathop{\min}_{|z|\leq 1} {\left( z - Y_{ij} \right)}^2$ yields the clip operator. Moreover, the diagonal constraint is enforced exactly. This operation is firmly non-expansive (see Lemma~\ref{lem:proj-firm}), costs $\mathcal{O}(n^2)$ arithmetic operations and can be implemented in-place (with $\Theta(n^2)$ storage for the matrix itself).

Both projectors are firmly non-expansive; their averaged composition is an averaged non-expansive operator. Alternating these maps yields a Fej\'er‑monotone sequence with respect to the intersection, guaranteeing convergence (see App.~\ref{app:ssec:AAP-theoretical-guarantees} for details).

\begin{remark}
If one omits entrywise clipping (i.e., uses only $\mathrm{diag}(\mathbf{S})=\mathbf{1}$ as the second constraint), it can be useful to additionally enforce $\|\mathbf{S}\|_F\le n$ to keep iterates uniformly bounded. When $\mathsf{proj}_{\mathcal{C}_{\mathrm{unit}}^{\,n}}$ includes $|S_{ij}|\le 1$, the bound $\|\mathbf{S}\|_F\le n$ holds automatically.
Therefore, in order to project onto $\mathcal{K}_{+}^{\,n} \cap \mathcal{B}_{F}^{\,n}$, where $\mathcal{B}_{F}^n  \coloneqq \{ \mathbf{S} \in \mathbb{R}^{n \times n}  \mid {\Vert \mathbf{S} \Vert}_{\mathrm{F}} \leq n\},$ we need an additional Frobenius-ball rescaling step. 
To enforce this additional constraint, define
\begin{equation}
t \coloneqq   {\Vert \boldsymbol{\lambda}^+ \Vert}_2= \Bigl(\sum_{k=1}^{n}(\lambda_{k}^{+})^{2}\Bigr)^{1/2},
\qquad \alpha \coloneqq \min\Bigl\{1,\;\frac{n}{t}\Bigr\}.
\end{equation}
%
%
The convex program $\mathop{\min}_{\mathbf{S} \succeq 0, \, {\Vert \mathbf{S} \Vert}_{\mathrm{F}} \leq n} {\Vert \mathbf{S} - \mathsf{proj}_{\mathcal{K}_{+}^{\,n}} \left( \mathbf{Y} \right)\Vert}_{\mathrm{F}}^2$ decouples in the eigen-basis and yields a \textit{radial projection}. Because the eigenvectors are orthogonal, the joint projection onto the intersection \(\mathcal{K}_{+}^{\,n}\cap\mathcal{B}_{F}^{\,n}\) is obtained by scaling the positive eigenvalues:
\begin{equation}\label{eq:PSD-ball-proj}
\mathsf{proj}_{\mathcal{K}_{+}^{\,n} \, \cap \, \mathcal{B}_{F}^{\,n}}(\mathbf{Y})
\;=\;
\mathbf{U} \mathsf{diag}(\alpha\lambda_{1}^{+},\dots,\alpha\lambda_{n}^{+})\,
\mathbf{U}^{\top}\, .
\end{equation}
\end{remark}

\begin{theorem}[DP guarantee for fast AAP all-pairs Gram release]
\label{thm:dp-allpairs-aap}
Let $f_{\mathsf{full}}(\mathbf{E})=\mathbf{S}=\mathbf{E}\mathbf{E}^\top \in \mathtt{S}^n$ be the all-pairs cosine Gram statistic.
Fix an adjacency relation $\sim$ on embedding matrices and assume the global Frobenius sensitivity
$\Delta_{\mathsf{full}}
\;\coloneqq\;
\sup_{\mathbf{E}\sim \mathbf{E}'} \bigl\|f_{\mathsf{full}}(\mathbf{E})-f_{\mathsf{full}}(\mathbf{E}')\bigr\|_{\mathrm{F}}$
is finite. Let $\varepsilon>0$, $\delta\in(0,1)$, set $c_{\varepsilon,\delta}\coloneqq 2\log(2/\delta)/\varepsilon^2$ and
$\sigma^2 \coloneqq c_{\varepsilon,\delta}\Delta_{\mathsf{full}}^2$.
Sample $\mathbf{W}\in\mathbb{R}^{n\times n}$ with i.i.d.\ entries $W_{ij}\sim\mathcal{N}(0,\sigma^2)$ and define the symmetric noise
$\mathbf{G}\coloneqq \tfrac12(\mathbf{W}+\mathbf{W}^\top)\in\mathtt{S}^n$.
Let $\mathsf{proj}_{\mathcal{C}_{\mathsf{coll}}}:\mathtt{S}^n\to\mathtt{S}^n$ denote the (deterministic) output of Algorithm~\ref{app:alg:fast_projection} run for either (i) a fixed number of iterations $T$, or
(ii) any stopping time that is measurable with respect to the noisy input $\mathbf{S}+\mathbf{G}$
(equivalently, depends only on the iterates / $\mathbf{S}+\mathbf{G}$, not on the raw data).
Release $\widehat{\mathbf{S}}\;\coloneqq\;\mathsf{proj}_{\mathcal{C}_{\mathsf{coll}}}\!\bigl(f_{\mathsf{full}}(\mathbf{E})+\mathbf{G}\bigr)$.
Then $\mathbf{E}\mapsto \widehat{\mathbf{S}}$ is $(\varepsilon,\delta)$-DP.
\end{theorem}

\begin{proof}
By Corollary~\ref{cor:gaussian-symmetric} (Gaussian mechanism with symmetric averaging), the intermediate release
$f_{\mathsf{full}}(\mathbf{E})+\mathbf{G}$ is $(\varepsilon,\delta)$--DP when $\sigma^2=c_{\varepsilon,\delta}\Delta_{\mathsf{full}}^2$.
The mapping $\mathsf{proj}_{\mathcal{C}_{\mathsf{coll}}}$ (including any stopping rule measurable w.r.t. its input) is a measurable
post-processing of $f_{\mathsf{full}}(\mathbf{E})+\mathbf{G}$, hence the output $\widehat{\mathbf{S}}$ remains
$(\varepsilon,\delta)$-DP by Lemma~\ref{lem:post-processing}.
\end{proof}

\vfill

\clearpage

\subsubsection{Fast Averaged Alternating Projection Algorithm}

Our algorithm is provided in Algorithm~\ref{app:alg:fast_projection}.

\vspace{15pt}
%
\begin{algorithm}[H]
\small
\caption{Fast Alternating Projection onto $\mathcal{C}_{\mathsf{coll}}$}
\label{app:alg:fast_projection}
\begin{algorithmic}[1]
  \State \textbf{Input:} $\mathbf{E} \in \mathbb{R}^{n \times d}$, $\varepsilon > 0$, $\delta \in (0, 1)$, sensitivity $\Delta>0$, tolerance $\tau>0$ \textbf{or} maximum iterations $T\in\mathbb{N}$
  \State \textbf{Output:} $\widehat{\mathbf{S}} \in \mathcal{C}_{\mathsf{coll}} \subseteq \mathbb{R}^{n \times n}$ with $(\varepsilon, \delta)$--DP guarantee \, s.t. \,
      $\bigl\|\widehat{\mathbf{S}} - \mathsf{proj}_{\mathcal{C}_{\mathsf{coll}}}(\mathbf{S} + \mathbf{W})\bigr\|_{\mathrm{F}}\le\tau$
\vspace{10pt}
  \State \textbf{Construct:} $\mathbf{S}\gets\mathbf{E}\mathbf{E}^\top$.
  \State \textbf{DP noise:} set $\sigma^2 \;\gets\; c_{\varepsilon, \delta} \Delta^2$ 
  \State Sample noise $\mathbf{W} \sim \mathcal{N}\left( \mathbf{0}, \sigma^2 \, \mathbf{I}_{n \times n }\right)$
  \State $\mathbf{S}' \gets \mathbf{S} + \frac{1}{2} \left( \mathbf{W} + \mathbf{W}^\top \right)$
  \State Initialize $\widehat{\mathbf{S}}_{(0)} \gets \mathbf{S}'$

  \For{$t=0$ \textbf{to} $T-1$}

      \State \textsf{/* symmetrize current iterate */}
      \State $\mathbf{Y}_t = \frac{1}{2} \big(\widehat{\mathbf{S}}_t  + \widehat{\mathbf{S}}_t^\top \big)$
      
      \State \textsf{/* projection onto PSD Cone $\mathcal{K}_{+}^{\,n}$ */}
      \State Compute eigen-decomposition $\mathbf{Y}_t = \mathbf{U} \mathsf{diag}(\lambda_1, \dots, \lambda_n) \mathbf{U}^{\top}$
      \State $\lambda^+_k \gets \max\{0,\lambda_k\}$, $\forall k \in [n]$
      \State $\mathbf{P}_t \gets \mathsf{proj}_{\mathcal{K}_{+}^{\,n}}(\mathbf{Y}_t) = \mathbf{U} \mathsf{diag}(\lambda^+_1, \dots, \lambda_n^+)\mathbf{U}^\top$

      \If{Frobenius-ball constraint is enforced and $\|\mathbf{P}_t\|_{\mathrm{F}} > n$}
        \State $\mathbf{P}_t \gets (n/\|\mathbf{P}_t\|_{\mathrm{F}})\,\mathbf{P}_t$  \hfill\quad\% radial projection onto $\mathcal{K}_+^n\cap\mathcal{B}_{\mathrm{F}}^n$
      \EndIf
      
      \State \textsf{/* projection onto Unit-Hyper-Cube $\displaystyle\mathcal{C}_{\mathrm{unit}}^{\,n}$*/} 
      \State  $\mathbf{Q}_{t} \gets \mathsf{proj}_{\mathcal{C}_{\mathrm{unit}}^{\,n}}(\mathbf{Y}_t)$  \hfill \quad\% $(\mathbf{Q}_t)_{ii}=1$;\ $(\mathbf{Q}_t)_{ij}=\mathrm{clip}((\mathbf{Y}_t)_{ij},-1,1)$ for $i\neq j$
      
      \State \textsf{/* averaged update */}
      \State $\widehat{\mathbf{S}}_{t+1} \gets \tfrac12\bigl(\mathbf{P}_t+\mathbf{Q}_t\bigr)$

      \State \textsf{/* stopping tests */}
      \State $r_{\mathrm{chg}}\gets \|\widehat{\mathbf{S}}_{t+1}-\widehat{\mathbf{S}}_t\|_{\mathrm{F}}$
      \State $r_{\mathrm{psd}}\gets \|\mathbf{Y}_t-\mathbf{P}_t\|_{\mathrm{F}}$;\quad $r_{\mathrm{box}}\gets \|\mathbf{Y}_t-\mathbf{Q}_t\|_{\mathrm{F}}$
      \If{$r_{\mathrm{chg}}\le\tau$ \textbf{and} $\max\{r_{\mathrm{psd}},r_{\mathrm{box}}\}\le\tau$}
         \State \textbf{break}
      \EndIf
  \EndFor
  
  \State \textsf{/* PSD on return */}
  \State $\mathbf{S}_{\mathrm{avg}} \gets \widehat{\mathbf{S}}_{t+1}$
  \State $\widehat{\mathbf{S}}\gets \mathsf{proj}_{\mathcal{K}_+^n}(\tfrac12(\mathbf{S}_{\mathrm{avg}}+\mathbf{S}_{\mathrm{avg}}^\top))$
         
  \State \textbf{if} $\min_i \widehat{S}_{ii}\le 0$ \textbf{then}  
       $\widehat{\mathbf{S}} \gets \widehat{\mathbf{S}} + \mu \, \mathbf{I}$ \text{ with small } $\mu>0$ \text{ (e.g.,} $\mu=10^{-8}\|\widehat{\mathbf{S}}\|_{\mathrm{F}}/n)$
  \State $\mathbf{D} \gets \mathsf{diag}(\, \widehat{\mathbf{S}} \,)^{1/2}$, \quad  
  $\widehat{\mathbf{S}}\gets \mathbf{D}^{-1}\,\widehat{\mathbf{S}}\,\mathbf{D}^{-1}$   
  
  \State \textbf{Output:} $\widehat{\mathbf{S}}$ 
\end{algorithmic}
\end{algorithm}


\clearpage

\appsubsection{Averaged Alternating Projection: Theoretical Guarantees}
\label{app:ssec:AAP-theoretical-guarantees}

\subsubsection{Foundational Definitions and Lemmas}

\begin{definition}[Bounded Linear Regularity]
\label{def:dist-blr}
Let $\mathcal{H}$ be a (finite-dimensional) Hilbert space with norm $\| \cdot\|$. For a nonempty set $\mathcal{A} \subset \mathcal{H}$, define the point--set distance $\mathsf{dist}(x,\mathcal{A})\;\coloneqq\; \inf_{a \in \mathcal{A}}\| x - a \|$. Let $\mathcal{C}_1, \mathcal{C}_2 \subset \mathcal{H}$ be nonempty, closed, and convex, and set $\mathcal{C}\coloneqq \mathcal{C}_1\cap\mathcal{C}_2\neq\varnothing $. For $R>0$, denote the closed ball $\mathbb{B}_R \coloneqq \{x\in\mathcal{H}:\ \|x\|\le R\}$. The pair $(\mathcal{C}_1,\mathcal{C}_2)$ is called \textit{boundedly linearly regular} if
for every $R>0$ there exists a constant $\kappa_R\ge 1$ such that
\begin{equation}\label{eq:blr-ball}
\mathsf{dist}(x,\mathcal{C})
\;\le\; \kappa_R\max\bigl\{\mathsf{dist}(x,\mathcal{C}_1),\ \mathsf{dist}(x,\mathcal{C}_2)\bigr\},
\qquad \forall\,x\in \mathbb{B}_R.
\end{equation}
\end{definition}

In our ScoreShield application, $\mathcal{H}=(\mathbb{R}^{n\times n},\langle\cdot,\cdot\rangle_F)$ and $\|\cdot\|=\|\cdot\|_F$.

\begin{definition}[Fej\'er Monotone Sequence]
\label{def:Fejer}
Let $( \mathcal{X} , \langle \cdot, \cdot \rangle)$ be a real Hilbert space, let $\|\cdot\|$ be its induced norm, and let $\mathcal{C} \subset \mathcal{X}$ be non–empty and closed\footnote{Convexity of $\mathcal{C}$ is not required for the definition itself, but almost every convergence theorem that uses Fej\'er monotonicity assumes $\mathcal{C}$ is closed and convex.}. A sequence $(\mathbf{S}_t)_{t\ge0} \subset \mathcal{X}$ is called \textit{Fej\'er monotone with respect to $\mathcal{C}$} if
\begin{equation}
\|\mathbf{S}_{t+1} - \mathbf{Z}\| \;\le\; \|\mathbf{S}_t - \mathbf{Z}\|,  \qquad \forall \, \mathbf{Z} \in \mathcal{C}, \; \,\forall t\ge 0 .
\end{equation}
\end{definition}

\begin{lemma}[Basic Properties of Fej\'er Monotone Sequences]
\label{lem:Fejer-properties}
Let $(\mathbf{S}_t)$ be Fej\'er monotone w.r.t. a non–empty, closed and convex set $\mathcal{C}$ in a Hilbert space. Then
\begin{enumerate}[label=\textup{(\alph*)}, leftmargin=*]
\item $\bigl(\| \mathbf{S}_t - \mathbf{Z} \|\bigr)_{t \ge 0}$ is non-increasing for each $\mathbf{Z}\in \mathcal{C}$ and therefore convergent;
\item The sequence $(\mathbf{S}_t)$ is bounded;
\item The distance sequence $d_t  \coloneqq  \mathsf{dist}(\mathbf{S}_t, \mathcal{C})=\inf_{\mathbf{Z}\in \mathcal{C}}\|\mathbf{S}_t - \mathbf{Z}\|$ is decreasing and convergent;
\item
If there exists a subsequence $(\mathbf{S}_{t_k})$ that converges in norm to some $\mathbf{Z}_\infty \in \mathcal{C}$, then $\mathbf{S}_t\to \mathbf{Z}_\infty$ in norm.
\end{enumerate}
\end{lemma}

\begin{proof}
Properties (a)–(c) follow directly from Definition~\ref{def:Fejer}. For (d),  assume $\mathbf{S}_{t_k}\to \mathbf{Z}_\infty\in\mathcal{C}$ in norm. By Fej\'er monotonicity, the sequence $a_t\coloneqq \|\mathbf{S}_t-\mathbf{Z}_\infty\|$ is nonincreasing, hence $\lim_{t\to\infty} a_t$ exists. But $a_{t_k}\to 0$, so $\lim_{t\to\infty} a_t=0$, i.e., $\mathbf{S}_t\to \mathbf{Z}_\infty$ (see also Lemma~\ref{lem:Opial-finite}).
\end{proof}

\begin{lemma}[Strong Quasi‑Nonexpansive for Averaged Operator $T_\lambda$]
\label{lem:sqne}
Let $\mathcal{H} = \left(\mathbb{R}^{n \times n}, {\langle \cdot, \cdot \rangle}_{\mathrm{F}} \right)$ be a real Hilbert space (in our case, $\mathbb{R}^{n\times n}$ with the Frobenius inner product), and let $\mathcal{C}_1, \mathcal{C}_2 \subset \mathcal{H}$ be nonempty, closed, convex sets. Denote by $\mathsf{proj}_{\mathcal{C}_i}$ the orthogonal projector onto $\mathcal{C}_i$, $i=1,2$. For $\lambda\in(0,1)$ define $T_\lambda \; \coloneqq \; \lambda \mathsf{proj}_{\mathcal{C}_1} + (1-\lambda) \mathsf{proj}_{\mathcal{C}_2}$. Then the following hold.
\begin{enumerate}[label=\textup{(\roman*)}, wide=0pt, leftmargin=*]
\item 
\textbf{Firm nonexpansiveness (averagedness).}  
Each $\mathsf{proj}_{\mathcal{C}_i}$ is firmly nonexpansive, hence $1/2$--averaged. A convex combination of $1/2$--averaged (resp. firmly nonexpansive) operators is again $1/2$--averaged (resp. firmly nonexpansive). Consequently,
\begin{equation}
\label{eq:fne}
\|T_\lambda \mathbf{S} - T_\lambda \mathbf{Y}\|^2 \;\le\; \langle T_\lambda \mathbf{S} - T_\lambda \mathbf{Y},\; \mathbf{S} - \mathbf{Y} \rangle,
\qquad \forall \, \mathbf{S} , \mathbf{Y} \in \mathcal{H},
\end{equation}
and $T_\lambda$ is $1/2$--averaged.
\item 
\textbf{Strong quasi--nonexpansiveness (SQNE).}  
For every $\mathbf{Z} \in \mathsf{Fix} (T_\lambda)$ and every $\mathbf{S} \in \mathcal{H}$,
\begin{equation}\label{eq:sqne}
\|T_\lambda \mathbf{S} - \mathbf{Z} \|^2
\;\le\; \| \mathbf{S} - \mathbf{Z}\|^2 \;-\; \|T_\lambda \mathbf{S} - \mathbf{S}\|^2.
\end{equation}
In particular, the Picard iterates $\mathbf{S}_{t+1}=T_\lambda\mathbf{S}_t$ are Fej\'er monotone w.r.t. $\mathsf{Fix}(T_\lambda)$ and satisfy $\sum_{t\ge0}\|\mathbf{S}_{t+1}-\mathbf{S}_t\|_F^2<\infty$ (hence $\|\mathbf{S}_{t+1}-\mathbf{S}_t\|_F\to 0$).
\end{enumerate}
\end{lemma}

\begin{proof}
\leavevmode

\noindent
\textbf{(i)}
Each projector $\mathsf{proj}_{\mathcal{C}_i}$ is firmly nonexpansive:
\begin{equation}
\|\mathsf{proj}_{\mathcal{C}_i} (\mathbf{S}) - \mathsf{proj}_{\mathcal{C}_i} (\mathbf{Y})\|^2 \;\le\; \langle \mathsf{proj}_{\mathcal{C}_i} (\mathbf{S}) - \mathsf{proj}_{\mathcal{C}_i} (\mathbf{Y}),\; \mathbf{S} - \mathbf{Y} \rangle,\qquad i=1,2.
\end{equation}
Let $\mathbf{A} \coloneqq \mathsf{proj}_{\mathcal{C}_1} (\mathbf{S}) - \mathsf{proj}_{\mathcal{C}_1} (\mathbf{Y})$ and $\mathbf{B} \coloneqq \mathsf{proj}_{\mathcal{C}_2} (\mathbf{S}) - \mathsf{proj}_{\mathcal{C}_2} (\mathbf{Y})$. Then $T_\lambda \mathbf{S} - T_\lambda \mathbf{Y} = \lambda \mathbf{A} + (1-\lambda) \mathbf{B}$. Using the convexity of the norm-square and dropping a nonpositive variance term,
\begin{equation}
\|\lambda \mathbf{A} + (1-\lambda)\mathbf{B}\|^2
\;\le\; \lambda\|\mathbf{A}\|^2 + (1-\lambda)\|\mathbf{B}\|^2.
\end{equation}
Applying firm nonexpansiveness of $\mathsf{proj}_{\mathcal{C}_1}$ and $\mathsf{proj}_{\mathcal{C}_2}$ to the RHS gives
\begin{equation}
\lambda\|\mathbf{A}\|^2 + (1-\lambda)\|\mathbf{B}\|^2
\;\le\; \lambda\langle \mathbf{A}, \mathbf{S}-\mathbf{Y}\rangle + (1-\lambda)\langle \mathbf{B}, \mathbf{S}-\mathbf{Y}\rangle
= \langle \lambda \mathbf{A}+(1-\lambda)\mathbf{B},\; \mathbf{S}-\mathbf{Y}\rangle.
\end{equation}
This is exactly Eq.~\ref{eq:fne}. Thus $T_\lambda$ is firmly nonexpansive, hence $\tfrac12$–averaged.

\noindent
\textbf{(ii)}
Let $\mathbf{Z} \in \mathsf{Fix}(T_\lambda)$ so that $T_\lambda \mathbf{Z}= \mathbf{Z}$. By (i), $T_\lambda$ is firmly nonexpansive, so
\begin{equation}
\|T_\lambda \mathbf{S} - T_\lambda \mathbf{Z}\|^2 \le \langle T_\lambda \mathbf{S} - T_\lambda \mathbf{Z},\; \mathbf{S}-\mathbf{Z}\rangle.
\end{equation}
But $T_\lambda \mathbf{Z}=\mathbf{Z}$, hence
\begin{equation}
\|T_\lambda \mathbf{S} - \mathbf{Z}\|^2 \le \langle T_\lambda \mathbf{S} - \mathbf{Z},\; \mathbf{S}-\mathbf{Z}\rangle.
\end{equation}
Expand the right-hand side using the polarization identity:
\begin{equation}
\langle T_\lambda \mathbf{S} - \mathbf{Z},\; \mathbf{S} - \mathbf{Z}\rangle
= \tfrac12\bigl( \|T_\lambda \mathbf{S} - \mathbf{Z}\|^2 + \|\mathbf{S}-\mathbf{Z}\|^2 - \|T_\lambda \mathbf{S} - \mathbf{S}\|^2 \bigr).
\end{equation}
Rearranging gives
\begin{equation}
\|T_\lambda \mathbf{S} - \mathbf{Z}\|^2
\le \|\mathbf{S}-\mathbf{Z}\|^2 - \|T_\lambda \mathbf{S} - \mathbf{S}\|^2,
\end{equation}
which is Eq.~\ref{eq:sqne}. This is the strong quasi–nonexpansiveness with parameter 1. 
Summing Eq.~\ref{eq:sqne} over $t$ gives $\sum_t\|\mathbf{S}_{t+1}-\mathbf{S}_t\|_F^2<\infty$.
\end{proof}

\begin{lemma}[Fixed Points of $T_\lambda$ via a Convex Potential]
\label{lem:Tlambda-fix}
Define the weighted proximity functional
\begin{equation}
J_\lambda(\mathbf{S}) \;\coloneqq\;
\frac{\lambda}{2}\,\mathsf{dist}(\mathbf{S},\mathcal{C}_1)^2
+\frac{1-\lambda}{2}\,\mathsf{dist}(\mathbf{S},\mathcal{C}_2)^2,
\qquad \mathbf{S}\in\mathcal{H},
\end{equation}
where $\mathsf{dist}(\mathbf{S},\mathcal{C})\coloneqq\|\mathbf{S}-\mathsf{proj}_{\mathcal{C}}(\mathbf{S})\|_F$. Then $J_\lambda$ is convex and Fr\'echet differentiable with
\begin{equation}
\nabla J_\lambda(\mathbf{S})=\mathbf{S} - T_\lambda(\mathbf{S}).   
\end{equation}
Consequently,
\begin{equation}
\mathsf{Fix}(T_\lambda)=\arg\min_{\mathbf{S}\in\mathcal{H}} J_\lambda(\mathbf{S}).    
\end{equation}
If moreover $\mathcal{C}_1\cap\mathcal{C}_2\neq\varnothing$, then $\min J_\lambda=0$ and
\begin{equation}
\mathsf{Fix}(T_\lambda)=\arg\min J_\lambda=\mathcal{C}_1\cap\mathcal{C}_2.   
\end{equation}
\end{lemma}

\begin{proof}
For any closed convex $\mathcal{C}$, the function $f(\mathbf{S})\coloneqq \tfrac12\mathsf{dist}(\mathbf{S},\mathcal{C})^2 =\min_{\mathbf{Z}\in \mathcal{C}}\tfrac12 \|\mathbf{S}-\mathbf{Z}\|_F^2$ is convex and differentiable with $\nabla f(\mathbf{S})=\mathbf{S}-\mathsf{proj}_{\mathcal{C}}(\mathbf{S})$. Therefore
\begin{equation}
\nabla J_\lambda(\mathbf{S})
=\lambda (\mathbf{S} - \mathsf{proj}_{\mathcal{C}_1}(\mathbf{S})) + (1-\lambda)(\mathbf{S}-\mathsf{proj}_{\mathcal{C}_2} (\mathbf{S}))
=\mathbf{S}-(\lambda \mathsf{proj}_{\mathcal{C}_1}(\mathbf{S}) + (1-\lambda)\mathsf{proj}_{\mathcal{C}_2}(\mathbf{S})) = \mathbf{S}-T_\lambda\mathbf{S}.
\end{equation}
Since $J_\lambda$ is convex differentiable, $\nabla J_\lambda(\mathbf{S})=0$ is equivalent to $\mathbf{S}\in\arg\min J_\lambda$. Thus $\mathsf{Fix}(T_\lambda)=\arg\min J_\lambda$. If $\mathcal{C}_1\cap\mathcal{C}_2\neq\varnothing$, then $J_\lambda(\mathbf{S})= 0$ iff $\mathbf{S}\in\mathcal{C}_1$ and $\mathbf{S}\in\mathcal{C}_2$, i.e., $\mathbf{S}\in\mathcal{C}_1\cap\mathcal{C}_2$.
\end{proof}

\begin{lemma}[Bounded BLR from relative-interior intersection]
\label{lem:blr-interior}
Let $\mathcal{H}$ be a finite-dimensional Hilbert space with norm $\|\cdot\|$, and let $\mathcal{C}_1,\mathcal{C}_2\subset\mathcal{H}$ be nonempty, closed, and convex with $\mathcal{C}\coloneqq \mathcal{C}_1\cap \mathcal{C}_2 \neq \varnothing$. Assume $\mathsf{ri}(\mathcal{C}_1)\cap \mathsf{ri}(\mathcal{C}_2)\neq\varnothing$. Then $(\mathcal{C}_1,\mathcal{C}_2)$ is boundedly linearly regular in the sense of
Definition~\ref{def:dist-blr}; i.e., for every $R>0$ there exists $\kappa_R\ge 1$ such that
\begin{equation}
\mathsf{dist}(x,\mathcal{C})\;\le\;\kappa_R\max\{\mathsf{dist}(x,\mathcal{C}_1),\mathsf{dist}(x,\mathcal{C}_2)\}, \qquad \forall\, x \in\mathbb{B}_R .
\end{equation}
\end{lemma}

\begin{proof}
Fix $R>0$ and consider the closed ball $\mathbb{B}_R=\{x\in\mathcal{H}:\|x\|\le R\}$. Under the qualification condition $\mathsf{ri}(\mathcal{C}_1)\cap \mathsf{ri}(\mathcal{C}_2)\neq \varnothing$, the pair $(\mathcal{C}_1,\mathcal{C}_2)$ satisfies a standard Slater-type constraint qualification for the
convex feasibility problem $\mathcal{C}_1 \cap \mathcal{C}_2$. A classical consequence is a \emph{metric inequality} (a.k.a.\ bounded linear regularity/error bound) on bounded sets: there exists a constant $\gamma_R>0$ such that
\begin{equation}\label{eq:metric-ineq-sumsq}
\mathsf{dist}(x,\mathcal{C})^2
\;\le\; \gamma_R\bigl(\mathsf{dist}(x,\mathcal{C}_1)^2+\mathsf{dist}(x,\mathcal{C}_2)^2\bigr),
\qquad \forall\,x \in\mathbb{B}_R ,
\end{equation}
see, e.g., \citep[Cor.~6]{bauschke1999strong} (and also \citep[Sec.~5]{bauschke1996projection}).
Now for any $x\in\mathbb{B}_R$,
\begin{equation}
\mathsf{dist}(x,\mathcal{C}_1)^2+\mathsf{dist}(x,\mathcal{C}_2)^2
\;\le\; 2\max\{\mathsf{dist}(x,\mathcal{C}_1),\mathsf{dist}(x,\mathcal{C}_2)\}^2 .
\end{equation}
Combine this with Eq.~\ref{eq:metric-ineq-sumsq} and take square-roots to obtain
\begin{equation}
\mathsf{dist}(x,\mathcal{C})
\;\le\; \sqrt{2\gamma_R}\,\max\{\mathsf{dist}(x,\mathcal{C}_1),\mathsf{dist}(x,\mathcal{C}_2)\},
\qquad \forall\,x\in\mathbb{B}_R .
\end{equation}
Thus Definition~\ref{def:dist-blr} holds with $\kappa_R\coloneqq \sqrt{2\gamma_R}$.
\end{proof}

\begin{remark}[Restricting BLR to the iterate region]
Suppose the iterates satisfy $\mathbf{S}_t\in\mathcal{B}$ for all $t$, where $\mathcal{B} \subset \mathcal{H}$ is bounded (e.g., a Fej\'er ball). Pick any $R>0$ such that $\mathcal{B}\subseteq \mathbb{B}_R$. If BLR holds on $\mathbb{B}_R$ with constant $\kappa_R$ (for instance, by Lemma~\ref{lem:blr-interior}), then the same inequality holds for all $x\in\mathcal{B}$ with the \emph{same} constant $\kappa_R$.
\end{remark}

\begin{lemma}[BLR for PSD cone and unit--diagonal box]
\label{lem:blr-corr}
Consider $\mathcal{H} = (\mathbb{R}^{n\times n}, \langle \cdot, \cdot \rangle_F)$ and set
$\mathcal{C}_1\coloneqq \mathcal{K}_+^{\,n}$ and $\mathcal{C}_2\coloneqq \mathcal{C}_{\mathrm{unit}}^{\,n}$. Then $(\mathcal{C}_1,\mathcal{C}_2)$ is boundedly linearly regular: for every $R>0$ there exists $\kappa_R\ge 1$ such that
\begin{equation}
\mathsf{dist}(S,\mathcal{C}_1\cap\mathcal{C}_2)
\;\le\; \kappa_R \max\{\mathsf{dist}(S,\mathcal{C}_1),\mathsf{dist}(S,\mathcal{C}_2)\},
\qquad \forall S\in\mathbb{B}_R.   
\end{equation}
\end{lemma}

\begin{proof}
$\mathbf{I}_n\in \mathsf{ri}(\mathcal{C}_1)\cap \mathsf{ri}(\mathcal{C}_2)$, hence Lemma~\ref{lem:blr-interior} applies.
\end{proof}

\begin{lemma}[Opial's Lemma in Finite Dimension {\citep{opial1967weak}}%
\footnote{See also the streamlined presentation in \citep{arakcheev2025opial}.}]
\label{lem:Opial-finite}
Let $\mathcal{H}$ be a finite-dimensional Hilbert space and let $\mathcal{C}\subset\mathcal{H}$ be nonempty, closed, and convex. Let $(\mathbf{S}_t)_{t\ge 0}\subset\mathcal{H}$ be a sequence such that:
\begin{enumerate}[label=\textup{(O\arabic*)},leftmargin=*]
\item[\bf(O1)] 
For every $\mathbf{Z}\in\mathcal{C}$, the limit $\lim_{t\to\infty}\|\mathbf{S}_t-\mathbf{Z}\|$ exists.
\item[\bf(O2)] 
Every norm-convergent subsequence $(\mathbf{S}_{t_k})$ has its limit in $\mathcal{C}$; i.e.,
if $\mathbf{S}_{t_k}\to \mathbf{S}_\infty$ then $\mathbf{S}_\infty\in\mathcal{C}$.
\end{enumerate}
Then $(\mathbf{S}_t)$ converges in norm to some point $\mathbf{S}_\star\in\mathcal{C}$.
\end{lemma}

\begin{proof}
Pick any $\mathbf{Z}_0\in\mathcal{C}$. By \textbf{(O1)}, the real sequence $r_t\coloneqq\|\mathbf{S}_t - \mathbf{Z}_0\|$ has a finite limit, hence is bounded. Therefore $(\mathbf{S}_t)$ is bounded. Since $\mathcal{H}$ is finite-dimensional, every bounded sequence has a norm-convergent subsequence \citep{rudin1976principles, bauschke2020correction}. Thus there exist indices $t_k\uparrow\infty$ and a point $\mathbf{S}_\infty\in\mathcal{H}$ such that $\mathbf{S}_{t_k}\to \mathbf{S}_\infty$ in norm. By \textbf{(O2)}, we have $\mathbf{S}_\infty\in\mathcal{C}$. Apply \textbf{(O1)} with $\mathbf{Z}=\mathbf{S}_\infty\in\mathcal{C}$ to conclude that the limit $L\coloneqq\lim_{t\to\infty}\|\mathbf{S}_t-\mathbf{S}_\infty\|$ exists. Along the subsequence $t_k$, $\|\mathbf{S}_{t_k}-\mathbf{S}_\infty\|\ \longrightarrow\ 0$, so necessarily $L=0$. Hence $\|\mathbf{S}_t-\mathbf{S}_\infty\|\to 0$, i.e., $\mathbf{S}_t\to\mathbf{S}_\infty$ in norm. Setting $\mathbf{S}_\star\coloneqq \mathbf{S}_\infty\in\mathcal{C}$ completes the proof.
\end{proof}

\subsubsection{Convergence and Linear Rates for AAP in ScoreShield Regime~(ii)}

\begin{theorem}[Convergence and Linear Feasibility Rates of AAP Under BLR]
\label{thm:aap}
Let $\mathcal{H} = \left( \mathbb{R}^{n \times n} , {\langle \cdot, \cdot \rangle}_{\mathrm{F}} \right)$ 
be the Hilbert space of real $n\times n$ matrices with Frobenius inner product.
Let $\mathcal{C}_1, \mathcal{C}_2 \subset \mathcal{H}$ be non-empty, closed, convex with $\mathcal{C}_1 \cap \mathcal{C}_2 \neq \varnothing$. 
For a relaxation parameter $\lambda \in (0, 1)$ define the averaged projection operator $T_{\lambda} = \lambda \, \mathsf{proj}_{\mathcal{C}_{1}} + (1 - \lambda) \mathsf{proj}_{\mathcal{C}_{2}}$\footnote{Note that our iteration uses a convex combination of projectors $T_\lambda=\lambda\,\mathsf{proj}_{\mathcal{C}_1}+(1-\lambda)\,\mathsf{proj}_{\mathcal{C}_2}$.
This is a Krasnosel'ski\u{\i}–Mann \citep{mann1953mean, ma1955two} averaged projector scheme, distinct from the classical
von Neumann alternating projections $\mathsf{proj}_{\mathcal{C}_2}( \mathsf{proj}_{\mathcal{C}_1} (\cdot))$ (composition) \citep{von1949rings}.}, where the orthogonal projector is defined as $\mathsf{proj}_{\mathcal{C}} (\mathbf{S})  \coloneqq  \arg \min_{\mathbf{Z} \in \mathcal{C}} \Vert \mathbf{S} - \mathbf{Z} \Vert_{\mathrm{F}}$. Define point–set distance as $\mathsf{dist} ( \mathbf{S}, \mathcal{C} )  \coloneqq  \Vert \mathbf{S} - \mathsf{proj}_{\mathcal{C}} (\mathbf{S}) \Vert_{\mathrm{F}}$. 
Given an initial matrix $\mathbf{S}_0 \in  \mathcal{H}$, iterate $\mathbf{S}_{t+1} = T_{\lambda} \mathbf{S}_t , \, t= 0, 1, \dots \; $.
\begin{enumerate}[label=\textup{(\roman*)}, wide=0pt, leftmargin=*]
\item 
\textbf{Convergence and Optimality.}
%
The sequence $(\mathbf{S}_t)_{t \ge 0}$ converges in the Frobenius norm to some $\mathbf{S}_\star \in \mathcal{C}_1\cap\mathcal{C}_2$. Moreover, 
%
\begin{equation}
\mathsf{Fix}(T_\lambda)=\mathcal{C}_1 \cap \mathcal{C}_2
=\arg\min_{\mathbf{S}\in\mathcal{H}} \ \mathcal{J}(\mathbf{S}),\qquad
    \mathcal{J}(\mathbf{S})  \coloneqq  \frac{\lambda}{2}  \mathsf{dist}^2 ( \mathbf{S}, \mathcal{C}_1 ) + \frac{1 - \lambda}{2} \mathsf{dist}^2 ( \mathbf{S}, \mathcal{C}_2 )
\end{equation}
and $\mathcal{J}(\mathbf{S}) = 0$ on $\mathcal{C}_1 \cap \mathcal{C}_2$.
If $\mathcal{C}_1 \cap \mathcal{C}_2$ contains more than one element, the particular minimizer reached may depend on $\mathbf{S}_0$.
\item 
\textbf{Linear Feasibility Decay on a Bounded Region $\mathsf{dist}(\mathbf{S}_t,\mathcal{C}_1 \cap \mathcal{C}_2)$ under BLR.}
%
Assume the pair $(\mathcal{C}_1,\mathcal{C}_2)$ is BLR. There exists $R > 0$ such that $\mathbf{S}_t \in \mathbb{B}_R, \forall t$ and that BLR hold on $\mathbb{B}_R$ with constant $\kappa_R$
\begin{equation}\label{eq:blr}
    \mathsf{dist} \bigl(\mathbf{S},\mathcal{C}_1\cap\mathcal{C}_2\bigr) \le
    \kappa_R \, \max \bigl\{ \mathsf{dist}(\mathbf{S},\mathcal{C}_1), \mathsf{dist}(\mathbf{S},\mathcal{C}_2) \bigr\},
    \quad \forall\, \mathbf{S} \in \mathbb{B}_R.
\end{equation}
%
%
Let $m  \coloneqq  \min\{\lambda,1-\lambda\}$. Then
\begin{equation}\label{eq:R-linear}
\mathsf{dist}(\mathbf{S}_t, \mathcal{C}_1 \cap \mathcal{C}_2)
\;\le\; \rho^{\,t}_R \, \mathsf{dist}(\mathbf{S}_0, \mathcal{C}_1 \cap \mathcal{C}_2),\qquad t=0,1,\dots\; , \qquad \rho_R \coloneqq  \sqrt{1-\frac{m}{\kappa^2_R}}\in(0,1).
\end{equation}

\item 
\textbf{Linear Decay in Norm under a Local Metric Error Bound.}
Let $\mathcal{B} \coloneqq  \bigl\{\mathbf{S}:\ \|\mathbf{S}-\mathbf{Z}\|_{\mathrm{F}}\le\|\mathbf{S}_0-\mathbf{Z}\|_{\mathrm{F}},\ \forall\mathbf{Z}\in\mathcal{C}\bigr\}$ be the Fejér ball containing all iterates. If there exist $\sigma>0$ and a (necessarily unique) $\mathbf{S}_\star\in  \mathcal{C}_1 \cap \mathcal{C}_2$ such that
\begin{equation}
\label{eq:metric-eb}
\|\mathbf{S}-\mathbf{S}_\star\|_{\mathrm{F}}\le\sigma\,\mathsf{dist}(\mathbf{S},  \mathcal{C}_1 \cap \mathcal{C}_2 ), \qquad \forall\,\mathbf{S} \in \mathcal{B},
\end{equation}
then
\begin{equation}\label{eq:norm-rate}
\|\mathbf{S}_t-\mathbf{S}_\star\|_{\mathrm{F}}\le
\sigma\,\rho^{\,t}_R \,\mathsf{dist}(\mathbf{S}_0,  \mathcal{C}_1 \cap \mathcal{C}_2)
\ \le\ \sigma\, \rho^{\,t}_R \,\|\mathbf{S}_0-\mathbf{S}_\star\|_{\mathrm{F}}.
\end{equation}
Thus $\|\mathbf{S}_t-\mathbf{S}_\star\|_{\mathrm{F}}$ also decays $R$-linearly.
\end{enumerate}
\end{theorem}


\begin{proof}
\leavevmode

\noindent
\textbf{(i)}
By Lemma~\ref{lem:sqne}, $T_\lambda$ is firmly nonexpansive and the Picard iterates are Fej\'er monotone w.r.t. $\mathsf{Fix}(T_\lambda)$, with $\sum_{t\ge 0}\|\mathbf{S}_{t+1}-\mathbf{S}_t\|_F^2<\infty$, hence
$\|\mathbf{S}_{t+1}-\mathbf{S}_t\|_F\to 0$. We verify Opial's conditions from Lemma~\ref{lem:Opial-finite} with $\mathcal{C}=\mathsf{Fix}(T_\lambda)$. For any $\mathbf{Z}\in\mathsf{Fix}(T_\lambda)$, Fej\'er monotonicity gives that $\|\mathbf{S}_t-\mathbf{Z}\|_F$ is nonincreasing, hence has a limit (O1).
For (O2), let $\mathbf{S}_{t_k}\to \mathbf{S}_\infty$ in norm. Then $\|T_\lambda\mathbf{S}_{t_k}-\mathbf{S}_{t_k}\|_F=\|\mathbf{S}_{t_k+1}-\mathbf{S}_{t_k}\|_F\to 0$. Since $T_\lambda$ is continuous (Lipschitz), $T_\lambda\mathbf{S}_{t_k}\to T_\lambda\mathbf{S}_\infty$, so $T_\lambda\mathbf{S}_\infty=\mathbf{S}_\infty$, i.e., $\mathbf{S}_\infty\in\mathsf{Fix}(T_\lambda)$. Opial's lemma yields $\mathbf{S}_t\to \mathbf{S}_\star\in\mathsf{Fix}(T_\lambda)$. Finally, Lemma~\ref{lem:Tlambda-fix} gives $\mathsf{Fix}(T_\lambda)=\arg\min \mathcal{J}$, and since $\mathcal{C}_1\cap\mathcal{C}_2\neq\varnothing$,
$\min\mathcal{J}=0$ and $\arg\min\mathcal{J}=\mathcal{C}_1\cap\mathcal{C}_2$, proving (i).

\noindent
\textbf{(ii)}
Fix $\mathbf{S}\in\mathbb{B}_R$ and note that $\mathsf{dist}(T_\lambda\mathbf{S},\mathcal{C})^2
= \min_{\mathbf{Z}\in\mathcal{C}}\|T_\lambda\mathbf{S}-\mathbf{Z}\|_{\mathrm{F}}^2 \le \|T_\lambda\mathbf{S} - \mathsf{proj}_{\mathcal{C}} (\mathbf{S})\|_{\mathrm{F}}^2$. 
By firm nonexpansiveness of each projector, convexity of $\|\cdot\|_{\mathrm{F}}^2$, and the projector Pythagorean inequality (see Lemma~\ref{cor:pythagoras}), we have
\begin{subequations}\label{eq:linear_rate_upperbound}
\begin{align}
\|T_\lambda \mathbf{S} &- \mathsf{proj}_{\mathcal{C}_1 \cap \mathcal{C}_2} (\mathbf{S})\|_{\mathrm{F}}^2\\
&= \|\lambda(\mathsf{proj}_{\mathcal{C}_1}(\mathbf{S})-\mathsf{proj}_{\mathcal{C}_1 \cap \mathcal{C}_2} (\mathbf{S}))+(1-\lambda)(\mathsf{proj}_{\mathcal{C}_2}(\mathbf{S})-\mathsf{proj}_{\mathcal{C}_1 \cap \mathcal{C}_2} (\mathbf{S}))\|_{\mathrm{F}}^2 \\
&\le \lambda\|\mathsf{proj}_{\mathcal{C}_1}(\mathbf{S})-\mathsf{proj}_{\mathcal{C}_1 \cap \mathcal{C}_2} (\mathbf{S})\|_{\mathrm{F}}^2
 +(1-\lambda)\|\mathsf{proj}_{\mathcal{C}_2}(\mathbf{S})-\mathsf{proj}_{\mathcal{C}_1 \cap \mathcal{C}_2} (\mathbf{S})\|_{\mathrm{F}}^2 \\
&\le \|\mathbf{S}-\mathsf{proj}_{\mathcal{C}_1 \cap \mathcal{C}_2} (\mathbf{S})\|_{\mathrm{F}}^2
 - \Bigl[\lambda\|\mathbf{S}-\mathsf{proj}_{\mathcal{C}_1}(\mathbf{S})\|_{\mathrm{F}}^2
+(1-\lambda)\|\mathbf{S}-\mathsf{proj}_{\mathcal{C}_2}(\mathbf{S})\|_{\mathrm{F}}^2\Bigr].
\end{align}
\end{subequations}
That is
\begin{equation}
\mathsf{dist}(T_\lambda\mathbf{S}, \mathcal{C}_1 \cap \mathcal{C}_2 )^2
\le \mathsf{dist}(\mathbf{S}, \mathcal{C}_1 \cap \mathcal{C}_2 )^2 -\Bigl[\lambda \mathsf{dist}(\mathbf{S},\mathcal{C}_1)^2 + (1-\lambda)\mathsf{dist}(\mathbf{S},\mathcal{C}_2)^2\Bigr].
\end{equation}

Using $m = \min\{\lambda,1-\lambda\}$ we have
\begin{subequations}
\begin{eqnarray}
    \lambda \mathsf{dist}(\mathbf{S},\mathcal{C}_1)^2+(1-\lambda)\mathsf{dist}(\mathbf{S},\mathcal{C}_2)^2 \;&\ge&\; m(\mathsf{dist}(\mathbf{S},\mathcal{C}_1)^2+\mathsf{dist}(\mathbf{S},\mathcal{C}_2)^2) \\
\;&\ge&\; m\max\{\mathsf{dist}(\mathbf{S},\mathcal{C}_1)^2,\mathsf{dist}(\mathbf{S},\mathcal{C}_2)^2\} \\
\;&\ge&\; \frac{m}{\kappa^2_R}\,{\mathsf{dist}(\mathbf{S} ,\mathcal{C}_1 \cap \mathcal{C}_2 )}^2,
\end{eqnarray}
\end{subequations}
where the last step uses BLR Eq.~\ref{eq:blr}.
Insert this lower bound in Eq.~\ref{eq:linear_rate_upperbound} to obtain
\begin{equation}
\mathsf{dist}(T_\lambda \mathbf{S}, \mathcal{C}_1 \cap \mathcal{C}_2 )^2
\le {\mathsf{dist}(\mathbf{S} ,\mathcal{C}_1 \cap \mathcal{C}_2 )}^2 - \frac{m}{\kappa^2_R} {\mathsf{dist}(\mathbf{S} ,\mathcal{C}_1 \cap \mathcal{C}_2 )}^2
= \Bigl(1-\frac{m}{\kappa^2_R}\Bigr){\mathsf{dist}(\mathbf{S} ,\mathcal{C}_1 \cap \mathcal{C}_2 )}^2.
\end{equation}
Hence
\begin{equation}
\mathsf{dist}(T_\lambda \mathbf{S},\mathcal{C}_1 \cap \mathcal{C}_2)
\le \rho_R \,\mathsf{dist}(\mathbf{S}, \mathcal{C}_1 \cap \mathcal{C}_2),
\qquad \rho \coloneqq \sqrt{1-\frac{m}{\kappa^2_R}} \in (0, 1).
\end{equation}
Iterating yields the geometric decrease Eq.~\ref{eq:R-linear}.

\textbf{(iii)} If Eq.~\ref{eq:metric-eb} holds on $\mathcal{B}$, then for all $t$
\(
\|\mathbf{S}_t-\mathbf{S}_\star\|_{\mathrm{F}}
\le \sigma\,\mathsf{dist}(\mathbf{S}_t, \mathcal{C}_1 \cap \mathcal{C}_2)
\le \sigma\,\rho^t\,\mathsf{dist}(\mathbf{S}_0,  \mathcal{C}_1 \cap \mathcal{C}_2)
\),
which gives Eq.~\ref{eq:norm-rate}. The last inequality in the statement follows since
$\mathsf{dist}(\mathbf{S}_0,\mathcal{C}_1 \cap \mathcal{C}_2)\le\|\mathbf{S}_0 - \mathbf{S}_\star\|_{\mathrm{F}}$.
\end{proof}

\begin{corollary}[Iteration Complexity]
\label{cor:iter}
Let $\mathbf{S}_{t+1}= T_\lambda \mathbf{S}_t$ and let $\tau>0$ be a target tolerance (i.e., $\mathsf{dist}(\mathbf{S}_t,\mathcal{C})\le \tau$ is desired).
Under the assumptions of Theorem~\ref{thm:aap},
\begin{equation}
\mathsf{dist}(\mathbf{S}_t,\mathcal{C})\le \rho^t\,\mathsf{dist}(\mathbf{S}_0, \mathcal{C})
\quad \Rightarrow\quad
t \;\ge\; \frac{\log\bigl(\mathsf{dist}(\mathbf{S}_0, \mathcal{C})/\tau\bigr)}{\log(1/\rho)}
\;=\; \mathcal{O}\bigl(\log(1/\tau)\bigr), \qquad \rho= \sqrt{1-\frac{m}{\kappa^2}}.
\end{equation}
\end{corollary}

\begin{remark}[Practical Expected Iteration]
Since $\mathbf{S}_0 = \mathbf{S} + \frac{1}{2} \left( \mathbf{W} + \mathbf{W}^\top \right)$ and $\mathbf{S} \in \mathcal{C}$, we have $\mathsf{dist} \left( \mathbf{S}_0 , \mathcal{C} \right) \leq {\Vert \frac{1}{2}  \left( \mathbf{W} + \mathbf{W}^\top \right)\Vert}_{\mathrm{F}}$. If $W_{ij} \overset{i.i.d.}{\sim} \mathcal{N} (0, \sigma^2)$, we have 
\begin{equation}
   \mathbb{E} \Big[ {\big\Vert \frac{1}{2}  \left( \mathbf{W} + \mathbf{W}^\top \right) \big\Vert}_{\mathrm{F}} \Big] \leq \sigma \sqrt{\frac{n (n+1)}{2}}. 
\end{equation}
With Gaussian mechanism $\sigma = \Delta \sqrt{2 \log (2/ \delta)}/ \varepsilon$ we have $\mathbb{E} \big[ {\Vert \frac{1}{2}  \left( \mathbf{W} + \mathbf{W}^\top \right) \Vert}_{\mathrm{F}} \big] \leq \frac{\Delta}{\varepsilon} \sqrt{n (n+1) \log (2/\delta)}$. Hence, a practical expected iteration bound is
\begin{equation}
t \;\gtrsim\;
\frac{\log\!\Big(\frac{\Delta}{\varepsilon\tau}\sqrt{n(n+1)\log(2/\delta)}\Big)}{\log(1/\rho)}.
\end{equation}
Therefore decreasing $\varepsilon$ or $\delta$ (more noise) increases the initial distance but does not change the geometric rate $\rho$, which depends only on the geometry.
%
\end{remark}

\appsubsection{Symmetrization Prior to PSD Projection}

Before each PSD projection we replace the iterate $\widehat{\mathbf{S}}_t$ by its symmetric part $\mathbf{Y} = \frac{1}{2} (\widehat{\mathbf{S}}_t + \widehat{\mathbf{S}}_t^\top)$.
This transformation serves three purposes. 
(i) It preserves the inherent symmetry of the target Gram matrix $\mathbf{S} = \mathbf{E} \mathbf{E}^\top$,  ensuring structural consistency throughout the alternating-projection procedure. 
(ii) It guarantees that the subsequent eigen-decomposition used to realize $\mathsf{proj}_{\mathcal{K}_{+}^{\,n}} (\cdot)$ is well-posed, thereby removes the skew-symmetric component which cannot be reduced by PSD projection and would otherwise inflate the distance. (iii) Because for any matrix the closest PSD point in Frobenius norm is obtained from its symmetric part, our explicit symmetrization merely makes an implicit step in the original perturb-and-project algorithm of Cohen-Addad \textit{et al.} \citep{cohen2024perturb} transparent without altering privacy or utility guarantees. The operation costs only $\mathcal{O} (n^2)$ flops, which is negligible relative to the $\mathcal{O} (n^3)$ eigen-decomposition that follows.

\begin{lemma}[Symmetrization Invariance of the PSD Projection]
\label{lem:symmetrization}
Let $\widehat{\mathbf{S}}_t \in  \mathbb{R}^{n\times n}$ be any (not necessarily symmetric) matrix and define its symmetric and skew-symmetric parts
\begin{equation}
\mathbf{Y}  \coloneqq  \tfrac12\bigl(\widehat{\mathbf{S}}_t + \widehat{\mathbf{S}}_t^\top\bigr),
\qquad
\mathbf{K}  \coloneqq  \tfrac12\bigl(\widehat{\mathbf{S}}_t - \widehat{\mathbf{S}}_t^\top\bigr).
\end{equation}
Then
\begin{equation}
\label{eq:symmetrization-identity}
\min_{\mathbf{X} \succeq 0} \|\mathbf{X} - \widehat{\mathbf{S}}_t\|_{\mathrm{F}}^{2}
\;=\; \min_{\mathbf{X} \succeq 0} \|\mathbf{X} - \mathbf{Y}\|_{\mathrm{F}}^{2}
\;+\; \tfrac{1}{4} \|\widehat{\mathbf{S}}_t - \widehat{\mathbf{S}}_t^\top\|_{\mathrm{F}}^{2},
\end{equation}
and every minimizer of the left-hand side is symmetric and coincides with a minimizer of the first term on the right-hand side. In particular,
\begin{equation}
\mathsf{proj}_{\mathcal{K}_{+}^{\,n}}(\widehat{\mathbf{S}}_t)
\;=\; \mathsf{proj}_{\mathcal{K}_{+}^{\,n}}\bigl(\mathbf{Y}\bigr).
\end{equation}
\end{lemma}

\begin{proof}
Write $\widehat{\mathbf{S}}_t = \mathbf{Y} + \mathbf{K}$ with $\mathbf{Y}^\top=\mathbf{Y}$ and $\mathbf{K}^\top=-\mathbf{K}$. Any feasible $\mathbf{X} \succeq 0$ is symmetric, so
\begin{equation}
\|\mathbf{X} - \widehat{\mathbf{S}}_t\|_{\mathrm{F}}^{2}
= \|\mathbf{X} - (\mathbf{Y} + \mathbf{K})\|_{\mathrm{F}}^{2}
= \|\mathbf{X} - \mathbf{Y}\|_{\mathrm{F}}^{2} + \|\mathbf{K}\|_{\mathrm{F}}^{2},
\end{equation}
because $\langle \mathbf{X} - \mathbf{Y}, \mathbf{K}\rangle_{\mathrm{F}} = 0$ (symmetric vs.\ skew–symmetric orthogonality). Since $\|\mathbf{K}\|_{\mathrm{F}}^{2} = \tfrac14\|\widehat{\mathbf{S}}_t - \widehat{\mathbf{S}}_t^\top\|_{\mathrm{F}}^{2}$ is independent of $\mathbf{X}$, minimizing over $\mathbf{X}\succeq 0$ yields Eq.~\ref{eq:symmetrization-identity}. The independence also shows any minimizer must minimize $\|\mathbf{X} - \mathbf{Y}\|_{\mathrm{F}}^{2}$ subject to $\mathbf{X}\succeq 0$, hence is symmetric and equal to $\mathsf{proj}_{\mathcal{K}_{+}^{\,n}}(\mathbf{Y})$.
\end{proof}

\begin{corollary}[Distance Decomposition]
\label{cor:symmetrization-distance}
Under our setup addressed in Theorem~\ref{thm:aap}, for any $\widehat{\mathbf{S}}_t$ we have
\begin{equation}
\mathsf{dist}\bigl(\widehat{\mathbf{S}}_t,\mathcal{K}_{+}^{\,n}\bigr)^{2}
= \mathsf{dist}\Bigl(\tfrac12(\widehat{\mathbf{S}}_t+\widehat{\mathbf{S}}_t^\top),
\mathcal{K}_{+}^{\,n}\Bigr)^{2} + \tfrac{1}{4}\|\widehat{\mathbf{S}}_t - \widehat{\mathbf{S}}_t^\top\|_{\mathrm{F}}^{2}.
\end{equation}
\end{corollary}

\begin{proof}
Immediate from Lemma~\ref{lem:symmetrization} by recognizing each distance as the objective value at the respective minimizer.
\end{proof}


\appsubsection{%
  Dykstra's Algorithm for the Metric Projection onto
  \texorpdfstring{%
    $\mathcal{K}_+^{\,n}\cap\mathcal{C}_{\mathrm{unit}}^{\,n}$%
  }{K plus n intersect C unit n}%
}
\label{app:ssec:dykstra}

Let consider the Hilbert space $\mathcal{H} \coloneqq (\mathtt{S}^n,\langle\cdot,\cdot\rangle_{\mathrm{F}})$ of real symmetric $n\times n$ matrices equipped with the Frobenius inner product. Define the closed convex sets
\begin{equation}
\mathcal{C}_1 \coloneqq \mathcal{K}_+^{\,n}=\{\mathbf{S}\in\mathtt{S}^n:\ \mathbf{S}\succeq 0\},
\qquad
\mathcal{C}_2 \coloneqq \mathcal{C}_{\mathrm{unit}}^{\,n}
=\{\mathbf{S}\in\mathtt{S}^n:\ S_{ii}=1,\ |S_{ij}|\le 1\ (i\neq j)\}. \nonumber 
\end{equation}
Given a (possibly non-symmetric) noisy matrix $\widetilde{\mathbf{S}}_0\in\mathbb{R}^{n\times n}$ arising from the DP perturbation step, we first symmetrize $\mathbf{S}_0 \;\coloneqq \; \tfrac12 \bigl( \widetilde{\mathbf{S}}_0 + \widetilde{\mathbf{S}}_0^\top \bigr)\in \mathtt{S}^n$. This symmetrization is a deterministic post-processing and, for the PSD projection, does not change the result (Lemma~\ref{lem:symmetrization}). The Euclidean (Frobenius) metric projection onto the cosine-Gram feasibility set is
\begin{equation}
\mathsf{proj}_{\mathcal{C}_1 \cap \mathcal{C}_2}(\mathbf{S}_0)
\;=\; \arg\min_{\mathbf{S} \in \mathcal{C}_1 \cap \mathcal{C}_2}\ \tfrac12 \| \mathbf{S} - \mathbf{S}_0 \|_{\mathrm{F}}^2.
\end{equation}

\noindent
\textbf{Feasibility Enforcement versus Metric Projection.}
Our fast projection method based on AAP is a feasibility method, i.e., it generates iterates whose distance to $\mathcal{C}_1 \cap \mathcal{C}_2$ decreases (and is $R$-linear under bounded linear regularity on the Fej\'er ball), but it does not, in general, return the metric projection of $\mathbf{S}_0$ onto $\mathcal{C}_1 \cap \mathcal{C}_2$. When the best-approximation property is required, one may instead use Dykstra's algorithm (see Algorithm~\ref{alg:dykstra}), which alternates the same two projectors with correction terms and converges to $\mathsf{proj}_{\mathcal{C}_1 \cap \mathcal{C}_2}(\mathbf{S}_0)$ \citep{dykstra1983algorithm, bauschke1994dykstra, boyle1986method, dykstra1985iterative}.
Per outer iteration, both methods apply (i) one PSD-cone projection (dominated by an eigendecomposition) and (ii) one entrywise projection onto $\mathcal{C}_2$. Dykstra additionally stores and updates two dense correction matrices.

\begin{algorithm}[t]
\caption{Dykstra's algorithm for $\mathsf{proj}_{\mathcal{C}_1\cap\mathcal{C}_2}(\mathbf{S}_0)$}
\label{alg:dykstra}
\begin{algorithmic}[1]
\State \textbf{Input:} $\mathbf{S}_0\in\mathtt{S}^n$, sets $\mathcal{C}_1,\mathcal{C}_2\subset\mathtt{S}^n$
\State Initialize $\mathbf{S}^{(0)}\gets \mathbf{S}_0$, $\mathbf{U}^{(0)}\gets \mathbf{0}$, $\mathbf{V}^{(0)}\gets \mathbf{0}$
\For{$k=0,1,2,\dots,K-1$} 
    \State $\mathbf{Y}^{(k)} \gets \mathsf{proj}_{\mathcal{C}_1}\!\big(\mathbf{S}^{(k)}+\mathbf{U}^{(k)}\big)$
    \State $\mathbf{U}^{(k+1)} \gets \mathbf{S}^{(k)}+\mathbf{U}^{(k)}-\mathbf{Y}^{(k)}$
    \State $\mathbf{S}^{(k+1)} \gets \mathsf{proj}_{\mathcal{C}_2}\!\big(\mathbf{Y}^{(k)}+\mathbf{V}^{(k)}\big)$
    \State $\mathbf{V}^{(k+1)} \gets \mathbf{Y}^{(k)}+\mathbf{V}^{(k)}-\mathbf{S}^{(k+1)}$
\EndFor
\State \textbf{Return:} $\mathbf{S}^{(K)}$
\end{algorithmic}
\end{algorithm}

\begin{theorem}[Convergence of Dykstra's algorithm to the metric projection]
\label{thm:dykstra}
Let $\mathcal{H}$ be a finite-dimensional Hilbert space and let
$\mathcal{C}_1,\mathcal{C}_2\subset\mathcal{H}$ be nonempty, closed, convex sets with
$\mathcal{C}_1\cap\mathcal{C}_2\neq\varnothing$.
Let $\mathbf S_0\in\mathcal H$, and let $(\mathbf S^{(k)})_{k\ge0}$ be the primal sequence produced by Algorithm~\ref{alg:dykstra}. Then
\[
    \mathbf S^{(k)}
    \longrightarrow
    \mathsf{proj}_{\mathcal C_1\cap\mathcal C_2}(\mathbf S_0)
\]
in norm as $k\to\infty$.
\end{theorem}

\begin{proof}
This is the classical convergence guarantee for Dykstra's algorithm for projecting onto the intersection of finitely many closed convex sets in finite-dimensional Hilbert spaces; see
\citep{dykstra1983algorithm,boyle1986method,dykstra1985iterative,bauschke1994dykstra}.
\end{proof}

\appsubsection{Feasible Sets and Projection Decompositions}

In this subsection we clarify the feasible set and projection decomposition used for regime~(ii), and contrasts it with the perturb-and-project construction of Cohen-Addad \textit{et al.}~\citep{cohen2024perturb}. Our projection uses the decomposition $A=\mathcal{K}_+^n$ and $B=\mathcal{C}_{\mathrm{unit}}^n$, where
\begin{equation}
\mathcal{K}_+^n = \{\mathbf{M}\in\mathbb S^n:\mathbf{M}\succeq0\},
\qquad 
\mathcal{C}_{\mathrm{unit}}^n = \{\mathbf{M}\in\mathbb S^n:M_{ii}=1,\ |M_{ij}|\le1,\ i\ne j\}.
\end{equation}
Projection onto $A$ is eigenvalue clipping, and projection onto $B$ sets the diagonal to one and clips the off-diagonal entries to $[-1,1]$.

Cohen-Addad \textit{et al.} instead use a decomposition of the form $A'=\mathcal{K}_+^n\cap \mathcal{B}_F^n$ and $B'=\mathcal{C}_{\max}^n$, where
\begin{equation}
\mathcal{B}_F^n=\{\mathbf{M}\in\mathbb S^n:\|\mathbf{M}\|_F\le n\},
\qquad
\mathcal{C}_{\max}^n = \{\mathbf{M}\in\mathbb S^n:\|\mathbf{M}\|_{\max}\le1\}.
\end{equation}
This decomposition does not enforce $M_{ii}=1$. Its Frobenius-ball step can radially rescale the positive spectrum, and this rescaling may change the diagonal entries.  Such a step is incompatible with exact preservation of the cosine self-similarity constraint $M_{ii}=1$.

In our setting the radius-$n$ Frobenius ball is unnecessary. If $\mathbf{M}\in\mathcal{C}_{\mathsf{coll}}$, then $|M_{ij}|\le1$ for all entries and therefore
\begin{equation}
\|\mathbf{M}\|_F^2 = \sum_{i,j}M_{ij}^2 \le n^2,
\qquad
\|\mathbf{M}\|_F\le n.
\end{equation}
Hence every feasible cosine Gram matrix already lies in $\mathcal{B}_F^n$. The unit-diagonal constraint gives compactness without adding a separate Frobenius-ball constraint.

\clearpage
\appsection{Supplementary Details for Regime (ii): Extended Experiments} 
\label{sec:supplementary-regime2-experiments}

\paragraph{Evaluation overview.}
We evaluate settings in which downstream algorithms access only a symmetric similarity matrix, or deterministic transformations of it.
The experiments assess three questions:
\begin{enumerate}[label=(\roman*), leftmargin=*]
\item How much utility is retained after enforcing the cosine-Gram feasibility constraints on a privatized Gram matrix?
\item How does the projected release compare with the corresponding non-private Gram baseline?
\item How does the proposed feasibility solver scale compared with an SDP baseline when both are tractable?
\end{enumerate}
Some experiments also report the intermediate noisy Gram matrix before projection. When this baseline is reported, it is labeled \textbf{Noisy}. For CIFAR-10 and CIFAR-100, the reported plots compare only the non-private Gram and the projected \textsc{ScoreShield} Gram; the unprojected noisy baseline is not included in those figures.

\appsubsection{Experimental Setup}

\paragraph{Implementation.}

Experiments are implemented in Python using PyTorch. Feature encoders are instantiated with \texttt{timm} and evaluated with deterministic preprocessing; no stochastic data augmentations are used. Computation uses a single CUDA GPU when available and otherwise runs on CPU. The AAP solver runs for at most $500$ iterations and terminates early when the relative Frobenius-norm change falls below $10^{-7}$. For benchmarks involving random subsampling, each configuration is repeated $R=5$ times with distinct seeds. We report mean $\pm$ standard deviation and plot mean curves with shaded $\pm$ standard-deviation bands.

\paragraph{Privacy parameters.}

We fix $\delta=10^{-5}$ and sweep $\varepsilon$ on a logarithmic grid (e.g., $\{0.1,1,10,100\}$) to cover the privacy–utility range. We perturb the Gram entrywise via the Gaussian mechanism $S'_{ij}\; = \;S_{ij} + Z_{ij}$, $Z_{ij} \stackrel{\mathrm{i.i.d.}}{\sim}\mathcal{N}(0, \sigma^2)$, with scale $\sigma = \sigma(\varepsilon,\delta,\Delta)$ calibrated from an $\ell_2$-sensitivity bound $\Delta$ for a single entry (classic sufficient vs. analytic calibration). We then enforce symmetry by $\mathbf{S}'\!\leftarrow\tfrac{1}{2}(\mathbf{S}'+ \mathbf{S}'^{\top})$. The same calibration is used across benchmarks.

\paragraph{Reported evaluation variants.}
Depending on the benchmark, we consider the following matrix variants:
\begin{enumerate}[leftmargin=*]
\item
\textbf{Non-private} ($\mathbf{S}$): construct the cosine Gram from $\ell_2$-normalized embeddings, or from the task-specific construction noted below. This serves as the non-private reference.

\item
\textbf{Noisy} ($\mathbf{S}'$): apply the Gaussian mechanism to $\mathbf{S}$ with scale $\sigma=\sigma(\varepsilon,\delta,\Delta)$ calibrated for the adopted adjacency model, and enforce symmetry by $\mathbf{S}'\leftarrow \frac{1}{2}(\mathbf{S}'+\mathbf{S}'^{\top})$.
The resulting matrix need not be positive semidefinite, need not have unit diagonal, and may have entries outside $[-1,1]$.

\item
\textbf{\textsc{ScoreShield}} ($\widehat{\mathbf{S}}$): apply a deterministic post-processing map to the noisy matrix to enforce the cosine-Gram constraints. In the large-scale experiments, this step is implemented by AAP over the PSD cone, the unit-diagonal affine set, and the entrywise box $[-1,1]$.
\end{enumerate}
The `Noisy' variant is not reported for every benchmark.

\paragraph{Stability of graph-based methods.}

From the cosine Gram $\mathbf{S}$ we form an affinity $\mathbf{A}=\tfrac{1}{2}(\mathbf{S} + \mathbf{1})$, clip to $[0,1]$, and set $\mathsf{diag}(\mathbf{A})= \mathbf{0}$. To control degrees and avoid isolates, we optionally sparsify with a symmetric $m$-nearest-neighbor graph (retain edge $(i,j)$ iff $i\!\in\!\mathrm{NN}_m(j)$ or $j\!\in\!\mathrm{NN}_m(i)$). Spectral clustering is applied to the symmetric normalized Laplacian $\mathbf{L}_{\mathrm{sym}}= \mathbf{I} - \mathbf{D}^{-1/2}\mathbf{A} \mathbf{D}^{-1/2}$. Unless otherwise stated, we use an oracle-$k$ setting where $k$ equals the number of ground-truth classes in the subsample (capped at $10$ for CIFAR-10); a fixed-$k$ variant is reported in ablations.

\paragraph{Backbones and datasets.}

Unless otherwise noted, we extract frozen image embeddings with DINOv2-B/14 \citep{oquab2023dinov2} (ImageNet-pretrained) using standard evaluation preprocessing (resize and center-crop). Vision benchmarks are CIFAR-10/100 \citep{krizhevsky2009cifar}, and Oxford-IIIT Pets \citep{parkhi2012pets}. Non-vision benchmarks are STS-B \citep{cer2017sts} and MovieLens-100K \citep{harper2015movielens}.

\appsubsection{Benchmark Suite and Metrics}

All tasks consume only the released Gram matrix $S$ and therefore directly assess the usability of a DP Gram release. Metrics are defined precisely below.

\paragraph{Unsupervised clustering (vision).}
 \leavevmode

\textit{Goal:} recover semantic clusters from $\mathbf{S}$ alone. \\
\textit{Construction:} for a random subsample of $n$ images, compute cosine similarities
$S_{ij}= \tfrac{\mathbf{x}_i^\top \mathbf{x}_j}{\|\mathbf{x}_i\|\,\|\mathbf{x}_j\|}$ from $\ell_2$-normalized embeddings and form an affinity $\mathbf{A}=\tfrac{1}{2}(\mathbf{S}+\mathbf{1})$ with $\mathrm{diag}(\mathbf{A})= \mathbf{0}$. Apply spectral clustering (optionally with a symmetric $m$-NN sparsification).\\
\textit{Metric:} Normalized Mutual Information (NMI) between predicted $\widehat{Y}$ and labels $Y$,
\begin{equation}
\mathrm{NMI}(Y,\widehat{Y}) = \frac{2 I(Y; \widehat{Y})}{H(Y) + H(\widehat{Y})} \in [0,1].    
\end{equation}
Report mean$\pm$std over $R=5$ subsamples per $(\varepsilon,n)$.

\paragraph{$k$-NN classification (vision).}
\leavevmode

\textit{Goal:} assign labels using neighborhoods induced by $\mathbf{S}$.\\
\textit{Construction:} with labels $\mathbf{y}\in\{1,\dots,C\}^n$, predict $\widehat{y}_i$ by majority vote over the $k$ largest off-diagonal entries in row $i$ of $\mathbf{S}$ (exclude $S_{ii}$).\\
\textit{Metric:} Top-1 accuracy $\frac{1}{n}\sum_{i=1}^n \mathbf{1}\{\widehat{y}_i = y_i\}$. Ties are broken by class frequency, then index.

\paragraph{Instance retrieval (vision).}
\leavevmode

\textit{Goal:} retrieve same-class instances using only $\mathbf{S}$.\\
\textit{Construction:} for each query $i$, rank $j \neq i$ by descending $S_{ij}$.\\
\textit{Metrics:} Recall@K and mAP. With $r_i(j)$ the $j$-th ranked neighbor and $M_i$ the number of relevant items,
\begin{equation}
\mathrm{R@K}= \frac{1}{n}\sum_{i=1}^n \mathbf{1}\{\exists\, j\le K:\, y_{r_i(j)}=y_i\},\qquad
\mathrm{mAP}=\frac{1}{n}\sum_{i=1}^n \frac{1}{M_i}\sum_{j=1}^{n-1} \mathrm{Prec}@j\cdot \mathbf{1}\{y_{r_i(j)}=y_i\}.    
\end{equation}

\paragraph{Same/different verification (vision).}
\leavevmode

\textit{Goal:} decide whether $(i,j)$ share a label using $S_{ij}$ as the score.\\
\textit{Metric:} ROC–AUC computed from $\{(S_{ij},\mathbf{1}\{y_i=y_j\})\}_{i<j}$ (diagonal excluded).

\paragraph{Semantic textual similarity (NLP).}
\leavevmode

\textit{Goal:} preserve sentence-pair similarity.\\
\textit{Construction:} embed STS-B sentences with SBERT and form a cosine Gram $\mathbf{S}$; for each labeled pair $(u,v)$, use $S_{uv}$.\\
\textit{Metric:} Spearman's rank correlation $\rho$ between $\{S_{uv}\}$ and human scores.

\paragraph{User-based collaborative filtering (recommender).}
\leavevmode

\textit{Goal:} predict ratings from user–user similarity.\\
\textit{Construction:} represent each user by a centered rating vector; $\mathbf{S}$ is the user–user Gram. Predict $\widehat{r}_{ui}$ by a similarity-weighted average over neighbors who rated item $i$.\\
\textit{Metric:} RMSE on a held-out test set,
\begin{equation}
\mathrm{RMSE} = \sqrt{\frac{1}{|\mathcal{T}|} \sum_{(u,i)\in\mathcal{T}} \big(\widehat{r}_{ui} - r_{ui}\big)^2 }.    
\end{equation}

\appsubsection{How the Gram is Consumed}

All downstream methods use only the Gram matrix $\mathbf{S} \in\mathbb{R}^{n\times n}$ (or deterministic transforms thereof); no raw features are accessed at inference time. For vision tasks, embeddings are $\ell_2$-normalized prior to forming cosine similarities, so $\mathbf{S}$ is a cosine Gram.

\noindent
- \textit{Clustering.} We construct an affinity $\mathbf{A}= \tfrac{1}{2}(\mathbf{S}+ \mathbf{1})$, set $\mathsf{diag}(\mathbf{A})=0$, optionally apply a symmetric $m$-nearest-neighbor sparsification for graph stability, and run spectral clustering with $k$ chosen either by oracle ($k=$ number of unique labels in the subsample) or fixed.

\noindent
- \textit{$k$-NN classification.} For each index $i$, we exclude self-similarity and assign $\widehat{y}_i$ by majority vote over the $k$ largest off-diagonal entries in row $i$ of $\mathbf{S}$; ties are resolved deterministically (class frequency, then index).

\noindent
- \textit{Retrieval and verification.} Retrieval ranks candidates $j\neq i$ by descending $S_{ij}$; metrics (e.g., Recall@K, mAP) are computed from these rankings. Verification uses $S_{ij}$ as the decision statistic to produce ROC–AUC or TMR@FMR.

\noindent
- \textit{Semantic similarity (STS-B).} For each labeled sentence pair $(u,v)$, the predicted similarity is the single Gram entry $S_{uv}$; utility is measured via Spearman's $\rho$ against human scores.

\noindent
- \textit{Recommender (MovieLens).} A user–user Gram is formed from centered rating vectors; predicted ratings are similarity-weighted averages over neighbors who rated the target item.

\noindent
- \textit{Link prediction (if used).} Candidate edges $(u,v)$ are scored directly by $S_{uv}$ and ranked accordingly.

In all cases, the identical pipeline is applied to non-private $\mathbf{S}$, noisy $\mathbf{S}'$, and ScoreShield projected $\widehat{\mathbf{S}}$.

\appsubsection{Aggregation and Visualization}

In stochastic settings with random $n$-subsampling, each configuration is repeated $R=5$ times; we report $\mu\pm\sigma$ across repeats and plot mean curves with translucent $\pm1\sigma$ bands. Deterministic settings with fixed $n$ report single values. Runtime summaries report measured wall-clock time for the AAP feasibility solver and, when the optimization terminates within the prescribed wall-time budget, for the SDP metric-projection baseline at the same $(\varepsilon,\delta,\Delta,n)$. If the SDP solver does not terminate within the budget, the corresponding entry is omitted rather than extrapolated.

\subsection{Runtime Comparison with an SDP Metric-Projection Baseline}

We compare AAP with an SDP metric-projection baseline for the nearest feasible cosine-Gram problem
\begin{equation}
\min_{\mathbf{S}}\ \|\mathbf{S}-\mathbf{S}'\|_F^2
\quad
\text{s.t.}\quad
\mathbf{S}\succeq 0,\quad
\mathrm{diag}(\mathbf{S})=\mathbf{1},\quad
|S_{ij}|\le 1\ \ (i\neq j).
\end{equation}
The SDP baseline is implemented in CVXPY using MOSEK, Clarabel, or SCS backends, subject to a fixed wall-time budget. AAP is not an exact metric-projection solver; it is the feasibility-enforcement post-processing used in the large-scale experiments. Its dominant per-iteration cost is the PSD projection, implemented through an eigen-decomposition. The SDP baseline solves the Frobenius metric-projection problem above and is included only for problem sizes where it terminates within the wall-time budget.

\begin{figure*}[!htbp]
  \centering
  \foreach \bigdval in {0.5,2}{%
    \foreach \smalldelta/\deltadisp in {1.00e-08/10^{-8},1.00e-05/10^{-5},0.01/10^{-2}}{%
      \begin{subfigure}[t]{0.32\textwidth}
        \includegraphics[
          trim={20pt 25pt 0pt 60pt}, 
          clip,
          width=\linewidth
        ]
        {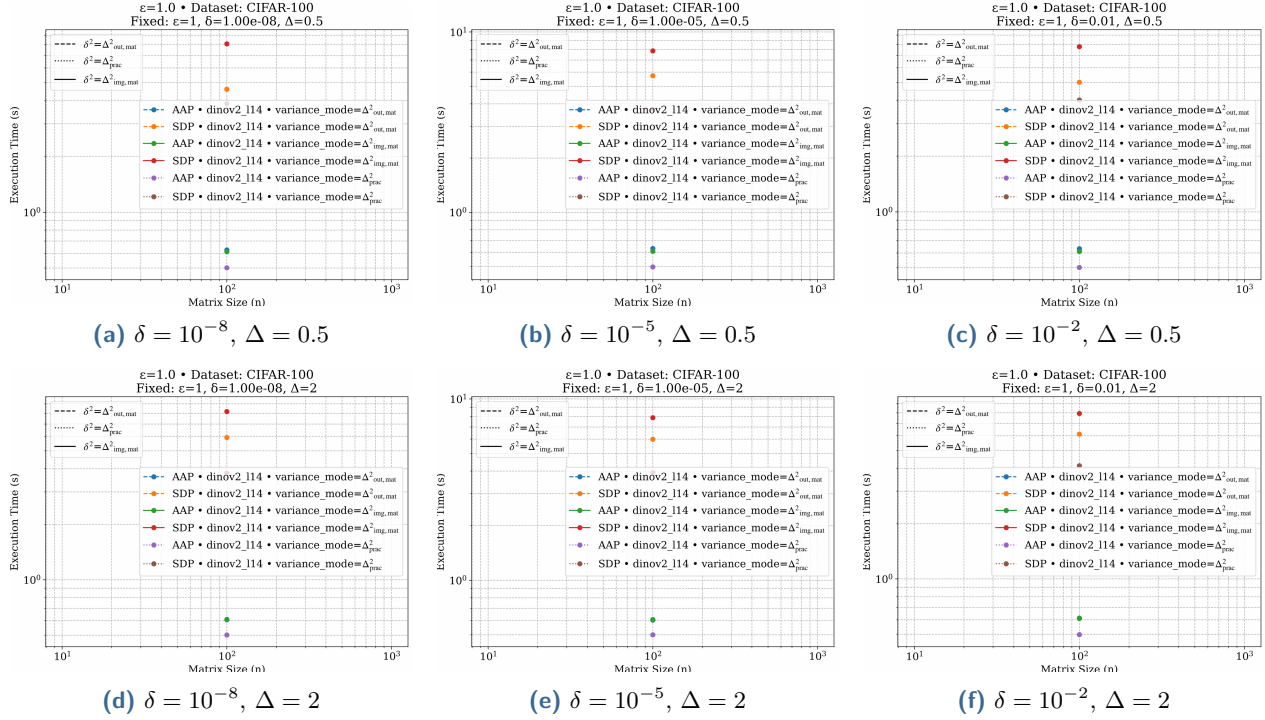}
        \caption{$\delta=\deltadisp$, $\Delta=\bigdval$}
      \end{subfigure}%
      \hfill
    }
    \par\medskip
  }
  \caption{\small CIFAR-100 clustering runtime comparison. Wall-clock execution time is reported in seconds on a logarithmic scale for the AAP feasibility solver and the SDP metric-projection baseline when the latter terminates within the prescribed wall-time budget.}
  \label{fig:cifar100_runtime}
\end{figure*}

Figure~\ref{fig:cifar100_runtime} reports a representative CIFAR-100 clustering runtime comparison at matched $(\varepsilon,\delta,\Delta)$. The vertical axis is logarithmic. In the displayed configurations, AAP has lower wall-clock runtime than the SDP metric-projection baseline whenever both are reported. These measurements are used only to document the runtime behavior of the implementation, not to claim that AAP computes the exact metric projection.



\appsubsection{Benchmarks}

Each grid fixes the sensitivity $\Delta$ by row and increases the failure probability $\delta$ by column. The horizontal axis is the privacy budget $\varepsilon$ (log scale). Curves compare: (i) the non-private Gram $\mathbf{S}$, (ii) the noisy Gram $\mathbf{S}'$ obtained by the Gaussian mechanism calibrated for $(\varepsilon,\delta,\Delta)$, and (iii) the projected Gram $\widehat{\mathbf{S}}$ produced by our ScoreShield fast alternating projection. An SDP projection baseline is included when numerically tractable. Shaded ribbons show mean~$\pm\,1\sigma$ over $R{=}5$ repeats (when subsampling is used).

\subsubsection{CIFAR-10}
\label{subsec:cifar10}
\begin{figure*}[!htbp]
  \centering
  \foreach \bigdval in {0.5,2}{%
    \foreach \smalldelta/\deltadisp in {1.00e-08/10^{-8},1.00e-05/10^{-5},0.01/10^{-2}}{%
      \begin{subfigure}[t]{0.32\textwidth}
        \includegraphics[
          trim={20pt 25pt 0pt 80pt}, 
          clip,
          width=\linewidth
        ]{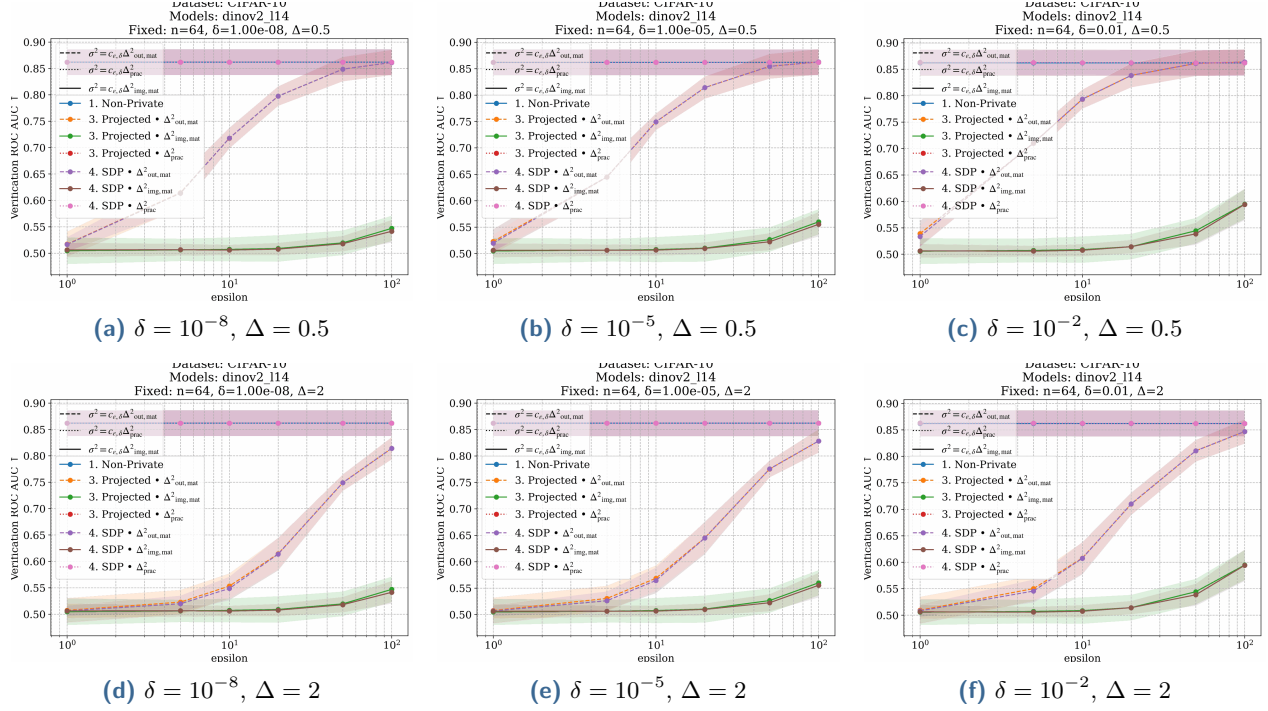}
        \caption{$\delta=\deltadisp$, $\Delta=\bigdval$}
      \end{subfigure}%
      \hfill
    }
    \par\medskip
  }
  \caption{\small CIFAR-10 verification AUC ($n=64$). Each row fixes $\Delta$; $\delta$ increases from left to right.
  The x-axis is $\varepsilon$ (log scale). Shaded regions indicate variability across runs.}
  \label{fig:cifar10_simple_auc_n64}
\end{figure*}

\begin{figure*}[!htbp]
  \centering
  \foreach \bigdval in {0.5,2}{%
    \foreach \smalldelta/\deltadisp in {1.00e-08/10^{-8},1.00e-05/10^{-5},0.01/10^{-2}}{%
      \begin{subfigure}[t]{0.32\textwidth}
        \includegraphics[
          trim={20pt 25pt 0pt 80pt}, 
          clip,
          width=\linewidth
        ]{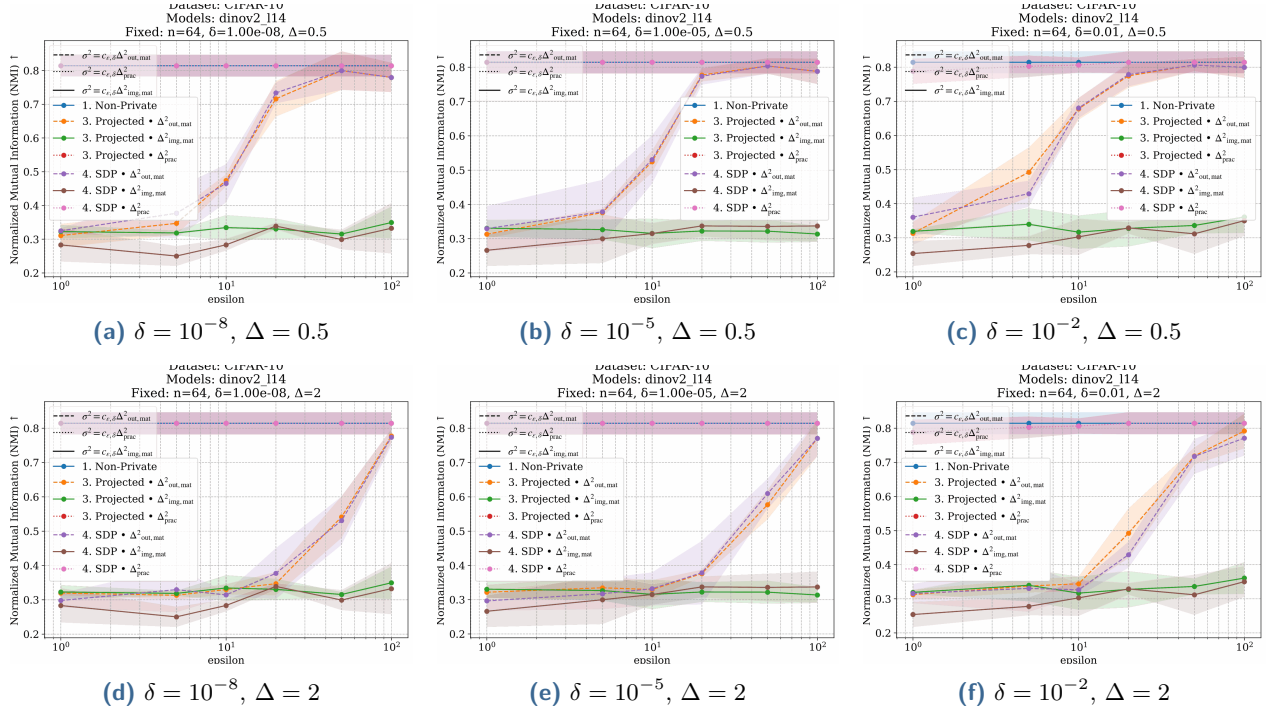}
        \caption{$\delta=\deltadisp$, $\Delta=\bigdval$}
      \end{subfigure}%
      \hfill
    }
    \par\medskip
  }
  \caption{\small CIFAR-10 clustering NMI ($n=64$). Each row fixes the sensitivity parameter $\Delta$; $\delta$; columns correspond to increasing $\delta$.
  The horizontal axis is $\varepsilon$ on a logarithmic scale. Shaded regions indicate variability across runs.}
  \label{fig:cifar10_simple_nmi_n64}
\end{figure*}

\textbf{Dataset and protocol.} CIFAR-10 has 10 classes and 60K images~\citep{krizhevsky2009cifar}. For each run we uniformly subsample $n{=}64$ images, extract frozen DINOv2-B/14 embeddings, form a cosine Gram $S$, and evaluate two tasks that consume only $S$: (i) pairwise same/different verification via ROC–AUC computed from $\{(S_{ij},\mathbf{1}\{y_i{=}y_j\})\}_{i<j}$; and (ii) spectral clustering scored by NMI with oracle $k{=}10$. This setting isolates the small-$n$ vision regime. Results are in Figs.~\ref{fig:cifar10_simple_auc_n64}–\ref{fig:cifar10_simple_nmi_n64}.

\clearpage

\subsubsection{CIFAR-100}
\label{subsubsec:cifar100}
\begin{figure}[!htbp]
  \centering

  \captionsetup{aboveskip=0pt, belowskip=0pt}
  \captionsetup[sub]{aboveskip=2pt, belowskip=2pt}

  \foreach \bigdval in {0.5,1,2}{%
    \foreach \smalldelta/\deltadisp [count=\col from 1]%
      in {1.00e-08/10^{-8},1.00e-05/10^{-5},0.01/10^{-2}}{%
      \begin{subfigure}[t]{0.32\linewidth}
        \includegraphics[
          trim=20pt 25pt 0pt 80pt, 
          clip,
          width=\linewidth
        ]{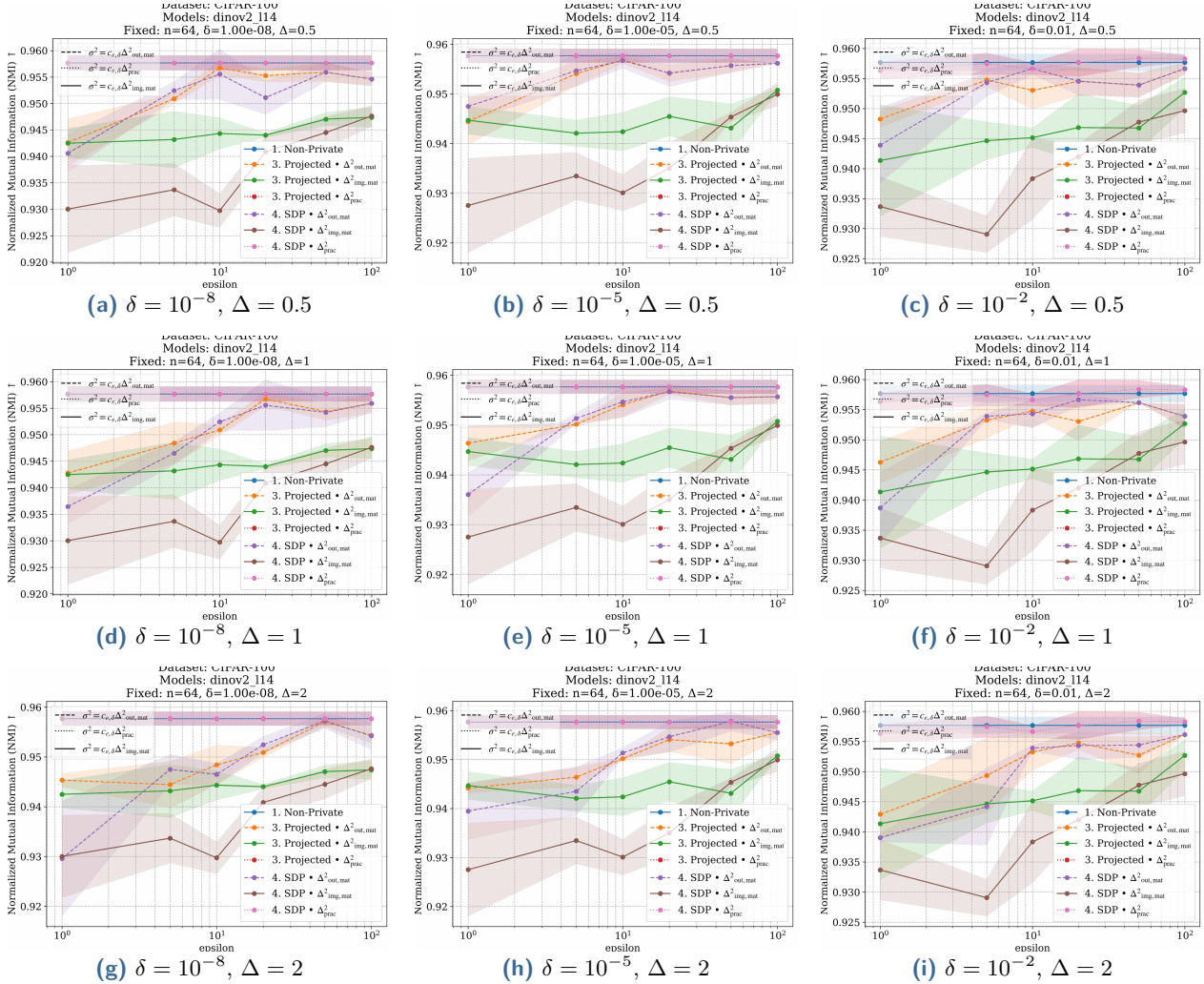}%
        \caption{$\delta=\deltadisp$, $\Delta=\bigdval$}
      \end{subfigure}%
      \ifnum\col<3\hfill\fi
    }%
    \par\vspace{0.3em}
  }%
  \caption{\small Results for CIFAR-100 utility NMI ($n=64$). Each row represents a fixed $\Delta$ with $\delta$ increasing from left to right. The horizontal axis is $\varepsilon$ on a logarithmic scale. Shaded bands show variability across repeated runs.}
  \label{fig:cifar100_nmi_n64}
\end{figure}

\begin{figure}[!htbp]
  \centering

  \captionsetup{aboveskip=0pt, belowskip=0pt}
  \captionsetup[sub]{aboveskip=2pt, belowskip=2pt}

  \foreach \bigdval in {0.5,1,2}{%
    \foreach \smalldelta/\deltadisp [count=\col from 1]%
      in {1.00e-08/10^{-8},1.00e-05/10^{-5},0.01/10^{-2}}{%
      \begin{subfigure}[t]{0.32\linewidth}
        \includegraphics[
          trim=20pt 25pt 0pt 80pt, 
          clip,
          width=\linewidth
        ]{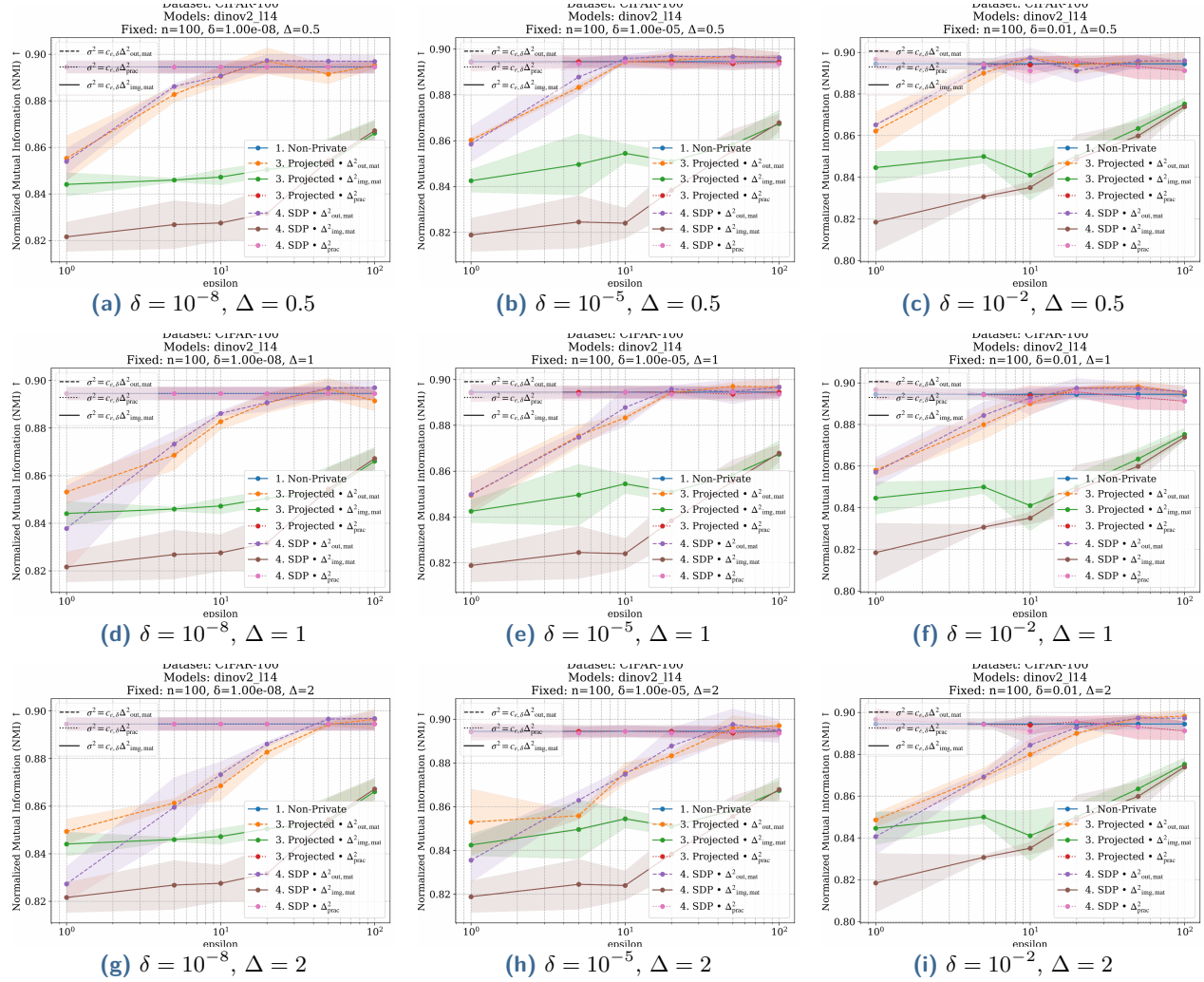}
        \caption{$\delta=\deltadisp$, $\Delta=\bigdval$}
      \end{subfigure}%
      \ifnum\col<3\hfill\fi
    }%
    \par\vspace{0.3em}
  }%

  \caption{\small Results for CIFAR-100 utility NMI ($n=100$). Each row represents a fixed $\Delta$ with $\delta$ increasing from left to right. The horizontal axis is $\varepsilon$ on a logarithmic scale. Shaded bands show variability across repeated runs.}
  \label{fig:cifar100_nmi_n100}
\end{figure}

\textbf{Dataset and protocol.} CIFAR-100 has 100 classes and 60K images. To enable comparison with the SDP projection baseline (which becomes memory-bound for larger $n$) we report spectral clustering NMI at $n\in\{64,100\}$. For clustering, $k$ is set to the number of distinct labels present in the subsample (oracle $k$). We omit $k$-NN and verification on CIFAR-100, as stable estimates in this setting require on the order of $10^3$ images (i.e., $\gtrsim$10 per class), which is outside the feasible range for the SDP baseline. See Figs.~\ref{fig:cifar100_nmi_n64} and \ref{fig:cifar100_nmi_n100}.

\vspace{10pt}

\clearpage
\subsubsection{STSBench}

\vspace{10pt}

\textbf{Semantic Textual Similarity (STS-B)}~\citep{cer2017sts}. 
We embed the $n{=}2{,}552$ unique sentences with SBERT, construct the sentence–sentence cosine Gram $S$, and evaluate utility as Spearman’s rank correlation $\rho$ between $\{S_{uv}\}_{(u,v)\in\mathcal{L}}$ and the human similarity scores on the labeled pairs $\mathcal{L}$ (diagonal excluded; higher is better).

\vspace{10pt}

\begin{figure}[!htbp]
  \centering

  \captionsetup{aboveskip=0pt, belowskip=0pt}
  \captionsetup[sub]{aboveskip=2pt, belowskip=2pt}

  \foreach \bigdval in {0.5,1,2}{%
    \foreach \smalldelta/\deltadisp [count=\col from 1]%
      in {1.00e-08/10^{-8},1.00e-05/10^{-5},0.01/10^{-2}}{%
      \begin{subfigure}[t]{0.32\linewidth}
        \includegraphics[
          trim=20pt 25pt 0pt 80pt, 
          clip,
          width=\linewidth
        ]{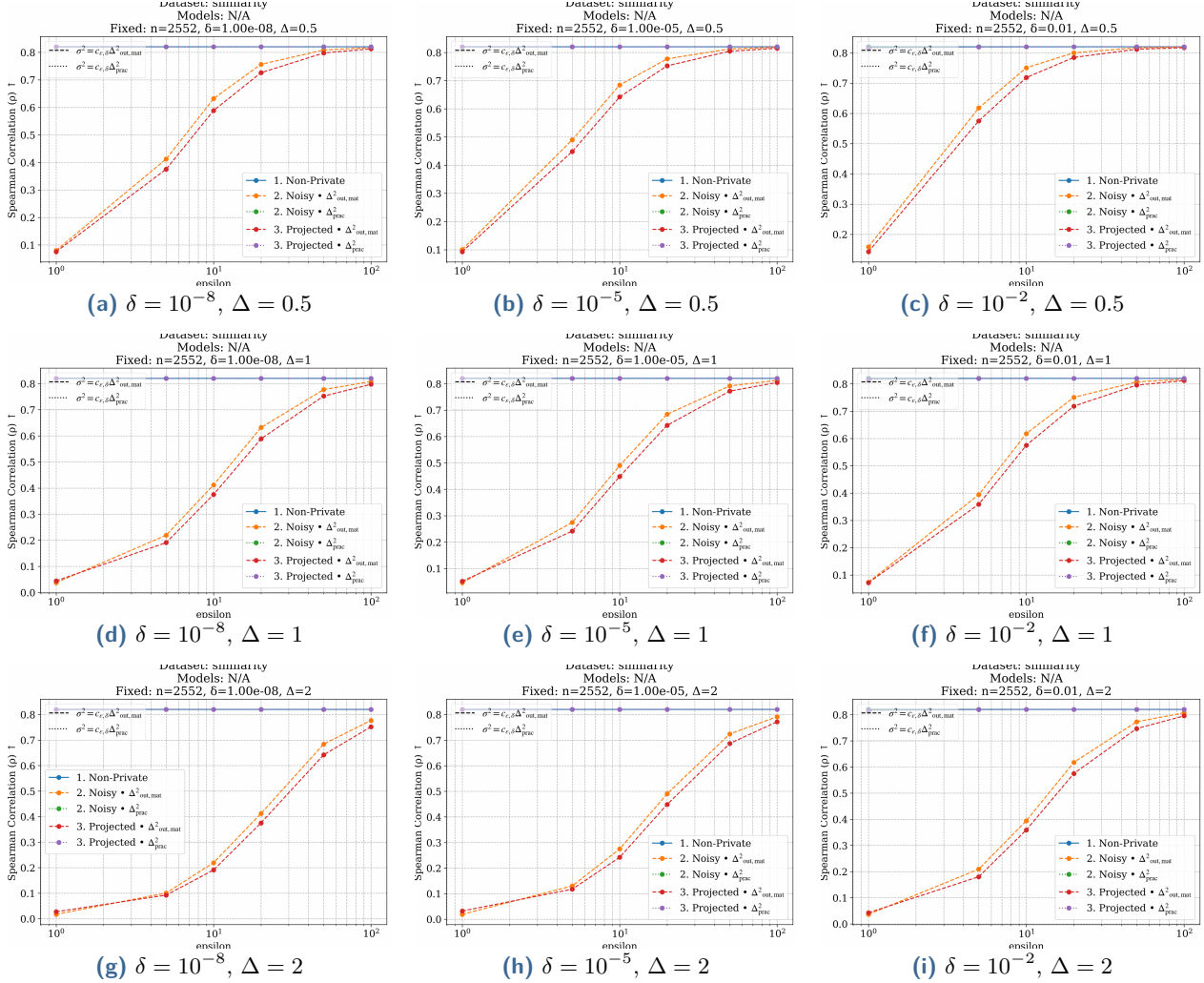}
        \caption{$\delta=\deltadisp$, $\Delta=\bigdval$}
      \end{subfigure}%
      \ifnum\col<3\hfill\fi
    }%
    \par\vspace{0.3em}
  }%

  \caption{\small Results for Spearman correlation ($\uparrow$) for STSBench. Each row corresponds to a fixed $\Delta$, with $\delta$ increasing from left to right. The x-axis shows $\varepsilon$ (log scale), and the color hues indicate variance across repeated runs. For large $n$, solving the SDP with linear system solvers (e.g., MOSEK) was computationally infeasible within the time or memory budget.}
  \label{fig:stsbench}
\end{figure}

\clearpage
\subsubsection{MovieLens}

\vspace{10pt}

\textbf{MovieLens–100K}~\citep{harper2015movielens}. 
We form a user–user cosine Gram $S$ over the $n{=}943$ users using user–mean–centered rating vectors, and evaluate a standard similarity-weighted neighborhood predictor on the official test split. Utility is reported as RMSE on held-out ratings. (The SDP baseline is omitted at this scale due to solver limitations.)

\vspace{10pt}

\begin{figure}[!htbp]
  \centering

  \captionsetup{aboveskip=0pt, belowskip=0pt}
  \captionsetup[sub]{aboveskip=2pt, belowskip=2pt}

  \foreach \bigdval in {0.5,1,2}{%
    \foreach \smalldelta/\deltadisp [count=\col from 1]%
      in {1.00e-08/10^{-8},1.00e-05/10^{-5},0.01/10^{-2}}{%
      \begin{subfigure}[t]{0.32\linewidth}
        \includegraphics[
          trim=20pt 25pt 0pt 80pt, 
          clip,
          width=\linewidth
        ]{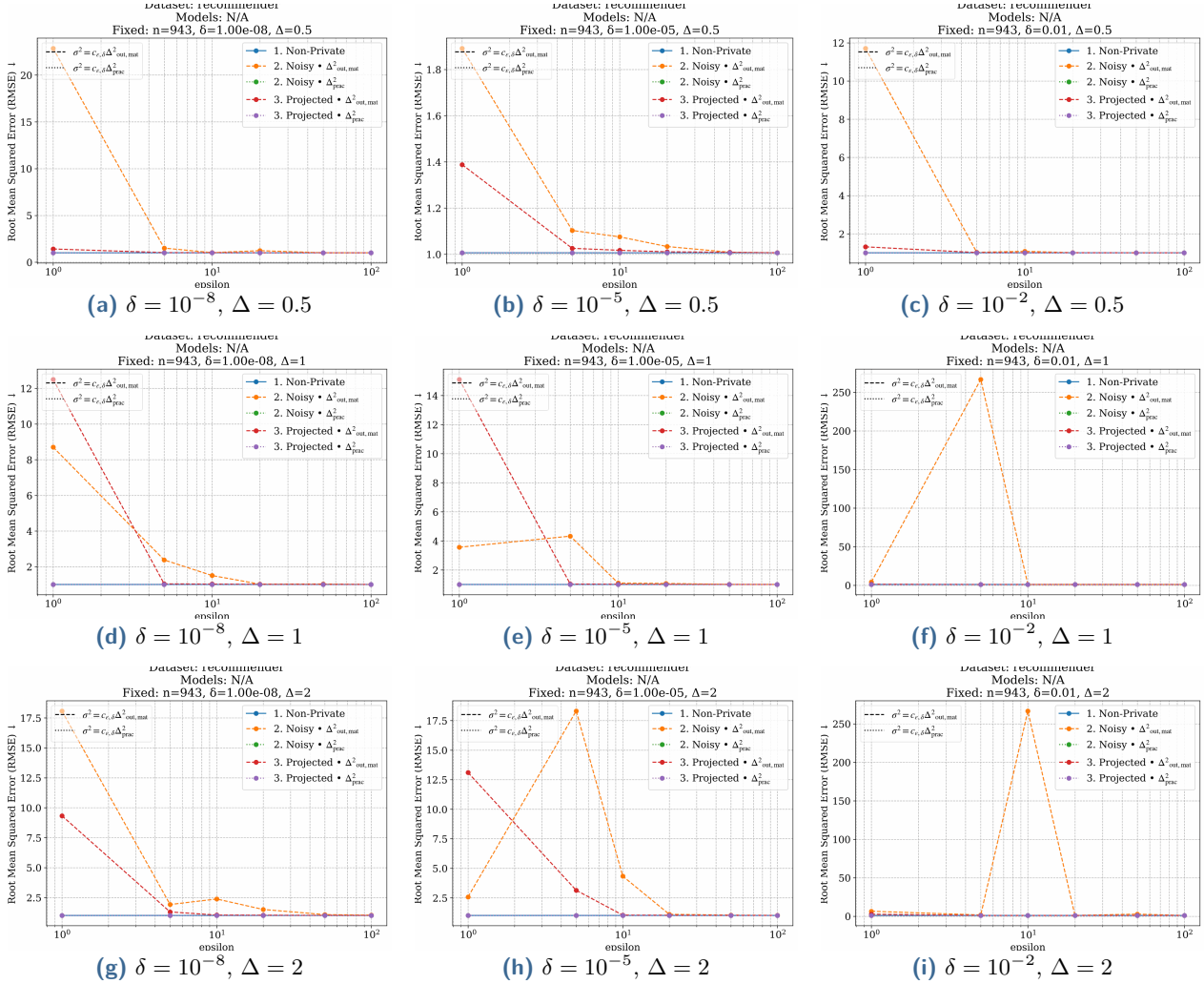}
        \caption{$\delta=\deltadisp$, $\Delta=\bigdval$}
      \end{subfigure}%
      \ifnum\col<3\hfill\fi
    }%
    \par\vspace{0.3em}
  }%

  \caption{\small Results for RMSE ($\downarrow$) for MovieLens Benchmark. Each row corresponds to a fixed $\Delta$, with $\delta$ increasing from left to right. The x-axis shows $\varepsilon$ (log scale), and the color hues indicate variance across repeated runs. For large $n$, solving the SDP with linear system solvers (e.g., MOSEK) was computationally infeasible within the time or memory budget.}
  \label{fig:movielens_rmse}
\end{figure}

\clearpage
\subsubsection{Oxford Pets}

\vspace{10pt}

\textbf{Oxford–IIIT Pets}~\citep{parkhi2012pets}: 37 breeds, $\sim$7.4K images. 
For each run, we sample $n{=}2048$ images, extract frozen DINOv2-B/14 embeddings, $\ell_2$-normalize, and form the cosine Gram $S$. 
We report three utilities: (i) ROC–AUC for same/different verification from pair scores $\{S_{ij}\}$ (diagonal excluded); (ii) $5$-NN top-1 accuracy using each Gram row as the neighborhood (self excluded); and (iii) instance retrieval Recall@1 from rankings induced by $S$.

\vspace{10pt}

\begin{figure}[!htbp]
  \centering

  \captionsetup{aboveskip=0pt, belowskip=0pt}
  \captionsetup[sub]{aboveskip=2pt, belowskip=2pt}

  \foreach \bigdval in {0.5,1,2}{%
    \foreach \smalldelta/\deltadisp [count=\col from 1]%
      in {1.00e-08/10^{-8},1.00e-05/10^{-5},0.01/10^{-2}}{%
      \begin{subfigure}[t]{0.32\linewidth}
        \includegraphics[
          trim=20pt 25pt 0pt 80pt, 
          clip,
          width=\linewidth
        ]{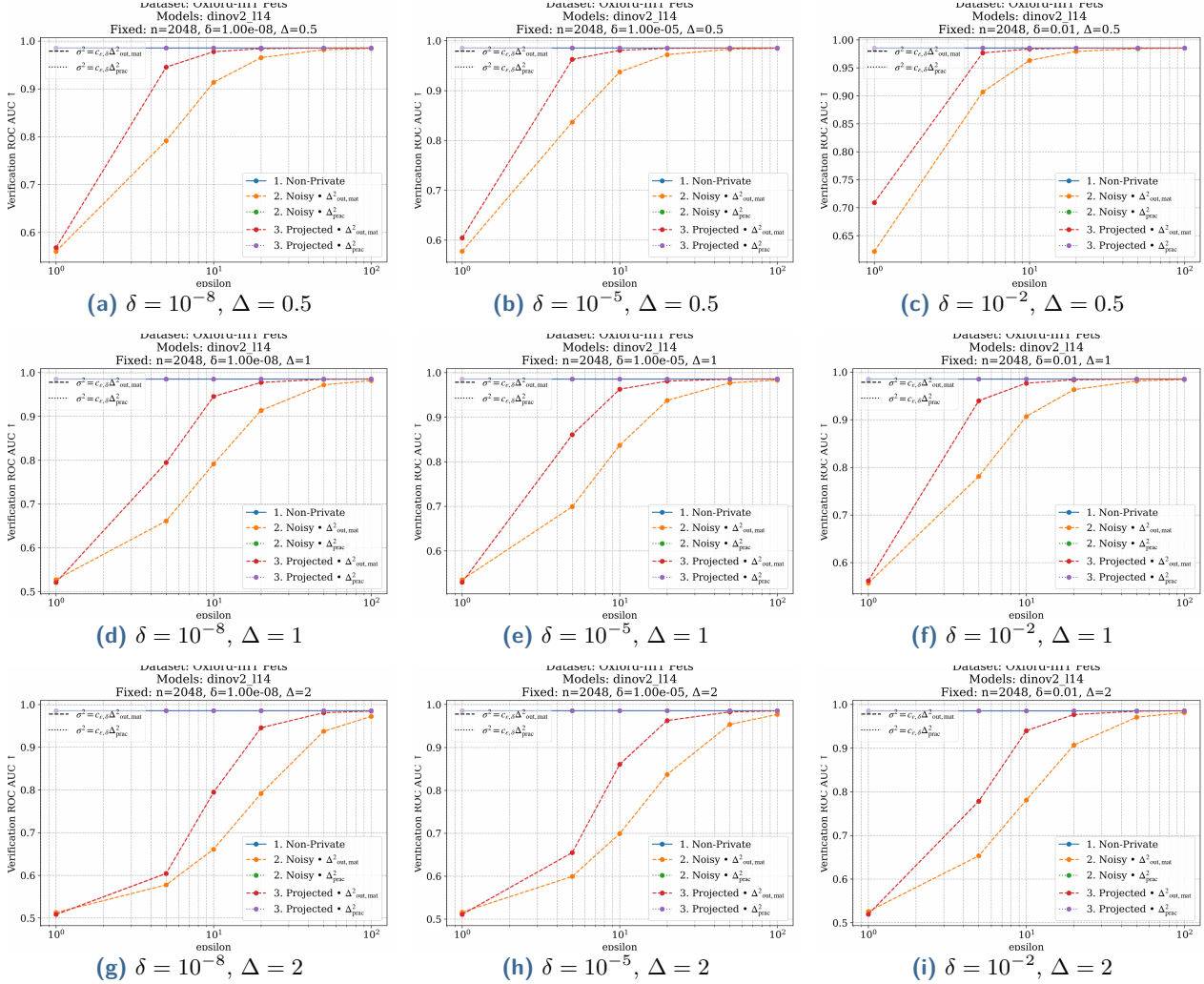}
        \caption{$\delta=\deltadisp$, $\Delta=\bigdval$}
      \end{subfigure}%
      \ifnum\col<3\hfill\fi
    }%
    \par\vspace{0.3em}
  }%
  \caption{\small 
Oxford-IIIT Pets same/different verification ROC--AUC ($\uparrow$) at $n=2048$.
Each row fixes the sensitivity parameter $\Delta$; columns correspond to increasing $\delta$.
The horizontal axis is $\varepsilon$ on a logarithmic scale. Shaded bands show variability across repeated runs.}
  \label{fig:oxfordpets_n2048}
\end{figure}

\begin{figure}[!htbp]
  \centering

  \captionsetup{aboveskip=0pt, belowskip=0pt}
  \captionsetup[sub]{aboveskip=2pt, belowskip=2pt}

  \foreach \bigdval in {0.5,1,2}{%
    \foreach \smalldelta/\deltadisp [count=\col from 1]%
      in {1.00e-08/10^{-8},1.00e-05/10^{-5},0.01/10^{-2}}{%
      \begin{subfigure}[t]{0.32\linewidth}
        \includegraphics[
          trim=20pt 25pt 0pt 80pt, 
          clip,
          width=\linewidth
        ]{appx_results/case2/oxford_pets/plot_utility_knn_acc_knn_oxford_pets_n=2048_delta=\smalldelta_bigd=\bigdval.png}
        \caption{$\delta=\deltadisp$, $\Delta=\bigdval$}
      \end{subfigure}%
      \ifnum\col<3\hfill\fi
    }%
    \par\vspace{0.3em}
  }%

  \caption{\small Results for KNN Accuracy ($\uparrow$) ($n=2048$). Each row represents a fixed $\Delta$ with $\delta$ increasing from left to right. The horizontal axis is $\varepsilon$ on a logarithmic scale. Shaded bands show variability across repeated runs.}
  \label{fig:oxfordpets_knn_n2048}
\end{figure}

\begin{figure}[!htbp]
  \centering

  \captionsetup{aboveskip=0pt, belowskip=0pt}
  \captionsetup[sub]{aboveskip=2pt, belowskip=2pt}

  \foreach \bigdval/\bigdfile in {0.5/0p5,1/1,2/2}{%
    \foreach \deltafile/\deltadisp [count=\col from 1]%
      in {1em08/10^{-8},1em05/10^{-5},0p01/10^{-2}}{%
      \begin{subfigure}[t]{0.32\linewidth}
        \includegraphics[
          trim=20pt 25pt 0pt 80pt, 
          clip,
          width=\linewidth
        ]{appx_results/case2/oxford_pets/plot_utility_recall1_oxford_pets_n2048_delta\deltafile_bigd\bigdfile.png}
        \caption{$\delta=\deltadisp$, $\Delta=\bigdval$}
      \end{subfigure}%
      \ifnum\col<3\hfill\fi
    }%
    \par\vspace{0.3em}
  }%

  \caption{\small Results for Retrieval Recall@1 ($\uparrow$) ($n=2048$). Each row represents a fixed $\Delta$ with $\delta$ increasing from left to right. The horizontal axis is $\varepsilon$ on a logarithmic scale. Shaded bands show variability across repeated runs.}
  \label{fig:oxfordpets_recall1_n2048}
\end{figure}

\clearpage
\renewcommand{\refname}{Appendix References}
\putbib
\end{bibunit}


\begin{thebibliography}{40}
\providecommand{\natexlab}[1]{#1}
\providecommand{\url}[1]{\texttt{#1}}
\expandafter\ifx\csname urlstyle\endcsname\relax
  \providecommand{\doi}[1]{doi: #1}\else
  \providecommand{\doi}{doi: \begingroup \urlstyle{rm}\Url}\fi

\bibitem[Abbas et~al.(2023)Abbas, Tirumala, Simig, Ganguli, and
  Morcos]{abbas2023semdedup}
Amro Kamal~Mohamed Abbas, Kushal Tirumala, Daniel Simig, Surya Ganguli, and
  Ari~S Morcos.
\newblock {SemDeDup}: Data-efficient learning at web-scale through semantic
  deduplication.
\newblock In \emph{ICLR 2023 Workshop on Mathematical and Empirical
  Understanding of Foundation Models}, 2023.

\bibitem[Anderson et~al.(2025)Anderson, Amit, and Goldsteen]{anderson2024my}
Maya Anderson, Guy Amit, and Abigail Goldsteen.
\newblock Is my data in your retrieval database? membership inference attacks
  against retrieval augmented generation.
\newblock In \emph{International Conference on Information Systems Security and
  Privacy}, volume~2, pp.\  474--485. Science and Technology Publications, Lda,
  2025.

\bibitem[Bai et~al.(2023)Bai, Zhang, Song, Shao, Wang, Cui, and
  Russello]{bai2023cryptomask}
Jianli Bai, Xiaowu Zhang, Xiangfu Song, Hang Shao, Qifan Wang, Shujie Cui, and
  Giovanni Russello.
\newblock Cryptomask: Privacy-preserving face recognition.
\newblock In \emph{International Conference on Information and Communications
  Security}, pp.\  333--350. Springer, 2023.

\bibitem[Blocki et~al.(2012)Blocki, Blum, Datta, and
  Sheffet]{blocki2012johnson}
Jeremiah Blocki, Avrim Blum, Anupam Datta, and Or~Sheffet.
\newblock The johnson-lindenstrauss transform itself preserves differential
  privacy.
\newblock In \emph{2012 IEEE 53rd Annual Symposium on Foundations of Computer
  Science}, pp.\  410--419. IEEE, 2012.

\bibitem[Chamikara et~al.(2020)Chamikara, Bertok, Khalil, Liu, and
  Camtepe]{chamikara2020privacy}
Mahawaga Arachchige~Pathum Chamikara, Peter Bertok, Ibrahim Khalil, Dongxi Liu,
  and Seyit Camtepe.
\newblock Privacy preserving face recognition utilizing differential privacy.
\newblock \emph{Computers \& Security}, 97, 2020.

\bibitem[Choi et~al.(2025)Choi, Park, Byun, Lee, and
  Park]{choi2025safeguarding}
Yujin Choi, Youngjoo Park, Junyoung Byun, Jaewook Lee, and Jinseong Park.
\newblock Safeguarding privacy of retrieval data against membership inference
  attacks: Is this query too close to home?
\newblock \emph{arXiv preprint arXiv:2505.22061}, 2025.

\bibitem[Chum et~al.(2007)Chum, Philbin, Sivic, Isard, and
  Zisserman]{chum2007total}
Ondrej Chum, James Philbin, Josef Sivic, Michael Isard, and Andrew Zisserman.
\newblock Total recall: Automatic query expansion with a generative feature
  model for object retrieval.
\newblock In \emph{2007 IEEE 11th international conference on computer vision},
  pp.\  1--8. IEEE, 2007.

\bibitem[Cohen-Addad et~al.(2024)Cohen-Addad, d'Orsi, Epasto, Mirrokni, and
  Zhong]{cohen2024perturb}
Vincent Cohen-Addad, Tommaso d'Orsi, Alessandro Epasto, Vahab Mirrokni, and
  Peilin Zhong.
\newblock Perturb-and-project: differentially private similarities and
  marginals.
\newblock In \emph{Proceedings of the 41st International Conference on Machine
  Learning}, pp.\  9161--9179, 2024.

\bibitem[Deng et~al.(2019)Deng, Guo, Xue, and Zafeiriou]{deng2019arcface}
Jiankang Deng, Jia Guo, Niannan Xue, and Stefanos Zafeiriou.
\newblock Arcface: Additive angular margin loss for deep face recognition.
\newblock In \emph{Proceedings of the IEEE/CVF conference on computer vision
  and pattern recognition}, pp.\  4690--4699, 2019.

\bibitem[Dwork et~al.(2014)Dwork, Roth, et~al.]{dwork2014algorithmic}
Cynthia Dwork, Aaron Roth, et~al.
\newblock The algorithmic foundations of differential privacy.
\newblock \emph{Foundations and trends{\textregistered} in theoretical computer
  science}, 9\penalty0 (3--4):\penalty0 211--407, 2014.

\bibitem[Galbally et~al.(2010)Galbally, McCool, Fierrez, Marcel, and
  Ortega-Garcia]{galbally2010vulnerability}
Javier Galbally, Chris McCool, Julian Fierrez, Sebastien Marcel, and Javier
  Ortega-Garcia.
\newblock On the vulnerability of face verification systems to hill-climbing
  attacks.
\newblock \emph{Pattern Recognition}, 43\penalty0 (3):\penalty0 1027--1038,
  2010.

\bibitem[Grislain(2025)]{grislain2025rag}
Nicolas Grislain.
\newblock Rag with differential privacy.
\newblock In \emph{2025 IEEE Conference on Artificial Intelligence (CAI)}, pp.\
   847--852. IEEE, 2025.

\bibitem[Ji et~al.(2022)Ji, Wang, Huang, Wu, Xu, Ding, Zhang, Cao, and
  Ji]{ji2022privacy}
Jiazhen Ji, Huan Wang, Yuge Huang, Jiaxiang Wu, Xingkun Xu, Shouhong Ding,
  ShengChuan Zhang, Liujuan Cao, and Rongrong Ji.
\newblock Privacy-preserving face recognition with learnable privacy budgets in
  frequency domain.
\newblock In \emph{European Conference on Computer Vision}, pp.\  475--491.
  Springer, 2022.

\bibitem[Ji \& Li(2024)Ji and Li]{ji2024less}
Tianxi Ji and Pan Li.
\newblock Less is more: Revisiting the gaussian mechanism for differential
  privacy.
\newblock In \emph{33rd USENIX Security Symposium (USENIX Security 24)}, pp.\
  937--954, 2024.

\bibitem[Johnson et~al.(2019)Johnson, Douze, and J{\'e}gou]{johnson2019billion}
Jeff Johnson, Matthijs Douze, and Herv{\'e} J{\'e}gou.
\newblock Billion-scale similarity search with gpus.
\newblock \emph{IEEE Transactions on Big Data}, 7\penalty0 (3):\penalty0
  535--547, 2019.

\bibitem[Kenthapadi et~al.(2013)Kenthapadi, Korolova, Mironov, and
  Mishra]{kenthapadi2012privacy}
Krishnaram Kenthapadi, Aleksandra Korolova, Ilya Mironov, and Nina Mishra.
\newblock Privacy via the johnson-lindenstrauss transform.
\newblock \emph{Journal of Privacy and Confidentiality}, 5\penalty0
  (1):\penalty0 39--71, 2013.

\bibitem[Koga et~al.(2024)Koga, Wu, Zhang, and Chaudhuri]{koga2024privacy}
Tatsuki Koga, Ruihan Wu, Zhiyuan Zhang, and Kamalika Chaudhuri.
\newblock Privacy-preserving retrieval-augmented generation with differential
  privacy.
\newblock \emph{arXiv preprint arXiv:2412.04697}, 2024.

\bibitem[Lewis et~al.(2020)Lewis, Perez, Piktus, Petroni, Karpukhin, Goyal,
  K{\"u}ttler, Lewis, Yih, Rockt{\"a}schel, et~al.]{lewis2020retrieval}
Patrick Lewis, Ethan Perez, Aleksandra Piktus, Fabio Petroni, Vladimir
  Karpukhin, Naman Goyal, Heinrich K{\"u}ttler, Mike Lewis, Wen-tau Yih, Tim
  Rockt{\"a}schel, et~al.
\newblock Retrieval-augmented generation for knowledge-intensive nlp tasks.
\newblock \emph{Advances in neural information processing systems},
  33:\penalty0 9459--9474, 2020.

\bibitem[Li et~al.(2025{\natexlab{a}})Li, He, Wang, Feng, Li, and
  Zhang]{li2025budgetleak}
Hao Li, Jiajun He, Guangshuo Wang, Dengguo Feng, Zheng Li, and Min Zhang.
\newblock Budgetleak: Membership inference attacks on rag systems via the
  generation budget side channel.
\newblock \emph{arXiv preprint arXiv:2511.12043}, 2025{\natexlab{a}}.

\bibitem[Li et~al.(2025{\natexlab{b}})Li, Liu, Wang, and
  Yang]{li2025generating}
Yuying Li, Gaoyang Liu, Chen Wang, and Yang Yang.
\newblock Generating is believing: Membership inference attacks against
  retrieval-augmented generation.
\newblock In \emph{ICASSP 2025-2025 IEEE International Conference on Acoustics,
  Speech and Signal Processing (ICASSP)}, pp.\  1--5. IEEE, 2025{\natexlab{b}}.

\bibitem[MA(1955)]{ma1955two}
KRASNOSEL'SKII MA.
\newblock Two comments on the method of successive approximations.
\newblock \emph{Usp. Math. Nauk}, 10:\penalty0 123--127, 1955.

\bibitem[Maiorana et~al.(2014)Maiorana, Hine, and Campisi]{maiorana2014hill}
Emanuele Maiorana, Gabriel~Emile Hine, and Patrizio Campisi.
\newblock Hill-climbing attacks on multibiometrics recognition systems.
\newblock \emph{IEEE Transactions on Information Forensics and Security},
  10\penalty0 (5):\penalty0 900--915, 2014.

\bibitem[Mann(1953)]{mann1953mean}
W~Robert Mann.
\newblock Mean value methods in iteration.
\newblock \emph{Proceedings of the American Mathematical Society}, 4\penalty0
  (3):\penalty0 506--510, 1953.

\bibitem[Maze et~al.(2018)Maze, Adams, Duncan, Kalka, Miller, Otto, Jain,
  Niggel, Anderson, Cheney, et~al.]{ijbc}
Brianna Maze, Jocelyn Adams, James~A Duncan, Nathan Kalka, Tim Miller, Charles
  Otto, Anil~K Jain, W~Tyler Niggel, Janet Anderson, Jordan Cheney, et~al.
\newblock Iarpa janus benchmark-c: Face dataset and protocol.
\newblock In \emph{2018 international conference on biometrics (ICB)}, pp.\
  158--165. IEEE, 2018.

\bibitem[Mehrabi et~al.(2021)Mehrabi, Morstatter, Saxena, Lerman, and
  Galstyan]{mehrabi2021survey}
Ninareh Mehrabi, Fred Morstatter, Nripsuta Saxena, Kristina Lerman, and Aram
  Galstyan.
\newblock A survey on bias and fairness in machine learning.
\newblock \emph{ACM computing surveys (CSUR)}, 54\penalty0 (6):\penalty0 1--35,
  2021.

\bibitem[Meng et~al.(2021)Meng, Zhao, Huang, and Zhou]{meng2021magface}
Qiang Meng, Shichao Zhao, Zhida Huang, and Feng Zhou.
\newblock Magface: A universal representation for face recognition and quality
  assessment.
\newblock In \emph{Proceedings of the IEEE/CVF conference on computer vision
  and pattern recognition}, pp.\  14225--14234, 2021.

\bibitem[Mori et~al.(2025)Mori, Kakizaki, Miyagawa, and
  Sakuma]{mori2025differentially}
Junki Mori, Kazuya Kakizaki, Taiki Miyagawa, and Jun Sakuma.
\newblock Differentially private synthetic text generation for
  retrieval-augmented generation (rag).
\newblock \emph{arXiv preprint arXiv:2510.06719}, 2025.

\bibitem[Muennighoff et~al.(2023)Muennighoff, Tazi, Magne, and
  Reimers]{muennighoff2022mteb}
Niklas Muennighoff, Nouamane Tazi, Lo{\"\i}c Magne, and Nils Reimers.
\newblock {MTEB}: Massive text embedding benchmark.
\newblock In \emph{Proceedings of the 17th Conference of the European Chapter
  of the Association for Computational Linguistics}, pp.\  2014--2037, 2023.

\bibitem[Naseh et~al.(2025)Naseh, Peng, Suri, Chaudhari, Oprea, and
  Houmansadr]{naseh2025riddle}
Ali Naseh, Yuefeng Peng, Anshuman Suri, Harsh Chaudhari, Alina Oprea, and Amir
  Houmansadr.
\newblock Riddle me this! stealthy membership inference for retrieval-augmented
  generation.
\newblock In \emph{Proceedings of the 2025 ACM SIGSAC Conference on Computer
  and Communications Security}, pp.\  1245--1259, 2025.

\bibitem[Ng et~al.(2001)Ng, Jordan, and Weiss]{ng2001spectral}
Andrew Ng, Michael Jordan, and Yair Weiss.
\newblock On spectral clustering: Analysis and an algorithm.
\newblock \emph{Advances in neural information processing systems}, 14, 2001.

\bibitem[Phan \& Nguyen(2022)Phan and Nguyen]{phan2022deepface}
Hai Phan and Anh Nguyen.
\newblock Deepface-emd: Re-ranking using patch-wise earth mover's distance
  improves out-of-distribution face identification.
\newblock In \emph{Proceedings of the IEEE/CVF Conference on Computer Vision
  and Pattern Recognition}, pp.\  20259--20269, 2022.

\bibitem[Reimers \& Gurevych(2019)Reimers and
  Gurevych]{reimers2019sentencebert}
Nils Reimers and Iryna Gurevych.
\newblock Sentence-bert: Sentence embeddings using siamese {BERT}-networks.
\newblock In \emph{Proceedings of EMNLP-IJCNLP}, 2019.

\bibitem[Shi \& Malik(2000)Shi and Malik]{shi2000normalized}
Jianbo Shi and Jitendra Malik.
\newblock Normalized cuts and image segmentation.
\newblock \emph{IEEE Transactions on pattern analysis and machine
  intelligence}, 22\penalty0 (8):\penalty0 888--905, 2000.

\bibitem[Slyman et~al.(2024)Slyman, Lee, Cohen, and Kafle]{slyman2024fairdedup}
Eric Slyman, Stefan Lee, Scott Cohen, and Kushal Kafle.
\newblock {FairDeDup}: Detecting and mitigating vision-language fairness
  disparities in semantic dataset deduplication.
\newblock In \emph{Proceedings of the IEEE/CVF Conference on Computer Vision
  and Pattern Recognition}, pp.\  13905--13916, 2024.

\bibitem[Whitelam et~al.(2017)Whitelam, Taborsky, Blanton, Maze, Adams, Miller,
  Kalka, Jain, Duncan, Allen, et~al.]{whitelam2017iarpaijbb}
Cameron Whitelam, Emma Taborsky, Austin Blanton, Brianna Maze, Jocelyn Adams,
  Tim Miller, Nathan Kalka, Anil~K Jain, James~A Duncan, Kristen Allen, et~al.
\newblock Iarpa janus benchmark-b face dataset.
\newblock In \emph{proceedings of the IEEE conference on computer vision and
  pattern recognition workshops}, pp.\  90--98, 2017.

\bibitem[Wu et~al.(2025{\natexlab{a}})Wu, Wang, and Wang]{wu2025beyond}
Ruihan Wu, Erchi Wang, and Yu-Xiang Wang.
\newblock Beyond per-question privacy: Multi-query differential privacy for rag
  systems.
\newblock In \emph{NeurIPS 2025 Workshop: Reliable ML from Unreliable Data},
  2025{\natexlab{a}}.

\bibitem[Wu et~al.(2025{\natexlab{b}})Wu, Wang, Zhang, and Wang]{wu2025private}
Ruihan Wu, Erchi Wang, Zhiyuan Zhang, and Yu-Xiang Wang.
\newblock Private-rag: Answering multiple queries with llms while keeping your
  data private.
\newblock \emph{arXiv preprint arXiv:2511.07637}, 2025{\natexlab{b}}.

\bibitem[Wu et~al.(2023)Wu, Ge, Luo, Liu, and Xu]{wu2023face}
Yang Wu, Zhiwei Ge, Yuhao Luo, Lin Liu, and Sulong Xu.
\newblock Face clustering via graph convolutional networks with confidence
  edges.
\newblock In \emph{Proceedings of the IEEE/CVF International Conference on
  Computer Vision}, pp.\  20990--20999, 2023.

\bibitem[Yang et~al.(2017)Yang, Zhu, Ma, Xiang, and Zhou]{yang2017privacy}
Mengmeng Yang, Tianqing Zhu, Lichuan Ma, Yang Xiang, and Wanlei Zhou.
\newblock Privacy preserving collaborative filtering via the
  johnson-lindenstrauss transform.
\newblock In \emph{2017 IEEE Trustcom/BigDataSE/ICESS}, pp.\  417--424. IEEE,
  2017.

\bibitem[Zhong et~al.(2017)Zhong, Zheng, Cao, and Li]{zhong2017re}
Zhun Zhong, Liang Zheng, Donglin Cao, and Shaozi Li.
\newblock Re-ranking person re-identification with k-reciprocal encoding.
\newblock In \emph{Proceedings of the IEEE conference on computer vision and
  pattern recognition}, pp.\  1318--1327, 2017.

\end{thebibliography}


\begin{thebibliography}{46}
\providecommand{\natexlab}[1]{#1}
\providecommand{\url}[1]{\texttt{#1}}
\expandafter\ifx\csname urlstyle\endcsname\relax
  \providecommand{\doi}[1]{doi: #1}\else
  \providecommand{\doi}{doi: \begingroup \urlstyle{rm}\Url}\fi

\bibitem[Adler \& Taylor(2007)Adler and Taylor]{adler2007random}
Robert~J Adler and Jonathan~E Taylor.
\newblock \emph{Random fields and geometry}.
\newblock Springer, 2007.

\bibitem[Arakcheev \& Bauschke(2025)Arakcheev and Bauschke]{arakcheev2025opial}
Aleksandr Arakcheev and Heinz~H Bauschke.
\newblock On opial's lemma.
\newblock \emph{arXiv preprint arXiv:2503.22004}, 2025.

\bibitem[Balle \& Wang(2018)Balle and Wang]{balle2018improving}
Borja Balle and Yu-Xiang Wang.
\newblock Improving the gaussian mechanism for differential privacy: Analytical
  calibration and optimal denoising.
\newblock In \emph{International conference on machine learning}, pp.\
  394--403. PMLR, 2018.

\bibitem[Bamber(1975)]{bamber1975area}
Donald Bamber.
\newblock The area above the ordinal dominance graph and the area below the
  receiver operating characteristic graph.
\newblock \emph{Journal of mathematical psychology}, 12\penalty0 (4):\penalty0
  387--415, 1975.

\bibitem[Bauschke \& Borwein(1994)Bauschke and Borwein]{bauschke1994dykstra}
Heinz~H Bauschke and Jonathan~M Borwein.
\newblock Dykstra' s alternating projection algorithm for two sets.
\newblock \emph{Journal of Approximation Theory}, 79\penalty0 (3):\penalty0
  418--443, 1994.

\bibitem[Bauschke \& Borwein(1996)Bauschke and Borwein]{bauschke1996projection}
Heinz~H Bauschke and Jonathan~M Borwein.
\newblock On projection algorithms for solving convex feasibility problems.
\newblock \emph{SIAM review}, 38\penalty0 (3):\penalty0 367--426, 1996.

\bibitem[Bauschke \& Combettes(2020)Bauschke and
  Combettes]{bauschke2020correction}
Heinz~H Bauschke and Patrick~L Combettes.
\newblock Correction to: convex analysis and monotone operator theory in
  hilbert spaces.
\newblock In \emph{Convex analysis and monotone operator theory in Hilbert
  spaces}, pp.\  C1--C4. Springer, 2020.

\bibitem[Bauschke et~al.(1999)Bauschke, Borwein, and Li]{bauschke1999strong}
Heinz~H Bauschke, Jonathan~M Borwein, and Wu~Li.
\newblock Strong conical hull intersection property, bounded linear regularity,
  jameson’s property (g), and error bounds in convex optimization.
\newblock \emph{Mathematical Programming}, 86\penalty0 (1):\penalty0 135--160,
  1999.

\bibitem[Blocki et~al.(2012)Blocki, Blum, Datta, and
  Sheffet]{blocki2012johnson}
Jeremiah Blocki, Avrim Blum, Anupam Datta, and Or~Sheffet.
\newblock The johnson-lindenstrauss transform itself preserves differential
  privacy.
\newblock In \emph{2012 IEEE 53rd Annual Symposium on Foundations of Computer
  Science}, pp.\  410--419. IEEE, 2012.

\bibitem[Boyle \& Dykstra(1986)Boyle and Dykstra]{boyle1986method}
James~P Boyle and Richard~L Dykstra.
\newblock A method for finding projections onto the intersection of convex sets
  in hilbert spaces.
\newblock In \emph{Advances in Order Restricted Statistical Inference:
  Proceedings of the Symposium on Order Restricted Statistical Inference held
  in Iowa City, Iowa, September 11--13, 1985}, pp.\  28--47. Springer, 1986.

\bibitem[Cer et~al.(2017)Cer, Diab, Agirre, Lopez-Gazpio, and
  Specia]{cer2017sts}
Daniel Cer, Mona Diab, Eneko Agirre, I{\~n}igo Lopez-Gazpio, and Lucia Specia.
\newblock {S}em{E}val-2017 task 1: Semantic textual similarity multilingual and
  cross-lingual focused evaluation.
\newblock In \emph{Proceedings of the 11th International Workshop on Semantic
  Evaluation (SemEval-2017)}, pp.\  1--14, 2017.

\bibitem[Chamikara et~al.(2020)Chamikara, Bertok, Khalil, Liu, and
  Camtepe]{chamikara2020privacy}
Mahawaga Arachchige~Pathum Chamikara, Peter Bertok, Ibrahim Khalil, Dongxi Liu,
  and Seyit Camtepe.
\newblock Privacy preserving face recognition utilizing differential privacy.
\newblock \emph{Computers \& Security}, 97, 2020.

\bibitem[Chu \& Raginsky(2025)Chu and Raginsky]{chu2025talagrand}
Yifeng Chu and Maxim Raginsky.
\newblock Talagrand meets talagrand: Upper and lower bounds on expected soft
  maxima of gaussian processes with finite index sets.
\newblock \emph{arXiv preprint arXiv:2502.06709}, 2025.

\bibitem[Cohen-Addad et~al.(2024)Cohen-Addad, d'Orsi, Epasto, Mirrokni, and
  Zhong]{cohen2024perturb}
Vincent Cohen-Addad, Tommaso d'Orsi, Alessandro Epasto, Vahab Mirrokni, and
  Peilin Zhong.
\newblock Perturb-and-project: differentially private similarities and
  marginals.
\newblock In \emph{Proceedings of the 41st International Conference on Machine
  Learning}, pp.\  9161--9179, 2024.

\bibitem[Dodd \& Pepe(2003)Dodd and Pepe]{dodd2003partial}
Lori~E Dodd and Margaret~S Pepe.
\newblock Partial auc estimation and regression.
\newblock \emph{Biometrics}, 59\penalty0 (3):\penalty0 614--623, 2003.

\bibitem[Dong et~al.(2022)Dong, Liang, and Yi]{dong2022differentially}
Wei Dong, Yuting Liang, and Ke~Yi.
\newblock Differentially private covariance revisited.
\newblock \emph{Advances in Neural Information Processing Systems},
  35:\penalty0 850--861, 2022.

\bibitem[Dykstra(1983)]{dykstra1983algorithm}
Richard~L Dykstra.
\newblock An algorithm for restricted least squares regression.
\newblock \emph{Journal of the American Statistical Association}, 78\penalty0
  (384):\penalty0 837--842, 1983.

\bibitem[Dykstra(1985)]{dykstra1985iterative}
Richard~L Dykstra.
\newblock An iterative procedure for obtaining i-projections onto the
  intersection of convex sets.
\newblock \emph{The annals of Probability}, pp.\  975--984, 1985.

\bibitem[{Gemma Team} et~al.(2025){Gemma Team}, Kamath, Ferret, Pathak,
  Vieillard, Merhej, Perrin, Matejovicova, Ramé, Rivière, Rouillard, Mesnard,
  Cideron, bastien Grill, Ramos, Yvinec, Casbon, Pot, Penchev, Liu, Visin,
  Kenealy, Beyer, Zhai, Tsitsulin, Busa-Fekete, Feng, Sachdeva, Coleman, Gao,
  Mustafa, Barr, Parisotto, Tian, Eyal, Cherry, Peter, Sinopalnikov,
  Bhupatiraju, Agarwal, Kazemi, Malkin, Kumar, Vilar, Brusilovsky, Luo,
  Steiner, Friesen, Sharma, Sharma, Gilady, Goedeckemeyer, Saade, Feng,
  Kolesnikov, Bendebury, Abdagic, Vadi, György, Pinto, Das, Bapna, Miech,
  Yang, Paterson, Shenoy, Chakrabarti, Piot, Wu, Shahriari, Petrini, Chen, Lan,
  Choquette-Choo, Carey, Brick, Deutsch, Eisenbud, Cattle, Cheng, Paparas,
  Sreepathihalli, Reid, Tran, Zelle, Noland, Huizenga, Kharitonov, Liu,
  Amirkhanyan, Cameron, Hashemi, Klimczak-Plucińska, Singh, Mehta, Lehri,
  Hazimeh, Ballantyne, Szpektor, Nardini, Pouget-Abadie, Chan, Stanton,
  Wieting, Lai, Orbay, Fernandez, Newlan, yeong Ji, Singh, Black, Yu, Hui,
  Vodrahalli, Greff, Qiu, Valentine, Coelho, Ritter, Hoffman, Watson,
  Chaturvedi, Moynihan, Ma, Babar, Noy, Byrd, Roy, Momchev, Chauhan, Sachdeva,
  Bunyan, Botarda, Caron, Rubenstein, Culliton, Schmid, Sessa, Xu, Stanczyk,
  Tafti, Shivanna, Wu, Pan, Rokni, Willoughby, Vallu, Mullins, Jerome, Smoot,
  Girgin, Iqbal, Reddy, Sheth, Põder, Bhatnagar, Panyam, Eiger, Zhang, Liu,
  Yacovone, Liechty, Kalra, Evci, Misra, Roseberry, Feinberg, Kolesnikov, Han,
  Kwon, Chen, Chow, Zhu, Wei, Egyed, Cotruta, Giang, Kirk, Rao, Black, Babar,
  Lo, Moreira, Martins, Sanseviero, Gonzalez, Gleicher, Warkentin, Mirrokni,
  Senter, Collins, Barral, Ghahramani, Hadsell, Matias, Sculley, Petrov,
  Fiedel, Shazeer, Vinyals, Dean, Hassabis, Kavukcuoglu, Farabet, Buchatskaya,
  Alayrac, Anil, Dmitry, Lepikhin, Borgeaud, Bachem, Joulin, Andreev, Hardin,
  Dadashi, and Hussenot]{gemma2025gemma3}
{Gemma Team}, Aishwarya Kamath, Johan Ferret, Shreya Pathak, Nino Vieillard,
  Ramona Merhej, Sarah Perrin, Tatiana Matejovicova, Alexandre Ramé, Morgane
  Rivière, Louis Rouillard, Thomas Mesnard, Geoffrey Cideron, Jean bastien
  Grill, Sabela Ramos, Edouard Yvinec, Michelle Casbon, Etienne Pot, Ivo
  Penchev, Gaël Liu, Francesco Visin, Kathleen Kenealy, Lucas Beyer, Xiaohai
  Zhai, Anton Tsitsulin, Robert Busa-Fekete, Alex Feng, Noveen Sachdeva,
  Benjamin Coleman, Yi~Gao, Basil Mustafa, Iain Barr, Emilio Parisotto, David
  Tian, Matan Eyal, Colin Cherry, Jan-Thorsten Peter, Danila Sinopalnikov,
  Surya Bhupatiraju, Rishabh Agarwal, Mehran Kazemi, Dan Malkin, Ravin Kumar,
  David Vilar, Idan Brusilovsky, Jiaming Luo, Andreas Steiner, Abe Friesen,
  Abhanshu Sharma, Abheesht Sharma, Adi~Mayrav Gilady, Adrian Goedeckemeyer,
  Alaa Saade, Alex Feng, Alexander Kolesnikov, Alexei Bendebury, Alvin Abdagic,
  Amit Vadi, András György, André~Susano Pinto, Anil Das, Ankur Bapna,
  Antoine Miech, Antoine Yang, Antonia Paterson, Ashish Shenoy, Ayan
  Chakrabarti, Bilal Piot, Bo~Wu, Bobak Shahriari, Bryce Petrini, Charlie Chen,
  Charline~Le Lan, Christopher~A. Choquette-Choo, CJ~Carey, Cormac Brick,
  Daniel Deutsch, Danielle Eisenbud, Dee Cattle, Derek Cheng, Dimitris Paparas,
  Divyashree~Shivakumar Sreepathihalli, Doug Reid, Dustin Tran, Dustin Zelle,
  Eric Noland, Erwin Huizenga, Eugene Kharitonov, Frederick Liu, Gagik
  Amirkhanyan, Glenn Cameron, Hadi Hashemi, Hanna Klimczak-Plucińska, Harman
  Singh, Harsh Mehta, Harshal~Tushar Lehri, Hussein Hazimeh, Ian Ballantyne,
  Idan Szpektor, Ivan Nardini, Jean Pouget-Abadie, Jetha Chan, Joe Stanton,
  John Wieting, Jonathan Lai, Jordi Orbay, Joseph Fernandez, Josh Newlan,
  Ju~yeong Ji, Jyotinder Singh, Kat Black, Kathy Yu, Kevin Hui, Kiran
  Vodrahalli, Klaus Greff, Linhai Qiu, Marcella Valentine, Marina Coelho,
  Marvin Ritter, Matt Hoffman, Matthew Watson, Mayank Chaturvedi, Michael
  Moynihan, Min Ma, Nabila Babar, Natasha Noy, Nathan Byrd, Nick Roy, Nikola
  Momchev, Nilay Chauhan, Noveen Sachdeva, Oskar Bunyan, Pankil Botarda, Paul
  Caron, Paul~Kishan Rubenstein, Phil Culliton, Philipp Schmid, Pier~Giuseppe
  Sessa, Pingmei Xu, Piotr Stanczyk, Pouya Tafti, Rakesh Shivanna, Renjie Wu,
  Renke Pan, Reza Rokni, Rob Willoughby, Rohith Vallu, Ryan Mullins, Sammy
  Jerome, Sara Smoot, Sertan Girgin, Shariq Iqbal, Shashir Reddy, Shruti Sheth,
  Siim Põder, Sijal Bhatnagar, Sindhu~Raghuram Panyam, Sivan Eiger, Susan
  Zhang, Tianqi Liu, Trevor Yacovone, Tyler Liechty, Uday Kalra, Utku Evci,
  Vedant Misra, Vincent Roseberry, Vlad Feinberg, Vlad Kolesnikov, Woohyun Han,
  Woosuk Kwon, Xi~Chen, Yinlam Chow, Yuvein Zhu, Zichuan Wei, Zoltan Egyed,
  Victor Cotruta, Minh Giang, Phoebe Kirk, Anand Rao, Kat Black, Nabila Babar,
  Jessica Lo, Erica Moreira, Luiz~Gustavo Martins, Omar Sanseviero, Lucas
  Gonzalez, Zach Gleicher, Tris Warkentin, Vahab Mirrokni, Evan Senter, Eli
  Collins, Joelle Barral, Zoubin Ghahramani, Raia Hadsell, Yossi Matias,
  D.~Sculley, Slav Petrov, Noah Fiedel, Noam Shazeer, Oriol Vinyals, Jeff Dean,
  Demis Hassabis, Koray Kavukcuoglu, Clement Farabet, Elena Buchatskaya,
  Jean-Baptiste Alayrac, Rohan Anil, Dmitry, Lepikhin, Sebastian Borgeaud,
  Olivier Bachem, Armand Joulin, Alek Andreev, Cassidy Hardin, Robert Dadashi,
  and Léonard Hussenot.
\newblock Gemma 3 technical report, 2025.
\newblock URL \url{https://arxiv.org/abs/2503.19786}.

\bibitem[Grislain(2025)]{grislain2025rag}
Nicolas Grislain.
\newblock Rag with differential privacy.
\newblock In \emph{2025 IEEE Conference on Artificial Intelligence (CAI)}, pp.\
   847--852. IEEE, 2025.

\bibitem[Harper \& Konstan(2015)Harper and Konstan]{harper2015movielens}
F~Maxwell Harper and Joseph~A Konstan.
\newblock The movielens datasets: History and context.
\newblock \emph{{ACM} transactions on interactive intelligent systems (TIIS)},
  5\penalty0 (4):\penalty0 1--19, 2015.

\bibitem[Ji et~al.(2022)Ji, Wang, Huang, Wu, Xu, Ding, Zhang, Cao, and
  Ji]{ji2022privacy}
Jiazhen Ji, Huan Wang, Yuge Huang, Jiaxiang Wu, Xingkun Xu, Shouhong Ding,
  ShengChuan Zhang, Liujuan Cao, and Rongrong Ji.
\newblock Privacy-preserving face recognition with learnable privacy budgets in
  frequency domain.
\newblock In \emph{European Conference on Computer Vision}, pp.\  475--491.
  Springer, 2022.

\bibitem[Ji \& Li(2024)Ji and Li]{ji2024less}
Tianxi Ji and Pan Li.
\newblock Less is more: Revisiting the gaussian mechanism for differential
  privacy.
\newblock In \emph{33rd USENIX Security Symposium (USENIX Security 24)}, pp.\
  937--954, 2024.

\bibitem[Koga et~al.(2024)Koga, Wu, Zhang, and Chaudhuri]{koga2024privacy}
Tatsuki Koga, Ruihan Wu, Zhiyuan Zhang, and Kamalika Chaudhuri.
\newblock Privacy-preserving retrieval-augmented generation with differential
  privacy.
\newblock \emph{arXiv preprint arXiv:2412.04697}, 2024.

\bibitem[Krishna et~al.(2025)Krishna, Krishna, Mohananey, Schwarcz, Stambler,
  Upadhyay, and Faruqui]{krishna2025fact}
Satyapriya Krishna, Kalpesh Krishna, Anhad Mohananey, Steven Schwarcz, Adam
  Stambler, Shyam Upadhyay, and Manaal Faruqui.
\newblock Fact, fetch, and reason: A unified evaluation of retrieval-augmented
  generation.
\newblock In \emph{Proceedings of the 2025 Conference of the Nations of the
  Americas Chapter of the Association for Computational Linguistics: Human
  Language Technologies (Volume 1: Long Papers)}, pp.\  4745--4759, 2025.
\newblock Dataset: google/frames-benchmark.

\bibitem[Krizhevsky(2009)]{krizhevsky2009cifar}
Alex Krizhevsky.
\newblock Learning multiple layers of features from tiny images.
\newblock Technical report, University of Toronto, 2009.
\newblock Tech Report.

\bibitem[Li et~al.(2026)Li, Zhang, Long, Keqin, Song, Bai, Yang, Xie, Yang,
  Liu, Zhou, and Lin]{qwen3vlembedding}
Mingxin Li, Yanzhao Zhang, Dingkun Long, Chen Keqin, Sibo Song, Shuai Bai,
  Zhibo Yang, Pengjun Xie, An~Yang, Dayiheng Liu, Jingren Zhou, and Junyang
  Lin.
\newblock Qwen3-vl-embedding and qwen3-vl-reranker: A unified framework for
  state-of-the-art multimodal retrieval and ranking.
\newblock \emph{arXiv preprint arXiv:2601.04720}, 2026.

\bibitem[MA(1955)]{ma1955two}
KRASNOSEL'SKII MA.
\newblock Two comments on the method of successive approximations.
\newblock \emph{Usp. Math. Nauk}, 10:\penalty0 123--127, 1955.

\bibitem[Mann(1953)]{mann1953mean}
W~Robert Mann.
\newblock Mean value methods in iteration.
\newblock \emph{Proceedings of the American Mathematical Society}, 4\penalty0
  (3):\penalty0 506--510, 1953.

\bibitem[Maze et~al.(2018)Maze, Adams, Duncan, Kalka, Miller, Otto, Jain,
  Niggel, Anderson, Cheney, et~al.]{ijbc}
Brianna Maze, Jocelyn Adams, James~A Duncan, Nathan Kalka, Tim Miller, Charles
  Otto, Anil~K Jain, W~Tyler Niggel, Janet Anderson, Jordan Cheney, et~al.
\newblock Iarpa janus benchmark-c: Face dataset and protocol.
\newblock In \emph{2018 international conference on biometrics (ICB)}, pp.\
  158--165. IEEE, 2018.

\bibitem[Moreau(1962)]{moreau1962decomposition}
Jean~Jacques Moreau.
\newblock D{\'e}composition orthogonale d'un espace hilbertien selon deux
  c{\^o}nes mutuellement polaires.
\newblock \emph{Comptes rendus hebdomadaires des s{\'e}ances de l'Acad{\'e}mie
  des sciences}, 255:\penalty0 238--240, 1962.

\bibitem[Mori et~al.(2025)Mori, Kakizaki, Miyagawa, and
  Sakuma]{mori2025differentially}
Junki Mori, Kazuya Kakizaki, Taiki Miyagawa, and Jun Sakuma.
\newblock Differentially private synthetic text generation for
  retrieval-augmented generation (rag).
\newblock \emph{arXiv preprint arXiv:2510.06719}, 2025.

\bibitem[Muschelli~III(2020)]{muschelli2020roc}
John Muschelli~III.
\newblock Roc and auc with a binary predictor: a potentially misleading metric.
\newblock \emph{Journal of classification}, 37\penalty0 (3):\penalty0 696--708,
  2020.

\bibitem[Opial(1967)]{opial1967weak}
Zdzis{\l}aw Opial.
\newblock Weak convergence of the sequence of successive approximations for
  nonexpansive mappings.
\newblock \emph{Bulletin of the American Mathematical Society}, 73\penalty0
  (4):\penalty0 591--597, 1967.

\bibitem[Oquab et~al.(2024)Oquab, Darcet, Moutakanni, Vo, Szafraniec, Khalidov,
  Fernandez, HAZIZA, Massa, El-Nouby, Assran, Ballas, Galuba, Howes, Huang, Li,
  Misra, Rabbat, Sharma, Synnaeve, Xu, Jegou, Mairal, Labatut, Joulin, and
  Bojanowski]{oquab2023dinov2}
Maxime Oquab, Timoth{\'e}e Darcet, Th{\'e}o Moutakanni, Huy~V. Vo, Marc
  Szafraniec, Vasil Khalidov, Pierre Fernandez, Daniel HAZIZA, Francisco Massa,
  Alaaeldin El-Nouby, Mido Assran, Nicolas Ballas, Wojciech Galuba, Russell
  Howes, Po-Yao Huang, Shang-Wen Li, Ishan Misra, Michael Rabbat, Vasu Sharma,
  Gabriel Synnaeve, Hu~Xu, Herve Jegou, Julien Mairal, Patrick Labatut, Armand
  Joulin, and Piotr Bojanowski.
\newblock {DINO}v2: Learning robust visual features without supervision.
\newblock \emph{Transactions on Machine Learning Research}, 2024.
\newblock Featured Certification.

\bibitem[Parkhi et~al.(2012)Parkhi, Vedaldi, Zisserman, and
  Jawahar]{parkhi2012pets}
Omkar~M Parkhi, Andrea Vedaldi, Andrew Zisserman, and CV~Jawahar.
\newblock Cats and dogs.
\newblock In \emph{2012 IEEE conference on computer vision and pattern
  recognition}, pp.\  3498--3505. IEEE, 2012.

\bibitem[Rudin(1976)]{rudin1976principles}
Walter Rudin.
\newblock Principles of mathematical analysis.
\newblock \emph{3rd ed.}, 1976.

\bibitem[Talagrand(1992)]{talagrand1992sudakov}
Michel Talagrand.
\newblock Sudakov-type minoration for gaussian chaos processes.
\newblock \emph{Israel Journal of Mathematics}, 79\penalty0 (2):\penalty0
  207--224, 1992.

\bibitem[Talwar et~al.(2015)Talwar, Guha~Thakurta, and Zhang]{talwar2015nearly}
Kunal Talwar, Abhradeep Guha~Thakurta, and Li~Zhang.
\newblock Nearly optimal private lasso.
\newblock \emph{Advances in Neural Information Processing Systems}, 28, 2015.

\bibitem[Vera et~al.(2025)Vera, Dua, Zhang, Salz, Mullins, Panyam, Smoot, Naim,
  Zou, Chen, et~al.]{embeddinggemma2025}
Henrique~Schechter Vera, Sahil Dua, Biao Zhang, Daniel Salz, Ryan Mullins,
  Sindhu~Raghuram Panyam, Sara Smoot, Iftekhar Naim, Joe Zou, Feiyang Chen,
  et~al.
\newblock Embeddinggemma: Powerful and lightweight text representations.
\newblock \emph{arXiv preprint arXiv:2509.20354}, 2025.
\newblock Model: google/embeddinggemma-300M.

\bibitem[Von~Neumann(1949)]{von1949rings}
John Von~Neumann.
\newblock On rings of operators. reduction theory.
\newblock \emph{Annals of Mathematics}, 50\penalty0 (2):\penalty0 401--485,
  1949.

\bibitem[Wu et~al.(2025{\natexlab{a}})Wu, Wang, and Wang]{wu2025beyond}
Ruihan Wu, Erchi Wang, and Yu-Xiang Wang.
\newblock Beyond per-question privacy: Multi-query differential privacy for rag
  systems.
\newblock In \emph{NeurIPS 2025 Workshop: Reliable ML from Unreliable Data},
  2025{\natexlab{a}}.

\bibitem[Wu et~al.(2025{\natexlab{b}})Wu, Wang, Zhang, and Wang]{wu2025private}
Ruihan Wu, Erchi Wang, Zhiyuan Zhang, and Yu-Xiang Wang.
\newblock Private-rag: Answering multiple queries with llms while keeping your
  data private.
\newblock \emph{arXiv preprint arXiv:2511.07637}, 2025{\natexlab{b}}.

\bibitem[Yang et~al.(2019)Yang, Lu, Lyu, and Hu]{yang2019two}
Hanfang Yang, Kun Lu, Xiang Lyu, and Feifang Hu.
\newblock Two-way partial auc and its properties.
\newblock \emph{Statistical methods in medical research}, 28\penalty0
  (1):\penalty0 184--195, 2019.

\bibitem[Yang et~al.(2017)Yang, Zhu, Ma, Xiang, and Zhou]{yang2017privacy}
Mengmeng Yang, Tianqing Zhu, Lichuan Ma, Yang Xiang, and Wanlei Zhou.
\newblock Privacy preserving collaborative filtering via the
  johnson-lindenstrauss transform.
\newblock In \emph{2017 IEEE Trustcom/BigDataSE/ICESS}, pp.\  417--424. IEEE,
  2017.

\bibitem[Yang et~al.(2021)Yang, Xu, Bao, He, Cao, and Huang]{yang2021all}
Zhiyong Yang, Qianqian Xu, Shilong Bao, Yuan He, Xiaochun Cao, and Qingming
  Huang.
\newblock When all we need is a piece of the pie: A generic framework for
  optimizing two-way partial auc.
\newblock In \emph{International Conference on Machine Learning}, pp.\
  11820--11829. PMLR, 2021.

\end{thebibliography}
\end{document}